\documentclass [11pt, twoside] {uwthesis}[2012/06/19]
 
\usepackage{url}
\usepackage{hyperref}
\usepackage{tabularx}
\usepackage{diagbox}
\usepackage{quantikz}
\usepackage{boldline}
\usepackage{physics}
\usepackage{makecell}
\usepackage{amsmath, amsthm, amssymb,slashed,url}
\usepackage{alltt}  %

\usepackage{bm}
\usepackage{graphicx}
\usepackage{tikz}
\usepackage{multirow} 
\usepackage{appendix} 

\newcommand{\inv}[1]{\frac{1}{#1}}
\newcommand{\mpi}{M_\pi}
\newcommand{\nn}{\nonumber \\}
\def\si{^1 \hskip -0.03in S _0}
\def\siii{^3 \hskip -0.025in S _1}

\def\sss{\scriptscriptstyle}

\def\onefourth{\textstyle{\frac{1}{4}} }

\def\oneht{\textstyle{\frac{1}{2} }}

\newcommand{\LOss}{\scriptscriptstyle{LO}}
\newcommand{\NLOss}{\scriptscriptstyle{NLO}}
\def\mapup#1{{\smash{\mathop{\llra}\limits^{#1}}}}
\def\llra{{\relbar\joinrel\longrightarrow}}
\newcolumntype{Y}{>{\centering\arraybackslash}X}
\newcommand{\nub}{\overline{\nu}}
\newcommand{\eb}{\overline{e}}
\newcommand{\ub}{\overline{u}}
\newcommand{\db}{\overline{d}}
\def\pslash{{ p\hskip-0.5em /}}
\newcommand{\infiL}{{\mathcal{I}_L}}
\def\Dslash{{ D\hskip-0.6em /}}
\newcommand\scalemath[2]{\scalebox{#1}{\mbox{\ensuremath{\displaystyle #2}}}}
\begin{document}
 
%
%

\prelimpages
 
%
%
\Title{High Energy Physics from Low Energy Physics}
\Author{Roland Carlos Farrell}
\Year{2024}
\Program{Physics}

\Chair{Silas Beane}{Professor}{Physics}
\Signature{Martin J. Savage}
\Signature{David B. Kaplan}

\copyrightpage

\titlepage

{\Degreetext{A dissertation%
\\
  submitted in partial fulfillment of the\\ requirements for the degree of}
 \def\thefootnote{\fnsymbol{footnote}}
 \let\footnoterule\relax
 \titlepage
 }
\setcounter{footnote}{0}

%
%

\abstract{
The separation between physics at low and high energies is essential for physics to have any utility;
the details of quantum gravity are not necessary to calculate the trajectory of a cannon ball.
However, physics at low and high energies are not completely independent, and this thesis explores two ways that they are related. 

The first is through a UV/IR symmetry that relates scattering processes at low and high energies.
This UV/IR symmetry manifests in geometrical properties of the $S$-matrix, and of the RG flow of the coupling constants in the corresponding effective field theory.
Low energy nuclear physics nearly realizes this UV/IR symmetry, providing an explanation for the smallness of shape parameters in the effective range expansion of nucleon-nucleon scattering, and inspiring a new way to organize the interactions between neutrons and protons.

The second is through the use of quantum computers to simulate lattice gauge theories.
Quantum simulations rely on the universality of the rules of quantum mechanics, which can be applied equally well to describe a (low energy) transmon qubit at 15 milli-Kelvin as a (high energy) 1 TeV quark.
This thesis presents the first simulations of one dimensional lattice quantum chromodynamics on a quantum computer, culminating in a real-time simulation of beta-decay.
Results from the first simulations of a lattice gauge theory on 100+ qubits of a quantum computer are also presented. 
The methods developed in this thesis for quantum simulation are ``physics-aware", and are guided by the symmetries and hierarchies in length scales of the systems being studied.
Without these physics-aware methods, 100+ qubit simulations of lattice gauge theories would not have been possible on the noisy quantum computers that are presently available.
}
 
%
%
\tableofcontents
\listoffigures
\listoftables

%
%
\acknowledgments{
  \noindent
  It takes a village to raise a child, and two cities to raise a PhD student.
  I have had the good fortune of doing both my undergraduate and PhD studies only 60 miles from where I grew up.
  I would like to thank my parents for providing a home that that always offered a cozy escape from the big city, and a place to heal after an elbow surgery at the beginning of my PhD and a knee surgery at the end.
  I also would like to thank them for instilling in me the unwavering belief that there is no limit to what I am capable of.
  In addition, I would like to thank my other friends and family from Mount Vernon who have patiently listened to my attempts at explaining quantum field theory, entanglement and quantum computing over the years: Olivia Farrell, Chris Perry, Peter Whidden, Casey Goodwin, Abe Nurkiewicz, Joe Ordoñez, Edie Granger, Eugene Kang, Stella Ordoñez, Sapphire Ordoñez and Ellen Gray.

I have made many great friends in Seattle who have helped make my life outside of research rich and vibrant. 
I would like to thank Richard Ellison for running a house with a revolving door of interesting people, teaching me how to cook and always sharing meals with me. 
I am grateful to my housemates Nikita Zemlevskiy, Chris Owen, William Marshall, Michael Dom, Hoang Nguyen, Murali Saravanan and Henry Froland for always making it fun to kill time.
Special thanks to Nikita, Murali and Henry for being people that I can rely on, and the help that they offered me as I was recovering from knee surgery.
  I am grateful to my fellow bulger hunters Dane Pollett, John Ferre and Zack Aemmer for the great adventures in the Washington wilderness, and to my other climbing and ski partners Tatsumi Nitta, Karla Diaz, Roel Ardiente, Ashlynn, Sasha Krassovsky and Michelle Yang for many great explorations of the mountains on the Hwy-2 and I-90 corridors (as well as the climbing gyms on the I-5 corridor).
  Special thanks to Sasha for subsidizing many meals, and always being excited about making code run fast.
  I am also grateful to Matthew Hsieh, Zach Oropesa, Francesco Cueto, Chris Liu, Victor Ho and Valentin Monfort for many great barbecues and hot pot dinners.

I would like to thank the members of my PhD cohort for the shared commiserations over TA duties, finding an advisor and research: John Goldak, Yiyun Dong, Wan Jin Yeo, John Ferre, Ramya Bhaskar, Zachary Draper, Teresa Lo, Chris Thomas, Michael Clancy, Ryan Lanzetta and Arnab Manna.
I would also like to thank the members of IQuS for fostering a great research environment that makes me excited to come into the office every day. 
I have especially benefited from collaboration with Marc Illa who has shown me the art of making beautiful figures, selecting text vertically and using high-performance and quantum computers.
 I would like to thank the professors/mentors I have had throughout my PhD, especially Andreas Karch, David Kaplan, Silas Beane, Martin Savage, Sanjay Reddy, Lukasz Fidkowski and Ann Nelson.
 Their passion and joy for theoretical physics was contagious and continues to fuel my desire for discovery.
 I am grateful to Catherine Provost for helping me get into the UW PhD program, and to both Catherine and Katie Hennessy for generally making all aspects of being a PhD student easier.

I would like to thank the participants of the 2022 DNP Summer School, the 2022 Talent Summer School and the 2023 Quantum Connections Summer School, with special thanks to  Rossie Jiang for her support and companionship.
Meeting so many young physicists from all over the world doing exciting research has really made me internalize that I am part of a thriving global physics community.
 I would like to thank the great friends and colleagues I met while in Bern, especially Matteo Traschel, Martina, Steven Waldvogel and Maike.
 The mantra that ``There are no strangers in my life, only friends" is something that I strive towards.
 
I would like to thank my advisor Silas Beane for guiding me on the leap between being a homework-solving student to a (quasi-)independent researcher. 
I really value that you were always available and excited to discuss research, that all of my ideas were taken seriously, and the emphasis that was put on creative thinking.
I hope to follow your example of always focusing on the problems that are the most interesting, independent of hype, status quo or inertia of learning something new.
I also appreciate the emphasis that was placed on taking advantage of all opportunities to travel and share my research. 
I would also like to thank Martin Savage for essentially being a second advisor during the last two years.
Your tenacity and enthusiasm is infectious and has pushed me to do my best work.
I appreciate that you never shoot down other people's ideas, and particularly value the occasional brutally honest feedback you have given me.
On multiple occasions this feedback has led to a complete rewiring of my thinking, and non-analytic jumps in my growth as a researcher.

}

%
%
\dedication{\begin{center}To my parents\end{center}}

%

%
%

\textpages

\chapter {Introduction}
\noindent
The new tools and fresh perspectives offered by quantum information are disrupting many areas of physics.  
Simulations of quantum many-body systems using quantum computers are close to surpassing the capabilities of classical computers, including exciting applications relevant to condensed matter~\cite{Kim:2023bwr,Shinjo:2024vci,Andersen:2024aob}, random circuit sampling~\cite{DeCross:2024tmi,Arute:2019zxq}, fault tolerance~\cite{Bluvstein:2023zmt} and nuclear physics~\cite{Farrell:2024fit,Farrell:2023fgd}. 
And the rate of progress shows no sign of slowing down, with demonstrations in the last year of large-scale 2D ion traps~\cite{Valentini:2024fly,Guo:2023xza}, Rydberg arrays with thousands of qubits~\cite{Manetsch:2024lwl}, the manipulation of 8 trapped ion ququarts~\cite{Zalivako:2024bjm} and the knitting together of multiple 100+ qubit superconducting Quantum Processing Units (QPUs)~\cite{Vazquez:2024qmo}.
The purely theoretical side of quantum information has also proven to be extremely valuable.
Studies of entanglement have revealed new ways of identifying topological order either through properties of their entanglement Hamiltonians~\cite{Li_2008} or bipartite entanglement~\cite{Kitaev:2005dm}, hint at explanations for ```accidental" symmetries in (hyper) nuclear~\cite{Beane:2018oxh,Low:2021ufv,Bai:2022hfv,Liu:2023bnr,Liu:2022grf,Kirchner:2023dvg} and particle~\cite{Carena:2023vjc,Beane:2021zvo,Aoude_2020} physics and have inspired powerful new numerical methods for simulating quantum many body systems using classical computers~\cite{PhysRevLett.69.2863,Verstraete_2008}.
These advancements have the potential to revolutionize many fields of physics; from quantum chemistry and materials~\cite{Bauer:2020epk} to hot and dense quantum chromodynamics (QCD)~\cite{Banuls:2019bmf,Bauer:2022hpo,Bauer:2023qgm}. 

A primary motivation for the work presented in this thesis is the unexplained puzzles and open questions present in the Standard Model of particle physics~\cite{Glashow:1961tr,Higgs:1964pj,Weinberg:1967tq,Salam:1968rm,Politzer:1973fx,Gross:1973id}, and its low energy manifestation in nuclear physics.
The Standard Model is a mathematical framework that describes the fundamental particles and their interactions.
The mathematical underpinning is a quantum field theory that has a tremendous amount of predictive power and has been validated by experiment to staggering precision~\cite{Fan:2022eto,ParticleDataGroup:2022pth,Tiesinga:2021myr}.
Despite having the equations that in principle can be used to predict the properties of matter in almost every setting, there are still many open questions.
These include the origin of ``accidental" symmetries or unexplained hierarchies that are present in the masses and interactions of the Standard Model e.g. why is the Higgs boson so light? Why is charge-conjugation and parity (CP) a good symmetry of the strong interaction? Why is the binding energy of the deuteron so small?
There are also many states of matter whose simulation is beyond the capabilities of the most powerful (classical) supercomputers that could ever be built~\cite{Troyer_2005}.
Lines of inquiry in this direction include:
What are the phases of matter beyond nuclear saturation density~\cite{Schafer:2005ff}?
How do nuclei fragment into partons during high energy collisions; and how do the fragments eventually re-hadronize~\cite{Andersson:1983ia,Fries:2003vb,Fries:2003kq}?
What are the mechanisms that drive strongly interacting matter to reach thermal equilibrium~\cite{Srednicki:1994mfb}?
How does the QCD vacuum respond in real-time to external probes?
Addressing these questions is crucial for understanding the mass distribution of neutron stars in our universe~\cite{Burgio:2021vgk}, for sharpening our inferences based on measurements in heavy ion collisions and how the conditions shortly after the big bang have led to the universe we see today.

Many of the above questions have been thought about for a long time, but still remain unresolved, and for good reason --they are very difficult problems!
However, the new tools and way of thinking coming from quantum information are providing new angles to approach these old problems.
At a high level, quantum information is about identifying and utilizing the correlations between quantum states that have no classical analog.
These correlations emerge from the foundational quantum mechanical features of superposition, particle indistinguishability, uncertainty and measurement.
For macroscopic objects, these quantum correlations are washed away, but at atomic and subatomic level they are absolutely essential.
Characteristics of these quantum correlations can often be used to predict physical properties of a quantum state, for example the connection between the entanglement spectrum and topological order in the quantum hall state~\cite{Li_2008}.
Additionally, the advantages of quantum computers over classical ones rests on their ability to efficiently manipulate and process these inherently quantum correlations.
This enables quantum computers to more efficiently solve ``quantum" problems such as the simulation of dynamics in strongly correlated quantum many-body systems.
Surprisingly, this also enables quantum computers to efficiently solve classes of problems with no obvious quantum structure, the most prominent being Shor's algorithm for factoring prime numbers~\cite{Shor:1994jg}.\footnote{Another surprising usage of quantum information is for provably secure methods of cryptography~\cite{Bennett:2014rmv}.}

There are many features of the Standard Model and its low energy effective field theories (EFTs) that, although mathematically consistent, beg for a deeper explanation.
As discussed above, this includes the empirical observations of approximate symmetries and unexplained hierarchies.
These peculiarities could truly be accidental, but it is also possible that they emerge from mechanisms that are not currently understood.
One possible explanation is that interactions are organized in terms of how much entanglement they generate. 
This would provide a new way to organize the interactions present in chiral and nuclear EFTs.
This conjecture was explored in the seminal work of Beane et. al~\cite{Beane:2018oxh} which looked at the spin entanglement generated by the scattering of (hyper)nucleons.
They obtained the striking result that the interaction chosen by nature leads to very little entanglement generated in scattering near threshold.
That is, out of all the possible values for the coupling constants that parameterize the low energy interaction, nature favors those which generate little spin entanglement.
Not only is entanglement suppressed, but the minimal entanglement solution also gives rise to an enhanced $SU(4)$ symmetry for nucleon-nucleon scattering, and a $SU(16)$ symmetry for hypernuclear scattering.
The approximate $SU(4)$ symmetry in the nucleon-nucleon interaction was first pointed out by Wigner~\cite{Wigner:1936dx,Wigner:1937zz,Wigner:1939zz}, and there is evidence of the $SU(16)$ symmetry in the hypernucleon-hypernucleon interaction from lattice QCD~\cite{NPLQCD:2020lxg,Wagman:2017tmp}.
This preliminary investigation motivated the conjecture that the confinement-deconfinement transition in QCD leads to emergent entanglement suppression in hadronic physics.

Chapter~\ref{chap:hadronEP} of this thesis explore this conjecture in the context of scattering between pions and nucleons. 
Entanglement suppression has also been explored in systems of light nuclei and nucleons~\cite{Bai:2022hfv,Kirchner:2023dvg}, hypernucleons~\cite{Liu:2023bnr,Liu:2022grf,Beane:2018oxh}, Higgs bosons~\cite{Carena:2023vjc} and black holes~\cite{Aoude_2020}.
Pions are pseudoscalar particles, with no spin to entangle, and instead entanglement in isospin space is considered.
Pions have isospin $I = 1$, while nucleons are $I=1/2$, and their internal states map onto qutrit and qubit Hilbert spaces respectively.
Using the highly accurate determination of the $\pi\pi$ and $\pi N$ scattering phase shifts, the isospin entanglement generated from scattering is determined across a wide range of center of mass (c.o.m) energies.
One interesting feature is a local minimum of the entanglement near c.o.m energies that excite the $\Delta$ resonance.
This is due to the rapid variation in the phase shift causing a corresponding rapid variation in the entanglement of $S$-matrix.
Additionally, the entanglement is determined analytically from chiral perturbation theory providing a set of low-energy theorems for isospin entanglement near threshold.
Unlike in the (hyper)nucleon-nucleon system, the only minimal entanglement solution consistent with the symmetries of the interactions is the trivial one --no scattering, with no enhanced symmetry.
However, this no scattering condition is almost satisfied for low-energy interactions involving pions. 
This is because pions are the (pseudo) Goldstone bosons of chiral symmetry breaking, and are therefore derivatively coupled, with an interaction that vanishes at low energies~\cite{PhysRev.124.246,PhysRev.122.345,PhysRevLett.4.380,PhysRev.117.648}. 
Indeed, this expectation is further reinforced by large $N_c$ that predicts non-interacting mesons~\cite{tHooft:1973alw,tHooft:1974pnl}.

The entanglement produced in scattering is further explored in  chapter~\ref{chap:SmatGeom} and motivates the development of a new geometric formulation of scattering.
A scattering process begins with an initial state of two  particles well separated in space.
The particles propagate towards each other, interact within some spatial volume, and then propagate away from each other and become well-separated again.
In the initial ``in" state, there are no correlations between the two particles and consequently no entanglement.
However, the final ``out" state, can exhibit entanglement which was generated from the interaction.
Once particles are entangled, they will have quantum correlations that can be detected no matter how far away they are by e.g. Bell measurements~\cite{RevModPhys.38.447}.
The observable consequences of scattering are encoded in the $S$-matrix that evolves the ``in" state to the ``out" state, $\hat{S}\vert\text{in}\rangle  = \vert\text{out}\rangle$.
Thus, the $S$-matrix encodes the capacity for the interaction to entangle the two particles.

The $S$-matrix is typically determined by solving an EFT describing particles interacting locally.
In this EFT-based approach, spacetime constraints like Galilean invariance and causality are encoded in the dependence of the scattering on the external kinematics e.g. the c.o.m. energy $E$.
The framework provided by EFTs is extremely powerful, and has enabled precision calculations of observables, with quantifiable uncertainties~\cite{Weinberg:1978kz,Beane:2000fx,Hammer:2019poc}.
However, keeping spacetime constraints manifest may obscure features of scattering that are non-local, such as entanglement.
Motivated by this, a new geometric formulation of scattering is developed and explored in chapter~\ref{chap:SmatGeom}.

\begin{figure}[!ht]
\centering
\includegraphics[width = 0.6\textwidth]{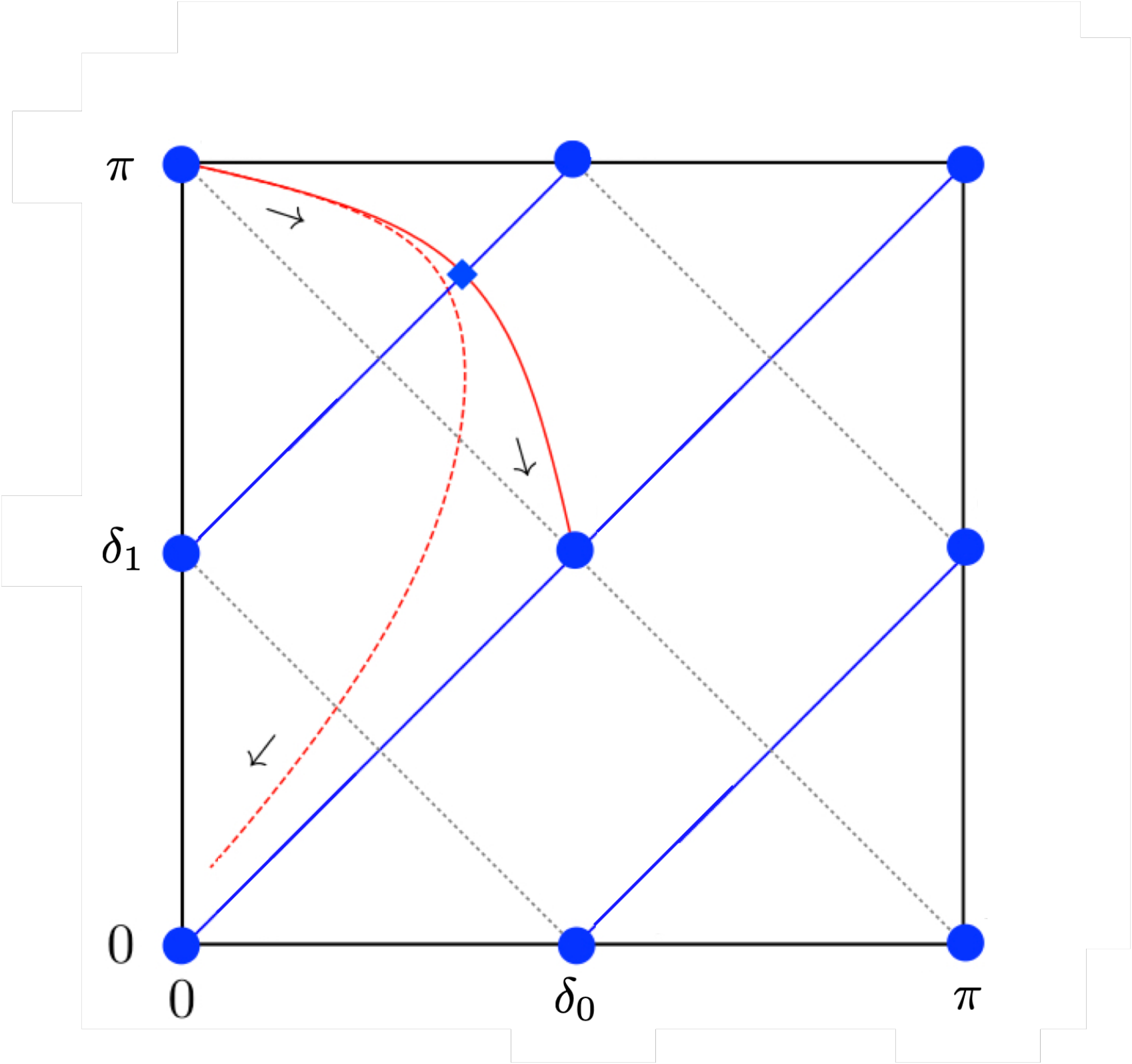}
\caption{The flat torus corresponding to all unitary $S$-matrices for two channel scattering.
The two phase shifts $\delta_0$ and $\delta_1$ can be parameterized by the c.o.m. energy i.e. $\delta(E)$, and give rise to trajectories that correspond to the $S$-matrix as a function of energy. 
The solid red curve is the example for phase shifts parameterized by the nucleon-nucleon scattering lengths, and the dashed red curve corresponds to the s-wave nucleon-nucleon scattering~\cite{NNOnline}, with $\delta_0$ ($\delta_1$) corresponding to scattering in the spin singlet (triplet) channel.
The arrows represent increasing $E$.
This figure is adapted from Ref.~\cite{Beane:2020wjl}.}
    \label{fig:flatTorus}
\end{figure}

Quantum mechanics is unitary, and the $S$-matrix can be parameterized by energy-dependent phase shifts $\delta(E)$ that characterize the strength of the interaction in the various scattering channels i.e. $\hat{S} = e^{2 i \delta(E)}$.
In non-relativistic scattering from a finite range potential it can be shown that the s-wave phase shift can be parameterized by the effective range expansion (ERE) as~\cite{PhysRev.76.38},
\begin{equation}
k\cot{\delta} \ = \ -\frac{1}{a} \ + \ \frac{r}{2}k^2  \ + \ {\mathcal O}(k^4) \ ,
\label{eq:ERE}
\end{equation}
where $a$ is the scattering length, $r$ is the effective range and $k = \sqrt{2ME}$ is the magnitude of the incoming momentum in the c.o.m. frame.
The coefficients parameterizing the ${\mathcal O}(k^4)$ and higher order terms are known as shape parameters.
In the geometric formulation these phase shifts form a basis for a space that encompasses all possible unitary $S$-matrices.
Due to the $\pi$ periodicity of the phase shift, this space has the topology of a flat torus.
For a given $S$ matrix, the phase shifts as a function of energy form trajectories on the flat torus.
The flat torus and example $S$-matrix trajectories are shown in Fig.~\ref{fig:flatTorus} for two-channel scattering.

The way that spacetime constraints such as causality and Galilean invariance manifest on the flat torus is explored throughout chapter~\ref{chap:SmatGeom}. 
It is shown that causality constrains the allowed tangent vectors to the $S$-matrix trajectories and Galilean invariance corresponds to the freedom to choose an (inaffine) parameterization of the trajectories.
This geometric way of viewing scattering also reveals a new UV/IR symmetry that relates scattering at low and high energies.
These symmetries are only present for $S$-matrices that have phase shifts parameterized by scattering lengths, or with effective ranges that are correlated with the scattering lengths.
Any higher order shape parameters in the effective range expansion necessarily break the UV/IR symmetry.
An example of a UV/IR symmetric $S$-matrix, corresponding to a reflection symmetric $S$-matrix trajectory, is shown as the solid red trajectory in Fig.~\ref{fig:flatTorus}.
This work demonstrated that the new outlook obtained from studying entanglement can be valuable in unexpected ways.
In this case, entanglement motivated the development of the geometric formulation of scattering, which in turn revealed a new symmetry that on the surface is completely unrelated to entanglement.

The implications of this UV/IR symmetry are further explored throughout chapters~\ref{chap:SmatUVIR} and~\ref{chap:Weinberg}.
In chapter~\ref{chap:SmatUVIR}, it is shown how this UV/IR symmetry manifests in the renormalization group (RG) running of couplings in the corresponding EFT of contact operators.
The focus is on the EFT that reproduces phase shifts parameterized only by scattering lengths.
This EFT contains only momentum-independent interactions --delta functions in position space, that are singular at high momentum, and consequently needs to be regularized and renormalized.
After renormalization, the coupling constants depends on the cutoff or RG scale that has been introduced to define the theory.
The coupling constants as a function of the RG scale trace out trajectories in ``coupling constant space".
These coupling constant trajectories also possess reflection symmetries that are now generated by a UV/IR transformation that interchanges low and high energy RG scales.
This is the fingerprint in the EFT of the reflection symmetries that the UV/IR symmetric $S$-matrix trajectories possess on the flat torus.\footnote{
Note that the UV divergence in this EFT is a linear divergence, which are conventionally thought to not contain ``physics", and are completely ignored when regulating with dimensional regularization and $\overline{\text{MS}}$.}
Further, it is shown that the assumption of a UV/IR symmetry constrains the RG running of the coupling constants, allowing the functional form to be determined without having to compute any loop integrals. 
The UV/IR symmetry also implies consistency relations for the RG scale dependence of the coupling constants of interactions that break the UV/IR symmetry.
This is used to determine the RG running of the coupling constant that generates effective range effects.

In chapter~\ref{chap:Weinberg}, the implications of the UV/IR symmetry for nuclear physics are considered.
At low energies, the interaction between nucleons can be being efficiently described by pionless EFT~\cite{Weinberg:1990rz}.
This EFT contains nucleons as the degree of freedoms, with all mesons and higher energy baryons effectively integrated out.
Scattering at the lowest energies is dominated by the $s$-wave, and the phase shifts in the spin singlet and triplet channels can be parameterized by the ERE of Eq.~\ref{eq:ERE}.
The ``natural" scale for this EFT is set by the lowest energy excitation that is not explicitly included in the theory.
For the case of pionless EFT, this is set by the pion mass $M_{\pi} \approx 1.5 \ \text{fm}^{-1}$.
In nuclear physics, the s-wave ERE parameters are~\cite{Kaplan:1998we,deSwart:1995ui},
\begin{align}
&a_{0} \ = \ -23.7 \ \text{fm} \ \ , \ \ a_{1} \ = \ 5.4 \ \text{fm} \nonumber \\
&r_{0} \ = \  2.7 \ \text{fm} \ \ , \ \ r_{1} \ = \ 1.7 \ \text{fm},
\end{align}
with the shape parameters very small, consistent with zero.
All of these parameters are larger than the naive breakdown scale of $M_{\pi}^{-1}$ and, in particular, the size of $a_0$ leads to the deuteron being very weakly bound.
Approximating the phase shift with only a scattering length and effective range reproduces the empirical measured nucleon-nucleon phase  up to $k\approx 160 \ \text{MeV}$.

As mentioned above, and shown in chapter~\ref{chap:SmatGeom}, an $S$-matrix with phase shifts parameterized by only scattering lengths and effective ranges is UV/IR symmetric, provided that the effective ranges are correlated with the scattering lengths.
If one assumes a UV/IR symmetry in the nucleon-nucleon interaction, emerging from some mechanism in QCD not currently understood, then this would explain the smallness of the shape parameters as UV/IR symmetry forbids shape corrections.
This UV/IR symmetric interaction forms the basis for a new EFT expansion, where LO treats scattering length and effective range to all orders, in such a way to preserve the UV/IR symmetry.
It is shown how this UV/IR symmetry implies a set of algebraic constraints on the two-body potential generated by the EFT.
These constraints are solved by a non-local potential, of a similar form to that proposed by Yamaguchi in 1954~\cite{Yamaguchi:1954mp}.
Higher order terms in this EFT break the UV/IR symmetry, either by shifting the effective range from being correlated with the scattering length or by introducing shape parameters.
This UV/IR symmetry provides motivation for a new way of organizing the nuclear interactions, and may lead to better convergence in many-body calculations.

The second half of this thesis explores the use of quantum computers to simulate lattice gauge theories.
Many problems relevant to nuclear and particle physics can only be addressed by solving QCD.
QCD is a quantum field theory that describes the interactions between quarks and gluons.
In a quantum field theory, the effective strength of an interaction is heavily influenced by quantum fluctuations.
These quantum fluctuations change with the energy scale of the interaction.
In asymptotically free theories like QCD~\cite{Gross:1973id,Politzer:1973fx}, the interaction strength is weak at high energies, and observables can be determined in perturbation theory.
However, as the relevant energies approaches the scale of confinement, around 1 GeV, the interaction strength becomes strong and perturbation theory breaks down.
This is the scale where the quarks and gluons become bound inside hadrons (neutrons, protons, pions etc.).
Observables at this scale can still be computed from the QCD path integral but require non-perturbative methods.
The only known framework for non-perturbatively defining QCD is through a powerful numerical method called lattice QCD.
In lattice QCD, spacetime is discretized on a lattice and different field configurations in the path integral are importance sampled.
This importance sampling relies on there being a well-defined probability distribution from which the different field configurations can be drawn.
This is satisfied for systems at zero density and in Euclidean space (imaginary time).
There has been tremendous success with this approach, and lattice QCD has been used to postdict, and in some cases predict meson decay rates and scattering parameters, hadronic masses, QCD phases at low density and high temperature and the anomalous magnetic moment of the muon~\cite{FlavourLatticeAveragingGroupFLAG:2021npn,Davoudi:2022bnl,USQCD:2022mmc,Davoudi:2020ngi,BMW:2008jgk,BMW:2014pzb}.
However, this method breaks down for observables involving real-time response or in systems at finite baryon density~\cite{Bauer:2022hpo,Catterall:2022wjq,Humble:2022klb}.
In these cases, the ``weight" of the different field configurations becomes complex, and the condition of an underlying probability distribution breaks down.
This is the sign-problem in lattice QCD, and it is believed to be NP-hard~\cite{Troyer_2005}.

Fortunately, over 40 years ago it was realized by Feynman and others that there is another route toward simulating quantum systems~\cite{Benioff1980,Feynman1982,Feynman1986,Lloyd1073}.
This is quantum simulation, where the target quantum theory is mapped onto another quantum system that can be well controlled in the laboratory.
Unlike for lattice QCD simulations using classical computers, it is believed that quantum simulation of real dynamics or systems at finite density are free of the sign problem~\cite{Bauer:2022hpo,Catterall:2022wjq,Humble:2022klb}.
The zeroth order step of quantum simulation is mapping the Hilbert space of the target theory to one that is natively available on the quantum simulator, usually in the form of a register of qubits or higher dimensional qudits.
A typical quantum simulation of a dynamical process proceeds by first preparing an initial, physically interesting state, evolving it with the time evolution operator $U(t) = e^{-i \hat{H} t}$ and then measuring observables in the final state.
Broadly speaking there are two types of quantum simulators; analog and digital.
An analog platform is capable of producing the unitary evolution $e^{-i \hat{H}'(\theta_i) t}$ for some class of Hamiltonians $\hat{H}'(\theta_i)$ with control over a set of parameters $\theta_i$.
If there exists a choice of $\theta_i$ that can reproduce the target Hamiltonian, then initial state preparation can often be done by adiabatically, and the time evolution operator can be reproduced natively\footnote{
In adiabatic state preparation the parameters are evolved as a function of an adiabatic parameter $\theta_i(\tau)$ with $\tau \in [0,1)$ such that the ground state of $\hat{H}'(\theta_i(0))$ is easy to prepare, and $\hat{H}'(\theta_i(1))$ is the target Hamiltonian.
This is also an efficient protocol for state preparation in digital quantum computing provided that the mass gap does not vanish during the adiabatic evolution~\cite{farhi2000quantum,van_Dam_2001}.}.
Analog quantum simulations often feature very high fidelities~\cite{Ebadi2021,Semeghini:2021wls}, but have limited range of application as the unitaries can only cover what is possible with the $\hat{H}'(\theta_i)$.

A digital quantum computer has the advantage that in principle it can implement the evolution under any unitary.
Digital quantum computers come equipped with a set of elementary unitary operations, a universal ``gate set", from which any unitary operation can be constructed.
This universal gate sets often consist of an arbitrary single qubit rotation, and an entangling two-qubit operation like a controlled-not (CNOT).
On current, Noisy Intermediate Scale Quantum (NISQ) era~\cite{Preskill:2018jim}, digital quantum computers it is the two-qubit gates that are the primary source of noise and errors.
In a quantum simulation, gates are arranged in such a way to form quantum circuits that prepare the desired initial state and implement the time evolution operator.

Quantum simulation of lattice gauge theories is still in its infancy, with the state-of-the-art focusing on toy models of QCD, often in fewer than three dimensions, with simpler gauge groups and/or without dynamical fermions~\cite{Klco:2018kyo,Mazzola:2021hma,deJong:2021wsd,Gong:2021bcp,Mildenberger:2022jqr,Charles:2023zbl,Pomarico:2023png,Lu:2018pjk,Mil:2019pbt,Yang:2020yer,Zhou:2021kdl,PhysRevResearch.5.023010,zhang2023observation,Farrell:2024fit,Farrell:2022vyh,Farrell:2023fgd,Farrell:2022wyt,Ciavarella:2024fzw}.
Chapter~\ref{chap:axial} establishes the foundation for quantum simulations of $1+1$D QCD, explores aspects of the theory with exact diagonalization and presents results from the first simulation of QCD in $1+1$D on a quantum computer.
This includes working out the zeroth order step of quantum simulation; mapping the Hilbert space of $1+1$D QCD to the Hilbert space of qubits. 
An in depth analysis of the spectrum of $1+1$D QCD reveals several interesting features.
With $N_f=2$ flavors of quarks there exists a bound state of two baryons, analogous to the deuteron in QCD.
It is found that this binding is due to the vacuum energy density per flavor of $N_f=2$ QCD being lower than in $N_f=1$ QCD. 
This result is intriguing as it indicates that the presence of a bound state can be deduced from the $N_f$-dependence of the vacuum energy.
Additionally, it is found that the structure of entanglement in the vacuum can be used to identify a phase transition.
As a function of the coupling constant, $g$, the vacuum transitions from being primarily composed of ``mesonic" excitations consisting of quark-antiquark pairs, to being primarily composed of ``baryonic" excitations consisting of baryon-antibaryon pairs.
This is due to a competition between the mass energy, which counts the occupation of quarks and antiquarks, and the energy in the chromoelectric field.
Mesonic excitations only contribute two occupation to the mass energy, but have a string of ${\bf 3}$ or $\overline{{\bf 3}}$ color flux that contributes chromoelectric energy.
Baryonic excitations on the other hand contribute six occupation to the mass energy, but are locally color singlets and do not excite the chromoelectric field.
Therefore, increasing $g$ with the quark mass fixed causes the low energy excitations to transition from being mesonic to baryonic.
At the critical coupling where this transition occurs, the bipartite entanglement between quarks and antiquarks spikes.
This is because there are contributions of both mesonic and baryonic states to the vacuum wavefunction, and consequently more pure states contributing to the reduced density matrix.
It is possible that such a rearrangement could occur in the QCD vacuum as the strong coupling constant increases near the confinement-deconfinement transition.

Chapter~\ref{chap:axial} also presents results from the first quantum simulations of QCD in $1+1$D.
To accomplish this, quantum circuits are developed for preparing the vacuum and implementing time evolution.
The time evolution circuits are executed on IBM's superconducting 7-qubit quantum computers {\tt ibm\_perth} and {\tt ibm\_jakarta} for $N_f=1$ QCD (corresponding to 6 qubits per spatial site).
The bare vacuum-to-vacuum amplitudes are measured, and found to agree with expectations, with statistical uncertainties below 1\%.
This work was featured in a podcast available on \href{https://www.youtube.com/watch?v=PS_8oRaqQRc}{YouTube}.

Chapter~\ref{chap:beta} extends these simulations to a single generation of Standard model fermions; $N_f=2$ QCD with $u$ and $d$ quarks as well as $e^-$ and $ \nu_e$ leptons.
This system requires 16 qubits per spatial lattice site, $6$ each for the $u$ and $d$ quarks, and $2$ each for the $e^-$ and $\nu_e$ leptons.
By coupling the quarks to the leptons with an effective 4-Fermi operator, weak decays in real-time are simulated.\footnote{Note that, as a chiral gauge theory, an {\it ab initio} treatment of the weak interaction on the lattice is currently not available.}
This simulation requires the preparation of an initial baryon state, in our case the $\Delta^-$ baryon as well as the lepton vacuum.
Time evolution leads to a non-zero decay rate corresponding to the process $\Delta^- \to \Delta^0 + e^- + \overline{\nu}_e$, that is detected by measuring the electric charge in the lepton sector.
The circuits that prepare the $\Delta^-$ baryon and evolve it forward in time are executed using Quantinuum's {\tt H1-1} 20-qubit trapped ion quantum computer~\cite{h1-1}.
The electric charge is measured with statistical uncertainties at the 5\% percent level, and the findings are consistent with classically computed expectations.
This work serves as a proof of concept, demonstrating the ability of quantum computers to simulate weak decays in real-time.
Work is currently underway to extend these simulations to neutrinoless double beta decay on multiple lattice sites.

\begin{figure}[!ht]
\centering
\includegraphics[width = \textwidth]{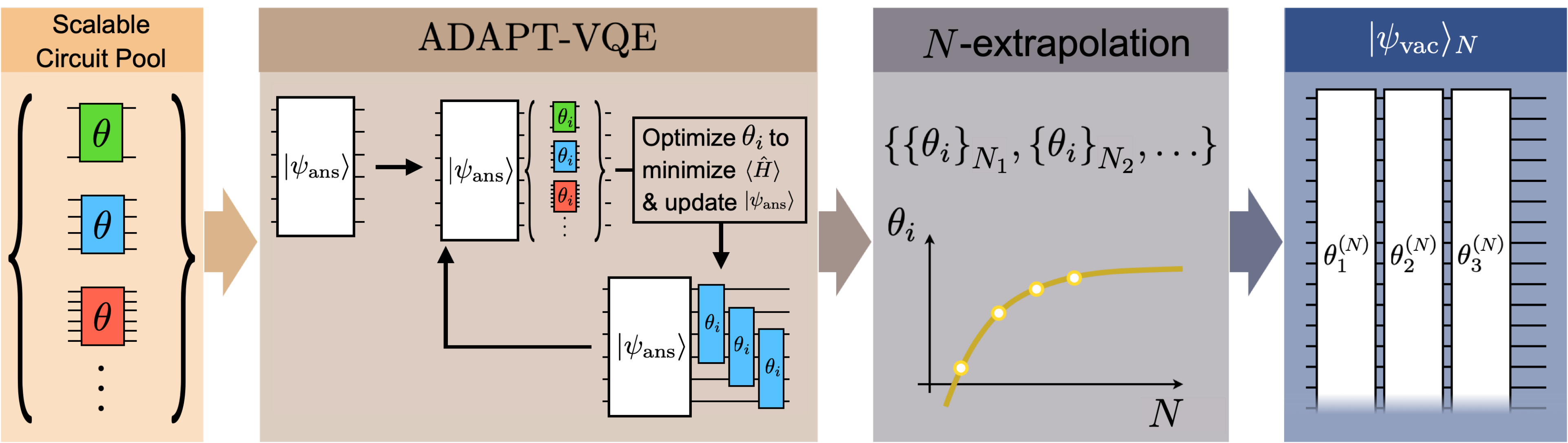}
\caption{The SC-ADAPT-VQE algorithm for state preparation on a quantum computer.
Left to right: A scalable pool of circuit elements that respects the symmetries of the desired state, and that build out correlations over a hierarchy of length scales.  
Parameterized circuits are built from the scalable circuit elements and optimized to minimize the energy using ADAPT-VQE~\cite{Grimsley_2019}.
This process is repeated for a sequence of system sizes $\{N_1,N_2,...\}$ on a classical computer, and the variational parameters $\{\theta_i\}$ that define the state preparation circuits are extrapolated to a desired $N$.
The extrapolated circuits are executed on a quantum computer to prepare the desired state.
This figure is adapted from Ref.~\cite{Farrell:2023fgd}}
    \label{fig:SCADAPTVQE}
\end{figure}
The work discussed in chapters~\ref{chap:axial} and~\ref{chap:beta} will be foundational for future quantum simulations of $1+1$D QCD and hopefully QCD in higher dimensions.
However, even the $N_f=2$ demonstration was limited to 16 qubits, which could be simulated by performing matrix multiplication of $2^{16}\times 2^{16}$ matrices ($2^{16}= 65,536$).
This can easily be handled on a laptop, which can handle up to $\sim 2^{26} \times 2^{26}$ dimensional matrices.
However, the exponential growth of Hilbert space causes exact methods to quickly hit a ceiling.
Indeed, even the most powerful supercomputers, with petabytes of memory, can only simulate up to 48 qubits using exact matrix methods~\cite{De_Raedt_2019}.
Surpassing this threshold of exact computation with classical computers was part of the motivation of the work in chapters~\ref{chap:SCADAPT} (quantum simulations using 100 qubits) and~\ref{chap:SchwingerTevol} (quantum simulations using 112 qubits).
To facilitate the quantum simulation of larger systems, these chapters focus on a simpler lattice gauge theory, the lattice Schwinger model, which is quantum electrodynamics in $1+1$D.
Like QCD, the Schwinger model is a confining gauge theory, has a chiral condensate and possesses composite, “hadronic”, particles that bind and form “nuclei”.

In chapter~\ref{chap:SCADAPT} the first step of a quantum simulation of the Schwinger model is addressed; state preparation on a quantum computer. 
A new state preparation algorithm, Scalable Circuits ADAPT-VQE (SC-ADAPT-VQE), is introduced, see Fig.~\ref{fig:SCADAPTVQE} for an illustration.
Physical systems have many properties that simplify state preparation: they are often invariant under spatial translations, locally interacting and possess a mass gap. 
SC-ADAPT-VQE uses these features to determine scalable quantum circuits for preparing states with localized correlations.
Scalability allows these state preparation circuits to be optimized on small and modest sized lattices using {\it classical computers}, and then robustly extrapolated to prepare the desired state on large systems using {\it quantum computers}.
The use of classical computers circumvents the challenge of optimizing parameterized quantum circuits on a noisy quantum computer, a task which is known to cause numerical instabilities like barren plateaus~\cite{Wang:2020yjh,Scriva:2023sgz}.
In chapter~\ref{chap:SCADAPT}, SC-ADAPT-VQE is applied to the preparation of the Schwinger model vacuum. 
By optimizing scalable state preparation circuits on systems of up to 28 qubits using classical computers, the ground state is prepared on 100 qubits of IBM’s quantum computer {\tt ibm\_cusco}. 
Measurements of the local chiral condensate and charge-charge correlators are found to be in excellent agreement with results obtained from matrix-product-state simulations.
The work in this chapter was featured in a  \href{https://www.youtube.com/watch?v=L7Kk_lR1Y2M&embeds_referring_euri=https%3A%2F%2Fresearch.ibm.com%2F&embeds_referring_origin=https%3A%2F%2Fresearch.ibm.com&source_ve_path=OTY3MTQ&feature=emb_imp_woyt}{\raisebox{-0.05\height} \ live YouTube Stream} and \href{https://research.ibm.com/blog/100-qubit-utility}{\raisebox{-0.05\height} \ blog post} by IBM.

The work in chapter~\ref{chap:SCADAPT} solved the first step for large scale quantum simulations of the Schwinger model.
Chapter~\ref{chap:SchwingerTevol} builds off this and presents results for quantum simulations of hadron dynamics in the lattice Schwinger model.
The first problem that is addressed is the preparation of a hadronic state that will be time evolved.
In this context, a hadron is an electron positron pair, bound together by the confining electromagnetic potential.
We choose to prepare a hadron wavepacket; a superposition of single hadron states that is localized in both position and momentum space.
This state has localized correlations, and can therefore be prepared with SC-ADAPT-VQE.
The SC-ADAPT-VQE circuits determined in chapter~\ref{chap:SCADAPT} are first used to prepare the vacuum everywhere on the lattice, and then SC-ADAPT-VQE is applied again to determine scalable circuits that excite a hadron wavepacket on top of the vacuum.

The next challenge addressed in chapter~\ref{chap:SchwingerTevol} is the implementation of the time evolution operator.
Fermions in the Schwinger model interact with each other through a linear Coulomb potential $\hat{H}_{el}$.
On a quantum computer this corresponds to an interaction that is all-to-all between every pair of qubits.
This all-to-all interaction is problematic to implement on a quantum device for two reasons.
First, the number of gates required to implement $e^{-i \hat{H}_{el} t}$ grows quadratically with the simulation volume.
For system sizes of 100+ qubits, corresponding to 50+ lattice sites, the necessary gate count surpasses the capabilities of present-day quantum devices.
Second, the required connectivity for efficient implementation is all-to-all.
Such connectivity is not available on IBM's superconducting architecture, which only has native nearest-neighbor interactions.
Motivated by the non-perturbative mechanism of confinement, we introduce a truncated interaction that removes interactions between sites separated by $\gtrsim$ the confinement length.
This interaction is systematically improvable by increasing the interaction range, and converges exponentially.
The truncated interaction allows $e^{-i \hat{H}_{el} t}$ to be implemented with a number of gates that scales linearly with the lattice size instead of quadratically.
Additionally, interactions only need to be engineered between qubits separated by approximately a confinement length instead of across the whole lattice.
This significantly reduces the two-qubit gate count, and makes quantum simulations of time evolution possible.

Our simulation of hadron dynamics proceeds by preparing the hadron wavepacket with SC-ADAPT-VQE, and then performing Trotterized time evolution using the Hamiltonian with the truncated electric interaction.
These simulations are performed on 112 qubits (56 lattice sites) for up to 14 Trotter steps.
The simulation of 14 Trotter steps uses 13,858 two-qubit gates with an associated two-qubit circuit depth of 370.
The results of the quantum simulation show a clear signal of a hadron propagating through the vacuum, and qualitative agreement is obtained with matrix product state simulations.
Vital to the success of these simulations was an incredible amount of statistics, 156 millions shots in total, that enabled powerful error mitigation techniques.
At the time (January 2024) this was the most complex digital quantum computation to be published on arXiv\footnote{Complexity is being measured by either the number of two-qubit gates or by the quantum volume $ = (\# \text{ of qubits}) \times (\text{CNOT depth})$.}.
This work has been featured in an IBM \href{https://www.ibm.com/quantum/blog/hadron-dynamics-simulations}{blogpost}.

The work in chapters~\ref{chap:SCADAPT} and~\ref{chap:SchwingerTevol} propelled the quantum simulation of lattice gauge theories beyond what is possible with exact methods using classical computers.
However, they are still susceptible to simulation using classical computers with (approximate) tensor-network techniques.
This is because simulations of the vacuum, and of the dynamics of a single hadron only populate relatively low-energy states, which can be efficiently represented with a Matrix Product State ansatz~\cite{Hastings_2007,arad2013area,Brand_o_2014}.
The work presented in chapter~\ref{chap:deDx} takes steps toward quantum simulations of fragmentation and hadronization at high energies, which will likely be beyond the capabilities of tensor network based techniques.
An open question relevant to understanding heavy-ion collisions at BNL~\cite{PHENIX:2004vcz,BRAHMS:2004adc,PHOBOS:2004zne,STAR:2005gfr,STAR:2010vob,STAR:2017sal} and the LHC~\cite{Loizides:2016tew,Foka:2016zdb,Foka:2016vta}, and QCD dynamics at high energies more generally, are the mechanisms underlying fragmentation and re-hadronization.
In a high energy collision between two heavy nuclei, the nucleons in the nucleus can be broken apart freeing the quarks and gluons from hadronic confinement.
The resulting state is believed to form an exotic phase of matter called a quark gluon plasma.
As the quark gluon plasma expands, the average energy of the partons decrease, and the quarks and gluons re-hadronize.
Questions related to the formation of quark gluon plasma, its subsequent expansion and thermalization and hadronization require the real-time simulation of out-of-equilibrium QCD dynamics to address.
Such simulations are impossible using classical computers, but will hopefully be possible using future quantum computers.

The Schwinger model is also a confining lattice gauge theory, and this process of fragmentation and hadronization can be simulated via the high energy collision of hadrons.
In chapter~\ref{chap:deDx} such simulations are performed using classical computers to prepare for future quantum simulations.
The background for our collisions is a dense medium of static hadrons on half of the lattice that are fixed in place by background charges.
Hadrons collide with this medium at high energies, and the energy deposition, entanglement and charge density are measured as a function of incident velocity.
The collision does not factorize as a product of individual collisions, providing clear evidence of quantum coherence between the constituents of the medium.
In addition, measurements of the charge distribution after the collision indicate that hadron are produced at sufficiently high collision energies.
A careful study of lattice artifacts reveals that entanglement, as a non-local quantity, is significantly more sensitive to lattice spacing effects than local observables.
With the goal of eventually performing these simulations on a quantum computer, it is shown how SC-ADAPT-VQE can be used to determine circuits that prepare the static hadrons that make up the medium.
The simulation strategy and state preparation circuits developed in this work are steps toward simulating fragmentation and hadronization in the Schwinger model, and eventually QCD, on a quantum computer.

\chapter{Entanglement minimization in hadronic scattering with pions}
\label{chap:hadronEP}
\noindent 
\textit{This chapter is associated with Ref.~\cite{Beane:2021zvo}: \\
``Entanglement minimization in hadronic scattering with pions" by Silas R. Beane, Roland C. Farrell and Mira Varma.}
  \section{Introduction}
\label{sec:intro}

\noindent It is of current interest to uncover implications of quantum
entanglement for the low-energy interactions of hadrons and nuclei\footnote{For a recent review, see Ref.~\cite{Klco:2021lap}.}. As
these interactions are profitably described by effective quantum field
theory (EFT), which is an expansion of the relevant effective action
in local operators, entanglement may have subtle implications for EFT
which are difficult to identify due to its intrinsic non-locality.
Ideally entanglement properties reveal themselves as regularities in
hadronic data and, possibly, as accidental approximate symmetries. In
addition to the non-local nature of entanglement, a difficulty lies
with parsing the distinction, if any, between entanglement effects and
generic quantum correlations which account for the deviation of QCD
path integral configurations from a classical path. For instance, if
one assumes that QCD with $N_c=3$ is near the large-$N_c$
limit~\cite{tHooft:1973alw,Witten:1979kh,Witten:1979pi,Kaiser:2000gs}, then one
might expect that it would be difficult to distinguish between
large-$N_c$ expectations and some fundamental underlying principle
that minimizes entanglement independent of the value of $N_c$.  To
make this more concrete, consider two local or non-local QCD operators
${\cal O}_1$ and ${\cal O}_2$. If the vacuum expectation value of the
product of these operators
obeys the factorization rule~\cite{tHooft:1973alw,Witten:1979kh,Witten:1979pi}
\begin{equation}
\langle {\cal O}_1 {\cal O}_2 \rangle \ =\ \langle {\cal O}_1 \rangle \langle {\cal O}_2 \rangle \ +\ O(\epsilon)
	\label{eq:factorgen}
\end{equation}
where $\epsilon$ is a small number, then the variance of any operator
vanishes in the limit $\epsilon\rightarrow 0$. 
The variance of an operator is related to the sampling of multiple field configurations in the path integral, and the vanishing of the variance often implies the existence of a master field solution.
A theory whose
operators obey this factorization behaves like a classical
theory,\footnote{Ordinarily one identifies the classical theory with
  the trivial $\hbar\rightarrow 0$ limit. However,
  Ref.~\cite{Yaffe:1981vf} has established a more general criterion
  for the classical limit.} and therefore has a small parameter
$\epsilon$ which measures quantum effects. Large-$N_c$ QCD is such a
theory, and indeed, at least for a class of QCD operators, one can
identify $\epsilon = 1/N_c$. The factorization property,
Eq.~(\ref{eq:factorgen}), is then easily deduced from Feynman diagrams
involving quarks and gluons and amounts to the dominance of
disconnected contributions in the path integral.

On the other hand, one might imagine that the factorization of
Eq.~(\ref{eq:factorgen}) arises as a property of the path integral,
rather than as a property of the local action (as in varying $N_c$ and
taking it large in QCD). It is not {\it a priori} unlikely that, at
least for a class of QCD operators, the path integral minimizes
quantum fluctuations via a mechanism that is not currently understood.
For instance, starting with QCD defined at short distances, the
procedure of sequentially integrating out short distance modes to
obtain low-energy hadronic scattering amplitudes may remove
highly-entangled states that arise from non-perturbative QCD dynamics,
leaving a low-energy EFT that is near a classical trajectory.  It is
intuitively sensible that the QCD confinement length acts as a natural
cutoff of entanglement in the low-energy EFT.  This notion can be
raised to the conjecture that QCD will minimize the entanglement in
low-energy hadronic interactions. Testing this conjecture relies on
finding hadronic systems where its consequences deviate from those
implied by large-$N_c$. And the success of the large-$N_c$ approximation
in describing the world renders this task challenging.  Evidence in
favor of this conjecture was found in Ref.~\cite{Beane:2018oxh} in a
study of baryon-baryon scattering systems (See also
Refs.~\cite{Beane:2020wjl} and \cite{Low:2021ufv}). This work relied
both on theoretical arguments and high-precision lattice QCD
simulations of baryon-baryon scattering systems with strangeness. In
this chapter, the conjecture of minimal
entanglement will be investigated in both $\pi\pi$ and $\pi N$ scattering.

Finding measures of the entanglement due to interaction is both
non-trivial and non-unique.  The most fundamental object in the
scattering process is the unitary $S$-matrix.  In a scattering process
in which the two in-state particles form a product state, the
$S$-matrix will entangle the in-state particles in a manner that is
dependent on the energy of the scattering event. A useful measure of
this entanglement is the entanglement power (EP) of the
$S$-matrix~\cite{PhysRevA.63.040304,mahdavi2011cross,Beane:2018oxh}. In
the case of nucleon-nucleon ($NN$) scattering, the EP was found for all
momenta below inelastic threshold~\cite{Beane:2018oxh}. However, the
most interesting phenomenological result is at threshold, where the
vanishing EP implies the vanishing of the leading-order spin
entangling operator, which in turn implies Wigner $SU(4)$
symmetry~\cite{Wigner:1936dx,Wigner:1937zz,Wigner:1939zz}.  As this
symmetry is a consequence of large-$N_c$
QCD~\cite{Kaplan:1995yg,Kaplan:1996rk,CalleCordon:2008cz}, the
minimization of entanglement and the large-$N_c$ approximation are
found to be indistinguishable in the two-flavor case. By contrast, in
the three-flavor case, minimization of the entanglement power in
baryon-baryon scattering implies an enhanced $SU(16)$ symmetry which
is not necessarily implied by large-$N_c$ and is realized in lattice
QCD simulations~\cite{Beane:2018oxh,Low:2021ufv}. Given that
baryon-baryon scattering exhibits entanglement minimization, it is of
interest to determine whether other low-energy QCD scattering systems
exhibit this property. In investigating the EP of scattering systems
involving pions, once again a crucial difficulty is distinguishing
consequences of entanglement minimization and the large-$N_c$
limit. In the $\pi\pi$ system the implications of entanglement minimization are found to
be indistinguishable from implications of large-$N_c$. In the $\pi N$
system the implications of entanglement minimization are distinct,
however the absence of an enhanced symmetry limits the predictive
power to simple scaling laws with no smoking-gun predictions.

This chapter is organized as follows. In Section~\ref{sec:pipiS}, the EP
of the $\pi\pi$ $S$-matrix is considered in detail.  After
introducting the standard definition and conventions of the $\pi\pi$
$S$-matrix, the $S$-matrix is formulated in a basis convenient for
calculation of the EP. Explicit expressions are derived for the EP of
the first few partial waves in terms of phase shifts and leading-order
expressions in chiral perturbation theory are provided.  Using the
highly-accurate Roy-equation solutions for the low-energy phase
shifts, the experimental EP for the first few partial waves are given
up to inelastic threshold. The consequences of minimizing the EP are
considered and compared to large-$N_c$ expectations.  In
Section~\ref{sec:piNS}, the same procedure is repeated for the $\pi N$
$S$-matrix. Finally, Section~\ref{sec:discuss} is a discussion of the
possible conclusions that can be drawn from the conjecture of minimal
entanglement.

\section{The \texorpdfstring{$\pi\pi$}{} System}
\label{sec:pipiS}

\noindent There are, of course, several important differences between baryon-baryon and pion-pion scattering.
Firstly, with pions there is no notion of spin entanglement.  However,
isospin (or generally flavor) entanglement is present and can be
quantified using the EP and it is not clear that there is any
meaningful distinction between these two kinds of
entanglement. Indeed, it is straightforward to see that the ``spin''
entanglement of Ref.~\cite{Beane:2018oxh} can be reformulated as
``isospin'' entanglement with identical consequences\footnote{At the
  level of the EFT, this is simply realized via Fierz
  identities.}. This is no surprise as entanglement is fundamentally a
property of a non-product state vector whose existence relies either on an internal or a spacetime symmetry. Secondly, the crucial distinction between
baryon-baryon scattering at very low-energies and the scattering of
pions is that pion scattering at low-energies is strongly constrained
by spontaneous chiral symmetry breaking in QCD. In particular, chiral
symmetry implies that low-energy pion scattering on an arbitrary
hadronic target is weak. The weak nature of the interaction is due to
the smallness of the light-quark masses relative to a characteristic
QCD scale. This translates to a chiral suppression of the EP at
low-energies. Chiral symmetry breaking at large-$N_c$ does involve
enhanced symmetry~\cite{Kaiser:2000gs}; for $N$ flavors, the QCD
chiral symmetries and their pattern of breaking are enhanced to
$U(N)\otimes U(N)\to U(N)$, as signaled by the presence of a new
Goldstone boson, $\eta'$, whose squared mass scales as $1/N_c$.
Intuitively, the anomaly, as an intrinsically quantum phenomenon, is a
strongly entangling effect which would generally vanish as quantum
fluctuations are suppressed. However, this is not assumed as the focus of this paper is
two-body scattering which does not access the anomaly.

\subsection{$S$-matrix definition}

\noindent The $S$-matrix is defined as
\begin{equation}
	S = 1 + i T 
	\label{eq:sbasic}
\end{equation}
where unity, corresponding to no interaction, has been separated out. This then defines the $T$-matrix.
The $S$-matrix element for the scattering process $\pi^i \pi^j \to \pi^k \pi^l$ is then given by
\begin{multline}
  \bra{\pi^k(p_3) \pi^l(p_4)} S \ket{\pi^i(p_1) \pi^j(p_2)} =   \langle \pi^k(p_3) \pi^l(p_4) |  \pi^i(p_1) \pi^j(p_2) \rangle
  \\ + \bra{\pi^k(p_3) \pi^l(p_4)} iT \ket{\pi^i(p_1) \pi^j(p_2)}
\end{multline}
where $i$, $j$, $k$, and $l$ are the isospin indices of the pion states.
The projection operators onto states of definite isospin are\footnote{For a detailed construction, see Ref.~\cite{lanz2018determination}.}
\begin{eqnarray}
	P_0^{kl,ij} &=& \inv{3} \delta^{kl} \delta^{ij} \ , \\
	P_1^{kl,ij} &=& \inv{2} \left( \delta^{ki} \delta^{lj} - \delta^{li} \delta^{kj} \right) \ , \\
	P_2^{kl,ij} &=& \inv{2} \left( \delta^{ki} \delta^{lj} + \delta^{li} \delta^{kj} \right) 
								- \inv{3}\delta^{kl} \delta^{ij} \ ,
	\label{eq:pionnIP}
\end{eqnarray}
where the subscript indicates the total isospin, $I$, of the $\pi\pi$ system.
Straightforward field-theoretic manipulations then give
\begin{multline}
  \bra{\pi^k(p_3) \pi^l(p_4)} S \ket{\pi^i(p_1) \pi^j(p_2)} \\ = 
  (2\pi)^4 \delta^4(p_1+p_2-p_3-p_4)\, \frac{16 \pi}{\sigma(s)}\,\sum_{\ell=0}^\infty (2\ell+1) P_\ell(\cos \theta)\, {\bf\cal S}_\ell^{kl,ij} \ ,
\end{multline}
where the $P_\ell$ are the Legendre polynomials, and
\begin{equation}
	\sigma(s) \equiv \sqrt{1-4 \mpi^2/s} \ ,
\end{equation}
with $s=4(q^2+M_\pi^2)$ and $q$ is the center-of-mass three-momentum of the pions.
The focus here will be on the $S$-matrices of definite partial wave:
\begin{equation}
{\bf\cal S}_\ell^{kl,ij} \equiv  e^{2i\delta_\ell^0} P_0^{kl,ij}+e^{2i\delta_\ell^1} P_1^{kl,ij}+e^{2i\delta_\ell^2} P_2^{kl,ij} \ ,
\end{equation}
    \label{eq:Smatcomp}
which satisfy the unitarity constraint
\begin{equation}
{\bf\cal S}_\ell^{kl,ij} {\bf\cal S}_\ell^{*ij,mn} \ =\ \delta^{km} \delta^{ln} \ .
\end{equation}
Since the pions obey Bose statistics, the angular momentum, $\ell$, is
even for the states with $I = 0$ or $2$ and odd for states with $I = 1$.

As the initial state in the scattering process is a product state of two pions, each in the ${\bf 3}$-dimensional
($I=1$) irrep of $SU(2)$ isospin, it is convenient to work in the direct-product  matrix basis.
The pion isospin matrices are the three-by-three matrices $\hat { t}_\alpha$ which satisfy
\begin{eqnarray}
{[\,\hat { t}_\alpha\,  ,\,\hat { t}_\beta\, ]}\, =\, i\,\epsilon_{\alpha\beta\gamma} \, \hat { t}_\gamma \ .
\end{eqnarray}
In the product Hilbert space ${\cal H}_1\otimes{\cal H}_2$, the total isospin of the two-pion system is $\hat {\bm t}_1\otimes {\cal I}_3 +{\cal I}_3\otimes \hat {\bm t}_2$, where
${\cal I}_3$ is the three-by-three unit matrix, which implies
\begin{eqnarray}
  \hat {\bm t}_1 \cdot \hat {\bm t}_2 & =&  \oneht \Big\lbrack I\left(I+1\right) \;-\; 4 \Big\rbrack \hat {\bf 1}\ =\ \hat {\bf 1}
  \begin{cases}
-2,\qquad I=0 \\
-1,\qquad I=1 \\
\ \; 1,\qquad I=2
  \end{cases}
    \label{eq:basic2}
\end{eqnarray}
where
$ \hat {\bf 1} = \hat {\cal I}_3\otimes  \hat {\cal I}_3$ and
$\hat {\bm t}_1 \cdot \hat {\bm t}_2 = \sum\limits_{\alpha=1}^3 \ \hat{ t}_1^\alpha \otimes \hat{ t}_2^\alpha$.
The $9\times 9$ dimensionality of the matrix is in correspondence with the dimensionality of the $SU(2)$ isospin product representation
${\bf 3}\otimes{\bf 3}= {\bf 1}\oplus {\bf 3}\oplus {\bf 5}$. There are now three invariants and three observables; one easily finds the $S$-matrix
in the direct-product matrix basis
\begin{eqnarray}
  \hat {\bf S}_\ell   & = &
e^{2i\delta_\ell^0} \hat {\bf P}_{0}+e^{2i\delta_\ell^1} \hat {\bf P}_{1}+e^{2i\delta_\ell^2} \hat {\bf P}_{2} \ ,
\label{eq:Sdefpipi}
\end{eqnarray}
where the three $9\times 9$ projection matrices are
\begin{eqnarray}
  \hat {\bf P}_{0}& =& -\frac{1}{3}\left(\hat {\bf 1}- \left( \hat  {\bm t}_1 \cdot   \hat  {\bm t}_2 \right)^2 \right) \ , \\
    \hat {\bf P}_{1}& =& \hat {\bf 1}-\frac{1}{2}\left( \left( \hat  {\bm t}_1 \cdot   \hat  {\bm t}_2 \right)+ \left( \hat  {\bm t}_1 \cdot   \hat  {\bm t}_2 \right)^2 \right) \ , \\
        \hat {\bf P}_{2}& =& \frac{1}{3}\left(\hat {\bf 1}+ \frac{3}{2} \left( \hat  {\bm t}_1 \cdot   \hat  {\bm t}_2 \right)+ \frac{1}{2} \left( \hat  {\bm t}_1 \cdot   \hat  {\bm t}_2 \right)^2 \right) \ .
\end{eqnarray}
It is readily checked that the $S$-matrix is unitary, and using the representation $({ t}_\gamma)_{\alpha\beta}=-i\epsilon_{\alpha\beta\gamma}$, it is straightforward to establish
equivalence with the component form, Eq.~(\ref{eq:Smatcomp}).
The trace is given by $e^{i 2 \delta_\ell^0} +3 e^{i 2 \delta_\ell^1} +5 e^{i 2 \delta_\ell^2}$ which correctly
counts the isospin multiplicity, and is in correspondence with the nine eigenvalues of $\hat {\bf S}$.

\subsection{Entanglement power}
\label{sec:pipiEP}

\noindent Consider the $\ell=1$ $S$-matrix. As this system can scatter only in the $I=1$ channel, it provides
a useful example of how the $S$-matrix entangles the initial two-pion state. From Eq.~(\ref{eq:Sdefpipi}) one
finds
\begin{eqnarray}
  \hat {\bf S}_1   & = & \frac{1}{2}\left( 1+ e^{i 2 \delta_1^1}\right) \hat   {\bf 1}
\ +\ \frac{1}{2}\left( 1- e^{i 2 \delta_1^1}  \right) {\cal P}_{12}
\label{eq:Spipiell1}
\end{eqnarray}
where the SWAP operator is given by
\begin{equation}
{\cal P}_{12} =\left( \hat  {\bm t}_1 \cdot   \hat  {\bm t}_2 \right)^2 + \hat  {\bm t}_1 \cdot   \hat  {\bm t}_2 - \hat {\bf 1} \ .
\end{equation}
As the SWAP operator interchanges the pions in the initial two-pion product state, leaving another two-pion product state, it
does not entangle. Therefore, the $S$-matrix has the two obvious non-entangling solutions $\delta_1^1=0$ (no interaction) and
$\delta_1^1=\pi/2$ (at resonance). One measure of $S$-matrix entanglement would then be the (absolute value squared of the) product of the coefficients of the non-entangling
solutions:
\begin{eqnarray}
\Big\lvert\;\left( 1+ e^{i 2 \delta_1^1}\right) \left( 1- e^{i 2 \delta_1^1}\right) \Big\lvert^2\; \sim \sin^2\left(2 \delta_1^1\right) \ .
\label{eq:EPnaive}
\end{eqnarray}

A state-independent measure of the entanglement generated by the
action of the $S$-matrix on the initial product state of two free
particles is the
EP~\cite{PhysRevA.63.040304,mahdavi2011cross,Beane:2018oxh}.  In order
to compute the EP an arbitrary initial product state should be
expressed in a general way that allows averaging over a given
probability distribution folded with the initial state. Recall that in
the $NN$ case, there are two spin states (a qubit) for each nucleon and
therefore the most general initial nucleon state involves two complex
parameters or four real parameters. Normalization gets rid of one
parameter and there is an overall irrelevant phase which finally
leaves two real parameters which parameterize the ${\bf CP}^1$
manifold, also known as the 2-sphere ${\bf S}^2$, or the Bloch sphere.
Now in the isospin-one case we have three isospin states (a qutrit)
which involves three complex parameters. Again normalization and
removal of the overall phase reduce this to four real parameters which
parameterize the ${\bf CP}^2$
manifold~\cite{Brody_2001,Bengtsson:2001yd,bengtsson_zyczkowski_2006}. Since
the $\pi\pi$ initial state is the product of two isospin-one states,
there will be eight parameters to integrate over to get the EP.

There are now two qutrits in the initial state, which live in the Hilbert spaces ${\cal H}_{1,2}$, each
spanned by the states $\lbrace |\,{\bf -1}_i \, \rangle , |\,{\bf 0}_i
\, \rangle , |\,{\bf 1}_i \, \rangle \rbrace$ with $i=1,2$.  It is of interest
to determine the EP of a given $S$-matrix operator, which is a measure of the entanglement of the scattered state averaged
over the ${\bf CP}^2$ manifolds on which the qutrits live.  Consider
an arbitrary initial product state of the qutrits
\begin{eqnarray}
|\,\Psi \, \rangle  \ =\  U\left(\alpha_1,\beta_1,\mu_1,\nu_1\right) | \, \rangle_1 \otimes U\left(\alpha_2,\beta_2,\mu_2,\nu_2\right) | \, \rangle_2 
\end{eqnarray}
with
\begin{eqnarray}
\hspace{-0.34in}U\left(\alpha_i,\beta_i,\mu_i,\nu_i\right) | \, \rangle_i  =
\cos\beta_i \sin\alpha_i|\,{\bf -1} \, \rangle_i  +  e^{i\mu_i}\sin\beta_i \sin\alpha_i|\,{\bf 0} \, \rangle_i  + e^{i\nu_i}\cos\alpha_i |\,{\bf 1} \, \rangle_i\, ,
\end{eqnarray}
where $0 \leq {\mu_i,\nu_i} < 2{\pi}$ and $0 \leq {\alpha_i,\beta_i} \leq {\pi}/2$.
The geometry of ${\bf CP}^2$ is described by the Fubini-Study (FS) line element~\cite{Brody_2001,Bengtsson:2001yd,bengtsson_zyczkowski_2006}
\begin{eqnarray} 
\begin{aligned}
ds_{\scriptstyle FS}^2 = \; &d\alpha^2+ \sin ^2(\alpha ) d\beta^2 +\left ( \sin ^2(\alpha ) \sin ^2(\beta ) - \sin ^4(\alpha ) \sin ^4(\beta ) \right ) d\mu^2 + \\
& \sin ^2(\alpha ) \cos ^2(\alpha ) d\nu^2 -2 \sin ^2(\alpha ) \cos ^2(\alpha ) \sin ^2(\beta ) d\mu  d\nu \ .
\end{aligned}
\label{FSmetric}
\end{eqnarray}
Of special interest here is the differential volume element which in these coordinates is
\begin{eqnarray}
  dV_{\scriptstyle FS} &=& \sqrt{det\,g_{\scriptstyle FS}}\, d\alpha\, d\beta\, d\mu\, d\nu  \nonumber \\
  &=&\cos\alpha\cos\beta\sin^3\alpha\sin\beta\, d\alpha\, d\beta\, d\mu\, d\nu \  
\end{eqnarray}
and the volume of the ${\bf CP}^2$ manifold is found to be,
\begin{eqnarray}
\int  dV_{\scriptstyle FS} & =& \frac{\pi^2}{2} \ .
\end{eqnarray}

The final state of the scattering process is obtained by acting with the unitary $S$-matrix of definite angular momentum
on the arbitrary initial product state:
\begin{eqnarray}
|\,\bar \Psi \, \rangle  \ =\  \hat {\bf S}_\ell |\, \Psi \, \rangle  \ .
\end{eqnarray}
The associated density matrix is
\begin{eqnarray}
\rho_{1,2}   \ =\  |\, \bar \Psi \, \rangle  \langle \, \bar \Psi  | \, ,
\end{eqnarray}
and tracing over the states in ${\cal H}_{2}$ gives the reduced density matrix
\begin{eqnarray}
\rho_{1}   \ =\  {\rm Tr}_2 \big\lbrack \rho_{1,2}  \big\rbrack .
\end{eqnarray}
The linear entropy of the $S$-matrix is then defined as\footnote{Note that this is related to the (exponential of the) second R\'enyi entropy.}
\begin{eqnarray}
E_{\hat {\bf S}_\ell}  \ =\  1 \ -\ {\rm Tr}_1 \big\lbrack \left( \rho_{1} \right)^2 \big\rbrack .
\end{eqnarray}
Finally, the linear entropy is integrated over the initial ${\bf CP}^2$ manifolds to form the average, and the entanglement power is
\begin{eqnarray}
  {\mathcal E}({\hat {\bf S}_\ell}) \ =\ \left(\frac{2}{\pi^2}\right)^2 \left(\prod_{i=1}^2\int dV^i_{\scriptstyle FS}\right) {\cal P} E_{\hat {\bf S}_\ell} 
\end{eqnarray}
where ${\cal P}$ is a probability distribution which here will be taken to be unity.
Evaluating this expression using Eq.~(\ref{eq:Sdefpipi}) yields the s-wave $\pi\pi$ EP:
\begin{eqnarray}
 {\mathcal E}({\hat {\bf S}_0}) & =& \frac{1}{648}\left( 156 - 6 \cos[4 \delta_0^0 ] - 65 \cos[2 (\delta_0^0 - \delta_0^2)] \right. \nn 
  &&\qquad\qquad\qquad\left.\hspace{-0.48in}- 10 \cos[4 (\delta_0^0 - \delta_0^2)] - 60 \cos[4 \delta_0^2] - 
 15 \cos[2 (\delta_0^0  + \delta_0^2)] \right)\ ,
\label{eq:EPswave}
\end{eqnarray}
and the p-wave $\pi\pi$ EP:
\begin{eqnarray}
  {\mathcal E}({\hat {\bf S}_1})  = \frac{1}{4}\sin^2\left(2 \delta_1^1\right) \ .
\label{eq:EPpwave}
\end{eqnarray}
Notice that this matches the intuitive construction which led to Eq.~(\ref{eq:EPnaive}).
The EPs have the non-entangling solutions:
\begin{eqnarray}
\delta_0^0 & =& \delta_0^2 \ =\ 0, \frac{\pi}{2} \ , \\
\delta_1^1 & =& 0, \frac{\pi}{2} \ .
\label{eq:noEPsols2bose}
\end{eqnarray}
Therefore, in the s-wave, entanglement minimization implies that {\it
  both} isospins are either non-interacting or at resonance, while in
the p-wave, entanglement minimization implies that the $I=1$ channel
is either non-interacting or at resonance. As no $I=2$ resonances are
observed in nature (and their coupling to pions is suppressed in
large-$N_c$ QCD~\cite{Weinberg:2013cfa}), the s-wave EP has a single minimum corresponding to
no interaction. By contrast, the $I=1$ channel will exhibit minima of
both types. It is worth considering the EP of a simple resonance model.
Consider the unitary $S$-matrix:
\begin{eqnarray}
  {\hat {\bf S}_1} \ =\ \frac{s-m_1^2 - i m_1 \Gamma_1}{s-m_1^2 + i m_1 \Gamma_1} \ ,
\end{eqnarray}
where $m_1$ ($\Gamma_1$) are the mass (width) of the resonance. The EP is
\begin{eqnarray}
  &&{\mathcal E}({\hat {\bf S}_1})  = \left( \frac{m_1 \Gamma_1 \left(s-m_1^2\right)}{\left(m_1 \Gamma_1\right)^2+ \left(s-m_1^2\right)^2}\right)^2 \ ,
\end{eqnarray}
which vanishes on resonance at $s=m_1^2$ and has maxima at $s=m_1(m_1\pm \Gamma_1)$. It is clear that the minimum corresponds to
$\hat {\bf S}\propto {\cal P}_{12}$. As the $\rho$-resonance dominates the $I=1$ channel at energies below $1~{\rm GeV}$, the EP in nature
will be approximately of this form.

\begin{figure}[!h]
\centering
\includegraphics[width = 0.81\textwidth]{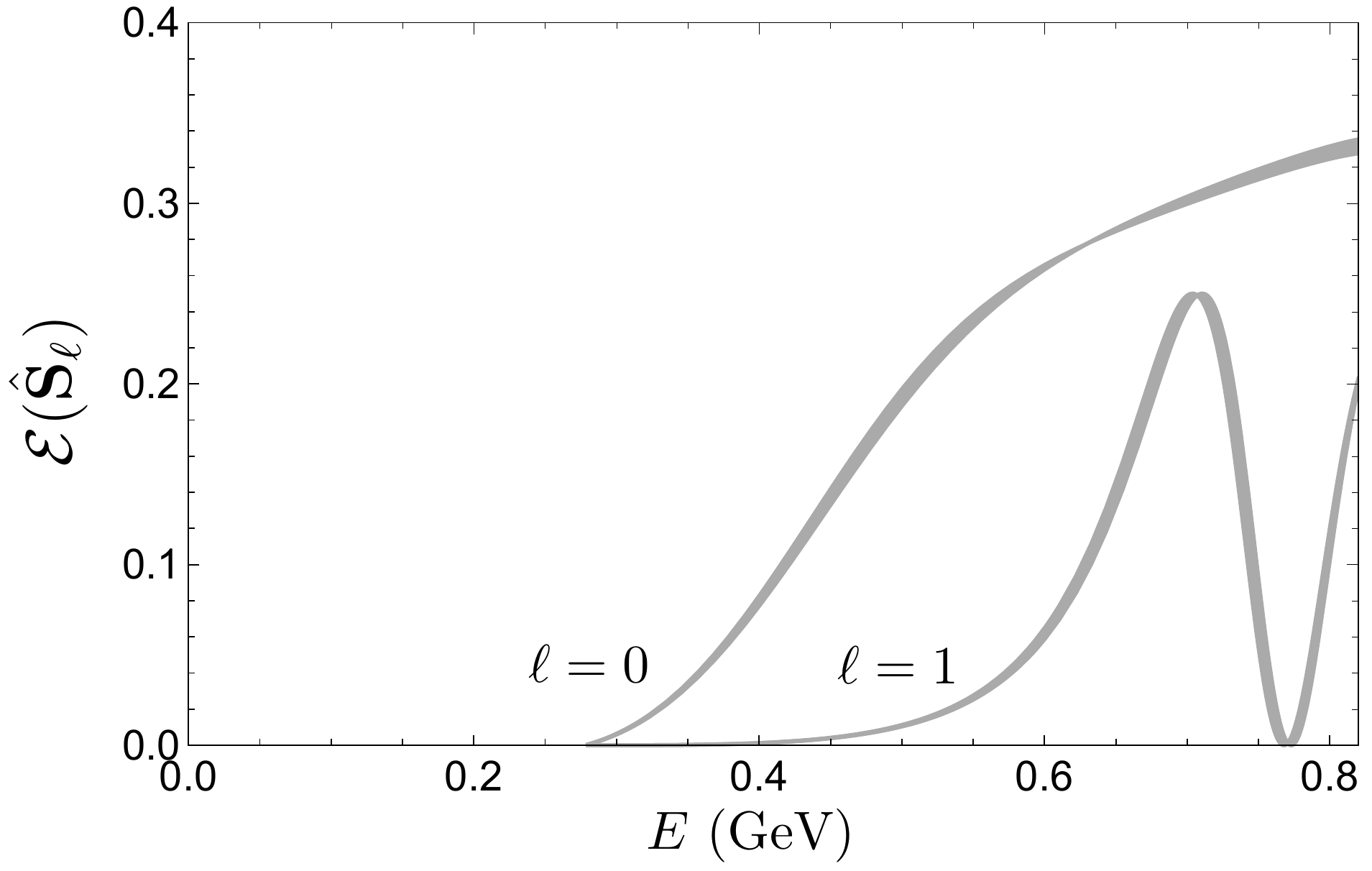}
\caption{Entanglement power of the $\pi\pi$ $S$-matrix for $\ell=0,1$ taken from Roy equation determinations (the bands represent an estimate of the uncertainties~\cite{Ananthanarayan:2000ht,Colangelo:2001df})
  of the $\pi\pi$ phase shifts.}
    \label{fig:EPpipiROY}
\end{figure}

The $\pi\pi$ phase shifts are the most accurately known of all hadronic $S$-matrices as the Roy equation constraints~\cite{Roy:1971tc}
come very close to a complete determination of the phase shifts~\cite{Ananthanarayan:2000ht,Colangelo:2001df}.
In Fig.~(\ref{fig:EPpipiROY}) the EPs for the first few partial waves are plotted using the Roy equation determinations of the $S$-matrix.

\subsection{Chiral perturbation theory}

\noindent Near threshold, the phase shift can be expressed in the effective range expansion as
\begin{equation}
	\delta_\ell^I(s) = \oneht \sin^{-1} \lbrace 2\sigma (s) q^{2\ell}\left( a^I_\ell \;+\; \mathcal{O}(q^2) \right)\rbrace \ ,
\end{equation}
where the scattering lengths, $a^I_\ell$, relevant to s-wave and p-wave scattering, are given at leading order in chiral perturbation theory by~\cite{Weinberg:1978kz,Gasser:1983yg}
\begin{equation}
a_0^0\ =\ \frac{7M_\pi^2}{32\pi F_\pi^2}  \ \ , \ \   a_0^2\ =\ -\frac{M_\pi^2}{16\pi F_\pi^2} \ \ , \ \ a_1^1\ =\ \frac{1}{24\pi F_\pi^2} \ ,
\end{equation}
where $F_\pi=93~{\rm MeV}$ is the pion decay constant. Near threshold the s-wave and p-wave EPs are given by
\begin{eqnarray}
&&{\mathcal E}({\hat {\bf S}_0}) \ =\ \frac{1}{9 M_\pi^2}\big\lbrack 4(a^0_0)^2 -5 (a^0_0 a^2_0)+10 (a^2_0)^2 \big\rbrack\; q^2 \ +\ \mathcal{O}(q^4)\ , \nn
&&{\mathcal E}({\hat {\bf S}_1}) \ =\ \frac{1}{M_\pi^2} (a^1_1)^2 \; q^6 \ +\ \mathcal{O}(q^{8})  \ .
\end{eqnarray}
As $a_0^0$ ($a_0^2$ ) is positive (negative) definite, the EP is trivially minimized with vanishing scattering lengths.
This then implies a bookkeeping where $F_\pi=\mathcal{O}(\epsilon^{-n})$ where $n$ is a positive number. 
Hence, in the limit of vanishing entanglement, the pions are non-interacting, and the dominant interaction is from tree diagrams; i.e. loops are suppressed by inverse powers
of $F_\pi$. In the large-$N_c$ limit, one finds  $\epsilon=1/N_c$ and $n=1/2$~\cite{tHooft:1973alw,Witten:1979kh,Witten:1979pi}.
Evidently the implications of vanishing entanglement for the $\pi\pi$ $S$-matrix are indistinguishable from large-$N_c$ expectations\footnote{We also studied the effect of explicit
chiral symmetry breaking on the entanglement power by varying the coefficients of operators with
insertions of the quark mass matrix in the effective action. No evidence of a connection between chiral symmetry breaking and the entanglement power was found. This aligns with large-$N_c$
expectations as the meson masses are independent of $N_c$. For an example of a relationship between
entanglement and chiral symmetry breaking see~\cite{Beane:2019loz}.}.

\section{The \texorpdfstring{$\pi N$}{} System}
\label{sec:piNS}

\noindent As baryons are formed from $N_c$ quarks, the baryon masses
and axial matrix elements grow with $N_c$.  The unitarity of the
$S$-matrix then places powerful constraints on baryon properties via
large-$N_c$ consistency
conditions~\cite{Dashen:1993as,Dashen:1993ac,Dashen:1993jt,Dashen:1994qi}.
At leading order in the large-$N_c$ expansion this yields predictions
that are equivalent in the two (three) flavor case to $SU(4)$
($SU(6)$) spin-flavor symmetry which place the ground-state baryon
spin states in the ${\bf 20}$ (${\bf 56}$) dimensional irrep together
with the delta (baryon decuplet). Therefore, the large-$N_c$ limit not
only predicts an enhanced symmetry but also alters the definition of
a baryon in a fundamental way. Moreover, any sensible effective
theory of $\pi N$ scattering in the large-$N_c$ limit must include the
delta resonance as an explicit degree of freedom. In what follows, the
consequences of entanglement minimization of the low-energy $S$-matrix
are considered for $N_c=3$ QCD.

\subsection{$S$-matrix definition}
The $S$-matrix element for the scattering process, $\pi^a(q_1) N(p_1) \to \pi^b(q_2) N(p_2)$, is given by
\begin{multline}
  \bra{\pi^b(q_2) N(p_2)} S \ket{\pi^a(q_1) N(p_1)} =   \langle \pi^b(q_2) N(p_2) |  \pi^a(q_1) N(p_1) \rangle
  \\ + \bra{\pi^b(q_2) N(p_2)} iT \ket{\pi^a(q_1) N(p_1)} ,
\end{multline}
where $a$ and $b$ label the isospin of the pion.
The $T$ matrix element in the center-of-mass system (cms) for the process may be parameterized as~\cite{Fettes_1998}
\begin{equation}
    \begin{split}
        T^{ba}_{\pi N} = \left( \frac{E+m}{ 2m} \right ) \bigg \{ \delta^{ba} &\left [ g^+(\omega,t) + i \vec{\sigma} \cdot (\vec{q_2} \times \vec{q_1}) h^+(\omega,t) \right ] \\
         + i\epsilon^{abc} \tau^c &\left [ g^-(\omega,t) + i \vec{\sigma} \cdot (\vec{q_2} \times \vec{q_1}) h^-(\omega,t) \right ] \bigg \}
    \end{split}
\end{equation}
where $E$ is the nucleon energy, $\omega$ is the pion energy, $m$ is
the nucleon mass and $t = (q_1 - q_2)^2$ is the square of the momentum
transfer. The $\sigma$($\tau$) matrices act on the spin(isospin) of
the incoming nucleon. This decomposition reduces the scattering
problem to calculating $g^{\pm}$, the isoscalar/isovector
non-spin-flip amplitude and $h^{\pm}$, the isoscalar/isovector
spin-flip amplitude. The amplitude can be further projected onto
partial waves by integrating against $P_{\ell}$, the relevant Legendre
polynomial:
\begin{equation}
    f_{\ell \pm}^{\pm}(s) = \frac{E + m}{16\pi \sqrt{s}} \int_{-1}^{+1} dz \left [ g^{\pm} P_{\ell}(z) + {\vec{q}}^{\, 2} h^{\pm} \big ( P_{\ell \pm 1}(z) -z P_{\ell}(z)\big ) \right ]\ .
\end{equation}
Here $z=\cos{\theta}$ is the cosine of the scattering angle, $s$ is
the cms energy squared and $\vec{q}^{\, 2} = \vec{q_1}^2 =
\vec{q_2}^2$. The subscript $\pm$ on the partial wave amplitude
indicates the total angular momentum $J = \ell \pm s$. The amplitudes
in the total isospin $I=\frac{1}{2}$ and $I=\frac{3}{2}$ can be
recovered via the identification:
\begin{equation}
    f_{\ell \pm}^{\frac{1}{2}} = f_{\ell \pm}^+ + 2 f_{\ell \pm}^- \ \ , \ \ f_{\ell \pm}^{\frac{3}{2}} = f_{\ell \pm}^+ - f_{\ell \pm}^- \ .
\end{equation}
Below inelastic threshold the scattering amplitude is related to a unitary $S$-matrix by
\begin{equation}
    S_{\ell \pm}^I(s) = 1 + 2 i \lvert \, \vec{q} \, \rvert f_{\ell \pm}^I(s) \ \ , \ \ S_{\ell \pm}^I(s) S_{\ell \pm}^I(s)^{\dagger} = 1 
\end{equation}
and the $S$-matrix can be parameterized in terms of phase shifts,
\begin{equation}
    S_{\ell \pm}^I(s) = e^{2 i \delta_{\ell \pm}^I(s)} \ .
    \label{eq:pinsmatrix}
\end{equation}
For a more detailed derivation of the $\pi N$ $S$-matrix
see~\cite{Fettes_1998,Yao_2016,Moj_i__1998,Scherer2012}. Scattering in
a given partial wave and total angular momentum channel leads to a
$S$-matrix which acts on the product Hilbert space of
the nucleon and pion isospin, $\mathcal{H}_{\pi} \otimes
\mathcal{H}_{N}$. The $S$-matrix can then be written in terms of total
isospin projection operators
\begin{equation}
    \hat{{\bf S}}_{\ell \pm} = e^{2 i \delta_{\ell \pm}^{3/2}} \hat{{\bf P}}_{3/2} + e^{2 i \delta_{\ell \pm}^{1/2}} \hat{{\bf P}}_{1/2}
\end{equation}
where the $6 \times 6$ projection matrices are
\begin{equation}
    \begin{split}
        &\hat{{\bf P}}_{3/2} = \frac{2}{3} \left (\hat{{\bf 1}} + \hat{{\bf t}}_{\pi} \cdot \hat{{\bf t}}_N \right ) \ , \\
        &\hat{{\bf P}}_{1/2} = \frac{1}{3} \left (\hat{{\bf 1}} - 2(\hat{{\bf t}}_{\pi} \cdot \hat{{\bf t}}_N) \right )\ .
    \end{split}
\end{equation}
The operators $\hat{{\bf t}}_N$ and $\hat{{\bf t}}_{\pi}$ are in the
$2$ and $3$ dimensional representations of $SU(2)$ isospin
respectively and $\hat {\bm t}_{\pi} \cdot \hat {\bm t}_N =
\sum\limits_{\alpha=1}^3 \ \hat{ t}_{\pi}^\alpha \otimes \hat{ t}_N^\alpha$.

\subsection{Entanglement power}
\begin{figure}
    \centering
    \includegraphics[width = 1 \textwidth]{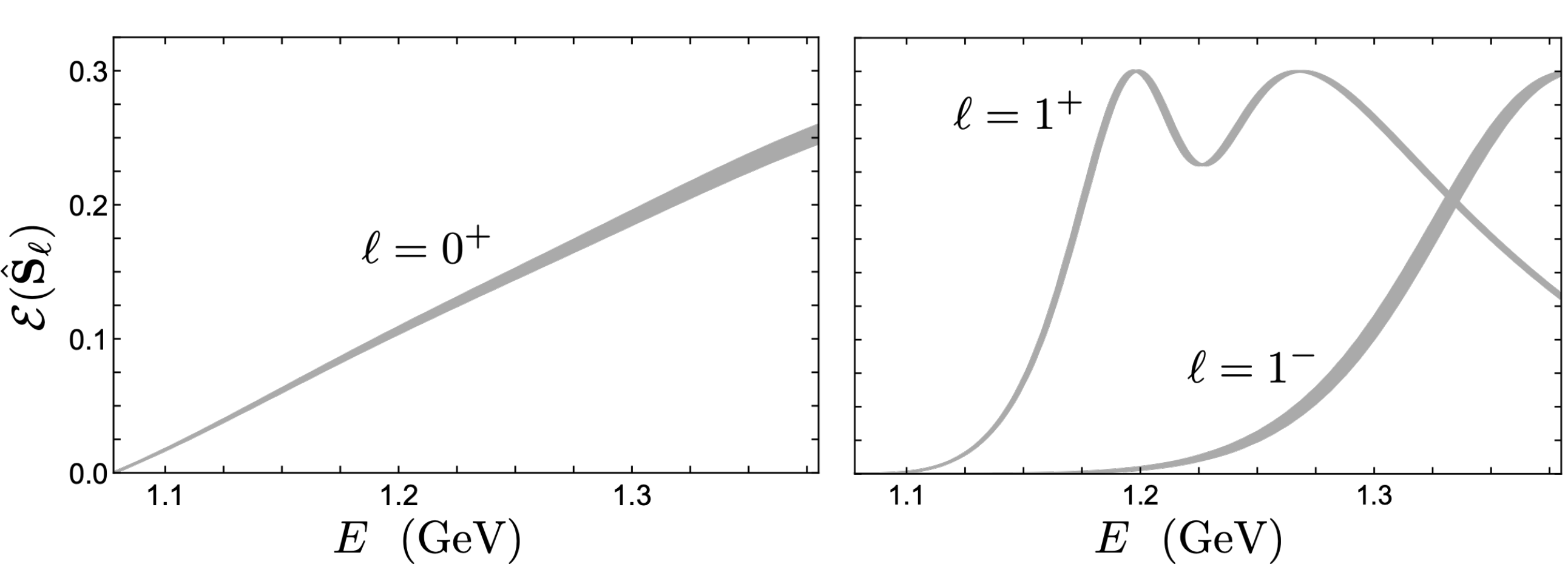}
    \caption{Entanglement power of the $\pi N$ $S$-matrix for $\ell = 0 , 1$ taken from Roy-Steiner equation determinations (the bands represent an estimate of the uncertainties~\cite{Hoferichter_2016}) of the $\pi N$ phase shifts.}
     \label{fig:piNEP}
\end{figure}

The entanglement power of the $\pi N$ $S$-matrix can be computed in a
similar manner as for the $\pi \pi$ EP. The incoming separable state
now maps to a point on the product manifold, ${\bf CP}^2 \times {\bf
  S}^2$. The construction of the reduced density matrix follows the
same steps as in section \ref{sec:pipiEP} and the entanglement power
is found to be,
\begin{equation}
\begin{split}
{\mathcal E}({\hat {\bf S}}_{\ell \pm}) &= \left ( \frac{2}{\pi^2}\frac{1}{4 \pi} \right ) \left (\int dV_{FS} d \Omega \right ) \mathcal{P} E_{\hat {\bf S}_{\ell \pm}} \\
&=
\frac{8}{243}  \bigg[ 17 + 10 \cos \left ( 2 \big ( \delta_{\ell \pm}^{{3/2}} - \delta_{\ell \pm}^{{1 / 2}}\big ) \right ) \bigg] \sin ^2\left (\delta_{\ell \pm}^{{3 / 2}} - \delta_{\ell \pm}^{{1 / 2}} \right )
\end{split}
\end{equation}
where $\mathcal{P}$ has been taken to be $1$.  Note that the two
particles are now distinguishable and so scattering in each partial
wave is no longer constrained by Bose/Fermi statistics. It follows
that the $S$-matrix is only non-entangling when it is proportional to
the identity which occurs when,
\begin{equation}
    \delta_{\ell \pm}^{{3 / 2}} = \delta_{\ell \pm}^{{1 / 2}} \ .
\end{equation}
Notice that the EP allows for interesting local minima when the difference in $I=3/2$ and $I=1/2$ phase shifts is $\pi/2$.
The $\pi N$ phase shifts are determined very accurately by the
Roy-Steiner equations up to a center-of-mass energy of 1.38
GeV~\cite{Hoferichter_2016} and the entanglement power for the first
couple partial waves is shown in Fig.~(\ref{fig:piNEP}).
There is a local minimum near the delta resonance position in the p-wave due to the rapid change of the
$I=3/2$ phase shift.

\subsection{Chiral perturbation theory}
Near threshold the phase shifts can be determined by the scattering lengths through the effective range expansion,
\begin{equation}
   \delta_{\ell \pm}^I = \cot^{-1}{\left \{ \frac{1}{\lvert \, \vec{q} \, \rvert ^{2 \ell +1} }\left ( \frac{1}{a_{\ell \pm}^I}  + \mathcal{O}(\vec{q}^{\, 2}) \right ) \right \} }\ .
\end{equation}
This leads to the threshold form of the entanglement power,
\begin{equation}
{\mathcal E}({\hat {\bf S}}_{\ell \pm}) = \frac{8}{9} \left (a^\frac{1}{2}_{\ell \pm}- a^\frac{3}{2}_{\ell \pm}\right )^2 \vec{q}^{\, 2+4l}
\end{equation}
which can only vanish if $a^\frac{1}{2}_{\ell \pm} = a^\frac{3}{2}_{\ell \pm}$.
The scattering lengths at leading order in heavy-baryon chiral perturbation theory, including the delta, are given by~\cite{Fettes_1998, Fettes_2001},
\begin{equation}
    \begin{split}
    &a^\frac{1}{2}_{0+} = \frac{2 M_{\pi} m}{8 \pi (m + {M_{\pi}}) F_{\pi}^2} \ \ , \ \ a^\frac{3}{2}_{0+} = \frac{- M_{\pi} m}{8 \pi (m + {M_{\pi}}) F_{\pi}^2} \\
    &a^\frac{1}{2}_{1-} = -\frac{m \left(9  g_A^2 \Delta +9 g_A^2 M_{\pi}-8 g_{\pi N \Delta}^2 M_{\pi}\right)}{54 \pi  F_{\pi}^2 M_{\pi} (\Delta+M_{\pi}) (m+M_{\pi})} \ \ , \ \ 
    a^\frac{3}{2}_{1-} =-\frac{m \left(9  g_A^2 \Delta +9 g_A^2 M_{\pi}-8 g_{\pi N \Delta}^2 M_{\pi}\right)}{216 \pi  F_{\pi}^2 M_{\pi} (\Delta+M_{\pi}) (m+M_{\pi})} \\
    &a^\frac{1}{2}_{1+} = \frac{m \left(-3  g_A^2 \Delta +3 g_A^2 M_{\pi}+8 g_{\pi N \Delta}^2 M_{\pi}\right)}{72 \pi  F_{\pi}^2 M_{\pi} (\Delta-M_{\pi}) (m+M_{\pi})} \ \ , \ \
    a^\frac{3}{2}_{1+} = \frac{m \left(-3 g_A^2 \Delta^2 -2  g_{\pi N \Delta}^2 M_{\pi} \Delta +3 g_A^2 M_{\pi}^2\right)}{36 \pi  F_{\pi}^2 M_{\pi} \left(M_{\pi}^2-\Delta^2\right) (m+M_{\pi})}
    \end{split}
\end{equation}
where $\Delta = m_{\Delta} - m_N$ is the delta-nucleon mass splitting. The corresponding EPs near threshold are,
\begin{equation}
\begin{split}
&{\mathcal E}({\hat {\bf S}}_{0+}) =  \frac{m^2 M_{\pi}^2}{8 \pi ^2 F_{\pi}^4 (m+M_{\pi})^2}\vec{q}^{\, 2} \\
&{\mathcal E}({\hat {\bf S}}_{1-}) = \frac{m^2  \left(9  g_A^2 \Delta +9 g_A^2 M_{\pi}-8 g_{\pi N \Delta}^2 M_{\pi}\right)^2}{5832 \pi ^2 f_{\pi}^4 M_{\pi}^2 (\Delta+M_{\pi})^2 (m+M_{\pi})^2} \vec{q}^{\, 6}\\
&{\mathcal E}({\hat {\bf S}}_{1+}) = \frac{m^2  \left(-9  g_A^2 \Delta^2 +4  g_{\pi N \Delta}^2 \Delta M_{\pi}+ \left(9 g_A^2+8 g_{\pi N \Delta}^2\right) M_{\pi}^2 \right)^2}{5832 \pi ^2 f_{\pi}^4 M_{\pi}^2 \left(M_{\pi}^2-\Delta^2\right)^2 (m+M_{\pi})^2} \vec{q}^{\, 6}\ .
\end{split}
\end{equation}
Once again the only non-entangling solution consistent with chiral
symmetry is no interaction, with the same scaling of $F_\pi$ as found in $\pi\pi$ scattering.
Unlike the large-$N_c$ limit, there is no reason to expect an enhancement of the axial couplings,
which in that case gives rise to the contracted spin-flavor symmetries~\cite{Dashen:1993as}.

\section{Discussion}
\label{sec:discuss}

\noindent In QCD the number of colors, $N_c$, is a parameter that
appears in the action and in some sense acts as a knob that dials the
amount of quantum correlation in the hadronic $S$-matrix.  The
simplifications, counting rules and enhanced symmetries implied by the
large-$N_c$ approximation have proved highly successful in explaining
regularity in the hadronic spectrum. Recent work in
Ref.~\cite{Beane:2018oxh} has conjectured that, independent of the
value of $N_c$, quantum entanglement is minimized in hadronic
$S$-matrices. Verifying this conjecture relies on finding consequences
of the conjecture that are distinct from large-$N_c$ predictions, and
indeed this has been found to be the case in baryon-baryon scattering.  In
particular, minimization of entanglement near threshold leads to
enhanced symmetry that is verified by lattice QCD simulations.  Here
this conjecture has been considered for $\pi\pi$ and $\pi N$
scattering.  As shown long ago by Weinberg, the scattering of soft
pions off any target is completely determined by chiral
symmetry~\cite{PhysRevLett.17.616} and is weak at low energies.  Here
it has been found that the only $\pi \pi$ or $\pi N$ $S$-matrix,
consistent with the low energy theorems, that does not entangle
isospin is the identity i.e. no scattering. In the context of chiral
perturbation theory this corresponds to $F_{\pi}$ being large when
entanglement is minimized, consistent with large $N_c$ scaling. Unlike
in the large $N_c$ limit, entanglement minimization of the $S$-matrix
says nothing about the scaling of the baryon masses and axial couplings and therefore
implies no new symmetries in the $\pi N$ sector.  Because of the
weakness of pion processes implied by chiral symmetry, it may be the
case that only systems without external Goldstone bosons (like $NN$)
give non-trivial constraints from entanglement minimization.
Considering general meson-nucleon scattering, it is clear that
scalar-isoscalar mesons have no spin or isospin to entangle.
Insofar as resonance saturation is effective, entanglement minimization would then predict the contribution to baryon-baryon scattering from
the exchange of non scalar-isoscalar resonances to sum together to give an equal
contribution to all spin-isospin channels~\cite{Epelbaum_2002}. This would then naturally lead to the $SU(16)$ symmetry seen in the three flavor baryon
sector~\cite{Beane:2018oxh}.

\chapter{Geometric Scattering Theory and UV/IR Symmetries of the S-matrix}
\label{chap:SmatGeom}
\noindent 
\textit{
This chapter is associated with ``Geometry and entanglement in the scattering matrix"~\cite{Beane:2020wjl} and
``Causality and dimensionality in geometric scattering"~\cite{Beane:2021B}
by Silas R. Beane and Roland C. Farrell.}
\section{Introduction}

\noindent The scattering matrix ($S$-matrix) encodes observable
consequences of the quantum mechanical interaction of two particles.
Typically, the $S$-matrix is determined by solving an effective quantum field theory (EFT) describing particles interacting through a local potential.
In this EFT-based approach, spacetime constraints like Galilean invariance and causality are encoded in the dependence of the scattering on the external kinematics e.g. the center of mass (c.o.m.) momentum ${\vec p}$ and energy $E$.
The framework provided by EFTs is extremely powerful, and has enabled precision calculations of observables, with quantifiable uncertainties.
However, keeping spacetime constraints manifest can obscure features of scattering that are non-local or mix different length scales. 
This chapter develops an alternative, geometric, formulation of scattering that is better suited for exploring these non-local aspects of scattering.
The geometric construction is applied to
the non-relativistic s-wave scattering of spin-1/2 fermions. 
Identifying these fermions with neutrons and protons describes the low-energy EFT of the Standard Model, and has widespread and important applications in nuclear physics.\footnote{See
  Ref.~\cite{Beane:2000fx,Hammer:2019poc} for a review.}
  
A new symmetry is found that relates scattering at low and high energies (a UV/UR symmetry).
This symmetry only exists for phase shifts parameterized by scattering lengths and effective ranges, and may explain why shape parameters in nuclear physics are anomalously small.\cite{Cohen:1998jr,deSwart:1995ui}
This geometric construction also provides a new perspective on how causality constrains a scattering process, and how entanglement is generated via scattering.

An elastic scattering process is naturally parameterized by (possibly multiple) momentum-dependent phase shift(s) $\delta_s(p)$ which ensures that the $S$-matrix is a unitary operator e.g. for single channel scattering
$\hat{S} = e^{i 2 \delta(p)}$.
In the geometric formulation of scattering, the phase shifts are used as a coordinate basis.
Due to the $2\pi$ periodicity of each $2\delta_s(p)$, this space has the topology of a torus.
The momentum-dependent $S$-matrix then emerges as a trajectory, parameterized by the momentum, through the torus.
In the geometric construction, these trajectories are generated from a combination of the intrinsic curvature of the space and ``external" forces that permeate the torus.

It will be shown that there exist special $S$-matrix trajectories that have a symmetry that maps low- and high-energy scattering processes into each other (a UV/IR symmetry).
These symmetries are hidden at the level of the EFT action (see Chapter~\ref{chap:SmatUVIR}), but manifest as reflection symmetries of the $S$-matrix trajectories.
  These UV/IR symmetries leave classes of
observables invariant and will also be referred to as a ``conformal symmetries" since they leave various combinations of the phase shifts unchanged. 
The UV/IR symmetries also allow for the determination of exact solutions for the external forces needed to generate the $S$-matrix trajectories.
These exact solutions provide a forum for the exploration of
spacetime constraints on the $S$-matrix in the geometric formulation
of scattering.

The $S$-matrix evolves an initial state from the boundary of space in
the infinite past, to the boundary of space in the infinite future,
and must do so in a manner consistent with causality and with
awareness of the number of spatial dimensions in which it is acting.
Constraints due to causality on non-relativistic scattering have
implications for the analytic structure of the
$S$-matrix~\cite{PhysRev.74.131,PhysRev.83.249,PhysRev.91.1267} and,
for systems arising from finite-range potentials, for the range of
allowed values of the scattering
parameters~\cite{Wigner:1955zz,Phillips:1996ae,Hammer:2009zh,Hammer:2010fw}.
These bounds, known as Wigner bounds, provide powerful constraints on
the exact $S$-matrix solutions implied by the UV/IR symmetries. In the
geometric theory it is found that these bounds manifest themselves as
constraints on the tangent vectors of $S$-matrix trajectories. 
 In addition, as quantum mechanics depends strongly on spatial
dimensionality, the differences between scattering in two and 
three dimensions are explored in the geometric formulation. 
The resulting $S$-matrix in two spatial dimensions again
has a solution implied by the UV/IR symmetries. Despite the strikingly
different physics that it gives rise to, the form of the
two-dimensional external potential differs from its three-dimensional
counterpart only by a change of coupling strength and phase.

This chapter is organized as follows.  Section~\ref{sec:smatth} sets up
the $S$-matrix framework, focusing on the properties of the most
general $S$-matrix consistent with finite-range forces. The $S$-matrix
is shown to allow conformal symmetries that are not manifest in the
EFT action, and which provide powerful geometric constraints. In
Section~\ref{sec:geom}, the $S$-matrix of contact forces is shown to
be the solution of a dynamical system which evolves the two-particle
state in a two-dimensional space defined by the two phase shifts and
bounded by unitarity. The conformal symmetries allow an exact
determination of the forces that determine the $S$-matrix in this
space. Section~\ref{sec:stcon} explores the manner in which spacetime features of
scattering manifest themselves in the geometric theory of
scattering. Constraints due to causality are considered, and the
dependence on spatial dimensionality is found by varying between three
and two dimensions.  
Finally, Section~\ref{sec:conc} summarizes and concludes.

\section{S-matrix theory}
\label{sec:smatth}

\subsection{S-matrix theory of contact forces}

\noindent It is a simple matter to write down the $S$-matrix without
reference to any underlying field theory by directly imposing general
physical principles and symmetries. Consider two species of
equal-mass, spin-$1/2$ fermions, which we label as neutrons and
protons (i.e. nucleons), that interact at low energy via forces that are
strictly of finite range. The spins of the two-body system can be
either aligned or anti-aligned. Therefore, near threshold the
$S$-matrix is dominated by the s-wave and can be written as~\cite{Beane:2018oxh}
\begin{eqnarray}
  \hat {\bf S}(p)
  & = &
\frac{1}{2}\left(  e^{i 2 \delta_1(p)} + e^{i 2 \delta_0(p)} \right)
 \hat   {\bf 1}
\ +\
\frac{1}{2}\left( e^{i 2 \delta_1(p)} - e^{i 2 \delta_0(p)} \right)
\hat{{\cal P}}_{12}
  \label{eq:Sdef1}
\end{eqnarray}
where the SWAP operator is
\begin{eqnarray}
\hat{{\cal P}}_{12} = \oneht \left({ \hat   {\bf 1}}+  \hat  {\bm \sigma} \cdot   \hat  {\bm \sigma} \right)
  \label{eq:swap}
\end{eqnarray}
and, in the direct-product space of the nucleon spins,
\begin{eqnarray}
 \hat {\bf 1} \equiv \hat {\cal I}_2\otimes  \hat {\cal I}_2 \ \ \ , \ \
  \hat {\bm \sigma} \cdot \hat {\bm \sigma} \equiv \sum\limits_{\alpha=1}^3 \
\hat{{\sigma}}^\alpha \otimes \hat{{ \sigma}}^\alpha \ ,
  \label{eq:Hil1}
\end{eqnarray}
where ${\cal I}_2$ is the $2\times 2$ unit matrix, and the $\hat{{
    \sigma}}^\alpha$ are the Pauli matrices. The $\delta_s$ are s-wave
phase shifts with $s=0$ corresponding to the spin-singlet ($\si$)
channel and $s=1$ corresponding to the
spin-triplet ($\siii$) channel.

The initial state for a scattering process between two particles is unentangled, and can be written as a tensor product.
When the $S$-matrix is the identity or SWAP operator it maps tensor product states into tensor product states, and therefore does not generate any spin entanglement. 
In general, the $S$-matrix is an
entangling operator as the total-spin basis which diagonalizes the interaction is
different from the single-particle basis which describes the initialstate.  Therefore, it is 
imperative to treat the $S$-matrix as the fundamental object of study,
rather than the EFT action or the scattering amplitude, when
addressing issues related to quantum entanglement.

The entangling character of the $S$-matrix can be captured by its
entanglement power (EP)~\cite{Beane:2018oxh,Beane:2020wjl,Low:2021ufv}
\begin{eqnarray}
{\mathcal E}({\hat {\bf S}})
& = &
\frac{1}{6}\ \sin^2\left(2(\delta_1-\delta_0)\right)
\ ,
  \label{eq:epSnf21}
\end{eqnarray}
which is a state-independent measure of the entanglement generated by
the $S$-matrix acting on an initial product state. Note that this
object manifestly couples the two spin states in a manner that is
quite distinct from the Lorentz-invariant, spin decoupled interactions
that are encoded in the EFT action. Indeed, when it vanishes there is
an enhanced $SU(4)$ spin-flavor symmetry (Wigner's supermultiplet
symmetry~\cite{Wigner:1936dx,Wigner:1937zz,Wigner:1939zz}) which
explicitly relates the singlet and triplet scattering channels.

The two angular degrees of freedom, the phase shifts $\delta_{0,1}$,
are characterized by the effective range
expansion (ERE)
\begin{eqnarray}
  p\cot\delta_s(p) & =& -\frac{1}{a_s} \ +\ \oneht r_s p^2  \ +\ v_{2;s} p^4 \ + \ {\mathcal O}(p^6)
  \label{eq:EFT1}
\end{eqnarray}
with ${\vec p}$ (with $p=|{\vec p}\,|$) chosen to be the center-of-mass (c.o.m.) momentum.
Here, $a_{0,1}$, $r_{0,1}$ and $v_{2;0,1}$ are the scattering length, effective range and first shape parameter for the spin singlet ($0$) and triple ($1$) channels.
The phase shift parameterizes the scattering amplitude
\begin{equation}
T_s(p)\ =\ -\frac{4 \pi}{M}
\left[ p\cot\delta_s(p) - i p \right]^{-1},
\label{eq:reamp}
\end{equation}
which is related to the $S$-matrix element by
\begin{equation}
  S_s(p)\ =\ e^{2 i \delta_s(p)}\ =\ 1 \ - \ i \frac{p M}{2\pi} T_s(p) \ .
   \label{eq:reampbS}
\end{equation}

Near threshold, the $S$-matrix can be written as
\begin{eqnarray}
  \hat {\bf S} & = & \frac{1}{2}\left( S_1+ S_0 \right) \;\hat {\bf 1} \;+\; \frac{1}{2}\left( S_1- S_0 \right)\;\hat{{\cal P}}_{12} \ ,
   \label{eq:Sdef1res}
\end{eqnarray}
where the $S$-matrix elements are 
\begin{eqnarray}
S_s &=& e^{2i\delta_s(p)} \ =\ \frac{1- i a_s(p) p}{1+ i a_s(p) p} \ ,
   \label{eq:Selementscatt}
\end{eqnarray}
and a momentum-dependent scattering length is defined as
\begin{eqnarray}
a_s(p) \ \equiv\ \frac{a_s}{1-\oneht a_s r_s p^2 + {\mathcal O}(p^4)} \ .
   \label{eq:mdepscatt1}
\end{eqnarray}
The momentum dependent scattering lengths are related to the phase shifts as
\begin{eqnarray}
\phi & \equiv& 2\delta_0\ =\ -2\tan^{-1}\!\left( a_0(p) p \right) \ \ \ , \ \ \ \theta\ \equiv\ 2\delta_1\ =\ -2\tan^{-1}\!\left( a_1(p) p \right) \ .
 \label{eq:confRmodSOL1}
\end{eqnarray}
Here the phase shifts have been expressed in terms of the angular variables $\phi\in [0,2\pi]$ and $\theta\in [0,2\pi]$.

\begin{figure}[!ht]
\centering
\includegraphics[width = 0.45\textwidth]{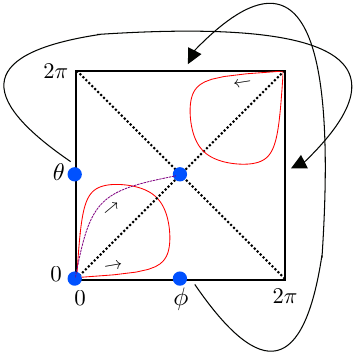}
\caption{The flat-torus manifold on which the phase shifts
  propagate. The blue dots correspond to the four fixed points of the
  RG. The dotted lines forming the diagonals of the square are axes of
  symmetry for $S$-matrix trajectories. The purple dashed curve is an example of an
  $S$-matrix trajectory in the scattering length approximation corresponding to the fourth
  row of Table~\ref{tab:LOsymclass} (with
  $a_1,a_0 >0$ and $a_1/a_0=5$). The red curves are examples of
  $S$-matrix trajectories with range parameter given in the fifth row
  of Table~\ref{tab:genconfrm}.  These trajectories leave
  $\phi+\theta$ invariant (with $a_1,a_0 <0$ (bottom left quadrant)
  and with $a_1,a_0 >0$ (top right quadrant), $a_0/a_1=15$ and
  $\lambda=0.01$).}
  \label{fig:ftbase}
\end{figure}

The $S$-matrix of Eq.~(\ref{eq:Sdef1res}) is specified by the two
angular variables $\phi(p)$ and $\theta(p)$ that are determined by the
Schr\"odinger equation once the finite-range quantum mechanical
potential is specified. As these variables are periodic,
the two-dimensional ``phase space'' that these variables define is a
flat torus manifold, illustrated in Fig.~(\ref{fig:ftbase}). The range
of values that $\phi(p)$ and $\theta(p)$ can take are bounded by
unitarity, with boundary values determined by the four renormalization group (RG) fixed points,
which occur at $\hat {\bf S} = \pm \hat {\bf 1}$ and $\pm \hat{{\cal
    P}}_{12}$ when the s-wave scattering lengths are either vanishing
or infinite (at unitarity)~\cite{Beane:2018oxh,Beane:2020wjl}.  Generally, in
effective range theory, the $S$-matrix trajectory on the flat torus
will originate at the trivial fixed point at scattering threshold and
trace out a curve that exits the flat-torus and enters a bulk space at
the first inelastic threshold~\cite{Beane:2020wjl}. In what follows,
all inelasticities will be pushed to infinite momentum and $S$-matrix
trajectories will begin and end at an RG fixed point.

\subsection{UV/IR symmetries of the $S$-matrix}
\label{sec:UVIRGeneral}

\subsubsection*{Out-state density matrix}

\noindent In this section $S$-matrices with a momentum inversion
symmetry that interchanges the IR and the UV will be studied. The $S$-matrix is
defined as the operator which evolves the incoming (``in'') state
before scattering into the outgoing (``out'') state after scattering,
i.e. $\hat {\bf S} \lvert \text{in} \rangle = \lvert \text{out}
\rangle$.  The criterion for the presence of a symmetry will be based on the
transformation properties of the density matrix of the ``out'' state
\begin{equation}
    \rho = \lvert \text{out} \rangle \langle \text{out} \rvert = \hat {\bf S} \lvert \text{in} \rangle \langle \text{in} \rvert \hat {\bf S}^{\dagger} \ .
\end{equation}
The first kind of symmetry transformation that will be considered is
$\rho \mapsto \rho$ which leaves all spin-observables invariant. In
the following sections it will be shown that there are non-trivial
instances of this symmetry where the $S$-matrix itself is not
invariant. The second kind of symmetry transformation that will be considered
is $\rho \mapsto \bar{\rho}$ where
\begin{equation}
    \bar{\rho} \equiv \hat {\bf S}^{*} \lvert \text{in} \rangle \langle \text{in} \rvert \hat {\bf S}^T \ .
\end{equation}
This is the density matrix that would be produced if all
phase shifts change sign, or, equivalently, if attractive
interactions are replaced with repulsive interactions of equal magnitude, and
vice-versa.

\subsubsection*{The scattering-length approximation}

\noindent The $S$-matrix takes constant values at the fixed points of
the RG, which implies that at both the trivial and unitary fixed
points, the underlying EFT has non-relativistic conformal invariance
(Sch\"odinger symmetry).  The scattering trajectories are bridges
which connect various conformal field theories. In addition to this
conformal invariance of the EFT action at the RG fixed points, there is a UV/IR symmetry
which acts directly on the $S$-matrix and which is present at finite
values of the scattering lengths. 

First, consider the scattering length approximation defined by $a_s\neq0$ and $r_s=0$.
The momentum inversion map
\begin{eqnarray}
p\mapsto \frac{1}{|a_1 a_0| p} \ ,
   \label{eq:moebius2}
\end{eqnarray}
transforms the phase shifts and density
matrices as shown in Table~\ref{tab:LOsymclass}.
As this takes $p=0$ to $p=\infty$ it is a transformation which interchanges the IR and UV.
This momentum inversion
is a conformal invariance that leaves the combination of angular
variables $\phi+\theta$ ($a_1 a_0 <0$) or $\phi-\theta$ ($a_1 a_0 >0$)
invariant and implies a reflection symmetry of the $S$-matrix
trajectory, as illustrated for a specific case in
Fig.~(\ref{fig:ftbase}) (purple curve). When $a_1 a_0 > 0$, the two
phase shifts conspire to leave the density matrix
unchanged despite neither phase shift separately being invariant.
This demonstrates that, in multi-channel scattering, there
exist symmetries which are not manifest at the level of the scattering
amplitude and appear as reflections of $S$-matrix trajectories.  It
is notable that the EP is invariant with respect to the
transformation of Eq.~(\ref{eq:moebius2})~\cite{Beane:2020wjl}.
\begin{table}[h!]
\centering
\begin{tabularx}{0.8\textwidth} { 
  | >{\raggedright\arraybackslash}X 
  | >{\centering\arraybackslash}X 
  | >{\centering\arraybackslash}X 
    | >{\centering\arraybackslash}X 
  | >{\raggedleft\arraybackslash}X | }
 \hline
  $\phi\mapsto$  & $\theta\mapsto$ & $\rho\mapsto$ & $a_0$ &  $a_1$  \\
\hlineB{3}
 $-{\pi} + \theta$  & ${\pi} + \phi$ & $\bar{\rho}$ & $+$ & $-$  \\
  \hline
${\pi} + \theta$  & $-{\pi} + \phi$ & $\bar{\rho}$ & $-$ & $+$ \\
\hline
${\pi} - \theta$  & ${\pi} - \phi$ & $\rho$ & $-$ & $-$ \\
 \hline
$-{\pi} - \theta$  & $-{\pi} - \phi$ & $\rho$ & $+$ & $+$ \\
\hlineB{2}
\end{tabularx}
  \caption{Action of the momentum inversion transformation in the scattering length approximation.}
 \label{tab:LOsymclass}
\end{table}

\subsubsection*{Including effective ranges}

\noindent The scattering length approximation is a part of
a larger class of UV/IR symmetric $S$-matrix models which include range effects
that are strictly correlated with the scattering lengths. These
UV/IR symmetries have a distinct character as the range effects
necessarily arise from derivative operators in the EFT.
Consider the general momentum inversion
\begin{eqnarray}
p\mapsto \frac{1}{\lambda |a_1 a_0| p} \ ,
   \label{eq:moebiusgen}
\end{eqnarray}
where the real parameter $\lambda >0$. One can ask: what is the most
general $S$-matrix for which this inversion symmetry gives rise to
interesting symmetries? This transformation rules out all
shape-parameter effects and correlates the effective ranges with
the scattering lengths in a specific way. Table~\ref{tab:genconfrm}
gives the effective-range parameters for all $S$-matrix models with
a symmetry under the momentum inversion of Eq.~(\ref{eq:moebiusgen}).
Note that the first four rows correlate the singlet
effective range with the triplet scattering length and vice-versa.
\begin{table}[h!]
\centering
\begin{tabularx}{0.8\textwidth} { 
  | >{\raggedright\arraybackslash}X 
  | >{\centering\arraybackslash}X 
  | >{\centering\arraybackslash}X 
    | >{\centering\arraybackslash}X 
  | >{\raggedleft\arraybackslash}X | }
 \hline
  $\phi\mapsto$  & $\theta\mapsto$ & $\rho\mapsto$ & $r_0$ & $r_1$ \\
\hlineB{3}
 $\phi$  & $\theta$ & $\rho$ & $-2\eta/a_0$ & $-2\eta/a_1$ \\
\hline
$\phi$  & $-\theta$ & $\rho_-  + \bar{\rho}_+$ & $-2\eta/a_0$ & $+2\eta/a_1$ \\
 \hline
$-\phi$  & $\theta$ & $\rho_+ + \bar{\rho}_- $ & $+2\eta/a_0$ & $-2\eta/a_1$ \\
 \hline
$-\phi$  & $-\theta$ & $\bar{\rho}$ & $+2\eta/a_0$ & $+2\eta/a_1$ \\
\hlineB{2}
$\theta$  & $\phi$ & $\bar{\rho}$ & $-2\eta/a_1$ & $-2\eta/a_0$ \\
 \hline
$-\theta$  & $-\phi$ & $\rho$ & $+2\eta/a_1$ & $+2\eta/a_0$ \\
 \hline
\end{tabularx}
  \caption{Action of the momentum-inversion transformation on models that have a non-zero effective range. Here $\eta \equiv \lambda \lvert a_0 a_1 \rvert$ and $\rho_{\pm}$ is the density matrix projected onto the total spin $\frac{1}{2} \left ( 1 \pm 1 \right )$ subspace.}
 \label{tab:genconfrm}
\end{table}

An interesting and well-known feature of the effective-range expansion
in nucleon-nucleon (NN) scattering is the smallness of the shape parameter
corrections (see, for instance, Ref.~\cite{Cohen:1998jr,deSwart:1995ui}) as compared
to the range of the interaction, which is given roughly by the Compton
wavelength of the pion. Vanishing shape corrections is a key signature
of an $S$-matrix with momentum-inversion symmetry. Indeed, as the
NN s-wave effective ranges are positive while the
scattering lengths have opposite sign, the model given in the second
row of Table~\ref{tab:genconfrm}, with $\lambda$ fitted to the data,
provides a description of the low-energy s-wave phase shifts that
improves upon the scattering-length approximation. As will be seen
below, models that arise from zero-range forces with exact
momentum-inversion symmetry and a positive effective range strictly
violate causality. However, relaxing the zero-range condition
can lead to interesting results for nuclear physics, as 
is considered in Chapter~\ref{chap:Weinberg}.

\section{Geometric scattering theory}
\label{sec:geom}

\subsection{Metric on the flat-torus}

\noindent As the space on which the two phase shifts, $\theta$ and $\phi$,  propagate is a two-dimensional flat space,
the line element should take the form $ds^2 \propto d\phi^2\;+\;  d\theta^2$. This metric
can be obtained formally by parameterizing the $S$-matrix of Eq.~(\ref{eq:Sdef1res}) as
\begin{eqnarray}
  \hat {\bf S}   & = & \big\lbrack x(p)\;+\; i\;y(p) \big\rbrack    \hat   {\bf 1}
\ +\
\big\lbrack z(p)\;+\; i\;w(p) \big\rbrack \hat{{\cal P}}_{12} \ ,
   \label{eq:Sdef1gen}
\end{eqnarray}
with
\begin{eqnarray}
  x & =&   \oneht\; \lbrack \cos(\phi) + \cos(\theta) \rbrack \ \ , \ \ 
  y \ =\   \oneht\; \lbrack \sin(\phi) + \sin(\theta) \rbrack\ \ , \ \ \nn \\
  z & =&   \oneht\; \lbrack -\cos(\phi) + \cos(\theta) \rbrack\ \ , \ \ 
  w \ =\   \oneht\; \lbrack -\sin(\phi) + \sin(\theta) \rbrack \ .
   \label{eq:PS1}
\end{eqnarray}
Then, as an embedding in $\mathbb{R}^4$, with line
element
\begin{eqnarray}\;
ds^2 \ =\ dx^2\;+\;dy^2\;+\;dz^2\;+\;dw^2 \ ,
   \label{eq:R4metricgen1}
\end{eqnarray}
one finds the flat two-dimensional Euclidean line element
\begin{eqnarray}\;
ds^2 \ =\ \oneht\;\left( d\phi^2\;+\;  d\theta^2\right)\ .
   \label{eq:T2inR41}
\end{eqnarray}
With $\phi$ and $\theta$ periodic, the corresponding metric describes the flat torus $\mathbb{T}^2\sim S^1 \times S^1\hookrightarrow \mathbb{R}^4$, where $S^1$
is the circle. From this line element, one can read off the flat-torus metric tensor $g_{ab}$, with $a,b=1,2$.
For a basis independent construction of this geometry see App.~\ref{app:geom}.

\subsection{Geometric action} \label{sec:geoact}

\noindent The action for a general parameterization of a curve on a space with
coordinates $\mathcal{X}^1=\phi$ and $\mathcal{X}^2=\theta$ and metric tensor $g_{ab}$
can be expressed as~\cite{garay2019classical}
 \begin{equation}
 \int L\left(\mathcal{X},\dot{\mathcal{X}} \right)d\sigma \ =\ \int \left({\bf{N}}^{-2}g_{ab} \dot{\mathcal{X}}^a \dot{\mathcal{X}}^b \ -\  \mathbb{V(\mathcal{X})} \right){\bf{N}}d\sigma \ ,
    \label{eq:action1}
 \end{equation}
 where $\sigma$ parameterizes the curve (affine or inaffine), $L$ is the Lagrangian,
 $\dot{\mathcal{X}}\equiv{d\mathcal{X}}/{d\sigma}$, and
 $\mathbb{V(\mathcal{X})}$ is an external geometric potential which is assumed
 to be a function of $\mathcal{X}$ only. The corresponding
 Euler-Lagrange equations give the trajectory equations
 \begin{equation}
\ddot{\mathcal{X}}^a \ +\ {}_g\Gamma^a_{\ bc} \dot{\mathcal{X}}^b \dot{\mathcal{X}}^c \ =\ \kappa(\sigma) \dot{\mathcal{X}}^a
   \ -\ \oneht {\bf{N}}^2 g^{ab}  \partial_b\mathbb{V(\mathcal{X})} \ ,
    \label{eq:exEOMfromact1}
 \end{equation}
 where ${}_g\Gamma^a_{\ bc}$ are the Christoffel symbols for the metric $g_{ab}$, and
 \begin{equation}
\kappa(\sigma) \ \equiv \ \frac{\dot{\bf{N}}}{\bf{N}} = \frac{d}{d\sigma}\ln \frac{d\tau}{d\sigma} \ .
    \label{eq:inaffinity1}
 \end{equation}
 Here $\kappa$ is the inaffinity~\cite{blau2020}, which vanishes when $\sigma=\tau$ with $\tau$ an affine parameter. 
 
 An interesting feature of the geometric construction of scattering is that the relative momentum
 that describes the motion of the center-of-mass is a non-affine parameter. For a constant geometric potential, the trajectory equations reduce to the geodesic equations
 which describe straight-line trajectories on the flat-torus. Any curvature indicates the presence of a non-constant geometric potential. Now, {\it a priori}, if a solution for $\phi$ and $\theta$ is specified,
 there are two equations of motion for three unknowns, the inaffinity and the two force components in the $\phi$ and $\theta$ directions. However, the presence of UV/IR symmetries can reduce the number of
 unknowns to two and thus allows an explicit construction of the geometric potential~\cite{Beane:2020wjl}.

\subsection{Solvable models}

\noindent In the scattering length approximation, the UV/IR symmetry of Eq.~(\ref{eq:moebius2}) determines the geometric potential exactly. It is given by~\cite{Beane:2020wjl}
\begin{eqnarray}
\mathbb{V}(\phi,\theta) & = &  \, \frac{|a_0 a_1|}{\left(|a_0|+|a_1| \right)^2c_1^2} \tan^2\left(\oneht(\phi+\epsilon\,\theta)\right) \ ,
\label{eq:entPOT1} 
\end{eqnarray}
where $\epsilon\,=-1$ for $a_1 a_0 >0$ and $\epsilon\,=+1$ for $a_1 a_0 <0$, and $c_1$ is an
integration constant. The inaffinity associated with a trajectory parameterized by the c.o.m momentum is constructed from
\begin{eqnarray}
{\bf{N}} &=& \frac{c_1}{p}\left(\sin\phi -\epsilon \sin\theta \right) \ .
\label{eq:Nfun} 
\end{eqnarray}
The $S$-matrix trajectory is independent of the parameterization and can be simply described
---with vanishing inaffinity and choice ${\bf{N}}={c_1}=1$--- by an affine parameter $\tau$ 
via the simple Lagrangian
\begin{eqnarray}
L &=& \oneht \left( \dot{\phi}\,+\,\dot{\theta} \right) \ -\ \mathbb{V}(\phi,\theta) \ ,
\label{eq:LagAFFLO1}
\end{eqnarray}
where the dots denote differentiation with respect to $\tau$. Of course $\tau$ has no interpretation as a momentum or energy in a scattering
process; such a parameter is not affine.

The conformal $S$-matrix models with the UV/IR symmetry of Eq.~(\ref{eq:moebiusgen}), and $\lambda$-dependent effective ranges, also lead to solvable geometric
potentials. The general solution is cumbersome, however in the special case where the effective ranges
are correlated to the scattering lengths as in the last two rows of Table~\ref{tab:genconfrm} ---and $\lambda=1/4$--- the geometric potential is identical to Eq.~(\ref{eq:entPOT1}) except for an overall factor of $1/2$ and a rescaled argument
\begin{eqnarray}
\oneht(\phi+\epsilon\,\theta) & \longrightarrow & \onefourth(\phi+\epsilon\,\theta) \ .
\label{eq:entPOT1lam14} 
\end{eqnarray}
It will be seen in section~\ref{sec:causalsing} why this case is special.

\section{Spacetime constraints}
\label{sec:stcon}

\subsection{Galilean invariance}

\noindent The parameter $p$, which labels the c.o.m. momentum of the scattering process,
is related to the total energy, $E$, in the system by $p =
\sqrt{M E}$. If the incoming particle momenta are ${\vec p}_1$ and
${\vec p}_2$ then in the c.o.m. frame, ${\vec p}_1 = - {\vec p}_2
\equiv \vec{\underline{p}}$ and $p = \lvert \, \vec{\underline{p}} \,
\rvert$. Other Galilean frames can be reached from the c.o.m. frame
via a combined rotation ${\bf{\cal R}}$ and boost by a velocity ${\vec
  v}$:
\begin{eqnarray}
  \vec{\underline{p}} \ && \longrightarrow \ {\bf{\cal R}} \, \vec{\underline{p}} + M {\vec v}
 \label{eq:Galilean}
\end{eqnarray}
which implies the transformation
\begin{eqnarray}
  p \ && \longrightarrow \  \lvert \, \vec{\underline{p}} \, \rvert \, \sqrt{1+{\bf x}} \ ,
 \label{eq:Galilean2}
\end{eqnarray}
with ${\bf x}\equiv {(M {\vec v\,})^2}/{{\vec {\underline{p}}}^2}$. Varying ${\bf x}$ between $0$ and $1$ corresponds to transforming
between the c.o.m. and laboratory frames. Hence, Galilean invariance allows arbitrary reparameterizations of the $S$-matrix
of the form $p \rightarrow \Omega \, p$ with $1\leq \Omega \leq\infty$ and $\Omega$ interpolating between the c.o.m at rest
and boosted to infinite momentum. As this is just a rescaling of $p$, changing to another inertial frame does not affect the inaffinity, Eq.~(\ref{eq:inaffinity1}).

\subsection{Causality bounds on zero-range scattering}

\subsubsection*{Wigner bounds}

\begin{figure}[!ht]
\centering
\includegraphics[width = 0.55\textwidth]{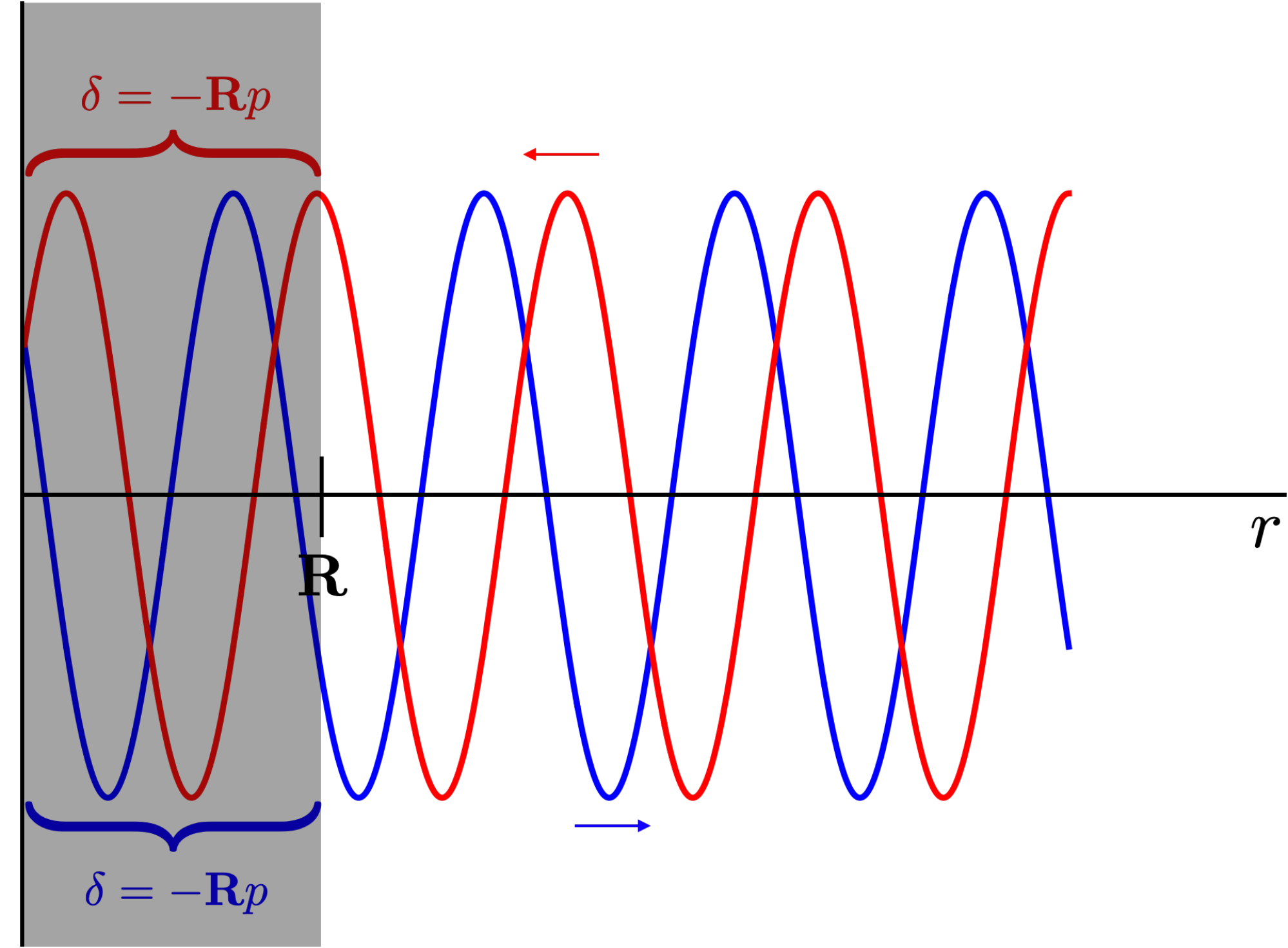}
\caption{An illustration of the minimum phase shift for scattering off a potential of range ${\bf R}$. The red (blue) curve represents the phase of the incoming (outgoing) s-wave spherical wave.}
  \label{fig:ClassWigBound}
\end{figure}
\noindent In non-relativistic scattering with finite-range forces, causality places bounds on
physical scattering parameters by way of Wigner
bounds~\cite{Wigner:1955zz,Phillips:1996ae,Hammer:2009zh,Hammer:2010fw}.
Consider a two-body s-wave wave function 
both free and in the presence of an interaction potential of range
${\bf R}$. The difference in phase between the scattered and free spherical wave is
defined to be twice the phase shift. The most negative phase shift is
obtained when the scattered wave does not penetrate the potential 
and reflects off the boundary at $r = {\bf R}$. This provides a
lower bound on the phase shift $\delta(p) \geq - {\bf R} p$, see
Fig.~(\ref{fig:ClassWigBound}). Now consider a plane wave at an
infinitesimally larger momentum, $\bar{p}$ with $\delta(\bar{p}) \geq -
{\bf R} \bar{p} $. The difference between $\delta(\bar{p})$ and $\delta(p)$
provides a semi-classical bound on the derivative of the phase shift with respect to
momentum
\begin{equation}
    \frac{d \delta}{dp} \geq - {\bf R} \ .
  \label{eq:WBclass}
\end{equation}
By time evolving the plane waves, the above becomes a bound on the time delay between the incident and scattered wave, $\Delta t \geq -{M {\bf R}}/{p}$. It is in this sense that causality constrains non-relativistic scattering.
A more careful derivation, which includes quantum mechanical effects,
induces a second term on the right hand side of Eq.~(\ref{eq:WBclass})
and leads to the bound~\cite{Wigner:1955zz}:
\begin{eqnarray}
  \frac{d \delta}{dp} \geq -{\bf R} \ +\ \frac{\sin \left(2\delta+2 p {\bf R}\right)}{2p}  \ .
  \label{eq:WB3dbp1}
\end{eqnarray}
Evaluated at threshold this becomes a constraint on the effective range parameter:
\begin{eqnarray}
r\ \leq \ 2\bigg\lbrack {\bf R}\, -\, \frac{{\bf R}^2}{a} \, +\, \frac{{\bf R}^3}{3 a^2} \bigg\rbrack \ .
  \label{eq:WB3da}
\end{eqnarray}

In the Wilsonian EFT paradigm, an $S$-matrix element derived from EFT
is dependent on a momentum cutoff, $\Lambda\sim 1/{\bf R}$, which is
kept finite and varied to ensure cutoff-independence to a given order
in the perturbative EFT expansion.  What occurs above this scale is
irrelevant to the infrared physics that is encoded by the
$S$-matrix and compared to experiment. An explicit calculation of the
Wigner bound in the EFT of contact operators with cutoff
regularization can be found in Ref.~\cite{Phillips:1996ae}.  As the
bound depends explicitly on the EFT cutoff, its relevance in
physical scenarios is somewhat ambiguous as the EFT can violate
causality bounds as long as the violations occur above the cutoff,
and the bound itself weakens as higher-order corrections in
the EFT expansion are included~\cite{Beck:2019abp}.
\begin{table}[h!]
\centering
\begin{tabularx}{0.9\textwidth} { 
  | >{\raggedright\arraybackslash}X 
  | >{\centering\arraybackslash}X 
  | >{\centering\arraybackslash}X 
    | >{\centering\arraybackslash}X 
  | >{\raggedleft\arraybackslash}X | }
 \hline
  $\phi\mapsto$  & $\theta\mapsto$  & $\rho \mapsto$ & $r_0$ & $r_1$ \\
\hlineB{3}
 $\phi$  & $\theta$ & $\rho$ &   $-2a_1\lambda$  \ \ ${\scriptstyle (a_1>0)}$ & $-2a_0\lambda$ \ \  ${\scriptstyle (a_0>0)}$\\
\hline
$\phi$  & $-\theta$ & $ \rho_- +  \bar{\rho}_+$ & $+2a_1\lambda$ \ \ ${\scriptstyle (a_1<0)}$& $-2a_0\lambda$ \ \ ${\scriptstyle (a_0>0)}$\\
 \hline
$-\phi$  & $\theta$ & $ \rho_+ +  \bar{\rho}_-$ & $-2a_1\lambda$ \ \ ${\scriptstyle (a_1>0)}$& $+2a_0\lambda$ \ \ ${\scriptstyle (a_0<0)}$\\
 \hline
$-\phi$  & $-\theta$ & $\bar{\rho}$ & $+2a_1\lambda$ \ \ ${\scriptstyle (a_1<0)}$& $+2a_0\lambda$ \ \ ${\scriptstyle (a_0<0)}$\\
\hlineB{2} 
$\theta$  & $\phi$ & $\bar{\rho}$ & $-2a_0\lambda$ \ \ ${\scriptstyle (a_0>0)}$& $-2a_1\lambda$ \ \ ${\scriptstyle (a_1>0)}$\\
 \hline
$-\theta$  & $-\phi$ & $\rho$& $+2a_0\lambda$ \ \ ${\scriptstyle (a_0<0)}$& $+2a_1\lambda$ \ \ ${\scriptstyle (a_1<0)}$\\
 \hline
\end{tabularx}
  \caption{Action of the momentum-inversion transformation on causal models that have a non-zero effective range.}
 \label{tab:genconfrmwb}
\end{table}

\begin{figure}[!ht]
\centering
\includegraphics[width = 0.8\textwidth]{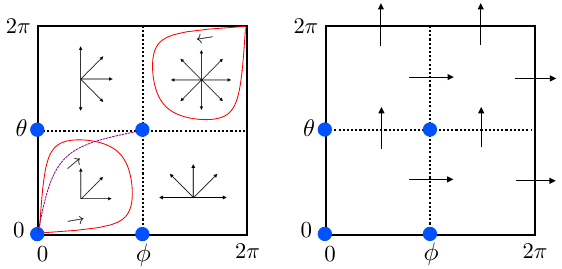}
\caption{Left panel: the range of tangent vectors on the flat
  torus allowed by the Wigner bound is superposed on Fig.~(\ref{fig:ftbase}).
  The purple trajectory in the bottom left quadrant has zero effective range
  and is seen to be consistent with the tangent-vector conditions.
  The red trajectory in the bottom left quadrant has positive effective ranges
  and is seen to violate the tangent-vector conditions, while the trajectory
  in the top right quadrant has negative effective ranges and is consistent
  with the conditions. Right panel: by matching the allowed tangent
  vectors at the boundaries of each quadrant it is found that
  $S$-matrix trajectories can only exit a quadrant via the upper or
  right edge.}
  \label{fig:WigTan}
\end{figure}

The $S$-matrix models with momentum-inversion symmetry can originate from
zero-range or finite-range forces. Here it will be assumed that the
underlying theory has strictly zero-range forces. This then implies
strong causality bounds whose geometric interpretation can be studied.
Explicitly, with zero-range forces, causality requires $r_s \leq 0$ and
the tangent vectors to $S$-matrix trajectories satisfy
\begin{eqnarray}
\dot{\phi}(p) \geq \frac{\sin\phi(p)}{p} \  \ \ , \ \ \ \dot{\theta}(p) \geq \frac{\sin\theta(p)}{p} \ ,
  \label{eq:WB3dx}
\end{eqnarray}
where dot represents differentiation with respect to momenta.  The
allowed tangent vectors clearly depend on the quadrant of the flat
torus in which they lie. In addition, by enforcing continuity of the
tangent vectors at the boundary of each quadrant, it is found that an
$S$-matrix trajectory can only exit a quadrant through the upper or
right edge. These various geometric constraints are illustrated in
Fig.~(\ref{fig:WigTan}) using the examples of Fig.~(\ref{fig:ftbase}).
It is notable that the large momentum behavior of any $S$-matrix curve
which ends at the trivial fixed point must be in the top-right
quadrant, which is also the only place where a trajectory can have
loops. Since the Wigner bound segregates by quadrant and indicates a
direction of preferred $S$-matrix evolution, causality introduces an
asymmetry which breaks the homogeneity and discrete isotropy of a
generic flat torus.
Applying the Wigner bound to the symmetric $S$-matrix models in
Table~\ref{tab:genconfrm} restricts the allowed signs of the
scattering lengths as shown in Table~\ref{tab:genconfrmwb}.

\subsubsection*{Causal singularities of the $S$-matrix}
\label{sec:causalsing}

\noindent In addition to Wigner bounds, causality in non-relativistic scattering is manifest in various constraints on the analytic structure
of the $S$-matrix in the complex-momentum plane~\cite{PhysRev.74.131,PhysRev.83.249,PhysRev.91.1267}. The simplicity of the $S$-matrix
models with momentum-inversion symmetry reveal these constraints and their relation with the Wigner bound in straightforward fashion.
The s-wave $S$-matrix elements with momentum inversion symmetry are ratios of polynomials of second degree and can thus be expressed as
\begin{eqnarray}
S_s &\equiv&\frac{\left(p+p_s^{\sss(1)}\right)\left(p+p_s^{\sss(2)}\right)}{\left(p-p_s^{\sss(1)}\right)\left(p-p_s^{\sss(2)}\right)} \ ,
   \label{eq:Selementres}
\end{eqnarray}
where
\begin{eqnarray}
p_s^{\sss(1,2)}&=& \frac{1}{r_s}\left(i\pm \sqrt{\frac{2r_s}{a_s}-1}\right) \ .
   \label{eq:polposres}
\end{eqnarray}
Consider the evolution of the singularities in the complex-$p$ plane as $\lambda$ is varied~\cite{Habashi:2020qgw}
for the causal model given in the last row of Table~\ref{tab:genconfrmwb}.
This model, with both scattering lengths negative, leaves $\phi-\theta$ invariant, and
has poles at
\begin{eqnarray}
p_s^{\sss(1,2)}&=& -\frac{1}{2|a_s|\lambda}\left(i\pm \sqrt{4\lambda-1}\right) \ .
   \label{eq:polposresmod}
\end{eqnarray}

\begin{figure}[!ht]
\centering
\includegraphics[width = 0.65\textwidth]{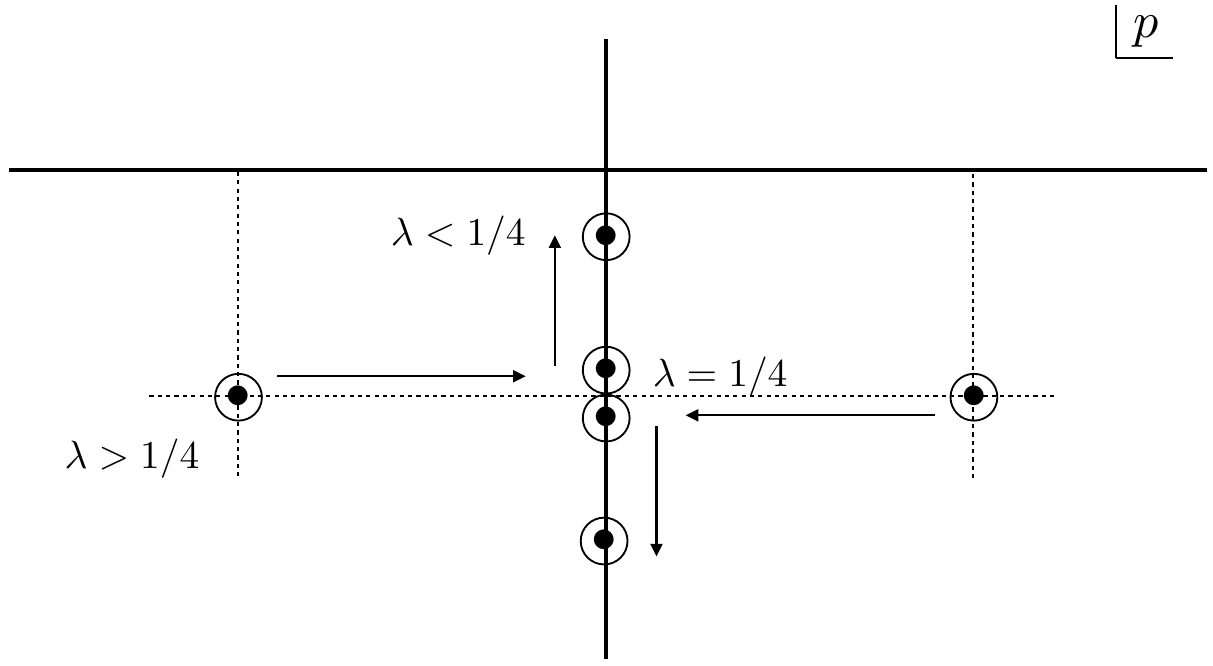}
\caption{Singularities of $S$-matrix elements in the complex-$p$ plane as $\lambda$ is varied. The arrows indicate direction of decreasing $\lambda$.}
  \label{fig:poles}
\end{figure}

There are three distinct cases, illustrated in Fig.~(\ref{fig:poles}).

\noindent $\underline{\lambda > 1/4}$: there are two resonance poles in the lower-half complex plane on
opposite sides of the imaginary axis. Dropping the partial-wave subscript on the scattering length,
\begin{eqnarray}
  p^{\sss(1)}&=& -p_R-i p_I \ \ \ , \ \ \
  p^{\sss(2)}= p_R-i p_I \ ,
   \label{eq:polposcase1a}
\end{eqnarray}
with
\begin{eqnarray}
  p_R &=& \frac{1}{2|a|\lambda}\sqrt{4\lambda-1}\ \ , \ \ p_I = \frac{1}{2|a|\lambda} \ ;
  \label{eq:polposcase1b}
\end{eqnarray}
$\underline{\lambda = 1/4}$: there is a double pole corresponding to a virtual state on the negative imaginary axis at
\begin{eqnarray}
  p^{\sss(1)} &=&  p^{\sss(2)}= -\frac{i}{2|a|\lambda} \ ;
   \label{eq:polposcase2}
\end{eqnarray}
$\underline{\lambda < 1/4}$: there are two poles corresponding to virtual states on the negative imaginary axis at
\begin{eqnarray}
  p^{\sss(1)}&=& -i p_- \ \ \ , \ \ \
  p^{\sss(2)}= -i p_+ \ ,
   \label{eq:polposcase3a}
\end{eqnarray}
with
\begin{eqnarray}
  p_\pm &=& \frac{1}{2|a_s|\lambda}\left(1\pm \sqrt{1-4\lambda}\right) \ .
  \label{eq:polposcase3b}
\end{eqnarray}
It is clear that the Wigner bound implies that the poles of the $S$-matrix elements lie in the lower-half of the complex-momentum plane as
one would expect of states that decay with time. Note that the special case of the double pole (and vanishing square root)
is in correspondence with the special case geometric potential which takes the simple form, Eq.~(\ref{eq:entPOT1lam14}).

\subsection{Spatial dependence of scattering}
\label{subsec:2dscat}

\noindent The $S$-matrix is clearly aware of the number of spatial
dimensions that it is acting in and one therefore expects that this is
reflected in the geometric theory of scattering via a modified
geometric potential. It is straightforward to carry out the analysis
that was done above in two spatial dimensions. In two dimensions, low energy
scattering arising from short-range forces is enhanced in the IR due to an apparent scaling symmetry of the Schr\"odinger equation. One consequence of this is that there is only a single fixed point $S$-matrix, the identity, which is reached at both zero and
infinite scattering length. Spin and particle statistics are also distinct in
two dimensions, however, for our purposes, all that will be required is a
scattering process with two independent low-energy channels. One way this could be achieved is by placing the three-dimensional scattering system in a strongly
anisotropic harmonic potential which effectively confines one of the
spatial dimensions~\cite{PhysRevA.64.012706,PhysRevA.76.063610,PhysRevA.85.061604,PhysRevA.98.051603,PhysRevLett.94.210401}. This allows
the two-fermion system to be continuously deformed from three to two
dimensions and provides a means of studying the dependence of
the geometric theory, constructed above, on spatial dimensionality.
A qualitatively equivalent and simpler way of achieving this reduction of
dimensionality is by periodically identifying and compactifying one of
the spatial dimensions~\cite{PhysRevA.93.063631,Beane:2018huc}, say in the
$z$-direction. 

Regardless of how the two dimensional system is obtained, the ERE is~\cite{Braaten:2004rn,Beane:2010ny,Kaplan:2005es}
\begin{eqnarray}
\cot{\delta_s(p)} & =& \frac{1}{\pi} \log({\textit{\textbf a}}^2_s p^2) + \sigma_{2, \, s} p^2 + \mathcal{O}(p^4)
 \label{eq:2DERE}
\end{eqnarray}
where the ${\textit{\textbf a}}_s$ and $\sigma_{2, \, s}$ are the two-dimensional scattering lengths and areas, respectively. The full $S$-matrix can be constructed from the phase shifts as in Eq.~(\ref{eq:Sdef1}). Retaining just the first term in the ERE gives rise to the scattering length approximation which, in terms of periodic variables on the flat torus, is
\begin{eqnarray}
\phi & =& 2\cot^{-1}\!\left(\frac{1}{\pi} \log({\textit{\textbf a}}^2_0 p^2)\right) \ \ \ , \ \ \ \theta\ =\ 2\cot^{-1}\!\left(\frac{1}{\pi}\log({\textit{\textbf a}}_1^2 p^2)\right) \ ,
 \label{eq:LOinaffineSOL2D}
\end{eqnarray}
where the higher order effective area and shape parameters have
been set to zero\footnote{This can be obtained from the
compactification of a spatial dimension if the $d=3$ effective range
parameters are functions of the compactification
radius~\cite{Beane:2018huc}.}. Note that there is an IR enhancement in two dimensions, as made evident by the logarithmic dependence on the c.o.m. momenta, and that there exists a bound state for either sign of coupling strength~\cite{Beane:2010ny,Kaplan:2005es,Beane:2021A}.
  
The momentum inversion transformation takes a similar form as in three dimensions,
$p\to ({\textit{\textbf a}}_1 {\textit{\textbf a}}_0\, p)^{-1}$, and the
phase shifts transform as
\begin{eqnarray} 
   \phi(p) \mapsto -\theta(p) \ \ & ,& \ \ \theta(p) \mapsto -\phi(p) \ ,
  \label{eq:confmoebiusiso22d}
\end{eqnarray}
which leaves the density matrix invariant.
In two spatial dimensions, the momentum-inversion symmetry implies that all effective area and
shape parameters must vanish\footnote{In $d=2$, causality bounds the effective area
parameter~\cite{Hammer:2009zh,Hammer:2010fw}
\begin{eqnarray}
\sigma_{2,\,s}\ \leq \ \frac{{\bf R}^2}{\pi} \bigg\lbrace \bigg\lbrack \log\left(\frac{\bf R}{2{\textit{\textbf a}}_s}\right)\, +\, {\gamma} \, -\, \frac{1}{2} \bigg\rbrack^2  \,+\,\frac{1}{4} \bigg\rbrace \ ,
  \label{eq:WB2da}
\end{eqnarray}
where $\gamma$ is the Euler-Mascheroni constant. Therefore, momentum-inversion implies that the Wigner bound is saturated with $\sigma_{2,\,s} = 0$.}.
The geometric potential on the flat torus which reproduces the phase shifts of Eq.~(\ref{eq:LOinaffineSOL2D}) is found to be
\begin{eqnarray}
\mathbb{V}(\phi,\theta) & = &  \, -\frac{\pi^2}{4\left(\log({{\textit{\textbf a}}_0}/{{\textit{\textbf a}}_1})\right)^2c_1^2} \tan^2\left(\oneht(\phi+\theta)+\frac{\pi}{2}\right) \ .
\label{eq:entPOT12d} 
\end{eqnarray}
Notice that the harmonic dependence is the same as in three spatial
dimensions, Eq.~(\ref{eq:entPOT1}), except for an additional phase of
$\pi/2$ which causes the geometric potential to diverge when both phase shifts
sum to zero. This can only occur at threshold, and can be attributed
to the infinite force needed to reproduce the singular behavior of the
phase shift derivatives at $p=0$.  Another property of the
geometric potential is the divergence of the prefactor when ${\textit{\textbf
    a}}_0 = {\textit{\textbf a}}_1$. At the end of
section~(\ref{sec:geoact}) it was pointed out that, for UV/IR symmetric
trajectories, there are two trajectory equations for two unknowns, the
inaffinity and the geometric potential. However, when the scattering
lengths are equal, and $\phi = \theta$, the two equations are no
longer linearly independent. In this case the trajectory is a geodesic
---a straight line--- on the flat torus, and no geometric potential is needed.

\section{Conclusion}
\label{sec:conc}

\noindent The $S$-matrix is a unitary operator that
evolves a state vector from the boundary of spacetime, into the
spacetime bulk to experience interaction, and then back to the
spacetime boundary.  In this view of scattering, all spacetime
features like causality and spatial dimensionality are bulk
properties, and, as the $S$-matrix is purely a function of kinematical
variables like momentum and energy, the bulk properties must be
imprinted in some way on these variables. In a general scattering
process, the $S$-matrix evolves an initial unentangled product state
into an entangled state which in general experiences non-local correlations.
In order to avoid the assumption of locality, which is intrinsic to
the EFT paradigm, a geometric formulation
of scattering for two species of spin-1/2 fermions interacting
at low-energies via finite-range interactions has been developed. 
In this geometric theory 
the $S$-matrix emerges, without direct reference to spacetime, as a trajectory 
in an abstract space that is defined by unitarity. 
These $S$-matrix trajectories 
are generated by an entangling harmonic force whose form is 
---in certain special cases--- determined exactly by a UV/IR symmetry. 
It should be noted that the generation of the $S$-matrix from an entangling force is strikingly similar to recent proposals of the emergence of spacetime from entanglement~\cite{VanRaamsdonk:2010pw,Pastawski:2015qua,Headrick:2014cta}.
The next chapter will demonstrate how the UV/IR symmetries of the $S$-matrix manifest in the EFT as reflection symmetries of the RG flow of the coupling constants.

\newpage

\begin{subappendices}
\section{Basis Independent Geometric Formulation}
\label{app:geom}
\noindent The analysis of Section~\ref{sec:geom} relies on 
choosing a specific isotropic coordinate system to study the geometry
of the $S$-matrix. As the $S$-matrix is an operator in the product
Hilbert space of nucleon spins, it is interesting to consider distance
measures in a basis-independent manner.  For this purpose it is
convenient to make use of the Hilbert-Schmidt (HS) distance.  The HS
distance measure is a natural extension of the Frobenius inner
product, $\langle \hat{\bf A},\hat{\bf B} \rangle = {\rm Tr}\lbrack
\hat{\bf{ A}^\dagger} \hat{\bf{ B}} \rbrack$. It can be defined as~\cite{bengtsson_zyczkowski_2006}
\begin{equation}
    D(\hat{\bf{A}},\hat{\bf{B}})^2  \equiv
    d_n{\rm Tr}\left[\; (\hat{\bf{A}}-\hat{\bf{B}})(\hat{\bf{A}}-\hat{\bf{B}})^\dagger\; \right]
    \label{eq:hsdist}
\end{equation}
with $d_n$ an arbitrary normalization constant that will be set to $\frac{1}{2}$.
The HS distance is independent of basis, positive semi-definite and
zero if and only if $\hat{\bf{A}} = \hat{\bf{B}}$. If the $S$-matrix is
parameterized by phase shifts, say $\phi$ and $\theta$, then
the HS distance induces a metric on the space of $S$-matrices. This
allows for the direct study of the geometry of the S-matrix. The HS
distance between two $S$-matrices with distinct phase shifts, $\hat{\bf
  S}(\phi, \theta)$ and $\hat{\bf S'}(\phi',\theta')$, is
\begin{equation}
    D(\hat{\bf S}, \hat{\bf S'})^2 = 
     \oneht{\rm Tr}\left[\; (\hat{\bf S} - \hat{\bf S'})(\hat{\bf S} - \hat{\bf S'})^\dagger\; \right] = 2 \Big(\sin^2\left(\oneht(\phi-\phi')\right) + 3\sin^2\left(\oneht(\theta-\theta')\right) \Big) \ .
    \label{eq:globalmet}
\end{equation}
The metric is obtained by looking at the infinitesimal differences, $d\phi = \phi' - \phi$ and $d\theta = \theta' - \theta$ and is found to be,
\begin{equation}
    ds^2 = \oneht\left(3\,d \theta^2 + d\phi^2\right) \ .
    \label{eq:localmet}
\end{equation}
The unitary $S$-matrix is determined by the two degrees of freedom, $\phi$ and $\theta$,
and therefore, locally, the $S$-matrix lives on the space defined by this two-dimensional Euclidean metric
that can be rescaled to remove the anisotropic spin weighting factor of the spin-triplet phase shift $\theta$.

The HS distance serves to obtain an operator definition and an alternate understanding of
the EP. Recall that the S-matrix is non-entangling when either $\phi = \theta $ or $\phi = \theta \pm
\pi$ \footnote{When $\phi = \theta \pm \pi$ the $S$-matrix acts as a
  swap gate on the incoming nucleon-nucleon state up to an overall
  phase. Likewise, when $\phi = \theta \pm \frac{\pi}{2}$, the
  $S$-matrix acts as a root-swap gate on the incoming nucleon-nucleon state
  up to an overall phase. }. Therefore, the non-entangling
$S$-matrices form a codimension-one subspace within the space of all
possible $S$-matrices. The EP of a given S-matrix, $\hat{\bf
  S}(\phi,\theta)$, is found to be,
\begin{equation}
    \mathcal{E}(\hat{\bf{S}}) = D(\hat{\bf S}(\phi,\theta), \hat{\bf{S}}(\theta,\theta))^2 \hspace{.1cm} D(\hat{\bf S}(\phi,\theta), \hat{\bf{S}}(\theta - \pi,\theta))^2 = N_p \sin^2{(\phi - \theta)} \ ,
    \label{eq:hsep}
\end{equation}
where the freedom in defining the HS norm has been used to set the normalization to $N_p$.
As both $\hat{\bf{S}}(\theta,\theta)$ and $\hat{\bf{S}}(\theta -
\pi,\theta)$ are non-entangling, the EP can be interpreted as a
measure of the distance from a given $S$-matrix to the two
non-entangling subspaces.  Using the HS distance
highlights the fact that the EP of an operator is a
state-independent measure of entanglement.
\end{subappendices}

\chapter{UV/IR Symmetries of the S-matrix and RG flow}
\label{chap:SmatUVIR}
\noindent 
\textit{This chapter is associated with Ref.~\cite{Beane:2021xrk}: \\
``UV/IR Symmetries of the S-matrix and RG flow" by Silas R. Beane and Roland C. Farrell.}
\section{Introduction}

\noindent Non-relativistic s-wave scattering with finite-range forces
exhibits two fixed points of the RG: the trivial fixed point
corresponding to no interaction, and the unitary fixed point where the
interaction strength takes the maximal value consistent with
unitarity~\cite{Weinberg:1991um} (for a review, see
Ref.~\cite{Braaten:2004rn}). 
As shown in the previous chapter, the
$S$-matrix also has interesting properties with respect to UV/IR
transformations that invert the momentum. 
In the scattering length approximation, where the effective range and all higher-order shape parameters vanish, the UV/IR transformations maps the trivial RG fixed point into
the unitary RG fixed point and vice versa.  As a result, the UV/IR
transformation does not act simply on the scattering amplitude which
vanishes at the trivial fixed point.  Instead, it is the $S$-matrix,
which accounts for the trivial fixed point via the unit operator, that
manifests the UV/IR symmetry.  
These novel symmetries provide a new perspective on EFT descriptions of the
nucleon-nucleon (NN) interaction and facilitate the development of new
EFTs which enhance the convergence\footnote{For recent work which aims
  to improve the convergence of NN EFT see
  Refs.~\cite{SanchezSanchez:2017tws,Peng:2021pvo,Mishra:2021luw,Ebert:2021epn,Habashi:2020qgw,Habashi:2020ofb}.}
of the description at very low energies (see Chapter~\ref{chap:Weinberg}).

This chapter shows how the UV/IR properties of the
$S$-matrix are reflected in the (scheme dependent) beta-functions which characterize the
RG scale dependence of the EFT couplings. This will be explored for the same system as in the previous chapter: non-relativistic fermions with an s-wave interaction in both two and three spatial dimensions. 
This chapter is organized as follows.
Section~\ref{sec:smatth1} introduces the $S$-matrix and its UV/IR
symmetries (whose details are treated in appendix~\ref{app:Ward}). In section~\ref{sec:eftrg}, the EFT which matches to
the $S$-matrix in the scattering length approximation is reviewed and the
UV/IR symmetry is shown to be present in the RG flow of the
coefficients of momentum-independent local operators.
Constraints on higher-order EFT operators which follow from considering
UV/IR symmetry breaking in the $S$-matrix are considered in section~\ref{sec:eftsb}.
Section~\ref{sec:conc1} summarizes and concludes.

\section{S-matrix theory and UV/IR symmetry}
\label{sec:smatth1}

\noindent 
The fixed points of the RG are determined by the flow
of coupling constants with a change of scale in the EFT which gives rise to Eq.~(\ref{eq:Sdef1}).
In non-relativistic scattering, the $S$-matrix takes special
constant values at these fixed points~\cite{Beane:2018oxh}.  For a given
channel, the fixed points of the RG occur when the phase shifts,
$\delta_0$ and $\delta_1$, vanish (trivial fixed point) or are at ${\pi}/2$
(unitary fixed point); i.e. when $S_s=\pm 1$. Therefore the fixed
points of the full $S$-matrix occur when the phase shifts both vanish,
$\delta_1=\delta_0=0$, both are at unitarity, $\delta_1=\delta_0={\pi}/2$, or
when $\delta_1=0$, $\delta_0={\pi}/2$ or $\delta_1={\pi}/2$, $\delta_0=0$. The
$S$-matrices at these four fixed points are $\pm \hat {\bf 1}$ and
$\pm \hat{{\cal P}}_{12}$.  In the ERE, the fixed
points of the RG are reached at $a_1= a_0=0$, $|a_1|=| a_0|=\infty$,
and at $a_1=0\,,\, |a_0|=\infty$, $|a_1|=\infty\,,\, a_0=0$, with
all effective range and shape parameters taken to be vanishing (the
scattering length approximation).  If all inelastic thresholds are
absent, and $p$ is defined on the interval $[0,\infty)$, then all four
  of the RG fixed points are accessible (in a limiting sense) via
  scattering. In the scattering length approximation this can be seen
  to be a consequence of the invariance of the $S$-matrix with respect
  to the scaling transformation
\begin{eqnarray}
p\mapsto e^\beta p \ \ \ , \ \ \ a_s\mapsto e^{-\beta} a_s \ ,
 \label{eq:dilinv}
\end{eqnarray}
with $\beta$ an arbitrary parameter. Operationally, keeping $a$ fixed and scaling $p$ with $\beta$
positive (negative) accesses the $S$-matrix of large (small)
scattering lengths.  Hence, with $a_0$ and $a_1$ finite, the $S$-matrix
is a trajectory from $\hat {\bf 1}$ to $-\hat {\bf 1}$. With $a_0$
finite (at unitarity) and $a_1$ at unitarity (finite), the trajectory
originates at $-\hat{{\cal P}}_{12}$ ($\hat{{\cal P}}_{12}$) and again ends at
$-\hat {\bf 1}$.

In addition to the scale invariance of Eq.~(\ref{eq:dilinv}), the individual $S$-matrix
elements, $S_s$, transform simply with respect to the momentum-inversion transformation
\begin{eqnarray}
p\mapsto \frac{1}{a_s^2\, p} \ .
   \label{eq:singlechanmominv}
\end{eqnarray}
As this transformation maps threshold to asymptotic infinity and vice-versa, it is a UV/IR transformation. (This transformation and its associated Ward identity are considered in detail in appendix~\ref{app:Ward}.) As a transformation on phase shifts, one finds that
Eq.~(\ref{eq:singlechanmominv}) implies
\begin{eqnarray}
  \delta_s(p) \mapsto -\delta_s(p) \pm {\pi}/2 \ \ &,& \ \ S_s \to -S^*_s  \ \ ,
\end{eqnarray}
where the sign of the shift by $\pi/2$ is determined by the sign of
the scattering length.  Therefore, considering scattering near
threshold, this momentum inversion transformation interchanges the trivial and unitary fixed points of the RG. It will
be seen below how this transformation is manifest in the
running coupling constants of the EFT.

\section{Effective field theory and RG flow}
\label{sec:eftrg}

\subsection*{EFT action and potential}

\noindent It is interesting to ask whether the UV/IR properties of the
$S$-matrix are reflected in the EFT. The $S$-matrix in the
scattering-length approximation can be derived from an EFT with
highly-singular, momentum-independent, contact interactions. These
give rise to the quantum mechanical potential that enters the
Lippmann-Schwinger equation, which in turn generates the $S$-matrix.
The $S$-matrix models with exact UV/IR symmetry are,
by definition, UV-complete. Therefore one might expect the UV/IR
symmetry to be reflected in the flow of the EFT potential between
fixed points of the RG.

The EFT which describes low-energy s-wave scattering of nucleons is
constrained by spin, isospin and Galilean invariance (and, for the
case we consider, parity and time-reversal invariance). The
leading-order (LO) interactions in the Lagrangian density
are~\cite{Weinberg:1991um,Weinberg:1990rz}
\begin{equation}
{\cal L}_{\rm LO}
=
-\frac{1}{2} C_S (N^\dagger N)^2
-\frac{1}{2} C_T \left(N^\dagger{\bm \sigma}N\right)\cdot \left(N^\dagger{\bm \sigma}N\right) \ ,
\label{eq:interaction2}
\end{equation}
where the field $N$ represents both spin states of the proton and neutron fields.  These interactions
can be re-expressed as contact interactions in the $\si$ and $\siii$
channels with couplings $C_0 = ( C_S-3 C_T) $ and $C_1 = (C_S+C_T)$
respectively, where the two couplings are fit to reproduce the $\si$
and $\siii$ scattering lengths.  The quantum-mechanical potential is
scheme dependent and can be read off from the Lagrangian density~\cite{Beane:2018oxh}
\begin{eqnarray}
{ V}(\mu)_\sigma
  & = &
\frac{1}{2}\left( { C}_{1}(\mu) +  { C}_{0}(\mu)  \right)  {\hat   {\bf 1}} +
\frac{1}{2}\left( { C}_{1}(\mu) -  { C}_{0}(\mu) \right) \hat{{\cal P}}_{12}
\ ,
   \label{eq:Vrescale}
\end{eqnarray}
where the $S$-matrix basis has been chosen.  In what follows
the flow of the potential with the RG scale, $\mu$, will be considered
in three and two spatial dimensions ($d=3,2$). 

\subsection*{RG flow in $d=3$}

\begin{figure}[!ht]
\centering
\includegraphics[width = 0.73\textwidth]{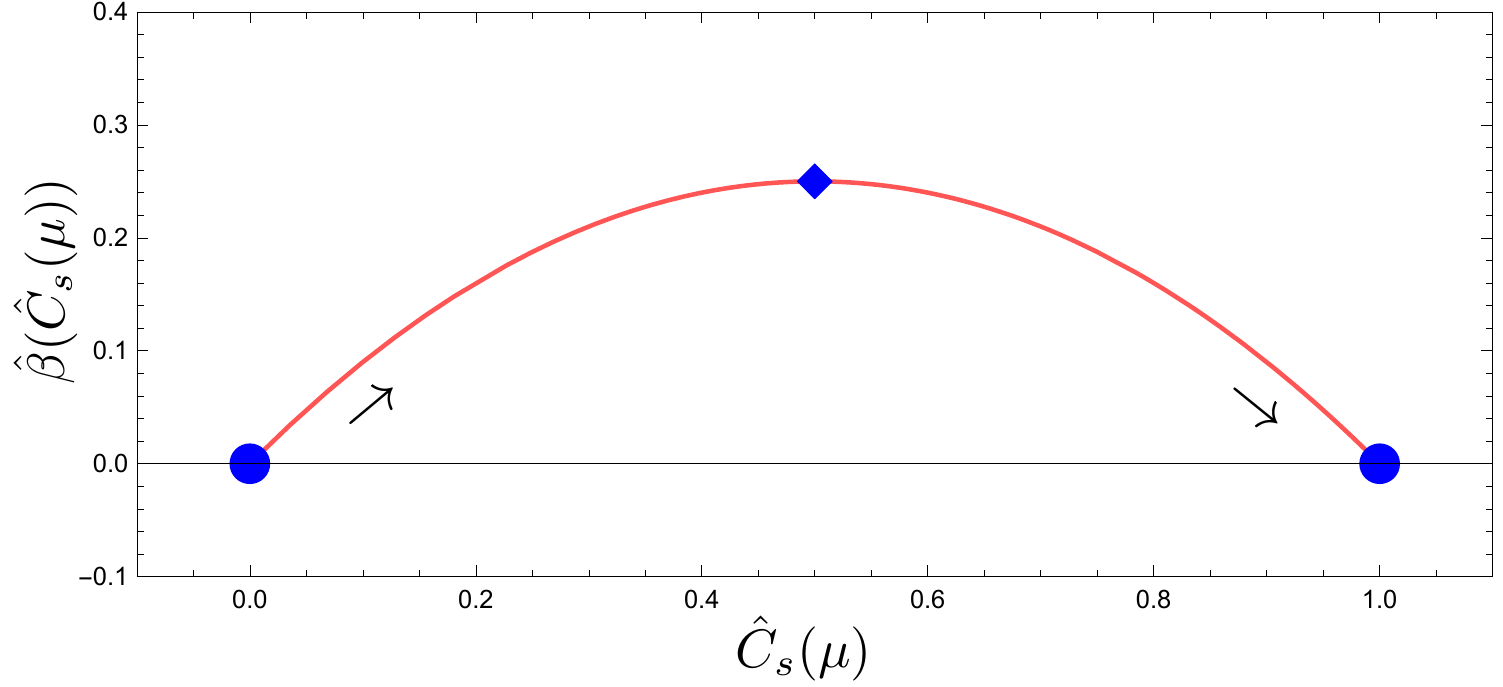}
\caption{The beta-function of Eq.~(\ref{eq:c0betafn}) plotted as a function of ${\hat C}_s(\mu)$. The blue dots are the RG fixed points.
  The beta-function curve evolves from $\mu=0$ (the trivial fixed point) to $\mu=\infty$ (the unitary fixed point). The diamond is the
fixed point of the scale-inversion symmetry, Eq.~(\ref{eq:singlechancutoffinv}), at ${\hat
  C}_s(\mu^\circ) = 1/2$.}
  \label{fig:eftc0beta}
\end{figure}

\noindent Solving the Lippmann-Schwinger equation with the potential
of Eq.~(\ref{eq:Vrescale}), or alternatively, summing to all orders
the loop diagrams with insertions of the operators in
Eq.~(\ref{eq:interaction2}), leads to the $d=3$ NN scattering amplitude.
In dimensional regularization (dim reg) with the power-divergence subtraction (PDS)
scheme~\cite{Kaplan:1998tg,Kaplan:1998we} the amplitude is
\begin{equation}
  i \mathcal{A} = \frac{-i V(\mu)}{1 + M V(\mu) \left( \mu + i p\right) /4\pi} \ \ \ ,
   \label{eq:AitoV}
\end{equation}
where $M$ is the nucleon mass and $V(\mu)$ is the projection of
${V}(\mu)_\sigma$ onto a particular scattering channel. The PDS scheme
offers a clean way of accounting for the linear
divergences which appear in loops. The PDS couplings also exhibit the trivial and unitary RG fixed points
and, for $\mu \sim p$, scale so as to justify their non-perturbative treatment (power counting is manifest).  
The relation between the $\mu$-dependent coefficients
and the phase shifts in the scattering-length
approximation follows from matching the scattering amplitude to the ERE and is
\begin{equation}
  p \cot \delta_s =  - \left( \frac{4 \pi}{M C_s} + \mu \right) = -\frac{1}{a_s}\ .
\end{equation}
Therefore, the running couplings in the PDS scheme are
\begin{equation}
  C_s(\mu) \ =\ \frac{4 \pi}{M} \frac{1}{{1}/{a_s}-{\mu}} \ .
  \label{eq:C0PDS}
\end{equation}
There is a fixed point at $C_s=0$, corresponding to free particles ($a_s=0$),
and a fixed point at $C_s=C_\star$ corresponding to a divergent scattering
length (unitarity). It is convenient to rescale the couplings to 
${\hat C}_s \equiv C_s/C_\star$. The beta-functions for the rescaled couplings
are then 
\begin{equation}
  {\hat\beta({\hat C}_s)}\ =\  \mu \frac{d}{d\mu} {\hat C}_s(\mu) \ =\ -{\hat C}_s(\mu)\left({\hat C}_s(\mu)-1\right) \ ,
   \label{eq:c0betafn}
\end{equation}
which has fixed points at ${\hat C}_s=0$ and $1$, as shown in Fig.~(\ref{fig:eftc0beta}). The coupling is near the
trivial fixed point for $\mu < 1/|a_s|$, and near the non-trivial fixed point for $\mu
> 1/ |a_s|$. The four fixed points in the NN system are at ${\hat C}_1= {\hat C}_0=0$, ${\hat C}_1={\hat C}_0=1$,
and at ${\hat C}_1=0\,,\,{\hat C}_0=1$, and ${\hat C}_1=1\,,\, {\hat C}_0=0$.
In the space of rescaled couplings, these four points,
$(0,0)$, $(1,1)$, $(0,1)$ and $(1,0)$ furnish a representation of the Klein
group~\cite{Beane:2018oxh}.

Now recall that the momentum inversion, $p \to {1 /(a_s^2 p)}$, has the
effect of interchanging the trivial and unitary fixed points of
the $S$-matrix elements in the scattering length approximation.
In the EFT, under an inversion of the PDS RG scale
\begin{eqnarray}
\mu \mapsto \frac{1}{a_s^2 \, \mu} \ ,
   \label{eq:singlechancutoffinv}
\end{eqnarray}
the couplings transform as
\begin{equation}
   {\hat C}_s(\mu) \mapsto 1 -{\hat C}_s(\mu)\ .
\end{equation}
This implies that the beta-function evaluated at two scales related by an inversion are
equal, i.e. ${\hat\beta({\hat C}_s)}\rvert_{\bar{\mu}} = {\hat\beta({\hat
C}_s)}\rvert_{1/(a_s^2 \bar{\mu})}$, for any $\bar{\mu}$. This scale-inversion
transformation maps the two RG fixed points to one another and the UV/IR
transformation property of the phase shifts is reflected in the
$\mu$ dependence of the coupling. In addition, the beta-function is reflection
symmetric about the fixed point of the inversion transformation, ${\hat
  C}_s(\mu^\circ) = 1/2$, which occurs at
$\mu^\circ=|a_s|^{-1}$, as shown in Fig.~(\ref{fig:eftc0beta}).

The analysis presented above holds for other renormalization schemes as well, provided that they preserve the UV/IR symmetry; i.e. reproduce the scattering length approximation.
For example, consider regulating with a hard-cutoff in momentum space. 
The scattering amplitude is
\begin{eqnarray} 
\label{eq:cutoffamp}
   {\cal A} = \frac{- C_s(\Lambda)}{1 + M C_s(\Lambda) \Big[\Lambda +  i p  \tan^{-1}\left(i \frac{\Lambda}{p} \right) \Big]/(2\pi^2)}
\end{eqnarray}
where the cutoff-dependent coupling is defined as
\begin{eqnarray}
   C_s(\Lambda) = \frac{4\pi}{M} \frac1{1/a_s - 2\Lambda/\pi}
   \label{eq:Cscutoff} \ .
\end{eqnarray}
For finite $\Lambda>p$, the expanded $\tan^{-1}{ (i \Lambda / p)} $ term generates cutoff-dependent contributions to the amplitude to all orders in the effective range expansion~\cite{vanKolck:1998bw,Birse:1998dk}. These higher-order
terms break the UV/IR symmetry and therefore, to preserve the symmetry, higher-dimensional operators must be added to the EFT action to cancel these symmetry-breaking effect. For instance, expanding Eq.~\ref{eq:cutoffamp} gives
\begin{eqnarray}
\mathcal{A} &=& -\frac{4 \pi}{M} \frac{1}{1/a_s + i p}\left (1 +\frac{2}{\pi\Lambda} \frac{1}{1/a_s + i p}p^2 +  \mathcal{O}(\Lambda^{-2}) \right)
\end{eqnarray}
which evidently requires a momentum dependent counterterm ---a shift in the  $C_{2 \, s}$ operator that appears at NLO in the EFT expansion--- that scales like $\mathcal{O}(\Lambda^{-1})$~\cite{Kaplan:1998tg,Kaplan:1998we,vanKolck:1998bw}. In addition, insertions of this counterterm in perturbation theory will generate new terms in the amplitude that scale like positive powers of the cutoff, and whose removal will in turn require even higher dimensional counterterms~\cite{vanKolck:1998bw}.
This procedure of choosing counterterms to reproduce the scattering length approximation is nothing new; the identical procedure must be carried out in order to renormalize the cutoff EFT in a manner that preserves the Schr\"odinger symmetry Ward identities~\cite{Mehen:1999nd}, in the unitary, $\lvert a_s \rvert  \to \infty$, limit.

A key observation is that the cutoff dependence of $C_s (\Lambda)$ does not change as more counterterms are added, and the RG flow of this coupling in the cutoff scheme is the same as in PDS, Eq.~(\ref{eq:C0PDS}), but with $\mu \mapsto \frac{2}{\pi}\Lambda$. Therefore, the RG scale-inversion symmetry of the leading-order beta function
is not an artifact of a particular scheme, but rather a manifestation of
a physical property of the system which is reflected in the RG evolution
of the EFT couplings. Note that after all the higher-dimensional symmetry-restoring operators have been added, the theory is valid for all momenta and varying the arbitrary cutoff sets the scale for the physical process. 

\begin{figure}[!ht]
\centering
\includegraphics[width = 0.8\textwidth]{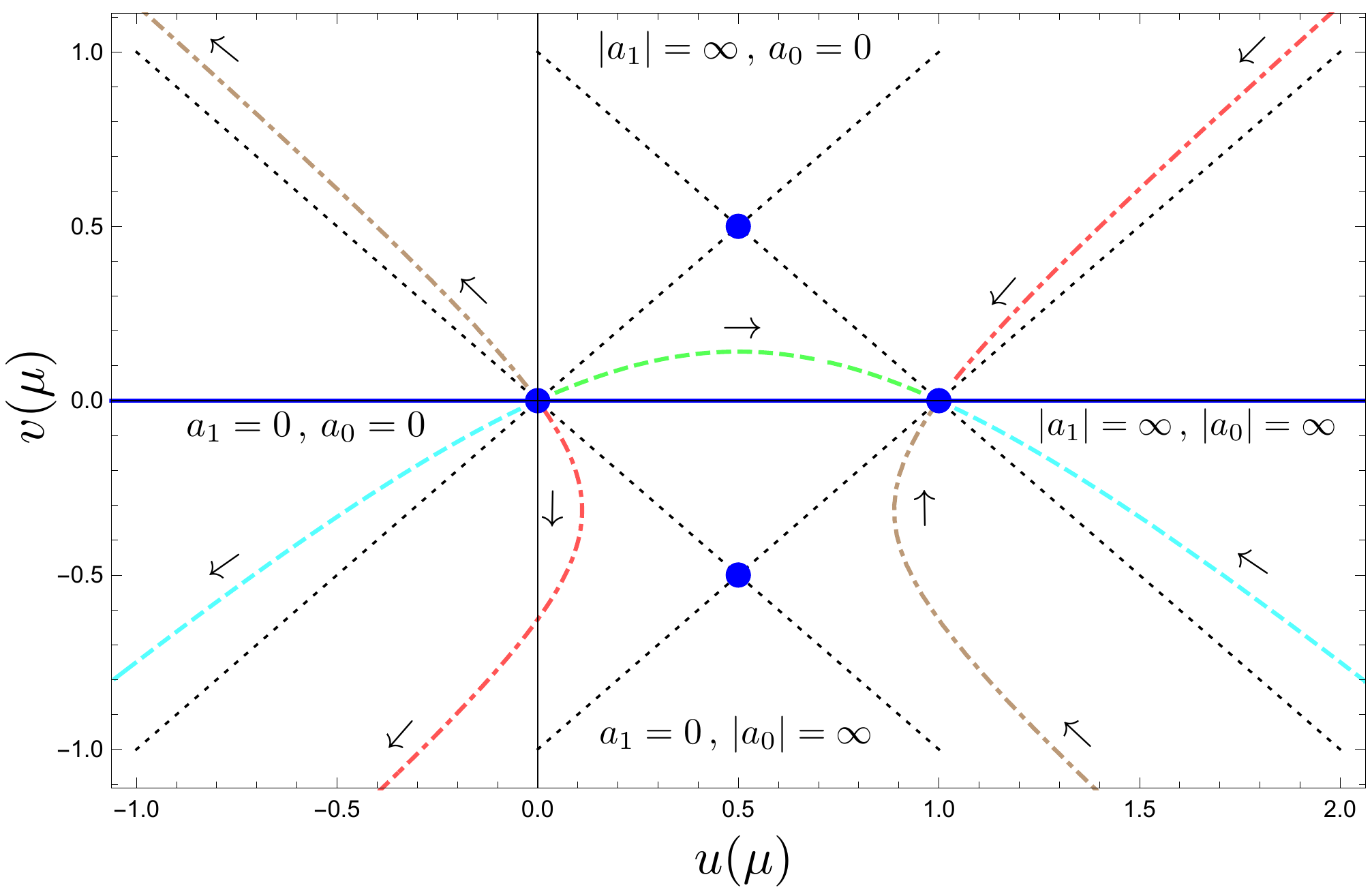}
\caption{The four fixed points of the RG in the $u-v$ plane in $d=3$. The
red-dashed line is the RG trajectory corresponding to the physical NN scattering lengths. The brown-dashed lines flip the signs of the physical scattering lengths.
The green (cyan) dashed curves are RG trajectories that exhibit the UV/IR symmetry and correspond to
both scattering lengths negative (positive). The arrows indicate direction of increasing $\mu$. Note that part of the cyan curve is not shown.}
  \label{fig:eftrhom}
\end{figure}

Returning to two-channel s-wave scattering, it is convenient to define the components of the re-scaled potential as
\begin{eqnarray}
u(\mu)  \; = \;  \frac{1}{2}\left( {\hat C}_{1}(\mu) +  {\hat C}_{0}(\mu)  \right) \ \ , \ \
v(\mu) \;=\; \frac{1}{2}\left( {\hat C}_{1}(\mu) -  {\hat C}_{0}(\mu) \right)  \ .
\label{eq:uvCs}
\end{eqnarray}
In the $u-v$ basis the RG fixed points of the rescaled potential are at $(0,0)$, $(1,0)$, $(1/2,1/2)$ and $(1/2,-1/2)$,
and the components of the potential flow with the RG according to the algebraic curve
\begin{eqnarray}
v\left(v-{\bar w}\right) \ = \ u\left( u-1\right) \ \ \ , \ \ \ {\bar w} \ \equiv \ \frac{a_1 + a_0}{a_1 - a_0}  \ .
   \label{eq:EFTconic}
\end{eqnarray}
The curves for all combinations of signs of scattering lengths are plotted in Fig.~(\ref{fig:eftrhom}). 

At the end of section~\ref{sec:smatth1} it was pointed out that when $a_1 a_0 > 0$ the momentum inversion symmetry is an exact symmetry of the
density matrix. In the figure this corresponds to the green and cyan curves which have a reflection symmetry about
the line $u = 1/2$. When $a_1 a_0 < 0$ the momentum inversion transformation maps the density matrix into one obtained from flipping the signs of the
scattering lengths. In the EFT this is seen through the
reflection about the line $u = 1/2$ mapping the trajectories with $a_0 > 0$ and $a_1 < 0$ (brown curve) and $a_0 < 0$ and $a_1 > 0$ (red curve) into each other.
In either case the reflection is generated by the scale-inversion 
\begin{eqnarray}
\mu \mapsto \frac{1}{\lvert a_1 a_0 \rvert \mu} \ .
   \label{eq:moebius4}
\end{eqnarray}
Therefore, in the EFT, the symmetry properties of the density matrix are encoded in
a geometric, reflection symmetry of the RG flow of the coupling constants. It
is curious that when the system is unbound in both s-wave channels, the RG flow is confined to the rhombus
formed by the four RG fixed points in much the same way that the $S$-matrix is confined via unitarity to the
flat torus defined by the two s-wave phase shifts~\cite{Beane:2020wjl}.

\subsection*{RG flow in $d=2$}

\noindent In order to further investigate the relation between UV/IR
symmetries of the $S$-matrix and RG flow, consider deforming the
scattering system to $d=2$ via an anisotropic harmonic
trap~\cite{PhysRevA.64.012706} or by compactifying a
dimension~\cite{PhysRevA.93.063631}. The $S$-matrix element in the
$d=2$ scattering length approximation becomes
\begin{eqnarray}
S_s &=& \frac{\log{\left ( {\textit{\textbf a}}_s^2 p^2 \right )} + i \pi}{\log{\left ( {\textit{\textbf a}}_s^2 p^2 \right )} - i \pi}  \ ,
   \label{eq:Selementscatt2d}
\end{eqnarray}
where ${\textit{\textbf a}}_s$ is the intrinsically positive $d=2$ scattering length.
Here the momentum inversion transformation, $p \mapsto 1/({\textit{\textbf a}}_s^2 p)$, maps $S_s\mapsto S_s^*$.

The scattering amplitude obtained in the EFT has a logarithmic divergence which requires regularization and
renormalization. Using dim reg with the $\overline{\text{MS}}$ scheme one finds that the couplings run
with the RG scale as~\cite{Kaplan:2005es,Beane:2010ny}
\begin{equation}
    C_s(\mu) = \frac{- 4 \pi}{M \log{\left ( {\textit{\textbf a}}_s^2 \mu^2 \right )}} \ .
\end{equation}
Notice that the coupling has a pole at $\mu = {\textit{\textbf a}}_s^{-1}$ where it changes sign. With $\mu > {\textit{\textbf a}}_s^{-1}$, 
$C_s(\mu)<0$ corresponding to attraction, and the coupling appears asymptotically free; i.e. flows to zero logarithmically.
With $\mu < {\textit{\textbf a}}_s^{-1}$,  $C_s(\mu)>0$ corresponding to repulsion, and the coupling runs into a Landau pole
in the UV at $\mu = {\textit{\textbf a}}_s^{-1}$.

\begin{figure}[!ht]
\centering
\includegraphics[width = 0.73\textwidth]{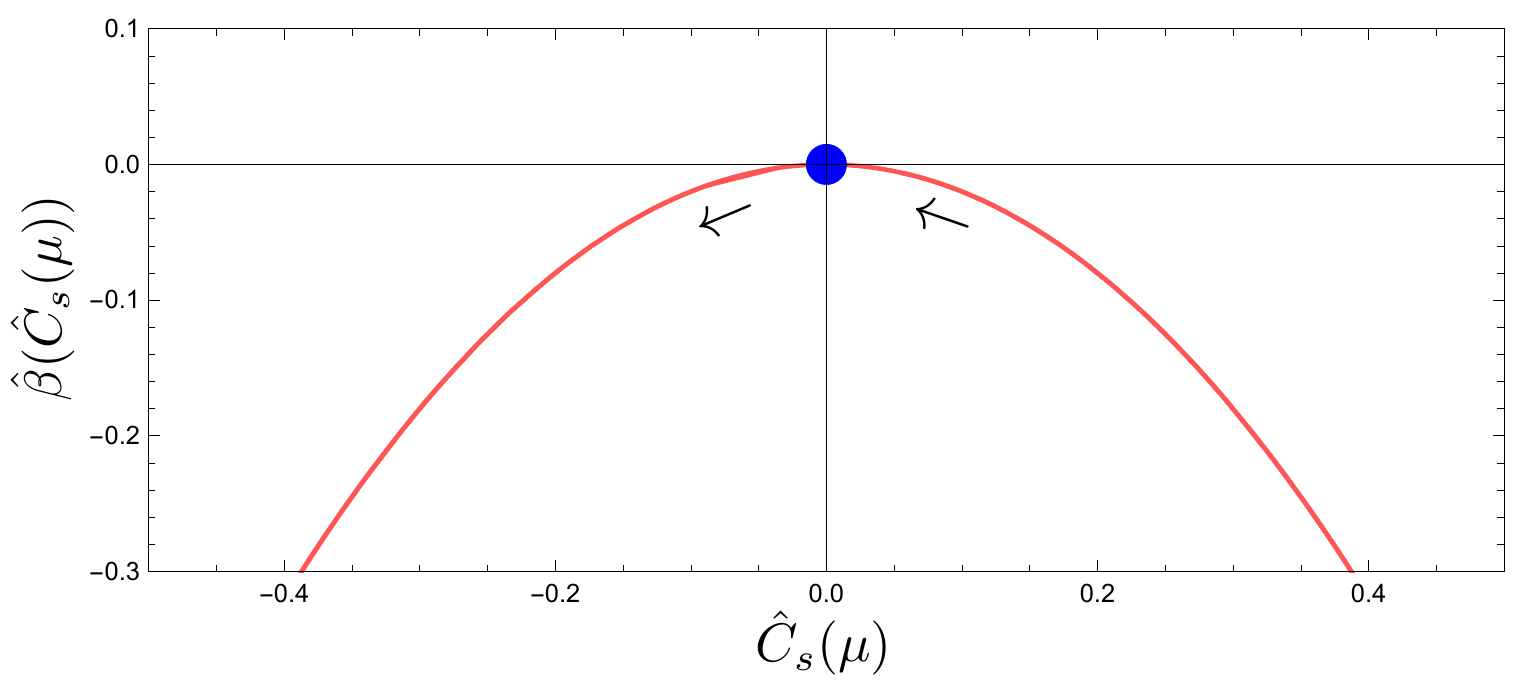}
\caption{The beta-function of Eq.~(\ref{eq:c0betafn2d}) plotted as a function of ${\hat C}_s(\mu)$. The blue dot is the RG fixed point.
  The beta-function curve evolves from $\mu=0$ (the trivial fixed point) to $\mu=\infty$ (again the trivial fixed point). The
fixed point of the scale-inversion symmetry, $\mu^\circ={\textit{\textbf a}}_s^{-1}$,  is at asymptotic infinity.}
  \label{fig:eftc0beta2d}
\end{figure}

It is convenient to define the dimensionless couplings, $\hat{C}_s(\mu) = {- M}C_s(\mu)/({4 \pi})$, whose beta-functions are
\begin{equation}
{\hat\beta({\hat C}_s)}\ =\    \mu \frac{d}{d\mu} \hat{C}_s(\mu) = -2 \hat{C}^2_s(\mu) \ .
  \label{eq:c0betafn2d}
\end{equation}
There is a single RG fixed point $\hat{C}_s = 0$, which is reached asymptotically at $\mu = 0$ and at $\mu = \infty$ as seen in
Fig.~(\ref{fig:eftc0beta2d}). Under an inversion of the RG scale,
$\mu \mapsto ({\textit{\textbf a}}_s^2 \mu)^{-1}$, the running couplings transform as
\begin{equation}
    \hat{C}_s(\mu) \mapsto -\hat{C}_s(\mu) \ ,
\end{equation}
which implies ${\hat\beta({\hat C}_s)}\rvert_{\bar{\mu}} = {\hat\beta({\hat
C}_s)}\rvert_{1/(a_s^2 \bar{\mu})}$ for any $\bar{\mu}$. The fixed point of the scale-inversion
transformation is at the Landau pole, $\mu^\circ = {\textit{\textbf a}}_s^{-1}$.

\begin{figure}[!ht]
\centering
\includegraphics[width = 0.7\textwidth]{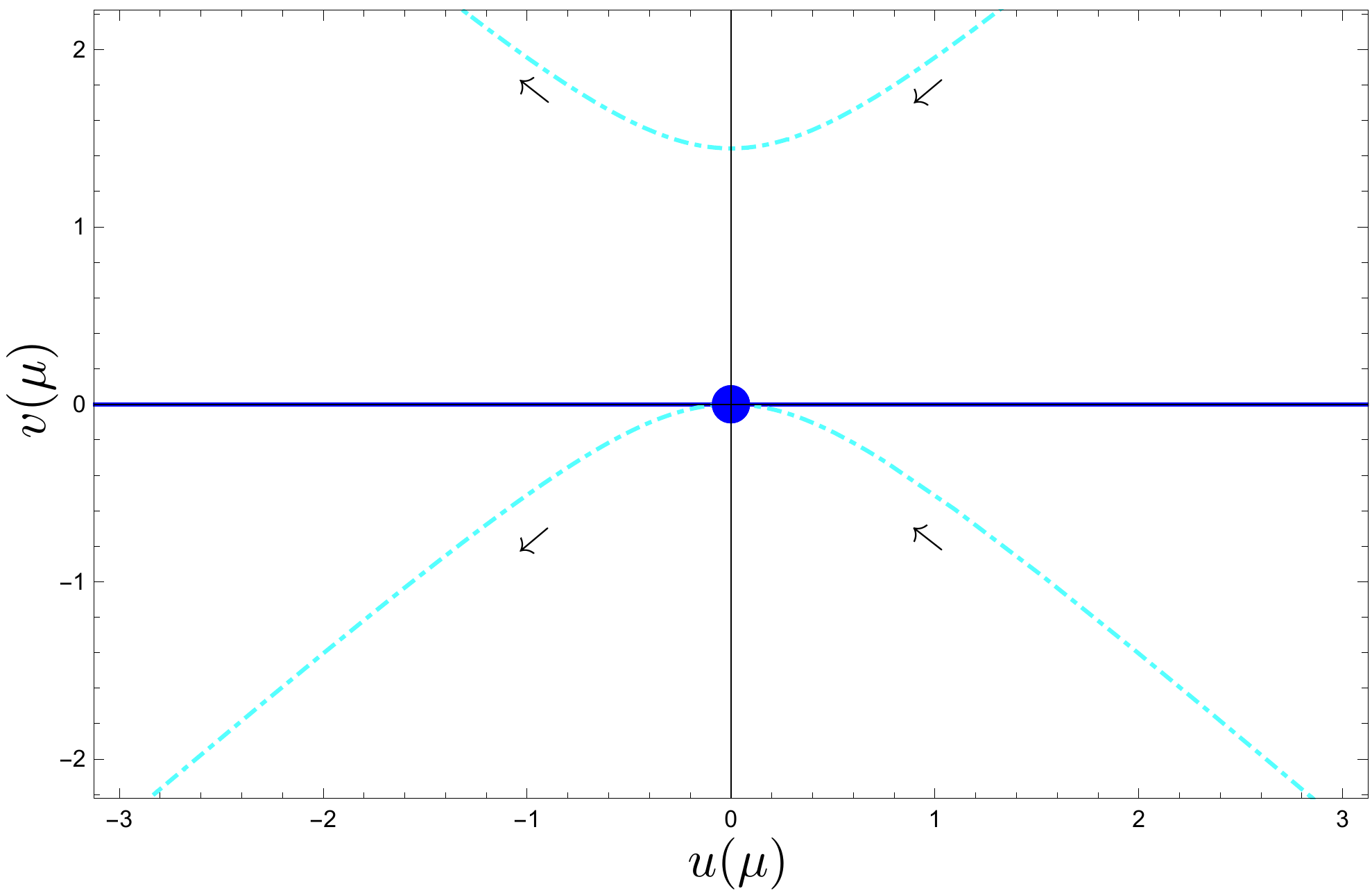}
\caption{The single fixed point of the RG in the $u-v$ plane in $d=2$.
The cyan dashed curve is the RG trajectory that exhibits the UV/IR symmetry. The trajectory begins and ends at the RG fixed point.  The arrows indicate direction of increasing $\mu$.}
  \label{fig:2dRG}
\end{figure}

Considering both s-wave scattering channels simultaneously, the momentum inversion
transformation of the $S$-matrix generalizes to $p \mapsto ({\textit{\textbf
    a}}_1 {\textit{\textbf a}}_0 \, p)^{-1}$ and leaves the
    density matrix invariant. Defining $u(\mu)$ and $v(\mu)$ as in $d = 3$, the components
of the potential flow with the RG according to the algebraic curve
\begin{eqnarray}
{\bar w}\left( v^2-u^2\right) = v\ \ \ , \ \ \ {\bar w} \ \equiv \  \log \left(\frac{ {\textit{\textbf a}}_1}{{\textit{\textbf a}}_0}   \right) \ .
   \label{eq:EFTconic2d}
\end{eqnarray}
The UV/IR transformation on the potential, via the inversion of
the scale, $\mu \mapsto ({\textit{\textbf a}}_1 {\textit{\textbf a}}_0 \,
\mu)^{-1}$, generates a reflection about the $v$-axis, $u \mapsto
-u$, $v \mapsto v$, in the $u-v$ plane. The fixed point of the scale-inversion
symmetry occurs at $\mu^{\circ} = ({\textit{\textbf a}}_1 {\textit{\textbf a}}_0)^{-1/2}$ where $\hat{C}_0 = -\hat{C}_1$. The RG
trajectory of the
potential is shown in Fig.~(\ref{fig:2dRG}). It is clear that, as in
three spatial dimensions, a symmetry of the density matrix is encoded in the
EFT as a reflection symmetry of the RG flow of the potential.
\section{UV/IR symmetry breaking}
\label{sec:eftsb}

\noindent The correspondence between the UV/IR transformation
properties of the S-matrix and the RG flow of EFT couplings is not
confined to the scattering length approximation. The UV/IR
transformation properties of the momentum-dependent corrections in the
effective range expansion are reflected in the EFT interaction
potential via constraints on the RG flow of the associated EFT
couplings.  This section will focus on corrections to the
single-channel example in $d=3$ considered above: the case of a large
s-wave scattering length with perturbative effective range
corrections~\cite{Kaplan:1998tg,Kaplan:1998we,vanKolck:1998bw,Birse:1998dk}.
The methods used in this section are similar to those used in
Ref.~\cite{Beane:2021C} where a distinct UV/IR symmetry was imposed to
constrain the EFT relevant for a non-perturbative treatment of the
effective range.  It will be shown that the UV/IR transformation
properties of the scattering-amplitude corrections reproduces key
features of the known RG flow of higher-order couplings in the EFT,
without explicit calculation of the higher-order effects.

Treating the scattering length to all orders with effective range and
shape parameter effects treated perturbatively, the $d=3$
s-wave\footnote{Note that in this section all spin labels are dropped
  and the results apply equally to the spin-singlet and spin-triplet
  channels.} scattering amplitude is given by the ERE which, up to
NLO,
is~\cite{Kaplan:1998tg,Kaplan:1998we,vanKolck:1998bw,Birse:1998dk}
\begin{eqnarray}
\mathcal{A} &=& -\frac{4 \pi}{M} \frac{1}{1/a + i p}\left (1+ \frac{r/2}{1/a + i p}p^2\right ) \equiv  \mathcal{A}_{-1} (p) + \mathcal{A}_{0}(p) 
\label{eq:pertamp}
\end{eqnarray}
where $r$ is the effective range.  Defining $\hat{p} \equiv a p$, at each order the scattering amplitude transforms simply under the UV/IR transformation, $\hat{p} \to 1/\hat{p}$,
\begin{eqnarray}
\mathcal{A}_{-1}(1/\hat{p}) &=& -i \hat{p}\,\mathcal{A}_{-1}(\hat{p})^* \ \ ,\ \ \mathcal{A}_{0}(1/\hat{p}) \, =\, -\hat{p}^{-2}\,\mathcal{A}_{0}(\hat{p})^* \ .
\label{eq:UVIRamp}
\end{eqnarray}
The scattering amplitude is generated from an EFT of contact interactions via a renormalized, on-shell, tree-level interaction potential
\begin{equation}
    V = C_0(\hat{\mu}) + C_2(\hat{\mu}) \hat{p}^2/a^2 \equiv V_{-1}(\hat{\mu}) + V_{0}(\hat{\mu}, \hat{p})
    \label{eq:pertpot}
\end{equation}
where $C_n$ is the coefficient of the four-fermion interaction with
$n$ derivatives and $\hat{\mu} = a \mu$.  Note that here the RG scale
$\mu$ can represent the PDS scale or a hard cutoff\,\footnote{For the case of cutoff regularization, we are omitting from the potential the counterterms necessary to exactly match onto Eq.~(\ref{eq:pertamp}).}, and therefore
Eq.~(\ref{eq:AitoV}) is not assumed. The potential relevant for
momentum $\hat{p}$ is obtained by setting $\hat{\mu} = \hat{p}$ and,
in a particular renormalization scheme, the UV/IR properties of the
potential should mirror those of the scattering amplitude that it
generates, i.e.  Eq.~(\ref{eq:UVIRamp}). Hermiticity implies that $V =
V^*$ and that there will be no imaginary phases in the UV/IR
transformation. The assumption that the interaction, as represented by
the potential, reflects the UV/IR transformation properties of the
amplitude at each order in the EFT expansion, suggests that by
imposing the UV/IR transformation, $\hat{\mu} \to 1/\hat{\mu}$ and
$\hat{p} \to 1/\hat{p}$, and setting $\hat{\mu} = \hat{p}$, the
potential should transform as
\begin{eqnarray}
    V_{-1}(1/\hat{p}) &=& \epsilon_{\scriptscriptstyle -1}\, \hat{p}\,V_{-1}(\hat{p}) \ \ ,\ \
    V_{0}(1/\hat{p}, 1/\hat{p}) \,=\, \epsilon_{\scriptscriptstyle 0}\, \hat{p}^{-2}\, V_{0}(\hat{p}, \hat{p}) 
    \label{eq:UVIRpot}
\end{eqnarray}
where $(\epsilon_{\scriptscriptstyle -1,0})^2=1$. 
This in turn implies that the renormalized couplings transform as
\begin{eqnarray}
    C_0(1/\hat{\mu}) &=& \epsilon_{\scriptscriptstyle -1}\, \hat{\mu}\,C_0(\hat{\mu}) \ \ ,\ \
    C_2(1/\hat{\mu}) \,=\, \epsilon_{\scriptscriptstyle 0}\, \hat{\mu}^2\, C_2(\hat{\mu}) \ .
    \label{eq:UVIRCn}
\end{eqnarray}
Hence, once $C_0$ is determined, the UV/IR transformation properties imply
$C_2\propto r (C_0)^2$, as confirmed by explicit calculation of higher
order loop effects in the EFT using both PDS and cutoff\,\footnote{Using cutoff 
regularization, the insertion of the $C_n$ operators in the EFT
calculation generates increasingly singular and non-linear cutoff dependence. However,
these singular contributions are
canceled by existing counterterms~\cite{vanKolck:1998bw}, as noted in section~\ref{sec:eftrg}.}
regularization~\cite{Kaplan:1998tg,Kaplan:1998we,vanKolck:1998bw,Birse:1998dk}.
This readily extends to higher orders in the EFT expansion; it is easy
to check that the UV/IR transformations of shape parameter and
higher-order amplitudes constrains $C_{2n}$ to scale as specific
powers of $C_0$.  Furthermore, from Eq.~(\ref{eq:pertamp}), it is
clear that the couplings should all vanish as $a \to 0$. Assuming
polynomial dependence on $a$ and $\mu$, dimensional analysis then
implies that $C_0(\hat{\mu}) = a/M \,f(\hat{\mu})$, and one class of
solutions to the UV/IR constraint is given by
\begin{equation}
    f(\hat{\mu}) = -\frac{c\left(\hat{\mu}+1\right)\hat{\mu}^n}{\hat{\mu}^{2n+2}+\epsilon_{\scriptscriptstyle -1}}
\end{equation}
where $c$ and $n$ are real constants. Setting
$\epsilon_{\scriptscriptstyle -1}=-1$, $n=0$, and $c = 4\pi$ recovers $C_0(\mu)$ in the PDS scheme~\cite{Kaplan:1998tg,Kaplan:1998we}.

\section{Summary and Conclusion}
\label{sec:conc1}

\noindent In both two and three spatial dimensions, the scattering
length approximation to the low-energy, s-wave $S$-matrix has a UV/IR
symmetry which leaves the density matrix of the ``out'' state
invariant. While this is not a symmetry of the scattering amplitude or
effective action in the sense of a transformation on the fields, the
UV/IR symmetry does appear as a symmetry of the RG evolution of the
EFT couplings. In a sense, the echo of the $S$-matrix symmetry in RG
evolution is a consistency check that indeed the $S$-matrix is
rendered UV complete by the symmetry. As the UV/IR transformation maps
threshold to asymptotic infinity and vice versa, its presence signals a
UV-complete description of the scattering event. That is, scattering
is well defined at all distance scales. Clearly then, the UV completeness
should be reflected in the interaction itself, which should also be
defined over all distance scales. In the EFT of contact operators,
this is reflected by the presence of the RG scale $\mu$ which can be
chosen to take any value, and in the UV/IR symmetry of the beta-function.

It was shown that the UV/IR symmetry has utility beyond the scattering
length approximation, where the simple RG evolution between two fixed
points breaks down. Indeed, the manner in which perturbations around
the scattering length approximation break the UV/IR symmetry strongly
constrains the RG flow of higher dimensional couplings in the
corresponding EFT. In addition, there is another class of UV/IR
symmetric $S$-matrices which include effective ranges that are
correlated to the scattering
lengths~\cite{Beane:2020wjl,Beane:2021B}. In that case, the symmetry
also appears in the quantum mechanical potential which gives rise to
the $S$-matrix, albeit in a different manner than was shown in this
paper~\cite{Beane:2021C}.

One of the important conclusions of this chapter is that there are
symmetries of a scattering process which are not manifest symmetries
of the scattering amplitude.  This arises because observables measured
on an ``out'' state depend on the full wave function after
scattering. That is, the contribution from the part of the wave
function which does not scatter (corresponding to the identity
operator in the $S$-matrix) is crucial in constructing the ``out''
state.  Furthermore, as the full wave function may decompose into many
scattering channels, each with their own scattering amplitude, there
can be symmetries which are only apparent if all scattering channels
are viewed holistically.  It would be interesting if such $S$-matrix
symmetries can arise in other contexts, perhaps unrelated to
momentum inversion.

One important issue that has not been addressed
is the spacetime nature of the UV/IR symmetry. As the UV/IR
transformation is an inversion of momentum, it necessarily involves a
scale transformation. Given the results of appendix~\ref{app:Ward}, it appears promising to investigate whether the
UV/IR symmetry could be understood as an extension of Schr\"odinger
symmetry~\cite{Hagen:1972pd,Niederer:1972zz,Mehen:1999nd,Nishida:2007pj}
to systems with finite scattering length.

\begin{subappendices}
\section{Momentum inversion Ward identity}
\label{app:Ward}
\noindent This appendix considers a generalization of the momentum inversion transformation of Eq.~(\ref{eq:singlechanmominv}) 
and derives the associated Ward identity. Consider the $S$-matrix of Eq.~(\ref{eq:Selementscatt}) in the scattering length approximation.
If one allows the momentum, $p$, to span the entire real line, then with respect to the real M\"obius transformation
\begin{eqnarray} \label{eq:mob1a}
p \mapsto \frac{\vartheta\,p+ 1/a_s}{ {\pm\left(a_s\,p-\vartheta \right)}} \ ,
   \end{eqnarray}
with $\vartheta$ an arbitrary real parameter, the $S$-matrix transforms as
\begin{eqnarray} \label{Smatmobius1a}
  S \mapsto \frac{\vartheta \pm i}{\vartheta\mp i}\,\begin{cases}
  S^* \ ,\\
  S\ .
\end{cases}
  \end{eqnarray}
That is, the $S$-matrix transforms to itself or its complex conjugate,
times a constant complex phase. Choosing the positive
sign in the transformation law, and $\vartheta=0$ recovers the UV/IR
transformation of Eq.~(\ref{eq:singlechanmominv}).

The M\"obius transformation is a general mapping of the momentum to itself and therefore
generally contains UV/IR transformations. In what follows, the Ward identity
for this symmetry will be derived. Let $\hat p\equiv a_s\, p$, and
consider the infinitesimal version of Eq.~(\ref{eq:mob1a}), with the
minus sign chosen in the transformation law. Since $\vartheta$ large
recovers the identity, take $\vartheta=1/\epsilon$ with $\epsilon$
infinitesimal. Then we see that Eq.~(\ref{eq:mob1a}) becomes
\begin{eqnarray} \label{eq:mob2a}
\hat p \mapsto \frac{\hat p+ \epsilon}{ {-\epsilon\,\hat p+1}}  \ .
   \end{eqnarray}
Now consider the infinitesimal translation
\begin{eqnarray} \label{eq:lm1a}
\hat p \mapsto  \hat p\;+\; \epsilon \ ,
\end{eqnarray}
the infinitesimal dilatation
\begin{eqnarray} \label{eq:l0a}
\hat p \mapsto  e^\epsilon \hat p = \hat p\;+\; \epsilon\,\hat p \;+\; \mathcal{O}(\epsilon^2) \ ,
\end{eqnarray}
and finally consider two inversions with a translation in between,
\begin{eqnarray} \label{eq:l1a}
\hat p \mapsto \left({\hat p}^{-1}-\epsilon\right)^{-1} = \frac{\hat p}{1-\epsilon\,\hat p} = \hat p\;+\; \epsilon\,\hat p^2 \;+\; \mathcal{O}(\epsilon^2) \ .
   \end{eqnarray}
This final step is critical in giving an infinitesimal description of momentum inversion.
These three transformations are generated by the differential operators $L_{-1}$, $L_{0}$,  and $L_{1}$, respectively, where
\begin{eqnarray} 
L_k\equiv - {\hat p}^{k+1}\frac{\partial}{\partial \hat p}  \ .
   \end{eqnarray}
These satisfy the ${\mathfrak sl}(2,\mathbb{R})$ algebra:
\begin{eqnarray} 
&& {[\, L_1\,  ,\, L_{-1}\, ]}\, =\, 2\,L_0 \ \ \ ,\ \ \  {[\, L_{\pm 1}\,  ,\, L_0\, ]}\, =\, \pm\,L_{\pm 1} \ .
\label{eq:sl2ralga}
\end{eqnarray}
Note that the general M\"obius transformation matrix of Eq.~(\ref{eq:mob1a}) has determinant $\mp(\vartheta^2+1)$
and therefore is an element of $PSL(2,\mathbb{R})$ only in the special case where the minus sign is chosen in the transformation law
and $\vartheta=0$\,\footnote{If $a_s = \pm 1$ then Eq.~(\ref{eq:mob1a}) is an element of the modular group, $PSL(2,\mathbb{Z})$.}. The general M\"obius transformation is an element of $PGL(2,\mathbb{R})$.

It is easy to check that the (infinitesimal) M\"obius transformation of Eq.~(\ref{eq:mob2a}) is constructed by 
successive transformations of Eq.~(\ref{eq:l1a}) and Eq.~(\ref{eq:lm1a}); that is, it is generated by $L_{1}$ and $L_{-1}$. Indeed the Ward identity is:
\begin{eqnarray} 
\left(L_{1} + L_{-1}-2 i \right)S\ =\ 0 \ .
   \end{eqnarray}
The general solution of this differential equation is Eq.~(\ref{eq:Selementscatt}), up to an overall complex coefficient.
Note that the (broken) on-shell dilatation Ward identity of the Schr\"odinger group~\cite{Mehen:1999nd} takes
the form~\cite{Beane:2020wjl}
\begin{eqnarray} 
  L_0 \: S&=& -\oneht\left( S^2-1\right)  \,
\end{eqnarray}
and, as expected, is respected at the RG fixed points, $S=\pm 1$. 

\end{subappendices}

\chapter{UV/IR Symmetric Effective Field Theories for the Nucleon-Nucleon Interaction}
\label{chap:Weinberg}
\noindent 
\textit{This chapter is associated with
``Symmetries of the Nucleon-Nucleon S-matrix and Effective Field Theory Expansions"~\cite{Beane:2021C} by Silas R. Beane and Roland C. Farrell.}
\section{Introduction}
\label{intro}

\noindent 
In this chapter the UV/IR symmetries explored in the previous chapters are used to construct a UV/IR symmetric EFT for the two nucleon system.
It is shown that the assumption of a UV/IR symmetry highly constrains the kinematical dependence of the interaction. 
These constraints manifest as a set of algebraic equations, that are solved by a ``Yamaguchi"-like potential~\cite{Yamaguchi:1954mp}.
This UV/IR symmetric interaction has the scattering length and effective ranges treated to all orders, with higher order shape parameters set to zero.
The UV/IR symmetric interaction constitutes LO in this new EFT, and it is shown how UV/IR symmetry breaking can be included as higher orders in perturbation theory.

\section{UV/IR symmetries of the NN s-wave S-matrix}
\label{sec:uvir}

\noindent It may seem that momentum inversion transformations are orthogonal to the idea of EFT since
they interchange the UV and the IR. Note however that at LO in the
EFT, one is considering scattering in a limit in which all
short-distance mass scales are taken to be very large. In this
limit, long-distance forces (like pion exchange) and inelastic
thresholds (like the pion-production threshold) are only probed as
momentum approaches infinity. Therefore it is reasonable, at LO, to consider
transformations of the momenta over the entire momentum half-line, $0<k<\infty$. 

Realistic NN scattering at low energies is not a single-channel system as the initial-state nucleons can be arranged into two distinct spin configurations. The NN $S$-matrix at very low energies is dominated by the s-wave, see Eq.~\ref{eq:Sdef1}.
The low-energy $S$-matrix can be well reproduced by the ERE truncated at the effective range, and below
the following physical effective range parameters will be used~\cite{Kaplan:1998we,deSwart:1995ui}: $a_{0}
= -23.714(13)$ fm, $a_{1} = 5.425(1)$ fm, $r_{0} = 2.73(3)$ fm, and
$r_{1} = 1.749(8)$ fm.

With effective range corrections included, the momentum inversion transformation
\begin{eqnarray}
p\mapsto \frac{1}{\lambda |a_1 a_0| p} \ ,
   \label{eq:moebiusgen3}
\end{eqnarray}
with the arbitrary real parameter $\lambda >0$, implies
\begin{eqnarray} 
&&   \delta_0(p) \mapsto \delta_0(p) \ \ \ , \ \ \ \ \ \, \delta_1(p) \mapsto -\delta_1(p) \ ,
  \label{eq:confmoebiusiso5}
\end{eqnarray}
but only in the special case\footnote{The general case is considered in Sec.~\ref{sec:UVIRGeneral}.} where the effective ranges are correlated with the scattering lengths 
as 
\begin{eqnarray} 
r_0&=&2\lambda a_1 \ \ \ ,\ \ \ r_1\,=\,-2\lambda a_0\ .
  \label{eq:confmoebiusiso6}
\end{eqnarray}

This UV/IR symmetry has interesting implications for
nuclear physics. The measured singlet NN scattering phase shift rises
steeply from zero due to the unnaturally large scattering length, and
then, as momenta approach inelastic threshold, the phase shift goes
through zero and becomes negative, indicating the fabled short-distance
repulsive core. While this impressionistic description assigns
physics to the potential, which is not an observable and indeed need
not be repulsive at short distances, the UV/IR symmetry directly
imposes the physically observed behavior of the phase shift. Even though the
singlet phase shift changes sign at momenta well beyond the range of
applicability of the pionless theory, ascribing this symmetry to the
LO results ---through the required presence of range corrections--- results
in a more accurate LO prediction than the usual pionless expansion~\cite{SanchezSanchez:2017tws}.

\section{EFT description: single-channel case}

\subsection{Potential and Lippmann-Schwinger equation}

\noindent It is convenient to introduce the generic UV scale ${\cal M}$ and the
generic IR scale $\aleph$. The EFT will describe physics for $k\sim \aleph \ll {\cal M}$.
Subleading corrections to the scattering amplitude in the EFT are expected to be parametrically
suppressed by powers of $k/{\cal M}$. For the NN system at very-low energies, described by the pionless EFT, the UV scale is ${\cal M}\sim M_\pi$.

The s-wave potential stripped from the most general effective Lagrangian of four-nucleon contact operators is~\cite{Kaplan:1996nv,Phillips:1997xu}
\begin{equation}
V(p',p)=C_0 + C_2 \left(p^2 + p'^2\right) + C_4 \left(p^4 + p'^4\right) + C_4' p^2 p'^2+\dots \ ,
\label{eq:Vexp}
\end{equation}
where the $C_n$ are the bare coefficients.
The scattering amplitude is obtained by solving the Lippmann-Schwinger (LS) equation with this potential
\begin{equation}
T(p',p;E)=V(p',p) + M\int \frac{d^3q}{(2 \pi)^3} \, V(p',q) 
\frac{1}{EM- {q^2}+i\epsilon} T(q,p;E) \ .
\label{eq:LSE}
\end{equation}
As the potential is separable to any order in the momentum expansion, the scattering amplitude can be
obtained in closed form to any desired order in the
potential~\cite{Phillips:1997xu}. Of course the singular nature of the
potential requires regularization and renormalization. The scattering
amplitude can be regulated using, for instance, dimensional regularization and its various schemes,  or by
simply imposing a hard UV cutoff, $\Lambda$, on the momentum integrals.

\subsection{Matching equations}

\noindent In the language of cutoff regularization, matching the solution of the LS equation with the ERE of
Eq.~(\ref{eq:reamp}), formally gives the all-orders matching equations
\begin{align}
a &= f_0(\Lambda; C_0,C_2,C_4,\ldots) \nn \ , \\
r  &= f_2(\Lambda; C_0,C_2,C_4,\ldots) \nn \ , \\
v_n  &= f_{2n}(\Lambda; C_0,C_2,C_4,\ldots) \ ,
  \label{eq:match}
\end{align}
where the $f_{2n}$ are non-linear functions determined by solving the
LS equation.  These equations can be inverted to find the now
cutoff-dependent coefficients $C_{2n}(\Lambda;a,r,v_2,\ldots)$.  To
obtain an EFT with predictive power, it is necessary to identify the
relative size of the effective range parameters. If they all scale as
powers of ${\cal M}^{-1}$, then the entire potential of
Eq.~(\ref{eq:Vexp}) can be treated in perturbation theory for momenta
$k\ll {\cal M}$. If there are big parts that instead scale as powers
of ${\aleph}^{-1}$, they will, via the interactions of Eq.~(\ref{eq:Vexp}), constitute the LO potential in the EFT
expansion.

\subsection{Scattering length approximation}

\noindent With $a\sim \aleph^{-1}$ and the effective range and shape parameters of 
natural size, $r\sim {\cal M}^{-1}$, $v_n\sim {\cal M}^{-2n+1}$, 
the amplitude can be expanded for $k\ll {\cal M}$ as
\begin{equation}
T(k)\ =\ -\frac{4 \pi}{M}\left( -\frac{1}{a} - i k \right)^{-1} \bigg\lbrack 1 + O(k) \bigg\rbrack \ .
\label{eq:reampb3}
\end{equation}
In order to generate this expansion in the EFT, the potential is written as
\begin{equation}
V(p',p)=V_{\LOss}+V_r(p',p) \ ,
\label{eq:Vexp2}
\end{equation}
where $V_{\LOss}=C_0$ is treated exactly in the LS equation and the residual potential, $V_r$, which includes range
and shape parameter corrections, is treated in perturbation theory.
Keeping the first term in the s-wave potential, the solution of the LS equation is
\begin{equation}
T_{\LOss}(k) =\left( \frac{1}{C_0} - {\mathbb{I}}(k) \right)^{-1},
\label{eq:sla1}
\end{equation}
where 
\begin{equation}
\mathbb{I}(k) \equiv  \left(\frac{\omega}{2}\right)^{3-d}  M\int \frac{d^dq}{(2 \pi)^{d}} \, \frac{1}
{{k^2}- {q^2}+i\epsilon}\ \  \mapup{\rm PDS}\ \ -\frac{M}{4\pi}\left(\omega +i k \right) \ ,
\label{eq:IEdef}
\end{equation}
has been evaluated in dimensional regularization with the PDS scheme~\cite{Kaplan:1998tg,Kaplan:1998we} and renormalized at the RG scale $\omega$.
The matching equations in this case are
\begin{eqnarray}
&& a = \left(\frac{4\pi}{M C_0} + \omega \right)^{-1}  \ ,\nn \\
&& r = v_n  = 0 \ .
  \label{eq:matchsla}
\end{eqnarray}
Inverting one finds
\begin{equation}
  C_0(\omega) \ =\ \frac{4 \pi}{M} \frac{1}{{1}/{a}-{\omega}} \ .
\end{equation}
The coupling at the unitary fixed point, $C_0=C_{0\star}$, corresponds to a divergent scattering
length (unitarity). A rescaled coupling can be defined as
${\hat C}_0 \equiv C_0/C_{0\star}$. The corresponding beta-function 
is then 
\begin{equation}
  {\hat\beta({\hat C}_0)}\ =\  \omega \frac{d}{d\omega} {\hat C}_0(\omega) \ =\ -{\hat C}_0(\omega)\left({\hat C}_0(\omega)-1\right) \ ,
   \label{eq:c0betafn3}
\end{equation}
which explicitly has fixed points at ${\hat C}_0=0$ and $1$. Note that the beta-function inherits the properties of the UV/IR symmetry as shown in Chapter~\ref{chap:SmatUVIR}.

\subsection{Range corrections with zero-range forces}

\noindent With $a, r\sim \aleph^{-1}$ and shape parameters of 
natural size, $v_n\sim {\cal M}^{-2n+1}$, 
the amplitude can be expanded for $k\ll {\cal M}$ as
\begin{equation}
T(k)\ =\ -\frac{4 \pi}{M}\left( -\frac{1}{a} + \frac{1}{2} r {k^2} - i k \right)^{-1} \bigg\lbrack 1 + O(k^3) \bigg\rbrack \ .
\label{eq:reampb2}
\end{equation}
In order to generate this expansion from the EFT perspective, one may choose $V_{\LOss}=C_0+C_2
\left(p^2 + p'^2\right)$. Due to the highly singular UV behavior of this potential, cutoff
regularization will be used to carefully account for the divergences that are generated when
solving the LS equation. The amplitude is found in closed form to be~\cite{Phillips:1997xu}
\begin{equation}
T_{\LOss}(k)=\left(\frac{(C_2 I_3 -1)^2}{C + C_2^2 I_5 + {k^2} C_2 (2 - C_2 I_3)} - {\mathbb I}(k)\right)^{-1} \ ,
\label{eq:Tonexp}
\end{equation}
where now ${\mathbb I}(k)$ is evaluated with cutoff regularization, and
\begin{equation}
I_n \equiv -M \int \frac{d^3q}{(2 \pi)^3} q^{n-3}\theta \left(\frac{\pi}{2}\Lambda-q\right)=-\frac{M \Lambda^n}{2^{1+n}\pi^{2-n}n}\ .
\label{eq:In}
\end{equation}
Matching the amplitude to the ERE gives
\begin{eqnarray}
a &=& \frac{M}{4 \pi}\left(\frac{(C_2 I_3 -1)^2}{C + C_2^2 I_5} - {I_1}\right)^{-1} \ ,\nn \\
r &=& \frac{8 \pi}{M}\left(\frac{(C_2 I_3 -1)^2}{C + C_2^2 I_5}\right)^2
\left[\frac{1}{(C_2 I_3 - 1)^2 I_3} - \frac{1}{I_3}\right]    \ ,   \nn \\
v_n  &=&  -\frac{4 \pi}{M}\frac{C_2^n(C_2 I_3 -2)^n(C_2 I_3 -1)^2}{(C + C_2^2 I_5)^{n+1}} \ .
  \label{eq:matchcrm}
\end{eqnarray}
Taking the limit $\Lambda \rightarrow \infty$ then recovers the ERE
with all shape parameter corrections vanishing, $v_n=0$. However, it
is straightforward to verify that in this limit, $r \leq 0$, as
required by the Wigner bound~\cite{Wigner:1955zz}. If s-wave NN scattering involved
negative effective ranges and resonances rather than bound states,
then this scheme, with the effective potential consisting of a finite
number of strictly delta-function potentials, would
suffice~\cite{Habashi:2020qgw}. However, as the s-wave NN effective ranges are
both positive, it is clear that $\Lambda$ must be kept
finite. In that case the higher-order terms in the bare
potential, even if neglected in the choice of $V_{\LOss}$, are generated
quantum mechanically at LO as evidenced by the non-vanishing shape
parameters. Therefore the higher order operators should be kept from the start. That is, one
is back to the general, formal statement of the matching conditions
given in Eq.~(\ref{eq:match}), where a renormalization scheme is
required which ensures that $v_n=0$, and a choice of $V_{\LOss}$ must be
found which achieves this while including all orders in the momentum
expansion.

\subsection{Range corrections with a finite-range scheme: LO}

\noindent The lesson provided by the Wigner bound is that if range corrections are treated exactly, then the LO potential
should, in general, include all orders in the momentum expansion. Therefore, the potential
can be written as in Eq.~(\ref{eq:Vexp2}) except now with a momentum dependent LO potential
\begin{equation}
V(p',p)=V_{\LOss}(p',p) + V_r(p',p) \ .
\label{eq:VexpEF}
\end{equation}
The residual potential, $V_r$, which accounts for NLO and higher effects in
perturbation theory, will be considered in detail below.
Now the potential is non-unique and there is no reason for the
separation into $V_{\LOss}$ and $V_r$ to be unique. Ideally one
finds a LO potential which identically gives the ERE with all
shape parameters vanishing, and indeed that is what will be achieved
by imposing the UV/IR symmetry at the level of the interaction.

As the s-wave EFT potential is non-local (non-diagonal in coordinate space) and separable to any order in the momentum expansion, it appears sensible to assume that the LO potential which generates the ERE with range corrections only is non-local and separable. However, such an assumption is not necessary: all potentials which generate the ERE truncated at the effective
range are, by definition, phase equivalent\footnote{Note that the potential
considered in the previous section is phase equivalent only in the limiting sense of
$\Lambda \to \infty$ and $r < 0$.}. Once one potential is found, others can be
obtained by unitary transformation. Here the simplest possibility
will be considered; that the LO potential is (rank-one) separable.

For a separable potential $V(p,p')$, the on-shell scattering amplitude
solves the LS equation algebraically to
\begin{equation}
T(k)\ =\ {V(k,k)}\left(1-M\int \frac{d^3q}{(2 \pi)^{3}} \, \frac{V(q,q)}
{{k^2}- {q^2}+i\epsilon}\right)^{-1} \ .
\label{sepsol}
\end{equation}
The momentum inversion symmetry of Eq.~(\ref{eq:singlechanmominv})
(assuming for simplicity $\eta\equiv a r/2 >0$) implies
\begin{equation}
  T(k)\mapsto T\left(\frac{1}{{\eta k}}\right) \ =\ -\eta k^2\, T^*(k) \ .
   \label{sepsol2}
\end{equation}
With the range of the potential taken to be
the momentum scale $\mu$, the potential can be taken to be the real function $V_{\LOss}(\mu;p,p')$. Constraints on the $S$-matrix concern the potential
$V_{\LOss}(\mu;p,p)\equiv V_{\LOss}(\mu;p)$. Therefore,
\begin{eqnarray}
  \hspace{-0.3in}T^*(k) & =&
{V_{\LOss}\left(\mu;k\right)}\left({1-M\int \frac{d^3q}{(2 \pi)^{3}} \, \frac{V_{\LOss}(\mu;q)} {{k^2}- {q^2}-i\epsilon}}\right)^{-1} \nn \\
     & =&
     \frac{-\frac{1}{\eta k^2} V_{\LOss}\left(\mu;\frac{1}{ {\eta k}}\right)}
     {\bigg\lbrack1-\frac{M}{2\pi^2}\int dq \, \left(-\frac{1}{\eta q^2}\right){V_{\LOss}(\mu;\frac{1}{{\eta q}})}\bigg\rbrack
-M\int \frac{d^3q}{(2 \pi)^{3}} \,  \left(-\frac{1}{\eta q^2}\right)\frac{V_{\LOss}(\mu;\frac{1}{{\eta q}})} {{k^2}- {q^2}-i\epsilon}}
       \label{eq:LSgen}
\end{eqnarray}
where the first line follows from Eq.~(\ref{sepsol}) and the second line follows from Eq.~(\ref{sepsol2}).
Now consider non-singular solutions of this equation of the form
\begin{equation}
V_{\LOss}\left(\mu;\frac{1}{{\eta q}}\right)\ = \ \epsilon\, \eta q^2\; V_{\LOss}(\nu;q)\ ,
\label{eq:yamatransfgen}
\end{equation}
where $\epsilon=\pm 1$, and the range of the potential has been allowed to vary under momentum inversion,
and so $\nu$ is an independent scale in correspondence with the range of the transformed potential. Integrating this equation gives
\begin{equation}
\int_0^\infty dq\, V_{\LOss}\left(\mu;q\right)\ = \ \epsilon\,\int_0^\infty dq\, V_{\LOss}(\nu;q)\ .
\label{eq:intcongen}
\end{equation}
With the choice $\epsilon=-1$ and $\mu=\nu$, this integral must vanish identically and there is clearly a formal solution
to Eq.~(\ref{eq:LSgen}). However, an explicit solution does not seem to exist for a real, finite-range potential.

What follows focuses on the case $\epsilon=1$ and $\mu\neq \nu$. In this case, Eq.~(\ref{eq:yamatransfgen}) is solved by
the ratio of polynomials
\begin{equation}
  V^n_{\LOss}\left(\mu;p\right)\ = \ \frac{N}{M\mu}\frac{1-\left(\frac{p^2}{\mu^2}\right)^{n+1}}{1-\left(\frac{p^2}{\mu^2}\right)^{n+2}} \ ,
\label{eq:gensolinv}
\end{equation}
where $n\in \mathbb{Z}: n\geq 0$,  $N$ is a dimensionless normalization constant and $\mu\,\nu=1/\eta$. One finds
\begin{equation}
\int_0^\infty dq\, V^n_{\LOss}\left(\mu;q\right)\ = \ \int_0^\infty dq\, V^n_{\LOss}\left(\nu;q\right)\ = \ \frac{\pi  N}{M (n+2)} \cot \left(\frac{\pi }{2 n+4}\right) \ .
\label{eq:intcongen2}
\end{equation}
With the postulated solution, Eq.~(\ref{eq:LSgen}) takes the form
\begin{eqnarray}
  \hspace{-0.3in}T^*(k) & =&
{V^n_{\LOss}\left(\mu;k\right)}\left({1-M\int \frac{d^3q}{(2 \pi)^{3}} \, \frac{V^n_{\LOss}(\mu;q)} {{k^2}- {q^2}-i\epsilon}}\right)^{-1} \nn \\
& =&
{{\bar V}^n_{\LOss}\left(\nu;k\right)}\left({1-M\int \frac{d^3q}{(2 \pi)^{3}} \, \frac{{\bar V^n}_{\LOss}(\nu;q)} {{k^2}- {q^2}-i\epsilon}}\right)^{-1} \ ,
       \label{eq:LSgenspec}
\end{eqnarray}
where
\begin{eqnarray}
  {\bar V}^n_{\LOss}\left(\nu;k\right)&=& -{V}^n_{\LOss}\left(\nu;k\right)\left( 1 + \frac{M}{2\pi^2}\int_0^\infty dq\,V^n_{\LOss}\left(\nu;q\right)\right)^{-1} \nn\\
&=& -\frac{N}{M\nu\left(1+\frac{  N}{2\pi (n+2)} \cot \left(\frac{\pi }{2 n+4}\right)\right)}\left(\frac{1-\left(\frac{k^2}{\nu^2}\right)^{n+1}}{1-\left(\frac{k^2}{\nu^2}\right)^{n+2}}\right) \ .
       \label{eq:PEpot}
\end{eqnarray}
Now all that is required is to show that $V^n_{\LOss}(\mu;k)$ and ${\bar V}^n_{\LOss}(\nu;k)$ are phase-equivalent potentials. Equating
the two sides of Eq.~(\ref{eq:LSgenspec}) at both $k=0$ and $k = \sqrt{\mu \nu}$
yields a single solution for $n$ and $N$:
\begin{equation}
n=0 \ \ \ , \ \ \ N= -4\pi\left(1 + \frac{\nu}{\mu}\right)\ .
\label{erepars2}
\end{equation}
Solving the LS equation with either phase-equivalent potential gives
\begin{equation}
a= \frac{1}{\mu}+\frac{1}{\nu}  \ \ , \ \ {r}=\frac{2}{\mu+\nu} \ \ ,\ \ v_n=0 \ .
\label{erepars}
\end{equation}
Finally, the potential takes the separable and non-local form~\cite{Yamaguchi:1954mp,Beane:1997pk,Phillips:1999bf}
\begin{equation}
  V_{\LOss}\left(\mu,\nu;p,p'\right)\ = \ -\frac{4\pi}{M\mu}\left(1 + \frac{\nu}{\mu}\right)\frac{1}{  \sqrt{1+\frac{p^2}{\mu^2}}\sqrt{1+\frac{p^{\prime 2}}{\mu^2}}   } \ ,
\label{eq:gensolinvyam}
\end{equation}
with phase-equivalent potential $\bar V_{\LOss}\left(\mu,\nu;p,p'\right)=V_{\LOss}\left(\nu,\mu;p,p'\right)$. 

Note that the potential of Eq.~(\ref{eq:gensolinvyam}) satisfies the scaling law of Eq.~(\ref{eq:yamatransfgen}) only if $N$ is held fixed. In some sense, the original potential that appears in the
LS equation may be viewed as a bare potential which is determined by requiring that Eq.~(\ref{eq:yamatransfgen}) leave the LS form invariant. The corresponding transformation of the $S$-matrix, $S \to S^*$, is seen only after solving the LS
equation, which fixes $N$ to its $\mu$ and $\nu$ dependent value in Eq.~(\ref{erepars2}). $N$ is therefore a kind of anomalous scaling factor which takes the bare potential to the renormalized form of Eq.~(\ref{eq:gensolinvyam}). Letting the UV/IR transformation act on both the momenta, $p$, and the scales, $\mu$ and
$\nu$, (i.e. $\mu \leftrightarrow \nu$), generates a transformation on $V_{\LOss}$ ($\bar V_{\LOss}$) which is augmented from that of Eq.~(\ref{eq:yamatransfgen}) by an anomalous
scaling factor of $\nu/\mu$ ($\mu/\nu$). One important observation is that, up to an anomalous scaling factor and a sign, $V_{\LOss}$ and $\bar V_{\LOss}$ transform in the same manner as the amplitude that they generate, Eq.~(\ref{sepsol2}). 

With $\mu$ and $\nu$ intrinsically positive, the general case, with scattering length and effective range of any sign
is obtained by taking $\zeta\mu\,\nu=1/\eta$, with $\zeta=1$ corresponding to $a r >0$ and $\zeta=-1$ corresponding to $a r <0$.
The general solution is then
\begin{equation}
  V_{\LOss}\left(\mu,\nu;p,p'\right)\ = \ -\frac{4\pi}{M\mu}\left(1 + \zeta\frac{\nu}{\mu}\right)\frac{1}{  \sqrt{1+\frac{p^2}{\mu^2}}\sqrt{1+\frac{p^{\prime 2}}{\mu^2}}   } \ ,
\label{eq:gensolinvyamanysing}
\end{equation}
with phase-equivalent potential  $\bar V_{\LOss}\left(\mu,\nu;p,p'\right)=\zeta V_{\LOss}\left(\nu,\mu;p,p'\right)$, and
\begin{equation}
a= \frac{1}{\mu}+\frac{1}{\zeta \nu}  \ \ , \ \ {r}=\frac{2}{\mu+ \zeta \nu} \ \ ,\ \ v_n=0 \ .
\label{ereparsgen}
\end{equation}

Having both $a$ and $r$ large as compared to the (inverse) UV scale
${\mathcal M}^{-1}\sim M^{-1}_\pi$ generally requires $\mu,\nu\sim \aleph$.
Expanding
$V_{\LOss}$ in powers of the momenta for $p,p'\ll\aleph$ and matching
onto the momentum expansion of Eq.~(\ref{eq:Vexp}) leads to the scaling
\begin{eqnarray}
C^{(\prime)}_{\scriptscriptstyle{LO}\,m} \ \sim \ \frac{4\pi}{M \aleph^{m+1}} \ .
\label{eq:bcpmodel3}
\end{eqnarray}
The coefficients of the residual potential 
are expected to be suppressed, in a manner to be determined below, by the UV scale. As
the potential is not unique, the decomposition into IR enhanced and UV
suppressed contributions is not unique.  Treating the expanded LO
potential as a renormalization scheme, then for momenta $p,p'\sim
\aleph$, all $C^{(\prime)}_{\scriptscriptstyle{LO}\,m}$ terms in the
potential should be summed into the LO potential to give
Eq.~(\ref{eq:gensolinvyamanysing}) which is treated exactly in the LS
equation, while the residual potential is treated in perturbation theory.
\begin{figure*}
 \centerline{\includegraphics[width=0.7\textwidth]{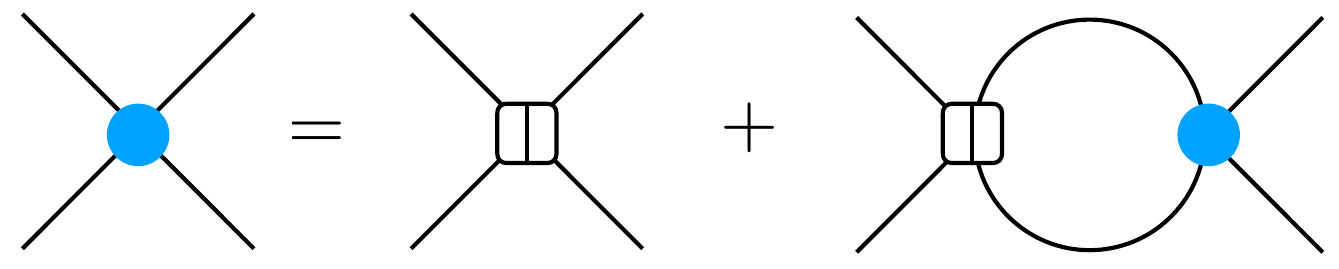}}
\caption{The LO scattering amplitude from the LS equation. The blue circle corresponds to the LO amplitude and the split square is an insertion of the LO separable potential.}
\label{fig:DiaLO}       
\end{figure*}

\subsection{Range corrections with a finite-range scheme: NLO}

\noindent Recall that treating $a,r\sim \aleph^{-1}$ and $v_n\sim {\mathcal M}^{-2n+1}$, for $k\ll {\mathcal M}$, the ERE of Eq.~(\ref{eq:reamp}) can be expanded to
give the NLO amplitude 
\begin{equation}
T_{\NLOss}(k)\ =\ \frac{4 \pi}{M}\left( -\frac{1}{a} + \frac{1}{2} r {k^2} - i k \right)^{-2} v_2 k^4 \ .
\label{eq:eftNLOrr}
\end{equation}
Note that it has been assumed in this expression that $r$ is close to its ``physical'' value.
More generally, and of greater utility when considering realistic NN scattering, one can decompose $r=r_{\LOss}+r_{\NLOss}$, where $r_{\LOss}\sim \aleph^{-1}$ and $r_{\NLOss}\sim {\mathcal M}^{-1}$. In this case
\begin{equation}
T_{\NLOss}(k)\ =\ \frac{4 \pi}{M}\left( -\frac{1}{a} + \frac{1}{2} r_{\LOss} {k^2} - i k \right)^{-2} \oneht r_{\NLOss} k^2
\label{eq:eftNLOrr2}
\end{equation}
so that shape-parameter corrections enter at NNLO (as a subleading contribution to range-squared effects).

\begin{figure*}
 \centerline{\includegraphics[width=0.97\textwidth]{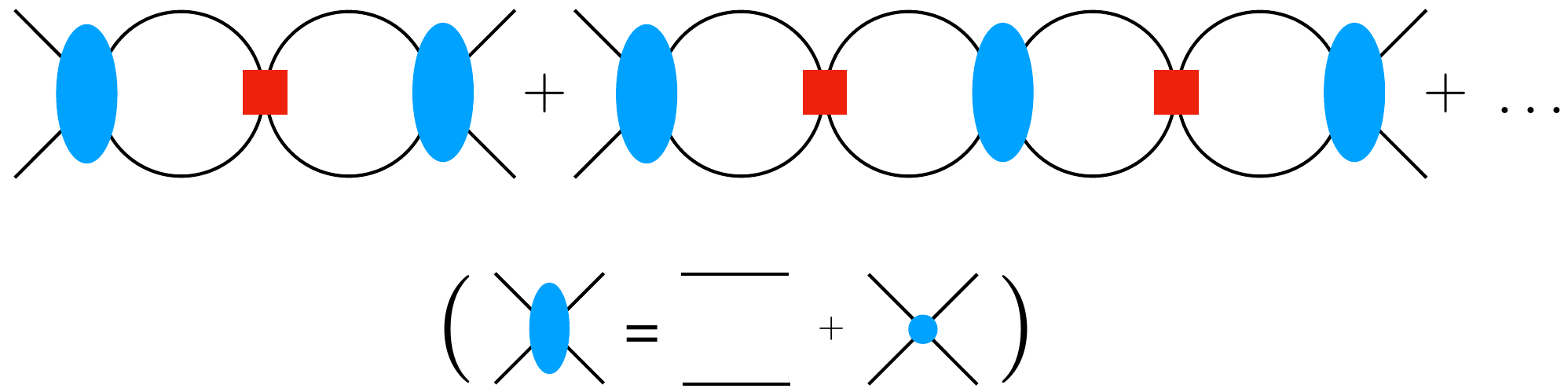}}
\caption{The NLO and beyond scattering amplitude in perturbation theory. The blue circle corresponds to the LO amplitude and the red square is an insertion of the NLO potential.}
\label{fig:DiaNLO}       
\end{figure*}

The goal in what follows is to generate the NLO amplitudes of Eq.~(\ref{eq:eftNLOrr2}) and Eq.~(\ref{eq:eftNLOrr}) (in that order) in the EFT.
The LS equation, Eq.~(\ref{eq:LSE}), for the full scattering amplitude is symbolically expressed as
\begin{equation}
T\ = \ V +  V\, G\, T \ ,
\label{eq:LSE2}
\end{equation}
where $G$ is the two-particle Green's function. Expanding the scattering amplitude and potential as
\begin{equation}
T=T_{\LOss}+T_{\NLOss}+\ldots \ \ \ ,\ \ \  V=V_{\LOss}+V_r
\label{eq:TandVexp}
\end{equation}
leads to $T_{\LOss}$ as an exact solution of the LS equation, as illustrated in Fig.~\ref{fig:DiaLO}, and the NLO and beyond amplitude
\begin{equation}
T_{\NLOss}\ +\ \ldots =V_r + V_r\, G\, T_{\LOss} + T_{\LOss}\, G\, V_r + T_{\LOss}\, G\, V_r\, G\, T_{\LOss} \ \ldots \ ,
\label{eq:LSE3}
\end{equation}
as illustrated diagrammatically to NLO via the Feynman diagrams in Fig.~\ref{fig:DiaNLO}. The form of
the bare EFT potential $V_r$ which matches to the expanded ERE is
straightforward to find using the UV/IR symmetry.  It is convenient to express the LO potential in the compact form
\begin{equation}
  V_{\LOss}\left(p,p'\right)\ = \ C_{\scriptscriptstyle{LO}\,0}\, {\mathcal G}(p) {\mathcal G}(p') \ .
\label{eq:LOcompact}
\end{equation}
where $C_{\scriptscriptstyle{LO}\,0}$ and ${\mathcal G}(p^{(\prime)})$ are defined by comparing with Eq.~(\ref{eq:gensolinvyamanysing}).
The LO amplitude is then
\begin{equation}
  T_{\LOss}\left(k\right)\ = \ C_{\scriptscriptstyle{LO}\,0}\, {\mathcal G}^2(k)\,{\bm Z}^{-1} \ ,
\label{eq:LOcompact2}
\end{equation}
with ${\bm Z} \equiv 1-C_{\scriptscriptstyle{LO}\,0}{\mathbb I}_2$ and the convergent integral 
\begin{equation}
{\mathbb I}_2 \ =\ M\int \frac{d^3q}{(2 \pi)^3} \frac{{\mathcal G}^2(q)}{k^2- {q^2}+i\epsilon} \ =\ -\frac{M}{4\pi}{\mathcal G}^2(k)\left(\mu+i k\right) .
\label{eq:LOcompact3}
\end{equation}

Now notice that the NLO amplitude of Eq.~(\ref{eq:eftNLOrr2}) transforms simply under the momentum inversion $k\mapsto 2/( \vert a r \vert k)= \mu\nu/k$ as $T_{\NLOss}\to
T^*_{\NLOss}$ $(a r > 0)$ or $T_{\NLOss}\to T_{\NLOss}$ $(a r <
0)$. Based on the discussion below Eq.~(\ref{eq:gensolinvyam}) it may be expected that the part of $V_r$ that generates
$T_{\NLOss}$ will be invariant under momentum inversion up to a sign and $\mu \leftrightarrow \nu$.
Consider the energy-dependent potential,
\begin{equation}
k^2  V_{\LOss}\left(k,k\right){\mathcal G}^2(k)\ = \ C_{\scriptscriptstyle{LO}\,0}k^2\, {\mathcal G}^4(k)
\label{eq:LOcompact4}
\end{equation}
which maps to (minus) itself with $\mu\leftrightarrow\nu$ for $a r$ positive (negative) and is therefore a candidate for the piece of $V_r$ which generates $T_{\NLOss}$. An energy-independent residual potential can then be defined as
\begin{equation}
V_r(p',p)= \big\lbrack c_0+ c_2 \left(p^2 + p'^2\right)+\dots \big\rbrack V_{\LOss}\left(p',p\right)\left({\mathcal G}^2(p)+{\mathcal G}^2(p')\right)\ ,
\label{eq:VexpEF2}
\end{equation}
where the $c^{(\prime)}_m$ coefficients are bare parameters, subject to renormalization. On-shell this potential has the desired UV/IR transformation properties. Note that the form of the potential reflects that only odd powers of ${\mathcal G}(p^{(\prime)})$ in $V_r$ will match to
a polynomial in $k$ for $k\sim \aleph$. Formally, 
the (bare) coefficients of the full potential can be expressed for $k\ll \aleph$ as 
$C^{(\prime)}_m=C^{(\prime)}_{\scriptscriptstyle{LO}\,m}+C^{(\prime)}_{\scriptscriptstyle{NLO}\,m}$, where $C^{(\prime)}_{\scriptscriptstyle{NLO}\,m}\equiv C_{\scriptscriptstyle{LO}\,0}\cdot c^{(\prime)}_m$.

Evaluating the diagrams of Fig.~\ref{fig:DiaNLO} with a single insertion of $V_r$ gives
\begin{eqnarray}
&& T_{\NLOss} = 2c_0 C_{\scriptscriptstyle{LO}\, 0} {\mathcal G}^2(k)\, \big\lbrack {\mathcal G}^2(k) + C_{\scriptscriptstyle{LO}\, 0}{\bm Z}^{-1}\left({\mathcal G}^2(k){\mathbb I}_2+{\mathbb I}_4\right) 
+ C^2_{\scriptscriptstyle{LO}\, 0}{\bm Z}^{-2}{\mathbb I}_2{\mathbb I}_4\big\rbrack \nn \\
&&+2c_2 C_{\scriptscriptstyle{LO}\, 0} {\mathcal G}^2(k)\, \big\lbrace 2{\mathcal G}^2(k) k^2
+ C_{\scriptscriptstyle{LO}\, 0}{\bm Z}^{-1}\lbrack   {\mathcal G}^2(k)\left(2k^2{\mathbb I}_2-{\mathbb J}_2\right)+2k^2{\mathbb I}_4-{\mathbb J}_4\rbrack \nn \\
&&+ C^2_{\scriptscriptstyle{LO}\, 0}{\bm Z}^{-2}\lbrack 2{\mathbb I}_2{\mathbb I}_4 k^2-{\mathbb I}_2{\mathbb J}_4 -{\mathbb J}_2{\mathbb I}_4 \rbrack
\big\rbrace
\label{eq:TNLObare}
\end{eqnarray}
where
\begin{eqnarray}
{\mathbb I}_4 & =& M\int \frac{d^3q}{(2 \pi)^3} \frac{{\mathcal G}^4(q)}{k^2- {q^2}+i\epsilon} \ =\ -\frac{M}{4\pi}{\mathcal G}^4(k)\left(\oneht\mu -\frac{k^2}{2\mu}+i k\right)  \ \ ,\nn \\
  {\mathbb J}_2 & =& \left(\frac{\omega}{2}\right)^{3-d}M\int \frac{d^dq}{(2 \pi)^d} {\mathcal G}^2(q) \ \  \mapup{\rm PDS}\ \  -\frac{M\mu^2}{4\pi}\left(\mu-\omega \right) \ \ , \nn \\
  {\mathbb J}_4 & =& M\int \frac{d^3q}{(2 \pi)^3} {\mathcal G}^4(q) \ =\ \frac{M\mu^3}{8\pi} \ .
\label{eq:divints}
\end{eqnarray}
Here the linearly divergent integral ${\mathbb J}_2$ has been evaluated in dimensional regularization with the PDS scheme~\cite{Kaplan:1998tg,Kaplan:1998we},
and $\omega$ is the renormalization scale\footnote{The linearly divergent integral ${\mathbb J}_2$ in the $\overline{MS}$ scheme can be obtained by setting $\omega = 0$. Similarly, cutoff regularization, as in Eq.~(\ref{eq:In}), is obtained by replacing $\omega$ with $\Lambda$ (for $\Lambda$ large).}.
In terms of renormalized parameters, the amplitude takes the form
\begin{eqnarray}
T_{\NLOss} &=&   C_{\scriptscriptstyle{LO}\, 0} \left( {\mathcal G}^2(k) {\bm Z}^{-1}\right)^2\big\lbrack c_0^R\left(1-\zeta \nu/\mu\right)+ c_2^R \left(3-\zeta \nu/\mu\right)k^2 \big\rbrack \ ,
\label{eq:TNLOrenorm}
\end{eqnarray}
where the renormalized parameters, $c^{(\prime)R}_m$,  are defined as
\begin{eqnarray}
c_2^R  &=&  c_2 \ \ \ , \ \ \ c_0^R  \,=\,  c_0(\omega) + {c_2^R}\left(\frac{\mu+ \zeta \nu}{\mu-\zeta \nu}\right)\big\lbrack \zeta \mu \nu +\omega\left( \mu-\zeta \nu\right)\big\rbrack \ .
\label{eq:renormpars}
\end{eqnarray}
Matching to the expanded ERE of Eq.~(\ref{eq:eftNLOrr2}) gives $c_0^R=0$\footnote{Note that one can also decompose $a=a_{\LOss}+a_{\NLOss}$ in which case this condition follows from $a_{\NLOss}=0$.} and 
\begin{eqnarray}
C_{\scriptscriptstyle{NLO}\,2}\,=\, \frac{M}{4\pi} C^2_{\scriptscriptstyle{LO}\,0}\left(3-\zeta \nu/\mu\right)^{-1}\oneht r_{\NLOss} \ ,
\label{eq:renormparsere}
\end{eqnarray}
with $C^{(\prime )}_{\scriptscriptstyle{NLO}\,m}  \,=\, C_{\scriptscriptstyle{LO}\,0}\cdot c^{(\prime )R}_m$.
This relation gives a subleading enhancement of the same form as the usual pionless EFT~\cite{vanKolck:1998bw,Kaplan:1998tg,Kaplan:1998we} up to the factor in parenthesis,
and results in the scaling
\begin{eqnarray}
c_2^R \ \sim \ \frac{1}{{\mathcal M} \aleph } \ \ \ , \ \ \ C_{\scriptscriptstyle{NLO}\,2} \sim \frac{4\pi}{M {\mathcal M}\aleph^2 } \ .
\label{eq:scaleNLO}
\end{eqnarray}

In similar fashion, the NLO amplitude of Eq.~(\ref{eq:eftNLOrr}) can be obtained via the 
energy-independent residual potential
\begin{eqnarray}
  V_r(p',p)&=& \big\lbrack c_0+ c_2 \left(p^2 + p'^2\right)+c_4 \left(p^4 + p'^4\right)+c'_4 p^2 p'^2+\dots \big\rbrack \nn \\
  &&\times V_{\LOss}\left(p',p\right)\left({\mathcal G}^2(p)+{\mathcal G}^2(p')\right)\ .
\label{eq:VexpEF2c4}
\end{eqnarray}
Working in dimensional regularization with $\overline{\textstyle{MS}}$,
the amplitude takes the form
\begin{eqnarray}
  T_{\NLOss} &=&   C_{\scriptscriptstyle{LO}\, 0} \left( {\mathcal G}^2(k) {\bm Z}^{-1}\right)^2\big\lbrack c_0^R\left(1-\zeta \nu/\mu\right)+ c_2^R \left(3-\zeta \nu/\mu\right)k^2 \nn \\
&&\qquad\qquad\qquad\qquad\qquad\qquad\qquad  + \lbrack 2 c_4^{\prime R} + c_4^R  \left(3- \zeta \nu/\mu\right)\rbrack k^4
  \big\rbrack \ ,
\label{eq:TNLOrenormcasev}
\end{eqnarray}
with the renormalized parameters 
\begin{eqnarray}
  c^{(\prime)R}_4  &=&  c^{(\prime)}_4  \ \ \ , \ \ \
  c_2^R \,=\,  c_2 - \mu \left(\mu+ \zeta \nu\right)\big\lbrack  c_4^R + \left(3- \zeta \nu/\mu\right)^{-1}c_4^{\prime R} \big\rbrack \ , \nn \\
  c_0^R  &=& 
  c_0 - \mu\left(\frac{\mu+ \zeta \nu}{\mu-\zeta \nu}\right)\big\lbrack
  -c_2 \nu \zeta+c_4^{R} \mu^2 (2 \mu+\nu \zeta)+c_4^{\prime R} \mu^2 (\mu+\nu \zeta)
  \big\rbrack \, .
  \label{eq:renormparscasev}
\end{eqnarray}
Matching to the expanded ERE of Eq.~(\ref{eq:eftNLOrr}) now gives $c_0^R=c_2^R=0$ and 
\begin{eqnarray}
C_{\scriptscriptstyle{NLO}\,4}+2C_{\scriptscriptstyle{NLO}\,4}'\left(3-\zeta \nu/\mu\right)^{-1} &=& \frac{M}{4\pi} C^2_{\scriptscriptstyle{LO}\,0} \left(3-\zeta \nu/\mu\right)^{-1} v_2 .
\label{eq:renormparserecasev}
\end{eqnarray}
The coefficients scale as
\begin{eqnarray}
c_4^{(\prime)R} \ \sim \ \frac{1}{{\mathcal M}^3 \aleph } \ \ \ , \ \ \ C^{(\prime)}_{\scriptscriptstyle{NLO}\,4} \sim \frac{4\pi}{M {\mathcal M}^3\aleph^2 } \ .
\label{eq:scaleNLOcasev}
\end{eqnarray}
This differs from the conventional pionless theory counting which has a nominally leading contribution to the
$C^{(\prime )}_{4}$ operators from effective range (squared) effects.

\section{EFT description: the NN s-wave phase shifts}

\begin{figure*}
 \centerline{\includegraphics[width=0.98\textwidth]{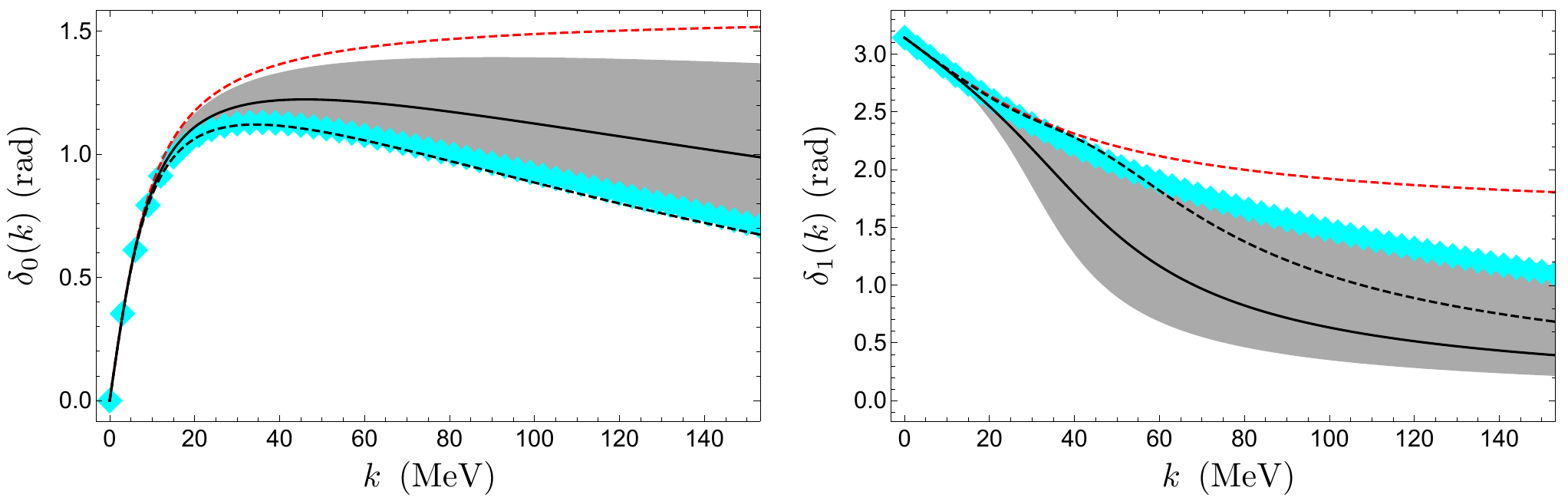}}
 \caption{Singlet (left panel) and triplet (right panel) s-wave NN
   phase shifts with the NLO amplitudes given by
   Eq.~(\ref{eq:eftNLOrr2}) plotted versus center-of-mass momentum
   $k$. The cyan polygons are the Nijmegen phase shift analysis
   (PWA93)~\protect\cite{NNOnline}. The red dashed curve is the LO
   phase shift in the pionless theory (scattering length only). The solid black
   curve is the central value of the LO phase shifts with exact UV/IR symmetry. The gray region represents an error band for the black curve and the dashed black
   curve is the NLO curve.}
\label{fig:LOsingtrip}       
\end{figure*}

\noindent The phase shifts to NLO in the EFT expansion are
\begin{eqnarray} 
\delta_s(k)&=&  \delta_{\LOss\, (s)}(k)\, +\,  \delta_{\NLOss\, (s)}(k) \ ,
  \label{eq:phaseshifts1}
\end{eqnarray}
with
\begin{eqnarray} 
  \delta_{\LOss\, (s)}(k)&=& -\frac{1}{2} i \ln\left(1 \ - \ i \frac{k M}{2\pi} T_{\LOss\, (s)}(k) \right) \ , \\
  \delta_{\NLOss\, (s)}(k)&=& -\frac{k M}{4\pi} T_{\NLOss\, (s)}(k) \left(1 \ - \ i \frac{k M}{2\pi}  T_{\LOss\, (s)}(k) \right)^{-1}  \ .
  \label{eq:phaseshifts2}
\end{eqnarray}
Recall from  section~\ref{sec:uvir} that in s-wave NN scattering, the UV/IR symmetry of the full $S$-matrix
requires range corrections that are correlated with the scattering lengths and treated exactly.
The physical effective ranges can therefore be expressed as $r_s=r_{\LOss\,(s)}+r_{\NLOss\,(s)}$ with
\begin{eqnarray} 
r_{\LOss\,(0)}&=&2\lambda a_1 \ \ \ ,\ \ \ r_{\LOss\,(1)}\,=\,-2\lambda a_0\ .
  \label{eq:confmoebiusiso4}
\end{eqnarray}
The singlet and triplet phase shifts with the NLO amplitude given by Eq.~(\ref{eq:eftNLOrr2}) are plotted in Fig.~(\ref{fig:LOsingtrip}).
At LO, there are three parameters given by the two s-wave scattering lengths and $\lambda$, which is fixed to
$\lambda=0.14\pm 0.11$, the range of values that exactly encompasses fits of $\lambda$ to each s-wave channel independently. This spread in $\lambda$
corresponds to the shaded gray region of the figures and is a conservative estimate of the LO uncertainty. The NLO curve has been generated
by tuning the $r_{\NLOss\,(s)}$ to give the ``physical'' effective ranges. A band on the NLO curves can easily be set by folding in the nominally
NNLO effect.

\begin{figure*}
 \centerline{\includegraphics[width=0.98\textwidth]{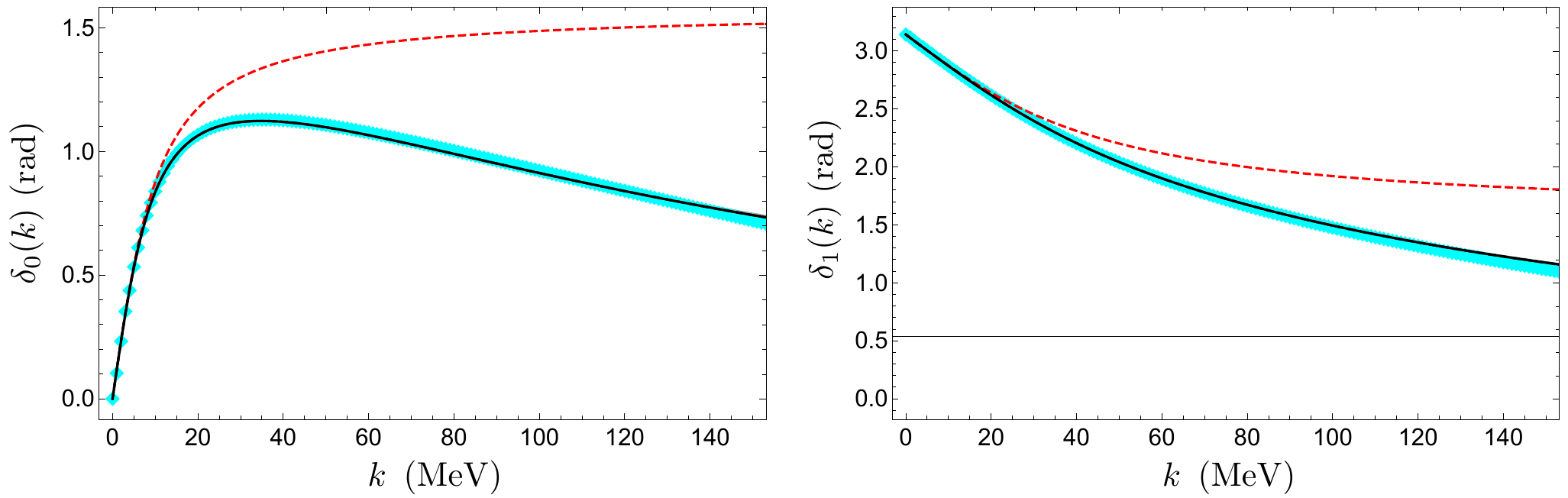}}
 \caption{
   Same as in Fig.~(\ref{fig:LOsingtrip}) but with the effective ranges treated exactly in each channel and the NLO amplitudes given by Eq.~(\ref{eq:eftNLOrr}).
   In both channels, the NLO curve overlaps with the central value of the LO phase shift.}
\label{fig:LOsingtripcasev}       
\end{figure*}

The singlet and triplet phase shifts with the NLO amplitude given by Eq.~(\ref{eq:eftNLOrr}) are plotted in Fig.~(\ref{fig:LOsingtripcasev}). As this treats both effective ranges exactly it provides an extremely accurate fit to the phase shifts due to the smallness of the shape parameters.

\section{Conclusion}

\noindent The s-wave NN $S$-matrix obtained from the ERE with
scattering length and effective range terms only, has interesting
UV/IR symmetries which are special inversions of the momenta. These
symmetries set the region of applicability of the EFT
descriptions. For instance, while it is common to view the EFT of
large scattering lengths as an expansion about the unitary fixed point
of the RG, it is, strictly speaking, an expansion about the fixed
point of a momentum inversion symmetry.  The UV/IR symmetries, while
not symmetries of the EFT action or of the scattering amplitude, are
present in the interaction. For instance, in the EFT of large
scattering lengths, the UV/IR symmetry is manifest in the RG flow of
the contact operator as an inversion symmetry of the RG scale which
interchanges the trivial and unitary RG fixed points and leaves the
beta function invariant~\cite{Beane:2021xrk}. When the effective range is also treated at LO in the EFT, the $S$-matrix has a (distinct) UV/IR symmetry which
effectively determines the LO potential, and constrains the form
of the perturbative NLO corrections.

There are many avenues to pursue with this new EFT. For instance, the softened asymptotic behavior of
the LO potential may resolve the issues of renormalization that arise when range corrections are
added to the integral equations that describe the three-nucleon system at very low energies. In addition, given the improved
convergence of LO in the EFT up to momenta beyond the range of validity of the pionless EFT, the perturbative pion paradigm may be worth revisiting in this scheme.
It may also be the case that the UV/IR
symmetries have interesting consequences for systems of many nucleons near
unitarity.

\chapter{Quantum Simulations of Quantum Chromodynamics in \texorpdfstring{$1+1$}{} Dimensions}
\label{chap:axial}
\noindent 
\textit{This chapter is associated with Ref.~\cite{Farrell:2022wyt}: \\
``Preparations for Quantum Simulations of Quantum Chromodynamics in \texorpdfstring{$1+1$}{} Dimensions: (I) Axial Gauge" by Roland C. Farrell, Ivan A. Chernyshev, Sarah J. M. Powell, Nikita A. Zemlevskiy, Marc Illa and Martin J. Savage.}

\section{Introduction}
\noindent
Simulations of the real-time dynamics of out-of-equilibrium, finite density quantum systems is a major goal of Standard Model (SM)~\cite{Glashow:1961tr,Higgs:1964pj,Weinberg:1967tq,Salam:1968rm,Politzer:1973fx,Gross:1973id} physics research and is expected to be computed efficiently~\cite{Lloyd1073} with ideal quantum computers~\cite{5392446,5391327,Benioff1980,Manin1980,Feynman1982,Fredkin1982,Feynman1986,doi:10.1063/1.881299,williamsNASAconference}.  
For recent reviews, see Refs.~\cite{Preskill:2021apy,Klco:2021lap,Bauer:2022hpo}.
Developing such capabilities would enable precision predictions of particle production and fragmentation in beam-beam collisions at the LHC and RHIC, of the matter-antimatter asymmetry production in the early universe, and of the structure and dynamics of dense matter in supernova and the neutrino flavor dynamics therein.
They would also play a role in better understanding protons and nuclei, particularly their entanglement structures and dynamics, and in exploring exotic strong-interaction phenomena such as color transparency. First steps are being taken toward simulating quantum field theories
(QFTs) using currently available, NISQ-era (Noisy Intermediate Scale Quantum) quantum devices~\cite{Preskill2018quantumcomputingin}, by studying low-dimensional and truncated many-body systems (see for example, Refs.~\cite{Hauke:2013jga,Banuls:2013jaa,Zohar:2016iic,Muschik:2016tws,Martinez:2016yna,Buyens:2016hhu,Banuls:2016lkq,Gonzalez-Cuadra:2017lvz,Dumitrescu:2018njn,PhysRevA.98.032331,Lu:2018pjk,Kaplan:2018vnj,Stryker:2018efp,Yeter-Aydeniz:2018mix,PhysRevD.101.074512,Avkhadiev:2019niu,Bauer:2019qxa,Klco:2019xro,Klco:2019yrb,Banuls:2019bmf,Luo:2019vmi,Funcke:2019zna,Davoudi:2019bhy,Magnifico:2019kyj,Mishra:2019xbh,Shehab:2019gfn,Yang_2020,Kharzeev:2020kgc,Shaw2020quantumalgorithms,PhysRevD.101.074512,PhysRevD.103.094501,Halimeh:2020ecg,Halimeh:2020djb,VanDamme:2020rur,Haase:2020kaj,Yeter-Aydeniz:2020jte,Davoudi:2021ney,ARahman:2021ktn,PhysRevLett.122.050403,Stryker:2021asy,aidelsburger2021cold,Bauer:2021gup,VanDamme:2021njp,Halimeh:2021lnv,Knaute:2021xna,Halimeh:2021vzf,Thompson:2021eze,Yeter-Aydeniz:2021mol,Yeter-Aydeniz:2021olz,Funcke:2021aps,Zhang:2021bjq,Rahman:2022rlg,Deliyannis:2022uyh,Bauer:2021gek,Illa:2022jqb,Mildenberger:2022jqr,Milsted:2020jmf}). 
These studies are permitting first quantum resource estimates to be made for more realistic simulations.

There has already been a number of quantum simulations of latticized $1+1$D quantum electrodynamics (QED, the lattice Schwinger model), starting with the pioneering work of Martinez {\it et al.}~\cite{Martinez:2016yna}.
The Schwinger model shares important features with quantum chromodynamics (QCD), such as charge screening, a non-zero fermion condensate, nontrivial topological charge sectors and a $\theta$-term.
Quantum simulations of the Schwinger model have been performed using quantum computers~\cite{Martinez:2016yna,Klco:2018kyo,Lu:2018pjk,Kokail:2018eiw,Nguyen:2021hyk,Thompson:2021eze}, and 
there is significant effort being made to extend this progress to higher dimensional QED~\cite{Zohar:2011cw,Zohar:2012ay,Tagliacozzo:2012vg,Zohar:2012ts,Wiese:2013uua,Marcos:2014lda,Kuno:2014npa,Bazavov:2015kka,Kasper:2015cca,Brennen:2015pgn,Kuno:2016xbf,Zohar:2016iic,Kasper:2016mzj,Gonzalez-Cuadra:2017lvz,Ott:2020ycj,Paulson:2020zjd,Kan:2021nyu,aidelsburger2021cold,Bauer:2021gek}.
These, of course, build upon far more extensive and detailed classical simulations of this model and analytic solutions of the continuum theory. 
There is also a rich portfolio of classical and analytic studies of $1+1$D $SU(N_c)$ gauge theories~\cite{Wilson:1994fk,Heinzl:1995jn,PhysRevD.31.2020,LIGTERINK2000983c,LIGTERINK2000215}, with some seminal papers preparing for quantum simulations~\cite{PhysRevD.11.395,PhysRevA.73.022328,Zohar:2012xf,PhysRevLett.110.125303,Tagliacozzo:2012vg,Banuls:2017ena,PhysRevD.101.074512,PhysRevD.103.094501,Paulson:2020zjd}, with the recent appearance of quantum simulations of a 1-flavor ($N_f=1$)  $1+1$D $SU(2)$ lattice gauge theory~\cite{Atas:2021ext}.
An attribute that makes such calculations attractive for early quantum simulations is that the gauge field(s) are uniquely constrained by Gauss's law at each lattice site.  
However, this is also a limitation for understanding higher dimensional theories where the gauge field is dynamical.  After pioneering theoretical works developing the formalism and also end-to-end  simulation protocols nearly a decade ago, it is only recently that first quantum simulations of the dynamics of a few plaquettes of gauge fields  have been performed~\cite{PhysRevD.101.074512,PhysRevD.103.094501,Atas:2021ext,Rahman:2022rlg}.

Due to its essential features, 
quantum simulations of the Schwinger model provide benchmarks for QFTs and quantum devices for the foreseeable future.  
Moving toward simulations of QCD requires including non-Abelian local gauge symmetry and multiple flavors of dynamical quarks. Low-energy, static and near-static observables in the continuum theory in $1+1$D
are well explored analytically and numerically, with remarkable results demonstrated, particularly in the 't Hooft model of large-$N_c$~\cite{tHooft:1973alw,tHooft:1974pnl} 
where the Bethe-Salpeter equation becomes exact. For a detailed discussion of $1+1$D $U(1)$ and $SU(N_c)$ gauge theories, see Refs.~\cite{Frishman:2010tc,Frishman2014book}.
Extending such calculations to inelastic scattering to predict, for instance, exclusive processes in high-energy hadronic collisions 
is a decadal challenge.

In $3+1$D QCD, the last 50 years have seen remarkable progress in using classical high-performance computing to provide robust numerical results using lattice QCD, e.g., Refs.~\cite{Joo:2019byq,Aoki:2021kgd}, 
where the quark and gluon fields are discretized in spacetime. Lattice QCD is providing complementary and synergistic results to those obtained in experimental facilities, moving beyond what is possible with analytic techniques alone. 
However, the scope of classical computations, even with beyond-exascale computing platforms~\cite{osti_1369223,Habib:2016sce,Joo:2019byq}, is limited by the use of a less fundamental theory (classical) to simulate a more fundamental theory (quantum).

Building upon theoretical progress in identifying 
candidate theories for early exploration (e.g., Ref.~\cite{Sala:2018dui}),
quantum simulations of $1+1$D non-Abelian gauge theories including matter were recently performed~\cite{Atas:2021ext} for a $N_c=2$ local gauge symmetry with one flavor of quark, $N_f=1$.
The Jordan-Wigner (JW) mapping~\cite{Jordan:1928wi} was used to define the lattice theory, and 
Variational Quantum Eigensolver (VQE)~\cite{Peruzzo_2014} quantum circuits were developed and used on IBM's quantum devices~\cite{IBMQ} 
to determine the vacuum energy, along with meson and baryon masses.  
Further, there have been previous quantum simulations of 
1- and 2-plaquette systems in $N_c=2,3$ Yang-Mills lattice gauge theories~\cite{Klco:2019evd,Ciavarella:2021nmj,Ciavarella:2021lel,ARahman:2021ktn,Illa:2022jqb} that did not include quarks.
Simulations of such systems are developing rapidly~\cite{Ciavarella:2021lel,Rahman:2022rlg} due to algorithmic and hardware advances. In addition, distinct mappings of these theories are being pursued~\cite{Brower:1997ha,Banerjee:2012xg,Tagliacozzo:2012df,Alexandru:2019nsa,Ji:2020kjk,Wiese:2021djl,Caspar:2022llo}.

This chapter focuses on the quantum simulation of $1+1$D $SU(N_c)$ lattice gauge theory for arbitrary $N_c$ and $N_f$.
Calculations are primarily done in $A^{(a)}_x=0$ axial (Arnowitt-Fickler) gauge,\footnote{For a discussion of Yang-Mills in axial gauge, see, for example, Ref.~\cite{Reinhardt:1996dy}.}
which leads to non-local interactions in order to define the chromo-electric field contributions to the energy density via Gauss's law.
This is in contrast to Weyl gauge, $A_t^{(a)}=0$, where contributions remain local.
The resource estimates for asymptotic quantum simulations of the Schwinger model in Weyl gauge have been recently performed~\cite{Shaw:2020udc}, and also for Yang-Mills gauge theory based upon the Byrnes-Yamamoto mapping~\cite{Kan:2021xfc}.
Here, the focus is on near-term, and hence non-asymptotic, quantum simulations to better assess the resource requirements for quantum simulations of non-Abelian gauge theories with multiple flavors of quarks. 
For concreteness, $N_f=2$ QCD is studied in detail, including the mass decomposition of the low-lying hadrons (the $\sigma$- and $\pi$-meson, the single baryon and the two-baryon bound state), color edge-states, entanglement structures within the hadrons and quantum circuits for time evolution.
Further, results are presented for the quantum simulation of a $N_f=1$, single-site system, using IBM's quantum computers~\cite{IBMQ}.
Such quantum simulations will play a critical role in evolving the functionality, protocols and workflows to be used in $3+1$D simulations of QCD, including the preparation of scattering states, time evolution and subsequent particle detection.
As a step in this direction, in a companion to the present paper, the results of this work have been applied to the quantum simulation of $\beta$-decay of a single baryon in $1+1$D QCD~\cite{PhysRevD.107.054513}.
Motivated by the recent successes in co-designing efficient multi-qubit operations in trapped-ion systems~\cite{Andrade:2021pil,Katz:2022czu}, 
additional multi-qubit or qudit operations are identified, 
specific to lattice gauge theories,
that would benefit from being native operations on quantum devices.

\section{QCD with Three Colors and Two flavors in \texorpdfstring{\boldmath$1+1$}{1+1}D}
\label{sec:Nc3Nf2}
\noindent
In $3+1$D, 
the low-lying spectrum of $N_f=2$ QCD is remarkably rich. 
The lightest hadrons are the $\pi$s, which are identified as the pseudo-Goldstone bosons associated with the spontaneous breaking of the approximate global $SU(2)_L\otimes SU(2)_R$ chiral symmetry, which becomes exact in the chiral limit where the $\pi$s are massless. At slightly higher mass are the broad $I=0$ spinless 
resonance, $\sigma$, and the narrow $I=0$, $\omega$, and $I=1$, $\rho$, vector resonances as well as the multi-meson continuum. 
The proton and neutron, which are degenerate in the isospin limit and the absence of electromagnetism,
are the lightest baryons, forming 
an $I=J=1/2$ iso-doublet.  
The next lightest baryons, which become degenerate with the nucleons in the large-$N_c$ limit (as part of a 
large-$N_c$ tower), are the four $I=J=3/2$ $\Delta$ resonances.
The nucleons bind together to form the periodic table of nuclei, the lightest being the deuteron, an $I=0$, $J=1$ neutron-proton bound state with a binding energy of $\sim 2.2~{\rm MeV}$, which is to be compared to the mass of the nucleon $M_N\sim 940~{\rm MeV}$.  
In nature, the low-energy two-nucleon systems have S-wave scattering lengths that are much larger than the range of their interactions, rendering them unnatural. Surprisingly, this unnaturalness persists for a sizable range of light-quark 
masses, e.g., Refs.~\cite{Beane:2002xf,Epelbaum:2012iu,Berengut:2013nh,Wagman:2017tmp,NPLQCD:2020lxg}. 
In addition, this unnaturalness, and the nearby renormalization-group fixed point~\cite{Kaplan:1998tg,Kaplan:1998we}, provides the starting point for a systematic effective field theory expansion about unitarity~\cite{Kaplan:1998tg,Kaplan:1998we,vanKolck:1998bw,Chen:1999tn}.
Much of this complexity is absent in a theory with only one flavor of quark.

As a first step toward $3+1$D QCD simulations of real-time dynamics of nucleons and nuclei, we will focus on preparing to carry out quantum simulations of $1+1$D QCD with $N_f=2$ flavors of quarks. While the isospin structure of the theory is the same as in $3+1$D, the lack of spin and orbital angular momentum significantly reduces the richness of the hadronic spectrum and S-matrix.
However, many of the relevant features and processes of $3+1$D QCD that are to be addressed by quantum simulation in the future are present in $1+1$D QCD.
Therefore, quantum simulations in $1+1$D are expected to provide inputs to the development of quantum simulations of QCD.

\subsection{Mapping \texorpdfstring{\boldmath$1+1$}{1+1}D QCD onto Qubits}
\noindent
The Hamiltonian describing non-Abelian lattice gauge field theories in arbitrary numbers of spatial dimensions was first given by Kogut and Susskind (KS) in the 1970s~\cite{Kogut:1974ag,Banks:1975gq}. For $1+1$D QCD with $N_f = 2$ discretized onto $L$ spatial lattice sites, which are mapped to 2L $q$, $\overline{q}$ sites to separately accommodate quarks and antiquarks,  the KS lattice Hamiltonian is
\begin{align}
    H_{\rm{KS}} 
     = & 
    \sum_{f=u,d}\left[
        \frac{1}{2 a} \sum_{n=0}^{2L-2} \left ( \phi_n^{(f)\dagger} U_n \phi_{n+1}^{(f)}
        \ +\ {\rm h.c.} \right ) 
    \: + \: 
    m_f \sum_{n=0}^{2L-1} (-1)^{n} \phi_n^{(f)\dagger} \phi_n^{(f)} 
    \right]
    \: + \: 
    \frac{a g^2}{2} 
    \sum_{n=0}^{2L-2} 
    \sum_{a=1}^{8}
    | {\bf E}^{(a)}_n|^2
    \nonumber\\
     & - \: \frac{\mu_B}{3} \sum_{f=u,d} \sum_{n=0}^{2L-1} \phi_n^{(f)\dagger} \phi^{(f)}_n 
    \ - \: 
    \frac{\mu_{I}}{2} \sum_{n=0}^{2L-1}\left(\phi_n^{(u)\dagger} \phi^{(u)}_n \ - \
    \phi_n^{(d)\dagger} \phi^{(d)}_n  \right)
    \ .
    \label{eq:KSHam}
\end{align}
The masses of the $u$- and $d$-quarks are $m_{u,d}$,
$g$ is the strong coupling constant at the spatial lattice spacing $a$,
$U_n$ is the spatial link operator in Weyl gauge
$A_t^{(a)}=0$,
$\phi^{(u,d)}_n$ are the $u$- and $d$-quark field operators which transform in the fundamental representation of $SU(3)$
and 
${\bf E}^{(a)}_n$ is the chromo-electric field associated with the $SU(3)$ generator,
$T^a$.
For convention, we write, for example, $\phi^{(u)}_n=(u_{n,r}, u_{n,g}, u_{n,b})^T$ to denote the $u$-quark field(s) at the $n^{\rm th}$ site in terms of 3 colors $r,g,b$.
With an eye toward simulations of dense matter systems, chemical potentials for baryon number, $\mu_B$, and the third component of isospin, $\mu_I$, are included.
For most of the results presented in
this work, the chemical potentials will be set to zero, $\mu_B=\mu_I = 0$,
and there will be exact isospin symmetry, $m_u=m_d \equiv m$.
In Weyl gauge and using the chromo-electric basis of the link operator $|{\bf R},\alpha,\beta\rangle_n$,
the contribution from the energy in the chromo-electric field from each basis state is proportional to the Casimir of the irrep ${\bf R}$.\footnote{
For an irrep, ${\bf R}$, represented by a tensor with $p$ upper indices and $q$ lower indices, $T^{a_1 \cdots a_p}_{b_1 \cdots b_q}$,
the Casimir provides
\begin{equation}
    \sum_{b=1}^{8}| {\bf E}^{(a)}_n|^2\ |{\bf R},\alpha,\beta\rangle_n = \frac{1}{3}\left( p^2+q^2+p q + 3 p + 3 q \right)\  |{\bf R},\alpha,\beta\rangle_n \ .
\label{eq:Casi}
\end{equation}
The indices $\alpha$ and $\beta$ specify the color state in the left (L) and right (R) link Hilbert spaces respectively. 
States of a color irrep {\bf R} are labelled by their total color isospin $T$, third component of color isospin $T^z$ and color hypercharge $Y$, i.e., $\alpha = (T_L, T^z_L, Y_L)$ and $\beta = (T_R, T^z_R, Y_R)$.
\label{foot:irrep}}
The fields have been latticized
such that the quarks reside on even-numbered sites, $n=0,2,4,6,\ldots$, and antiquarks reside on odd-numbered sites, $n=1,3,5,\ldots$.
Open boundary conditions (OBCs) are employed in the spatial direction, 
with a vanishing background chromo-electric field.
For simplicity, 
the lattice spacing will be set equal to $1$.

The KS Hamiltonian in Eq.~(\ref{eq:KSHam}) 
is constructed in Weyl gauge.
A unitary transformation can be performed on Eq.~(\ref{eq:KSHam}) to eliminate the gauge links~\cite{Sala:2018dui}, with Gauss's Law 
uniquely providing the energy in the chromo-electric field in terms of a non-local sum of products of charges, i.e., the Coulomb energy. 
This is equivalent to formulating the system in axial gauge~\cite{PhysRev.127.1821,weinberg1995quantum}, $A^{(a)}_x = 0$, from the outset.
The Hamiltonian in Eq.~(\ref{eq:KSHam}), when formulated with $A^{(a)}_x = 0$, becomes
\begin{align}
    H 
    = & 
    \sum_{f=u,d}\left[ 
        \frac{1}{2} \sum_{n=0}^{2L-2} \left ( \phi_n^{(f)\dagger} \phi_{n+1}^{(f)}
        \ +\ {\rm h.c.} \right ) 
    \: + \: 
    m_f \sum_{n=0}^{2L-1} (-1)^{n} \phi_n^{(f)\dagger} \phi_n^{(f)} 
    \right]
    \: + \: 
    \frac{g^2}{2}
    \sum_{n=0}^{2L-2} 
    \sum_{a=1}^{8}
    \left ( \sum_{m \leq n} Q^{(a)}_m \right ) ^2    \nonumber\\
     & - \: \frac{\mu_B}{3} \sum_{f=u,d} \sum_{n=0}^{2L-1} \phi_n^{(f)\dagger} \phi^{(f)}_n  
    \ - \:
    \frac{\mu_{I}}{2} \sum_{n=0}^{2L-1}\left(\phi_n^{(u)\dagger} \phi^{(u)}_n \ - \
    \phi_n^{(d)\dagger} \phi^{(d)}_n  \right)
    \ ,
    \label{eq:GFHam}
\end{align}
where the color charge operators on a given lattice site are the sum of contributions from the $u$- and $d$-quarks,
\begin{equation}
    Q^{(a)}_m \ =\ 
    \phi^{(u) \dagger}_m T^a \phi_m^{(u)}\ +\ 
    \phi^{(d) \dagger}_m T^a \phi_m^{(d)}
    \ .
    \label{eq:SU3charges}
\end{equation}
To define the fields, 
boundary conditions with $A_0^{(a)}(x)=0$ at spatial infinity and zero background chromo-electric fields are used, with Gauss's law sufficient to determine them at all other points on the lattice,
\begin{equation}
   {\bf E}^{(a)}_n  = \sum_{m\leq n} Q^{(a)}_m \ .
\end{equation}
In this construction, a state is completely specified by the fermionic occupation at each site. This is to be contrasted with the Weyl
gauge construction where both fermionic occupation and the $SU(3)$ multiplet defining the chromo-electric field are required.

\begin{figure}[!ht]
    \centering
    \includegraphics[width=15cm]{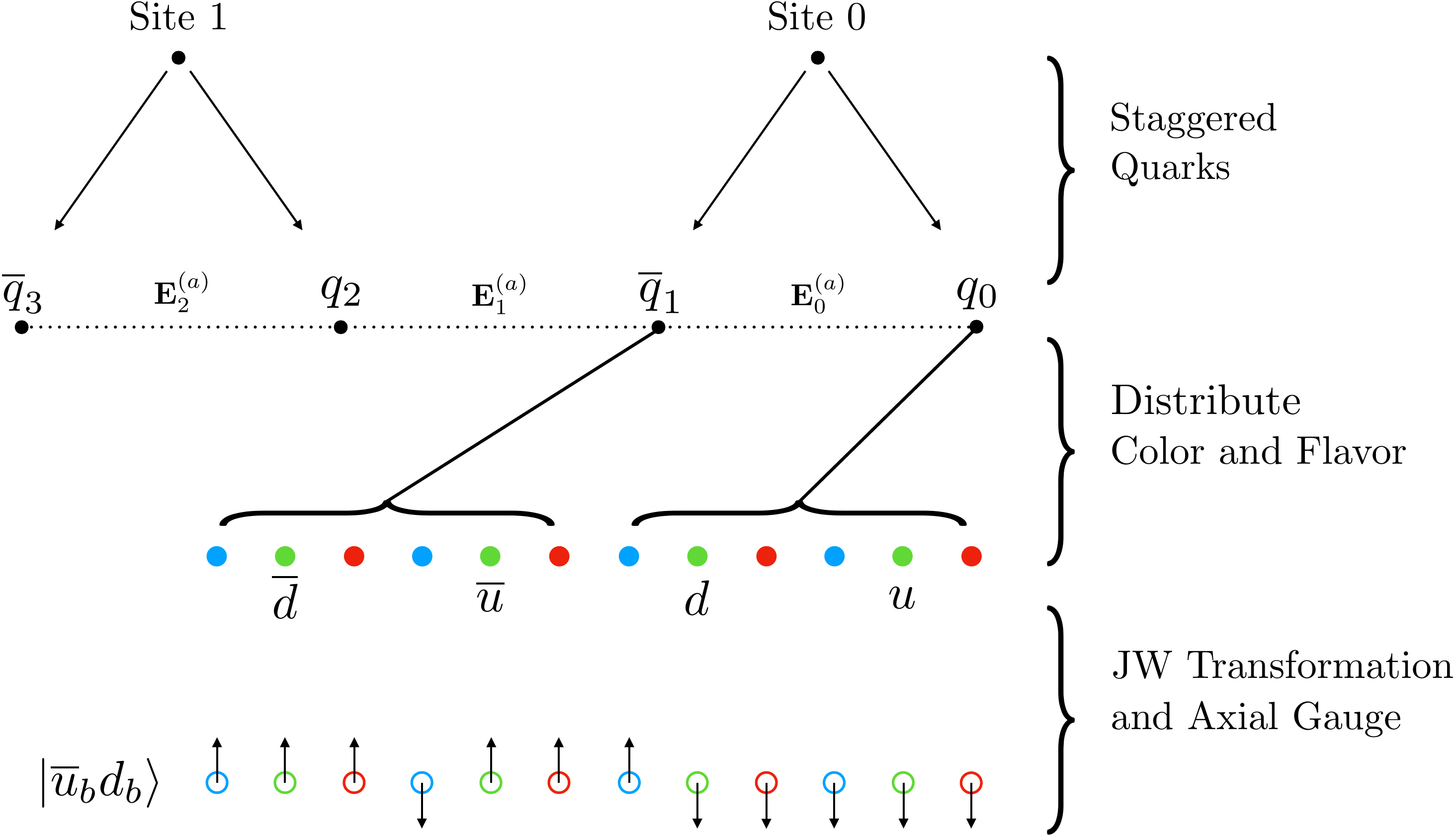}
    \caption{
    The encoding of $N_f=2$ QCD onto a lattice of spins describing $L=2$ spatial sites. 
    Staggering is used to discretize the quark fields, which doubles the number of lattice sites, with (anti)quarks on (odd) even sites.
    The chromo-electric field resides on the links between quarks and antiquarks. 
    Color and flavor degrees of freedom of each quark and antiquark site are distributed over six qubits with a JW mapping, and axial gauge along with Gauss's law are used to remove the chromo-electric fields. 
    A quark (antiquark) site is occupied if it is spin up (down), and the example spin configuration corresponds to the state $\ket{\overline{u}_b \, d_b}$.}
    \label{fig:2flavLayout}
\end{figure}

There are a number of ways that this system,
with the Hamiltonian given in Eq.~(\ref{eq:GFHam}), could be mapped
onto the register of a quantum computer.
In this work, both a staggered discretization and a JW transformation~\cite{1928ZPhy...47..631J} are chosen to map the $N_c=3$ and $N_f=2$
quarks to 6 qubits, with ordering $d_b, d_g, d_r, u_b, u_g, u_r$,
and the antiquarks associated with the same spatial site adjacent with ordering 
$\overline{d}_b, \overline{d}_g, \overline{d}_r, \overline{u}_b, \overline{u}_g, \overline{u}_r$.
This is illustrated in Fig.~\ref{fig:2flavLayout} and
requires a total of 12 qubits per spatial lattice site (see App.~\ref{app:hamConst} for more details). 
The resulting JW-mapped Hamiltonian is the sum of the following five terms:
\begin{subequations}
    \label{eq:H2flav}
    \begin{align}
    H = & \ H_{kin}\ +\ H_m\ +\ H_{el} \ +\ 
    H_{\mu_B}\ +\ H_{\mu_I} \ ,\\[4pt]
    H_{kin} = & \ -\frac{1}{2} \sum_{n=0}^{2L-2} \sum_{f=0}^{1} \sum_{c=0}^{2} \left[ \sigma^+_{6n+3f+c} \left ( \bigotimes_{i=1}^{5}\sigma^z_{6n+3f+c+i} \right )\sigma^-_{6(n+1)+3f+c} +\rm{h.c.} \right]\ ,
        \label{eq:Hkin2flav}\\[4pt]
    H_m = & \ \frac{1}{2} \sum_{n=0}^{2L-1} \sum_{f=0}^{1} \sum_{c=0}^{2} m_f\left[ (-1)^{n} \sigma_{6n + 3f + c}^z + 1\right]\ ,
        \label{eq:Hm2flav}\\[4pt]
    H_{el} = & \ \frac{g^2}{2} \sum_{n=0}^{2L-2}(2L-1-n)\left( \sum_{f=0}^{1} Q_{n,f}^{(a)} \, Q_{n,f}^{(a)} \ + \
        2 Q_{n,0}^{(a)} \, Q_{n,1}^{(a)}
         \right)   \nonumber \\[4pt]
        & + g^2 \sum_{n=0}^{2L-3} \sum_{m=n+1}^{2L-2}(2L-1-m) \sum_{f=0}^1 \sum_{f'=0}^1 Q_{n,f}^{(a)} \, Q_{m,f'}^{(a)} \ ,
         \label{eq:Hel2flav}\\[4pt]
    H_{\mu_B}  = & \ -\frac{\mu_B}{6} \sum_{n=0}^{2L-1} \sum_{f=0}^{1} \sum_{c=0}^{2}  \sigma_{6n + 3f + c}^z \ ,
        \label{eq:HmuB2flav}\\[4pt]
    H_{\mu_I} = & \ -\frac{\mu_I}{4} \sum_{n=0}^{2L-1} \sum_{f=0}^{1} \sum_{c=0}^{2} (-1)^{f} \sigma_{6n + 3f + c}^z  \ ,
        \label{eq:HmuI2flav}
    \end{align}
\end{subequations}
where now repeated adjoint color indices, $(a)$, are summed over,
the flavor indices, $f=0,1$, correspond to $u$- and $d$-quark flavors and $\sigma^\pm = (\sigma^x \pm i \sigma^y)/2$.
Products of charges are given in terms of spin operators as
\begin{align}
    &Q_{n,f}^{(a)} \, Q_{n,f}^{(a)} = \frac{1}{3}(3 - \sigma^z_{6n+3f} \sigma^z_{6n+3f+1} - \sigma^z_{6n+3f} \sigma^z_{6n+3f+2} - \sigma^z_{6n+3f+1} \sigma^z_{6n+3f+2}) \ ,  \nonumber \\[4pt]
    &Q_{n,f}^{(a)} \, Q_{m,f'}^{(a)} = \frac{1}{4}\bigg [2\big (\sigma^+_{6n+3f}\sigma^-_{6n+3f+1}\sigma^-_{6m+3f'}\sigma^+_{6m+3f'+1} \nonumber \\[4pt]
    & + \sigma^+_{6n+3f}\sigma^z_{6n+3f+1}\sigma^-_{6n+3f+2}\sigma^-_{6m+3f'}\sigma^z_{6m+3f'+1}\sigma^+_{6m+3f'+2}\nonumber \\[4pt] &+\sigma^+_{6n+3f+1}\sigma^-_{6n+3f+2}\sigma^-_{6m+3f'+1}\sigma^+_{6m+3f'+2} + { \rm h.c.}\big ) \nonumber \\[4pt]
    &+ \frac{1}{6}\sum_{c=0}^{2} \sum_{c'=0}^2( 3 \delta_{c c'} - 1 ) \sigma^z_{6n+3f+c}\sigma^z_{6m+3f'+c'} \bigg ] \ .
    \label{eq:QnfQmfp}
\end{align}
A constant has been added to $H_m$ to ensure that all basis states contribute positive mass. The Hamiltonian for $SU(N_c)$ gauge theory with $N_f$ flavors in the fundamental representation is presented in Sec.~\ref{sec:NcNf}. 
Note that choosing $A^{(a)}_x = 0$ gauge and enforcing Gauss's law has resulted in all-to-all interactions, the double lattice sum in $H_{el}$.

For any finite lattice system, there are color non-singlet states in the spectrum, which are unphysical and have infinite energy in the continuum and infinite-volume limits.
For a large but finite system, OBCs can also support finite-energy color non-singlet states which are localized to the end of the lattice (color edge-states).\footnote{Low-energy edge-states that have global charge in a confining theory can also be found in the simpler setting of the Schwinger model. 
Through exact and approximate tensor methods, we have verified that these states exist on lattices up to length $L=13$, and they are expected to persist for larger $L$.} 
The existence of such states in the spectrum is independent of the choice of gauge or fermion mapping.
The naive ways to systematically examine basis states and preclude such configurations is found to be impractical due to the non-Abelian nature of the gauge charges
and the resulting entanglement between states required for color neutrality.
A practical way to deal with this problem is to add a term to the Hamiltonian that 
raises the energy of color non-singlet states.
This can be accomplished by including the energy density in the chromo-electric field beyond the end of the lattice with a large coefficient $h$.  
This effectively adds the energy density in a finite chromo-electric field over a large spatial extent beyond the end of the lattice.
In the limit $h\rightarrow\infty$, only states with a vanishing chromo-electric field beyond the end of the lattice remain at finite energy, rendering the system within the lattice to be a color singlet.
This new term in the Hamiltonian is
\begin{equation}
    H_{\bf 1} = \frac{h^2}{2} \sum_{n=0}^{2L-1}
    \left( \sum_{f=0}^{1} Q_{n,f}^{(a)} \, Q_{n,f}^{(a)} \ + \  
    2 Q_{n,0}^{(a)} \, Q_{n,1}^{(a)}
     \right) \ + \  h^2 \sum_{n=0}^{2L-2} \sum_{m=n+1}^{2L-1}\sum_{f=0}^1 \sum_{f'=0}^1 Q_{n,f}^{(a)} \, Q_{m,f'}^{(a)} \ ,
     \label{eq:Hpen}
\end{equation}
which makes a vanishing contribution when the sum of charges over the whole lattice is zero; otherwise, it makes a contribution $\sim h^2$.

\subsection{Spectra for \texorpdfstring{\boldmath$L=1, 2$}{L=1,2} Spatial Sites}
\label{sec:Exact}
\noindent
The spectra and wavefunctions of systems with a small number of lattice sites can be determined by diagonalization of the Hamiltonian.
In terms of spin operators, the $N_f=2$ Hamiltonian in Eq.~(\ref{eq:H2flav}) decomposes into sums of tensor products of Pauli matrices. The tensor product factorization can be exploited to perform an exact diagonalization relatively efficiently. 
This is accomplished by first constructing a basis
by projecting onto states with specific quantum numbers, and then building the Hamiltonian in that subspace. 
There are four mutually commuting symmetry generators that allow states to be labelled by $(r,g,b,I_3)$: redness, greenness, blueness and the third component of isospin.
In the computational (occupation) basis, states are represented by bit strings of $0$s and $1$s. For example, the $L=1$ state with no occupation is $\ket{000000111111}$.\footnote{Qubits are read from right to left, e.g., $\ket{q_{11} \, q_{10}\, \ldots \, q_{1}\, q_{0}}$. Spin up is $\ket{0}$ and spin down is $\ket{1}$.} Projecting onto eigenstates of $(r,g,b,I_3)$ amounts to fixing the total number of $1$s in a substring of a state.
The Hamiltonian is formed by evaluating matrix elements of Pauli strings between states in the basis, and only involves $2\times 2$ matrix multiplication.
The Hamiltonian matrix is found to be sparse, as expected, and the low energy eigenvalues and eigenstates can be found straightforwardly.
As the dimension of the Hamiltonian grows exponentially with the spatial extent of the lattice, this method becomes intractable for large system sizes, as is well known.

\subsubsection{Exact Diagonalizations, Color Edge-States and Mass Decompositions of the Hadrons}
\noindent
For small enough systems, an exact diagonalization of the Hamiltonian matrix in the previously described basis can be performed.
Without chiral symmetry and its spontaneous breaking, the energy spectrum in $1+1$D does not contain a massless isovector state (corresponding to the QCD pion) in the limit of vanishing quark masses.
In the absence of 
chemical potentials for baryon number, $\mu_B=0$, or isospin, $\mu_I=0$,
the vacuum, $\ket{\Omega}$, has $B=0$
(baryon number zero) and $I=0$ (zero total isospin). 
The $I=0$
$\sigma$-meson is the lightest meson,
while the $I=1$ $\pi$-meson is the next lightest.
The lowest-lying eigenstates in the 
$B=0$ spectra for $L=1,2$
(obtained from
exact diagonalization of the Hamiltonian) 
are given in Table~\ref{tab:specB0}. 
The masses are defined by their energy gap to the vacuum, 
and all results in this section are for $m_u=m_d=m=1$.
\begin{table}[!ht]
\renewcommand{\arraystretch}{1.2}
\begin{tabularx}{0.4\textwidth}{||c | Y | Y | Y ||} 
\hline
\multicolumn{4}{||c||}{$L=1 $} \\
 \hline
 $g^2$ & $E_{\Omega}$ & $M_{\sigma}$ & $M_{\pi}$ \\
 \hline\hline
 8 & -0.205 & 5.73 & 5.82 \\ 
 \hline
 4 & -0.321 & 4.37 & 4.47\\
 \hline
 2 & -0.445 & 3.26 & 3.30 \\
 \hline
 1 & -0.549 & 2.73 & 2.74\\
 \hline
 1/2 & -0.619 & 2.48 & 2.48\\
 \hline
 1/4 & -0.661 & 2.35 & 2.36 \\
 \hline
 1/8 & -0.684 &  2.29 & 2.30 \\
 \hline
\end{tabularx}
\renewcommand{\arraystretch}{1}
\qquad\qquad\qquad
\renewcommand{\arraystretch}{1.2}
\begin{tabularx}{0.4\textwidth}{||c | Y | Y | Y ||} 
\hline
\multicolumn{4}{||c||}{$L=2 $} \\
 \hline
 $g^2$ & $E_{\Omega}$ & $M_{\sigma}$ & $M_{\pi}$\\
 \hline\hline
 8 & -0.611 & 5.82 & 5.92 \\ 
 \hline
 4 & -0.949 & 4.41 & 4.49 \\
 \hline
 2 & -1.30 & 3.27 & 3.31 \\
 \hline
 1 & -1.58 & 2.72 & 2.74 \\
 \hline
 1/2 & -1.77 & 2.45 & 2.46 \\
 \hline
 1/4 & -1.88 & 2.30 & 2.31 \\
 \hline
 1/8 & -1.94 & 2.22 & 2.22 \\
 \hline
\end{tabularx}
\renewcommand{\arraystretch}{1}
\caption{
The vacuum energy and the masses of the $\sigma$- and $\pi$-mesons for $1+1$D QCD with $N_f=2$ for systems with $L=1,2$
spatial sites. These results are insensitive to $h$ as they are color singlets.}
\label{tab:specB0}
\end{table}
By examining the vacuum energy density 
$E_{\Omega}/L$, it is clear that, as expected, this number of lattice sites is insufficient to fully contain hadronic correlation lengths.  
While Table~\ref{tab:specB0} shows the energies of color-singlet states, there are also non-singlet states in the spectra with similar masses,
which become increasingly localized near the end of the lattice, as discussed in the previous section.

\begin{figure}[!ht]
    \centering
    \includegraphics[width=14cm]{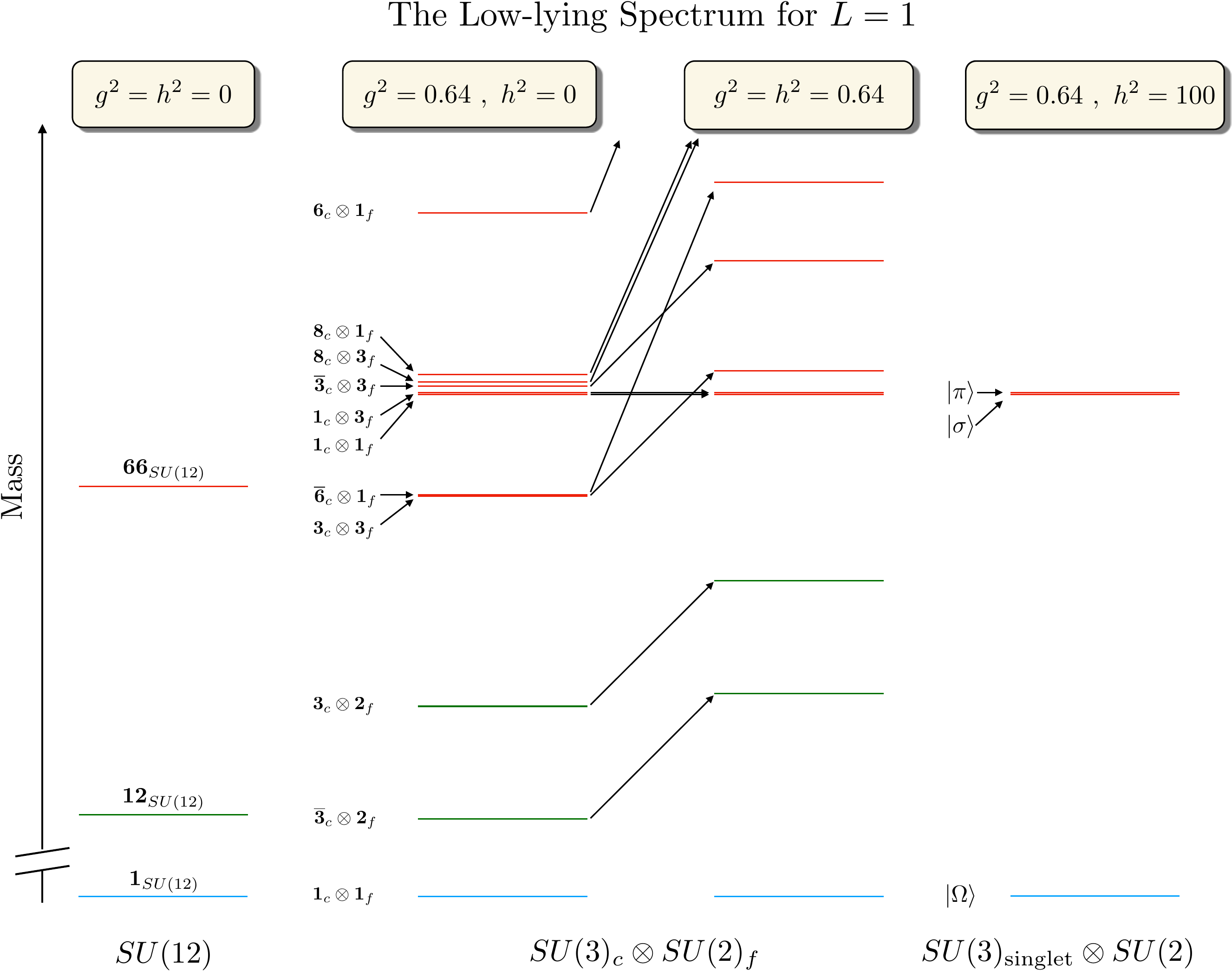}
    \caption{
    The spectrum of the Hamiltonian as the couplings $g$ and $h$ increase. 
    For $g=h=0$ there is an exact $SU(12)$ symmetry and the color-singlet 
    $\sigma$- and $\pi$-mesons are a part of the antisymmetric ${\bf 66}$ irrep. 
    When $g>0$ and $h=0$, the spectrum splits into irreps of global $SU(3)_c \otimes SU(2)_f$ with 
    color non-singlet states among the low-lying states.
    Increasing $h>0$ pushes non-singlet color states out of the low-lying spectrum. 
    Notice that the $\sigma$ and $\pi$ masses are insensitive to $h$, as expected.}
    \label{fig:specDegenh}
\end{figure}

It is informative to examine the spectrum of the $L=1$ system as both $g$ and $h$ are slowly increased and, in particular, take note of the relevant symmetries. For $g=h=0$,
with contributions from only the hopping and mass terms,
the system exhibits a global $SU(12)$ symmetry 
where the spectrum is that of free quasi-particles; see App.~\ref{app:freeSym}.
The enhanced global symmetry at this special point restricts the structure of the spectrum to the ${\bf 1}$ and ${\bf 12}$ of $SU(12)$ as well as the antisymmetric combinations of fundamental irreps, ${\bf 66}, {\bf 220}, \ldots$.  
For $g>0$, these $SU(12)$ irreps split into irreps of color $SU(3)_c$ and flavor $SU(2)_f$. 
The ${\bf 12}$ corresponds to single quark ($q$) or antiquark ($\overline{q}$) excitations
(with fractional baryon number), and splits into ${\bf 3}_c\otimes {\bf 2}_f$ for quarks and $\overline{\bf 3}_c\otimes
{\bf 2}_f$ for antiquarks.  In the absence of OBCs, these states would remain degenerate, but the boundary condition of vanishing background
chromo-electric field is not invariant under 
$q\leftrightarrow \overline{q}$ and the quarks get pushed to higher mass. As there is no chromo-electric energy associated with exciting an
antiquark at the end of the lattice in this mapping, the $\overline{\bf 3}_c\otimes {\bf 2}_f$ states remains low in the spectrum until $h\gg0$.
The ${\bf 66}$ corresponds to two-particle excitations, and contains all combinations of $qq$, $\overline{q}q$ and 
$\overline{q} \overline{q}$ excitations. 
The mixed color symmetry (i.e., neither symmetric or antisymmetric) of $\overline{q}q$ excitations allows for states with
${\bf 1}_c\otimes {\bf 1}_f
\oplus
{\bf 1}_c\otimes {\bf 3}_f
\oplus
{\bf 8}_c\otimes {\bf 1}_f
\oplus
{\bf 8}_c\otimes {\bf 3}_f
$,
while the $qq$ excitations with definite color symmetry allow for 
${\bf 6}_c\otimes {\bf 1}_f
\oplus \overline{\bf 3}_c\otimes {\bf 3}_f
$
and
$\overline{q}\overline{q}$ excitations allow for
$\overline{\bf 6}_c\otimes {\bf 1}_f
\oplus {\bf 3}_c\otimes {\bf 3}_f$,
saturating the $66$ states in the multiplet.
When $g>0$, these different configurations split in energy, and when
$h\gg0$, only color-singlet states are left in the low-lying spectrum. Figure~\ref{fig:specDegenh} shows the evolution of the spectrum as 
$g$ and $h$ increase.
The increase in mass of non-singlet color states  with $h$ is proportional to the Casimir of the $SU(3)_c$ representation which is evident in Fig.~\ref{fig:specDegenh} where, for example, the increase in the mass of the ${\bf 3}_c$s and $\overline{{\bf 3}}_c$s between $h^2 = 0$ and $h^2=0.64$ are the same.

\begin{figure}[!ht]
    \centering
    \includegraphics[width=12cm]{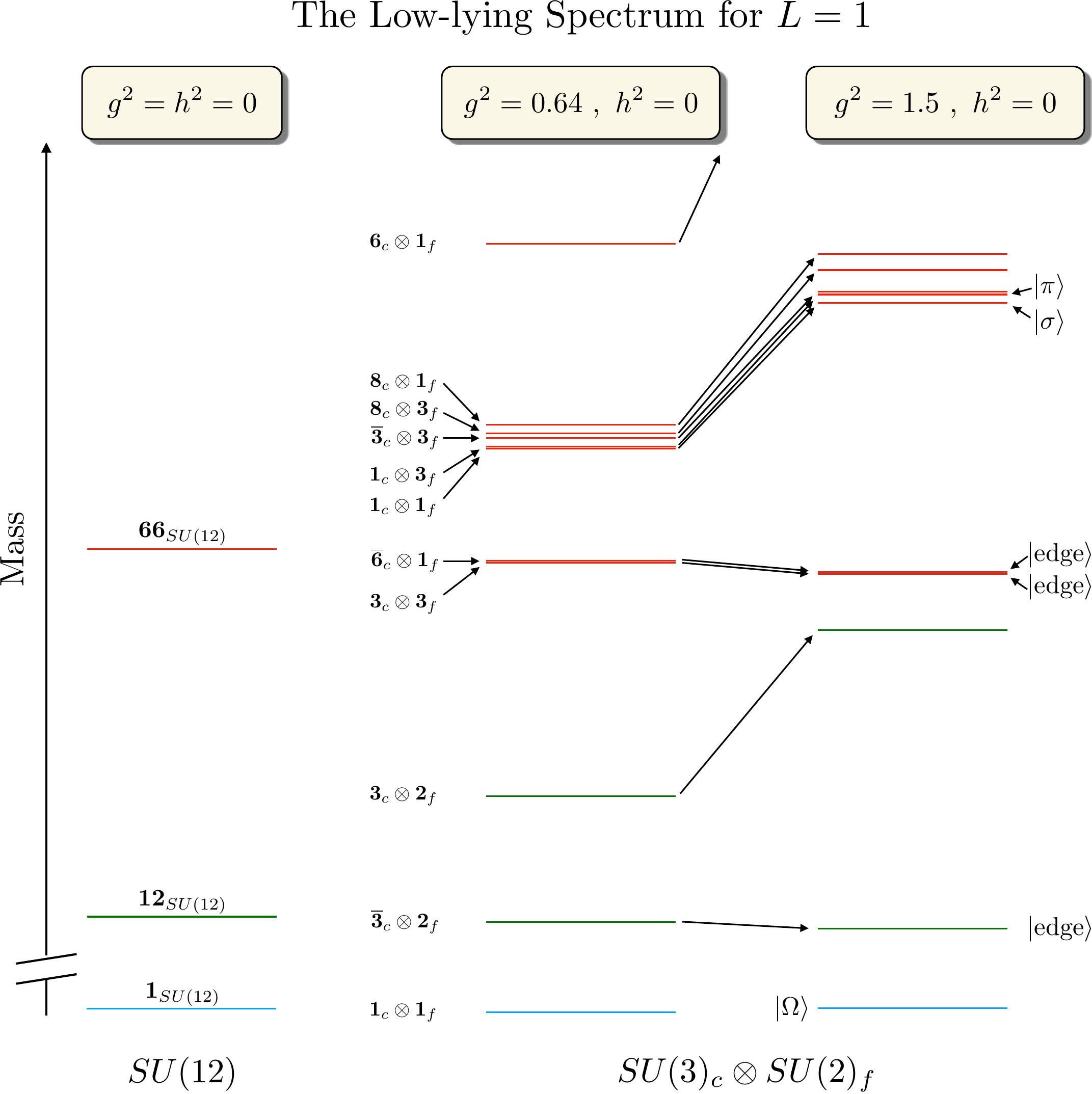}
    \caption{
    The spectrum of the Hamiltonian as $g$ increases for $h=0$. 
    When $g=h=0$ there is an exact $SU(12)$ symmetry and the $\sigma$- and $\pi$-mesons are a part of the antisymmetric ${\bf 66}$ irrep. 
    When $g>0$  but $h=0$ the spectrum splits into irreps of global $SU(3)_c \otimes SU(2)_f$, and non-singlet color states remain in the low-lying spectrum. 
    Increasing $g$ shifts all but the antiquark $\ket{\text{edge}}$ (states) to higher mass.
    }
    \label{fig:specDegeng}
\end{figure}

The antiquark states are particularly interesting as they correspond to edge states that are not ``penalized" in energy by the chromo-electric field when $h=0$. 
These states have an approximate
$SU(6)$ symmetry where the $6$ antiquarks transform in the fundamental.  
This is evident in the spectrum shown in 
Fig.~\ref{fig:specDegeng}
by the presence of a $\overline{{\bf 3}}_c \otimes {\bf 2}_f$ 
and nearly degenerate 
$\overline{\bf 6}_c\otimes {\bf 1}_f$ 
and 
${\bf 3}_c\otimes {\bf 3}_f$
which are identified as states of a 
${\bf 15}$ 
(an antisymmetric irrep of $SU(6)$)
that do not increase in mass as $g$ increases.
This edge-state $SU(6)$ symmetry is not exact 
due to interactions from the hopping term that couple the edge $\overline{q}$s to the rest of the lattice. 
These colored edge states are artifacts of OBCs and will persist in the low-lying spectrum for larger lattices.

Figures~\ref{fig:specDegenh} and \ref{fig:specDegeng}
reveal the near-degeneracy of the  $\sigma$- and $\pi$-mesons throughout the range of couplings $g$ and $h$, suggesting another approximate symmetry, which
can be understood in the small and large $g$ limits.
For small $g^2$, the effect of $H_{el} = \frac{g^2}{2}(Q_{0,u}^{(a)} + Q_{0,d}^{(a)})^2$ on the the $SU(12)$-symmetric spectrum can be obtained through perturbation theory.
To first order in $g^2$, the shift in the energy of any state is equal to the expectation value of $H_{el}$.
The $\sigma$- and $\pi$-meson states are both quark-antiquark states in the {\bf 66} irrep of $SU(12)$, and therefore, both have a ${\bf 3}_c$ color charge on the quark site and receive the same mass shift.\footnote{This also explains why
there are three other states nearly degenerate with the mesons, as seen in Fig.~\ref{fig:specDegenh}. 
Each of these states carry a ${\bf 3}_c$ or $\overline{{\bf 3}}_c$ color charge on the quark site and consequently have the same energy at first order in perturbation theory.
}
For large $g^2$, the only finite-energy excitations of the trivial vacuum (all sites unoccupied) are bare baryons and antibaryons,
and the spectrum is one of non-interacting color-singlet baryons.
Each quark (antiquark) site hosts $4$ distinct baryons (antibaryons) in correspondence with the multiplicity of the $I=3/2$ irrep.
As a result, the $\sigma$, $\pi$, $I=2,3$ mesons, deuteron and antideuteron are all degenerate.

\begin{figure}[!ht]
    \centering
    \includegraphics[width=14cm]{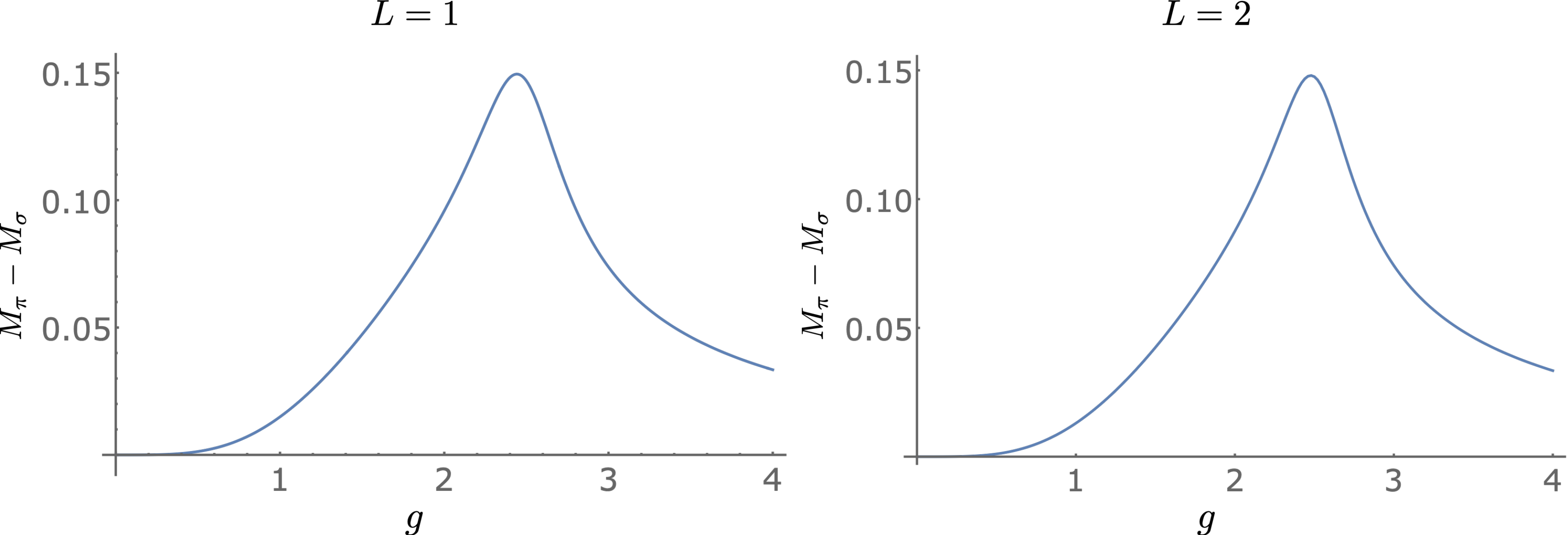}
    \caption{
    The mass splitting between the $\sigma$- and $\pi$-mesons for $L=1$ (left panel) and $L=2$ (right panel).}
    \label{fig:PiSigSplit}
\end{figure}

The $\sigma$- and $\pi$-meson mass splitting is shown in Fig.~\ref{fig:PiSigSplit} and has a clear maxima for $g \sim 2.4$. 
Intriguingly, this corresponds to the maximum of the linear entropy between quark and antiquarks (as discussed in Sec.~\ref{sec:eigenent}),
and suggests a connection between symmetry, via degeneracies in the spectrum, and entanglement.
This shares similarities with the correspondence between Wigner's $SU(4)$ spin-flavor
symmetry~\cite{PhysRev.51.106,PhysRev.51.947,PhysRev.56.519},
which becomes manifest in low-energy nuclear forces in the large-$N_c$ limit of QCD~\cite{Kaplan:1995yg,Kaplan:1996rk},
and entanglement suppression in
nucleon-nucleon scattering found in Ref.~\cite{Beane:2018oxh} (see also Refs.~\cite{Beane:2021zvo,Low:2021ufv,Beane:2020wjl}).

Color singlet baryons are also present in this system, formed by contracting the color indices of three quarks with a Levi-Civita tensor (and antibaryons are formed from three antiquarks).
A baryon is composed of three $I=1/2$ quarks in the (symmetric) $I=3/2$ configuration and in a (antisymmetric) color singlet.
It will be referred to as the $\Delta$, highlighting its similarity to the $\Delta$-resonance in $3+1$D QCD. 
Interestingly, there is an isoscalar $\Delta\Delta$ bound state, which will be referred to as the deuteron. 
The existence of a deuteron makes this system valuable from the standpoint of quantum simulations of the formation of nuclei in a model of reduced complexity.
The mass of the $\Delta$, $M_{\Delta}$, and the binding energy of the deuteron, $B_{\Delta \Delta} = 2 M_{\Delta} - M_{\Delta \Delta}$, are shown in Table~\ref{tab:specBneq0} for a range of strong couplings.
\begin{table}[!ht]
\renewcommand{\arraystretch}{1.2}
\begin{tabularx}{0.4\textwidth}{||c | Y | Y ||} 
\hline
\multicolumn{3}{||c||}{$L=1 $} \\
 \hline
 $g^2$ & $M_{\Delta}$ & $B_{\Delta \Delta}$\\
 \hline\hline
 8 & 3.10 & $2.61\times 10^{-4}$\\ 
 \hline
 4 & 3.16 & $5.48\times 10^{-4}$\\
 \hline
 2 & 3.22 & $6.12\times 10^{-4}$\\
 \hline
 1 & 3.27 & $3.84\times 10^{-4}$\\
 \hline
 1/2 & 3.31 & $1.61\times 10^{-4}$\\
 \hline
 1/4 & 3.33 & $5.27\times 10^{-5}$\\
 \hline
 1/8 & 3.34 & $1.52\times 10^{-5}$\\
 \hline
\end{tabularx}
\renewcommand{\arraystretch}{1}
\qquad\qquad\qquad
\renewcommand{\arraystretch}{1.2}
\begin{tabularx}{0.4\textwidth}{|| c | Y | Y ||} 
\hline
\multicolumn{3}{||c||}{$L=2 $} \\
 \hline
 $g^2$ & $M_{\Delta}$ & $B_{\Delta \Delta}$\\
 \hline\hline
 8  & 3.10 & $2.50\times 10^{-4}$\\ 
 \hline
 4  & 3.16 & $4.95\times 10^{-4}$\\
 \hline
 2  & 3.21 & $5.07\times 10^{-4}$\\
 \hline
 1  & 3.24 & $4.60\times 10^{-4}$\\
 \hline
 1/2  & 3.25 & $1.53\times 10^{-3}$\\
 \hline
 1/4 & 3.23 & $3.91\times 10^{-3}$\\
 \hline
 1/8 & 3.20 & $3.35\times 10^{-3}$\\
 \hline
\end{tabularx}
\renewcommand{\arraystretch}{1}
\caption{
The mass of the $\Delta$ and the binding energy of the deuteron 
in $1+1$D QCD with $N_f=2$ for systems with $L=1,2$ spatial sites.}
\label{tab:specBneq0}
\end{table}

\begin{figure}[!ht]
    \centering
    \includegraphics[width=\columnwidth]{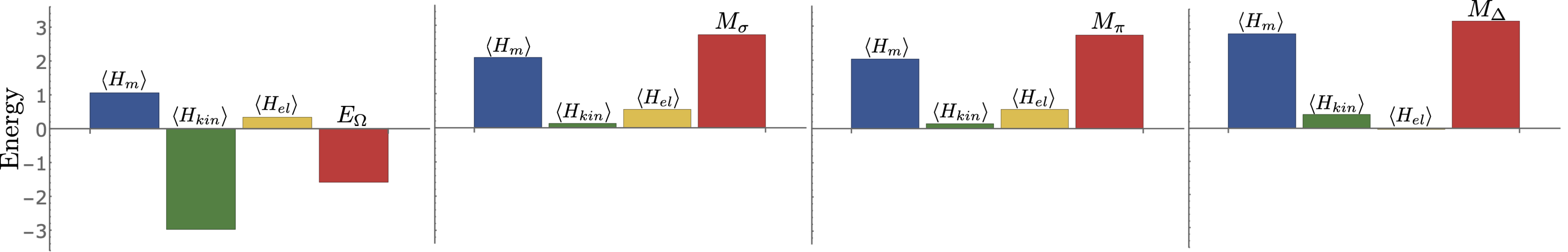}
    \caption{
    The decomposition of vacuum energy ($E_{\Omega}$) and the masses of the lightest hadrons ($M_{\sigma}$, $M_{\pi}$ and $M_{\Delta}$) 
    into contributions from the mass, the kinetic and the chromo-electric field terms in the Hamiltonian, defined in axial gauge, for $1+1$D QCD with $N_f=L=2$ and $m=g=1$.}
    \label{fig:massdeco}
\end{figure}

Understanding and quantifying the structure of the lowest-lying hadrons is a priority for nuclear physics research~\cite{LongRangePlan}.
Great progress has been made, experimentally, analytically and computationally,
in dissecting the mass and angular momentum of the proton (see, for example, Refs.~\cite{deFlorian:2009vb,Nocera:2014gqa,COMPASS:2015mhb,Yang:2018nqn,Alexandrou:2020sml,Ji:2021mtz,Wang:2021vqy,Lorce:2021xku}). 
This provides, in part, the foundation for anticipated precision studies at the future electron-ion collider (EIC)~\cite{Boer:2011fh,Accardi:2012qut} at Brookhaven National Laboratory.
Decompositions of the vacuum energy and the masses of the $\sigma$, $\pi$ and $\Delta$ are shown in Fig.~\ref{fig:massdeco} where, for example, the chromo-electric contribution to the
$\sigma$ is $\langle H_{el} \rangle = \bra{\sigma}  H_{el} \ket{\sigma} - \bra{\Omega}  H_{el} \ket{\Omega}$.
These calculations demonstrate the potential of future quantum simulations in being able to quantify decompositions of properties of the nucleon,
including in dense matter.
For the baryon states, it is $H_{el}$ that is responsible for the system coalescing into localized color singlets in order to minimize the energy in the chromo-electric field (between spatial sites).

\begin{figure}[!ht]
    \centering
    \includegraphics[width=14cm]{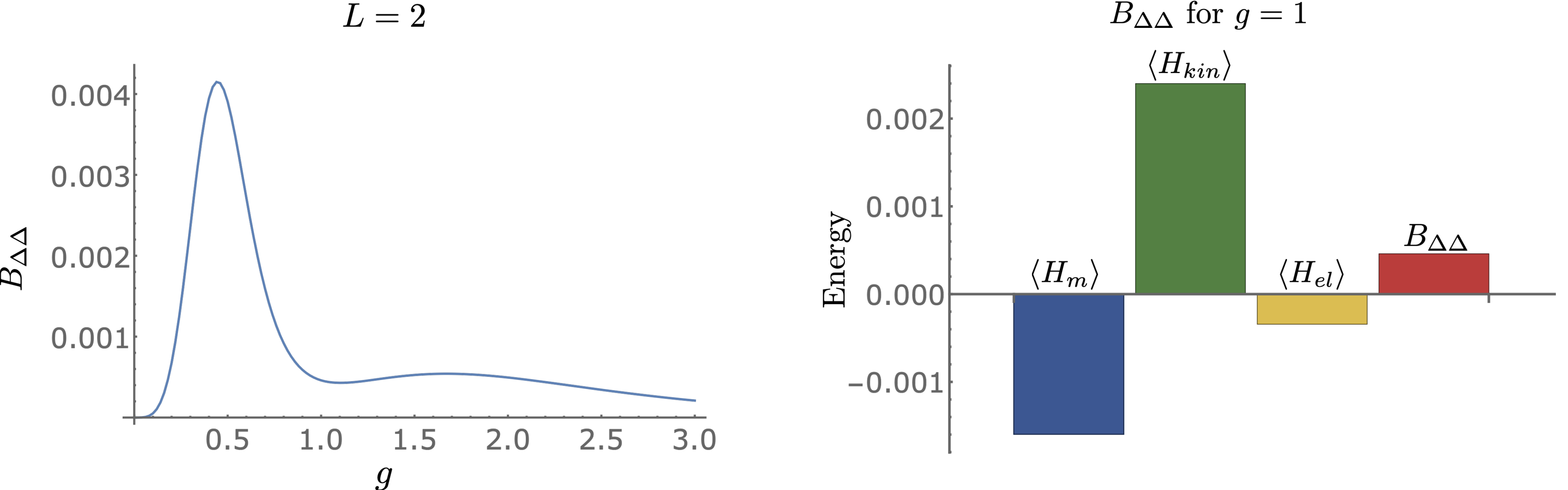}
    \caption{The left panel shows the deuteron binding energy, $B_{\Delta \Delta}$, for $m=1$ and $L=2$. The right panel shows the decomposition of $B_{\Delta \Delta}$ into contributions from the Hamiltonian for $g=1$.}
    \label{fig:deutBE}
\end{figure}

The deuteron binding energy is shown in the left panel of Fig.~\ref{fig:deutBE} as a function of $g$.
While the deuteron is unbound at $g=0$ for obvious reasons, it is also unbound at large $g$ because the spectrum is that of non-interacting color-singlet (anti)baryons.
Therefore, the non-trivial aspects of deuteron binding for these systems is for intermediate 
values of $g$. The decomposition of $B_{\Delta \Delta}$ is shown in the right panel of Fig.~\ref{fig:deutBE}, where, for example, the chromo-electric contribution is
\begin{equation}
    \langle H_{el} \rangle = 2 \big ( \bra{\Delta}  H_{el} \ket{\Delta} - \bra{\Omega}  H_{el} \ket{\Omega} \big ) - \big (\bra{\Delta \Delta}  H_{el} \ket{\Delta \Delta} - \bra{\Omega}  H_{el} \ket{\Omega} \big ) \ .
    \label{eq:deutBEdec}
\end{equation}
The largest contribution to the binding energy is $\langle H_{kin} \rangle$, which is the term responsible for creating $q \overline{q}$ pairs.
This suggests that meson-exchange may play a significant role in the attraction between baryons,
as is the case in $3+1$D QCD, but larger systems will need to be studied before 
definitive conclusions can be drawn.
One consequence of the lightest baryon 
being $I=3/2$ is that, for $L=1$, 
the $I_3=+3/2$ state completely occupies the up-quark sites.
Thus the system factorizes into an inert up-quark sector and a dynamic down-quark sector, and the absolute energy of the lowest-lying baryon state can be written as 
$E_{\Delta} =
M_{\Delta} + E_{\Omega}^{2 f} = 
3m + E_{\Omega}^{1 f}$,
where
$E_{\Omega}^{1,2 f}$ is the
vacuum energy of the 
$N_f=1,2$ flavor systems.
Analogously, the deuteron absolute energy is 
$E_{\Delta \Delta}=6m$, and therefore the 
deuteron binding energy can be written as
$B_{\Delta \Delta}= 2(3m+E_{\Omega}^{1 f}-E_{\Omega}^{2 f}) - (6m-E_{\Omega}^{2 f})
= 2E_{\Omega}^{1 f}-E_{\Omega}^{2 f}$.
This is quite a remarkable result because, in this system, the deuteron binding energy depends only on the difference between
the $N_f=1$ and $N_f=2$ vacuum energies, being bound when $2 E_{\Omega}^{1 f} - E_{\Omega}^{2 f} > 0$.
As has been discussed previously, it is the 
$q\overline{q}$ contribution from this difference that dominates the binding.
%

\subsubsection{The Low-Lying Spectrum Using D-Wave's Quantum Annealers}
\label{sec:dwave_spectrum}
\noindent
The low-lying spectrum of this system can also be determined through annealing by using
D-Wave's quantum annealer (QA) {\tt Advantage}~\cite{DwaveLeap}, 
a device with 5627 superconducting flux qubits, with a 15-way qubit connectivity via Josephson junctions rf-SQUID couplers~\cite{PhysRevB.80.052506}. 
Not only did this enable the determination of the energies of low-lying states, but it also assessed the ability of this quantum device to isolate nearly degenerate states.
The time-dependent Hamiltonian
of the device, which our systems are to be mapped,
are of the form of an Ising model, with the freedom to specify the single- and two-qubit coefficients. Alternatively, the Ising model can be rewritten in a quadratic unconstrained binary
optimization (QUBO) form, $f_Q(x)=\sum_{ij} Q_{ij}x_i x_j$, 
where $x_i$ are binary variables 
and $Q_{ij}$ is a QUBO matrix, which contains the coefficients of single-qubit ($i=j$) and two-qubit ($i\neq j$) terms. 
The QUBO matrix is the input that is submitted to 
{\tt Advantage}, with the output being a bit-string that minimizes $f_Q$.
Due to the qubit connectivity of {\tt Advantage},
multiple physical qubits are chained together to recover the required connectivity, limiting the system size that can be annealed.

The QA {\tt Advantage} was used to
determine the lowest three states in the $B=0$ sector of the $L=1$ system, with $m=g=1$ and $h=2$, following techniques presented in Ref.~\cite{Illa:2022jqb}. 
In that work, the objective function to be minimized is defined as $F=\langle \Psi \rvert \tilde{H} \lvert \Psi \rangle -\eta \langle \Psi| \Psi \rangle$~\cite{doi:10.1021/acs.jctc.9b00402}, where $\eta$ is a parameter that is included to avoid the null solution, and its optimal value 
can be iteratively tuned to be as close to the ground-state energy as possible. 
The wavefunction is expanded in a finite dimensional orthonormal basis $\psi_{\alpha}$, $\lvert \Psi \rangle =\sum^{n_s}_{\alpha} a_\alpha |\psi_{\alpha}\rangle$, which in this case reduces the dimensionality of $H$ to $88$, defining $\tilde{H}$, thus making it feasible to study with {\tt Advantage}. 
The procedure to write the objective function in a QUBO form can be found in Ref.~\cite{Illa:2022jqb} (and briefly described in App.~\ref{app:dwave}), where the coefficients $a_{\alpha}$ are
digitized using $K$ binary variables~\cite{doi:10.1021/acs.jctc.9b00402}, and the adaptive QA eigenvalue solver is implemented by using the zooming method~\cite{Chang:2019,ARahman:2021ktn}. To reduce the uncertainty in the resulting energy and wavefunction, due to the noisy nature of this QA, the iterative procedure
described in Ref.~\cite{Illa:2022jqb} was used, where the (low-precision) solution obtained from the machine after several zooming steps constituted the starting point of a new anneal. This led to a reduction of the uncertainty by an order of magnitude (while effectively only doubling the resources used).

\begin{figure}[!ht]
    \centering
    \includegraphics[width=\columnwidth]{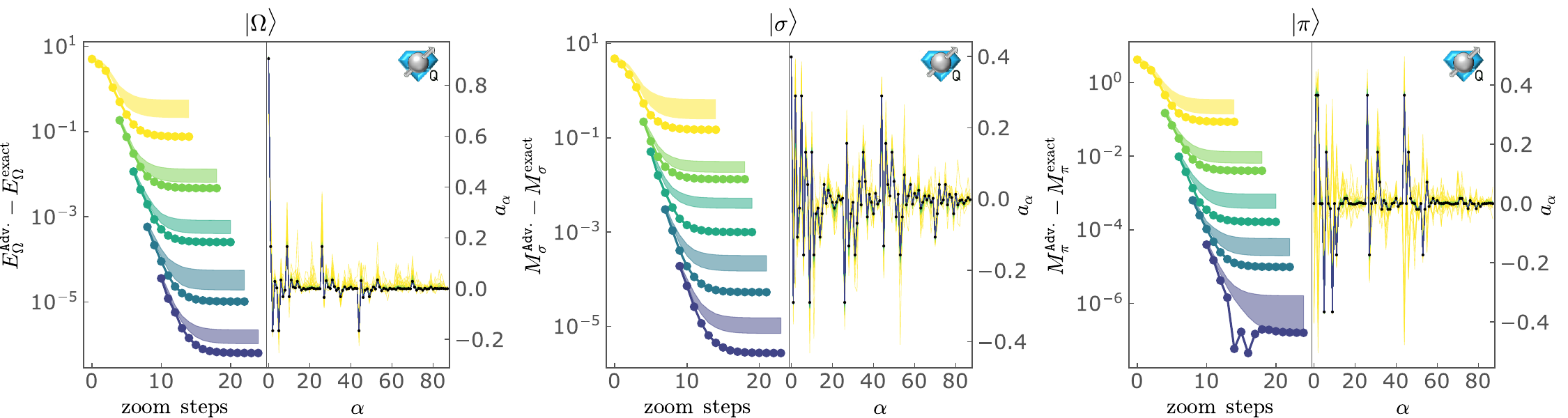}
    \caption{
    Iterative convergence of the energy, masses and wavefunctions for the three lowest-lying states in the $B=0$ sector of $1+1$D QCD with $N_f=2$ and $m=g=L=1$: vacuum (left), $\sigma$-meson (center) and $\pi$-meson (right). 
    The different colors correspond to different steps of the iterative procedure that is described in the main text. 
    The oscillatory behavior seen in the right panel around the 15th zoom step is discussed in App.~\ref{app:dwave}.
    The blue icons in the upper right indicate that this calculation was done on a quantum device~\cite{Klco:2019xro}.}
    \label{fig:QAresults}
\end{figure}

Results obtained using {\tt Advantage} are shown in Fig.~\ref{fig:QAresults}, where the three panels show the convergence of the energy of the 
vacuum state (left), the mass of the $\sigma$-meson (center) and the mass of the $\pi$-meson (right) as a function of zoom steps, as well as comparisons to the exact wavefunctions. The bands in the plot correspond to 68\% confidence intervals determined from 20 independent runs of the annealing workflow, where each corresponds to $10^3$ anneals with an annealing time of $t_A=20$ $\mu$s, and the points correspond to the lowest energy found by the QA. The parameter $K$ in the digitization of $a_{\alpha}$ is set to $K=2$. The parameter $\eta$ is first set close enough to the corresponding energy (e.g., $\eta=0$ for the ground-state), and for the subsequent iterative steps it is set to the lowest energy found in the previous step. The first two excited states are nearly degenerate, and after projecting out the ground state, {\tt Advantage} finds both states in the first step of the iterative procedure (as shown by the yellow lines in the $\pi$ wavefunction of Fig.~\ref{fig:QAresults}). 
However, after one iterative step, the QA converges to one of the two excited states. 
It first finds the second excited state (the $\pi$-meson), and once this state is known with sufficient precision, it can be projected out to study the other excited state. 
The converged values for the energies and masses of these states are shown in Table~\ref{tab:QAresults},
along with the exact results. The uncertainties in these values should be understood as uncertainties on an upper bound of the energy (as they result from a variational calculation). For more details see App.~\ref{app:dwave}.
\begin{table}[!ht]
\renewcommand{\arraystretch}{1.2}
\begin{tabularx}{\textwidth}{|| c | c | Y | Y | Y ||} 
\hline
\multicolumn{5}{||c||}{$L=1$} \\
 \hline
 \multicolumn{2}{||c|}{} & $\ket{\Omega}$ & $\ket{\sigma}$ & $\ket{\pi}$ \\
 \hline\hline
 \multirow{2}{*}{Exact} & Energy & $-0.5491067$ & $2.177749$ & $2.1926786$ \\ 
 \cline{2-5}
 & Mass & - & $2.726855$ & $2.7417853$\\
 \hline
 \multirow{2}{*}{\tt Advantage} &Energy & $-0.5491051(6)$ & $2.177760(4)$ & $2.1926809(7)$ \\ 
 \cline{2-5}
 & Mass & - & $2.726865(4)$ & $2.7417860(9)$\\
 \hline
\end{tabularx}
\renewcommand{\arraystretch}{1}
\caption{Energies and masses of the three lowest-lying states in the $B=0$ sector of $1+1$D QCD with $N_f=2$ and $m=g=L=1$. Shown are the exact results from diagonalization of the Hamiltonian matrix and those obtained from D-Wave's {\tt Advantage}.}
\label{tab:QAresults}
\end{table}
%

\subsubsection{Quark-Antiquark Entanglement in the Spectra via Exact Diagonalization}
\label{sec:eigenent}
\begin{figure}[!ht]
    \centering
    \includegraphics[width=14cm]{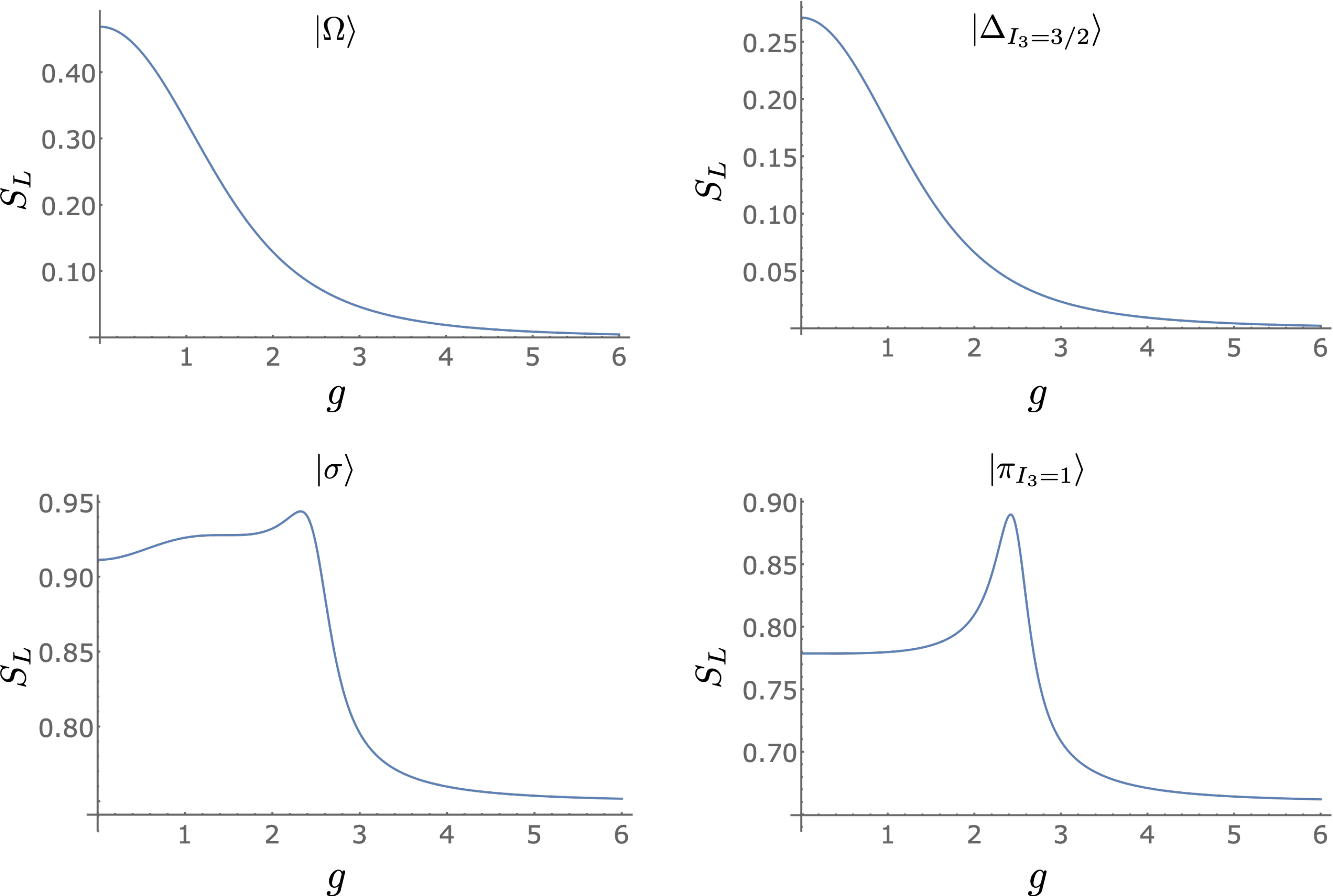}
    \caption{The linear entropy between quarks and antiquarks in $\ket{\Omega}$, $\ket{\Delta_{I_3=3/2}}$, $\ket{\sigma}$ and $\ket{\pi_{I_3=1}}$ for $m=L=1$.}
    \label{fig:m1N2Linent}
\end{figure}

\noindent
With $h \gg g$, the eigenstates of the Hamiltonian are color singlets and irreps of isospin. 
As these are global quantum numbers (summed over the lattice) the eigenstates are generically entangled among the color and isospin components at each lattice site.  With the hope of gaining insight into $3+1$D QCD, aspects of the entanglement structure of the $L=1$ wavefunctions are explored via exact methods.
An interesting measure of entanglement for these systems is the linear entropy between quarks and antiquarks, defined as
\begin{equation}
    S_L = 1 - \Tr [\rho_q^2]
    \ ,
\end{equation}
where $\rho_q =  \Tr_{\overline{q}} [\rho]$ and $\rho$ is a density matrix of the system. Shown in Fig.~\ref{fig:m1N2Linent} is the linear entropy between quarks and antiquarks in
$\ket{\Omega}$, $\ket{\sigma}$, $\ket{\pi_{I_3=1}}$ and $\ket{\Delta_{I_3=3/2}}$ as a function of $g$.
The deuteron is not shown as there is only one basis state contributing for $L=1$.

\begin{figure}
    \centering
    \includegraphics[width=14cm]{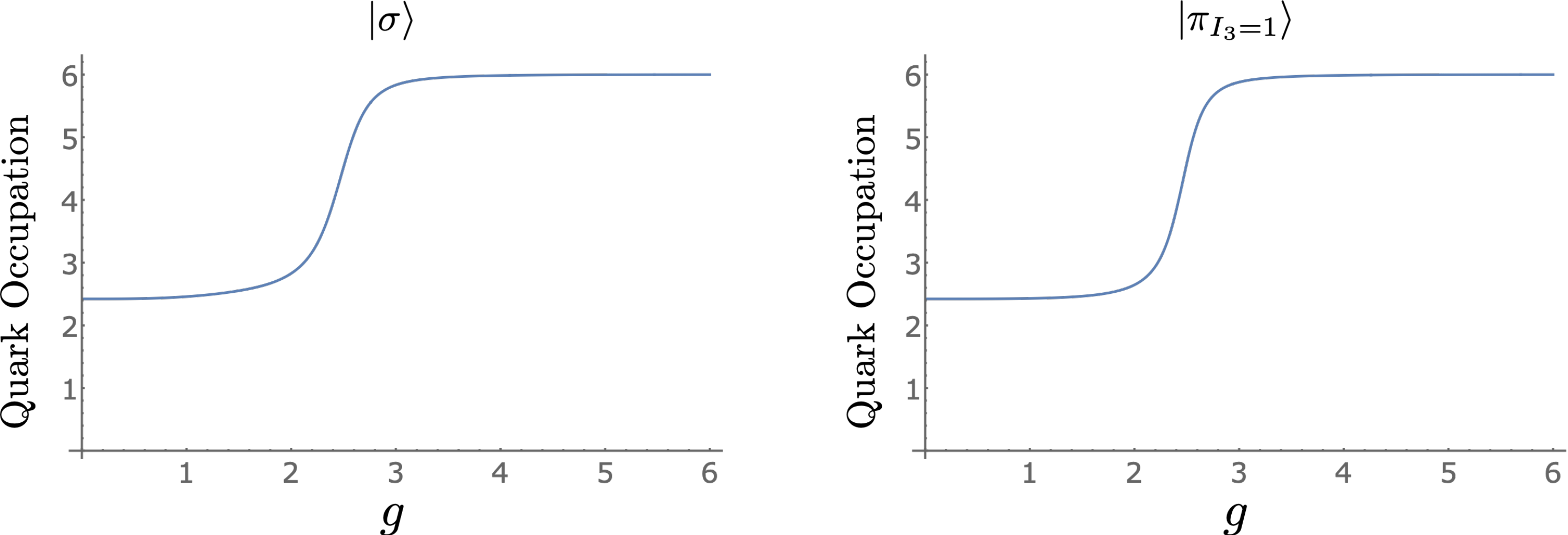}
    \caption{The expectation value of quark occupation in the $\ket{\sigma}$ and $\ket{\pi_{I_3 = 1}}$ for $m=L=1$.}
    \label{fig:m1N2Occ}
\end{figure}

The scaling of the linear entropy in the vacuum and baryon with $g$ can be understood as follows. 
As $g$ increases, color singlets on each site have the least energy density.
The vacuum becomes dominated by the unoccupied state and the $\Delta$ becomes dominated by the ``bare" $\Delta$ with all three quarks located on one site in a color singlet. 
As the entropy generically scales with the number of available states, 
the vacuum and baryon have decreasing entropy for increasing $g$.
The situation for the $\pi$ and $\sigma$ is somewhat more interesting. 
For small $g$, their wavefunctions are dominated by $q \overline{q}$ excitations on top of the trivial vacuum,  which minimizes the contributions from the mass term. 
However, color singlets are preferred as $g$ increases, 
and the mesons become primarily composed of baryon-antibaryon ($B \overline{B}$) excitations.
There are more $q \overline{q}$ states than 
$B \overline{B}$ states with a given $I_3$, 
and therefore there is more entropy at small $g$ than large $g$. 
The peak at intermediate $g$ occurs at the crossover between these two regimes where the meson has a sizable contribution from both $q \overline{q}$ and $B \overline{B}$ excitations. 
To illustrate this, 
the expectation value of total 
quark occupation (number of quarks plus the number of antiquarks) is shown in Fig.~\ref{fig:m1N2Occ}. 
For small $g$, the occupation is near $2$ since the state is mostly composed of $q \overline{q}$,
while for large $g$ it approaches $6$ as the state mostly consists of $B \overline{B}$. 
This is a transition from the excitations being 
``color-flux tubes" between quark and antiquark of the same color to bound states of color-singlet baryons and antibaryons.
%

\subsection{Digital Quantum Circuits}
\label{sec:Circuits}
\noindent
The Hamiltonian for $1+1$D QCD with arbitrary $N_c$ and $N_f$, when written in terms of spin operators, can be naturally mapped onto a quantum device with qubit registers. In this section the time evolution for systems with $N_c = 3$ and $N_f=2$ are developed.

\subsubsection{Time Evolution}
\begin{figure}[!ht]
    \centering
    \includegraphics[height=0.3\textheight]{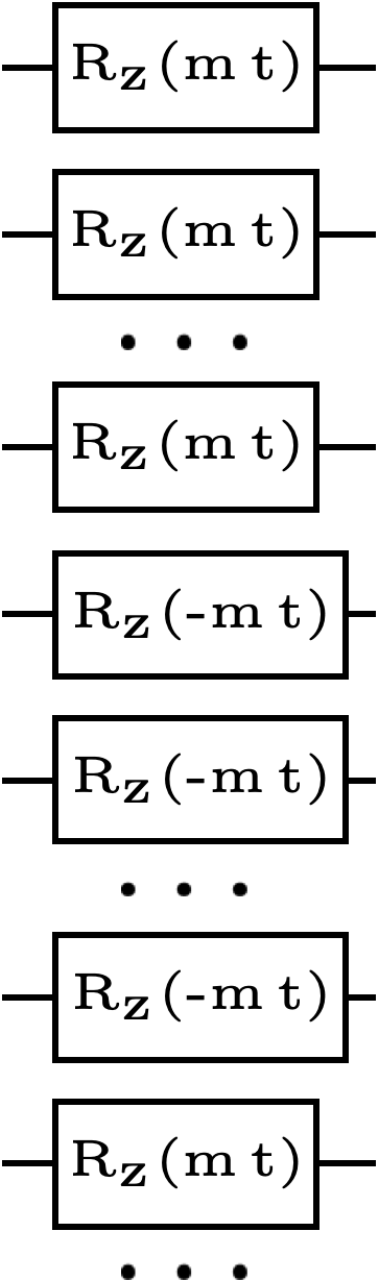}
    \caption{The quantum circuit that implements
    time evolution by the mass term,
    $U_m(t) = \exp(- i H_m t)$.}
    \label{circ:Um}
\end{figure}

\noindent
To perform time evolution on a quantum computer, the operator $U(t) = \exp(-i H t)$ is reproduced by a sequence of gates applied to the qubit register.
Generally, a Hamiltonian cannot be directly mapped to such a sequence efficiently, but each of the elements in a Trotter decomposition can, with systematically reducible errors.
Typically, the Hamiltonian is divided into Pauli strings whose unitary evolution can be implemented with quantum circuits that are readily constructed.
For a Trotter step of size $t$, the circuit that implements the time evolution from the mass term, $U_m(t) = \exp(- i H_m t)$, is shown in Fig.~\ref{circ:Um}.
The staggered mass leads to quarks being rotated by a positive angle and antiquarks being rotated by a negative angle. 
Only single qubit rotations about the z-axis are required for its implementation, with 
$R_Z(\theta) = \exp(-i \theta Z/2)$.
The circuit that implements
the evolution from the baryon chemical potential, $\mu_B$,
$U_{\mu_B}(t) = \exp(- i H_{\mu_B} t)$, 
is similar to $U_m(t)$  with 
$m \to \mu_B/3$, and with both quarks and antiquarks rotated by the same angle. 
Similarly, the circuit that implements the evolution from
the isospin chemical potential, $\mu_I$,
$U_{\mu_I}(t) = \exp(- i H_{\mu_I} t)$, 
is similar to $U_m(t)$ with $m \to
\mu_I/2$ and up (down) quarks rotated by a negative (positive) angle.

\begin{figure}[!ht]
    \centering
    \includegraphics[width=\textwidth]{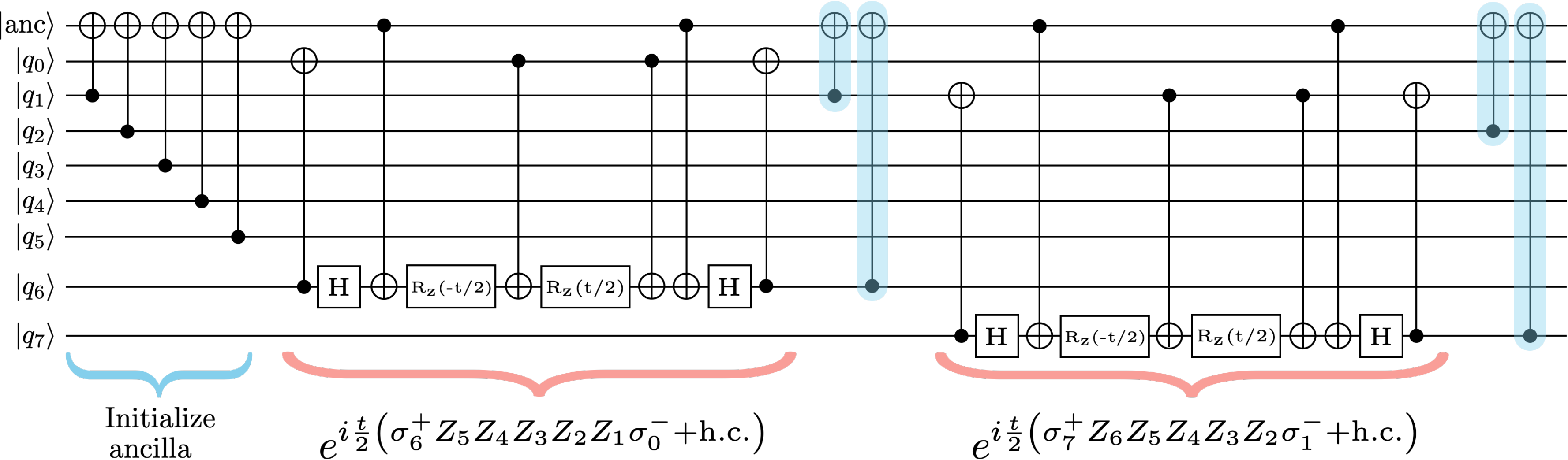}
    \caption{
    A circuit that implements the time evolution from two sequential hopping terms.
    Implementing $\exp(-i H_{kin} t)$ in Eq.~(\ref{eq:Hkin2flav}) is a straightforward extension of this circuit.}
    \label{circ:UkinAnc}
\end{figure}

The kinetic piece of the Hamiltonian, Eq.~(\ref{eq:Hkin2flav}), is composed of hopping terms of the form
\begin{equation}
H_{kin} \ \sim \ 
    \sigma^+ ZZZZZ \sigma^- + \rm{h.c.} \ .
    \label{eq:hop}
\end{equation}
The $\sigma^+$ and $\sigma^-$ operators enable quarks and antiquarks to move between sites with the same color and flavor 
(create $\overline{q}^\alpha_i q_\alpha^i$ pairs)
and the string of $Z$ operators 
incorporates the signs from Pauli statistics.
The circuits for Trotterizing these terms are 
based on circuits in Ref.~\cite{Stetina:2020abi}. We introduce an ancilla to 
accumulate the parity of the JW string of $Z$s.
This provides a mechanism for the 
different hopping terms to re-use 
previously computed
(partial-)parity.\footnote{An ancilla was used similarly in Ref.~\cite{Qchem2014}.}
The circuit for the first two hopping terms is shown in Fig.~\ref{circ:UkinAnc}.
The first circuit operations initialize 
the ancilla to store the parity of the string of $Z$s between the first and last qubit of the string. Next, the system is evolved
by the exponential of the hopping term. After the exponential of each hopping term, the ancilla is modified for the parity of the subsequent hopping term 
(the CNOTs highlighted in blue).
Note that the hopping of quarks, or antiquarks, of different flavors and colors commute, and the Trotter decomposition is exact (without Trotterization errors) over a single spatial site.

\begin{figure}[!ht]
    \centering
    \includegraphics[width=10cm]{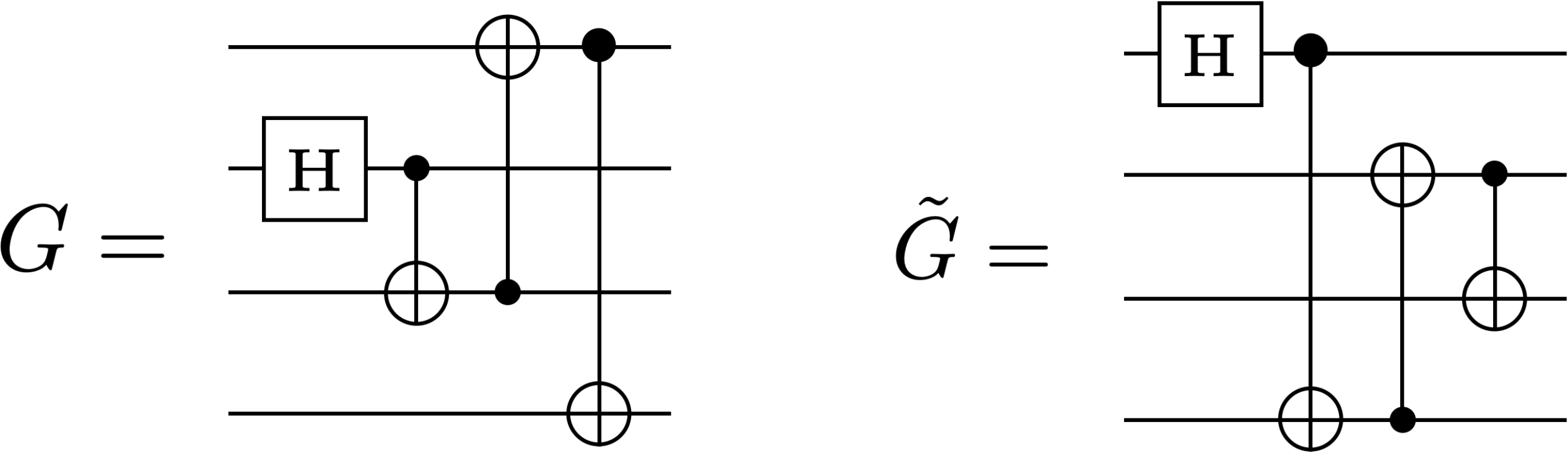}
    \caption{Two GHZ state-preparation circuits.}
    \label{circ:GHZ}
\end{figure}

\begin{figure}[!ht]
    \centering
    \includegraphics[width=15cm]{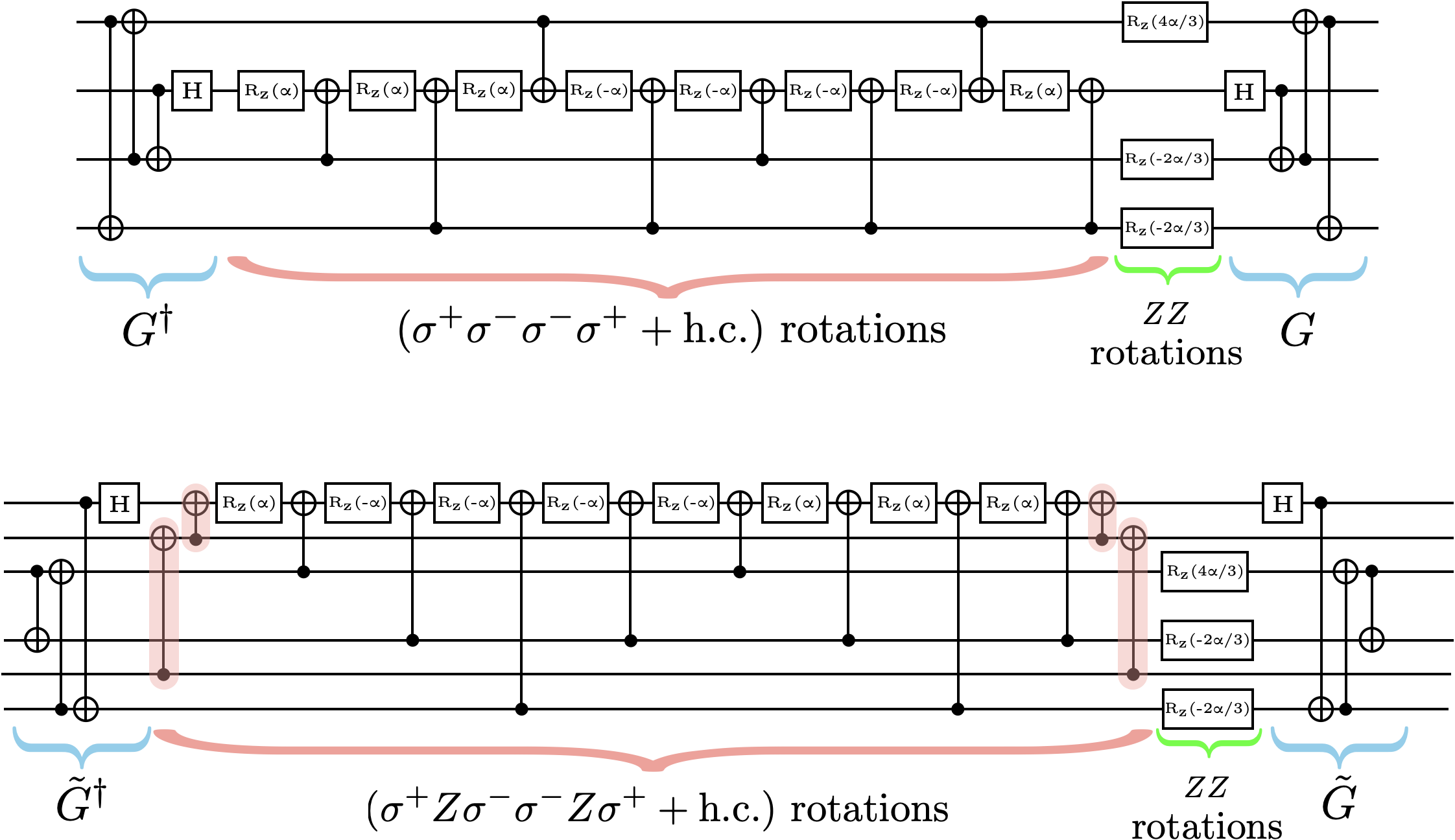}
    \caption{The circuits that implement the time evolution of ${\exp}(-8 i \alpha Q_{n,f}^{(a)} \, Q_{m,f'}^{(a)})$. 
    Specifically, the upper circuit implements\\
    $\exp{-i 4 \alpha [ (\sigma^+\sigma^-\sigma^-\sigma^+ + {\rm h.c.}) +  \frac{1}{12}(2 IZIZ -IZZI - ZIIZ) ]}$,\\ 
    while the lower circuit implements\\
    $\exp{-i 4 \alpha[ (\sigma^+ Z \sigma^-\sigma^- Z \sigma^- + {\rm h.c.}) +  \frac{1}{12}(2 ZIIZII -IIZZII - ZIIIIZ) ]}$.\\
    The CNOTs highlighted in red account for the $Z$s in $\sigma^+ Z \sigma^- \sigma^- Z \sigma^+$.
    For 
    $SU(3)$ with $N_f=2$ and 
    $L=1$, the required evolution operators have $\alpha = t g^2 /8$.}
    \label{circ:UpmmpZZ}
\end{figure}

Implementation of the time-evolution 
induced by the energy density in the 
chromo-electric field, $H_{el}$, 
given in Eq.~(\ref{eq:QnfQmfp}),
is the most challenging due to its 
inherent non-locality in axial gauge.
There are two distinct types of contributions: One is from same-site interactions and the other from interactions between different sites.
For the same-site interactions, the operator is the product of charges 
$Q_{n,f}^{(a)} \, Q_{n,f}^{(a)}$, which contains only $ZZ$ operators, and is digitized with the standard two CNOT circuit.\footnote{Using the native $ZX$ gate on IBM's devices allows this to be done with a single two-qubit entangling gate~\cite{Kim2021ScalableEM}.}
The $Q_{n,f}^{(a)} \, Q_{m,f'}^{(a)}$ operators contain 4-qubit interactions of the form 
$(\sigma^+ \sigma^- \sigma^- \sigma^+ + \ {\rm h.c.})$
and 
6-qubit interactions of the form 
$(\sigma^+ Z \sigma^- \sigma^- Z \sigma^+ + \ {\rm h.c.})$,
in addition to $ZZ$ contributions.
The manipulations required to implement the 6-qubit operators parallel those required for the 4-qubit operators, and here only the latter is discussed in detail.
These operators can be decomposed into eight mutually commuting terms,
\begin{align}
\sigma^+ \sigma^- \sigma^- \sigma^+ + {\rm h.c.} = \frac{1}{8}(&XXXX + YYXX + YXYX - YXXY - XYYX + XYXY \nonumber \\
&+ XXYY + YYYY) \ .
\label{eq:pmmp}
\end{align}
The strategy for identifying the corresponding time evolution circuit is to first apply a unitary that diagonalizes every term, apply the diagonal rotations, and finally, act with the inverse unitary to return to the computational basis. 
By only applying diagonal rotations, 
many of the CNOTs can be arranged to cancel.
Each of the eight Pauli strings
in Eq.~(\ref{eq:pmmp})
takes a state in the computational basis to the corresponding bit-flipped state (up to a phase). 
This suggests that the desired eigenbasis
pairs together states with their bit-flipped counterpart, which is an inherent property of the GHZ basis~\cite{Stetina:2020abi}. 
In fact, any permutation of the GHZ state-preparation circuit diagonalizes the interaction. 
The two that will be used,
denoted by $G$ and $\tilde G$,
are shown in Fig.~\ref{circ:GHZ}.
In the diagonal bases, the Pauli strings
in Eq.~(\ref{eq:pmmp}) become
\begin{align}
    G^{\dagger}\ (\sigma^+ \sigma^- \sigma^- \sigma^+ + {\rm h.c.})\ G =\ \frac{1}{8} (&IIZI - ZIZZ - ZZZZ + ZIZI + IZZI - IIZZ \nonumber \\
    &- IZZZ + ZZZI ) \ , \nonumber \\
    \tilde{G}^{\dagger}\ (\sigma^+ \sigma^- \sigma^- \sigma^+ + {\rm h.c.})\ \tilde{G} =\ \frac{1}{8} (&IIIZ - IZZZ - IIZZ + ZIIZ + IZIZ - ZZZZ \nonumber \\
    &- ZIZZ + ZZIZ) \ .
    \label{eq:pmmpdiag}
\end{align}
Another simplification comes from the fact that 
$ZZ$ in the computational basis becomes  
a single $Z$ in a GHZ basis if the GHZ state-preparation circuit has a CNOT connecting the two $Z$s. 
For the case at hand, this implies
\begin{align}
    G^{\dagger}\ (IZZI + IZIZ + ZIIZ)\ G =& \ IZII + IIIZ + ZIII \ , \nonumber \\[4pt]
    \tilde{G}^{\dagger}\ (ZIZI + IZZI + ZIIZ)\ \tilde{G} =& \ IIZI + IZII + ZIII  \ .
    \label{eq:ZZGHZ}
\end{align}
As a consequence, all nine $ZZ$ terms in $Q_{n,f}^{(a)} \, Q_{m,f'}^{(a)}$ 
become single $Z$s in a GHZ basis, thus requiring no additional CNOT gates to implement. 
Central elements of the circuits 
required to implement time evolution of the chromo-electric energy density
are shown in Fig.~\ref{circ:UpmmpZZ},
which extends the circuit presented in Fig.~4 of Ref.~\cite{Stetina:2020abi} to non-Abelian gauge theories.
More details on these circuits can be found in App.~\ref{app:circ}.

\subsubsection{Trotterization, Color Symmetry and Color Twirling}
\label{sec:colorBreak}
\noindent
After fixing the gauge, the Hamiltonian is no longer manifestly invariant under local  $SU(3)$ gauge transformations. 
However, as is well known, observables of the theory are correctly computed from such a gauge-fixed Hamiltonian, which possesses a remnant global $SU(3)$ symmetry.
This section addresses the extent to which this symmetry is preserved by Trotterization of the time-evolution operator. 
The focus will be on 
the $N_f=1$ theory as including additional flavors
does not introduce new complications.

Trotterization of the mass and kinetic parts of the Hamiltonian,
while having  non-zero commutators between some terms, preserves the global $SU(3)$ symmetry.
The time evolution of $Q_{n}^{(a)} \, Q_{n}^{(a)}$
can be implemented in a unitary operator without Trotter errors, and, therefore, does not break $SU(3)$. 
On the other hand,
the time evolution induced by 
$Q_{n}^{(a)} \, Q_{m}^{(a)}$
is implemented by the operator being divided into 
four terms:
$(Q^{(1)}_n \, Q^{(1)}_{m} + Q^{(2)}_n \, Q^{(2)}_m)$,  $(Q^{(4)}_n \, Q^{(4)}_m + Q^{(5)}_n \, Q^{(5)}_m)$, $(Q^{(6)}_n \, Q^{(6)}_m + Q^{(7)}_n \, Q^{(7)}_m)$ and $(Q^{(3)}_n \, Q^{(3)}_m + Q^{(8)}_n \, Q^{(8)}_m)$. In order for global $SU(3)$ to be unbroken, 
the sum over the entire lattice
of each of the 8 gauge charges must be unchanged under time evolution. 
Therefore, 
the object of interest is the commutator
\begin{equation}
\mathcal{C} = \left [ \sum_{n=0}^{2L-1}Q^{(a)}_{n} \ , \ Q^{(\tilde{b})}_{m}\cdot Q^{(\tilde{b})}_{l} \right ] \ ,
\label{eq:Qcomm}
\end{equation}
where $\tilde{b}$ is summed over the elements of one of the pairs in $\{(1,2),\,(4,5),\,(6,7),\,(3,8)\}$. 
It is found that this commutator only vanishes if $a=3$ or $a=8$, or if $\tilde{b}$ is summed over all $8$ values (as is the case for the exact time evolution operator). 
Therefore, Trotter time evolution does not preserve the global off-diagonal $SU(3)$ charges and, for example, color singlets can evolve into non-color singlets.
Equivalently, the Trotterized time evolution operator is not in the trivial representation of $SU(3)$.
To understand this point in more detail,
consider the transformation of 
$\left(T^a\right)^i_j  \ \left(T^a\right)^k_l$ for any given $a$. 
Because of the symmetry of this product of operators, each transforming as an ${\bf 8}$, the product must decompose into ${\bf 1}\oplus {\bf 8}\oplus {\bf 27} $,
where the elements of each of the irreps can be found from
\begin{align}
\left(T^a\right)^i_j  \ \left(T^a\right)^k_l = 
\left(\hat {\cal O}_{27}^a\right)^{ik}_{jl} 
-\frac{2}{5}\left[ \delta^i_j \left(\hat {\cal O}^a_8\right)^k_l + \delta^k_l \left(\hat {\cal O}^a_8\right)^i_j \right]
+&\frac{3}{5}
\left[\delta^i_l \left(\hat {\cal O}^a_8\right)^k_j + \delta^k_j \left(\hat {\cal O}^a_8\right)^i_l \right]
+ \nonumber \\
&\frac{1}{8} \left( \delta^i_l \delta^k_j\ -\ \frac{1}{3} \delta^i_j \delta^k_l \right) \hat {\cal O}^a_1
\  ,
\end{align}
where
\begin{align}
\left(\hat {\cal O}^a_{27}\right)^{ik}_{jl} 
 = & \
\frac{1}{2}
\left[ \left(T^a\right)^i_j \left(T^a\right)^k_l + \left(T^a\right)^i_l \left(T^a\right)^k_j \right]
-
\frac{1}{10}\left[ 
\delta^i_j \left(\hat {\cal O}^a_8\right)^k_l 
+ \delta^i_l \left(\hat {\cal O}^a_8\right)^k_j 
+ \delta^k_j \left(\hat {\cal O}^a_8\right)^i_l 
+ \delta^k_l \left(\hat {\cal O}^a_8\right)^i_j \right] \nonumber\\
& -\frac{1}{24} \left( \delta^i_j \delta^k_l + \delta^i_l \delta^k_j \right) \hat {\cal O}^a_1 \ ,
\nonumber\\
\left(\hat {\cal O}^a_8\right)^i_j 
 = & \
\left(T^a\right)^i_\beta \left(T^a\right)_j^\beta\ -\ \frac{1}{3} \delta^i_j \hat {\cal O}^a_1
\ ,\ \ 
\hat {\cal O}^a_1 \ =\  \left(T^a\right)^\alpha_\beta \left(T^a\right)_\alpha^\beta \ =\ \frac{1}{2}
\  .
\end{align}
When summed over $a=1,\ldots,8$, the contributions from the ${\bf 8}$ and ${\bf 27}$ vanish, leaving the familiar contribution from the ${\bf 1}$.
When only partials sums are available, as is the situation with individual contributions to the Trotterized evolution, 
each of the contributions is the exponential of 
${\bf 1}\oplus {\bf 8}\oplus 
{\bf 27} $, with only the singlet contributions leaving the lattice a color singlet.
The leading term in the expansion of the product of the four pairs of Trotterized evolution operators sum to leave only the singlet contribution. 
In contrast, higher-order terms do not cancel and 
non-singlet contributions are present.

This is a generic problem 
that will be encountered when satisfying Gauss's law
leads to non-local charge-charge interactions. 
This is not a problem for $U(1)$, and 
surprisingly, is not a problem for 
$SU(2)$ because $(Q^{(1)}_n \, Q^{(1)}_m, Q^{(2)}_n \, Q^{(2)}_m, Q^{(3)}_n \, Q^{(3)}_m )$ are in the Cartan sub-algebra of $SU(4)$ and therefore mutually commuting.  However, it is a problem for $N_c>2$.
One way around the breaking of global $SU(N_c)$ 
is through the co-design 
of unitaries that directly (natively) implement
${\exp}( i \alpha Q^{(a)}_n \, Q^{(a)}_m)$; see Sec.~\ref{sec:codesign}.
Without such a native unitary,
the breaking of $SU(N_c)$ appears as any other Trotter error, and can be systematically reduced in the same way. A potential caveat to this is if the time evolution operator took the system into a different phase, but our studies of $L=1$ show no evidence of this.

It is interesting to note that the terms generated by the Trotter commutators form a closed algebra.
In principle, a finite number of terms could be
included to define an effective Hamiltonian whose Trotterization exactly maps onto the desired evolution operator (without the extra terms).
It is straightforward to work out the terms generated order-by-order in the Baker-Campbell-Hausdorff formula.
Aside from re-normalizing the existing charges, there are $9$ new operator structures produced. 
For example, the leading-order commutators generate the three operators, ${\cal O}_i$, in Eq.~(\ref{eq:BCHOp}),
\begin{align}
{\cal O}_i = 
\begin{cases} 
      (\sigma^+ I \sigma^- \sigma^- Z \sigma^+ - \sigma^+ Z \sigma^- \sigma^- I \sigma^+) - {\rm h.c.}  \ ,\\
      (I \sigma^- \sigma^+ Z \sigma^+ \sigma^- - Z \sigma^- \sigma^+ I \sigma^+ \sigma^-) - {\rm h.c.}  \ , \\
       (\sigma^+ \sigma^- Z \sigma^- \sigma^+ I - \sigma^+ \sigma^- I \sigma^- \sigma^+ Z) - {\rm h.c.} \ .
\end{cases}
\label{eq:BCHOp}
\end{align}
In general, additional operators are constrained only by (anti)hermiticity, 
symmetry with respect to $n \leftrightarrow m$ and preservation of $(r,g,b)$, and should generically be included in the same spirit as terms in the  Symanzik-action~\cite{Symanzik:1983dc,Symanzik:1983gh} for lattice QCD.

With Trotterization of the gauge field introducing violations of gauge symmetry, and the presence of bit- and phase-flip errors within the device register, it is worth briefly considering a potential mitigation strategy. A single
bit-flip error will change isospin by $|\Delta I_3|=1/2$ and color charge by one unit of red or green or blue.
After each Trotter step on a real quantum device, such errors will be encountered and a mitigation or correction scheme is required.
Without the explicit gauge-field degrees of freedom and local charge conservation checks enabled by Gauss's law, such errors can only be detected globally, and hence, cannot be actively corrected during the evolution.\footnote{When local gauge fields are present,
previous works have found that including a quadratic ``penalty-term" in the Hamiltonian is effective in mitigating violation of Gauss's law~\cite{Hauke:2013jga,Zohar:2015hwa,Dalmonte:2016alw,Halimeh:2019svu}. See also Refs.~\cite{PhysRevLett.112.120406,Kasper:2020owz}.}
Motivated by this, consider introducing a twirling phase factor into the evolution, $\exp(-i \theta^a {\cal Q}^{(a)})$, where ${\cal Q}^{(a)}$ is the total charge on the lattice.
If applied after each Trotter step, with a randomly selected set of eight angles, $\theta^a$, 
the phases of color-nonsinglet states become random for each member of an ensemble, mitigating errors in some observables. 
Similar twirling phase factors could be included for the other charges that are conserved or approximately conserved.

\subsubsection{Quantum Resource Requirements for Time Evolution}
\noindent
It is straightforward to extend the circuits presented in the previous section to arbitrary $N_c$ and $N_f$. The quantum
resources required for time evolution can be quantified
for small, modest and asymptotically large systems. As discussed previously, a quantum register with $N_q=2 L N_c N_f$ qubits\footnote{The inclusion of an ancilla for the kinetic term increases the qubit requirement to $N_q = 2L N_c N_f + 1$.} is required to encode one-dimensional $SU(N_c)$ gauge theory
with $N_f$ flavors on $L$ spatial lattice sites using the JW transformation. For $SU(3)$ gauge theory, this leads to, for example, $N_q = 6L$ with only $u$-quarks and $N_q = 18L$ with $u,d,s$-quarks.
The five distinct contributions to the resource requirements, 
corresponding to application of the unitary operators providing 
a single Trotter step associated with the quark mass, $U_m$, the baryon chemical potential, $U_{\mu_B}$, the isospin chemical potential, $U_{\mu_I}$, the kinetic term, $U_{kin}$, and the chromo-electric field, $U_{el}$, are 
given in terms of the number of 
single-qubit rotations, denoted by ``$R_Z$'', the number of Hadamard gates, denoted by ``Hadamard'', and the number of CNOT gates, denoted by ``CNOT''.
It is found that\footnote{
For $N_c = 2$ only three of the $ZZ$ terms can be combined into $Q_{n,f}^{(a)} \, Q_{m,f'}^{(a)}$ and the number of CNOTs for one Trotter step of $U_{el}$ is 
\begin{equation}
U_{el}  \ : \ (2 L-1) N_f [9 (2 L-1) N_f-7] \ \ \ | \ \text{CNOT} \ .
\label{eq:Nc2CNOT}
\end{equation}
Additionally, for $N_c N_f < 4$, the Trotterization of $U_{{\rm kin}}$ is more efficient without an ancilla and the number of CNOTs required is
\begin{equation}
U_{ kin}  \ : \ 2 (2 L-1) N_c (N_c + 1) \ \ \ | \ \text{CNOT} \ .
\label{eq:UkinNoanc}
\end{equation}
The construction of the circuit that implements the time evolution of the hopping term for $N_c=3$ 
and $N_f=1$
is shown in Fig.~\ref{fig:Ukin1flavTrot}.
}
\begin{align}
   U_m  \ :& \ \ 2 N_c N_f L \ \ \ | \ R_Z \ ,\nonumber \\[5pt]
   U_{\mu_B}  \ :& \ \ 2 N_c N_f L \ \ \ | \ R_Z \ ,\nonumber \\[5pt]
   U_{\mu_I}  \ :& \ \ 2 N_c N_f L \ \ \ | \ R_Z \ ,\nonumber \\[5pt]
   U_{kin}  \ :& \ \ 2 N_c N_f(2L-1) \ \ \ | \ R_Z \ ,\nonumber \\
                & \ \ 2 N_c N_f (2L-1) \ \ \ | \ \text{Hadamard} \ ,\nonumber \\
                & \ \ 2 N_c N_f (8L-3) -4 \ \ \ | \ \text{CNOT} \ ,\nonumber \\[5pt]
    U_{el}  \ :& \ \ \frac{1}{2}(2L-1)N_c N_f\left [3-4N_c+N_f(2L-1)(5N_c-4)\right ] \ \ \ | \ R_Z \ ,\nonumber \\
                & \ \ \frac{1}{2}(2L-1)(N_c-1) N_c N_f\left [N_f(2L-1)-1\right ] \ \ \ | \ \text{Hadamard} \ ,\nonumber \\
                & \ \ \frac{1}{6} (2 L -1) (N_c-1) N_c N_f [(2 L-1) (2 N_c+17) N_f-2 N_c-11] \ \ \ | \ \text{CNOT} \ .
\label{eq:RHCN}
\end{align}

It is interesting to note the scaling of each of the contributions.  The mass, chemical potential and kinetic terms scale as ${\cal O}(L^1)$, while the non-local gauge-field contribution is ${\cal O}(L^2)$.
As anticipated from the outset, using Gauss's law to constrain the energy in the gauge field via the quark occupation has given rise to circuit depths that scale quadratically with the lattice extent, naively violating one of the criteria for quantum simulations at scale~\cite{Feynman:1981tf,DiVincenzo2000ThePI}.
This volume-scaling is absent for formulations that explicitly include the 
gauge-field locally, 
but with the trade-off of requiring a volume-scaling increase in the number of 
qubits or qudits or bosonic modes.\footnote{
The local basis on each link is spanned by the possible color irreps 
and the states of the left and right Hilbert spaces (see footnote~\ref{foot:irrep}). 
The possible irreps are built from the charges of the preceding fermion sites, 
and therefore the dimension of the link basis grows polynomially in $L$. 
This can be encoded in $\mathcal{O}(\log L)$ qubits per link and 
$\mathcal{O}(L \log L)$ qubits in total. 
The hopping and chromo-electric terms in the Hamiltonian are local, 
and therefore one Trotter step will require $\mathcal{O}(L)$ gate operations up to logarithmic corrections.} 
We expect that the architecture of quantum devices used for simulation 
and the resource requirements for the local construction will determine 
the selection of local versus non-local implementations.

For QCD with $N_f=2$, the total requirements are
\begin{align}
   R_Z  \ :& \ \ (2L-1)\left( 132 L -63 \right)+18 \ ,\nonumber \\
   \text{Hadamard} \ :& \ \ (2L-1)\left( 24L - 6 \right)  \ ,\nonumber \\
    \text{CNOT} \ :& \ \ (2L-1)\left( 184L - 78 \right) + 8 \ ,
\end{align}
and further, the CNOT requirements for a single Trotter step 
of $SU(2)$ and $SU(3)$ for $N_f = 1,2,3$ are shown in Table~\ref{tab:cnotA}.
\begin{table}[!ht]
\renewcommand{\arraystretch}{1.2}
\begin{tabularx}{0.48\textwidth}{||c | Y | Y | Y ||}
\hline
\multicolumn{4}{||c||}{\# CNOTs for one Trotter step of $SU(2)$} \\
 \hline
 $L$ & $N_f=1$ & $N_f=2$ & $N_f=3$ \\
 \hline\hline
 1 & 14 & 58 & 116 \\ 
 \hline
 2 & 96 & 382 & 818\\
 \hline
 5 & 774 & 3,082 & 6,812 \\
 \hline
 10 & 3,344 & 13,342 & 29,762 \\
 \hline
 100 & 357,404 & 1,429,222 & 3,213,062 \\
 \hline
\end{tabularx}
\renewcommand{\arraystretch}{1}
\ \
\renewcommand{\arraystretch}{1.2}
\begin{tabularx}{0.48\textwidth}{||c | Y | Y | Y ||}
\hline
\multicolumn{4}{||c||}{\# CNOTs for one Trotter step of $SU(3)$ } \\
 \hline
 $L$ & $N_f=1$ & $N_f=2$ & $N_f=3$ \\
 \hline\hline
 1 & 30 & 114 & 242 \\ 
 \hline
 2 & 228 & 878 & 1,940\\
 \hline
 5 & 1,926 & 7,586 & 16,970 \\
 \hline
 10 & 8,436 & 33,486 & 75,140 \\
 \hline
 100 & 912,216 & 3,646,086 & 8,201,600 \\
 \hline
\end{tabularx}
\renewcommand{\arraystretch}{1}
\caption{The CNOT requirements to perform one Trotter step of time evolution for a selection of simulation parameters.}
 \label{tab:cnotA}
 \end{table}
These resource requirements suggest that systems with up to $L=5$ could be simulated, with appropriate error mitigation protocols, using this non-local framework in the near future. Simulations beyond $L=5$ appear to present a challenge in the near term.

The resource requirements in Table~\ref{tab:cnotA} do not include those for a gauge-link beyond the end of the lattice.  As discussed previously, such additions to the time evolution could be used to move color-nonsinglet contributions to high frequency, allowing the possibility that they are filtered from observables.
Such terms contribute further to the quadratic volume scaling of resources.
Including chemical potentials in the time evolution does not increase the number of required entangling gates per Trotter step.  Their impact upon resource requirements may arise in preparing the initial state of the system.

\subsubsection{Elements for Future Co-Design Efforts}
\label{sec:codesign}
\noindent
Recent work has shown the capability of creating many-body entangling gates natively~\cite{Andrade:2021pil,Katz:2022czu} which have similar fidelity to two qubit gates.
This has multiple benefits. First, it allows for (effectively) deeper circuits to be run within coherence times. 
Second, it can eliminate some of the Trotter errors due to non-commuting terms. 
The possibility of using native gates for these calculations is particularly interesting from the standpoint of eliminating or mitigating the Trotter errors that violate the global $SU(3)$ symmetry, as discussed in Sec.~\ref{sec:colorBreak}.
Specifically, 
it would be advantageous to have a ``black box" unitary operation of the form,
\begin{align}
   e^{-i \alpha Q_n^{(a)} \, Q_m^{(a)}} =& \ \exp \bigg \{-i \frac{\alpha}{2} \bigg [\sigma^+_{n} \sigma^-_{n+1}\sigma^-_{m}\sigma^+_{m+1} + \sigma^-_{n}\sigma^+_{n+1}\sigma^+_{m}\sigma^-_{m+1} +  \sigma^+_{n+1}\sigma^-_{n+2}\sigma^-_{m+1}\sigma^+_{m+2} 
   \nonumber \\
   &+ \sigma^-_{n+1}\sigma^+_{n+2}\sigma^+_{m+1}\sigma^-_{m+2} + \sigma^+_{n}\sigma^z_{n+1}\sigma^-_{n+2}\sigma^-_{m}\sigma^z_{m+1}\sigma^+_{m+2}\nonumber \\
   & + \sigma^-_{n}\sigma^z_{n+1}\sigma^+_{n+2}\sigma^+_{m}\sigma^z_{m+1}\sigma^-_{m+2} + \frac{1}{6}(\sigma^z_n \sigma^z_m + \sigma^z_{n+1} \sigma^z_{m+1} + \sigma^z_{n+2} \sigma^z_{m+2})\nonumber \\
   &- \frac{1}{12}(\sigma^z_n \sigma^z_{m+1} + \sigma^z_n \sigma^z_{m+2} + \sigma^z_{n+1} \sigma^z_m + \sigma^z_{n+1} \sigma^z_{m+2} + \sigma^z_{n+2} \sigma^z_m + \sigma^z_{n+2 }\sigma^z_{m+1})
   \bigg ] \bigg \} \ ,
\end{align}
for arbitrary $\alpha$ and pairs of sites, $n$ and $m$ (sum on $a$ is implied).
A more detailed discussion of co-designing 
interactions for quantum simulations of these theories is clearly warranted.

\subsection{Results from Quantum Simulators}
\begin{figure}[!ht]
    \centering
    \includegraphics[width=\columnwidth]{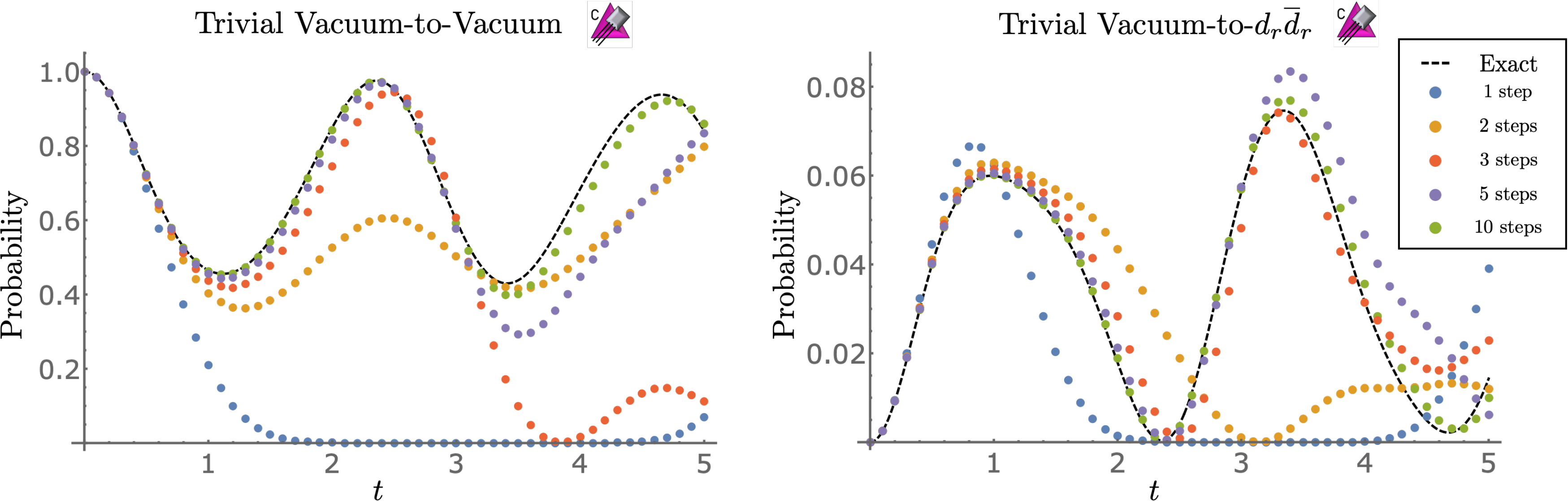}
    \caption{
    The trivial vacuum-to-vacuum 
    and trivial vacuum-to-$d_r \overline{d}_r$ probabilities in QCD with $N_f=2$
    for $m=g=L=1$. 
    Shown are the results obtained from exact exponentiation of the Hamiltonian (dashed black curve) and from the Trotterized implementation with $1$, $2$, $3$, $5$ and $10$ Trotter steps using the (classical) quantum simulators in {\tt cirq} and {\tt qiskit} (denoted by the purple icons~\cite{Klco:2019xro}).
    }
    \label{fig:VacTo}
\end{figure}

\noindent
The circuits laid out in Sec.~\ref{sec:Circuits} are too deep to be executed on currently available quantum devices,  
but can be readily implemented with quantum simulators such as {\tt cirq} and {\tt qiskit}.
This allows for an estimate of the number of Trotter steps required to achieve a desired precision in the determination of any given observable as a function of time.
Figure~\ref{fig:VacTo} shows results for the 
trivial vacuum-to-vacuum and trivial vacuum-to-$d_r \overline{d}_r$ probabilities as a function of time for $L=1$. See App.~\ref{app:Nf1SU3circs} for the full circuit which
implements a single Trotter step, and App.~\ref{app:MDTD} for the decomposition of the energy starting in the trivial vacuum.

The number of Trotter steps, 
$N_{\rm Trott}$, required to evolve out to a given $t$ within a specified (systematic) error, 
$\epsilon_{\rm Trott}$, was also investigated. 
$\epsilon_{\rm Trott}$ is defined as the 
maximum fractional error between the 
Trotterized and exact time evolution in two quantities, the vacuum-to-vacuum persistence probability 
and the vacuum-to-$d_r\overline{d}_r$ transition probability. For demonstrative purposes, an analysis at leading order in the Trotter expansion is sufficient.
Naive expectations based upon global properties of the Hamiltonian defining the evolution operators indicate that an upper bound for $\epsilon_{\rm Trott}$ scales as 
\begin{equation}
\Big\lvert \Big\lvert e^{-i H t} - \left[ U_1\left (\frac{t}{N_{\rm Trott}}\right ) \right]^{N_{\rm Trott}} \Big\rvert \Big\rvert \ \le \ \frac{1}{2} \sum_i \sum_{j>i} \Big\lvert \Big\lvert \left[ H_i , H_j \right] \Big\rvert \Big\rvert \frac{t^2}{N_{\rm Trott}} \ ,
\label{eq:LOTrottbound}
\end{equation}
where the Hamiltonian has been divided into sets of mutually commuting terms, $H = \sum_i H_i$. This upper bound indicates that the required number of Trotter steps to maintain a fixed error scales as  $N_{\rm Trott}\sim t^2$~\cite{Childs_2021}.

\begin{figure}[!ht]
    \centering
    \includegraphics[width=\columnwidth]{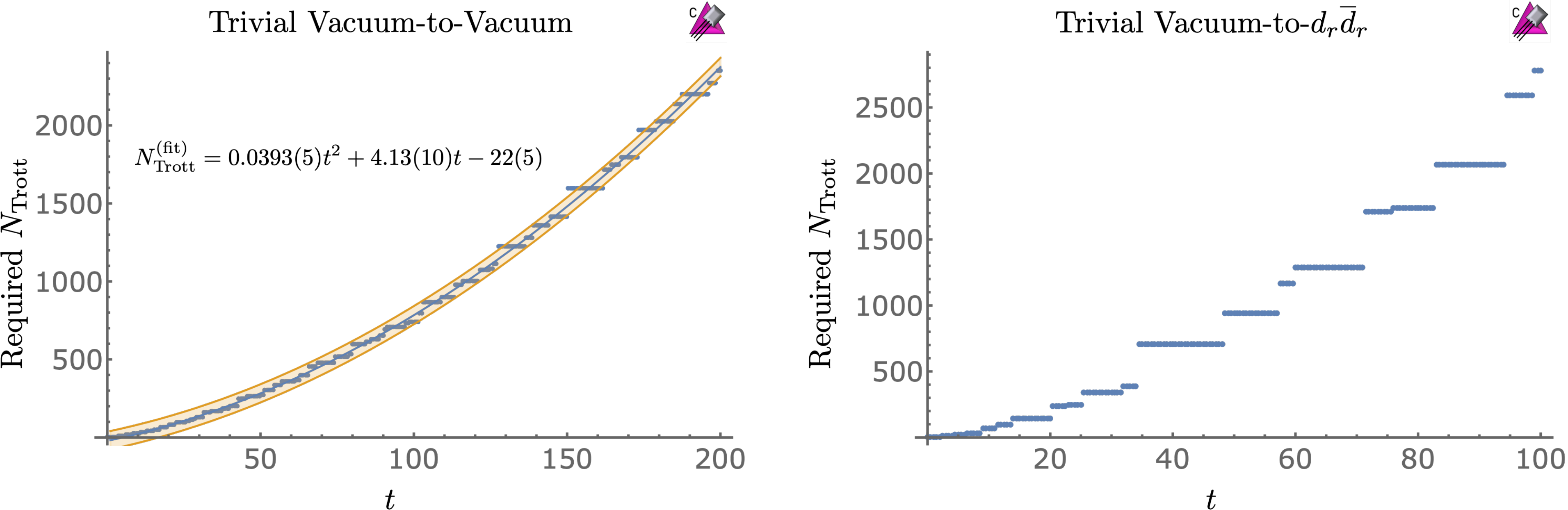}
    \caption{
    The number of Trotter steps, 
    $N_{\rm Trott}$,
   required to 
   achieve a systematic fractional error of 
   $\epsilon_{{\rm Trott}}\le 0.1$
    at time $t$ in the trivial vacuum-to-vacuum  probability (left panel) and the trivial vacuum-to-$d_r\overline{d}_r$ probability
    (right panel) for QCD with $N_f=2$ and
$m=g=L=1$. The blue points are results obtained by direct calculation.
   }
    \label{fig:TrotErrorB}
\end{figure}

To explore the resource requirements for simulation based upon explicit calculations between exclusive states, as opposed to upper bounds for inclusive processes, given in Eq.~(\ref{eq:LOTrottbound}), 
a series of calculations was performed requiring $\epsilon_{\rm Trott}\le0.1$ for a range of times, $t$.
Figure~\ref{fig:TrotErrorB} shows the required 
$N_{\rm Trott}$ as a function of $t$ for $m=g=L=1$.
The plateaus 
observed in Fig.~\ref{fig:TrotErrorB} arise from
resolving upper bounds from oscillating functions, 
and introduce a limitation in fitting to extract scaling behavior.  This is less of a limitation 
for the larger vacuum-to-vacuum probabilities which are fit well by a quadratic polynomial, starting from $t=1$, with coefficients,
\begin{equation}
    N_{{\rm Trott}} = 0.0393(5) t^2 + 4.13(10) t - 22(5) \ .
    \label{eq:TrotErrorFit}
\end{equation}
The uncertainty represents a 95\% confidence interval in the fit parameters and corresponds to the shaded orange region in
Fig.~\ref{fig:TrotErrorB}. The weak quadratic scaling with $t$ implies that, even out to $t \sim 100$, the number of Trotter
steps scales approximately linearly, and a constant error in the observables can be achieved with a fixed Trotter step size. 
We have been unable to distinguish between fits with and without logarithmic terms.

These results can be contrasted with those obtained for the Schwinger model in Weyl gauge. The authors of Ref.~\cite{Shaw:2020udc} estimate a resource
requirement, as quantified by the number of $T$-gates, that scales as $\sim (L t)^{3/2}\log L t$, increasing to
$\sim L^{5/2} t^{3/2}\log L t \log L$ if the maximal value of the gauge fields is accommodated within the Hilbert space.
The results obtained in this section suggest that resource requirements in axial gauge, as quantified by the number of CNOTs,
effectively scale as $\sim L^2 t$ up to intermediate times and as $\sim L^2 t^2$ asymptotically. In a scattering process with localized
wave-packets, it is appropriate to take $L\sim t$ 
(for the speed of light taken to be $c=1$),
as the relevant non-trivial time evolution is bounded by the light cone. 
This suggests that the required resources scale asymptotically as $\sim t^4$, independent of the chosen gauge to define the simulation. 
This could have been
anticipated at the outset by assuming that the minimum change in complexity for a process has physical meaning~\cite{https://doi.org/10.48550/arxiv.quant-ph/0701004,doi:10.1126/science.1121541,https://doi.org/10.48550/arxiv.quant-ph/0502070,Jefferson:2017sdb}.

\section{Simulating \texorpdfstring{\boldmath$1+1$}{1+1}D QCD with \texorpdfstring{\boldmath$N_f=1$}{Nf=1} and \texorpdfstring{\boldmath$L=1$}{L=1}}
\label{sec:Nc3Nf1}
\noindent
With the advances in quantum devices, algorithms and mitigation strategies, quantum simulations of $1+1$D QCD can now begin, and this section presents results for $N_f=1$ and $L=1$. Both state preparation and time evolution will be discussed.

\subsection{State Preparation with VQE}
\begin{figure}[!ht]
    \centering
    \includegraphics[width=6cm]{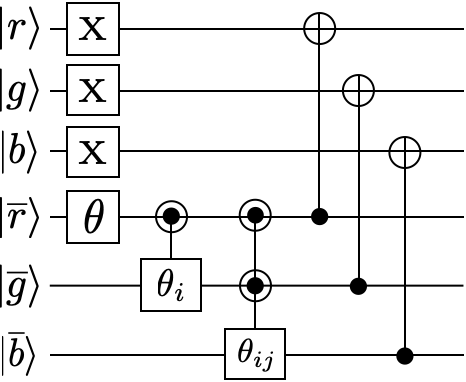}
    \caption{
    Building upon the trivial vacuum, 
    this circuit initializes the 
    most general real wavefunction 
    (with 7 independent rotation angles for 3 qubits)
    in the $\overline{q}$-sector, which is subsequently
    mirrored into the $q$-sector by 3 CNOTs. Gates labelled by ``$\theta$" are shorthand for $R_Y(\theta)$ and half-filled circles denote a control on $0$ 
    and a different control on $1$.
    }
    \label{circ:GeneralVQE}
\end{figure}

\noindent
Restricting the states of the lattice to be color singlets reduces the complexity of state preparation significantly.  
Transformations in the quark sector are mirrored in the antiquark sector.
A circuit that
prepares the most general state with $r=g=b=0$ is shown in Fig.~\ref{circ:GeneralVQE}. 
The (multiply-)controlled $\theta$ gates are short-hand for (multiply-)controlled $R_Y(\theta)$ gates with half-filled circles denoting a control on $0$ 
and a different control on $1$.
The subscripts on $\theta_{ij}$ signify that there are different angles for each controlled rotation. For example,
$\theta_{i}$ has two components, $\theta_{0}$ and $\theta_{1}$, corresponding to a rotation controlled on $0$ and $1$, respectively. 
The CNOTs at the end of the circuit
enforce that there are equal numbers of quarks and antiquarks with the same color,
i.e., that $r=g=b=0$. 
This circuit can be further simplified by constraining the angles to only parameterize color singlet states. The color singlet subspace is spanned by\footnote{
The apparent asymmetry between $q_r,q_g,q_b$ is due to the charge operators generating hops over different numbers of quarks or antiquarks. 
For example, $Q^{(1)}$ hops $q_r$ to $q_g$ without passing over any intermediate quarks, but $Q^{(4)}$ hops $q_r$ to $q_b$ passing over $q_g$. 
Also note that when $m=0$ the $\mathbb{Z}_2$ spin-flip symmetry reduces the space of states to be two-dimensional.}
\begin{align}
    \ket{{\Omega_0}} \ &, \ \ \frac{1}{\sqrt{3}}\left (\ket{q_r \overline{q}_r} - \ket{q_g \overline{q}_g} + \ket{q_b \overline{q}_b}\right ) \ , \nonumber \\
    \ket{q_r \overline{q}_r \, q_g \overline{q}_g \, q_b \overline{q}_b} \ &, \ \  \frac{1}{\sqrt{3}} \left (\ket{q_r \overline{q}_r \, q_g\overline{q}_g} - \ket{q_r \overline{q}_r \, q_b\overline{q}_b} + \ket{q_g \overline{q}_g \, q_b \overline{q}_b}\right ) \ , 
    \label{eq:vacBasis}
\end{align}
where $\ket{{\Omega_0}} = \ket{{000111}}$ is the trivial vacuum.
This leads to the following relations between angles,
\begin{align}
    \theta_{10} &= \theta_{01}\ ,  & \theta_{00} &= -2 \sin^{-1}\left[ \tan(\theta_{0}/2) \, \cos(\theta_{01}/2) \right] \ ,\nonumber \\
    \theta_{01} &= -2 \sin^{-1}\left[ \cos(\theta_{11}/2)\, \tan(\theta_{1}/2) \right]\ , & \theta_0 &= -2 \sin^{-1} \left[ \tan(\theta/2) \, \cos(\theta_{1}/2) \right] \ .
    \label{eq:angleconst}
\end{align}

The circuit in Fig.~\ref{circ:GeneralVQE} can utilize the strategy outlined in Ref.~\cite{Atas:2021ext}  to
separate into a ``variational" part and a ``static" part. 
If the VQE circuit can be written as 
$U_{var}(\theta) U_s$, 
where $U_s$ is independent
of the variational parameters, 
then $U_s$ can be absorbed by a redefinition of
the Hamiltonian. 
Specifically, matrix elements of the Hamiltonian can be written as 
\begin{equation}
    \bra{{\Omega_0}} U_{var}^{\dagger}(\theta) \tilde{H} U_{var}(\theta) \ket{{\Omega_0}}
    \ ,
\end{equation}
where $\tilde{H}= U_s^{\dagger} H U_s$.
Table~\ref{tab:PUHU} shows the 
transformations of various Pauli strings under conjugation by a CNOT controlled on the smaller index qubit. 
Note that the $\mathbb{Z}_2$ nature of this transformation is manifest.

\begin{figure}[!ht]
    \centering
    \includegraphics[width=\columnwidth]{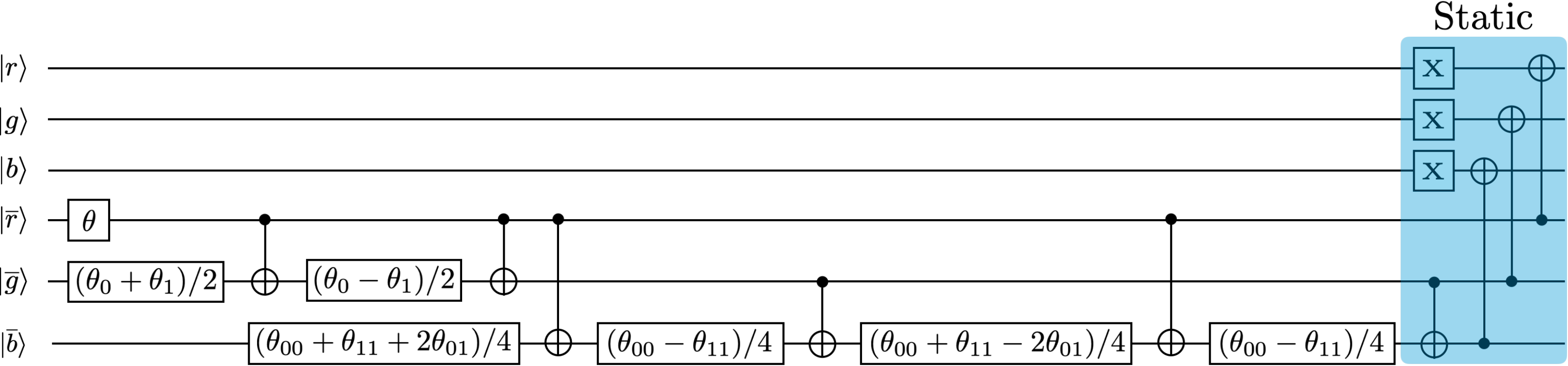}
    \caption{
    A circuit that initializes the most general 
    $B=0$ color singlet state for $N_f=1$ and $L=1$. Gates labelled by ``$\theta$" are shorthand for $R_Y(\theta)$
    and the $X$s at the end are to initialize the trivial vacuum. The color singlet constraint, $\theta_{10} = \theta_{01}$, has been used and the other angles are related by Eq.~(\ref{eq:angleconst}).}
    \label{circ:VQEImp}
\end{figure}

\begin{table}[!ht]
\renewcommand{\arraystretch}{1}
\resizebox{\textwidth}{!}{\begin{tabular}{||c | c | c | c | c | c | c | c ||} 
 \hline
 $XX \to IX$ & $XY \to IY$ & $YX \to ZY$ & $XZ \to XZ$ & $ZZ \to ZI$ & $YZ \to YI$ & $ZY \to YX$ & $YY \to (-)ZX$ \\
 \hline
 $IX \to XX$ & $IY \to XY$ & $XI \to XI$  & $IZ \to IZ$ & $ZI \to ZZ$  & $YI \to YZ$ &  $ZX \to (-)YY$  & $II \to II$\\
 \hline
\end{tabular}}
\renewcommand{\arraystretch}{1}
\caption{The transformation of Pauli strings under conjugation by a CNOT controlled on the smaller index qubit.}
\label{tab:PUHU}
\end{table}
In essence, entanglement is
traded for a larger number of  correlated measurements.  
Applying the techniques in Ref.~\cite{Klco:2019xro}, the VQE circuit of Fig.~\ref{circ:GeneralVQE} can be put into the form of Fig.~\ref{circ:VQEImp},
which requires $5$ CNOTs along with all-to-all connectivity between the three $\overline{q}$s.
%

\subsection{Time Evolution Using IBM's 7-Qubit Quantum Computers}
\begin{figure}[!ht]
    \centering
    \includegraphics[height=0.25\textheight]{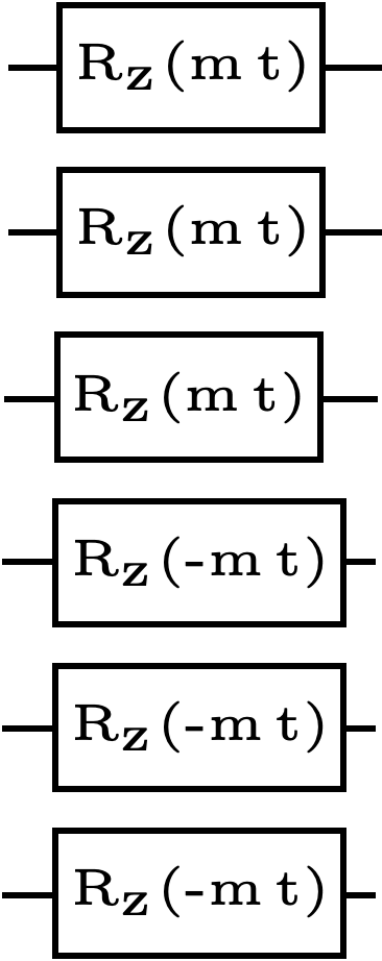}
    \caption{A circuit that implements 
    $U_m(t) = \exp(- i H_m t)$ for $N_f=1$ and $L=1$.}
    \label{circ:Um1flav}
\end{figure}

\begin{figure}[!ht]
    \centering
    \includegraphics[width=12cm]{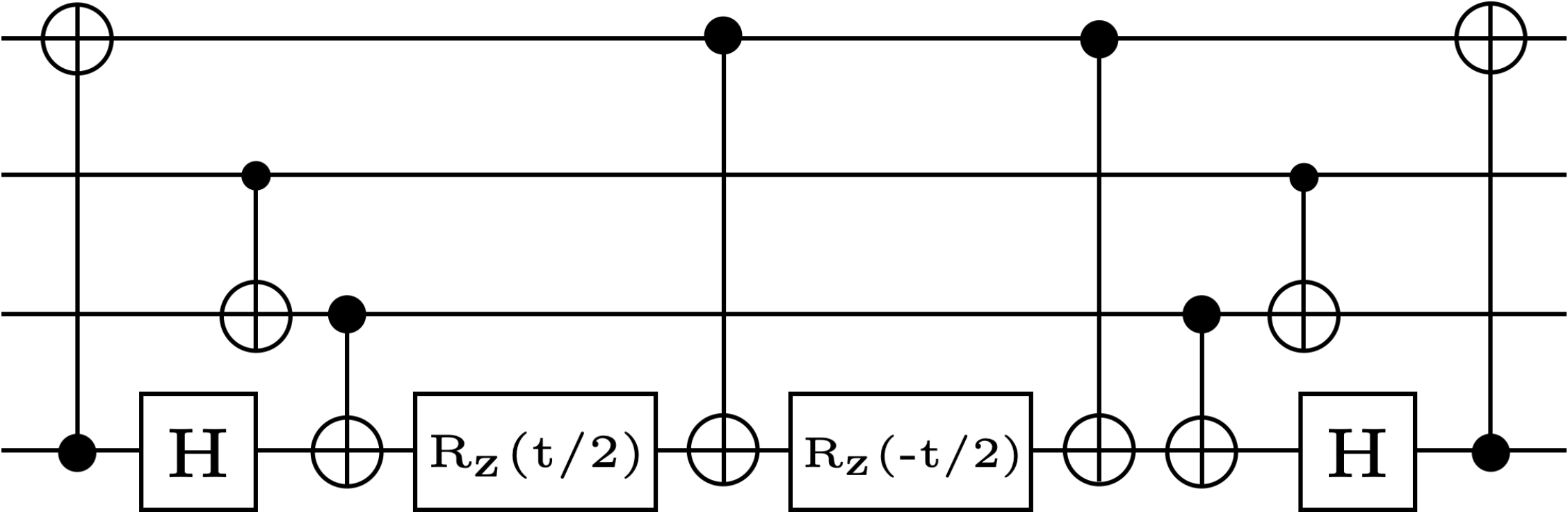}
    \caption{A circuit that implements $\exp[-i \frac{t}{2} (\sigma^+ Z Z \sigma^- + {\rm h.c.})]$.}
    \label{fig:Ukin1flavTrot}
\end{figure}

\noindent
A single leading-order Trotter step of $N_f=1$ QCD with $L=1$ requires 28 CNOTs.\footnote{By evolving with $U_{el}$ before $U_{kin}$ in the Trotterized time evolution, two of the CNOTs
become adjacent in the circuit and can be canceled.}
A circuit that implements one Trotter step of the mass term is shown in Fig.~\ref{circ:Um1flav}.
As discussed around Eq.~(\ref{eq:UkinNoanc}), it is more efficient to not use an ancilla qubit in the Trotterization of the kinetic part of the Hamiltonian. 
A circuit that implements one Trotter step of a single hopping term is shown in Fig.~\ref{fig:Ukin1flavTrot}~\cite{Stetina:2020abi}.
Similarly, for this system,
the only contribution to $H_{el}$ is $Q^{(a)}_n \, Q^{(a)}_n$, which contains three $ZZ$ terms that are Trotterized using the standard two CNOT implementation.
The complete set of circuits required for Trotterized time evolution are given in App.~\ref{app:Nf1SU3circs}.

\begin{figure}[!ht]
    \centering
    \includegraphics[width=12cm]{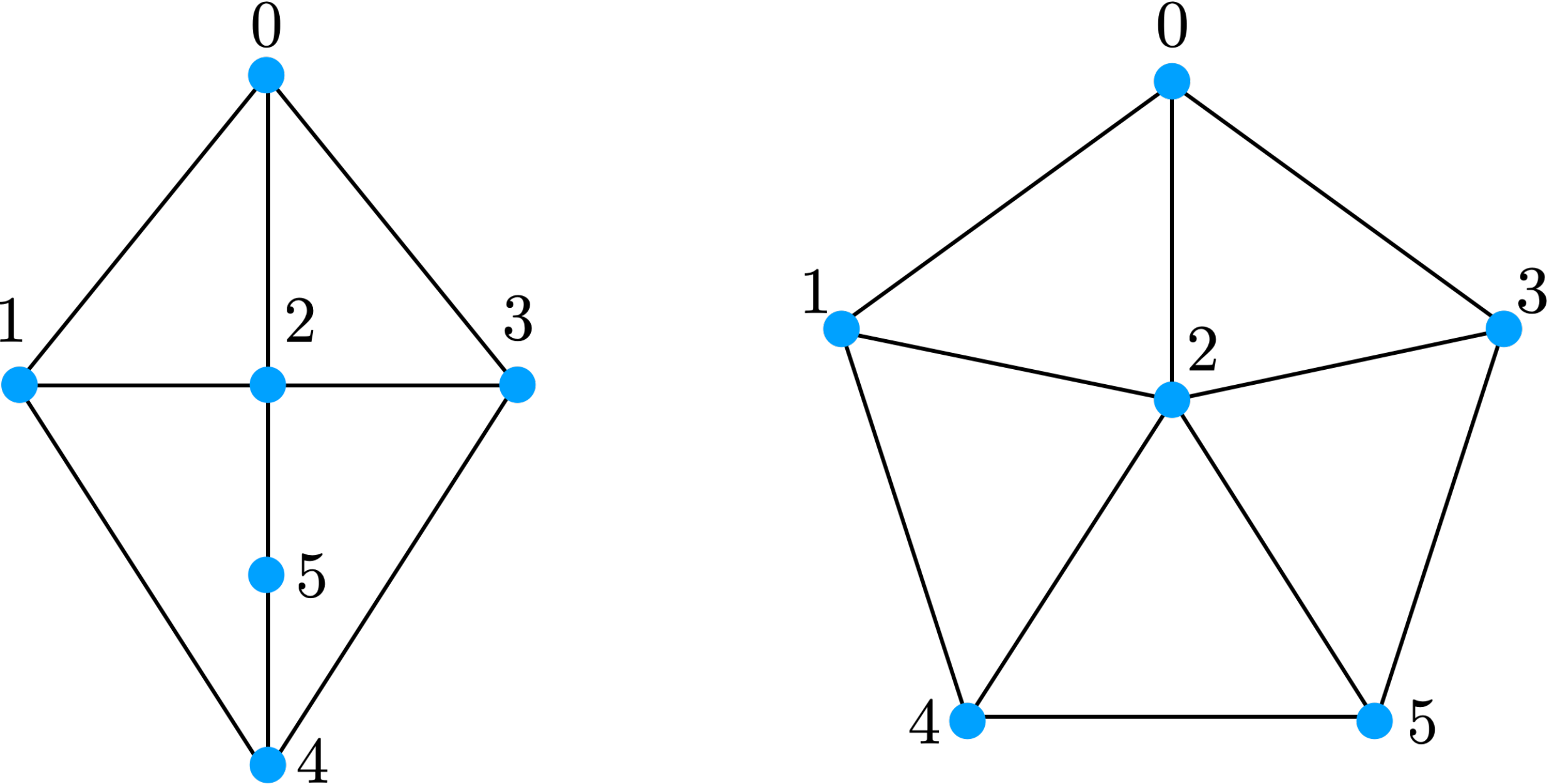}
    \caption{
    Two potential quantum device topologies for the  implementation of Trotterized time evolution.
    }
    \label{fig:TrotTopo}
\end{figure}

To map the system onto a quantum device, it is necessary to understand the required connectivity for efficient simulation. 
Together, the hopping and chromo-electric terms require connectivity between nearest neighbors as well as between $q_r$ and $q_b$ and
$q$s and $\overline{q}$s of the same color. 
The required device topology is planar and two embedding options are
shown in Fig.~\ref{fig:TrotTopo}.
The ``kite'' topology follows from the above circuits, 
while the ``wagon wheel'' topology makes use of the identities $CX(q_a,q_b) \cdot CX(q_b,q_c) = CX(q_a,q_c) \cdot CX(q_b,q_c) = CX(q_b,q_a) \cdot CX(q_a,q_c)$ where $CX(q_a,q_b)$ denotes a CNOT controlled on qubit $q_a$.
Both topologies can be employed on devices with all-to-all connectivity, such as trapped-ion systems, but
neither topology exists natively on available superconducting-qubit devices.

We performed leading-order Trotter evolution to study the trivial vacuum persistence and transition probability using IBM's quantum computers {\tt ibmq\_jakarta} and {\tt ibm\_perth}, each a {\tt r5.11H} quantum processor with 7 qubits and ``H"-connectivity.
The circuits developed for this system require a higher degree of connectivity than available with these devices, and so SWAP-gates were necessary for implementation.   
The IBM {\tt transpiler} was used to first compile the circuit for the H-connectivity and then again to compile the Pauli twirling (discussed next).
An efficient use of SWAP-gates allows for a single Trotter step to be executed with 34 CNOTs.

A number of error-mitigation techniques were employed to minimize associated systematic uncertainties in our calculations: randomized compiling of the CNOTs (Pauli twirling)~\cite{PhysRevA.94.052325} combined with decoherence renormalization~\cite{Urbanek:2021oej,Rahman:2022rlg}, measurement error mitigation, post-selecting on physical states and dynamical decoupling~\cite{PhysRevA.58.2733,DUAN1999139,ZANARDI199977,PhysRevLett.82.2417}.\footnote{A recent detailed study of the stability of some of IBM's quantum devices using a system of physical interest can be found in Ref.~\cite{Yeter-Aydeniz:2022vuy}.}
The circuits were randomly complied with each CNOT Pauli-twirled as a mechanism to transform coherent errors in the CNOT gates into statistical noise in the ensemble.
This has been shown to be effective in improving the quality of results in other simulations, for example, Refs.~\cite{Kim2021ScalableEM,Rahman:2022rlg}.
Pauli twirling involves multiplying the right side of each CNOT by a randomly chosen
element of the two-qubit Pauli group, $G_2$,
and the left side by $G'_2$ 
such that $G'_2 \, CX \, G_2 = CX$ (up to a phase). 
For an ideal CNOT gate, this would have no effect on the circuit. 
A table of required CNOT identities is given,
for example,
in an appendix in Ref.~\cite{Rahman:2022rlg}.
Randomized Pauli-twirling is combined with performing measurements of a ``non-physics", mitigation circuit, which is the time evolution circuit evaluated at $t=0$, and is the identity in the absence of noise.
Assuming that the randomized-compiling of the Pauli-twirled CNOTs transforms coherent noise into depolarizing noise, 
the fractional deviation of the noiseless and computed results 
from the asymptotic limit of complete decoherence
are expected to be approximately equal for both the physics and mitigation ensembles. Assuming linearity, it follows that
\begin{equation}
\left ( P_{\text{pred}}^{(\text{phys})}-\frac{1}{8} \right ) = \left ( P_{\text{meas}}^{(\text{phys})}-\frac{1}{8} \right ) \times \left ( \frac{1-\frac{1}{8}}{ P_{\text{meas}}^{(\text{mit})}-\frac{1}{8} } \right )\ ,
\label{eq:mit}
\end{equation}
where $P_{\text{meas}}^{(\text{phys})}$ and $P_{\text{meas}}^{(\text{mit})}$ are post-processed probabilities and
$P_{\text{pred}}^{(\text{phys})}$ is an estimate of the probability once the effects of depolarizing noise have been removed. 
The ``$\frac{1}{8}$" represents the fully decohered probability after post-selecting on physical states (described next) and the ``$1$" is the probability of measuring the initial state from the mitigation circuit in the absence of noise. 

The computational basis of 6 qubits contains $2^6$ states but time evolution only connects those with the same $r$, $g$ and $b$. Starting from the trivial vacuum, this
implies that only the $8$ states with $r=g=b=0$ are accessible through time evolution.
The results off the quantum computer were post-processed to only select events that populated 1 of the 8 physically
allowed states, discarding outcomes that were unphysical. Typically, this resulted in a retention rate of $\sim 30\%$. The
workflow interspersed physics and mitigation circuits to provide a correlated calibration of the quantum devices. This enabled the detection (and removal) of
out-of-specs device performance during post-processing. We explored using the same twirling sequences for both physics and
mitigation circuits and found that it had no significant impact. 
The impact of dynamical decoupling of idle qubits using {\tt qiskit}'s built in functionality was also investigated and found to have little effect. 
The results of each run were corrected for measurement error using IBM's available function, {\tt TensoredMeasFitter}, and associated downstream operations.

\begin{figure}[!ht]
    \centering
    \includegraphics[width=\columnwidth]{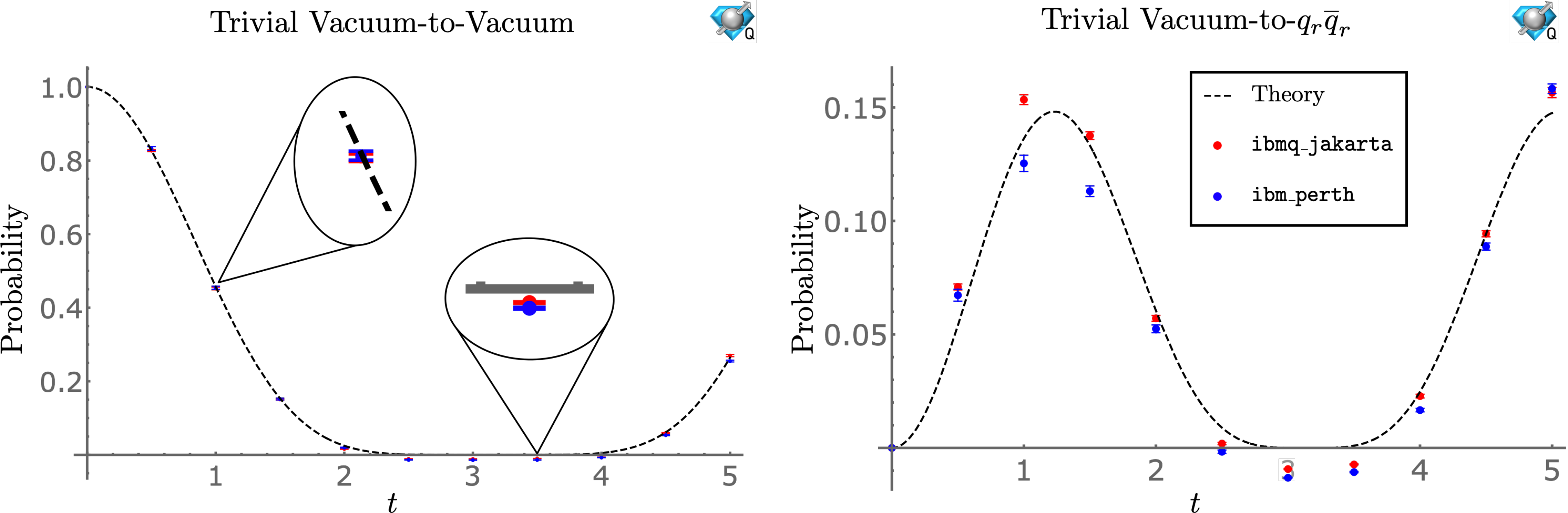}
    \caption{
    The trivial vacuum-to-vacuum (left panel) and trivial vacuum-to-$q_r \overline{q}_r$ (right panel) probabilities for $N_f=1$ QCD and $m=g=L=1$. 
    The dashed-black curve shows the expected result from one step of leading-order Trotter evolution.
    The results, given in Tables~\ref{tab:IBMvacresultsLO} and~\ref{tab:IBMrrbresultsLO}, were obtained by using $10^3$ shots for 447 Pauli-twirled circuits using IBM's quantum computers {\tt ibmq\_jakarta} (red) and {\tt ibm\_perth} (blue).
    }
    \label{fig:IBMresults}
\end{figure}

The results obtained for the trivial vacuum-to-vacuum and trivial vacuum-to-$q_r \overline{q}_r$ probabilities from one step of leading-order Trotter time evolution are shown in Fig.~\ref{fig:IBMresults}.
For each time, 447 Pauli-twirled physics circuits 
and 447 differently twirled circuits with zeroed angles (mitigation) were analyzed using $10^3$ shots on both {\tt ibmq\_jakarta} and {\tt ibm\_perth} (to estimate device systematics).
After post-selecting on physical states, correlated Bootstrap Resampling was used to form the final result.\footnote{As the mitigation and physics circuits were executed as adjacent jobs on the devices, the same Bootstrap sample was used to select results from both ensembles to account for temporal correlations.}
Tables~\ref{tab:IBMvacresultsLO} and~\ref{tab:IBMrrbresultsLO} display the results of the calculations performed using {\tt ibmq\_jakarta} and {\tt ibm\_perth} quantum computers.
The same mitigation data was used for both the trivial vacuum-to-vacuum and trivial vacuum-to-$q_r\overline{q}_r$ calculations, and is provided in columns 2 and 4 of Table~\ref{tab:IBMvacresultsLO}. 
See App.~\ref{app:LOTrott} for an extended discussion of leading-order Trotter. 
Note that the negative probabilities seen in Fig.~\ref{fig:IBMresults} indicate that additional non-linear terms are needed in Eq.~(\ref{eq:mit}).
\begin{table}[!ht]
\renewcommand{\arraystretch}{1.2}
\begin{tabularx}{\textwidth}{||c | Y | Y | Y | Y | Y | Y | Y ||}
\hline
\multicolumn{8}{||c||}{Vac-to-Vac Probabilities for $N_f=1$ QCD from IBM's {\tt ibmq\_jakarta} and {\tt ibm\_perth}} \\
 \hline
 $t$ & 
 \makecell{Mitigation \\ {\tt jakarta}} & 
 \makecell{Physics \\ {\tt jakarta}} & 
 \makecell{Mitigation \\ {\tt perth}} & 
 \makecell{Physics \\ {\tt perth}} & 
 \makecell{Results \\ {\tt jakarta}} & 
 \makecell{Results \\ {\tt perth}} & 
 Theory
 \\
 \hline\hline
 0 & - & - & - & - & - & - & 1  \\
 \hline
 0.5 & 
 0.9176(10) & 0.7607(24)  &
 0.8744(23) & 0.7310(42)  & 
 0.8268(27) & 0.8326(52) & 0.8274 \\
 \hline
 1.0 & 
 0.9059(12) & 0.4171(32) & 
 0.9118(16) & 0.4211(39) & 
 0.4523(36) & 0.4543(43) & 0.4568 \\
 \hline
 1.5 & 
 0.9180(12) & 0.1483(16) & 
 0.9077(17) & 0.1489(23) &  
 0.1507(17)  & 0.1518(25) & 0.1534 \\
 \hline
 2.0 & 
 0.8953(15) & 0.0292(08) & 
 0.8953(21) & 0.0324(10) &  
 0.0162(09)  &  0.0198(11) & 0.0249 \\
 \hline
 2.5 & 
 0.9169(12) & 0.0020(01) &
 0.8938(21) & 0.0032(02) & 
 -0.0109(03)   &  -0.0136(04) & 0.0010 \\
 \hline
 3.0 &
 0.9282(13) & 0.00010(2) &
 0.9100(13) & 0.00017(3) &  
 -0.0111(02)  &  -0.0140(02) & $1.3\times 10^{-7}$ \\
 \hline
 3.5 &
 0.9357(10)& 0.00017(3) & 
 0.9109(14) & 0.00037(4)& 
 -0.0097(02) &  -0.0138(02) & $3.2\times 10^{-5}$ \\
 \hline
 4.0 & 
 0.9267(13) & 0.0081(03) & 
 0.9023(14) & 0.0076(03) & 
 -0.0026(04) &  -0.0072(04) & 0.0052 \\
 \hline
 4.5 & 
 0.9213(12) & 0.0653(10) &
 0.8995(16) & 0.0619(11) & 
 0.0594(11)  & 0.0537(13) & 0.0614 \\
 \hline
 5.0 & 
 0.9105(12) & 0.2550(26) & 
 0.9031(14) & 0.2405(21) &  
 0.2698(29)  & 0.2550(23) & 0.2644 \\
 \hline
\end{tabularx}
\renewcommand{\arraystretch}{1}
\caption{
The trivial vacuum-to-vacuum probabilities for $m=g=L=1$ using {\tt ibmq\_jakarta} and {\tt ibm\_perth}, the underlying distributions of which are displayed in Fig.~\ref{fig:IBMhistos}.
The 2nd through 5th columns are the results after selecting only physical states and columns $6$ and $7$ are the results after using the mitigation circuit to account for depolarizing noise.
}
\label{tab:IBMvacresultsLO}
\end{table}
\begin{table}[!ht]
\renewcommand{\arraystretch}{1.2}
\begin{tabularx}{\textwidth}{||c | Y | Y | Y | Y | Y ||}
\hline
\multicolumn{6}{||c||}{Vac-to-$q_r\overline{q}_r$ Probabilities for $N_f=1$ QCD from IBM's {\tt ibmq\_jakarta} and {\tt ibm\_perth}} \\
 \hline
 $t$ & 
 \makecell{Physics \\ {\tt jakarta}} & 
 \makecell{Physics \\ {\tt perth}} & 
 \makecell{Results \\ {\tt jakarta}} & 
 \makecell{Results \\ {\tt perth}} & 
 Theory
 \\
 \hline\hline
 0 & - & - & - & - & 0  \\
 \hline
 0.5 & 
 0.0760(12) & 0.0756(22)  & 
 0.0709(13) & 0.0673(26) & 0.0539 \\
 \hline
 1.0 & 
 0.1504(19) & 0.1253(32) & 
 0.1534(22) & 0.1254(36) & 0.1363 \\
 \hline
 1.5 & 
 0.1364(15) & 0.1144(21) &  
 0.1376(17)  & 0.1131(23) & 0.1332 \\
 \hline
 2.0 & 
 0.0652(11) & 0.0611(15) &  
 0.0571(13)  &  0.0525(17) & 0.0603 \\
 \hline
 2.5 & 
 0.0136(04) & 0.0137(06) & 
 0.0019(05)   &  -0.0017(07) & 0.0089 \\
 \hline
 3.0 &
 0.0017(01) & 0.0011(01) &  
 -0.0093(02)  &  -0.0132(02) & $2.5\times 10^{-5}$ \\
 \hline
 3.5 &
 0.0024(01) & 0.0032(02)& 
 -0.0073(02) &  -0.0107(03) & 0.0010 \\
 \hline
 4.0 & 
 0.0314(07) & 0.0288(07) & 
 0.0228(08) &  0.0167(08) & 0.0248 \\
 \hline
 4.5 & 
 0.0971(12) & 0.0929(14) & 
 0.0943(13)  & 0.0887(16) & 0.0943 \\
 \hline
 5.0 & 
 0.1534(20) & 0.1546(19) &  
 0.1566(22)  & 0.1583(21) & 0.1475 \\
 \hline
\end{tabularx}
\renewcommand{\arraystretch}{1}
\caption{
The trivial vacuum-to-$q_r \overline{q}_r$ probabilities for $m=g=L=1$ using {\tt ibmq\_jakarta} and {\tt ibm\_perth}.
The 2nd and 3rd columns are the results after selecting only physical states and columns 4 and 5 are the results after using the mitigation circuit to account for depolarizing noise.
}
\label{tab:IBMrrbresultsLO}
\end{table}

\begin{figure}[!ht]
    \centering
    \includegraphics[width=0.9 \columnwidth]{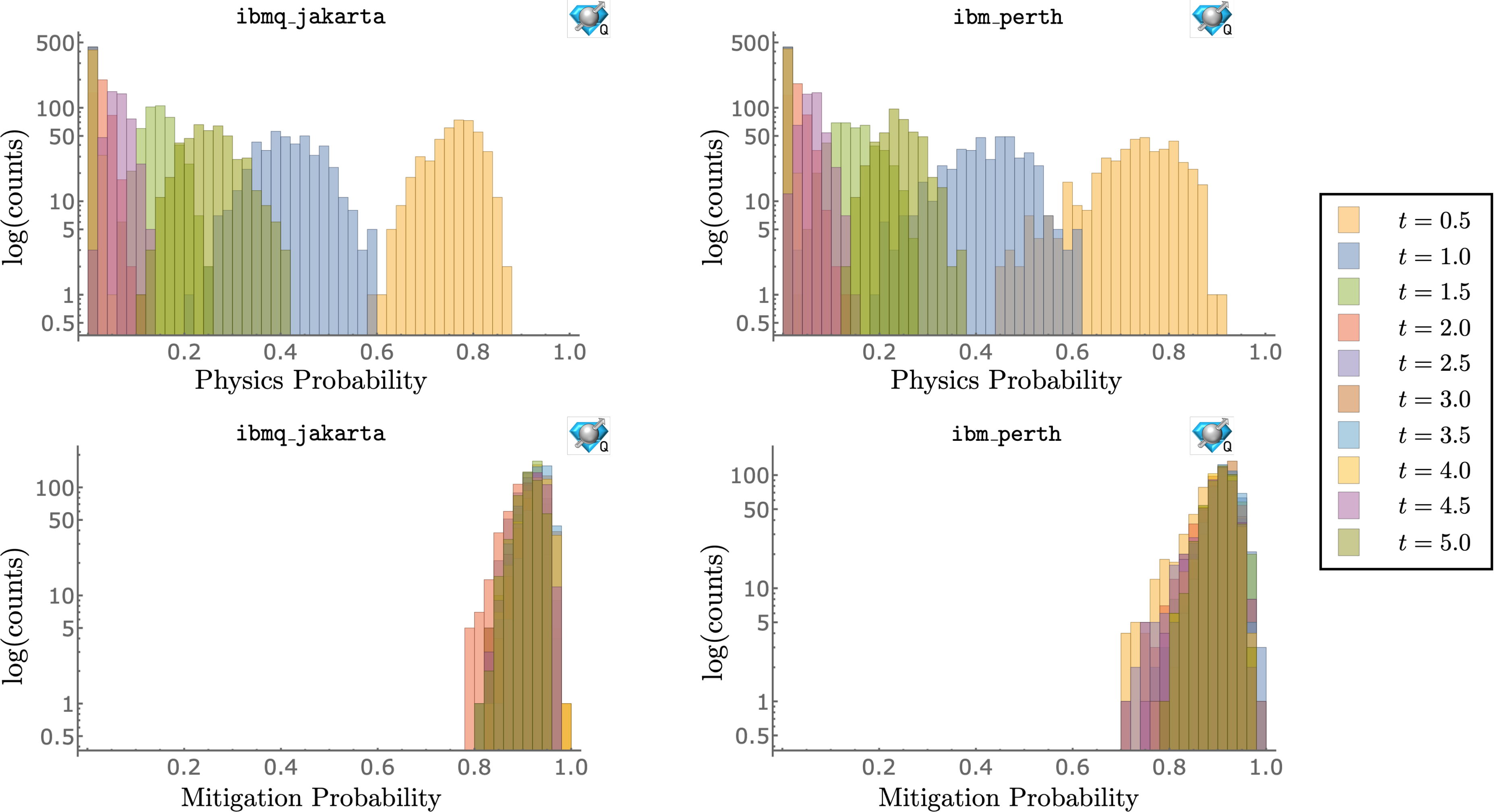}
    \caption{
    Histograms of the post-processed vacuum-to-vacuum results obtained using {\tt ibmq\_jakarta} and {\tt ibm\_perth}.
    The horizontal axes show the value of the vacuum-to-vacuum probability, and the vertical axes show bin counts on a log-scale.
    The top panels display the results obtained from the physics circuits for the range of evolution times and the bottom panels display the results obtained for the corresponding mitigation circuits.
    }
    \label{fig:IBMhistos}
\end{figure}

It is interesting to consider the distributions of events obtained from the Pauli-twirled circuits, as shown in Fig.~\ref{fig:IBMhistos}.
The distributions are not Gaussian and, in a number of instances, exhibit heavy tails particularly near the boundaries.\footnote{For a study of heavy-tailed distributions in Euclidean-space lattice QCD calculations, see Refs.~\cite{Wagman:2016bam,Wagman:2017gqi}.}
The spread of the distributions, associated with non-ideal CNOT gates, is seen to reach a maximum of $\sim 0.4$, but with a full-width at half-max that is $\sim 0.2$. These distributions are already broad with a 34 CNOT circuit, and we probed the limit
of these devices by time-evolving with two first-order Trotter steps,\footnote{Under a particular ordering of terms, two steps of first- and second-order Trotter time evolution are equivalent.}
which requires 91 CNOTs after accounting for SWAPs. 
Using the aforementioned techniques, this was found to be beyond the capabilities of {\tt ibmq\_jakarta}, {\tt ibmq\_lagos} and {\tt ibm\_perth}.

\section{Arbitrary \texorpdfstring{\boldmath$N_c$}{Nc} and \texorpdfstring{\boldmath$N_f$}{Nf}}
\label{sec:NcNf}
\noindent
In this section, the structure of the Hamiltonian for $N_f$ flavors of quarks in the fundamental representation of $SU(N_c)$ is developed. 
The mapping to spins has the same structure as for 
$N_f=2$ QCD, but now, there are $N_c\times N_f$ $q$s and $N_c\times N_f$ $\overline{q}$s per spatial lattice site.
While the mass and kinetic terms generalize straightforwardly, the energy in the chromo-electric field is more tricky.
After enforcing Gauss's law, it is 
\begin{equation}
    H_{el} = \frac{g^2}{2} \sum_{n=0}^{2L-2} \left ( \sum_{m \leq n} Q^{(a)}_m \right ) ^2
    \ ,\ \ 
    Q^{(a)}_m = \phi^{\dagger}_m T^a \phi_m
    \ ,
\end{equation}
where $T^a$ are now the generators of $SU(N_c)$.
The Hamiltonian,
including chemical potentials for baryon number (chemical potentials for other flavor combinations can be included as needed), is found to be
\begin{subequations}
    \label{eq:HNfNc}
    \begin{align}
        H = & \ H_{kin}\ +\ H_m\ +\ H_{el} \ +\ H_{\mu_B}  \ , \\[4pt]
        H_{kin} =& \ \frac{1}{2}\sum_{n=0}^{2L-2}\sum_{f=0}^{N_f-1}\sum_{c=0}^{N_c-1} \left[ \sigma_{i(n,f,c)}^+ \left ( \bigotimes_{j=1}^{N_cN_f-1}(-\sigma_{i(n,f,c)+j}^z ) \right ) \sigma_{i(n,f,c) + N_c N_f}^- +\rm{h.c.} \right] \ ,
        \label{eq:HkinN}\\[4pt]
        H_m =& \ \frac{1}{2} \sum_{n=0}^{2L-1} \sum_{f=0}^{N_f-1}\sum_{c=0}^{N_c-1} m_f \left[ (-1)^n \sigma^z_{i(n,f,c)} + 1 \right] \ ,
        \label{eq:HmN}\\[4pt]
        H_{el} =& \ \frac{g^2}{2} \sum_{n=0}^{2L-2}(2L-1-n)\left( \sum_{f=0}^{N_f-1} Q_{n,f}^{(a)} \, Q_{n,f}^{(a)} \ \ + \ \  
            2 \sum_{f=0}^{N_f-2} \sum_{f'=f+1}^{N_f-1}Q_{n,f}^{(a)} \, Q_{n,f'}^{(a)}
             \right)   \nonumber \\[4pt]
            & + g^2 \sum_{n=0}^{2L-3} \sum_{m=n+1}^{2L-2}(2L-1-m) \sum_{f=0}^{N_f-1} \sum_{f'=0}^{N_f-1} Q_{n,f}^{(a)} \, Q_{m,f'}^{(a)}  \ ,
        \label{eq:HelN}\\[4pt]
        H_{\mu_B} =& \ -\frac{\mu_B}{2 N_c} \sum_{n=0}^{2L-1} \sum_{f=0}^{N_f-1} \sum_{c=0}^{N_c-1} \sigma^z_{i(n,f,c)}   \ ,
        \label{eq:HmuBN}
    \end{align}
\end{subequations}
where, $i(n,f,c) = (N_c N_f n + N_c f + c)$, 
and the products of the charges are
\begin{align}
    4 Q_{n,f}^{(a)} \, Q_{n,f}^{(a)} =& \ \frac{N_c^2-1}{2} - \left (1+\frac{1}{N_c} \right )\sum_{c=0}^{N_c-2}\sum_{c' = c+1}^{N_c-1} \sigma^z_{i(n,f,c)} \sigma^z_{i(n,f,c')} \ ,  \nonumber \\[4pt]
    8 Q_{n,f}^{(a)} \, Q_{m,f'}^{(a)} =& \ 4 \sum_{c=0}^{N_c-2}\sum_{c'=c+1}^{N_c-1} \left[ \sigma^+_{i(n,f,c)} \ Z_{(n,f,c,c')} \ \sigma^-_{i(n,f,c')} \sigma^-_{i(m,f',c)} \ Z_{(m,f',c,c')} \ \sigma^+_{i(m,f',c')} + {\rm h.c.}
    \right] \  \nonumber \\
    &+ \sum_{c=0}^{N_c-1} \sum_{c'=0}^{N_c-1}\left (\delta_{cc'} - \frac{1}{N_c} \right)\sigma^z_{i(n,f,c)}\sigma^z_{i(m,f',c')}\ , \nonumber \\
     Z_{(n,f,c,c')}  \equiv & \ \bigotimes_{k=1}^{c'-c-1} \sigma^z_{i(n,f,c)+k}
    \ .
    \label{eq:QnfQmfpN}
\end{align}
The resource requirements for implementing Trotterized time evolution 
using generalizations of the circuits in Sec.~\ref{sec:Circuits} are given in Eq.~(\ref{eq:RHCN}). 

It is interesting to consider the large-$N_c$ limit of the Hamiltonian, 
where quark loops are parametrically suppressed and 
the system can be described semi-classically~\cite{tHooft:1973alw,tHooft:1974pnl,Witten:1979kh,RevModPhys.54.407}.
Unitarity requires rescaling the strong coupling, 
$g^2 \to g^2/N_c$ and leading terms in the Hamiltonian scale as $\mathcal{O}(N_c)$.
The leading order contribution to the product of charges is
\begin{align}
    4 Q_{n,f}^{(a)} \, Q_{n,f}^{(a)} =&\ \sum_{c=0}^{N_c-2}\sum_{c' = c+1}^{N_c-1} \left (1 - \sigma^z_{i(n,f,c)} \sigma^z_{i(n,f,c')}\right ) \ ,  \nonumber \\[4pt]
    8 Q_{n,f}^{(a)} \, Q_{m,f'}^{(a)} =&\ 4 \sum_{c=0}^{N_c-2}\sum_{c'=c+1}^{N_c-1} 
    \left[ \sigma^+_{i(n,f,c)} \ 
    Z_{(n,f,c,c')} \ 
    \sigma^-_{i(n,f,c')} \sigma^-_{i(m,f',c)} \ 
    Z_{(m,f',c,c')} \ 
    \sigma^+_{i(m,f',c')} + {\rm h.c.}
    \right] \ .
\end{align}
Assuming that the number of $q\overline{q}$ pairs that contribute to the meson wavefunctions do not scale with $N_c$, 
as expected in the large-$N_c$ limit,
$H_{el} \propto N_c$ 
and mesons are non-interacting, a well known consequence of the large-$N_c$ limit~\cite{tHooft:1973alw,tHooft:1974pnl}.
Baryons on the other hand are expected to have strong interactions at leading order in $N_c$~\cite{Witten:1979kh}. This is a semi-classical limit and we expect that there exists a basis
where states factorize into localized tensor products, and the time evolution operator is non-entangling.
The latter result has been observed in the large-$N_c$ limit of hadronic scattering~\cite{Beane:2018oxh,Beane:2021zvo,Low:2021ufv,Aoude:2020mlg,Cervera-Lierta:2017tdt}.

\section{Summary and Discussion}
\label{sec:SandC}
\noindent
Important for future quantum simulations of processes that can be meaningfully compared to experiment, the real-time dynamics of strongly-interacting systems are predicted to be efficiently computable with quantum computers of sufficient capability.
Building upon foundational work in quantum chemistry and in low-dimensional $U(1)$ and $SU(2)$ gauge theories, this work has developed the tools necessary for the quantum simulation of
$1+1$D QCD (in axial gauge) using open boundary conditions, with arbitrary numbers of quark flavors and colors and including chemical potentials for baryon number and isospin.
Focusing largely on QCD with $N_f=2$, which shares many of the complexities of QCD in $3+1$D, we have performed a detailed analysis of the required quantum resources for simulation of real-time dynamics, including efficient quantum circuits and associated gate counts, and the scaling of the number of Trotter steps for a fixed-precision time evolution.
The structure and dynamics of small systems, with $L=1,2$ for $N_c=3$ and $N_f=1,2$ have been detailed using classical computation, quantum simulators, D-Wave's {\tt Advantage} and IBM's 7-qubit devices {\tt ibmq\_jakarta} and {\tt ibm\_perth}. Using recently developed error mitigation strategies, relatively small uncertainties were obtained for a single Trotter step with $34$ CNOT gates after transpilation onto the QPU connectivity.

Through a detailed study of the low-lying spectrum, both the relevant symmetries and the color-singlets in the mesonic and baryonic sectors, including a bound two-baryon nucleus, have been identified.  
Open boundary conditions also permit low-lying color edge-states that penetrate into the lattice volume by a distance set by the confinement scale.
By examining quark entanglement in the hadrons, a transition from the mesons being primarily composed of quark-antiquarks to baryon-antibaryons was found.
We have presented the relative contributions of each of the terms in the Hamiltonian to the energy of the vacuum, mesons and baryons.  

This chapter has provided an estimate for the number of CNOT-gates required to implement one Trotter step in $N_f=2$, $1+1$D axial-gauge QCD. For $L = 10$ spatial sites, $\sim 3 \times 10^4$ CNOTs
are required, while $\sim 4 \times 10^6$ CNOTs are required for $L = 100$.
Realistically, quantum simulations with $L=10$ are a beginning toward providing results with a complete quantification of uncertainties, including lattice-spacing and finite-volume
artifacts, and $L=100$ will likely yield high-precision results. It was found that, in the axial-gauge formulation, resources for time evolution effectively scale as $L^2 t$ for intermediate times and $L^2 t^2$ for
asymptotic times. With $L\sim t$, this asymptotic scaling is the same as in the Schwinger model, suggesting no differences in scaling between Weyl and axial gauges.

\clearpage
\begin{subappendices}

\section{Mapping to Qubits}
\label{app:hamConst}
\noindent 
This appendix outlines how the qubit Hamiltonian in Eq.~(\ref{eq:H2flav}) is obtained from the lattice Hamiltonian in Eq.~(\ref{eq:GFHam}). 
For this system, 
the constraint of Gauss's law is sufficient to uniquely determine the chromo-electric field carried by the links between lattice sites in terms of a background chromo-electric field and the distribution of color charges.  The difference between adjacent chromo-electric fields at a site with charge 
$Q^{(a)}$ 
is
\begin{equation}
    {\bf E}^{(a)}_{n+1} - {\bf E}^{(a)}_n = Q^{(a)}_n \ ,
    \label{eq:GaussLaw}
\end{equation}
for $a=1$ to $8$, resulting in a
chromo-electric field 
\begin{equation}
    {\bf E}^{(a)}_{n} = {\bf F}^{(a)}
    \: + \: \sum_{i\leq n} Q^{(a)}_i \  .
    \label{eq:GaussLawSol}
\end{equation}
In general, there can be a non-zero background chromo-electric field, ${\bf F}^{(a)}$,
which in this paper has been set to zero.
Inserting the chromo-electric field in terms of the charges into Eq.~(\ref{eq:KSHam}) yields Eq.~(\ref{eq:GFHam}). 

The color and flavor degrees of freedom of each $q$ and $\overline{q}$ are then distributed over 
$6$ ($=N_c N_f$) sites as illustrated in Fig.~(\ref{fig:2flavLayout}). 
There are now creation and annihilation operators for each quark, and the Hamiltonian is
\begin{align}
   H =&\sum_{n=0}^{2L-1} \sum_{f=0}^1 \sum_{c=0}^2  \left [ \left ( m_f (-1)^n \: - \: \frac{\mu_B}{3} \: - \: \frac{\mu_I}{2}(-1)^f\right ) \psi^{\dagger}_{6n+3f+c} \psi_{6n+3f+c}  \right ] \nonumber \\[4pt]
   &+  \frac{1}{2} \sum_{n=0}^{2L-2}\sum_{f=0}^1 \sum_{c=0}^2  \left (\psi^{\dagger}_{6n+3f+c} \psi_{6(n+1)+3f+c} + \: {\rm h.c.} \right ) \: + \: \frac{g^2}{2} \sum_{n=0}^{2L-2} \left ( \sum_{m\leq n} \sum_{f=0}^1 Q^{(a)}_{m,f} \right ) ^2 \ ,
    \label{eq:FockHam}
\end{align}
where the color charge is evaluated over three $(r,g,b)$ occupation sites with the same flavor,
\begin{equation}
   Q_{m,f}^{(a)} = \sum_{c=0}^{2}\sum_{c'=0}^2 \psi^{\dagger}_{6m+3f+c} \ T^a_{cc'}\  \psi_{6m+3f+c'} \ ,
\end{equation}
and the $T^a$ are the eight generators of $SU(3)$. 
The fermionic  operators in Fock space are mapped onto spin operators via the JW transformation,
\begin{equation}
    \psi_n =  \bigotimes_{l<n}( -\sigma^z_l ) \sigma^-_n \ , \ \ \psi_n^{\dagger} =  \bigotimes_{l<n}( -\sigma^z_l ) \sigma^+_n \ .
    \label{eq:JW}
\end{equation}
In terms of spins, the eight $SU(3)$ charge operators become\footnote{Calculations of quadratics of the gauge charges are simplified by the Fierz identity,
\begin{equation}
    \left ( T^{(a)} \right )^{\alpha}_{\beta} \, \left (T^{(a)}\right )^{\gamma}_{\delta} = \frac{1}{2} (\delta^{\alpha}_{\delta} \delta^{\gamma}_{\beta} - \frac{1}{N_c} \delta^{\alpha}_{\beta}\delta^{\gamma}_{\delta}) \ .
    \label{eq:Fierz}
\end{equation}}
\begin{align}
    Q_{m,f}^{(1)} = & \ \frac{1}{2}\sigma^+_{6m+3f} \sigma^-_{6m+3f+1} + \rm{h.c.} \ , \nonumber \\
    Q_{m,f}^{(2)} = & \ -\frac{i}{2}\sigma^+_{6m+3f} \sigma^-_{6m+3f+1} + \rm{h.c.} \ , \nonumber \\
    Q_{m,f}^{(3)} = & \ \frac{1}{4}(\sigma^z_{6m+3f} - \sigma^z_{6m+3f+1}) \ , \nonumber \\
    Q_{m,f}^{(4)} = & \ -\frac{1}{2}\sigma^+_{6m+3f} \sigma^z_{6m+3f+1} \sigma^-_{6m+3f+2} + \rm{h.c.} \ , \nonumber \\
    Q_{m,f}^{(5)} = & \ \frac{i}{2}\sigma^+_{6m+3f} \sigma^z_{6m+3f+1} \sigma^-_{6m+3f+2} + \rm{h.c.} \ , \nonumber \\
    Q_{m,f}^{(6)} = & \ \frac{1}{2}\sigma^+_{6m+3f+1} \sigma^-_{6m+3f+2} + \rm{h.c.} \ , \nonumber \\
    Q_{m,f}^{(7)} = & \ -\frac{i}{2}\sigma^+_{6m+3f+1} \sigma^-_{6m+3f+2} + \rm{h.c.} \ , \nonumber \\
    Q_{m,f}^{(8)} = & \ \frac{1}{4 \sqrt{3}}(\sigma^z_{6m+3f} + \sigma^z_{6m+3f+1} - 2\sigma^z_{6m+3f+2})  \ .
    \label{eq:SU3chargesFull}
\end{align}
Substituting Eqs.~(\ref{eq:JW}) and~(\ref{eq:SU3chargesFull}) into Eq.~(\ref{eq:FockHam}) gives the Hamiltonian in Eq.~(\ref{eq:H2flav}). For reference, the expanded Hamiltonian for $L=1$ is
\begin{subequations}
    \label{eq:H2flavL1}
    \begin{align}
    H = & \ H_{kin}\ +\ H_m\ +\ H_{el} \ +\ 
    H_{\mu_B}\ +\ H_{\mu_I}\ ,\\[4pt]
    H_{kin} = & \ -\frac{1}{2} (\sigma^+_6 \sigma^z_5 \sigma^z_4 \sigma^z_3 \sigma^z_2 \sigma^z_1 \sigma^-_0 + \sigma^-_6 \sigma^z_5 \sigma^z_4 \sigma^z_3 \sigma^z_2 \sigma^z_1 \sigma^+_0 + \sigma^+_7 \sigma^z_6 \sigma^z_5 \sigma^z_4 \sigma^z_3 \sigma^z_2 \sigma^-_1 + \sigma^-_7 \sigma^z_6 \sigma^z_5 \sigma^z_4 \sigma^z_3 \sigma^z_2 \sigma^+_1  \nonumber \\
    &+\, \sigma^+_8 \sigma^z_7 \sigma^z_6 \sigma^z_5 \sigma^z_4 \sigma^z_3 \sigma^-_2 + \sigma^-_8 \sigma^z_7 \sigma^z_6 \sigma^z_5 \sigma^z_4 \sigma^z_3 \sigma^+_2 + \sigma^+_9 \sigma^z_8 \sigma^z_7 \sigma^z_6 \sigma^z_5 \sigma^z_4 \sigma^-_3 + \sigma^-_9 \sigma^z_8 \sigma^z_7 \sigma^z_6 \sigma^z_5 \sigma^z_4 \sigma^+_3 \nonumber \\
    &+\, \sigma^+_{10} \sigma^z_9 \sigma^z_8 \sigma^z_7 \sigma^z_6 \sigma^z_5 \sigma^-_4 + \sigma^-_{10} \sigma^z_9 \sigma^z_8 \sigma^z_7 \sigma^z_6 \sigma^z_5 \sigma^+_4 + \sigma^+_{11} \sigma^z_{10} \sigma^z_9 \sigma^z_8 \sigma^z_7 \sigma^z_6 \sigma^-_5 + \sigma^-_{11} \sigma^z_{10} \sigma^z_9 \sigma^z_8 \sigma^z_7 \sigma^z_6 \sigma^+_5 ) \ ,
        \label{eq:Hkin2flavL1}\\[4pt]
    H_m = & \ \frac{1}{2} \left [ m_u\left (\sigma^z_0 + \sigma^z_1 + \sigma^z_2 -\sigma^z_6 - \sigma^z_7 - \sigma^z_8 + 6\right )+ m_d\left (\sigma^z_3 + \sigma^z_4 + \sigma^z_5 -\sigma^z_9 - \sigma^z_{10} - \sigma^z_{11} + 6\right ) \right ]\ ,
        \label{eq:Hm2flavL1}\\[4pt]
    H_{el} = & \ \frac{g^2}{2} \bigg [ \frac{1}{3}(6 - \sigma^z_1 \sigma^z_0 - \sigma^z_2 \sigma^z_0 - \sigma^z_2 \sigma^z_1 - \sigma^z_4 \sigma^z_3 - \sigma^z_5 \sigma^z_3 - \sigma^z_5 \sigma^z_4) + \sigma^+_4\sigma^-_3\sigma^-_1\sigma^+_0  + \sigma^-_4\sigma^+_3\sigma^+_1\sigma^-_0  \nonumber \\ 
    &+ \, \sigma^+_5\sigma^z_4\sigma^-_3\sigma^-_2\sigma^z_1\sigma^+_0  
    \, \sigma^-_5\sigma^z_4\sigma^+_3\sigma^+_2\sigma^z_1\sigma^-_0 \, + \, \sigma^+_5\sigma^-_4\sigma^-_2\sigma^+_1 + \sigma^-_5\sigma^+_4\sigma^+_2\sigma^-_1  \nonumber \\
    & +\,\frac{1}{12}\left (2 \sigma^z_3 \sigma^z_0 + 2\sigma^z_4 \sigma^z_1 + 2\sigma^z_5 \sigma^z_2 - \sigma^z_5 \sigma^z_0 - \sigma^z_5 \sigma^z_1 - \sigma^z_4 \sigma^z_2 - \sigma^z_4 \sigma^z_0 - \sigma^z_3 \sigma^z_1  - \sigma^z_3 \sigma^z_2 \right ) \bigg ] \ ,
    \label{eq:Hel2flavL1}\\[4pt] 
    H_{\mu_B} = & \ -\frac{\mu_B}{6} \left ( \sigma^z_0 + \sigma^z_1 + \sigma^z_2 + \sigma^z_3 + \sigma^z_4 + \sigma^z_5
    - \sigma^z_6 + \sigma^z_7 + \sigma^z_8 + \sigma^z_9 + \sigma^z_{10} + \sigma^z_{11} \right )\ ,
        \label{eq:HmuB2flavL1}\\[4pt]
    H_{\mu_I} = & \ -\frac{\mu_I}{4} \left ( \sigma^z_0 + \sigma^z_1 + \sigma^z_2 - \sigma^z_3 - \sigma^z_4 - \sigma^z_5
    + \sigma^z_6 + \sigma^z_7 + \sigma^z_8 - \sigma^z_9 - \sigma^z_{10} - \sigma^z_{11} \right ) \ .
        \label{eq:HmuI2flavL1}
    \end{align}
\end{subequations}
%

\section{Symmetries of the Free-Quark Hamiltonian}
\label{app:freeSym}
\noindent 
Here the symmetries of the free-quark Hamiltonian are identified to better understand the degeneracies observed in the spectrum of $1+1$D QCD with $N_f=2$ and $L=1$ as displayed in Figs.~\ref{fig:specDegenh} and~\ref{fig:specDegeng}.
Specifically, the Hamiltonian with $g=h=\mu_B=\mu_I=0$, leaving only the hopping and mass terms ($m = m_u = m_d$), is
\begin{equation}
   H =   \sum_{f=0}^1 \sum_{c=0}^2  \left [ m \sum_{n=0}^{2L-1} (-1)^n \psi^{\dagger}_{6n+3f+c} \psi_{6n+3f+c} \: + \: \frac{1}{2} \sum_{n=0}^{2L-2} \left (\psi^{\dagger}_{6n+3f+c} \psi_{6(n+1)+3f+c} + \: {\rm h.c.} \right ) \right ] \ .
\end{equation}
The mapping of degrees of freedom is taken to be as shown in Fig.~\ref{fig:2flavLayout}, but it will be convenient to work with Fock-space quark operators instead of spin operators. 
In what follows the focus will be on $L=1$, and larger systems follow similarly.

The creation operators can be assembled into a 12-component vector, 
$\Psi^{\dagger}_i = (\psi_0^\dagger, \, \psi_1^\dagger, \ldots ,\psi_{10}^{\dagger}, \, \psi_{11}^{\dagger})$, 
in terms of which the Hamiltonian  becomes
\begin{equation}
    H = \Psi^{\dagger}_i M_{ij} \Psi_j \ ,
\end{equation}
where $M$ is a $12 \times 12$ block matrix of the form,
\begin{equation}
   M  = 
   \left[
   \begin{array}{c|c}
       m  & 1/2 \\
       \hline
    1/2  & -m 
\end{array}
\right]
\ ,
\end{equation}
with each block a $6 \times 6$  diagonal matrix. 
Diagonalizing $M$, gives rise to 
\begin{equation}
   \tilde M = 
   \left[
   \begin{array}{c|c}
       \lambda  & 0 \\
       \hline
    0  & -\lambda 
\end{array}
\right]
\ ,\ \ 
\lambda = \frac{1}{2}\sqrt{1+4m^2} \ ,
\end{equation}
with associated eigenvectors,
\begin{equation}
    \tilde{\psi}_i = \frac{1}{\sqrt{2}}\left (\sqrt{1+\frac{\lambda}{m}}\, \psi_i \ + \ \sqrt{1-\frac{\lambda}{m}}\, \psi_{6+i} \right ) \ , \ \tilde{\psi}_{6+i} = \frac{1}{\sqrt{2}}\left (-\sqrt{1-\frac{\lambda}{m}}\,\psi_i \ + \ \sqrt{1+\frac{\lambda}{m}}\, \psi_{6+i} \right )
    \ ,
\end{equation}
where $\tilde{\psi}_i$ ($\tilde{\psi}_{6+i}$) corresponds to the positive (negative) eigenvalue
and the index $i$ takes values $0$ to $5$.
These eigenvectors create superpositions of quarks and antiquarks with the same color and flavor, which are the OBC analogs of momentum plane-waves. 
In this basis, the Hamiltonian becomes
\begin{equation}
    H = \sum_{i=0}^{5} \lambda\left ( \tilde{\psi}^{\dagger}_i \tilde{\psi}_i - \tilde{\psi}^{\dagger}_{6+i} \tilde{\psi}_{6+i} \right )
    \ ,
    \label{eq:Hamifree0}
\end{equation}
which has a vacuum state,
\begin{equation}
    \lvert \Omega_0 \rangle = \prod_{i=0}^{i=5}\tilde{\psi}^{\dagger}_{6+i} \ket{\omega_0} \ ,
\end{equation}
where $\ket{\omega_0}$ is the unoccupied state,
and
$\lvert \Omega_0 \rangle$ corresponds to 
$\lvert 000000111111 \rangle$ (in binary)
in this transformed basis.
Excited states are formed by acting with either $\tilde{\psi}^{\dagger}_i$ or $\tilde{\psi}_{6+i}$ on 
$\lvert \Omega_0 \rangle$ which raises the energy of the system by $\lambda$. 
A further transformation is required for the $SU(12)$ symmetry to be manifest.
In terms of the 12-component vector, $\tilde{\Psi}^{\dagger} = (\tilde{\psi}^{\dagger}_0, \, \ldots, \, \tilde{\psi}^{\dagger}_5, \, \tilde{\psi}_6, \, \ldots, \, \tilde{\psi}_{11})$, the Hamiltonian in Eq.~(\ref{eq:Hamifree0}) becomes,
\begin{equation}
    H = 
    \sum_{i=0}^{5} \lambda\left ( \tilde{\psi}^{\dagger}_i \tilde{\psi}_i - \tilde{\psi}^{\dagger}_{6+i} \tilde{\psi}_{6+i} \right )
    \ =\ 
    \lambda\left( 
    \tilde{\Psi}^{\dagger} \tilde{\Psi} - 6 
    \right)
    \ ,
\end{equation}
where the canonical anticommutation relations have been used to obtain the final equality.
This is invariant under a $SU(12)$ symmetry, where $\tilde{\Psi}$ transforms in the fundamental representation. 
The free-quark spectrum ($g=h=0$) is therefore described by states with degeneracies corresponding to the ${\bf 1}$ and ${\bf 12}$ of $SU(12)$ as well as
the antisymmetric combinations of fundamental irreps, ${\bf 66}, {\bf 220}, \ldots$ as illustrated in Figs.~\ref{fig:specDegenh} and~\ref{fig:specDegeng}.
The vacuum state corresponds to the singlet of $SU(12)$. The lowest-lying {\bf 12} corresponds to single quark or antiquark excitations, which are color ${\bf 3}_c$s for quarks and $\overline{\bf 3}_c$s for antiquarks and will each appear as isodoublets, i.e., ${\bf 12}\rightarrow {\bf 3}_c\otimes {\bf 2}_f \oplus \overline{\bf 3}_c\otimes {\bf 2}_f$.
The {\bf 66} arises from double excitations of quarks and antiquarks.  The possible color-isospin configurations are, based upon totally-antisymmetric wavefunctions for $qq$, $\overline{q}\overline{q}$ and $\overline{q}q$,  
${\bf 66} =
{\bf 1}_c\otimes {\bf 1}_f
\oplus
{\bf 1}_c\otimes {\bf 3}_f
\oplus
{\bf 8}_c\otimes {\bf 1}_f
\oplus
{\bf 8}_c\otimes {\bf 3}_f
\oplus
{\bf 6}_c\otimes {\bf 1}_f
\oplus
\overline{\bf 6}_c\otimes {\bf 1}_f
\oplus
{\bf 3}_c\otimes {\bf 3}_f
\oplus
\overline{\bf 3}_c\otimes {\bf 3}_f
$.
The OBCs split the naive symmetry between quarks and antiquarks and, for $g\ne 0$, the lowest-lying color edge-states are from the antiquark sector with degeneracies $6$ from a single excitation and $6,9$ from double excitations. 
Larger lattices possess an analogous global 
$SU(12)$ symmetry, coupled between spatial sites by the hopping term, and the spectrum is again one of non-interacting quasi-particles.

\section{Details of the D-Wave Implementations}
\label{app:dwave}
\noindent
In this appendix, additional details are provided on the procedure used in Sec.~\ref{sec:dwave_spectrum} to extract the lowest three eigenstates and corresponding energies using D-Wave's {\tt Advantage}, (a more complete description can be found in Ref.~\cite{Illa:2022jqb}). The objective function $F$ to be minimized can be written in terms of binary variables and put into QUBO form. Defining $F=\langle \Psi \rvert \tilde{H} \lvert \Psi \rangle -\eta \langle \Psi| \Psi \rangle$~\cite{doi:10.1021/acs.jctc.9b00402}, and expanding the wavefunction with a finite dimensional orthonormal basis $\psi_{\alpha}$, $\lvert \Psi \rangle =\sum^{n_s}_{\alpha} a_\alpha |\psi_{\alpha}\rangle$, it is found
\begin{equation}
    F=\langle \Psi \rvert \tilde{H} \lvert \Psi \rangle -\eta \langle \Psi| \Psi \rangle = \sum_{\alpha\beta}^{n_s} a_\alpha a_\beta[\langle \psi_\alpha \rvert \tilde{H} \lvert \psi_\beta \rangle -\eta \langle \psi_\alpha| \psi_\beta \rangle] =\sum_{\alpha\beta}^{n_s} a_\alpha a_\beta (\tilde{H}_{\alpha\beta} -\eta \delta_{\alpha\beta})=\sum_{\alpha\beta}^{n_s} a_\alpha a_\beta h_{\alpha\beta}\ ,
\end{equation}
where $h_{\alpha\beta}$ are the matrix elements of the Hamiltonian that can be computed classically. The coefficients $a_\alpha$ are then expanded in a fixed-point representation using $K$ bits~\cite{doi:10.1021/acs.jctc.9b00402,Chang:2019,ARahman:2021ktn}, 
\begin{equation}
    a^{(z+1)}_\alpha=a^{(z)}_\alpha+\sum_{i=1}^{K}2^{i-K-z}(-1)^{\delta_{iK}}q^{\alpha}_i \ ,
\end{equation}
where $z$ is the zoom parameter. The starting point is $a_\alpha^{(z=0)}=0$, and for each consecutive value of $z$, the range of values that $a_\alpha^{(z+1)}$ is allowed to explore is reduced by a factor of $2$, centered around the previous solution $a_\alpha^{(z)}$. Now $F$ takes the following form,
\begin{equation}
    F=\sum_{\alpha,\beta}^{n_s}\sum_{i,j}^K Q_{\alpha,i;\beta,j} q^{\alpha}_i q^{\beta}_j\ , \ Q_{\alpha,i;\beta,j}=2^{i+j-2K-2z} (-1)^{\delta_{iK}+\delta_{jK}} h_{\alpha\beta} + 2 \delta_{\alpha\beta} \delta_{ij} 2^{i-K-z} (-1)^{\delta_{iK}} \sum_\gamma^{n_s} a^{(z)}_\gamma h_{\gamma\beta} \ .
\end{equation}
The iterative procedure used to improve the precision of the results is based on the value $a^{(z)}_\alpha$ obtained after $14$ zoom steps (starting from $a_\alpha^{(z_0=0)}=0$), and then launching a new annealing workflow with $z_1 \neq 0$ (e.g., $z_1=4$), with $a^{(z=z_0+14)}_\alpha$ as the starting point. After another 14 zoom steps, the final value $a^{(z=z_1+14)}_\alpha$ can be used as the new starting point for $a^{(z=z_2)}_\alpha$, with $z_2 > z_1$. This process can be repeated until no further improvement is seen in the convergence of the energy and wavefunction.

In Table~\ref{tab:QAresults2}, the difference between the exact energy of the vacuum and masses of the $\sigma$- and $\pi$-mesons and the ones computed with the QA, for each iteration of this procedure after 14 zoom steps, are given, together with the overlap of the wavefunctions $1-|\langle \Psi^{\rm exact}| \Psi^{\tt Adv.}\rangle|^2$. See also Fig.~\ref{fig:QAresults}.
\begin{table}[!ht]
\renewcommand{\arraystretch}{1}
\resizebox{\textwidth}{!}{\begin{tabular}{|| c | r | r@{\ \ \ \ \ } | r | r@{\ \ \ \ \ } | r | r@{\ \ \ \ \ } ||} 
\hline
  & \multicolumn{2}{c|}{$\ket{\Omega}$} & \multicolumn{2}{c|}{$\ket{\sigma}$} & \multicolumn{2}{c|}{$\ket{\pi}$} \\ \hline
  Step & \multicolumn{1}{c|}{$\delta E_\Omega$} & \multicolumn{1}{c|}{$1-|\langle \Psi_\Omega^{\rm exact}| \Psi_\Omega^{\tt Adv.}\rangle|^2$} & \multicolumn{1}{c|}{$\delta M_\sigma$} & \multicolumn{1}{c|}{$1-|\langle \Psi_\sigma^{\rm exact}| \Psi_\sigma^{\tt Adv.}\rangle|^2$} & \multicolumn{1}{c|}{$\delta M_\pi$} & \multicolumn{1}{c|}{$1-|\langle \Psi_\pi^{\rm exact}| \Psi_\pi^{\tt Adv.}\rangle|^2$} \\
 \hline \hline 
 0 & $4^{\,+2}_{\,-2}\times 10^{-1}$ & $10^{\,+3}_{\,-5}\times 10^{-2}$ & $4^{\,+2}_{\,-2}\times 10^{-1}$ & $11^{\,+7}_{\,-5}\times 10^{-2}$& $3^{\,+1}_{\,-1}\times 10^{-1}$ & $11^{\,+71}_{\,-4}\times 10^{-2}$\\[0.2ex]
 1 & $9^{\,+4}_{\,-3}\times 10^{-3}$ & $2^{\,+6}_{\,-5}\times 10^{-3}$  & $3^{\,+1}_{\,-1}\times 10^{-2}$ & $7^{\,+2}_{\,-2}\times 10^{-3}$ & $9^{\,+4}_{\,-3}\times 10^{-3}$ & $3^{\,+3}_{\,-1}\times 10^{-3}$\\[0.2ex]
 2 & $6^{\,+2}_{\,-2}\times 10^{-4}$ & $12^{\,+3}_{\,-5}\times 10^{-5}$ & $4^{\,+1}_{\,-1}\times 10^{-3}$ & $12^{\,+3}_{\,-4}\times 10^{-4}$& $7^{\,+2}_{\,-3}\times 10^{-4}$ & $3^{\,+2}_{\,-2}\times 10^{-4}$\\[0.2ex]
 3 & $4^{\,+1}_{\,-2}\times 10^{-5}$ & $9^{\,+3}_{\,-4}\times 10^{-6}$  & $2^{\,+1}_{\,-1}\times 10^{-4}$ & $6^{\,+1}_{\,-2}\times 10^{-5}$ & $4^{\,+2}_{\,-2}\times 10^{-5}$ & $12^{\,+6}_{\,-3}\times 10^{-6}$\\[0.2ex]
 4 & $16^{\,+6}_{\,-6}\times 10^{-7}$& $3^{\,+2}_{\,-1}\times 10^{-7}$  & $10^{\,+6}_{\,-3}\times 10^{-6}$& $9^{\,+1}_{\,-1}\times 10^{-6}$ & $7^{\,+9}_{\,-5}\times 10^{-7}$ & $8^{\,+2}_{\,-2}\times 10^{-6}$\\
 \hline
\end{tabular}}
\renewcommand{\arraystretch}{1}
\caption{Convergence of the energy, masses and wavefunctions of the three lowest-lying states in the $B=0$ sector of $1+1$D QCD with $N_f=2$ and $m=g=L=1$, between exact results from diagonalization of the Hamiltonian and those 
obtained from D-Wave's {\tt Advantage}.}
\label{tab:QAresults2}
\end{table}

\begin{figure}[!ht]
    \centering
    \includegraphics[width=\columnwidth]{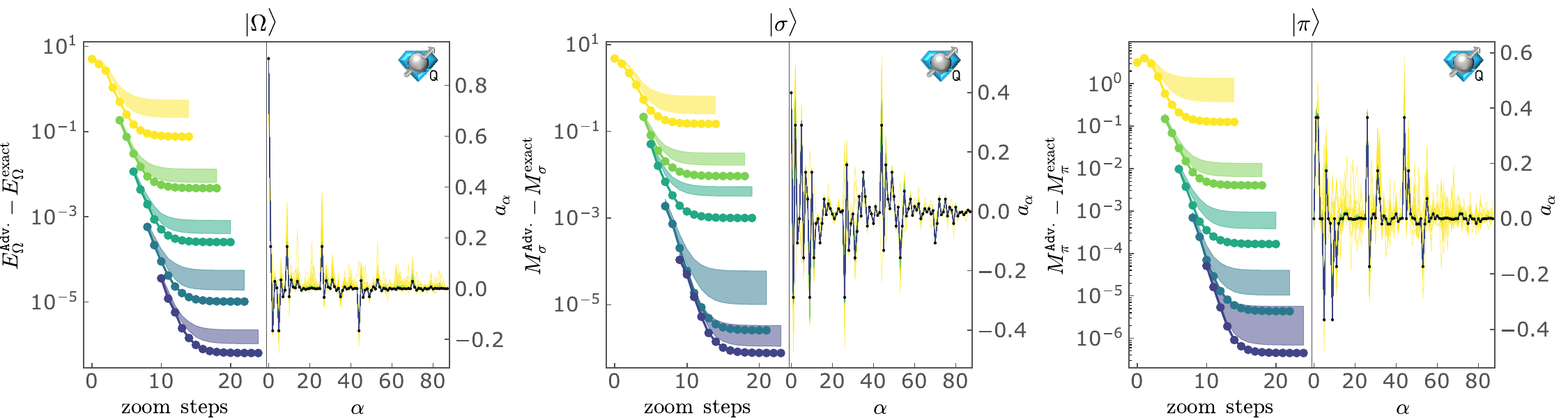}
    \caption{
    Iterative convergence of the energy, masses and wavefunctions for the three lowest-lying states in the $B=0$ sector of $1+1$D QCD with $N_f=2$ and $m=g=L=1$: vacuum (left), $\sigma$-meson (center) and $\pi$-meson (right). Compared to Fig.~\ref{fig:QAresults}, the exact wavefunctions are used when projecting the Hamiltonian to study the excited states.}
    \label{fig:QAresults_exact}
\end{figure}

Focusing on the lowest line of the last panel of Fig.~\ref{fig:QAresults}, which shows the convergence as a function of zoom steps for the pion mass, it can be seen that it displays some oscillatory behavior compared to the rest, which are smooth. This is expected, since the wavefunctions used to project out the lower eigenstates from the Hamiltonian are known with a finite precision (obtained from previous runs). For example, the vacuum state is extracted at the $10^{-6}$ precision level. Then, when looking at the excited states with increased precision (like for the pion, around $10^{-7}$), the variational principle might not hold, and the computed energy level might be below the ``true'' one (and not above). To support this argument, the same calculation has been pursued, but using the exact wavefunctions when projecting the Hamiltonian to study the excited states (instead of the ones computed using {\tt Advantage}), and no oscillatory behavior is observed, as displayed in Fig.~\ref{fig:QAresults_exact}.

\section{Quantum Circuits Required for Time Evolution by the Gauge-Field Interaction}
\label{app:circ}
\begin{figure}[!ht]
    \centering
    \includegraphics[width=8cm]{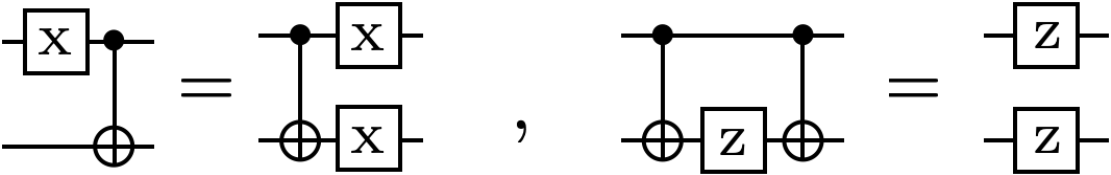}
    \caption{The $X$ and $Z$ circuit identities.}
    \label{circ:XZError}
\end{figure}

\begin{figure}
    \centering
    \includegraphics[width=\textwidth]{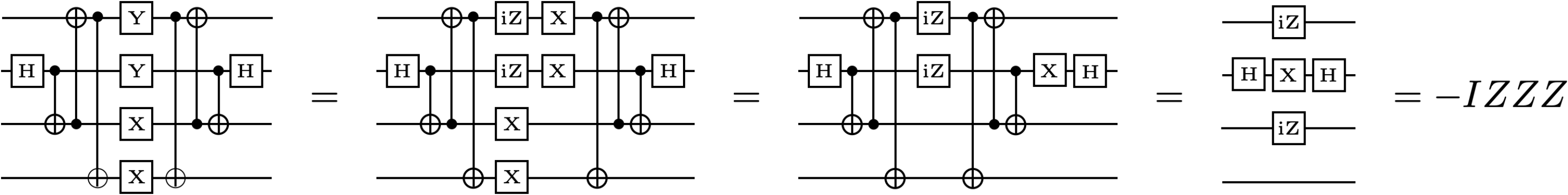}
    \caption{The diagonalization of $XXYY$ via a GHZ state-preparation circuit.}
    \label{fig:XXYYDiag}
\end{figure}

\noindent
This appendix provides more detail about the construction of the quantum circuits which implement the Trotterized time evolution of the chromo-electric terms of the Hamiltonian. 
It closely follows the presentation in the appendix of Ref.~\cite{Stetina:2020abi}.
The four-qubit interaction in $H_{el}$ has the form
\begin{equation}
\sigma^+ \sigma^- \sigma^- \sigma^+ + {\rm h.c.} = \frac{1}{8}(XXXX + XXYY + XYXY - XYYX + YXYX - YXXY +YYXX + YYYY) \ .
\label{eq:pmmpApp}
\end{equation}
Since the 8 Pauli strings are mutually commuting, they can be simultaneously diagonalized by a unitary transformation. The strategy for
identifying the quantum circuit(s) to implement this term will be to first change to a basis where every term is diagonal, then apply the diagonal unitaries and finally
return back to the computational basis.
The GHZ state-preparation circuits,
shown in Fig.~\ref{circ:GHZ}, 
diagonalize all 8 of the Pauli strings, for example,
\begin{align}
G^{\dagger} \ &( XXXX + YYXX + YXYX - YXXY - XYYX + XYXY + XXYY + YYYY) \ G \nonumber \\[4pt]
 &= \ IIZI - ZIZZ - ZZZZ + ZIZI + IZZI - IIZZ - IZZZ + ZZZI  \ .
\label{eq:GHZDiag}
\end{align}
This can be verified by using the identities that are shown in Fig.~\ref{circ:XZError} to simplify the circuits formed by conjugating each Pauli string by $G$. 
As an example, the diagonalization of $XXYY$ is displayed in Fig.~\ref{fig:XXYYDiag}. 
The first equality uses $Y = i Z X$ and the second equality uses the $X$
circuit identity to move all $X$s past the CNOTs. The third equality moves the $Z$s past
the controls of the CNOTs and uses the $Z$ circuit identity. The other Pauli strings are diagonalized in a similar manner.

It is also straightforward to show that, for example, 
\begin{equation}
    G^{\dagger}(IZZI + IZIZ + ZIIZ)G = IZII + IIIZ + ZIII \ .
    \label{eq:ZZGHZG}
\end{equation}
In general, a $ZZ$ in the computational basis becomes a single $Z$ in the GHZ basis if the state-preparation circuit has a CNOT that connects the
original two $Z$s. The two GHZ state-preparation circuits, $G$ and $\tilde{G}$, were chosen so that all $9$ of the $ZZ$ terms in Eq.~(\ref{eq:QnfQmfp}) are mapped to single qubit rotations.
Once in the GHZ basis, the diagonal unitaries are performed, e.g., $\exp(-i IZZZ)$. 
They are arranged to minimize the number of CNOTs required, and the optimal circuit layouts are shown in Fig.~\ref{circ:UpmmpZZ}.

\section{Complete Circuits for \texorpdfstring{\boldmath$N_f=1,2$}{Nf=1,2} QCD with \texorpdfstring{\boldmath$L=1$}{L=1}}
\label{app:Nf1SU3circs}
\begin{figure}[!ht]
    \centering
    \includegraphics[width=\columnwidth]{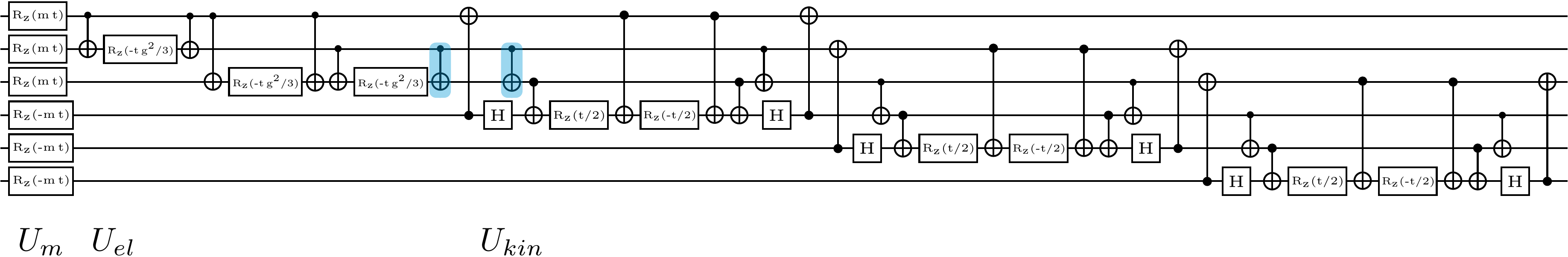}
    \caption{The complete circuit that implements a single Trotter step for $N_f=1$ QCD with $L=1$.
    }
    \label{fig:Nf1Trot}
\end{figure}

\begin{figure}[!ht]
    \centering
    \includegraphics[width=\columnwidth]{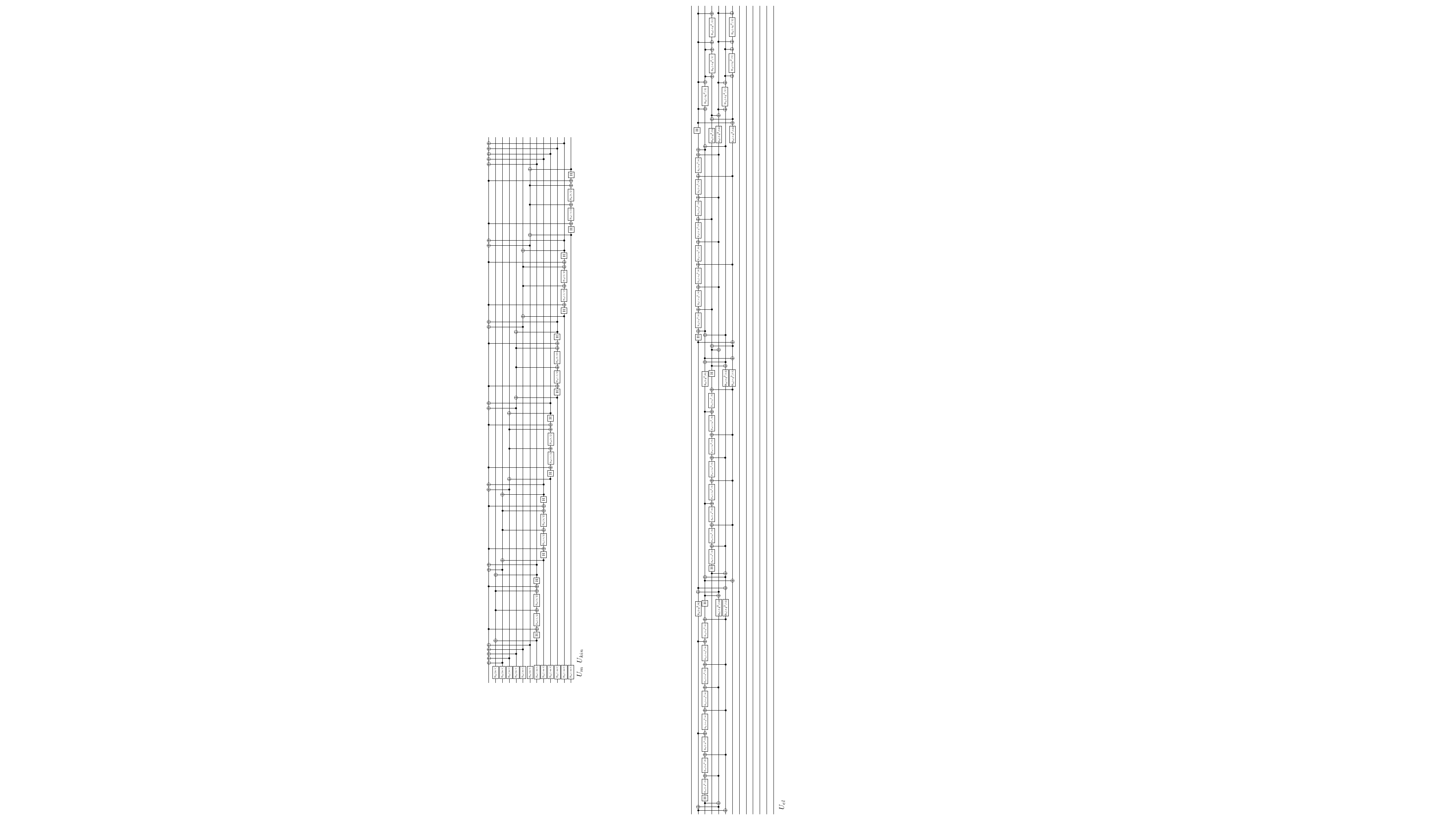}
    \caption{The complete circuit that implements a single Trotter step for $N_f=2$ QCD with $L=1$.
    }
    \label{fig:Nf2Trot}
\end{figure}

\noindent 
This appendix provides the complete set of circuits required to
implement one Trotter step for 
$N_f=1$ and $N_f=2$ QCD with $L=1$. 
The composite circuit for $N_f=1$ is shown in 
Fig.~\ref{fig:Nf1Trot} where, by ordering $U_{el}$ before $U_{kin}$, the CNOTs highlighted in blue cancel. The composite circuit for $N_f=2$ is shown in 
Fig.~\ref{fig:Nf2Trot}, 
where the ordering in the Trotterization
is $U_m$
followed by $U_{kin}$ 
and then by $U_{el}$.

%

\section{Energy Decomposition Associated with Time Evolution from the Trivial Vacuum}
\label{app:MDTD}
\begin{figure}[!ht]
    \centering
    \includegraphics[width=0.9\columnwidth]{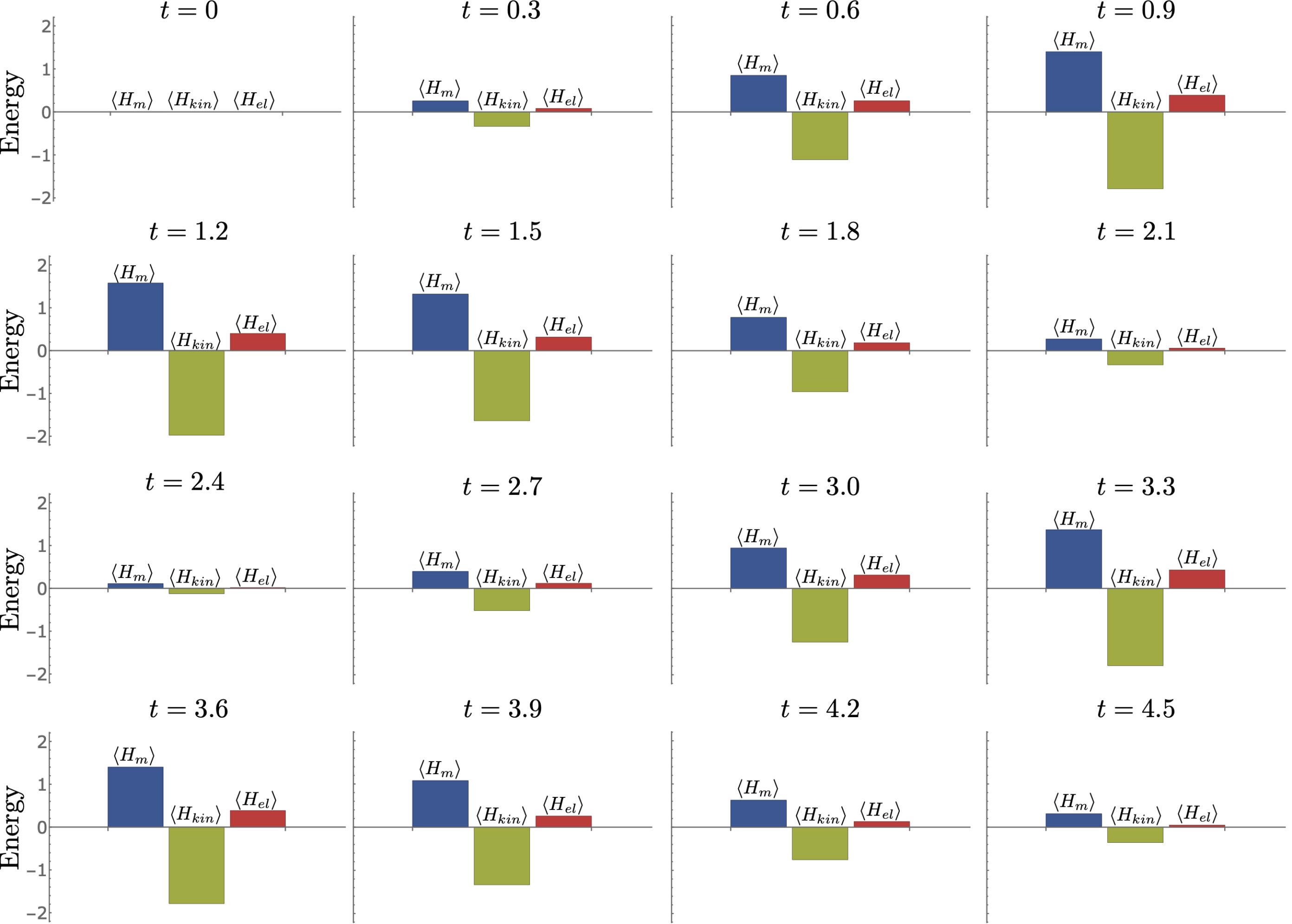}
    \caption{The time evolution of the decomposition of the energy starting from the trivial vacuum starting at $t=0$ for $N_f=2$ QCD with $m=g=L=1$.
    }
    \label{fig:Hanim}
\end{figure}

\noindent 
This appendix shows,
in Fig.~\ref{fig:Hanim}, the time evolution of the decomposition of the expectation value of the Hamiltonian starting with the trivial vacuum at $t=0$ for $N_f=2$ QCD with $m=g=L=1$. 
Notice that the sum of all three terms equals zero for all times as required by energy conservation and that the period of oscillations is the same as the period of the persistence amplitude shown in Fig.~\ref{fig:VacTo}. 
%

\section{Details on One First-Order Trotter Step of \texorpdfstring{\boldmath$N_f=1$}{Nf=1} QCD with \texorpdfstring{\boldmath$L=1$}{L=1}}
\label{app:LOTrott}
\begin{figure}[!ht]
    \centering
    \includegraphics[width=0.6\columnwidth]{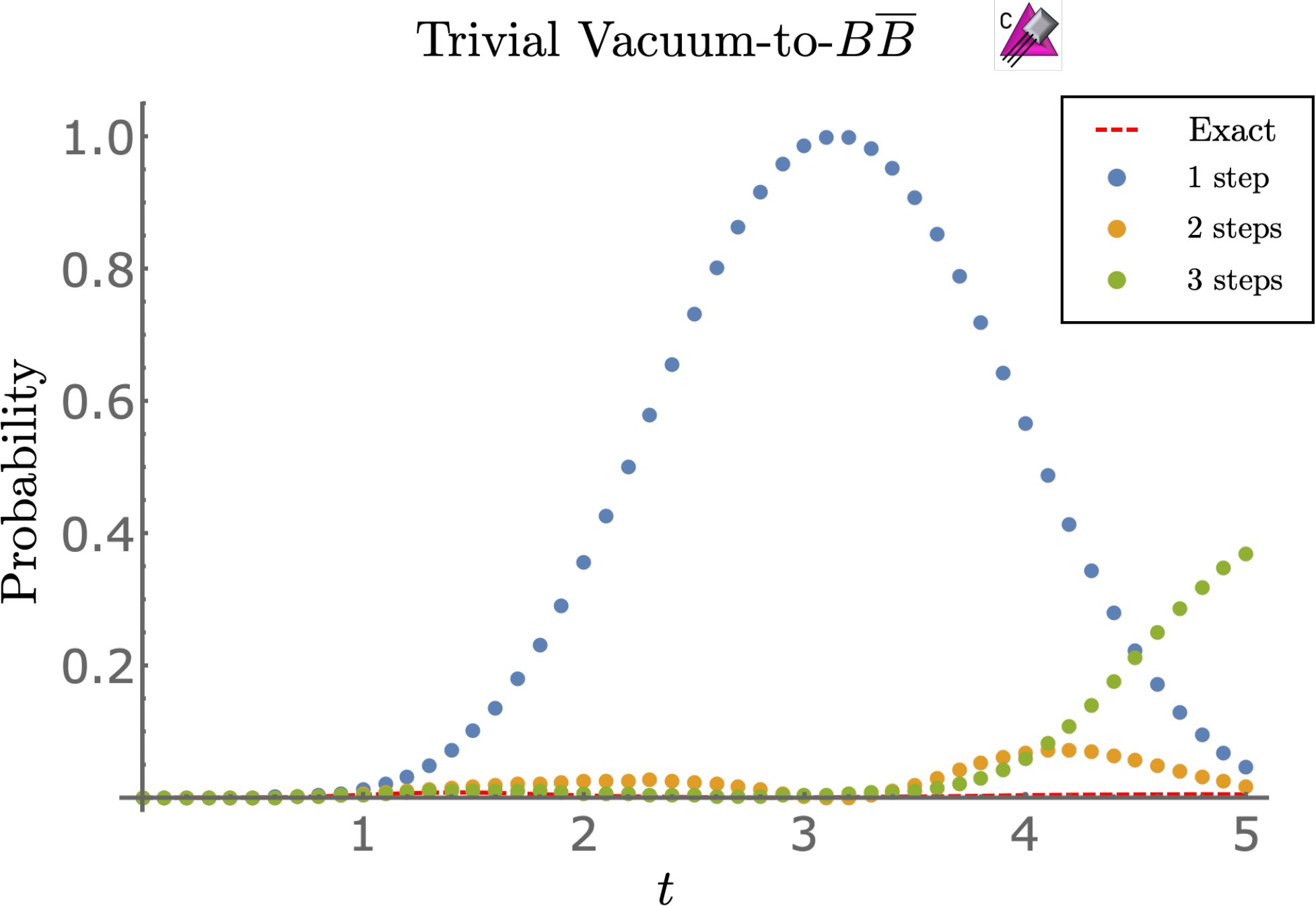}
    \caption{The trivial vacuum-to-$B\overline{B}$ probability for $1+1$D QCD with $m=g=L=1$. Shown are the results obtained from exact exponentiation of the Hamiltonian (dashed red curve) and from the Trotterized implementation with $1$, $2$ and $3$ Trotter steps.}
    \label{fig:VacToBBbar}
\end{figure}

\noindent 
This appendix discusses the theoretical expectations for one step of first-order Trotter time evolution for $N_f=1$ QCD with $L=1$.
The time evolution operator 
is decomposed
into $U_1(t) = U_{kin}(t) U_{el}(t) U_m(t)$ where the subscript ``$1$'' is to denote first-order Trotter. Both the trivial vacuum-to-vacuum and trivial vacuum-to-$q_r\overline{q}_r$ probabilities involve measurements in the computational basis where $U_m(t)$ and $U_{el}(t)$ are diagonal and have no effect. 
Thus, the time-evolution operator is effectively $U_1(t) = U_{kin}(t)$, which is exact (no Trotter errors) over a single spatial site. The trivial vacuum-to-vacuum, trivial vacuum-to-$q_r \overline{q}_r$ and trivial vacuum-to-$B \overline{B}$ probabilities are found to be,
\begin{align}
&\lvert \langle \Omega_0 \rvert e^{-i H_{kin} t} \lvert \Omega_0 \rangle\rvert ^2 = \cos^6(t/2) \ , \nonumber \\ 
&\lvert\langle q_r \overline{q}_r \rvert e^{-i H_{kin} t} \lvert \Omega_0 \rangle\rvert ^2 = \cos^4(t/2)\sin^2(t/2) \ , \nonumber \\
&\lvert\langle B \overline{B} \rvert e^{-i H_{kin} t} \lvert \Omega_0 \rangle\rvert ^2 = \sin^6(t/2) \ .
\end{align}
For large periods of the evolution, the wavefunction is dominated by $B\overline{B}$ as shown in Fig.~\ref{fig:VacToBBbar}. Exact time evolution, on the other hand, has a  small probability of $B\overline{B}$
which suggests that detecting  
$B \overline{B}$ could lead to an additional way to mitigate Trotter errors.
It is interesting that the kinetic term alone favors transitioning the trivial vacuum into color singlets on each site. This same behavior holds
for $N_f=2$ where the dominant transition is to $\Delta \Delta \overline{\Delta} \overline{\Delta}$.
\end{subappendices}

\chapter{Quantum Simulations of Weak decay in \texorpdfstring{$1+1$}{} Dimensional Quantum Chromodyanmics}
\label{chap:beta}
\noindent
\textit{This chapter is associated with Ref.~\cite{Farrell:2022vyh}: \\
``Preparations for Quantum Simulations of Quantum Chromodynamics in \texorpdfstring{$1+1$}{} Dimensions: (II) Single-Baryon \texorpdfstring{$\beta$}{}-Decay in Real Time" by Roland C. Farrell, Ivan A. Chernyshev, Sarah J. M. Powell, Nikita A. Zemlevskiy, Marc Illa and Martin J. Savage.}

 \section{Introduction}
\noindent
A quantitative exploration of hadronic decays and nuclear reaction dynamics resolved 
at very short time scales using quantum simulations will provide a new window into 
strong-interaction processes that lies beyond the capabilities of experiment.
In chemistry, the development of
femtosecond laser-pulse imaging in the 1980s~\cite{Zewail:1999}, allowed for reaction pathways to be studied in real time (for an overview, see Ref.~\cite{Durrani:2020}). 
Although a similar experimental procedure is not available for strong processes, it is expected that quantum simulations will provide analogous insight into hadronic dynamics.
Perhaps the simplest non-trivial class of such reactions to begin exploring is the
$\beta$-decay of low-lying hadrons and nuclei.
Single $\beta$-decay rates of nuclei have played a central role in defining the 
Standard Model (SM) of strong and electroweak processes~\cite{Glashow:1961tr,Higgs:1964pj,Weinberg:1967tq,Salam:1968rm}. They 
initially provided evidence that the weak (charged-current) 
quark eigenstates differ from the strong eigenstates, and, more recently, are
providing stringent tests of the unitarity of the Cabibbo-Kobayashi-Maskawa (CKM) matrix~\cite{Cabibbo:1963yz,Kobayashi:1973fv}.
For recent reviews of $\beta$-decay, see, e.g.,  Refs.~\cite{Gonz_lez_Alonso_2019,Hassan:2020hrj,Algora:2021,PhysRevC.102.045501}.
The four-Fermi operators responsible for $\beta$-decay~\cite{Feynman:1958ty} in the SM
emerge from operator production expansions (OPEs) 
of the non-local operators coming from the exchange of a charged-gauge boson ($W^-$) between quarks and leptons.
Of relevance to this work is the four-Fermi operator, which gives rise to the flavor changing quark process $d\rightarrow u e^-\overline{\nu}$.
In the absence of higher-order electroweak processes, including electromagnetism,
matrix elements of these operators factorize between the hadronic and leptonic sectors. This leaves, for example, a non-perturbative evaluation of $n\rightarrow p e^-\overline{\nu}$ for neutron decay, which is constrained significantly by the approximate global flavor symmetries of QCD.
Only recently have the observed systematics of $\beta$-decay rates of nuclei been understood without the need for phenomenological re-scalings of the axial coupling constant, 
$g_A$. 
As has long been anticipated, the correct decay rates are recovered when two-nucleon and higher-body interactions are included within the
effective field theories (EFTs) (or meson-exchange currents)~\cite{Baroni_2016,Krebs:2016rqz,Gysbers:2019uyb,Baroni_2021}. 
This was preceded by successes of EFTs in describing electroweak processes of few-nucleon systems through the inclusion of higher-body electroweak operators (not constrained by strong interactions alone), 
e.g., Refs.~\cite{Chen:1999tn,Butler:1999sv,Butler:2002cw,Baroni:2016xll,Li:2017udr,Baroni:2018fdn}.
The EFT framework describing nuclear $\beta$-decays involves contributions from ``potential-pion" and ``radiation-pion" exchanges~\cite{Kaplan:1998tg,Kaplan:1998we} 
(an artifact of a system of relativistic and non-relativistic particles~\cite{Grinstein:1997gv,Luke:1997ys})
and real-time simulations of these processes are expected to be able to isolate these distinct contributions.
Recently, 
the first Euclidean-space lattice QCD calculations of Gamow-Teller matrix elements in light nuclei 
(at unphysical light quark masses and without fully-quantified uncertainties) 
have been performed~\cite{Parreno:2021ovq}, 
finding results that are consistent with nature.

While $\beta$-decay is a well-studied and foundational area of sub-atomic physics, 
the double-$\beta$-decay of nuclei continues to present a theoretical challenge in the 
the search for physics beyond the SM.
For a recent review of the ``status and prospects" of $\beta\beta$-decay, see Ref.~\cite{Dolinski_2019}.
Although $2\nu\beta\beta$-decay is allowed in the SM, and is 
a second order $\beta$-decay process, 
$0\nu\beta\beta$-decay requires the violation of lepton number.
Strong interactions clearly play an essential role 
in the experimental detection of the $\beta\beta$-decay of nuclei, but
such contributions are non-perturbative and complex, and, for example, 
the EFT descriptions involve contributions from two- and higher-body correlated operators~\cite{Savage:1998yh,Shanahan:2017bgi,Tiburzi:2017iux,Cirigliano:2018hja,Cirigliano:2019vdj}.
The ability to study the real-time dynamics of such decay process in nuclei would likely 
provide valuable insight into the underlying strong-interaction mechanisms, and potentially offer first principles constraints beyond those from Euclidean-space lattice QCD.\footnote{
For discussions of the potential of lattice QCD to impact $\beta\beta$-decay, see, e.g., Refs.~\cite{Shanahan:2017bgi,Tiburzi:2017iux,Monge-Camacho:2019nby,Davoudi:2021noh,Cirigliano:2022oqy,USQCD:2022mmc,Detmold:2022jwu,Cirigliano:2022rmf}.}

This chapter is an extension of the previous chapter
to include flavor-changing 
weak interactions via a four-Fermi operator that generates the $\beta$-decay 
of hadrons and nuclei. The terms in the lattice Hamiltonian that generate a Majorana mass for the neutrinos are also given, although not included in the simulations.
Applying the JW mapping, it is found that a single generation of the SM (quarks and leptons) maps onto $16$ qubits per spatial lattice site. 
Using Quantinuum's {\tt H1-1} 20-qubit trapped ion quantum computer, the initial state of a baryon is both prepared and evolved with one and two Trotter steps on a single lattice site. 
Despite only employing a minimal amount of error mitigation, results at the 
$\sim 5\%$-level are obtained, consistent with the expectations.
Finally, we briefly comment on the potential of such hierarchical dynamics for error-correction purposes in quantum simulations.

\section{The \texorpdfstring{$\beta$}{Beta}-Decay Hamiltonian for Quantum Simulations in 1+1 Dimensions}
\noindent
In nature, the $\beta$-decays of neutrons and nuclei involve energy and momentum transfers related to the energy scales of nuclear forces and of isospin breaking. 
As these are much below the electroweak scale,
$\beta$-decay rates are well reproduced by matrix elements of 
four-Fermi effective interactions with $V-A$ structure~\cite{Feynman:1958ty,Sudarshan:1958vf}, of the form
\begin{equation}
    {\cal H}_\beta  = 
    \frac{G_F}{\sqrt{2}} \ V_{ud} \ 
    \overline{\psi}_u\gamma^\mu (1-\gamma_5)\psi_d\ 
    \overline{\psi}_e\gamma_\mu (1-\gamma_5)\psi_{\nu_e} 
    \ +\ {\rm h.c.}
    \ ,
    \label{eq:HbetaC}
\end{equation}
where $V_{ud}$ is the element of the CKM matrix for $d\rightarrow u$ transitions,
and $G_F$ is Fermi's coupling constant that is 
measured to be $G_F=1.1663787 (6) \times 10^{-5}~{\rm GeV}^{-2}$~\cite{Tiesinga:2021myr}.
This is the leading order (LO) SM result, obtained by matching amplitudes at 
tree-level, 
where $G_F/\sqrt{2} = g_2^2/(8 M_W^2)$ 
with $M_W$ the mass of the $W^\pm$ gauge boson
and $g_2$ the SU(2)$_L$ coupling constant.
Toward simulating the SM in $3+1$D, we consider
$1+1$D QCD containing $u$-quarks, $d$-quarks, electrons and electron neutrinos. 
For simplicity,
we model $\beta$-decay through a vector-like four-Fermi operator,
\begin{equation}
    {\cal H}_\beta^{1+1} =
    \frac{G}{\sqrt{2}} \
    \overline{\psi}_u\gamma^\mu \psi_d\ 
    \overline{\psi}_e\gamma_\mu \mathcal{C} \psi_{\nu}  
        \ +\ {\rm h.c.}
    \ ,
    \label{eq:HbetaC1}
\end{equation}
where $\mathcal{C} = \gamma_1$ is the charge-conjugation operator 
whose purpose will become clear. 
Appendices~\ref{app:betaSM} and~\ref{app:beta1p1} provide details on
calculating the single-baryon $\beta$-decay rates in the
infinite volume and continuum limits in the SM and in the $1+1$D model considered here.

The strong and weak interactions can be mapped
onto the finite-dimensional Hilbert space provided by a quantum computer
by using the Kogut-Susskind (KS) Hamiltonian formulation of
lattice gauge theory~\cite{Kogut:1974ag,Banks:1975gq}. 
The KS discretization of the fields is such that
$L$ spatial lattice sites 
are split into  $2L$ fermion sites 
that separately accommodate 
fermions (even sites) and anti-fermions (odd sites).
For the $\beta$-decay of baryons, the strong and the weak KS Hamiltonian (in axial gauge) 
has the form~\cite{Banuls:2017ena,Atas:2021ext,Farrell:2022wyt,Atas:2022dqm}
\begin{equation}
    H = H_{{\rm quarks}} + H_{{\rm leptons}} +  H_{{\rm glue}} + H_{\beta}
    \ ,
\end{equation}
where
\begin{align}
    H_{\rm{quarks}} 
    =&\ 
    \sum_{f=u,d}\left[
        \frac{1}{2 a} \sum_{n=0}^{2L-2} \left ( \phi_n^{(f)\dagger} \phi_{n+1}^{(f)}
        \ +\ {\rm h.c.} \right ) 
    \: + \: 
    m_f \sum_{n=0}^{2L-1} (-1)^{n} \phi_n^{(f)\dagger} \phi_n^{(f)} 
    \right] \ ,
    \nonumber\\
    H_{\rm{leptons}} 
    =&\ 
        \sum_{f=e,\nu}\left[
    \frac{1}{2 a} \sum_{n=0}^{2L-2} \left ( \chi_n^{(f)\dagger} \chi_{n+1}^{(f)}
        \ +\ {\rm h.c.} \right ) 
    \: + \: 
    m_f \sum_{n=0}^{2L-1} (-1)^{n} \chi_n^{(f)\dagger} \chi_n^{(f)} 
\right]  \ ,
    \nonumber\\
    H_{{\rm glue}}
    =&\ 
    \frac{a g^2}{2} 
    \sum_{n=0}^{2L-2} 
    \sum_{a=1}^8
    \left ( \sum_{m\leq n} Q^{(a)}_m \right )^2 \ ,
    \nonumber\\
    H_{{\rm \beta}}
    =&\
    \frac{G}{a \sqrt{2}} 
    \sum_{l=0}^{L-1} \bigg [
    \left (\phi_{2l}^{(u)\dagger} \phi_{2l}^{(d)} + \phi_{2l+1}^{(u)\dagger} \phi_{2l+1}^{(d)} \right ) \left (\chi_{2l}^{(e)\dagger} \chi_{2l+1}^{(\nu)} - \chi_{2l+1}^{(e)\dagger} \chi_{2l}^{(\nu)}\right ) 
    \nonumber\\
    &+
    \left ( \phi_{2l}^{(u)\dagger} \phi_{2l+1}^{(d)} + \phi_{2l+1}^{(u)\dagger} \phi_{2l}^{(d)} \right )
    \left (\chi_{2l}^{(e)\dagger} \chi_{2l}^{(\nu)} - \chi_{2l+1}^{(e)\dagger} \chi_{2l+1}^{(\nu)}\right )+
    {\rm h.c.} \bigg ] \ .
    \label{eq:KSHam1}
\end{align}
The masses of the $u$-, $d$-quarks, electron and neutrino (Dirac) are $m_{u,d,e,\nu}$,
and the strong and weak coupling constants are $g$ and $G$. The $SU(3)$ charges are
$Q_m^{(a)}$, and
$\phi^{(u,d)}_n$ are the $u$- and $d$-quark field operators (which both transform in the fundamental representation of $SU(3)$, and hence the sum over color indices has been suppressed). The electron and  neutrino field operators are
$\chi^{(e,\nu)}_n$, and for the remainder of this paper the lattice spacing, $a$, will be set to unity.
We emphasize that the absence of gluon fields is due to the choice of axial gauge, whereas the lack of weak gauge fields is due to the
consideration of a low energy effective theory in which the heavy weak gauge bosons have been integrated out. 
This results in, for example, the absence of parallel transporters in the fermion kinetic terms.
\begin{figure}[!t]
    \centering
    \includegraphics[width=15cm]{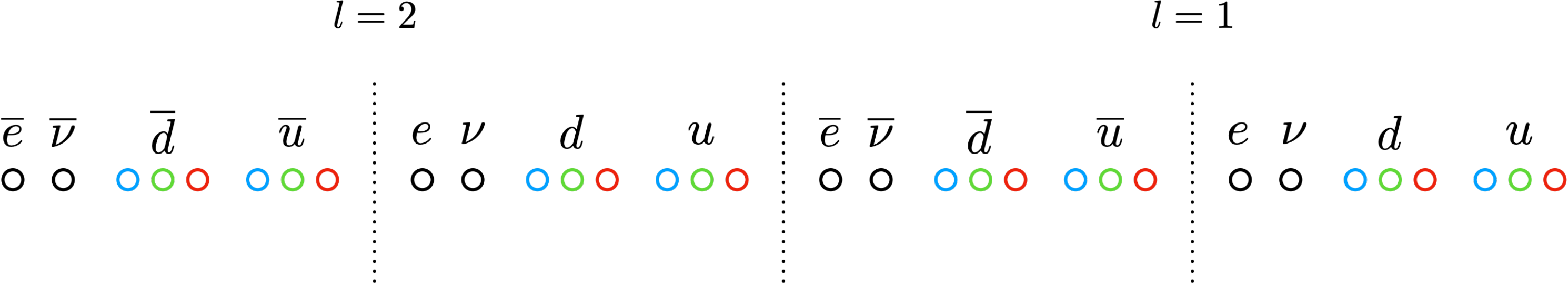}
    \caption{
    The qubit layout of a $L=2$ lattice,
    where fermions and anti-fermions are grouped together (which will be preferred if electromagnetism is included).  This layout extends straightforwardly to $L>2$.
}
    \label{fig:L1layout}
\end{figure}

The JW mapping of the Hamiltonian in Eq.~(\ref{eq:KSHam1}) to qubits, 
arranged as shown in Fig.~\ref{fig:L1layout},
is given by 
\begin{align}
    H_{\rm{quarks}} 
     \rightarrow &\
    \frac{1}{2} \sum_{l=0}^{L-1} \sum_{f=u,d}\sum_{c=0}^{2}  m_f\left ( Z_{l,f,c} - Z_{l,\overline{f},c} + 2\right )  \nonumber \\
    & -\frac{1}{2} \sum_{l=0}^{L-1} \sum_{f=u,d} \sum_{c=0}^{2}\left [ \sigma^+_{l,f,c} Z^7 \sigma^-_{l,\overline{f},c}  + (1-\delta_{l,L-1})  \sigma^+_{l,\overline{f},c} Z^7 \sigma^-_{l+1,f,c} + {\rm h.c.} \right ]\ , \nonumber \\[4pt]
    H_{\rm{leptons}} 
    \rightarrow  &\ \frac{1}{2} \sum_{l=0}^{L-1} \sum_{f=e,\nu}  m_f\left ( Z_{l,f} - Z_{l,\overline{f}} + 2\right )  \nonumber \\[4pt]
    &- \: \frac{1}{2} \sum_{l=0}^{L-1} \sum_{f=e,\nu} \left [ \sigma^+_{l,f} Z^7 \sigma^-_{l,\overline{f}}  + (1-\delta_{l,L-1})  \sigma^+_{l,\overline{f}} Z^7 \sigma^-_{l+1,f} + {\rm h.c.} \right ] \ ,
    \nonumber\\[4pt]
    H_{{\rm glue}}
    \rightarrow  &\ \frac{g^2}{2} \sum_{n=0}^{2L-2}(2L-1-n)\left( \sum_{f=u,d} Q_{n,f}^{(a)} \, Q_{n,f}^{(a)} \ + \
        2 Q_{n,u}^{(a)} \, Q_{n,d}^{(a)}
         \right)    \nonumber \\
         &+ \: g^2 \sum_{n=0}^{2L-3} \sum_{m=n+1}^{2L-2}(2L-1-m) \sum_{f=u,d} \sum_{f'=u,d} Q_{n,f}^{(a)} \, Q_{m,f'}^{(a)} \ ,
    \nonumber\\[4pt]
    H_{{\rm \beta}}
    \rightarrow  &\
    \frac{G}{\sqrt{2}}\sum_{l = 0}^{L-1}\sum_{c=0}^2\bigg ( \sigma^-_{l,\nub} Z^6 \sigma^+_{l,e} \sigma^-_{l,d,c} Z^2 \sigma^+_{l,u,c} \: - \: \sigma^+_{l,\eb} Z^8 \sigma_{l,\nu}^- \sigma_{l,d,c}^- Z^2 \sigma^+_{l,u,c} \nonumber \\
    &- \: \sigma^-_{l,\nub} Z^{2-c} \sigma^-_{l,\db,c} \sigma^+_{l,\ub,c} Z^c \sigma^+_{l,e} \: + \: \sigma^+_{l,\eb}Z^{3-c}\sigma^-_{l,\db,c} \sigma^+_{l,\ub,c} Z^{1+c} \sigma^-_{l,\nu} \nonumber \\
    &- \: \sigma^-_{l,\db,c} Z^{3+c} \sigma^+_{l,e} \sigma^-_{l,\nu} Z^{5-c} \sigma^+_{l,u,c} \: - \: \sigma^+_{l,\eb} \sigma^-_{l,\nub} \sigma^-_{l,\db,c}Z^{10}\sigma^+_{l,u,c} \nonumber \\ 
    & - \: \sigma^+_{l,\ub,c} Z^c \sigma^+_{l,e} \sigma^-_{l,\nu} Z^{2-c} \sigma^-_{l,d,c} \: - \: \sigma^+_{l,\eb} \sigma^-_{l,\nub} \sigma^+_{l,\ub,c} Z^4 \sigma^-_{l,d,c} \: + \: {\rm h.c.}\bigg )
     \ ,
\label{eq:KSHamJWmap}
\end{align}
where the sums of products of color charges are given by Eq.~\eqref{eq:QnfQmfp}.

\subsection{Efficiently Mapping the \texorpdfstring{$L=1$}{L=1} Hamiltonian to Qubits}
\begin{figure}[!t]
    \centering
    \includegraphics[width=15cm]{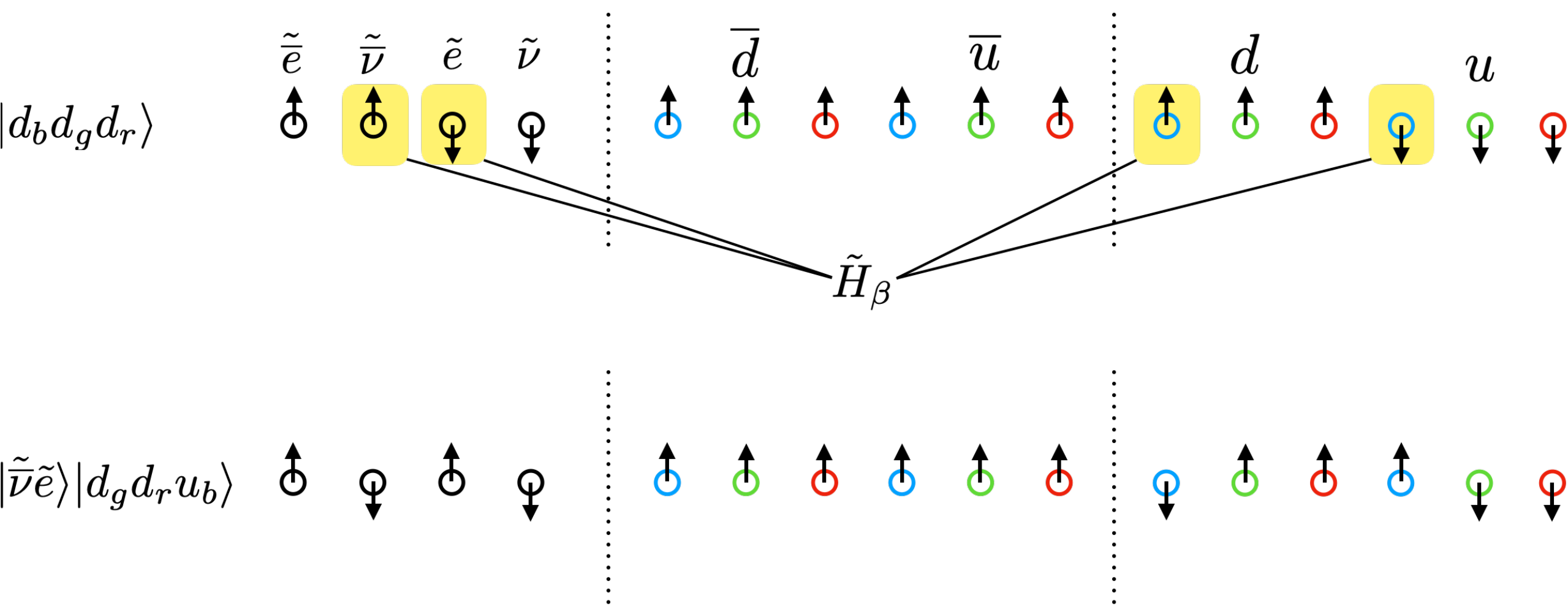}
    \caption{
    The $L=1$ lattice qubit layout of one generation of the SM that is used in this paper for quantum simulation. 
    Fermion (anti-fermion) sites are occupied when the spin is up (down), and the spins at the lepton sites represent occupation in the tilde basis.
    Specifically, 
    the example of $\ket{d_b d_g d_r}$ (upper lattice) decaying to $\ket{ \tilde{\overline{\nu}} \tilde{e}}\ket{d_g d_r u_b}$ (lower lattice) through one application of $\tilde{H}_{\beta}$ 
    in Eq.~(\ref{eq:tildeBeta}) is shown.
    }
    \label{fig:EWlayoutTilde}
\end{figure}

\noindent
To accommodate the capabilities of current devices, 
the quantum simulations performed in this work involve only a single spatial site, $L=1$, 
where the structure of the Hamiltonian can be simplified.
In particular, without interactions between leptons, it is convenient to work with field operators that create and annihilate eigenstates of the free lepton Hamiltonian, $H_{{\rm leptons}}$.
These are denoted by  ``tilde operators"~\cite{Farrell:2022wyt},
which create the open-boundary-condition (OBC) analogs of plane waves. 
In the tilde basis with the JW mapping, the lepton Hamiltonian is diagonal and becomes
\begin{equation}
    \tilde{H}_{{\rm leptons}} = \lambda_{\nu}(\tilde{\chi}^{(\nu) \dagger}_0 \tilde{\chi}^{(\nu)}_0-\tilde{\chi}^{(\nu) \dagger}_1 \tilde{\chi}^{(\nu)}_1)  + \lambda_{e}(\tilde{\chi}^{(e) \dagger}_0 \tilde{\chi}^{(e)}_0-\tilde{\chi}^{(e) \dagger}_1 \tilde{\chi}^{(e)}_1) \ \rightarrow \ \frac{\lambda_{\nu}}{2}(Z_{\nu} - Z_{\overline{\nu}})  + \frac{\lambda_{e}}{2}(Z_{e} - Z_{\overline{e}}) 
    \ ,
    \label{eq:tildeLep}
\end{equation}
where $\lambda_{\nu,e} = \frac{1}{2}\sqrt{1+4m_{\nu,e}^2}$. 
The $\beta$-decay operator in Eq.~(\ref{eq:KSHam1}) becomes
\begin{alignat}{2}
    \tilde{H}_{\beta} = \frac{G}{\sqrt{2}}\Bigg \{ \left ( \phi_0^{(u)\dagger} \phi_{0}^{(d)}  +  \phi_{1}^{(u)\dagger} \phi_{1}^{(d)} \right ) \bigg [&\frac{1}{2}(s_+^e s_-^{\nu} \ - \ s_-^e s_+^{\nu})\left (\tilde{\chi}_0^{(e)\dagger} \tilde{\chi}_{0}^{(\nu)}  + \tilde{\chi}_1^{(e)\dagger} \tilde{\chi}_1^{(\nu)}\right ) \nonumber \\
     + \ &\frac{1}{2}( s_+^e s_+^{\nu} \ + \ s_-^e s_-^{\nu})\left (\tilde{\chi}_0^{(e)\dagger} \tilde{\chi}_{1}^{(\nu)} - \tilde{\chi}_1^{(e)\dagger} \tilde{\chi}_0^{(\nu)}\right )\bigg] \nonumber \\
    + \ \left ( \phi_0^{(u)\dagger} \phi_{1}^{(d)}  + \phi_{1}^{(u)\dagger} \phi_{0}^{(d)} \right )\bigg [
    &\frac{1}{2}(s_+^e s_+^{\nu} \ -\ s_-^e s_-^{\nu})\left (\tilde{\chi}_0^{(e)\dagger} \tilde{\chi}_{0}^{(\nu)}  - \tilde{\chi}_1^{(e)\dagger} \tilde{\chi}_1^{(\nu)}\right ) \nonumber \\ 
    - \ &\frac{1}{2}(s_+^e s_-^{\nu} \ +\  s_-^e s_+^{\nu})\left (\tilde{\chi}_0^{(e)\dagger} \tilde{\chi}_{1}^{(\nu)} +  \tilde{\chi}_1^{(e)\dagger} \tilde{\chi}_0^{(\nu)}\right ) \bigg ] \ + \ {\rm h.c.} \Bigg \}  \ ,
\label{eq:HbetaTilNoJW}
\end{alignat}
where $s^{\nu,e}_\pm = \sqrt{1\pm m_{\nu,e}/\lambda_{\nu,e}}$. 
In our simulations, the initial state of the quark-lepton 
system is prepared in a strong eigenstate with baryon number $B=+1$ in the quark sector 
and the vacuum, $\lvert \Omega \rangle_{{\rm lepton}}$, in the lepton sector.
One of the benefits of working in the tilde basis is that the vacuum satisfies $\tilde{\chi}^{(e,v)}_0\lvert \Omega \rangle_{{\rm lepton}} = \tilde{\chi}^{(e,v) \dagger}_1 \lvert \Omega \rangle_{{\rm lepton}} = 0$, and the terms in the first and third lines of Eq.~(\ref{eq:HbetaTilNoJW}) do not contribute to $\beta$-decay. For the processes we are interested in, this results in an effective $\beta$-decay operator of the form
\begin{alignat}{2}
    \tilde{H}_{\beta} = \frac{G}{\sqrt{2}}\Bigg \{ \left ( \phi_0^{(u)\dagger} \phi_{0}^{(d)}  +  \phi_{1}^{(u)\dagger} \phi_{1}^{(d)} \right ) \bigg [&\frac{1}{2}( s_+^e s_+^{\nu} \ + \ s_-^e s_-^{\nu})\left (\tilde{\chi}_0^{(e)\dagger} \tilde{\chi}_{1}^{(\nu)}  - \tilde{\chi}_1^{(e)\dagger} \tilde{\chi}_0^{(\nu)}\right )\bigg] \nonumber \\
    - \ \left ( \phi_0^{(u)\dagger} \phi_{1}^{(d)} + \phi_{1}^{(u)\dagger} \phi_{0}^{(d)} \right )\bigg [  &\frac{1}{2}(s_+^e s_-^{\nu} \ +\  s_-^e s_+^{\nu})\left (\tilde{\chi}_0^{(e)\dagger} \tilde{\chi}_{1}^{(\nu)} +  \tilde{\chi}_1^{(e)\dagger} \tilde{\chi}_0^{(\nu)}\right ) \bigg ] \ + \ {\rm h.c.} \Bigg \}  \ .
\label{eq:HbetaTilNoVac}
\end{alignat}
The insertion of 
the charge-conjugation matrix,
$\mathcal{C}$, in the continuum operator, Eq.~(\ref{eq:HbetaC1}),
is
necessary to obtain a $\beta$-decay operator that does not annihilate the lepton vacuum.
To minimize the length of the string of $Z$s in the JW mapping, the lattice layout in Fig.~\ref{fig:EWlayoutTilde} is used.
In this layout, the hopping piece of $H_{{\rm quarks}}$ has only 5 $Z$s between the quark and antiquark raising and lowering operators and the $\beta$-decay operator is
\begin{align}
\!\!\!\!\!\! \tilde{H}_{\beta }
    \rightarrow
    \frac{G}{\sqrt{2}}
      \sum_{c=r,g,b} \bigg [&\frac{1}{2}( s_+^e s_+^{\nu} \ + \ s_-^e s_-^{\nu})\left ( \sigma^-_{\overline{\nu}} \sigma^+_e  \ - \  \sigma^+_{\overline{e}} Z^2 \sigma^-_{\nu} \right )\left (\sigma^-_{d,c}Z^2 \sigma^+_{u,c} \: + \: \sigma^-_{\overline{d},c} Z^2 \sigma^+_{\overline{u},c}\right ) \nonumber \\
     - \ &\frac{1}{2}(s_+^e s_-^{\nu} \ +\  s_-^e s_+^{\nu})\left (\sigma^-_{\overline{\nu}} \sigma^+_e \ + \  \sigma^+_{\overline{e}} Z^2 \sigma^-_{\nu} \right )\left ( \sigma^-_{\overline{d},c} Z^8 \sigma^+_{u,c} \: + \: \sigma^+_{\overline{u},c} Z^2 \sigma^-_{d,c} \right ) \ + \  {\rm h.c.} \bigg ]  \ .
     \label{eq:tildeBeta}
\end{align}
In total, the $L=1$ system requires $16$ ($12$ quark and $4$ lepton) 
qubits. See App.~\ref{app:fullHam} for the complete $L=1$ Hamiltonian in terms of qubits.

\subsection{A Majorana Mass for the Neutrino}
\noindent
Although not relevant to the simulations performed in Sec.~\ref{sec:BetaSim}, it is of current interest to consider the inclusion of a Majorana mass term for the neutrinos.
A Majorana mass requires and induces the violation of lepton number by
$|\Delta L| = 2$, and is not present in the minimal SM, defined by dim-4 operators.
However, the Weinberg operator~\cite{Weinberg:1979sa} enters at dim-5 and generates an effective Majorana mass for the neutrinos,
\begin{align}
\!\!\!\!\!\!\!\!\!\!\!\!
{\cal L}^{{\rm Weinberg}} 
&= \frac{1}{ 2\Lambda} 
\left( \overline{L}^c \epsilon \phi \right)
\left(\phi^T \epsilon L \right)
    \ +\ {\rm h.c.}
\ ,\ \ 
L = \left( \nu , e \right)^T_L
\ ,\ \ 
\phi = \left( \phi^+ , \phi^0 \right)^T
\ ,\ \ 
\langle \phi \rangle = \left( 0 , v/\sqrt{2} \right)^T
\ ,\ \ \epsilon = i\sigma_2
\ \ ,
\nonumber\\
&\rightarrow
-\frac{v^2}{4\Lambda}  \overline{\nu}^c_L \nu_L
    \ +\ {\rm h.c.}
\ +\ .... 
\label{eq:maja}
\end{align}
where $\phi$ is the Higgs doublet, 
$L^c$ denotes the charge-conjugated left-handed lepton doublet,
$v$ is the Higgs vacuum expectation value and $\Lambda$ is a high energy scale characterizing physics beyond the SM. 
The ellipsis denote interaction terms involving components of the Higgs doublet fields and the leptons.
This is the leading contribution beyond the minimal SM, 
but does not preclude contributions from other sources.
On a $1+1$D lattice there is only a single 
$\lvert \Delta L \rvert = 2$ local operator 
with the structure of a mass term
and, using the JW mapping along with the qubit layout in Fig.~\ref{fig:L1layout}, is of the form
\begin{equation}
H_{\rm Majorana} =
\frac{1}{2} m_M 
     \sum_{n={\rm even}}^{2L-2} 
\left( 
\chi_n^{(\nu)}
\chi_{n+1}^{(\nu)} \:
+ \: {\rm h.c.}
\right)
 \ \rightarrow
 \frac{1}{2} m_M  \sum_{l = 0}^{L-1} 
\left( 
\sigma^+_{l,\nu}\ 
Z^7
\sigma^+_{l,\nub} \: + \: {\rm h.c.}
\right)
\ .
\label{eq:majaHami}
\end{equation}
While the operator has support on a single spatial lattice site, it does not contribute to 
$0\nu\beta \beta$-decay on a lattice with only a single spatial site.   
This is because the processes that it could potentially induce, such as
$\Delta^-\Delta^-\rightarrow \Delta^0\Delta^0 e^- e^-$,
are Pauli-blocked by the single electron site.  
At least two spatial sites are required for any such process producing two electrons in the final state.

\section{Quantum Simulations of the \texorpdfstring{$\beta$}{Beta}-Decay of One Baryon on One Lattice Site}
\label{sec:BetaSim}
\noindent
In this section, quantum simulations of the $\beta$-decay of a single baryon are performed
in $N_f=2$ flavor QCD with $L=1$ spatial lattice site.
The required quantum circuits to perform one and two Trotter steps of time evolution were developed and run on the Quantinuum {\tt H1-1} $20$ qubit trapped ion quantum computer and its simulator {\tt H1-1E}~\cite{QuantHoney,h1-1e}.

\subsection{Preparing to Simulate \texorpdfstring{$\beta$}{Beta}-Decay}
\label{sec:BetaSimA}
\noindent
It is well known that, because of confinement, the  energy eigenstates (asymptotic states) of QCD 
are color-singlet hadrons, which are composite objects of quarks and gluons.
On the other hand, 
the operators responsible for $\beta$-decay, given in Eq.~(\ref{eq:tildeBeta}), 
generate transitions between $d$- and $u$-quarks.
As a result,  observable effects of $\tilde{H}_{\beta}$, in part, 
are found in transitions between 
hadronic states whose matrix elements depend on the distribution of the quarks within. 
Toward quantum simulations of the $\beta$-decay of neutrons and nuclei more generally, 
the present work focuses on the decay of a single baryon.

Generically, three elements are required for real-time quantum simulations of 
the $\beta$-decay of baryons:
\begin{enumerate}
    \item 
    Prepare the initial hadronic state that will subsequently undergo $\beta$-decay. 
    In this work, this is one of the single-baryon states (appropriately selected in the spectrum) 
    that is an eigenstate of the strong Hamiltonian alone, 
    i.e., the weak coupling constant is set equal to $G=0$.
    \item 
    Perform (Trotterized) time-evolution using the full ($G\neq0$) Hamiltonian.
    \item 
    Measure one or more of the lepton qubits. 
    If leptons are detected, then $\beta$-decay has occurred.
\end{enumerate}
In $1+1$D, Fermi statistics 
preclude the existence of a light isospin $I=1/2$ nucleon,
and the lightest baryons are in an $I=3/2$ multiplet 
$(\Delta^{++}, \Delta^+, \Delta^0, \Delta^-)$
(using the standard electric charge assignments of the up and down quarks). 
We have  chosen to simulate the decay 
$\Delta^- \to \Delta^0 + e + \overline{\nu}$, which, at the quark level, involves 
baryon-interpolating operators with the quantum numbers of
$ddd\rightarrow udd$.

In order for $\beta$-decay to be kinematically allowed, 
the input-parameters of the theory must be such that 
$M_{\Delta^-} > M_{\Delta^0} + M_{\overline{\nu}} + M_{e}$. 
This is accomplished through tuning the parameters of the Hamiltonian.
The degeneracy in the iso-multiplet is lifted 
by using different values for the  up and down quark masses. 
It is found that the choice of parameters, $m_u=0.9$, $m_d=2.1$, $g=2$ and $m_{e,\nu} = 0$ 
results in the desired hierarchy of baryon and lepton masses. 
The relevant part of the spectrum, obtained from an exact diagonalization of the Hamiltonian, 
is shown in Table~\ref{tab:betaMass}. Although kinematically allowed, multiple instances of $\beta$-decay cannot occur for $L=1$ as there can be at most one of each (anti)lepton.
\begin{table}[!ht]
\renewcommand{\arraystretch}{1.2}
\begin{tabularx}{0.7\textwidth}{||c | Y ||} 
\hline
\multicolumn{2}{||c||}{Energy of states relevant for $\beta$-decay (above the vacuum)} \\
 \hline
 State & Energy Gap\\
 \hline\hline
 $\Delta^{++}$ & 2.868 \\ 
 \hline
 $\Delta^{++}$ + $2 l$ & 3.868\\
 \hline
 $\Delta^+$ & 4.048 \\
 \hline
 $\Delta^{++}$ + $4l$  & 4.868\\
 \hline
 $\Delta^{+}$ + $2l$  & 5.048\\
 \hline
 $\Delta^0$ & 5.229 \\
 \hline
 $\Delta^{+}$ + $4l$& 6.048\\
 \hline
 $\Delta^0$ + $2l$ & 6.229 \\
 \hline
 $\Delta^-$ & 6.409 \\
 \hline
\end{tabularx}
\caption{
The energy gap above the vacuum of states relevant for $\beta$-decays of single baryons 
with $m_u = 0.9$, $m_d = 2.1$, $g=2$ and $m_{e,\nu} = 0$. The leptons are degenerate in energy and collectively denoted by $l$.}
\renewcommand{\arraystretch}{1}
\label{tab:betaMass}
\end{table}
Note that even though $m_{e,\nu} = 0$, 
the electron and neutrino are gapped due to the finite spatial volume.

\begin{figure}[!ht]
    \centering
    \includegraphics[width=\textwidth]{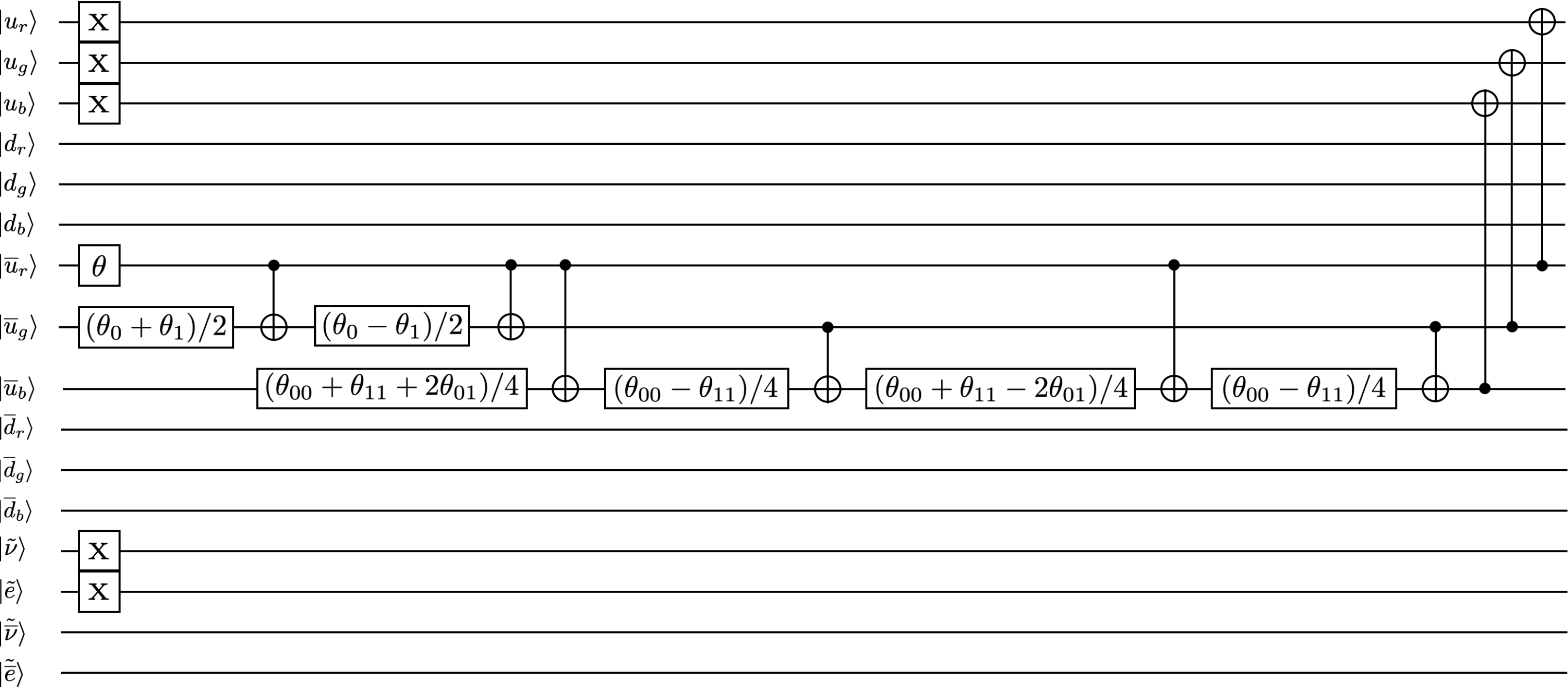}
    \caption{
    A quantum circuit that can be used to prepare the $\Delta^-$-baryon on $L=1$ spatial site.}
    \label{fig:DMVQE}
\end{figure}

To prepare the $\Delta^-$ initial state, 
we exploit the observation made in the previous chapter,
that the stretched-isospin eigenstates of the $\Delta$-baryons, 
with third component of isospin $I_3 = \pm 3/2$,
factorize between the $u$ and $d$ flavor sectors for $L=1$. 
Therefore, the previously developed 
Variational Quantum Eigensolver (VQE)~\cite{Peruzzo_2014} 
circuit~\cite{Farrell:2022wyt} used to prepare the one-flavor vacuum can be used to initialize the two-flavor $\Delta^-$ wave function.
This is done by initializing the vacuum in the lepton sector, 
preparing the state $\ket{d_r d_g d_b}$ in the $d$-sector, 
and applying the VQE circuit to produce the $u$-sector vacuum. 
In the tilde basis, the lepton vacuum is the unoccupied state (trivial vacuum), 
and the complete state-preparation circuit is shown in Fig.~\ref{fig:DMVQE}, 
where $\theta$ is shorthand for RY($\theta$). 
The rotation angles are related by
\begin{align}
    &\theta_0 = -2 \sin^{-1} \left[ \tan(\theta/2) \, \cos(\theta_{1}/2)  \right] \ , \nonumber \\ 
    &\theta_{00} = -2 \sin^{-1}\left[ \tan(\theta_{0}/2) \, \cos(\theta_{01}/2) \right] \ , \nonumber \\ 
    &\theta_{01} = -2 \sin^{-1}\left[ \cos(\theta_{11}/2)\, \tan(\theta_{1}/2) \right]
    \label{eq:angleconst1}
\end{align}
and, for $m_u = 0.9$ and $g=2$,\footnote{
The $\overline{u}$ and $u$ parts of the lattice are separated by a fully packed $d$ sector which implies that the part of the wavefunctions with odd numbers of anti-up quarks have relative minus signs compared to the one-flavor vacuum wavefunction.
}
\begin{equation}
    \theta = 0.2256 \ \ , \ \ \theta_1 = 0.4794 \ \ , \ \ \theta_{11} = 0.3265 \ .
\end{equation}
In total, state preparation requires the application of $9$ CNOT gates.

\begin{figure}[!ht]
    \centering
    \includegraphics[width=12cm]{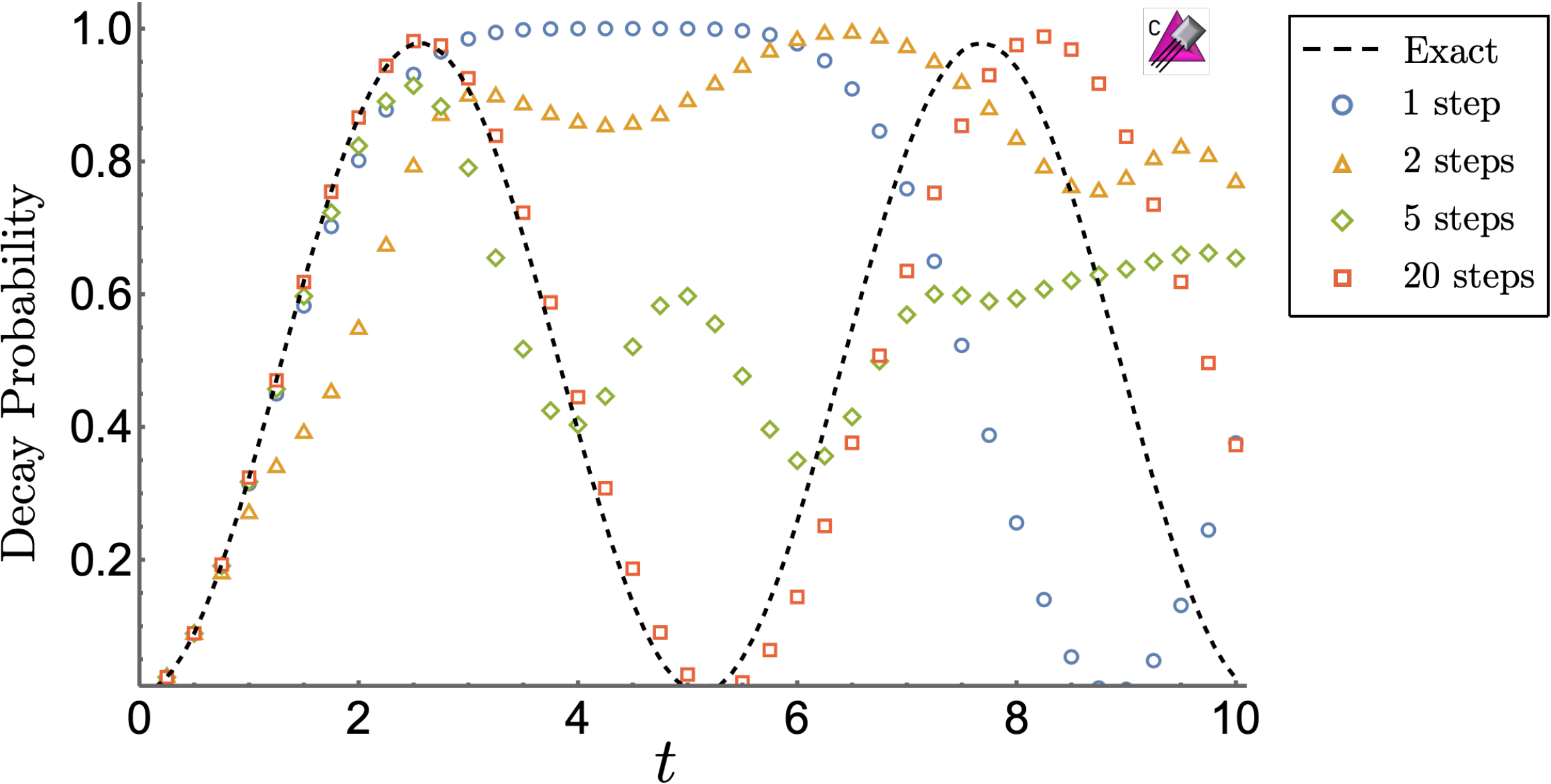}
    \caption{
The probability of $\beta$-decay, $\Delta^- \to \Delta^{0} + e + \overline{\nu}$, with $m_u = 0.9$, $m_d=2.1$, $m_{e,{\nu}} = 0$, $g=2$ and $G=0.5$ computed via exact diagonalization (dotted black line) and on the {\tt qiskit} quantum simulator~\cite{gadi_aleksandrowicz_2019_2562111} using $1,2,5,20$ Trotter steps.}
    \label{fig:BetaDecay}
\end{figure}

Once the $\Delta^-$ baryon state has been initialized 
on the register of qubits, it is then evolved in time with the full Hamiltonian.
The quantum circuits that implement the Trotterized time-evolution 
induced by $H_{{\rm quarks}}$ and $H_{{\rm glue}}$ were previously developed in
Ref.~\cite{Farrell:2022wyt}, 
where it was found that, by using an ancilla, each Trotter step 
can be implemented using 114 CNOTs. 
The lepton Hamiltonian, $\tilde{H}_{{\rm leptons}}$, has just single $Z$s which are Trotterized with single qubit rotations.
The circuits required to implement a Trotter step from $\tilde{H}_{\beta}$ are similar to those developed in Ref.~\cite{Farrell:2022wyt}, 
and their construction is outlined in App.~\ref{app:BetaCircuits}.
For the present choice of parameters, 
the main contribution to the initial ($\Delta^-$) wave function 
is $\lvert d_b d_g d_r \rangle$, 
i.e., the quark configuration associated with the ``bare" baryon in the d-sector and the trivial vacuum in the u-sector.
This implies that the dominant contribution to the $\beta$-decay 
is from the $\phi_0^{(u)\dagger} \phi_0^{(d)} \tilde{\chi}_0^{(e)\dagger} \tilde{\chi}_{1}^{(\nu)}$ 
term\footnote{Note that the $\phi_0^{(u)\dagger} \phi_0^{(d)} \tilde{\chi}_1^{(e)\dagger} \tilde{\chi}_{0}^{(\nu)}$ term is suppressed since the lepton vacuum in the tilde basis satisfies $\tilde{\chi}_1^{(e,\nu)\dagger} \lvert \Omega\rangle_{{\rm lep}} =  \tilde{\chi}_{0}^{(e,\nu)} \lvert \Omega\rangle_{{\rm lep}} = 0$.}  in Eq.~(\ref{eq:HbetaTilNoJW}), 
which acts only on valence quarks, and the $\beta$-decay operator can be approximated by 
\begin{equation}
    \tilde{H}_{\beta }^{{\rm val}}
    =
    \frac{G}{\sqrt{2}}
     \left (\sigma^-_{\overline{\nu}} \sigma^+_e  \sum_{c=r,g,b}\sigma^-_{d,c}Z^2 \sigma^+_{u,c}  + {\rm h.c.} \right  ) \ ,
     \label{eq:tildeBetaRed}
\end{equation}
for these parameter values. See App.~\ref{app:betaFull} for details on the validity of this approximation.
All of the results presented in this section implement this interaction, 
the Trotterization of which requires 50 CNOTs. 
Notice that, if the Trotterization of $\tilde{H}_{\beta}^{{\rm val}}$ is placed at the end of the first Trotter step, then\\
${U(t) = \exp(-i \tilde{H}_{\beta}^{{\rm val}} t) \times \exp \left [ -i (\tilde{H}_{{\rm leptons}} + H_{{\rm quarks}} + H_{{\rm glue}})t\right ]}$ and the initial exponential (corresponding to strong-interaction evolution) 
can be omitted as it acts on an eigenstate (the $\Delta^-$). 
This reduces the CNOTs required for one and two Trotter steps to $50$ and  $214$, respectively.
For an estimate of the number of CNOTs required to time evolve with the $\beta$-decay Hamiltonian on larger lattices see App.~\ref{app:LongJW}.
The probability of $\beta$-decay, 
as computed both through exact diagonalization of the Hamiltonian 
and through Trotterized time-evolution using the {\tt qiskit} classical simulator~\cite{gadi_aleksandrowicz_2019_2562111}, 
is shown in Fig.~\ref{fig:BetaDecay}.
The periodic structure is a finite volume effect, and the probability of 
$\beta$-decay is expected to tend to an exponential in time as $L$ increases, 
see App.~\ref{app:beta1p1aL}.

\begin{figure}[!ht]
    \centering
    \includegraphics[width=10cm]{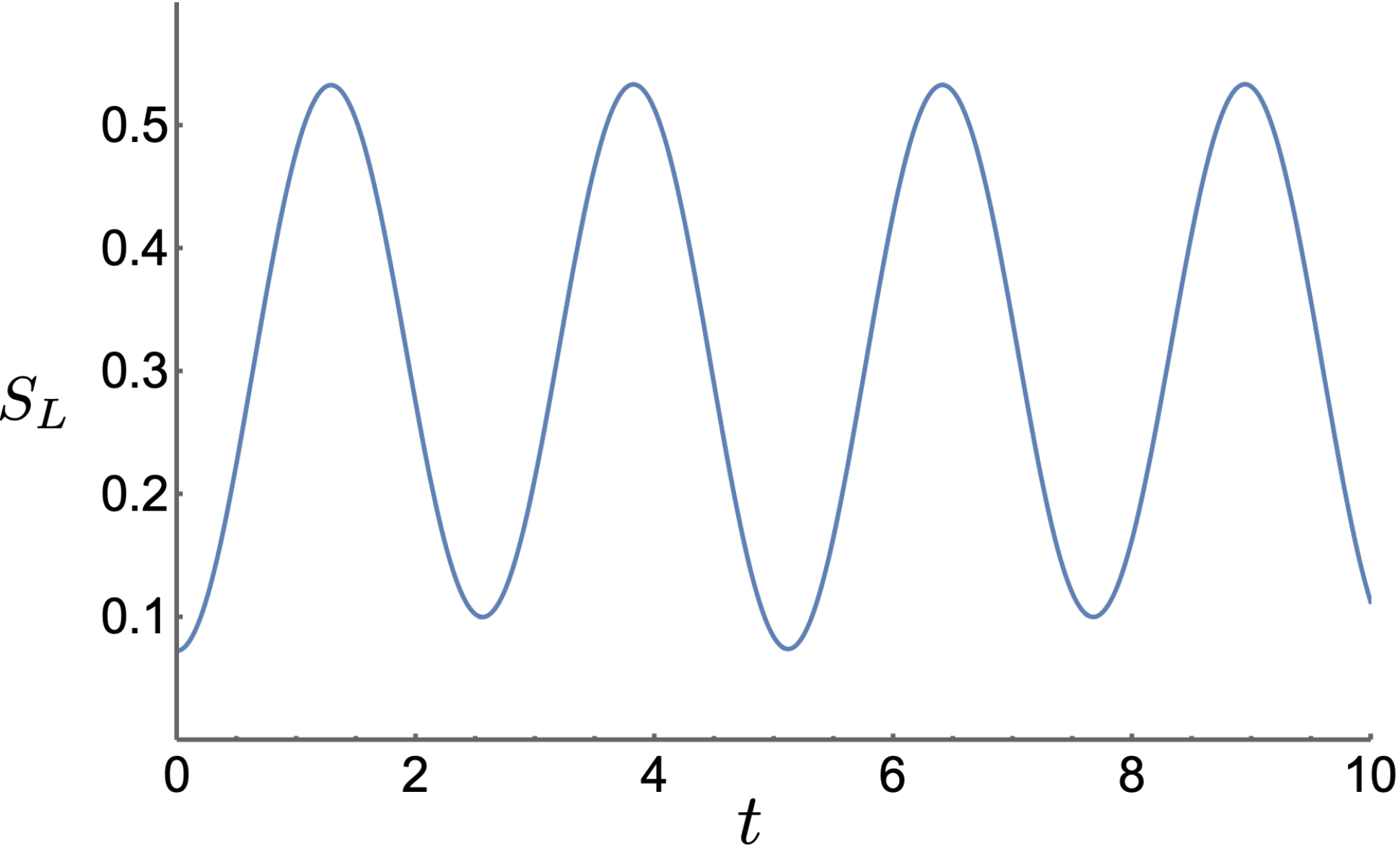}
    \caption{
    The linear entanglement entropy, $S_L$, between quarks and antiquarks plus leptons during the $\beta$-decay of an initial $\Delta^-$-baryon.}
    \label{fig:linEnt}
\end{figure}

Entanglement in quantum simulations of lattice gauge theories 
is a growing area of focus, 
see, e.g., Refs.~\cite{Ghosh:2015iwa,Soni:2015yga,Panizza:2022gvd,Rigobello:2021fxw}, and 
it is interesting to examine the evolution of entanglement during the $\beta$-decay process.
Before the decay, the quarks and antiquarks are together in a pure state as the leptons are in the vacuum, and subsequent time evolution of the state introduces components into the wavefunction that have non-zero population of the lepton states.
One measure of entanglement is the linear entropy,  
\begin{equation}
    S_L = 1 - \Tr[\rho_q^2]
    \ ,
\end{equation}
between the quarks and antiquarks plus leptons.
It is constructed by  tracing the full density matrix, $\rho$, 
over the antiquark and lepton sector to form the reduced density matrix 
$\rho_q = \Tr_{\overline{q}, {\rm leptons}} [\rho]$.
Figure~\ref{fig:linEnt} shows the linear entropy computed through exact diagonalization 
of the Hamiltonian with the parameters discussed previously in the text. 
By comparing with the persistence probability in Fig.~\ref{fig:BetaDecay}, 
it is seen that the entanglement entropy evolves at twice the frequency 
of the $\beta$-decay probability. 
This is because $\beta$-decay primarily transitions the baryon between the ground state of the $\Delta^-$ and $\Delta^0$. 
It is expected that these states will have a comparable amount of entanglement, 
and so the entanglement is approximately the same when the decay probabilities are $0$ and $1$. 
While this makes this particular example somewhat uninteresting, it does demonstrate that when multiple final states are accessible, the time-dependence of the entanglement structure might be revealing.

\subsection{Simulations Using Quantinuum's {\tt H1-1} 20 Qubit Trapped Ion Quantum Computer}
\label{sec:BetaSimB}
\noindent
Both the initial state preparation and one and two steps of Trotterized time evolution were executed
using Quantinuum's {\tt H1-1} 20 qubit trapped ion quantum computer~\cite{QuantHoney} and its simulator {\tt H1-1E}\footnote{The classical simulator {\tt H1-1E} includes depolarizing gate noise, leakage errors, crosstalk noise and dephasing noise due to transport and qubit idling~\cite{h1-1e}.} (for details on the specifications of {\tt H1-1}, see App.~\ref{app:H1specs}).
After transpilation onto the native gate set of {\tt H1-1}, a single Trotter step requires 59 $ZZ$ gates, while two Trotter steps requires 212 $ZZ$ gates.\footnote{The number of $ZZ$ gates could be further reduced by 5 by not resetting the ancilla.}
By post-selecting results on ``physical" states with baryon number $B=1$ and lepton number $L=0$ 
to mitigate single-qubit errors (e.g., Ref.~\cite{Klco:2019evd}),
approximately 90\% (50\%) of the total events from the one (two) Trotter step circuit remained. Additionally, for the two Trotter step circuit, results were selected where the ancilla qubit was in the $|0\rangle$ state (around 95\%).\footnote{For this type of error, the mid-circuit measurement and re-initialization option available for {\tt H1-1} could have been used to identify the case where the bit-flip occurred after the ancilla was used and the error had no effect on the final results.}

\begin{figure}[!t]
    \centering
    \includegraphics[width=\textwidth]{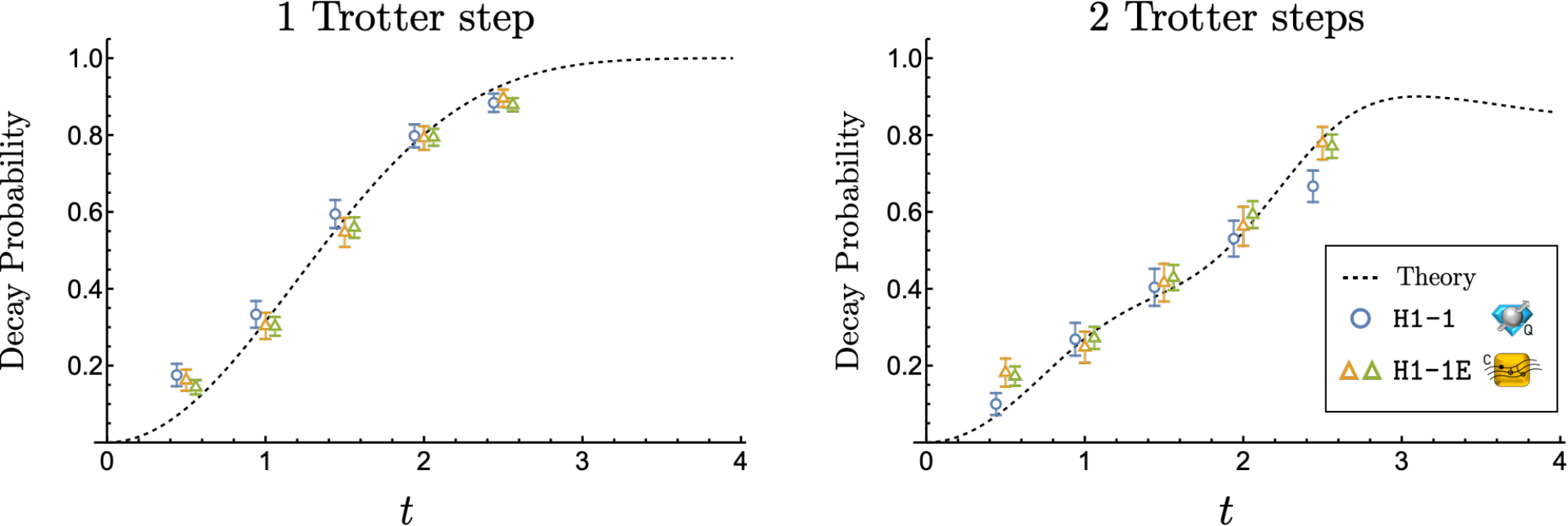}
    \caption{
    The probability of $\beta$-decay, $\Delta^- \to \Delta^{0} + e + \overline{\nu}$, with $m_u = 0.9$, $m_d=2.1$, $m_{e,{\nu}} = 0$, $g=2$ and $G=0.5$, using one (left panel) and two (right panel) Trotter steps (requiring 59 and 212 $ZZ$ gates, respectively), as given in Table~\ref{tab:H1results}.
    The dashed-black curves show the expected result from Trotterized time evolution, corresponding to the blue circles (orange triangles) in Fig.~\ref{fig:BetaDecay} for one (two) Trotter steps.
    The blue circles correspond to the data obtained on the {\tt H1-1} machine, 
    and the orange (green) triangles to the {\tt H1-1E} emulator, 
    each obtained from 200 shots (400 shots). The points have been shifted slightly along the $t$-axis for clarity.
    Error mitigation beyond physical-state post-selection has not been performed.
    The weak Hamiltonian in the time-evolution responsible for the decay is given in Eq.~(\ref{eq:tildeBetaRed}).
    }
    \label{fig:BetaDecayH1}
\end{figure}

The results of the simulations 
are shown in Fig.~\ref{fig:BetaDecayH1} and given in Table~\ref{tab:H1results}. 
By comparing the results from {\tt H1-1} and {\tt H1-1E} (using 200 shots) it is seen that the simulator is able to faithfully reproduce the behavior of the quantum computer.
The emulator was also run with 400 shots and clearly shows convergence to the expected value, verifying that the agreement between data and theory was not an artifact due to low statistics (and large error bars). 
Compared with the results presented in Ref.~\cite{Farrell:2022wyt} that were performed using IBM's {\tt ibmq$\_$jakarta} and {\tt ibm$\_$perth}, 
error mitigation techniques were not 
applied to the present simulations due to the overhead in resource requirements.
Specifically, Pauli twirling, dynamical decoupling, decoherence renormalization and measurement error mitigation 
were not performed. This is practical because the two-qubit gate, state preparation and measurement (SPAM) errors are an order of magnitude smaller on Quantinuum's trapped ion system 
compared to those of IBM's superconducting qubit systems (and a similar error rate on the single-qubit gates)~\cite{Pelofske:2022vyy}. 
\begin{table}[!ht]
\renewcommand{\arraystretch}{1.2}
\resizebox{\textwidth}{!}{\begin{tabularx}{1.1\textwidth}{||c | Y | Y | Y | Y | Y | Y | Y | Y ||}
\hline
\multicolumn{9}{||c||}{
Single-Baryon Decay Probabilities using Quantinuum's {\tt H1-1} and {\tt H1-1E}} \\
\hline
 & \multicolumn{4}{c|}{1 Trotter step} & \multicolumn{4}{c||}{2 Trotter steps} \\
 \hline
 $t$ & 
 {\tt H1-1} & {\tt H1-1E} & \makecell{{\tt H1-1E} \\($\times 2$ stats)} & Theory & 
 {\tt H1-1} & {\tt H1-1E} & \makecell{{\tt H1-1E} \\($\times 2$ stats)} & Theory
 \\
 \hline
 0.5 & 
 0.175(29) & 0.162(28) & 0.144(19) & 0.089 & 
 0.100(29) & 0.182(37) & 0.173(25) & 0.088\\
 \hline
 1.0 & 
 0.333(35) & 0.303(34) & 0.302(25) & 0.315 & 
 0.269(43) & 0.248(41) & 0.272(29) & 0.270 \\
 \hline
 1.5 & 
 0.594(37) & 0.547(38) & 0.559(27) & 0.582 & 
 0.404(48) & 0.416(49) & 0.429(33)  & 0.391 \\
 \hline
 2.0 & 
 0.798(30) & 0.792(30) & 0.794(22) & 0.801 & 
 0.530(47) & 0.563(51) & 0.593(35) & 0.547 \\
 \hline
 2.5 & 
 0.884(24) & 0.896(23) & 0.879(17) & 0.931 & 
 0.667(41) & 0.779(43) & 0.771(30) & 0.792 \\
 \hline
\end{tabularx}}
\renewcommand{\arraystretch}{1}
\caption{
The probability of $\beta$-decay, 
$\Delta^- \to \Delta^{0} + e + \overline{\nu}$, on $L=1$ spatial lattice with $m_u = 0.9$, $m_d=2.1$, $m_{e,{\nu}} = 0$, $g=2$ and $G=0.5$.
These simulations
were performed using Quantinuum's {\tt H1-1} and {\tt H1-1E} and included the initial state preparation and subsequent time evolution under 1 and 2 Trotter steps. The results are displayed in Fig.~\ref{fig:BetaDecayH1}. 
The columns labeled ($\times 2$ stats) were obtained using 400 shots, compared to the rest, that used 200 shots, and
uncertainties were computed assuming the results follow a binomial distribution.
}
\label{tab:H1results}
\end{table}
%

\section{Speculation about Quantum Simulations with a Hierarchy of Length Scales}
\noindent
It is interesting to consider how a hierarchy of length scales, 
as present in the SM, may be helpful in error correction.
In the system we have examined, the low energy strong sector is composed of mesons, baryons and nuclei, with both color singlet and non-singlet excitations (existing at higher energies).
As observed in Ref.~\cite{Farrell:2022wyt}, OBCs allow for 
relatively low-energy colored ``edge" states to exist near the boundary of the lattice.
The energy of a color non-singlet grows linearly with its distance 
from the boundary, leading to a force on colored objects.
This will cause colored errors in the bulk to migrate to the edge of the lattice where they could be detected and possibly removed.
This is one benefit of using axial gauge, where Gauss's law is automatically enforced, 
and a colored ``error" in the bulk generates a color flux tube that extends to the boundary.

Localized two-bit-flip errors  can create color-singlet 
excitations that do not experience a force toward the boundary, but which are
vulnerable to weak decay. 
For sufficiently large lattices,  color singlet excitations will decay weakly down to stable states
enabled by the near continuum of lepton states. 
In many ways, this resembles the
quantum imaginary-time evolution (QITE)~\cite{Kamakari:2021nmf,Hubisz:2020vhx,Turro:2021vbk}
algorithm, which is a special case of coupling to open systems,
where quantum systems are driven into their ground state by embedding them in a larger system that acts as a heat reservoir.
One can speculate that, in the future, quantum simulations of QCD 
will benefit from also including electroweak interactions as a mechanism to cool the strongly-interacting sector from particular classes of errors.

This particular line of investigation is currently at a ``schematic'' level, and significantly more work is required to quantify its utility.
Given the quantum resource requirements, it is likely that the Schwinger model will 
provide a suitable system to explore such scenarios.

\section{Summary and Conclusions}
\noindent
Quantum simulations of SM physics is in its infancy and, for practical reasons, has been previously
limited to either QCD or QED in one or two spatial dimensions. 
In this chapter, we have started the integration of the electroweak sector into quantum simulations of QCD by examining the time-evolution of the $\beta$-decay of one baryon.
In addition to the general framework that allows for 
simulations of arbitrary numbers of lattice sites in one dimension, 
we present results for $L=1$ spatial lattice site, which requires 16 qubits.
Explicitly, this work considered quantum simulations of 
$\Delta^-\rightarrow\Delta^0 e \overline{\nu}$
in two flavor $1+1$D QCD for $L=1$ spatial lattice site.
Simulations were performed using Quantinuum's {\tt H1-1} 20-qubit trapped ion quantum computer
and classical simulator {\tt H1-1E}, 
requiring 17 (16 system and 1 ancilla) qubits. 
Results were presented for both one and two Trotter steps, including the state preparation of the initial baryon, requiring 59 and 212 two-qubit gates respectively.
Even with 212 two-qubit gates, {\tt H1-1} provided results that 
are consistent with the expected results, within uncertainties, without error-mitigation beyond physical-state post selection.  
While not representative of $\beta$-decay in the continuum, 
these results demonstrate the potential of quantum simulations to determine 
the real-time evolution of decay and reaction processes in nuclear and 
high-energy processes.
High temporal-resolution studies of the evolution of the quarks and gluons
during hadronic decays and nuclear reactions
are expected to provide new insights into the mechanisms responsible for these processes, 
and lead to new strategies for further reducing systematic errors in their prediction.

\clearpage
\begin{subappendices}

\section{The Complete Spin Hamiltonian for \texorpdfstring{$L=1$}{L=1}}
\label{app:fullHam}
\noindent
After the JW mapping of the Hamiltonian to qubits, and using the tilde-basis for the leptons, 
the four contributing terms are
\begin{subequations}
    \label{eq:H2flavL11}
    \begin{align}
    H = & \ H_{{\rm quarks}}\ +\ \tilde{H}_{{\rm leptons}}\ +\ H_{{\rm glue}} \ +\ \tilde{H}_{\beta} ,\\[4pt]
    H_{{\rm quarks}}= & \ \frac{1}{2} \left [ m_u\left (Z_0 + Z_1 + Z_2 -Z_6 - Z_7 - Z_8 + 6\right )+ m_d\left (Z_3 + Z_4 + Z_5 -Z_9 - Z_{10} - Z_{11} + 6\right ) \right ] \nonumber \\
    &-\, \frac{1}{2} (\sigma^+_6 Z_5 Z_4 Z_3 Z_2 Z_1 \sigma^-_0 + \sigma^-_6 Z_5 Z_4 Z_3 Z_2 Z_1 \sigma^+_0 + \sigma^+_7 Z_6 Z_5 Z_4 Z_3 Z_2 \sigma^-_1 + \sigma^-_7 Z_6 Z_5 Z_4 Z_3 Z_2 \sigma^+_1  \nonumber \\
    &+\, \sigma^+_8 Z_7 Z_6 Z_5 Z_4 Z_3 \sigma^-_2 + \sigma^-_8 Z_7 Z_6 Z_5 Z_4 Z_3 \sigma^+_2 + \sigma^+_9 Z_8 Z_7 Z_6 Z_5 Z_4 \sigma^-_3 + \sigma^-_9 Z_8 Z_7 Z_6 Z_5 Z_4 \sigma^+_3 \nonumber \\
    &+\, \sigma^+_{10} Z_9 Z_8 Z_7 Z_6 Z_5 \sigma^-_4 + \sigma^-_{10} Z_9 Z_8 Z_7 Z_6 Z_5 \sigma^+_4 + \sigma^+_{11} Z_{10} Z_9 Z_8 Z_7 Z_6 \sigma^-_5 + \sigma^-_{11} Z_{10} Z_9 Z_8 Z_7 Z_6 \sigma^+_5 ) \ ,
        \label{eq:Hkin2flavL11}\\[4pt]
    \tilde{H}_{{\rm leptons}} = & \ \frac{1}{4}\sqrt{1 +4 m_e^2}(Z_{13} - Z_{15}) + \frac{1}{4}\sqrt{1 +4 m_{\nu}^2}(Z_{12} - Z_{14}) \nonumber \\[4pt]
    H_{{\rm glue}} = & \ \frac{g^2}{2} \bigg [ \frac{1}{3}(3 - Z_1 Z_0 - Z_2 Z_0 - Z_2 Z_1) + \sigma^+_4\sigma^-_3\sigma^-_1\sigma^+_0  + \sigma^-_4\sigma^+_3\sigma^+_1\sigma^-_0   \nonumber \\
    & + \, \sigma^+_5Z_4\sigma^-_3\sigma^-_2Z_1\sigma^+_0 + \sigma^-_5Z_4\sigma^+_3\sigma^+_2Z_1\sigma^-_0 \, +\, \sigma^+_5\sigma^-_4\sigma^-_2\sigma^+_1 + \sigma^-_5\sigma^+_4\sigma^+_2\sigma^-_1  \nonumber \\
    & +\,\frac{1}{12}\left (2 Z_3 Z_0 + 2Z_4 Z_1 + 2Z_5 Z_2 - Z_5 Z_0 - Z_5 Z_1 - Z_4 Z_2 - Z_4 Z_0 - Z_3 Z_1  - Z_3 Z_2 \right ) \bigg ] \ ,
    \label{eq:Hel2flavL11}\\[4pt] 
    \tilde{H}_{\beta} = & \ \frac{G}{\sqrt{2}}  \bigg \{ \frac{1}{2}( s_+^e s_+^{\nu} + s_-^e s_-^{\nu})\big [(\textcolor{blue}{\sigma^-_{14} \sigma^+_{13}} -  \sigma^+_{15} Z_{14} Z_{13}\sigma^-_{12})\big (\textcolor{blue}{\sigma^-_{3} Z_2 Z_1 \sigma^+_0 + \sigma^-_{4} Z_3 Z_2 \sigma^+_1 + \sigma^-_5 Z_4 Z_3 \sigma^+_2}   \nonumber \\
    &+ \sigma^-_{9} Z_8 Z_7 \sigma^+_6+ \ \sigma^-_{10} Z_9 Z_8 \sigma^+_7 + \sigma^-_{11} Z_{10} Z_9 \sigma^+_8) \ + \ (\textcolor{blue}{\sigma^+_{14} \sigma^-_{13}} - \sigma^-_{15} Z_{14} Z_{13}\sigma^+_{12})\big (\textcolor{blue}{\sigma^+_{3} Z_2 Z_1 \sigma^-_0 }  \nonumber \\
    & + \textcolor{blue}{\sigma^+_{4} Z_3 Z_2 \sigma^-_1 + \sigma^+_5 Z_4 Z_3 \sigma^-_2 }+ \ \sigma^+_{9} Z_8 Z_7 \sigma^-_6 + \sigma^+_{10} Z_9 Z_8 \sigma^-_7  + \sigma^+_{11} Z_{10} Z_9 \sigma^-_8)\big ]  \nonumber \\ 
    &- \ \frac{1}{2}(s_+^e s_-^{\nu}  +  s_-^e s_+^{\nu}) \big [ (\sigma^-_{14} \sigma^+_{13} +  \sigma^+_{15} Z_{14} Z_{13}\sigma^-_{12}) \big( \sigma^-_{9} Z_8 Z_7 Z_6 Z_5 Z_4 Z_3 Z_2 Z_1 \sigma^+_0  \nonumber \\
    &+ \sigma^-_{10} Z_9 Z_8 Z_7 Z_6 Z_5 Z_4 Z_3 Z_2 \sigma^+_1+ \ \sigma^-_{11} Z_{10} Z_9 Z_8 Z_7 Z_6 Z_5 Z_4 Z_3 \sigma^+_2 + \sigma^+_{6} Z_5 Z_4 \sigma^-_3 \nonumber \\
    &+ \sigma^+_{7} Z_6 Z_5 \sigma^-_4 + \sigma^+_{8} Z_7 Z_6 \sigma^-_5 \big ) \nonumber \\
    &+ \ (\sigma^+_{14} \sigma^-_{13} + \sigma^-_{15} Z_{14} Z_{13}\sigma^+_{12}) \big( \sigma^+_{9} Z_8 Z_7 Z_6 Z_5 Z_4 Z_3 Z_2 Z_1 \sigma^-_0 + \sigma^+_{10} Z_9 Z_8 Z_7 Z_6 Z_5 Z_4 Z_3 Z_2 \sigma^-_1 \nonumber \\
    &+ \  \sigma^+_{11} Z_{10} Z_9 Z_8 Z_7 Z_6 Z_5 Z_4 Z_3 \sigma^-_2 + \sigma^-_{6} Z_5 Z_4 \sigma^+_3 + \sigma^-_{7} Z_6 Z_5 \sigma^+_4 + \sigma^-_{8} Z_7 Z_6 \sigma^+_5 \big ) \big ] \bigg \} \ .
    \end{align}
\end{subequations}
In the mapping, the qubits are indexed right-to-left and, 
for example, qubit zero (one) corresponds to a red (green) up-quark.
The terms highlighted in blue provide the leading contribution to the $\beta$-decay process 
for the parameters used in the text and make up the operator used for the simulations performed in Sec.~\ref{sec:BetaSim}.

\section{\texorpdfstring{$\beta$}{Beta}-Decay in the Standard Model}
\label{app:betaSM}
\noindent 
To put our simulations in $1+1$D into context, 
it is helpful to  outline relevant aspects of single-hadron $\beta$-decays 
in the SM in $3+1$D.
Far below the electroweak symmetry-breaking scale,  
charged-current interactions can be included as an infinite 
set of effective operators in a systematic EFT description, ordered by their contributions in powers of low-energy scales divided by appropriate powers of $M_W$.
For instance, $\beta$-decay rates between hadrons scale as 
$\sim \Lambda (G_F \Lambda^2 )^2 (\Lambda / M_W )^n$, 
where $\Lambda$ denotes the low-energy scales,
$\frac{G_F}{\sqrt{2}} = \frac{g_2^2}{8 M_W^2}$ is Fermi's constant and 
LO (in $\Lambda / M_W$)
corresponds to $n=0$.
By matching operators at LO in SM interactions, the $\beta$-decay of the neutron is induced by an effective Hamiltonian density of the form~\cite{Feynman:1958ty,Sudarshan:1958vf}
\begin{equation}
    {\cal H}_\beta = 
    \frac{G_F}{\sqrt{2}} \ V_{ud} \ 
    \overline{\psi}_u\gamma^\mu (1-\gamma_5)\psi_d\ 
    \overline{\psi}_e\gamma_\mu (1-\gamma_5)\psi_{\nu_e}  
    \ +\ {\rm h.c.}
    \ ,
    \label{eq:betaHamiSM}
\end{equation}
where $V_{ud}$ is the element of the CKM matrix for $d\rightarrow u$ transitions.
As ${\cal H}_\beta $ factors into contributions from lepton and quark operators, the matrix element factorizes into a plane-wave lepton contribution and a non-perturbative hadronic component requiring matrix elements of the quark operator between hadronic states. 
With the mass hierarchies and symmetries in nature, 
there are two dominant form factors, so that,
\begin{equation}
    \langle p(p_p) | \overline{\psi}_u\gamma^\mu (1-\gamma_5)\psi_d | n(p_n)\rangle = 
    \overline{U}_p \left[\ g_V(q^2) \gamma^\mu - g_A(q^2) \gamma^\mu \gamma_5\ \right] U_n \ ,
\end{equation}
where $q$ is the four-momentum transfer of the process, $g_V(0) = 1$ in the isospin limit and $g_A(0)=1.2754(13)$~\cite{Workman:2022ynf} as measured in experiment.
The matrix element for $n\rightarrow p e^- \overline{\nu}_e$
calculated from the Hamiltonian in Eq.~(\ref{eq:betaHamiSM})
is
\begin{equation}
    \lvert \mathcal{M} \rvert^2 = 16 G_F^2 \lvert V_{ud}\rvert^2 M_n M_p (g_V^2 + 3 g_A^2)(E_{\nu}E_e + \frac{g_V^2-g_A^2}{g_V^2 + 3 g_A^2} {\bf p}_e \cdot {\bf p}_{\nub}) \ ,
\end{equation}
which leads to a neutron width of 
(at LO in $(M_n-M_p)/M_n$ and $m_e/M_n$)
\begin{equation}
    \Gamma_n =
    \frac{G_F^2 |V_{ud}|^2 (M_n-M_p)^5}{60\pi^3}
    \ \left( g_V^2 + 3 g_A^2 \right)\ f^\prime(y) \ ,
\end{equation}
where $f^\prime(y)$ is a phase-space factor, 
\begin{equation}
    f^\prime (y) = \sqrt{1-y^2}\left(1 - \frac{9}{2}y^2 - 4 y^4\right)
    \ -\ \frac{15}{2}y^4 \log\left[ \frac{y}{\sqrt{1-y^2}+1}\right] \ ,
\end{equation}
and $y=m_e/(M_n-M_p)$.
Radiative effects, recoil effects and other higher-order contributions have been neglected.

\section{\texorpdfstring{$\beta$}{Beta}-Decay in \texorpdfstring{$1+1$}{1+1} Dimensions: The \texorpdfstring{$L=\infty$}{L to infinity} and Continuum Limits}
\label{app:beta1p1}
\noindent 
In $1+1$D, the fermion field has dimensions 
$\left[\psi\right] = \frac{1}{2}$, 
and a four-Fermi operator has dimension 
$[\hat \theta ] = 2$.  
Therefore, while in $3+1$D 
$\left[G_F \right] = -2$, 
in $1+1$D, the coupling has dimension $\left[G \right] = 0$.
For our purposes, to describe the $\beta-$decay of a $\Delta^-$-baryon in $1+1$D,
we have chosen to work with an effective Hamiltonian of the form
\begin{equation}
    {\cal H}_\beta^{1+1} \ = \ 
    \frac{G}{\sqrt{2}} \
    \overline{\psi}_u\gamma^\mu \psi_d\ 
    \overline{\psi}_e\gamma_\mu\psi_{\overline{\nu}} 
    \ +\ {\rm h.c.}
    \ =\ 
    \frac{G}{\sqrt{2}} \
    \overline{\psi}_u\gamma^\mu \psi_d\ 
    \overline{\psi}_e\gamma_\mu \mathcal{C} \psi_{\nu} 
        \ +\ {\rm h.c.}
\ ,
\label{eq:betaHami1p1}
\end{equation}
where we have chosen the basis
\begin{equation}
    \gamma_0 \ = \ 
    \left(
    \begin{array}{cc}
    1&0 \\ 0&-1
    \end{array}
    \right)
    \ \ ,\ \ 
    \gamma_1 \ = \ 
    \left(
    \begin{array}{cc}
    0&1 \\ -1&0
    \end{array}
    \right)
    \ =\ \mathcal{C}
    \ ,\ 
    \gamma_0\gamma_\mu^\dagger \gamma_0\ =\ \gamma_\mu
    \ ,\ 
    \gamma_0 \mathcal{C} ^\dagger \gamma_0\ =\ \mathcal{C}
    \ \ ,\ \ 
    \{\gamma_\mu , \gamma_\nu \} \ =\ 2 g_{\mu\nu}
    \ .
\label{eq:gammaMats1p1}
\end{equation}
For simplicity, the CKM matrix element is set equal to unity 
as only one generation of particles is considered.

In the limit of exact isospin symmetry, which we assume to be approximately valid in this appendix, 
the four $\Delta$ baryons form an isospin quartet 
and can be embedded in a tensor $T^{abc}$ (as is the case for the $\Delta$ resonances in nature)
as 
$T^{111}=\Delta^{++}$,
$T^{112}=T^{121}=T^{211}=\Delta^{+}/\sqrt{3}$,
$T^{122}=T^{221}=T^{212}=\Delta^{0}/\sqrt{3}$,
$T^{222}=\Delta^{-}$.
Matrix elements of the isospin generators
are reproduced by an effective operator of the form
\begin{equation}
    \overline\psi_q \gamma^\mu \tau^\alpha \psi_q \ \rightarrow \
    3 \overline{T}_{abc} \gamma^\mu \left(\tau^\alpha \right)^c_d T^{abd} \ ,
\label{eq:DeltaIgens}
\end{equation}
which provides a Clebsch-Gordan coefficient for isospin raising operators,
\begin{equation}
    \overline\psi_q \gamma^\mu \tau^+ \psi_q \ \rightarrow \
    \sqrt{3}\  \overline{\Delta^{++}}\gamma^\mu\Delta^+ 
    \ +\ 2 \ \overline{\Delta^{+}}\gamma^\mu\Delta^0
    \ +\ 
    \sqrt{3}\  \overline{\Delta^{0}}\gamma^\mu\Delta^- \ .
\label{eq:DeltaIgens2}
\end{equation}

The matrix element for $\beta$-decay factorizes at LO in the electroweak interactions.
The hadronic component of the matrix element is given by 
\begin{align}
\langle \Delta^0(p_0) | \overline{\psi}_u\gamma^\alpha \psi_d | \Delta^-(p_-)\rangle & = \sqrt{3} g_V(q^2) \ \overline{U}_{\Delta^0}  \gamma^\alpha  U_{\Delta^-} \ =\ H^\alpha \ , \nonumber\\
H^\alpha H^{\beta\ \dagger} & = 3 |g_V(q^2) |^2 {\rm Tr}\left[ \gamma^\alpha \left( \pslash_- + M_{\Delta^-}\right) \gamma^\beta 
\left( \pslash_0 + M_{\Delta^0}\right) \right] \nonumber\\
& = 6 |g_V(q^2) |^2  
\left[p_-^\alpha p_0^\beta  + p_0^\alpha p_-^\beta  - g^{\alpha\beta} (p_-\cdot p_0)
\ +\ M_{\Delta^-} M_{\Delta^0} g^{\alpha\beta}\right] \ =\ H^{\alpha\beta} \ ,
\end{align}
and the leptonic component of the matrix element is given by, assuming that the electron and neutrino are massless, 
\begin{align}
\langle e^- \overline{\nu}_e | \overline{\psi}_e \gamma^\alpha C \psi_{\nu}  | 0\rangle 
& = \overline{U}_e\gamma^\alpha C V_\nu\ =\ L^\alpha \ ,
\nonumber\\
L^\alpha L^{\beta\ \dagger} & = {\rm Tr}\left[\ 
\gamma^\alpha C \pslash_\nu C \gamma^\beta \pslash_e \right] \ =\  {\rm Tr}\left[\ \gamma^\alpha \overline{\pslash}_\nu \gamma^\beta \pslash_e \right] \nonumber\\
& = 
2 \left[ \overline{p}_\nu^\alpha p_e^\beta + \overline{p}_\nu^\beta p_e^\alpha  -  g^{\alpha\beta} (\overline{p}_\nu\cdot p_e) \right] \ =\ L^{\alpha\beta}\ ,
\end{align}
where $p = (p^0, +p^1)$ and 
$\overline{p}=(p^0, -p^1)$.
Therefore, the squared matrix element of the process is
\begin{equation}
    |{\cal M}|^2 \ = \
    \frac{G^2}{2}
    H^{\alpha\beta} L_{\alpha\beta} \ = \ 
    12 G^2 g_V^2  M_{\Delta^-} 
    \left( M_{\Delta^-} - 2 E_{\overline{\nu}}\right)
    \left( E_e E_{\overline{\nu}} - {\bf p}_e\cdot {\bf p}_{\overline{\nu}} \right) \ ,
\end{equation}
from which
the delta decay width can be determined by standard methods,
\begin{align}
\Gamma_{\Delta^-} & =
\frac{1}{2M_{\Delta^-}}
\int 
\frac{d{\bf p}_e}{4\pi E_e}
\frac{d{\bf p}_{\overline{\nu}}}{4\pi E_{\overline{\nu}}}
\frac{d{\bf p}_0}{4\pi E_0} (2\pi)^2 \delta^2(p_- - p_0 - p_e - p_{\overline{\nu}})
|{\cal M}|^2 
\nonumber\\
& =
3 \frac{G^2 g_V^2}{2\pi}
\int\  dE_e\  dE_{\overline{\nu}}\ \delta(Q - E_e - E_{\overline{\nu}})
\ +\ {\cal O}\left(Q^n/M_\Delta^n\right)
\nonumber\\
& =
3 \frac{G^2 g_V^2 Q}{2\pi}
\ +\ {\cal O}\left(Q^n/M_\Delta^n\right)
\ ,
\label{eq:1p1decaywidth}
\end{align}
where $Q= M_{\Delta^-} - M_{\Delta^0}$ and we have retained only the leading terms in
an expansion in $Q/M_\Delta$ and evaluated the vector form factor at $g_V(q^2=0) \equiv g_V$. 
The electron and neutrino masses have been set to zero, and the inclusion of non-zero masses will lead to a phase-space factor, $f_1$, 
reducing the width shown in Eq.~(\ref{eq:1p1decaywidth}), 
and which becomes $f_1=1$ in the massless limit.

\section{\texorpdfstring{$\beta$}{Beta}-Decay in \texorpdfstring{$1+1$}{1+1} Dimensions: Finite \texorpdfstring{$L$}{L} and Non-zero Spatial Lattice Spacing}
\label{app:beta1p1aL}
\noindent 
The previous appendix computed the $\beta$-decay rate 
in $1+1$D in infinite volume and in the continuum.
However, lattice  calculations of such processes will necessarily be performed with a non-zero lattice spacing and a finite number of lattice points.
For calculations done on a Euclidean-space lattice, significant work has been done to develop the machinery used to extract physically meaningful results. 
This formalism was initially pioneered by L\"{u}scher~\cite{Luscher:1985dn,Luscher:1986pf,Luscher:1990ux} for hadron masses and two-particle scattering, and has been extended to more complex systems relevant to electroweak processes (Lellouch-L\"{u}scher)~\cite{Lellouch:2000pv,Detmold:2004qn,Christ:2005gi,Kim:2005gf,Hansen:2012tf,Meyer:2011um,Briceno:2012yi,Feng:2014gba,Briceno:2012yi,Meyer:2012wk,Bernard:2012bi,Agadjanov:2014kha,Briceno:2014uqa,Briceno:2015csa,Briceno:2015tza,Briceno:2019opb,Briceno:2020vgp} and 
to nuclear physics~\cite{Beane:2003da,Detmold:2004qn,Beane:2007qr,Beane:2007es,Luu:2010hw,Luu:2011ep,Davoudi:2011md,Meyer:2012wk,Briceno:2012rv,Briceno:2013lba,Briceno:2013bda,Briceno:2013hya,Briceno:2014oea,Grabowska:2021qqz}.   
L\"{u}scher's method was originally derived from an analysis of Hamiltonian dynamics in Euclidean space and later from a field theoretic point of view directly from correlation functions. 
The challenge is working around the Maiani-Testa theorem~\cite{Maiani:1990ca} and reliably determining Minkowski-space matrix elements from Euclidean-space observables.
This formalism has been used successfully for a number of important quantities, and continues to be the workhorse for Euclidean-space computations.

As quantum simulations provide observables directly in Minkowski space,
understanding the finite-volume and non-zero lattice spacing artifacts requires a similar but different analysis than in Euclidean space.\footnote{Estimates of such effects in model 1+1 dimensional simulations can be found in Ref.~\cite{PhysRevD.103.014506}.}  
While the method used in Euclidean space of determining S-matrix elements for scattering processes from energy eigenvalues can still be applied, Minkowski space simulations will also allow for a direct evaluation of scattering processes, removing some of the modeling that remains in Euclidean-space calculations.\footnote{
For example, the energies of states in different volumes are different, 
and so the elements of the scattering matrix are constrained over 
a range of energies and not at one single energy,
and {\it a priori} unknown interpolations are modeled.
}
Neglecting electroweak interactions beyond $\beta$-decay means that the final state leptons are non-interacting (plane-waves when using periodic boundary conditions),
and therefore the modifications to the density of states due to interactions, as encapsulated within the L\"{u}scher formalism, are absent.

With Hamiltonian evolution of a system described within a finite-dimensional Hilbert space, the persistence amplitude of the initial state coupled to final states via the weak Hamiltonian will be determined by the sum over oscillatory amplitudes.  
For a small number of final states, the amplitude will return to unity after some finite period of time.  
As the density of final states near the energy of the initial state becomes large, there will be cancellations among the oscillatory amplitudes, and the persistence probability will begin to approximate the ``classic" exponential decay over some time interval.  
This time interval will extend  to infinity as the density of states tends to a continuous spectrum. 
It is important to understand how to reliably extract an estimate of the decay rate, with a quantification of systematic errors, from the amplitudes measured in a quantum simulation.  
This is the subject of future work, but here a simple model will be used to demonstrate some of the relevant issues.

\begin{figure}[!ht]
    \centering
    \includegraphics[width=12cm]{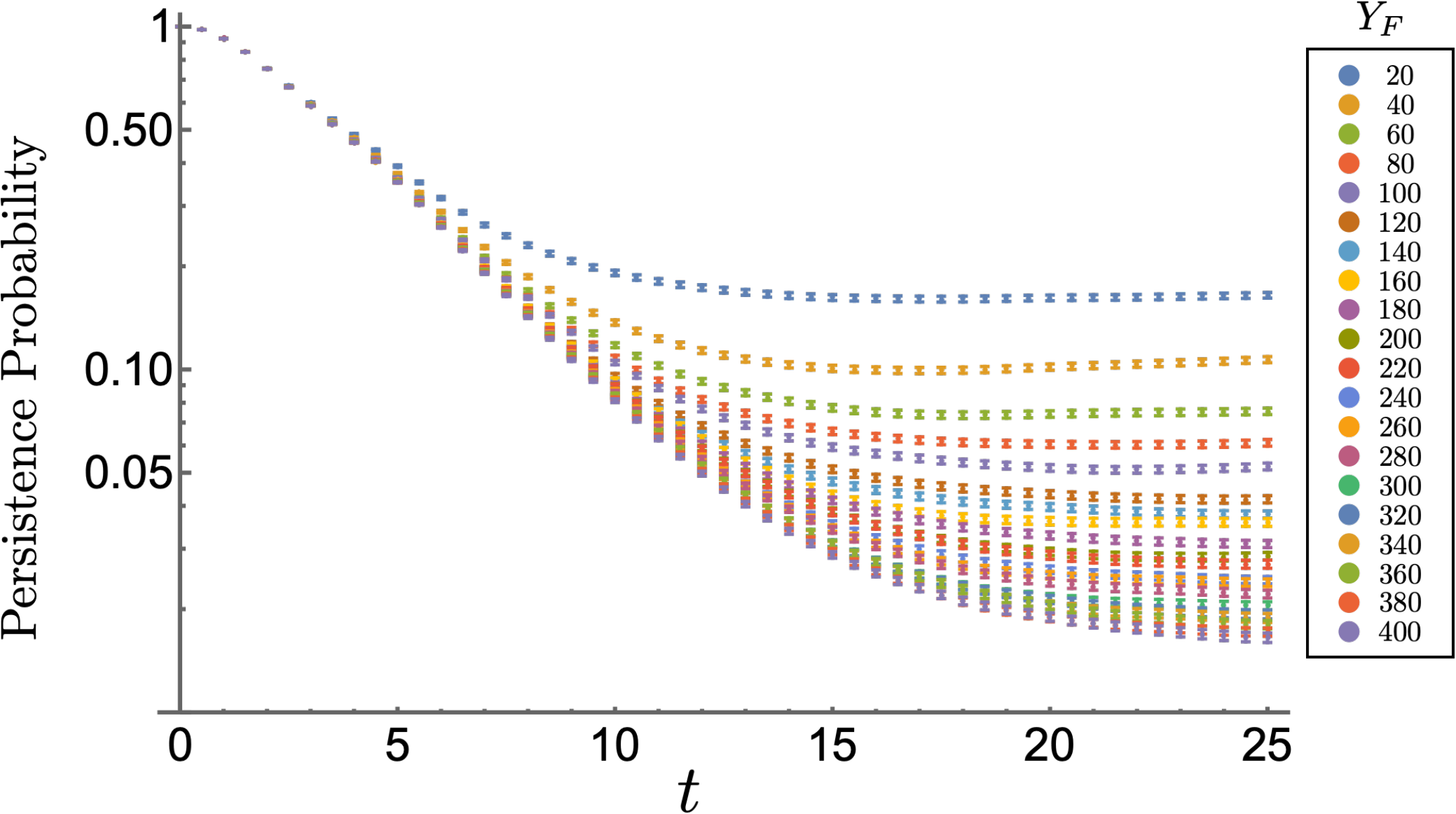}
    \caption{
    Ensemble averages (over 2000 random samples) of the persistence probability of an initial state in one sector of a strong Hamiltonian undergoing weak decay to states in a different sector, as described in this appendix. 
    The different colored points are results from calculations with an increasing number of final states, $Y_F$. 
    The weak coupling scales so that the decay probability converges to a well-defined value as the density of final states tends to a continuum.
    }
    \label{fig:ExpDecay}
\end{figure}

Consider the weak decay of a strong eigenstate in one sector to a strong eigenstate in a different sector (a sector is defined by its strong quantum numbers).
For this demonstration, 
we calculate the persistence probability of the initial state, averaged over random weak and strong Hamiltonians and initial states, as the number of states  below a given energy increases (i.e. increasing density of states).
Concretely, the energy eigenvalues of the initial strong sector range from 0 to 1.1, and 10 are selected randomly within this interval.
The initial state is chosen to be the one with the fifth lowest energy.
The eigenvalues in the final strong sector range between 0 and 2.03, and $Y_f = $ 20 to 400 are selected.
The weak Hamiltonian that induces transitions between the 10 initial states to the $Y_f$ final states is a dense matrix with each element selected randomly.
The weak coupling constant is scaled so that
$G^2\rho_f$ is independent of the number of states, where $\rho_f$ is the density of states.
This allows for a well-defined persistence probability as $Y_f \to \infty$.
For this example, the elements of the weak Hamiltonian were chosen between $\pm w_f$, 
where $w_f = 1/(2 \sqrt{Y_F})$.
Figure~\ref{fig:ExpDecay} shows the emergence of the expected exponential decay as the number of available final states tends toward a continuous spectrum.
In a quantum simulation of a lattice theory, the density of states increases with $L$, and the late-time deviation from exponential decay will exhibit oscillatory behavior, as opposed to the plateaus found in this statistically averaged model.
The very early time behavior of the probability is interesting to note, and exhibits a well-known behavior, e.g., Refs.~\cite{Urbanowski_2017,Giacosa_2017}.  
It is, as expected, not falling exponentially, which sets in over time scales set by the energy spectrum of final states.

Only small lattices are practical for near-term simulation and lattice artifacts will be important to quantify. Relative to the continuum, a finite lattice spacing modifies the energy-momentum relation and introduce a momentum cut-off on the spectra.  
However, if the initial particle has a mass that is much less than the cut-off, these effects should be minimal as the energy of each final state particle is bounded above by the mass of the initial particle.
As has been shown in this appendix, working on a small lattice with its associated sparse number of final states, will lead to significant systematic errors when extracting the decay rates directly from the persistence probabilities.
Further work will be necessary to determine how to reliably estimate these errors.

\section{\texorpdfstring{$\beta$}{Beta}-Decay Circuits}
\label{app:BetaCircuits}
\begin{figure}[!ht]
    \centering
    \includegraphics[width=10cm]{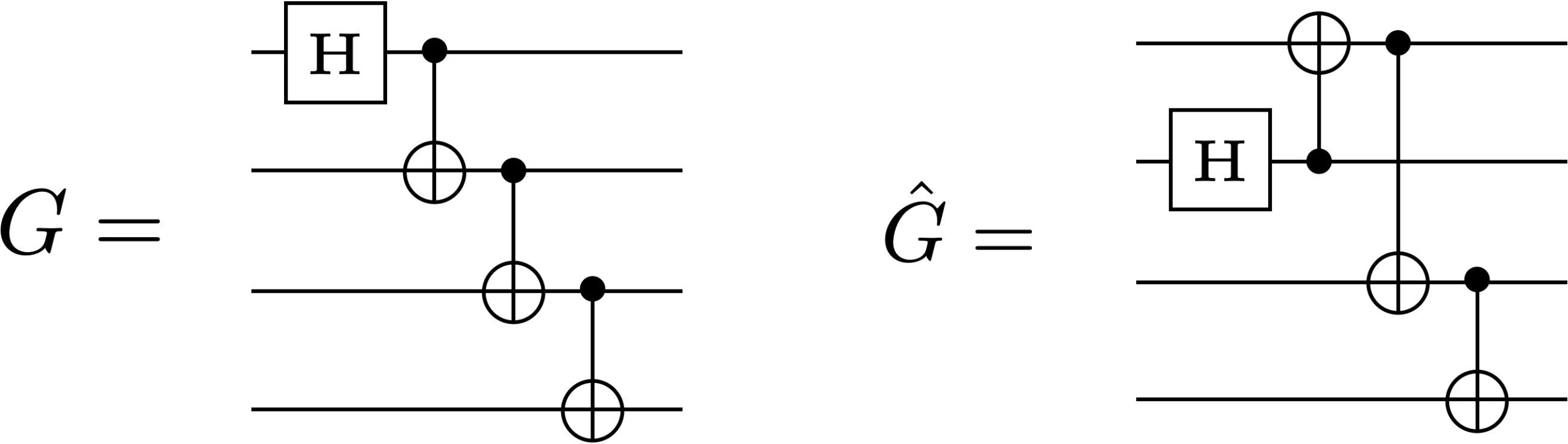}
    \caption{
    Two GHZ state preparation circuits.}
    \label{fig:GHZCirc}
\end{figure}

\begin{figure}[!ht]
    \centering
    \includegraphics[width=\columnwidth]{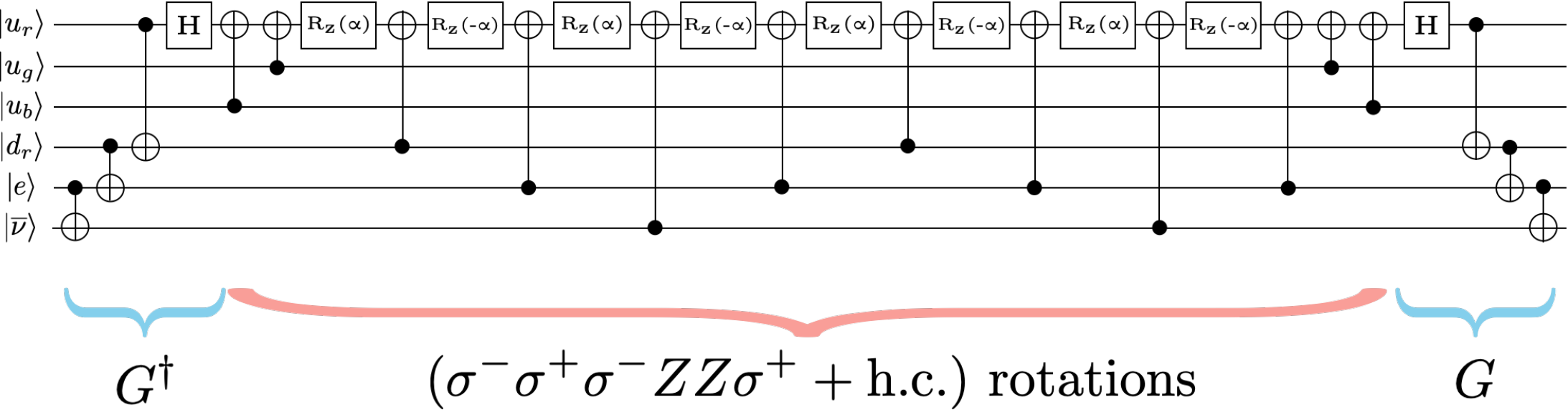}
    \caption{
    A quantum circuit that provides the time evolution associated with the 
    $\sigma^-_{\overline{\nu}} \sigma^+_e \sigma^-_{d,r} Z_{u,b} Z_{u,g} \sigma^+_{u,r}$ operator in the $\beta$-decay Hamiltonian, with 
    $\alpha = \sqrt{2} G t/8$.
    }
    \label{fig:BetaCirc}
\end{figure}

\noindent
The quantum circuits that implement the Trotterized time-evolution of the 
$\beta$-decay Hamiltonian are  similar to those
presented in the previous chapter,
and here the differences between the two will be highlighted.
The $\beta$-decay Hamiltonian in both the standard and tilde layouts, Eqs.~(\ref{eq:KSHamJWmap}) and~(\ref{eq:tildeBeta}), contains terms of the form
\begin{align}
    H_{\beta} \sim&\ (\sigma^- \sigma^+ \sigma^- \sigma^+ + {\rm h.c.}) + (\sigma^- \sigma^+ \sigma^+ \sigma^- + {\rm h.c.}) \nonumber \\
    =&\ \frac{1}{8}(XXXX+YYXX-YXYX+YXXY+XYYX-XYXY+XXYY+YYYY) \nonumber \\
    &+ \frac{1}{8}(XXXX + YYXX + YXYX - YXXY - XYYX + XYXY + XXYY + YYYY)
    \ ,
\end{align}
which can be diagonalized by the GHZ state-preparation circuits, $G$ and $\hat{G}$, shown in Fig.~\ref{fig:GHZCirc}. 
In the GHZ basis, it is found that
\begin{align}
    &G^{\dagger}(XXXX+YYXX-YXYX+YXXY+XYYX-XYXY+XXYY+YYYY)G \nonumber \\ 
    &= IIIZ - ZIIZ + ZZIZ - ZZZZ -IZIZ + IZZZ - IIZZ + ZIZZ 
    \ ,
\end{align}
and 
\begin{align}
    &\hat{G}^{\dagger}(XXXX + YYXX + YXYX - YXXY - XYYX + XYXY + XXYY + YYYY)\hat{G} \nonumber \\ 
    &= IIZI - ZIZI -ZZZZ+ZZZI+IZZZ-IZZI-IIZZ+ZIZZ
    \ .
\end{align}
Once diagonalized the circuit is a product of diagonal rotations, see Fig.~\ref{fig:BetaCirc} for an example of the quantum circuit that provides the time evolution associated with
$\sigma^-_{\overline{\nu}} \sigma^+_e \sigma^-_{d,r} Z_{u,b} Z_{u,g} \sigma^+_{u,r}$.
By diagonalizing with both $G$ and $\hat{G}$ and arranging terms in the Trotterization so that operators that act on the same quarks are next to each other, 
many of the CNOTs can be made to cancel.
Also, an ancilla can be used to efficiently store the parity of the string of $Z$s between the $\sigma^{\pm}$.
%

\section{Resource Estimates for Simulating \texorpdfstring{$\beta$}{Beta}-Decay Dynamics}
\label{app:LongJW}
\begin{figure}[!ht]
    \centering
    \includegraphics[width=15cm]{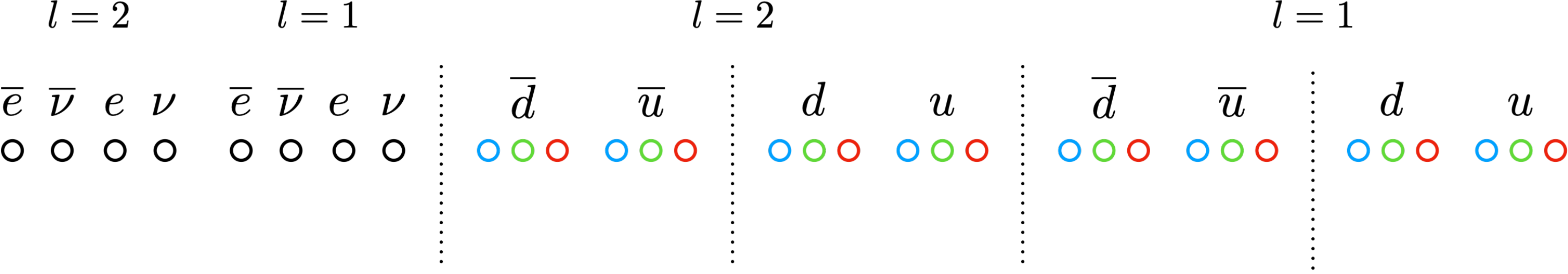}
    \caption{
    A qubit layout that is efficient for the simulation of $\beta$-decay. Shown is an example for $L=2$.}
    \label{fig:L2BetaLayout}
\end{figure}

\noindent 
For multiple lattice sites, it is inefficient to work with leptons in the tilde basis. 
This is due to the mismatch between the local four-Fermi interaction 
and the non-local tilde basis eigenstates. 
As a result, the number of terms in the $\beta$-decay component of the Hamiltonian will scale as $\mathcal{O}(L^2)$ in the tilde basis, 
as opposed to $\mathcal{O}(L)$ in the local occupation basis.
This appendix explores a layout different from the one in Fig.~\ref{fig:L1layout}, which is optimized for the simulation of $\beta$-decay on larger lattices.
To minimize the length of JW $Z$ strings, all leptons are placed at the end of the lattice, see Fig.~\ref{fig:L2BetaLayout}.
After applying the JW mapping, the $\beta$-decay operator becomes
\begin{align}
   H_{\beta} \rightarrow \frac{G}{\sqrt{2}}\sum_{l = 0}^{L-1}\sum_{c=0}^2\bigg (&\sigma^-_{l,\nub} \sigma^+_{l,e} \sigma^-_{l,d,c} Z^2 \sigma^+_{l,u,c} \: - \: \sigma^+_{l,\eb}Z^2\sigma^-_{l,\nu} \sigma^-_{l,d,c} Z^2 \sigma^+_{l,u,c} \: + \: \sigma^-_{l,\nub} \sigma^+_{l,e} \sigma^-_{l,\db,c} Z^2 \sigma^+_{l,\ub,c}  \nonumber \\ 
   &- \: \sigma^+_{l,\eb} Z^2 \sigma^-_{l,\nu} \sigma^-_{l,\db,c} Z^2 \sigma^+_{l,\ub,c} \: + \: \sigma^+_{l,e} \sigma^-_{l,\nu} \sigma^-_{l,\db,c} Z^8 \sigma^+_{l,u,c} \: - \: \sigma^+_{l,\eb} \sigma^-_{l,\nub} \sigma^-_{l,\db,c}Z^8 \sigma^+_{l,u,c} \nonumber \\
   &+ \: \sigma^+_{l,e} \sigma^-_{l,\nu} \sigma^+_{l,\ub,c} Z^2 \sigma^-_{l,d,c} \: - \: \sigma^+_{l,\eb} \sigma^-_{l,\nub} \sigma^+_{l,\ub,c} Z^2 \sigma^-_{l,d,c} \: + \: {\rm h.c.} \bigg ) \ .
   \label{eq:HBlow}
\end{align}

Using the techniques outlined in App.~\ref{app:BetaCircuits} 
to construct the relevant quantum circuits, 
the resources required per Trotter 
step of 
$H_{\beta}$ are estimated to be
\begin{align}
    R_Z  \ :& \ \ 192L \ ,\nonumber \\
   \text{Hadamard} \ :& \ \ 48L  \ ,\nonumber \\
    \text{CNOT} \ :& \ \ 436 L \ .
\end{align}
For small lattices, $L\lesssim 5$, it is expected that use of the tilde basis will be more efficient and these estimates should be taken as an upper bound.
Combining this with the resources required to time evolve with the rest of the Hamiltonian, see Ref.~\cite{Farrell:2022wyt}, the total resource requirements per Trotter step are estimated to be
\begin{align}
    R_Z  \ :& \ \ 264L^2 -54L +77 \ ,\nonumber \\
   \text{Hadamard} \ :& \ \ 48L^2 + 20L +2 \ ,\nonumber \\
    \text{CNOT} \ :& \ \ 368L^2 + 120L+74 \ .
\end{align}
It is important to note that the addition of $H_{\beta}$ does not contribute to the quadratic scaling of resources as it is a local operator. 
Recently, the capability to produce multi-qubit gates natively with similar fidelities to two-qubit gates has also been demonstrated~\cite{Katz:2022ajk,Katz:2022czu,Andrade:2021pil}.
This could lead to dramatic reductions in the resources required and, for example, the number of multi-qubit terms in the Hamiltonian scales as
\begin{equation}
    \text{Multi-qubit terms} \ : \ \ 96 L^2 -68L+22 \ .
\end{equation}
The required number of CNOTs and, for comparison, the number of multi-qubit terms in the Hamiltonian, for a selection of different lattice sizes are given in Table~\ref{tab:cnot}. 
\begin{table}[!ht]
\renewcommand{\arraystretch}{1.2}
\begin{tabularx}{0.48\textwidth}{||c | Y | Y ||}
 \hline
 $L$ & CNOTS & Multi-Qubit Terms \\
 \hline\hline
 5 & 9874 & 2082 \\
 \hline
 10 & 38,074 & 8942\\
 \hline
 50 & 926,074 & 236,622\\
 \hline
 100 & 3,692,074 & 953,222\\
 \hline
\end{tabularx}
\renewcommand{\arraystretch}{1}
\caption{The CNOT-gate requirements to perform one Trotter step of time evolution 
of $\beta$-decay for a selection of lattice sizes. For comparison, the number of multi-qubit terms in the Hamiltonian is also given.}
\label{tab:cnot}
\end{table}
Note that these estimates do not include the resources required to prepare the initial state.

\section{Technical Details on the Quantinuum {\tt H1-1} Quantum Computer}
\label{app:H1specs}
\noindent 
For completeness, this appendix contains a brief description of Quantinuum's {\tt H1-1} 20 
trapped ion quantum computer (more details can be found in~\cite{h1-1}). 
The {\tt H1-1} system uses the System Model {\tt H1} design, 
where unitary operations act on a single line of ${}^{172}$Y${}^+$ ions induced by lasers. 
The qubits are defined as the two hyperfine clock states in the ${}^2S_{1/2}$ ground state of ${}^{172}$Y${}^+$. 
Since the physical position of the ions can be modified, 
it is possible to apply two-qubit gates to any pair of qubits, 
endowing the device with all-to-all connectivity. 
Moreover, there are five different physical regions where these gates can be applied in parallel. Although we did not use this feature, it is also possible to perform a mid-circuit measurement of a qubit, i.e., initialize it and reuse it (if necessary).

The native gate set for {\tt H1-1} is the following,
\begin{equation}
    U_{1q}(\theta,\phi)=e^{-i\frac{\theta}{2}[\cos(\phi) X+\sin(\phi) Y]}\ , \quad R_Z(\lambda)=e^{-i\frac{\lambda}{2}Z}\ , \quad ZZ=e^{-i\frac{\pi}{4}ZZ} \ ,
\end{equation}
where $\theta$ in $U_{1q}(\theta,\phi)$
can only take the values $\{\frac{\pi}{2},\pi\}$, 
and arbitrary values of $\theta$ can be obtained by combining several single-qubit gates, $\tilde{U}_{1q}(\theta,\phi)=U_{1q}(\frac{\pi}{2},\phi+\frac{\pi}{2}) . R_Z(\theta) . U_{1q}(\frac{\pi}{2},\phi-\frac{\pi}{2})$. 
Translations between the gates used in the circuits shown in the main text and appendices to the native ones are performed automatically by {\tt pytket}~\cite{sivarajah2020t}.
The infidelity of the single- and two-qubit gates, as well as the error of the SPAM operations, are shown in Table~\ref{tab:H1err}.
\begin{table}[!ht]
\renewcommand{\arraystretch}{1.2}
\begin{tabularx}{0.7\textwidth}{||r | Y | Y | Y ||}
 \hline
  & Min & Average & Max \\
 \hline\hline
 Single-qubit infidelity & $2\times 10^{-5}$ & $5\times 10^{-5}$ & $3\times 10^{-4}$ \\
 \hline
 Two-qubit infidelity & $2\times 10^{-3}$ & $3\times 10^{-3}$ & $5\times 10^{-3}$ \\
 \hline
 SPAM error & $2\times 10^{-3}$ & $3\times 10^{-3}$ & $5\times 10^{-3}$ \\
 \hline
\end{tabularx}
\renewcommand{\arraystretch}{1}
\caption{
Errors on the single-qubit, two-qubit and SPAM operations, with their minimum, average and maximum values.
}
\label{tab:H1err}
\end{table}
%

\section{Time Evolution Under the Full \texorpdfstring{$\beta$}{Beta}-Decay Operator}
\label{app:betaFull}
\begin{figure}[!ht]
    \centering
    \includegraphics[width=12cm]{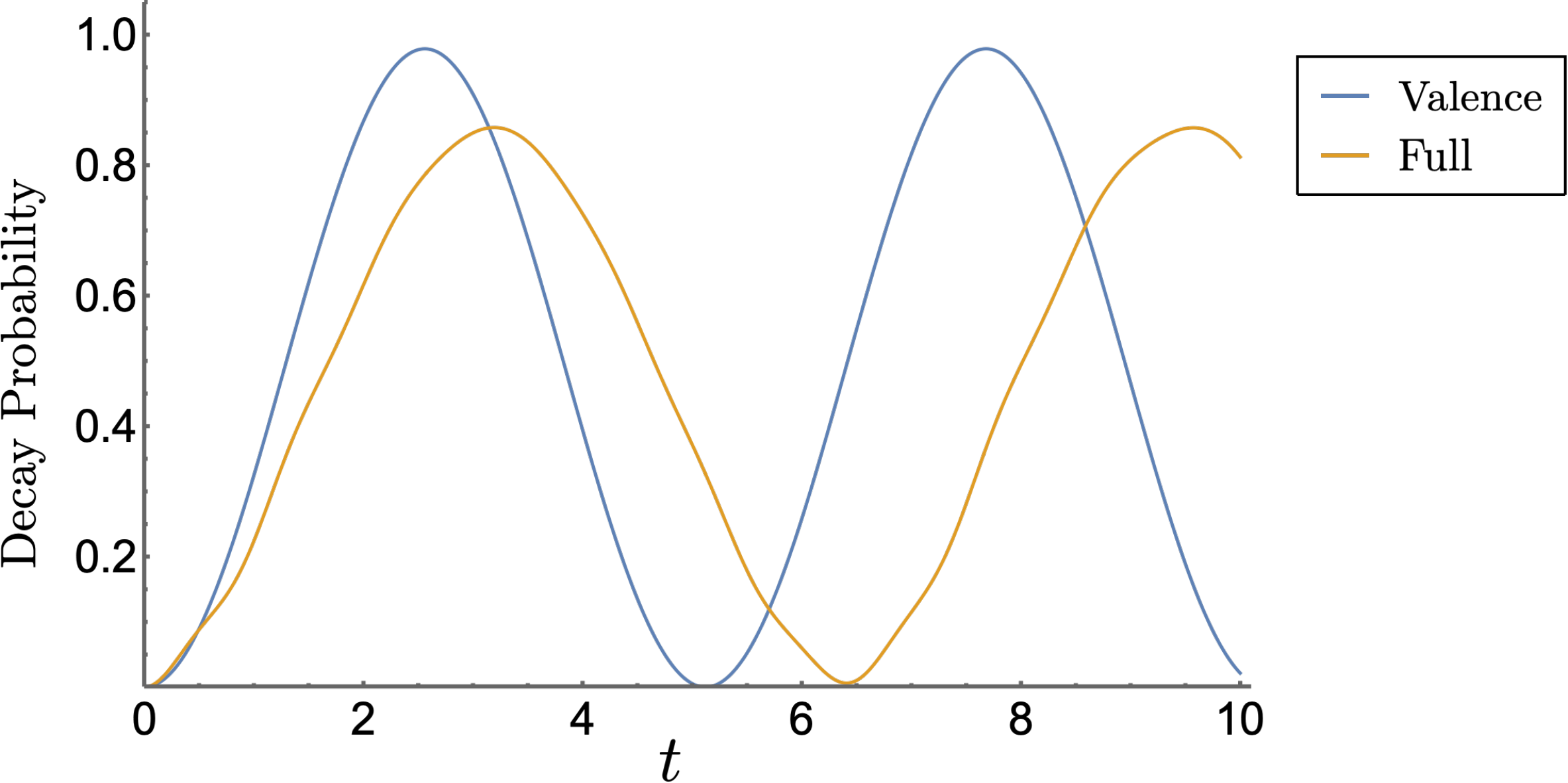}
    \caption{The probability of $\beta$-decay using both the approximate $\beta$-decay operator which only acts on valence quarks (blue) and the full operator (orange).}
    \label{fig:valfull}
\end{figure}

\noindent
The simulations performed in Sec.~\ref{sec:BetaSim} kept only the terms in the $\beta$-decay Hamiltonian which act on valence quarks, see Eq.~(\ref{eq:tildeBetaRed}). 
This appendix examines how well this valence quark $\beta$-decay operator approximates the full operator, Eq.~(\ref{eq:tildeBeta}), for the parameters used in the main text.
Shown in Fig.~\ref{fig:valfull} is the decay probability when evolved with both the approximate and full operator as calculated through exact diagonalization of the Hamiltonian.
The full $\beta$-decay operator has multiple terms that can interfere leading to a more jagged decay probability.
The simulations ran on {\tt H1-1} only went out to $t=2.5$ where the error of the approximate operator is $\sim 20\%$.
\end{subappendices}

\chapter{Quantum simulations of the Schwinger model vacuum on 100 Qubits}
\label{chap:SCADAPT}
\noindent
\textit{This chapter is associated with Ref.~\cite{Farrell:2023fgd}: \\
``Scalable Circuits for Preparing Ground State on Digital Quantum Computers: The Schwinger Model Vacuum on 100 Qubits" by Roland C. Farrell, Anthony N. Ciavarella, Marc Illa and Martin J. Savage. }
\section{Introduction}
\label{sec:I}
\noindent
Quantum simulations of physical systems described by the Standard Model~\cite{Weinberg:1967tq,Glashow:1961tr,Salam:1968rm,Politzer:1973fx,Gross:1973id,Higgs:1964pj}, 
and descendant effective field theories (EFT),
are anticipated to provide qualitatively new predictions about matter under extreme conditions; from the dynamics of matter in the early universe, 
to properties of the exotic phases of quantum chromodynamics (QCD) produced at the LHC and RHIC (for overviews and reviews, see Refs.~\cite{Banuls:2019bmf,Guan:2020bdl,Klco:2021lap,Delgado:2022tpc,Bauer:2022hpo,Humble:2022klb,Beck:2023xhh,Bauer:2023qgm,DiMeglio:2023nsa}).
One of the major challenges facing quantum simulations of physical systems is 
the preparation of initial states on quantum computers
that can be used to determine important quantities that are inaccessible to  
classical high-performance computing (HPC),
i.e., the problem of state preparation.
While simulating the dynamics of any given initial state 
is known to be efficient for an ideal quantum computer~\cite{Lloyd1073},
residing in the {\bf BQP} complexity class,
preparing an arbitrary state generally requires 
quantum resources that asymptotically scale 
super-polynomially  with increasing system size~\cite{10.1007/978-3-540-30538-5_31},
residing in the {\bf QMA} complexity class.\footnote{Note that adiabatic state preparation resides within {\bf BQP} when there is a path through parameter space in which the system remains gapped~\cite{farhi2000quantum,van_Dam_2001}.
However, even in gapped systems the gate count required for adiabatic preparation can be daunting; e.g., see Ref.~\cite{Chakraborty:2020uhf} where adiabatic preparation of the Schwinger model vacuum on 16 qubits was estimated to require $2.7\times10^5$ two-qubit gates.}
However, states of physical systems are not the general case, and are often constrained by both local and global symmetries.\footnote{Systems of importance to nuclear physics and high-energy physics are constrained by
a number of local, exact global and approximately global symmetries, 
some of which are emergent from the mechanisms of confinement 
and spontaneous symmetry breaking.} 
In some instances, these symmetries
allow observables to be computed by perturbing around states that can be efficiently 
initialized~\cite{Klco:2021lap}.
In the foreseeable future, quantum simulations  will be far from asymptotic in both system size and evolution time, and the resources required for both time evolution and state preparation
will be estimated by direct construction and extrapolations thereof.
Furthermore, successful quantum simulations will require specialized quantum circuits and workflows that are optimized for specific quantum hardware.

The development of algorithms for preparing 
non-trivial initial states on quantum computers, including the ground states of quantum field theories (QFTs), is an active area of research.
Even with many advances, 
algorithms remain limited in capability, and generally do not scale favorably to modest or large-scale simulations of quantum many-body systems.
Consequently, quantum simulations of small model systems are currently 
being performed across an array of science domains,
generally studying dynamics starting from tensor-product initial states.
While being the simplest gauge theory based on a continuous group, 
the Schwinger model~\cite{Schwinger:1962tp} (quantum electrodynamics in 1+1D)  
possesses many features of interest to both the 
quantum chromodynamics and quantum information science (QIS)
communities.
These include
the presence of a mass gap, charge screening, a chiral condensate, few-body bound states (``hadrons'' and ``nuclei''), and  a topological $\theta$-term.
It has emerged as a popular test bed for developing quantum simulation techniques
for  lattice gauge theories,
and has been explored using a variety of platforms, including trapped ions~\cite{Martinez:2016yna,Kokail:2018eiw,Nguyen:2021hyk,Mueller:2022xbg}, superconducting qubits~\cite{Klco:2018kyo,Mazzola:2021hma,deJong:2021wsd,Gong:2021bcp,Mildenberger:2022jqr,Charles:2023zbl,Pomarico:2023png}, photonic systems~\cite{Lu:2018pjk}, Rydberg atoms~\cite{Surace:2019dtp}, ultracold atoms~\cite{Mil:2019pbt,Yang:2020yer,Zhou:2021kdl,PhysRevResearch.5.023010,zhang2023observation} and classical electric circuits~\cite{Riechert:2021ink}, together with classical simulations~\cite{Banerjee:2012pg,Marcos:2013aya,Yang:2016hjn,Muschik:2016tws,Davoudi:2019bhy,Luo:2019vmi,Chakraborty:2020uhf,Ferguson:2020qyf,Davoudi:2021ney,Yamamoto:2021vxp,Honda:2021aum,Bennewitz:2021jqi,Andrade:2021pil,Honda:2021ovk,Halimeh:2022pkw,Xie:2022jgj,Davoudi:2022uzo,Nagano:2023uaq,Popov:2023xft,Nagano:2023kge}, calculations~\cite{Hauke:2013jga,Kasper:2016mzj,Notarnicola:2019wzb,Tran:2020azk,Shen:2021zrg,Jensen:2022hyu,Florio:2023dke,Ikeda:2023zil} 
and tensor-networks~\cite{Byrnes:2002nv,Banuls:2013jaa,Rico:2013qya,Buyens:2013yza,Kuhn:2014rha,Banuls:2015sta,Pichler:2015yqa,Buyens:2015tea,Banuls:2016lkq,Buyens:2016ecr,Buyens:2016hhu,Zapp:2017fcr,Ercolessi:2017jbi,Sala:2018dui,Funcke:2019zna,Magnifico:2019kyj,Butt:2019uul,Rigobello:2021fxw,Okuda:2022hsq,Honda:2022edn,Desaules:2023ghs,Angelides:2023bme,Belyansky:2023rgh} (for reviews on this last topic, see, e.g., Refs.~\cite{Banuls:2019rao,Banuls:2019bmf}).
There has also been pioneering work on quantum simulations of low-dimensional 
non-Abelian gauge theories, both with~\cite{Atas:2021ext,Atas:2022dqm,Farrell:2022vyh,Farrell:2022wyt,Ciavarella:2023mfc,Kuhn:2015zqa} 
and without~\cite{Klco:2019evd,Ciavarella:2021nmj,Ciavarella:2021lel,ARahman:2021ktn,Illa:2022jqb,Ciavarella:2022zhe,ARahman:2022tkr} matter.
While these are important benchmarks, more sophisticated simulations requiring the preparation of eigenstates or scattering states have so far been too demanding for 
NISQ-era quantum computers, and until now have been limited to 20 qubits~\cite{Kokail:2018eiw,Mildenberger:2022jqr}.

Many systems of physical interest, including QCD, 
have translational symmetry and possess an energy (mass) gap $\Lambda$ between the unique ground state and first excited state.
The  gap defines a characteristic length scale of the system $\xi = 1/\Lambda$, and parameterizes the decay of the longest distance 
correlations in the ground state wavefunction, falling as 
$\sim e^{- r/\xi}/r^\alpha$ for regions separated by $r\gg\xi$, for some $\alpha$.
A natural way to encode a lattice QFT onto a register of a digital quantum computer is by identifying subsets of qubits (or qudits) with spatial points of the lattice 
that align with the connectivity of the quantum computer.
A realization of the ground state on the register of a quantum computer
should reflect the localized correlations
between these subsets of 
qubits
separated by $r\gg\xi$~\cite{Klco:2019yrb,Klco:2020aud}.
In the absence of topological order, one way to establish the ground state is to initialize the quantum register in a 
state without correlations between qubits, e.g., a tensor product state,
and 
then systematically introduce correlations 
through the action of  quantum circuits.
A crucial point is that the localized correlations imply that the state preparation circuits need to have structure only for qubits spatially separated by $r \lesssim \xi$~\cite{Klco:2019yrb,Klco:2020aud}.
This is sufficient to obtain exponentially converged accuracy in the prepared state.
Due to translational invariance, the ground state for an arbitrarily large lattice can be prepared by repeating these circuits across the entire register.

To study the dynamics of physically relevant systems in a quantitative way,
with a complete quantification of uncertainties,
simulations of large volumes of spacetime are typically required.
Motivated by the discussion in the previous paragraph, we introduce Scalable Circuits ADAPT-VQE (SC-ADAPT-VQE), a new method for quantum state preparation that uses the hierarchies of length scales present in physical systems; see Fig.~\ref{fig:SCADAPTVQE} for an illustration.
In SC-ADAPT-VQE, 
quantum circuits that (efficiently) prepare a given state to a specified level of precision
are determined on modest-sized lattices that are large enough to contain the longest correlation lengths.
As long as $\xi$ is not too large, these circuits can be determined using {\it classical} computers.
This avoids the challenging task of optimizing circuits on a quantum computer with both statistical uncertainty and device noise~\cite{Wang:2020yjh,Scriva:2023sgz}.
Once determined, (discrete) translation invariance is used to scale these circuits up to the full lattice.
Since the quality of the prepared state becomes 
independent of the spatial lattice length $L$, 
with ${\mathcal O}(e^{-\xi/L})$ corrections,
this is a potential path toward 
quantum simulations of lattice QFTs
that are beyond the capabilities of HPC.

In this chapter, SC-ADAPT-VQE is applied to the Schwinger model and is used to prepare the vacuum on up to 100 qubits on IBM's {\tt Eagle} quantum processors.
Underlying the development is the ADAPT-VQE algorithm~\cite{Grimsley_2019} for quantum state preparation, which is modified to generate scalable circuits.
After the necessary Trotterized circuits have been built, SC-ADAPT-VQE is performed using the {\tt qiskit} classical simulator on 
system sizes up to $L=14$ (28 qubits).
It is found that both the energy density and the chiral condensate 
converge exponentially with circuit depth to the exact results.
Importantly, both the quality of the prepared state and the structure of the associated circuits are found to converge with system size.
This allows the state preparation circuits, determined on small lattices using classical computing, to be extrapolated to much larger lattices, with a quality that becomes independent of $L$.
The scaled circuits are used to prepare the $L\le 500$ vacua using {\tt qiskit}'s Matrix Product State (MPS) circuit simulator, 
and to prepare the  $L \le 50$ (100 qubits) vacua on the registers of 
IBM's superconducting-qubit quantum computers {\tt ibm\_brisbane} and {\tt ibm\_cusco}.
An improved and unbiased error mitigation technique, Operator Decoherence Renormalization (ODR), is developed and applied to the quantum simulations to estimate error-free observables.
The results obtained from both the MPS circuit simulator and  IBM's quantum computers
are found to be in excellent agreement with Density Matrix Renormalization Group (DMRG) calculations.

\section{The Lattice Schwinger Model}
\label{sec:SchwingerHam}
\noindent
The Schwinger model~\cite{Schwinger:1962tp} has a long history of study 
in the continuum and using numerical lattice techniques.
 In the continuum, it is described by the Lagrange density
 \begin{align}
 {\cal L}
 & =  \overline{\psi}\left( i \Dslash - m_\psi \right) \psi - \frac{1}{4} F^{\mu\nu} F_{\mu\nu}
 \ .
 \label{eq:LagSM}
 \end{align}
Electrically-charged fermions are described by the field operator $\psi$ with mass $m_\psi$,
the electromagnetic gauge field by $A_\mu$ with field tensor $F_{\mu\nu}$, 
and the covariant derivative is defined as $D_\mu = \partial_\mu - i e A_\mu$.
It is the Hamiltonian lattice formulation,
first developed and studied by Banks, Kogut and Susskind~\cite{Kogut:1974ag,Banks:1975gq},
that is relevant for quantum simulations. 
One feature of gauge theories in $1+1$D, which distinguishes 
them from theories in higher dimensions,
is that the gauge field is completely constrained 
by the distribution of fermion charges through Gauss's law.
In axial gauge, the spatial gauge field is absent, and the effects of the time-component of the gauge field
are included by non-local (Coulomb) interactions~\cite{Sala:2018dui,Farrell:2022wyt}.
With  open boundary conditions (OBCs),  
using the staggered fermion discretization~\cite{Kogut:1974ag} of the electron field,
and applying the Jordan Wigner (JW) transformation to map fermion field operators to spins, 
the Schwinger model Hamiltonian is (for a derivation, see, e.g., Ref.~\cite{Kokail:2018eiw})
\begin{align}
\hat H & \ =\  \hat H_m + \hat H_{kin} + \hat H_{el} \ = \nonumber \\
&\ \frac{m}{ 2}\sum_{j=0}^{2L-1}\ \left[ (-1)^j \hat Z_j + \hat{I} \right] \ + \ \frac{1}{2}\sum_{j=0}^{2L-2}\ \left( \hat \sigma^+_j \hat\sigma^-_{j+1} + {\rm h.c.} \right) \ + \ \frac{g^2}{ 2}\sum_{j=0}^{2L-2}\bigg (\sum_{k\leq j} \hat Q_k \bigg )^2 
\ ,
\nonumber \\ 
\hat Q_k & \ = \ -\frac{1}{2}\left[ \hat Z_k + (-1)^k\hat{I} \right] \ .
\label{eq:Hgf}
\end{align}
Here, $L$ is the number of  spatial lattice sites, 
corresponding to $2L$ staggered (fermion) sites, $m$ and $g$ are the (bare) electron mass and charge, respectively, and the staggered lattice spacing $a$ has been set to one.\footnote{For faster convergence to the continuum, an ${\mathcal O}(a)$ improvement to the mass term can be performed to restore a discrete remnant of chiral symmetry in the $m\to0$ limit~\cite{Dempsey:2022nys}.} 
``Physical'' quantities are 
derived from the corresponding dimensionless quantities by restoring factors of the 
spatial lattice spacing.
Even (odd) sites correspond to electrons (positrons), as reflected in the staggered mass term and charge operator.\footnote{The convention is that even fermion-sites correspond to electrons, 
such that $\hat Q \lvert\downarrow\rangle = 0$ and 
$ \hat Q \lvert\uparrow\rangle = -\lvert\uparrow\rangle$,
while
the odd fermion-sites correspond to positrons, such that 
$ \hat Q \lvert\uparrow\rangle = 0$ and 
$\hat Q \lvert\downarrow\rangle = +\lvert\downarrow\rangle$.
}
A background electric field can be included straightforwardly, 
equivalent to a  $\theta$-term,  but
will be set to zero in this work.
Due to the confinement, the low energy excitations are hadrons and the mass gap is given by $\Lambda = m_{\text{hadron}}$. For our purposes, $m_{\text{hadron}}$ is defined to be the energy difference in the $Q=0$ sector between the first excited state (single hadron at rest) and the vacuum.

\subsection{Infinite Volume Extrapolations of Local Observables}
\label{sec:infVolExtra}
\noindent
Central to the development of state preparation circuits is the scaling of expectation values of local observables in the ground state, 
with both the correlation length $\xi = 1/m_{{\rm hadron}}$, and the volume $L$.
Due to the exponential suppression of correlations in the ground state 
between regions separated by $r>\xi$, it is expected that, locally, the wavefunction has converged to its infinite volume form, with corrections of ${\mathcal O}(e^{-\xi/L})$.
As a result, expectation values of local observables will 
be exponentially converged to their infinite volume values.
However, near the boundaries of the lattice, the wavefunction is perturbed over a depth proportional to $\xi$, causing local observables to deviate from their infinite volume values.
Equivalently, boundary effects cause deviations in volume averages of local observables that are ${\mathcal O}(\xi/L)$.
This scaling of observables is responsible for the SC-ADAPT-VQE prepared vacuum converging exponentially in circuit depth, and enables the circuits to be systematically extrapolated to larger system sizes.

\begin{figure}[hbt!]
    \centering
    \includegraphics[width=\columnwidth]{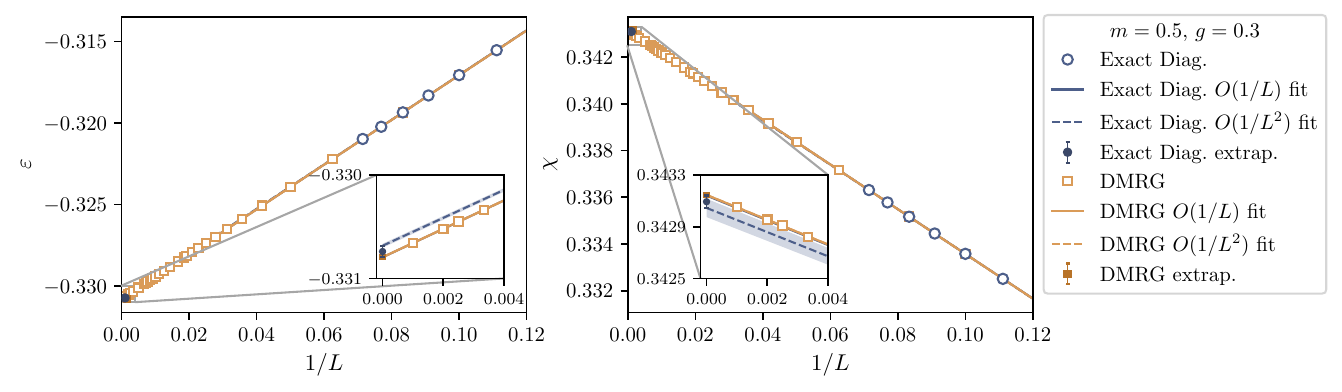}
    \caption{
    $L$-extrapolations of the vacuum energy density $\varepsilon$ (left panel)
and chiral condensate $\chi$ (right panel) for $m=0.5$ and $g=0.3$.
The results of exact diagonalization calculations for $L\ge 9$ (blue circles) given in 
Table~\ref{tab:Echifmp5gp3} and 
DMRG calculations (orange squares) given in Table~\ref{tab:DMRGresults}
are extrapolated to $L\rightarrow\infty$, as shown by the darker points, respectively.
The solid lines correspond to linear extrapolations and the dashed lines correspond to quadratic extrapolations, and are found to overlap (see inset).   
The difference between the $L\rightarrow\infty$ values of these two extrapolations defines the (fitting) uncertainties associated with the darker points. 
}
    \label{fig:EdChiLextrap53}
\end{figure}

Two quantities associated with the ground-state wavefunction (vacuum) 
that we focus on are the chiral condensate $\chi$, and the energy density
$\varepsilon$.
The chiral condensate\footnote{In the continuum, 
the chiral condensate is defined as 
$\chi_{\text{cont}}=\langle \overline{\psi}\psi \rangle$, which on the lattice becomes $\chi_{\text{lat}} = \frac{1}{L}\sum_j\langle \overline{\psi}_j\psi_j \rangle$, 
where $j$ labels the spatial site. 
To have a positive quantity, we have added a constant to the definition of $\chi$, 
$\chi \equiv \chi_{\text{lat}} + 1$. 
This counts the average number of electrons and positrons on a spatial site.} 
is an order parameter of chiral symmetry breaking, 
and in the JW mapping is
\begin{equation}
\chi \ = \ \frac{1}{2L}  \sum_{j=0}^{2L-1} \left\langle (-1)^j \hat Z_j + \hat I   \right\rangle \ \equiv \ \frac{1}{2L}  \sum_{j=0}^{2L-1}\chi_j
\ .
\label{eq:chiralCond}
\end{equation} 
The energy density is defined as 
$\varepsilon = \langle \hat{H} \rangle / L$, 
and 
in axial gauge is not a local observable because the contribution from the 
electric-field term in the Hamiltonian, $\hat{H}_{el}$, involves all-to-all couplings.
However, this is an artifact of using axial gauge and enforcing Gauss's law.
In Weyl gauge, with explicit (local) gauge degrees of freedom, 
the Hamiltonian is manifestly local, and therefore the energy density is a local observable.
These quantities are computed for $m=0.5,g=0.3$ using exact diagonalization for $L\le 14$ 
(Table~\ref{tab:Echifmp5gp3})
and DMRG for $L\gg 14$ (Table~\ref{tab:DMRGresults}).
As anticipated,
a linear extrapolation in $1/L$ is found to be consistent with 
these results, as seen in Fig.~\ref{fig:EdChiLextrap53}.
Additional details,
along with results for $m=0.1$  with $g=0.3$ and $g=0.8$, 
can be found in App.~\ref{app:LinvScaling}.

\section{SC-ADAPT-VQE for the Lattice Schwinger Model}
\label{sec:II}
\noindent
Underlying SC-ADAPT-VQE is ADAPT-VQE~\cite{Grimsley_2019},  
a quantum algorithm for state preparation that has been applied to spin models~\cite{VanDyke:2022ffj}, systems in quantum chemistry~\cite{Grimsley_2019,Tang:2021PRX,Yordanov:2021,Shkolnikov:2021btx,Bertels:2022oga,Anastasiou:2022swg,Feniou:2023gvo} and nuclear structure~\cite{Romero:2022blx,Perez-Obiol:2023vod}.
It builds upon the Variational Quantum Eigensolver (VQE)~\cite{Peruzzo_2014}, 
in which parameterized quantum circuits are optimized  
to minimize the expectation value of a Hamiltonian.
The parameterized circuits are constructed 
step-wise (or equivalently in layers), 
where the incrementally-improved ansatz states converge to the ground state with successive iterations.
At each step, 
the unitary transformation that maximally decreases the energy of the ansatz state is identified from a pre-defined set (``pool") of unitaries.
The quantum circuit corresponding to this unitary is then
appended to the state preparation circuit. 
The (initial) state from which the algorithm starts
will often be chosen to be a tensor product
or an entangled state that can be efficiently prepared on a quantum computer, 
such as a GHZ-state.
If the target state is the ground state of a confining gauge theory, e.g., the Schwinger model, 
the strong-coupling (trivial) vacuum,
\begin{equation}
|\Omega_0\rangle = 
\lvert\uparrow\downarrow\uparrow\downarrow \ldots\uparrow\downarrow\rangle
\ ,
\label{eq:SCvac}
\end{equation} 
can be a good choice for such an initial state as it has the correct long-distance 
structure in the gauge fields.
The ADAPT-VQE algorithm can be summarized as follows:
\begin{itemize}
    \item[1.] Define a pool of operators $\{ \hat{O} \}$ that are 
    constrained to respect some or all of the symmetries of the system. 
    \item[2.] Initialize the register of the quantum computer to a strategically selected state, 
    $|\psi_{{\rm ansatz}} \rangle$,  with the desired quantum numbers
   and symmetries of the target wavefunction.
    \item[3.] Measure the expectation value of the commutator of the Hamiltonian with each operator in the pool, 
    $\langle \psi_{{\rm ansatz}} \lvert [\hat{H}, \hat{O}_i] \lvert \psi_{{\rm ansatz}} \rangle$. 
    These are estimators of the associated decrease in energy from 
    transforming the  ansatz wavefunction by 
    $\lvert \psi_{{\rm ansatz}} \rangle \to e^{i \theta_i \hat{O}_i}\lvert \psi_{{\rm ansatz}} \rangle$,
    for an arbitrary parameter $\theta_i$.
    \item[4.] 
    Identify the operator, $\hat{O}_n$, with the largest magnitude commutator with the Hamiltonian.
    If the absolute value of this commutator is below some pre-determined threshold, terminate the algorithm. 
    If it is above the threshold, 
    update the ansatz with the parameterized evolution of the operator $\lvert \psi_{{\rm ansatz}} \rangle \to e^{i \theta_n \hat{O}_n}\lvert \psi_{{\rm ansatz}} \rangle$.
    \item[5.] Use VQE to find the values of the variational parameters that minimize the energy,  \\
    $\langle\psi_{{\rm ansatz}} (\theta_{1}, \theta_{2},..., \theta_n)\rvert \hat{H}\lvert\psi_{{\rm ansatz}} (\theta_{1}, \theta_{2},..., \theta_n)\rangle$.
    The previously optimized values for $\theta_{1,2,...,n-1}$
    and $\theta_n=0$, are used as initial conditions. 
    If the optimal value of the newest parameter, $\theta_n$, is below some pre-determined threshold, terminate the algorithm.
    \item[6.] Return to step 3.
\end{itemize}
For a given pool of operators, it is {\it a priori} 
unknown if this algorithm will furnish a wavefunction that satisfies the 
pre-determined threshold
for the observable(s) of interest,
but it is expected that the  pool can be expanded on the fly to 
achieve the desired threshold.
The systems that have been explored with this algorithm show,
for a fixed pool,
exponential convergence with increasing numbers of iterations~\cite{Grimsley_2019,Tang:2021PRX,Bertels:2022oga,Romero:2022blx,Feniou:2023gvo}.

Generally,
different terms contributing to operators
in the pool 
do not commute with each other.
Constructing quantum circuits that exactly implement the exponential of a sum of non-commuting terms is challenging, and in practice approximations such as first-order Trotterization are used.
This introduces (higher-order) systematic deviations from the target unitary operator in each case, and 
defines the pool of unitary operators,
\begin{align}
     \{ \hat{U}_i \} & = \{ {\exp}(i \theta_i \hat{O}_i) \}\rightarrow \left \{ \prod\limits_t \hat{U}_i^{(t)} \right \}
    \ .
    \label{eq:TrotOp}
\end{align}
These Trotterized unitary operators correspond to the quantum circuits 
that are implemented in state preparation.
In optimization of the quality of the state prepared on a given quantum computer, 
particularly a NISQ-era device,
there are tradeoffs between the gate-depth of a particular circuit implementation, 
the coherence time, 
the errors associated with gate operations, 
and the associated Trotter errors.
This is explored in App.~\ref{app:TrotterErrors}.

Typically, ADAPT-VQE is a hybrid classical-quantum algorithm that evaluates matrix elements of the Hamiltonian 
in trial wavefunctions on a quantum computer, with parameters that are optimized classically.
One disadvantage of this is that the evaluation of expectation values of the Hamiltonian requires a large number of measurements (shots) on quantum computers.
A novel part of SC-ADAPT-VQE is the use of a {\it classical} simulator to determine the ADAPT-VQE state preparation circuits.
As shown in Sec.~\ref{sec:vacum_class_quan}, these circuits can be scaled and used to prepare the vacuum on arbitrarily large lattices.

\subsection{A Scalable Operator Pool for the Lattice Schwinger Model}
\label{sec:CQAdaptVQESM}
\noindent
A successful application of SC-ADAPT-VQE to the preparation of the lattice Schwinger model vacuum requires choosing an efficient and scalable pool of operators.
These operators are
used to systematically improve the ansatz vacuum wavefunction, 
and are (only) constrained to be
charge neutral, 
symmetric under charge-conjugation and parity (CP) and, as a consequence of the CPT theorem~\cite{Bell:1955djs,Schwinger:1951xk,Luders:1954zz}, invariant under time reversal.\footnote{In the total charge $Q=0$ sector, there is a CP symmetry corresponding to the composition of a reflection through the mid-point of the lattice, exchanging {\it spatial} sites $n \leftrightarrow L-1-n$, and an interchange of an electron and a positron on each spatial site. 
In terms of spins on {\it staggered} sites this is realized as $\hat{\sigma}^i_n \leftrightarrow \hat{\sigma}_{2L-1-n}^i$ followed by $\hat{\sigma}_n^i \leftrightarrow \hat{X}_n\hat{\sigma}_n^i\hat{X}_n$, where $\hat{\sigma}^i$ with $i=1,2,3$ are the Pauli matrices.
For example, under a CP transformation, the following $L=4$ state becomes
\begin{align}
&
|\uparrow\downarrow\ \uparrow\uparrow\ \downarrow\downarrow\ \downarrow\uparrow\rangle \ = \ |. .\ \ . e^-\ \ e^+ .\ \ e^+ e^-\rangle \ \xrightarrow[ ]{\text{CP}} 
\ |\downarrow\uparrow\ \uparrow\uparrow\ \downarrow\downarrow\ \uparrow\downarrow\rangle 
\ =\ 
\ |e^+ e^-\ \ . e^-\ \ e^+ .\ \ . .\rangle
\ .
\label{eq:CP}
\end{align}
}
Ideally one wants to find the smallest pool of operators 
that is expressive enough to converge rapidly toward the vacuum.
For a lattice with OBCs, 
the system has translational symmetry in the volume that is broken by the boundaries (surface).
In the vacuum, 
the effects of the boundaries are expected to be localized, 
with penetration depths set by the mass gap.
Therefore, 
the pool of operators should contain 
translationally invariant ``volume" operators, 
and ``surface" operators that have support only near the boundaries.
In addition, a hierarchy is anticipated in which one-body operators 
are more important 
than two-body operators, 
two-body more important than three-body, and so on.\footnote{An $n$-body operator involves $n$ fermionic creation and $n$ fermionic annihilation operators.}
Note that because wavefunctions are evolved with ${\exp}(i \theta_i \hat{O}_i)$, 
arbitrarily high-body correlations are built from $n$-body operators 
(analogous to connected vs disconnected Feynman diagrams).
For the Schwinger model, we observe that one-body operators are sufficient.

With the above discussion as guidance, 
it is convenient to define two classes of one-body operators, 
one containing volume operators, 
and the other containing surface operators: 
\begin{align}
\hat{\Theta}_m^V &=   \frac{1}{2}\sum_{n=0}^{2L-1} (-1)^{n} \hat Z_n \ , \nonumber \\
\hat{\Theta}_{h}^V(d) &= \frac{1}{4}\sum_{n=0}^{2L-1-d} \left (\hat X_n \hat Z^{d-1} \hat X_{n+d} 
+ \hat Y_n \hat Z^{d-1} \hat Y_{n+d} \right ) \ ,  \nonumber \\
\hat{\Theta}_{m}^S(d) &= (-1)^d \frac{1}{2}\left ( \hat Z_d - \hat Z_{2L-1-d} \right ) 
\ , \nonumber \\
\hat{\Theta}_{h}^S(d) &= \frac{1}{4}\left (\hat X_1\hat Z^{d-1}\hat X_{d+1} + \hat Y_1\hat Z^{d-1}\hat Y_{d+1} 
\ + \  \hat X_{2L-2-d}\hat Z^{d-1}\hat X_{2L-2} + \hat Y_{2L-2-d}\hat Z^{d-1}\hat Y_{2L-2}\right )
\ .
\label{eq:PoolJW}
\end{align}
Unlabelled $\hat Z$s 
act on the qubits between the
leftmost and rightmost operators
(e.g., $\hat X_0 \hat Z^2 \hat X_3 = \hat X_0 \hat Z_1 \hat Z_2 \hat X_3$).
The first two operators in Eq.~(\ref{eq:PoolJW}) are translationally invariant,
$\hat{\Theta}_m^V$ is the mass term in the Hamiltonian,
and 
$\hat{\Theta}_{h}^V(d)$ is a generalized hopping term that spans an odd-number of fermion sites, $d$, connecting electrons and positrons at spatial sites separated by $\Delta L = (d-1)/2$. 
Only $d$-odd operators are retained, as the $d$-even operators break CP.
The second two operators in Eq.~(\ref{eq:PoolJW})
correspond to surface terms, of the form of a mass-density 
and of a hopping-density at and near the boundaries.
For $\hat{\Theta}_h^V(d)$, $d\in \{1,3,\ldots 2L-3\}$, and for  $\hat{\Theta}_{h}^S(d)$, $d\in \{1,3,\ldots 2L-5\}$,
preventing hopping between boundaries (which is found to improve convergence).

Time reversal symmetry implies that the vacuum wavefunction 
can be made real up to an overall phase. 
The SC-ADAPT-VQE ansatz is built from unitaries of the form $e^{i \theta_i \hat{O}_i}$, and furnishing a real wavefunction requires that the 
operators in the pool
are imaginary and anti-symmetric.
The operators in Eq.~(\ref{eq:PoolJW}) are real and are therefore disqualified from being members of the pool.
Instead, consider a pool comprised
of their commutators,\footnote{The commutators of $\hat{\Theta}$ operators not included in the pool are linear combinations of those that are.}
\begin{align}
\{ \hat{O} \} &= \left \{ \hat O_{mh}^{V}(d) \ , \ \hat O_{mh}^{S}(0,d) \ , \ \hat O_{mh}^{S}(1,d)
\right \}
\ ,\nonumber\\
 \hat O_{mh}^{V}(d) & \equiv i\left [\hat{\Theta}_m^V, \hat{\Theta}_{h}^V(d)\right ]
 = 
\frac{1}{2}\sum_{n=0}^{2L-1-d}(-1)^n\left (
\hat X_n\hat Z^{d-1}\hat Y_{n+d} - 
\hat Y_n\hat Z^{d-1}\hat X_{n+d} 
\right )
\ ,\nonumber\\
\hat O_{mh}^{S}(0,d) & \equiv i\left [\hat{\Theta}_{m}^S(0), \hat{\Theta}_{h}^V(d) \right ] \nonumber \\
 &= 
\frac{1}{4}\left (\hat X_0\hat Z^{d-1}\hat Y_{d} - \hat Y_0\hat Z^{d-1}\hat X_{d} 
- \hat Y_{2L-1-d}\hat Z^{d-1}\hat X_{2L-1} + \hat X_{2L-1-d}\hat Z^{d-1}\hat Y_{2L-1}\right ) 
\ ,\nonumber \\
 \hat O_{mh}^{S}(1,d) & \equiv i\left [\hat{\Theta}_{m}^S(1), \hat{\Theta}_{h}^S(d) \right ] \nonumber \\
 &= 
\frac{1}{4}\left (\hat Y_1\hat Z^{d-1}\hat X_{d+1} - \hat X_1\hat Z^{d-1}\hat Y_{d+1} 
+ \hat Y_{2L-2-d}\hat Z^{d-1}\hat X_{2L-2} - \hat X_{2L-2-d}\hat Z^{d-1}\hat Y_{2L-2} \right )
\ . 
\label{eq:poolComm}
\end{align}
While the contributions to extensive quantities from the 
volume operators,
$\hat O^{V}$,
typically scale as ${\cal O}(L)$,  the surface operators, $\hat O^{S}$,
make ${\cal O}(1)$ contributions as they are constrained to regions near the boundaries.\footnote{For the range of $m$ and $g$ we have considered, it was only necessary to consider $\hat{\Theta}_{m}^S(d)$ with $d=0,1$ in the pool.
Taking the continuum limit, where the correlation length diverges, will likely require keeping terms with $d>1$.} 
When acting on the strong-coupling vacuum, 
the exponential of an operator in the pool creates and annihilates $e^+ e^-$ pairs separated 
by distance $d$.
As the operators that are being considered are one-body, 
the variational algorithm is essentially building a coupled cluster singles (CCS) state (see, e.g., Refs.~\cite{RevModPhys.79.291,Hagen:2013nca}).

\section{Scalable Quantum Circuits from Classical Computing}
\label{sec:ClassSim}
\noindent
Integral to the application of SC-ADAPT-VQE 
is performing ADAPT-VQE on a series of systems that are large enough to enable a robust scaling of the parameterized circuits.
These scalable circuits can either be determined with classical computing, or by use of a smaller partition of a larger quantum computer.
In this section, 
SC-ADAPT-VQE is implemented using the ${\tt qiskit}$ noiseless classical simulator~\cite{IBMQ,qiskit}.

\subsection{Trotterized Quantum Circuits for the Scalable Operator Pool}
\noindent
As  discussed above, implementing the unitary operators in the pool, 
i.e., Eq.~\eqref{eq:TrotOp},
on classical simulators or quantum computers
requires mapping them to sequences of quantum gates.
For the individual terms in 
Eq.~\eqref{eq:poolComm}, 
we have chosen to do this using Trotterization.
The optimal gate decomposition 
is less important for implementation using a classical simulator,
but is crucial for successful simulations on a quantum computer.
With the goal of 
using IBM's superconducting-qubit quantum computers~\cite{IBMQ,qiskit},
our circuit designs aim to minimize two qubit gate count and circuit depth and require only nearest-neighbor connectivity.

\begin{figure}[t!]
    \centering
    \includegraphics[width=0.75\columnwidth]{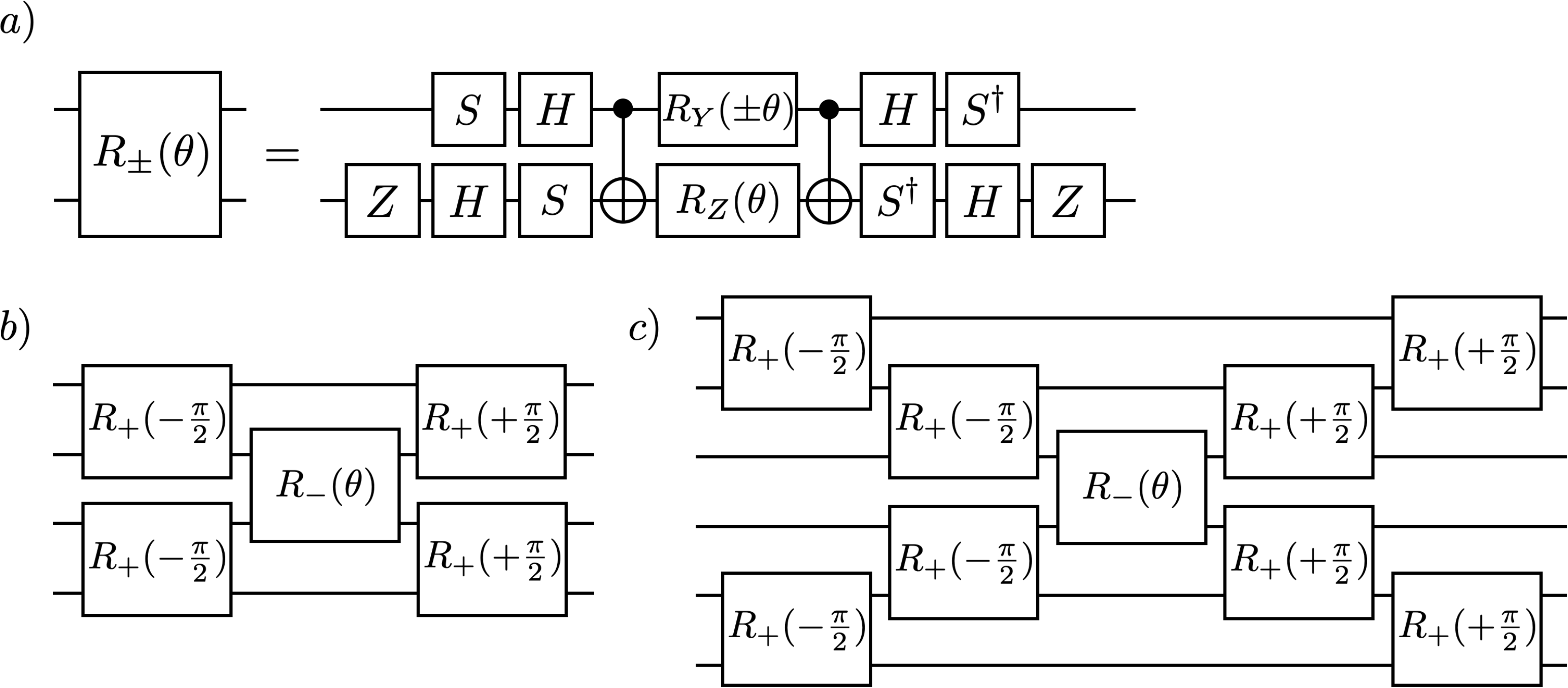}
    \caption{(a) Definition of the $R_{\pm}(\theta)$ gate, which
    implements ${\rm exp}[-i\theta/2 (\hat{Y}\hat{X}\pm \hat{X}\hat{Y})]$. 
    The $R_{\pm}(\theta)$ gate is used to implement 
    (b) ${\exp}[-i \theta/2 (\hat X\hat Z^2\hat Y - \hat Y\hat Z^2\hat X)]$ 
    and (c) ${\exp}[i \theta/2 (\hat X\hat Z^4\hat Y - \hat Y\hat Z^4\hat X)]$ (note the change in sign).}
    \label{fig:RpmOh35circ}
\end{figure}

\begin{figure}[t!]
    \centering
    \includegraphics[width=0.8\columnwidth]{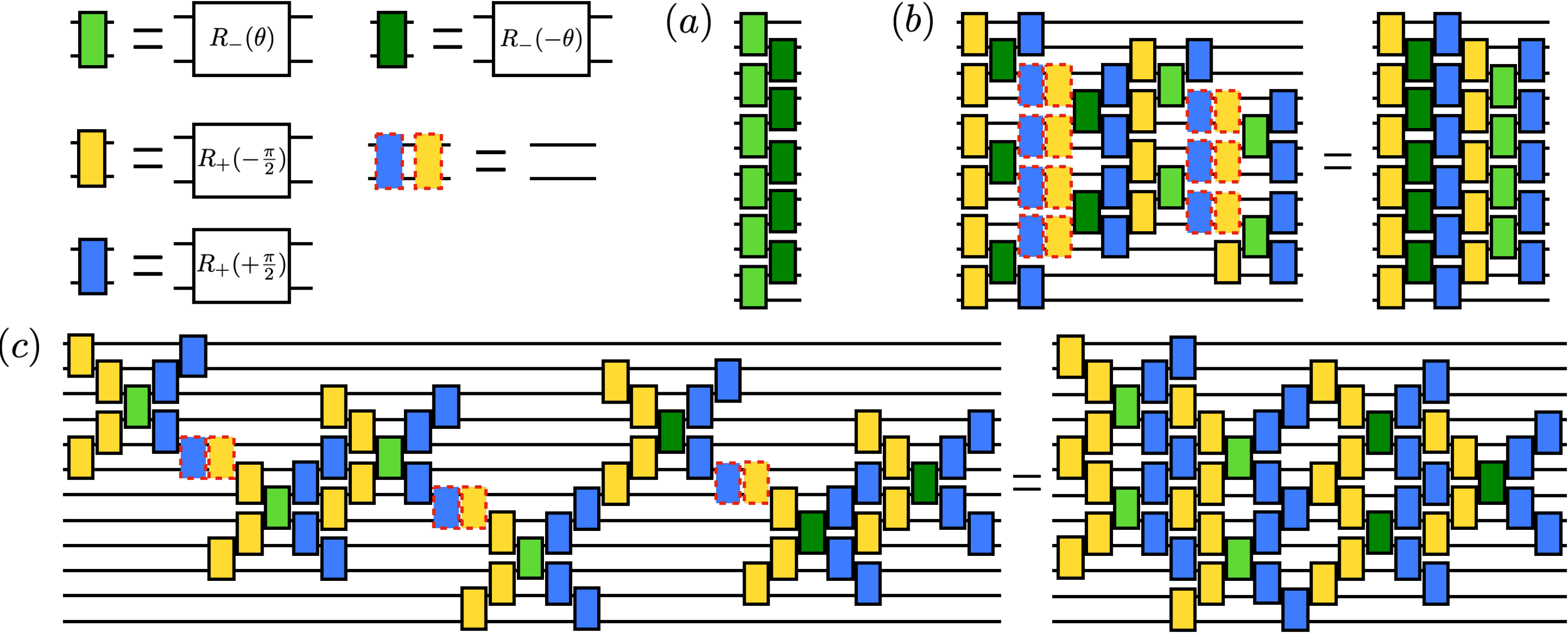}
    \caption{Simplifications of quantum circuits for the Trotterized unitaries corresponding to
    (a) $\hat O_{mh}^{V}(1)$, 
    (b) $\hat O_{mh}^{V}(3)$, 
    and (c) $\hat O_{mh}^{V}(5)$ for $L=6$, as explained in the main text. 
    Cancellations between $R_{+}(\pm \tfrac{\pi}{2})$ are highlighted with red-dash-outlined boxes.}
    \label{fig:ohd35multicirc}
\end{figure}

As can be seen in Eq.~\eqref{eq:poolComm}, 
each term in a given operator in the pool is of the form 
$(\hat{X} \hat{Z}^{d-1} \hat{Y} - \hat{Y} \hat{Z}^{d-1} \hat{X})$ 
for some odd value of $d$.
The construction of circuits implementing the corresponding unitary operators  follows the strategy outlined in Ref.~\cite{Algaba:2023enr}.
First, consider the Trotterization of terms with $d=1$, 
i.e., constructing a circuit corresponding to 
$e^{i \theta/2 (\hat{X}\hat{Y} \pm \hat{Y}\hat{X})} \equiv R_{\pm}(\theta) $.
There is a known 2-CNOT realization of this unitary operator~\cite{Algaba:2023enr}, 
shown in Fig.~\ref{fig:RpmOh35circ}a.
For terms with $d>1$, this circuit can be extended in an ``X" pattern as shown in Fig.~\ref{fig:RpmOh35circ}b and~\ref{fig:RpmOh35circ}c for $d=3$ and $d=5$, respectively.\footnote{
These circuits have been verified by comparison with 
Trotterized exponentials of fermionic operators.}
Terms with larger $d$ are constructed by extension of the legs of the ``X".
Compared with the traditional CNOT staircase-based circuits, 
there is a reduction by two CNOTs, 
and a reduction by a factor $\times 2$ in CNOT-depth.\footnote{The staircase circuit can be modified into an X-shaped one, reducing the depth, but with the same number of CNOTs~\cite{Cowtan:2019}.}
However, the primary advantage of these circuits is that they allow for an efficient arrangement of terms leading to cancellations among neighboring $R_+(\pm\tfrac{\pi}{2})$ gates.
As depicted in Fig.~\ref{fig:ohd35multicirc}, this is made possible by arranging the circuit elements so that sequential terms are offset by $d-1$ qubits, 
i.e., start on qubit $\{0,d-1,2(d-1),\ldots\}$.
This allows the outermost gates to cancel (using the identity in the upper left of Fig.~\ref{fig:ohd35multicirc}). 
Also, for $d\geq 5$, the next layer should start $(d-1)/2$ qubits below the previous one, 
as the circuit depth can be reduced
by interleaving the legs of the ``X".
Further optimizations are possible by noting that distinct orderings of terms, while equivalent up to higher order Trotter errors, can have different convergence properties; see App.~\ref{app:TrotterErrors}.
%

\subsection{
Building Scalable State Preparation Quantum Circuits using SC-ADAPT-VQE with Classical Computing
}
\noindent
In this section, SC-ADAPT-VQE is used to prepare approximations to the vacuum of the lattice Schwinger model on up to $L=14$ spatial sites (28 qubits) using classical simulations of the quantum circuits developed in the previous section (second step in Fig.~\ref{fig:SCADAPTVQE}).

%
\begin{table}[!ht]
\renewcommand{\arraystretch}{1.2}
\resizebox{\textwidth}{!}{\begin{tabularx}{1.3\textwidth}{|c || Y | Y || Y | Y || Y ||  Y |}
 \hline
 $L$ 
 & $\varepsilon^{\rm (aVQE)}$ 
 & $\varepsilon^{\rm (exact)}$ 
 & $\chi^{\rm (aVQE)}$ 
 & $\chi^{\rm (exact)}$ 
 & $\infiL$ & \text{\# CNOTs/qubit} 
  \\
 \hline\hline
 6 & -0.30772& -0.30791  & 0.32626& 0.32720 & 0.00010 & 31.2\\
 \hline
7 & -0.31097& -0.31117  & 0.32847& 0.32947 & 0.00011 & 33.6\\
\hline
8 & -0.31348& -0.31363  & 0.33036& 0.33118 & 0.00008 & 35.8\\
\hline
9 & -0.31539& -0.31553  & 0.33171& 0.33251 & 0.00008 & 37.1\\
\hline
10 & -0.31691& -0.31706  & 0.33279& 0.33358 & 0.00008 & 38.2\\
\hline
11 & -0.31816& -0.31831  & 0.33367& 0.33445 & 0.00008 & 39.1\\
\hline
12 & -0.31920& -0.31935  & 0.33441& 0.33517 & 0.00008 & 39.8\\
\hline
13 & -0.32008& -0.32023 &  0.33504& 0.33578 & 0.00008 & 40.5\\
\hline
14 & -0.32084 & -0.32098 & 0.33557 & 0.33631 &0.00008 & 41.0\\
 \hline
\end{tabularx}}
\caption{
Energy density, chiral condensate and wavefunction infidelity 
for the vacuum of the Schwinger model with $m=0.5,g=0.3$.
Both the results obtained from 7 steps of the SC-ADAPT-VQE (aVQE) algorithm using {\tt qiskit}'s classical simulator and the exact values
are given.
The last column shows the number of CNOTs/qubit in the state preparation circuit.}
 \label{tab:Echifmp5gp3}
\end{table}
\begin{figure}[!ht]
    \centering
    \includegraphics[width=\columnwidth]{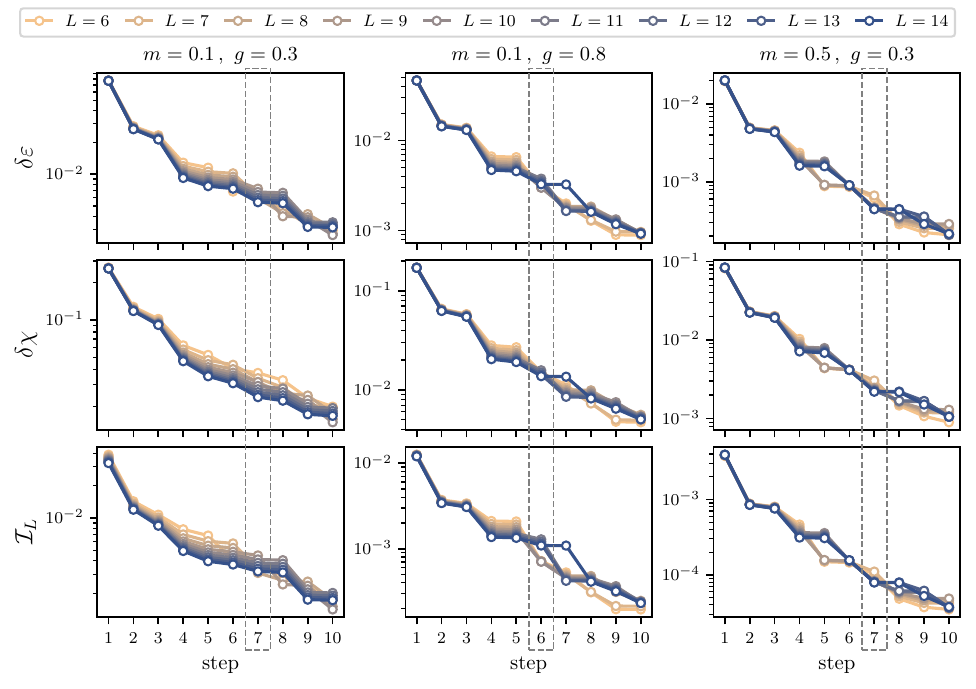}
    \caption{Deviations from the exact values of the energy density $\delta \varepsilon$, chiral condensate $\delta \chi$, and 
    wavefunction infidelity density $\infiL$ 
    obtained with SC-ADAPT-VQE. 
    The deviation in quantity ``$x$'' is defined 
    as $\delta x = |(x^{(\rm aVQE)}-x^{(\rm exact)})/x^{(\rm exact)}|$,
    where $x^{(\rm exact)}$ denotes the exactly calculated value at the same $L$. 
    Results are shown for $L=6$ to $L=14$   
    as a function of step number for $m=0.1, g=0.3$ (left column),  
    $m=0.1, g=0.8$ (center column) 
    and $m=0.5, g=0.3$ (right column).
    The numerical values for $m=0.5, g=0.3$ for the $7^{\rm th}$ step (highlighted with the dashed box) are given in Table~\ref{tab:Echifmp5gp3}, and the sequencing  of the corresponding Trotterized operators 
    and the variational parameters are given in Table~\ref{tab:AnglesXCircmp5gp3}.
    The corresponding results for $m=0.1,g=0.3$ ($7^{\rm th}$ step) and $m=0.1, g=0.8$ ($6^{\rm th}$ step) can be found in  App.~\ref{app:classData}.}
    \label{fig:ClassConvXCirc}
\end{figure}

In addition to the energy density and chiral condensate introduced in Sec.~\ref{sec:infVolExtra}, the infidelity density,
\begin{equation}
\infiL = \frac{1}{L} \left( 1 - 
|\langle\psi_{\rm ansatz}|\psi_{\rm exact}\rangle|^2 \right)
\ ,
\label{eq:fiddens}
\end{equation} 
is also studied, where $|\psi_{\rm exact}\rangle$ is the exact vacuum wavefunction on a lattice with $L$ spatial sites.
An infidelity density that is constant in $L$ corresponds to constant deviations in local observables evaluated in the prepared state.

To investigate the interplay between $L$ and 
$\xi=1/m_{\text{hadron}}$, 
three sets of parameters are considered: 
$m=0.1, g=0.3$ ($\xi_{L=14} = 2.6$), 
$m=0.1, g=0.8$  ($\xi_{L=14} = 1.3$) 
and $m=0.5, g=0.3$  ($\xi_{L=14} = 0.9$).
The $\xi$ are determined with exact diagonalization, and are found to weakly depend on $L$.
Note that increasing either $m$ or $g$ decreases the correlation length.
To make systematically improvable predictions 
of observables from the QFT that emerges from a given lattice model, 
extrapolations to the continuum (lattice spacing to zero) and infinite-volume ($L \to \infty)$ limits must be performed.
This requires that the relevant correlation length(s) are all much greater than the lattice spacing, $\xi \gg 1$ in lattice units, but are well contained in the lattice volume, $L\gg\xi$.
We primarily focus on extrapolation to large lattices, and therefore only require $L\gg \xi$.
As a result, the parameter set $m=0.5, g=0.3$ is used as the primary example throughout this work.

The values of $\varepsilon$, $\chi$ and $\infiL$ obtained at 
the $7^{\rm th}$ step of SC-ADAPT-VQE with $m=0.5, g=0.3$ are given in 
Table~\ref{tab:Echifmp5gp3},
while their deviations from the exact values 
are shown in Fig.~\ref{fig:ClassConvXCirc},
as a function of increasing number of SC-ADAPT-VQE steps.
The corresponding numerical values obtained from the other parameter sets are presented in App.~\ref{app:classData}.\footnote{The $6^{\rm th}$ and $7^{\rm th}$ steps were chosen for study in detail as the operator ordering has stabilized for $L\leq 14$. This allows the operator structure to be displayed in a single table, and enables the systematic extrapolation of parameters. 
The available classical computing resources limited the maximum number of steps of SC-ADAPT-VQE to 10.}
As seen by their approximately linear behavior in the log-plots in Fig.~\ref{fig:ClassConvXCirc}, 
the error in each of these quantities decreases exponentially with algorithm step,
indicating convergence to the target wavefunction.
This exponential trend is demonstrated out to 10 steps, reaching a convergence comparable to the systematic errors introduced in the $L$-extrapolations below.
This provides evidence that this choice of initial state and operator pool does not suffer from ``barren plateaus" or local minima.
For a given step in the algorithm, 
the error is seen to become independent of system size.
This indicates that 
extrapolations of the circuits to arbitrarily large systems will have errors that are {\it independent of $L$}.
As discussed above, it is expected that SC-ADAPT-VQE will converge more rapidly for systems with smaller correlation lengths.
This is indeed seen in Fig.~\ref{fig:ClassConvXCirc}, 
where the correlation length decreases from left to right, while the convergence improves.
Also included in Table~\ref{tab:Echifmp5gp3} is the number of CNOTs per qubit in the SC-ADAPT-VQE circuit. 
It is seen to  scale as a constant plus a subleading ${\mathcal O}(1/L)$ term,
leading to an asymptotic value of 48 CNOTs per qubit.
This scaling is due to there being $(2L-d)$ terms in each volume operator.

The structure of the SC-ADAPT-VQE state preparation circuit and the corresponding variational parameters for $m=0.5$ and $g=0.3$ 
are given in Table~\ref{tab:AnglesXCircmp5gp3}.
Notice that initially localized operators are added to the wavefunction (small $d$), 
followed by increasingly longer-range ones, as well as surface operators. Systems with longer correlation lengths require larger $d$ operators (e.g., compare Table~\ref{tab:AnglesXCircmp5gp3} and Table~\ref{tab:AnglesXCircmp1gp3}), 
in line with previous discussions on the exponential decay of correlations for $d > \xi$.
It is also seen that the surface operators become less important 
(appear later in the ansatz structure) for larger lattices.
For example, as shown in Table~\ref{tab:AnglesXCircmp5gp3}, the $5^{{\rm th}}$ step of SC-ADAPT-VQE transitions from being a surface to a volume operator at $L=10$ (causing the jump in convergence at the fifth step in the right column of Fig.~\ref{fig:ClassConvXCirc}).
This is expected as they contribute ${\mathcal O}(1/L)$ 
to the energy density, whereas volume operators contribute ${\mathcal O}(1)$.

\begin{figure}[!b]
    \centering
    \includegraphics[width=0.65\columnwidth]{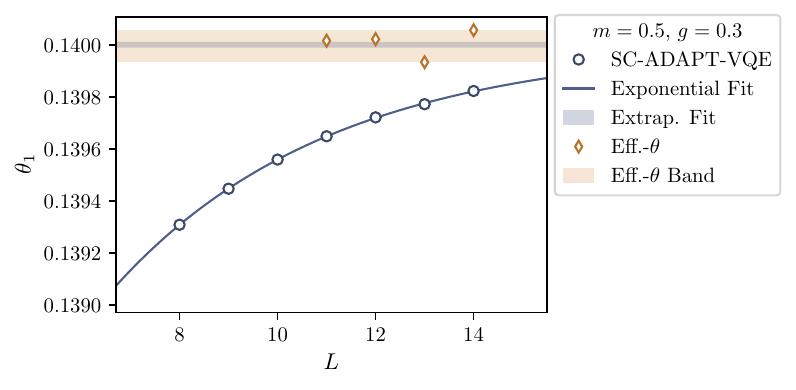}
    \caption{
    Example of fitting the asymptotic $L$-dependence of a parameter defining the SC-ADAPT-VQE state-preparation circuit.
    The results for $\theta_1$, corresponding to evolving by $\hat O_{mh}^{V}(1)$ (blue circles),   determined from classical simulations,
    for 
    $m=0.5$ and $g=0.3$ given in Table~\ref{tab:AnglesXCircmp5gp3},
    are extrapolated to $L=\infty$ by 
i) use of a 3-parameter fit given in Eq.~\eqref{eq:thetaextrap}, 
as shown by the blue line,
with an asymptotic value shown by the blue region,
and by
ii) the forming of effective-$\theta$ (orange diamonds) defined in Eq.~\eqref{eq:LdifsB},
with the maximum and minimum values shown as the orange shaded region.
}
    \label{fig:ThetaLextrap53}
\end{figure}
Importantly, Table~\ref{tab:AnglesXCircmp5gp3} shows that the order of operators, 
and the corresponding variational parameters are converging with increasing system size (third step in Fig.~\ref{fig:SCADAPTVQE}). 
This is due to exponentially decaying correlations for $d \gg \xi$, 
and it is expected that the variational parameters will also converge exponentially,
once $L$ is sufficiently large to contain $\xi$, and we assume the following form:
\begin{equation}
\theta_i \ = \ \theta_i^{L=\infty} \ + \  c_1\,e^{-c_2 \, L} 
\ .
\label{eq:thetaextrap}
\end{equation}
Table~\ref{tab:AnglesXCircmp5gp3} shows that this convergence sets in 
for $L>7$,\footnote{The ordering of operators changes at $L=10$ but the operator content is unchanged, so it is still possible to use $L=8,9$ in the extrapolation.} 
and the variational parameters extrapolated to $L=\infty$ are given in the last row of Table~\ref{tab:AnglesXCircmp5gp3}.
These are used in the next section to initialize the vacuum on 
lattices up to $L=500$.
An example of 
extrapolating the variational parameters
is shown in Fig.~\ref{fig:ThetaLextrap53} for 
the parameter $\theta_1$,
associated with
$\hat O_{mh}^{V}(1)$.
The exact results obtained for $L\le 14$ are well reproduced and 
extrapolated with the exponential functional form in Eq.~(\ref{eq:thetaextrap}).\footnote{
One could imagine generating the $\theta^{L=\infty}_i$ for a variety of $m$ and $g$, and then machine learning the variational parameters for all $m$ and $g$.
This could be particularly useful for $m$ and $g$ that approach the continuum limit, where the correlation length can no longer be contained within lattice volumes accessible to classical simulators.}
A more complete discussion of the parameter extrapolations, along with
examples for $m=0.1$ and $g=0.3$ and for $m=0.1$ and $g=0.8$, can be found in App.~\ref{app:ThetaScaling}.

\begin{table}[!t]
\renewcommand{\arraystretch}{1.4}
\resizebox{\textwidth}{!}{\begin{tabularx}{1.3\textwidth}{|c || Y | Y | Y | Y || Y |  Y| Y | Y | Y |}
 \hline
 \diagbox[height=23pt]{$L$}{$\theta_i$} & $\hat O_{mh}^{V}(1)$
 & $\hat O_{mh}^{V}(3)$ & $\hat O_{mh}^{V}(5)$ & $\hat O_{mh}^{V}(1)$ & $\hat O_{mh}^{V}(7)$ &  $\hat O_{mh}^{S}(0,1)$ & 
 $\hat O_{mh}^{V}(7)$  &
 $\hat O_{mh}^{V}(1)$ &
 $\hat O_{mh}^{S}(0,3)$ \\
 \hline\hline
 6 & 0.18426 & -0.03540 & 0.00731 & 0.11866 & -- & 0.06895 & -0.00182 & -- & -0.03145\\
 \hline
 7 & 0.18440 & -0.03574 & 0.00729 & 0.11864 & -- & 0.06867 & -0.00177 & -- & -0.03066\\
 \hline
 8 & 0.13931 & -0.03727 & 0.00760 & 0.08870 & -- & 0.06925 & -0.00183 & 0.07457 & --\\
 \hline
 9 & 0.13945 & -0.03714 & 0.00755 & 0.08849 & -- & 0.06904 & -0.00180 & 0.07473 & --\\
 \hline
 10 & 0.13956 & -0.03703 & 0.00752 & 0.08832 & -0.00178 & 0.06888 & -- & 0.07485 & --\\
 \hline
 11 & 0.13965 & -0.03695 & 0.00749 & 0.08819 & -0.00177 & 0.06875 & -- & 0.07494  & --\\
 \hline
 12 & 0.13972 & -0.03688 & 0.00747 & 0.08808 & -0.00176 & 0.06865 & -- & 0.07502 & -- \\
 \hline
 13 & 0.13977 & -0.03683 & 0.00745 & 0.08800 & -0.00175 & 0.06856 & -- & 0.07508 & --\\
 \hline
14&0.13982& -0.03678&0.00744&0.08793&-0.00174&0.06849&--&0.07513&--\\
\hline
 \hline
  $\infty$ &0.1400 & -0.0366 & 0.0074 & 0.0877 & -0.0017 & 0.0682 & -- & 0.0753 & --\\
 \hline
\end{tabularx}}
\caption{Structure of the ansatz wavefunction with $m=0.5$ and $g=0.3$ 
through 7 steps of the SC-ADAPT-VQE algorithm 
obtained from a classical simulation using {\tt qiskit}. 
For a given $L$, the order that the operators are added to the ansatz is displayed from
left to right, with the associated parameter, $\theta_i$, given as the entry in the table.
The  operators,
$\hat O_{mh}^{V}(d_h)$ and $\hat O_{mh}^{S}(d_m,d_h)$,
are defined in Eq.~(\ref{eq:poolComm}).
An entry of  `` -- '' means that the operator does not contribute.
The bottom row corresponds to an extrapolation to $L=\infty$ as detailed in Eq.~\eqref{eq:thetaextrap}.
}
 \label{tab:AnglesXCircmp5gp3}
\end{table}
%

\section{Preparing the Vacuum of the Schwinger Model on Large Lattices}
\label{sec:vacum_class_quan}
\noindent
The vacuum preparation circuits, determined for $L\leq14$ with SC-ADAPT-VQE using an exact (statevector) classical simulator, are scaled to prepare the vacuum on much larger lattices. 
These scaled circuits are used to prepare the vacuum on lattices of up to $L=500$ (1000 qubits) using a classical MPS circuit simulator
and up to $L=50$ (100 qubits) using IBM's {\tt Eagle}-processor quantum computers (fourth step in Fig.~\ref{fig:SCADAPTVQE}).
We emphasize that this scaling requires no further optimization of the circuits.
The chiral condensate and energy density determined from the classical simulator
are found to be consistent with DMRG calculations.
On the quantum computers, the chiral condensate and charge-charge correlators 
are measured
to probe the quality of one- and two-qubit observables.
The results are in agreement with those from the classical MPS simulator, within statistical uncertainties.

\subsection{
Classical Simulation}
\label{sec:scaling}
\noindent
Very large quantum circuits that do not generate long-range entanglement can be efficiently simulated using the 
{\tt qiskit} {\tt matrix\_product\_state} classical simulator.
Here it is used to simulate the preparation of the Schwinger model vacuum 
on $L\gg 14$ lattices,
applying the scalable circuits 
determined in the previous section
from 7 steps of SC-ADAPT-VQE on $L\le 14$ lattices.
The values obtained for the chiral condensate and energy density up to $L=500$ are compared with DMRG results, and are presented in Table~\ref{tab:SQCResults}.
The deviations in the energy density ($\sim 1\times 10^{-4}$) and chiral condensate ($\sim 1\times 10^{-3}$) are in good agreement with what was found for smaller $L$; 
see Table~\ref{tab:Echifmp5gp3}.
This demonstrates that the systematic errors in the vacuum wavefunctions prepared with the scaled quantum circuits are (approximately) independent of $L$ over this range of lattice volumes.\footnote{The $6^{{\rm th}}$ operator in the extrapolation is a surface operator, whose contribution to the energy density scales as $1/L$. 
Therefore, if SC-ADAPT-VQE could be performed on, for example, $L=500$, this operator would likely not be in the ansatz.
Evidently the ``error" introduced by extrapolating the ansatz with a surface operator is small since the deviation of observables for large $L$ is the same as for $L\leq14$.}
The scaled circuits corresponding to $m=0.1, g=0.3$ and $m=0.1, g=0.8$ have also been used to successfully prepare the vacuum.
However, due to the larger correlation lengths, 
MPS calculations with $L\gtrsim 100$ required excessive classical resources, and were not performed.  See App.~\ref{app:classData} for more details.
\begin{table}[!t]
\renewcommand{\arraystretch}{1.4}
\begin{tabularx}{\textwidth}{|c || Y | Y || Y | Y |}
 \hline
 $L$ 
 & $\varepsilon^{\rm (SC-MPS)}$ 
 & $\varepsilon^{\rm (DMRG)}$ 
 & $\chi^{\rm (SC-MPS)}$  
 & $\chi^{\rm (DMRG)}$ \\
 \hline\hline
 50 & -0.32790 & -0.32805 &0.34044 & 0.34123\\
 \hline
 100 & -0.32928 & -0.32942 & 0.34135 & 0.34219\\
 \hline
200  & -0.32996 & -0.33011 & 0.34181 & 0.34267 \\
\hline
300 & -0.33019 & -0.33034 & 0.34196 & 0.34282 \\
\hline
400 & -0.33031 & -0.33045 & 0.34204 &0.34291\\
\hline
500 & -0.33038 & -0.33052 & 0.34209 & 0.34296 \\
 \hline
\end{tabularx}
\caption{
Results for large lattices with $m=0.5, g=0.3$ through 7 steps of SC-ADAPT-VQE 
using circuits scaled from $L\leq14$.
The superscript ``SC-MPS'' denotes the results obtained 
from the scaled circuits
using the {\tt qiskit} MPS classical simulator,
and the superscript ``DMRG'' denotes results obtained from DMRG calculations.
}
 \label{tab:SQCResults}
\end{table}

It is worth summarizing what has been accomplished 
in this work with classical simulations:
\begin{itemize}
    \item In Sec.~\ref{sec:SchwingerHam}, the vacuum energy density and chiral condensate were determined exactly for 
    $L\le 14$ (28 staggered lattice sites) using exact diagonalization, and for $L\le 10^3$ using DMRG.
    The results for $L\ge 9$ were (consistently) extrapolated to $L\rightarrow\infty$, with $1/L$ scaling.
    \item In Sec.~\ref{sec:ClassSim}, SC-ADAPT-VQE, based on the scalable operator pool determined in Sec.~\ref{sec:II}, was performed on $L\leq 14$ lattices.
    Intensive quantities were found to converge exponentially with circuit depth, and the errors in these quantities, as well as the structure of the state preparation circuits, were found to become independent of $L$.
    This enabled the variational parameters defining the state preparation circuits to be consistently extrapolated to arbitrarily large $L$.
    \item 
    In this section,
    the quantum circuits corresponding to 7 steps of SC-ADAPT-VQE were scaled and applied to large lattices using the {\tt qiskit} MPS circuit simulator.  
    The deviations of the energy density and chiral condensate computed from these wavefunctions were found to be independent of $L$, i.e., consistent with $L\le 14$.
\end{itemize}
These main points indicate that the quantum circuits determined classically with SC-ADAPT-VQE can be used to prepare the vacuum 
of the Schwinger model 
on quantum computers at scale with a precision that is independent of system size.

\subsection{Quantum Simulations on 100 Qubits using IBM's Quantum Computers}
\label{sec:SCQuSim}
\noindent
The quantum circuits determined via classical simulation on $L\le 14$ lattices  
are now scaled to larger $L$ to
prepare the vacuum of the Schwinger model 
on up to 100 qubits of 
IBM's 127 superconducting-qubit 
{\tt Eagle} quantum computers with heavy-hexagonal communication fabric.
Hamiltonian parameters $m=0.5, g=0.3$ with $L=14,20,30,40,50$,
and 
state preparation circuits scaled from 2 steps of SC-ADAPT-VQE 
(compared to 7 steps in the previous section), are used.
Fewer steps equates to shallower circuits, and a preliminary study of the performance of the computer with more steps can be found in App.~\ref{app:qusimDetail}.
The variational parameters extrapolated to the chosen range of $L$ for 2 steps of SC-ADAPT-VQE are given in 
Table~\ref{tab:angles_ibm} in
App.~\ref{app:qusimDetail}.

The large number of qubits and two-qubit gates involved in these 
simulations make error mitigation essential to obtain reliable estimates of observables.
Specifically, this work uses readout-error mitigation (REM), dynamical decoupling (DD), Pauli twirling (PT), and decoherence renormalization. 
The {\tt qiskit} Runtime Sampler primitive is used to obtain readout-corrected quasi-distributions via the matrix-free measurement mitigation (M3) from Ref.~\cite{Nation:2021kye}. 
Also included in the primitive is DD, which is used to suppress crosstalk and idling errors~\cite{Viola:1998dsd,2012RSPTA.370.4748S,Ezzell:2022uat}.
Crucial to the error mitigation is decoherence renormalization~\cite{Urbanek_2021,ARahman:2022tkr,Farrell:2022wyt,Ciavarella:2023mfc}, 
modified in this work for simulations on a large number of qubits, which we call {\it Operator Decoherence Renormalization} (ODR).
Underpinning decoherence renormalization is PT~\cite{Wallman:2016nts}, which turns coherent two-qubit gate errors into incoherent errors, which can be inverted to recover error-free expectation values.
Unlike previous applications of decoherence renormalization, 
which assume a constant decoherence across the device, 
ODR estimates the decoherence separately for each operator.
This is done by running a mitigation circuit, which has the same 
operator structure as the one used to extract the observables, but with the noise-free result being known {\it a priori}. 
We choose the state preparation circuits with the variational parameters 
set to zero for mitigation,
and 
in the absence of noise this prepares the strong coupling vacuum, $\lvert \Omega_0 \rangle$ in Eq.~\eqref{eq:SCvac}.
Naively, it could be expected that post-selecting results on states with 
total charge $Q = 0$ would eliminate the leading bit-flip errors~\cite{Klco:2019evd}.
However, when post-selection is combined with ODR, which accommodates single-qubit decoherence, undesirable correlations between qubits are introduced.
We find that performing both mitigation techniques 
(post-selection and ODR)
degrades the quality of two-qubit observables, 
and post selection is not used in this work as it is found to be less effective.
More details about ODR and post-selection can be found in App.~\ref{app:QSimError}.

\begin{figure}[!t]
    \centering
    \includegraphics[width=\columnwidth]{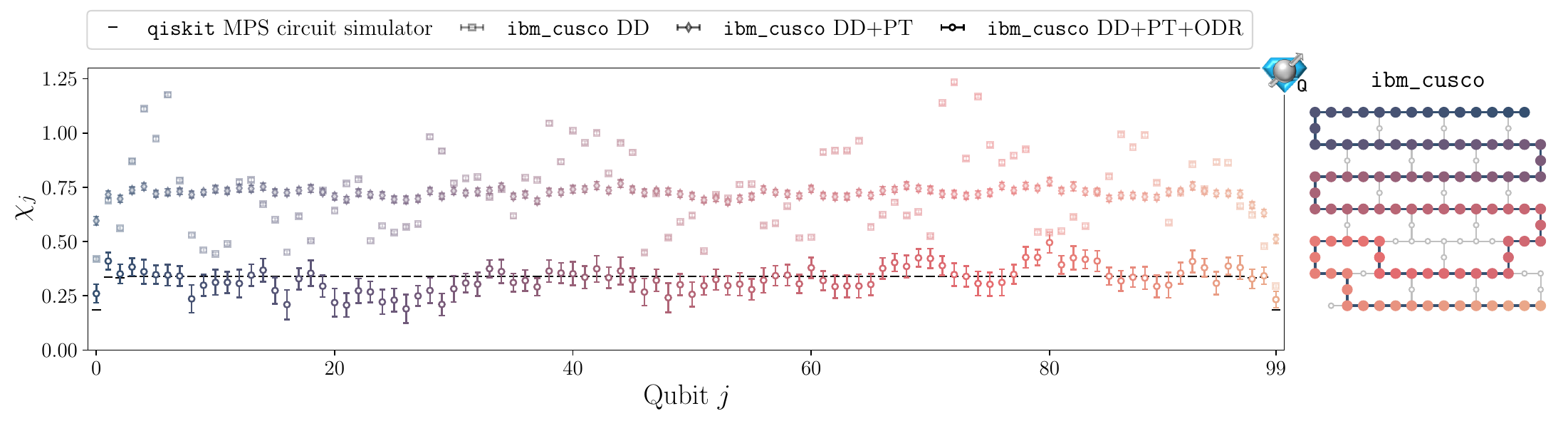}
    \caption{Local chiral condensate $\chi_j$ for $L=50$, as obtained from 
    IBM's Eagle-processor quantum computer
    {\tt ibm\_cusco} after different steps of error mitigation: DD (squares), PT (diamonds), and ODR (circles).
    This is compared with the expected results obtained from the {\tt qiskit} MPS circuit simulator (black dashes). 
    Averaging $\chi_j$ over all of the  
    qubits (including at the boundaries) gives the chiral condensate presented in Table~\ref{tab:SQCResults_ibm}.
    The layout of the qubits used on the processor is shown on the right.
    These results were obtained by performing 150 Pauli twirls, each involving $8\times 10^3$ shots
    for the physics circuits and the corresponding mitigation circuits.
    The blue icon in the upper right indicates that this calculation was done on a quantum device~\cite{Klco:2019xro}.}
    \label{fig:chi_50}
\end{figure}

The local chiral condensate, $ \chi_j $ in Eq.~(\ref{eq:chiralCond}), 
obtained from {\tt ibm\_cusco} for $L=50$ is shown in Fig.~\ref{fig:chi_50},
where the subscript ``$j$" denotes the qubit index.\footnote{
For all of the results presented in this work, 
correlated bootstrap re-sampling was used to estimate statistical (shot) uncertainties.
The circuits used for $L\leq 40$ were executed on {\tt ibm\_brisbane} with 
40 Pauli-twirled instances for both the mitigation and the physics circuits, 
each with $8\times10^3$ shots. 
For $L=50$, the M3 method was not applied due to the large overhead in classical computing, 
and production was executed on {\tt ibm\_cusco} with 150 Pauli-twirled instances.
Additional details can be found in App.~\ref{app:qusimDetail}.}
Three different sets of results (in different stages of error mitigation) are shown: with only DD applied (squares), with DD and PT applied (diamonds), and after ODR (circles). Looking at the results with only DD (squares), it is seen that the noise is not uniform across the device, signaling a significant contribution of coherent noise. After PT (diamonds), this coherent noise is averaged out, and is transformed into incoherent (depolarizing) noise, seen by the almost-constant shift of the results compared with the MPS simulation. Finally, ODR removes this shift by mitigating the effects of depolarizing noise. More details on the interplay between these methods can be found in App.~\ref{app:qusimDetail}.

With the statistics and twirlings gathered,
the $1\sigma$ uncertainties in each point are $\sim 15\%$ of their mean, 
and each $ \chi_j $ is within $3 \sigma$ of the MPS simulator result (the individual values of $\chi_j$ can be CP averaged to reduce the uncertainty, as shown in Fig.~\ref{fig:chi_14-50_CP} in App.~\ref{app:qusimDetail}).
It is expected that these uncertainties will reduce with increased statistics 
and twirlings.
Notice that the expected values of 
$ \chi_j $ deviate from the volume average 
for only a few qubits near the boundaries.
This is because the boundaries  perturb the wavefunction only over a few correlation lengths, leaving the rest of the volume unaffected.
The chiral condensates for $L=14,20,30,40$ and 50 are given in Table~\ref{tab:SQCResults_ibm}.
This is an average over the whole lattice, Eq.~\eqref{eq:chiralCond}, and therefore
the uncertainty decreases with increasing $L$
due to increased sampling.
Despite having smaller uncertainties, 
the results remain within $3\sigma$ of the MPS simulator result.
Also given in Table~\ref{tab:SQCResults_ibm} is the number of two-qubit CNOT gates.
The number of CNOTs is seen to grow linearly with $L$, without affecting the quality of the result, and 788 CNOTs over 100 qubits is well within the capabilities of the quantum computer.
This is in line with other quantum simulations that have been performed with large numbers of qubits and CNOTs using IBM's quantum computers~\cite{Yu:2022ivm,Kim:2023bwr,Shtanko:2023tjn}.
\begin{table}[!t]
\renewcommand{\arraystretch}{1.4}
\begin{tabularx}{\textwidth}{|c |c |c || Y | Y | Y | Y |}
 \hline
 $L$ & Qubits & CNOTs
 & $\chi^{\rm (SC-{\tt IBM})}$ before ODR 
 & $\chi^{\rm (SC-{\tt IBM})}$  after ODR
 & $\chi^{\rm (SC-MPS)}$
 & $\chi^{\rm (DMRG)}$ \\
 \hline\hline
 14 & 28 & 212 & 0.491(4)  & 0.332(8) & 0.32879 & 0.33631 \\
 \hline
 20 & 40 & 308 & 0.504(3) & 0.324(5) & 0.33105 & 0.33836 \\
 \hline
 30 & 60 & 468 & 0.513(2) & 0.328(4) & 0.33319 & 0.33996 \\
\hline
 40 & 80 & 628 & 0.532(2) & 0.334(3) & 0.33444 & 0.34075 \\
\hline
 50 & 100 & 788 & 0.721(2) & 0.326(3) & 0.33524 & 0.34123 \\
 \hline
\end{tabularx}
\caption{
Chiral condensate in the Schwinger model vacuum 
obtained from {\tt ibm\_brisbane} ($L\leq 40$) and {\tt ibm\_cusco} ($L=50$) for large lattices with $m=0.5, g=0.3$ using the scaled circuits from two steps of SC-ADAPT-VQE.
The values before and after application of ODR are given in columns four and five. 
Column six gives results obtained from running the two step SC-ADAPT-VQE circuits on an MPS classical simulator (the noiseless result), while column seven gives the results from DMRG calculations.}
\label{tab:SQCResults_ibm}
\end{table}
This highlights the fact that it is not the total number of CNOT gates in the
quantum circuit that is limiting the scale of simulations, but rather it is the number of CNOT gates per qubit. 
This, of course, assumes that the CNOT gates in a single layer of the circuit can be enacted in parallel.
Due to this, increasing $L$ actually improves volume-averaged quantities by $\sim 1/\sqrt{L}$ due to statistical averaging.
In a similar vein, since scalable circuits repeat structures of size $\xi$ many times over the whole lattice, 
the number of Pauli-twirls being sampled is effectively multiplied by $L/\xi$.

\begin{figure}[!t]
    \centering
    \includegraphics[width=0.95\columnwidth]{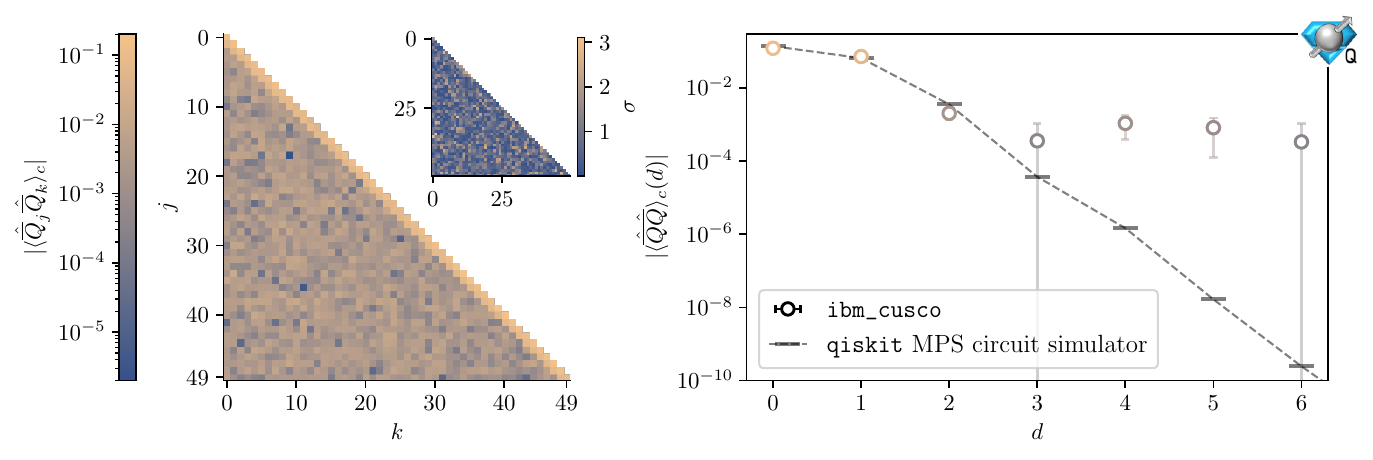}
    \caption{
    Left: Connected contributions to the spatial charge-charge correlation functions, 
    $\langle \hat{\overline{Q}}_{j} \hat{\overline{Q}}_{k} \rangle_c$, 
    for $L=50$ 
    (the inset shows the number of standard deviations by which the results obtained from {\tt ibm\_cusco} deviate from the MPS simulator results).
    Right: Volume averaged correlation functions as a function of distance $d$, 
    $\langle \hat{\overline{Q}} \hat{\overline{Q}} \rangle_c (d)$, with the points following the same color map as in the left main panel (error bars show $1\sigma$ standard deviations).}
    \label{fig:qiqj_40}
\end{figure}

To further probe the quality of the prepared wavefunctions, correlations between electric charges on the spatial sites are considered.
The charge on a spatial site is defined 
to be the sum of charges on the two associated staggered sites, 
$\hat{\overline{Q}}_{k} = \hat{Q}_{2k} + \hat{Q}_{2k+1}$,
where $k$ is an integer corresponding to the spatial site.
Of particular interest are connected correlation functions between spatial charges,\footnote{For periodic boundary conditions, $\langle\hat{ \overline{Q}}_k\rangle  = 0$, but for OBCs $\langle\hat{ \overline{Q}}_k\rangle$ decays exponentially away from the boundaries; see App.~\ref{app:LinvScaling}.} defined as
\begin{equation}
\langle \hat{\overline{Q}}_{j} \hat{\overline{Q}}_{k} \rangle_c \ = \ \langle \hat{\overline{Q}}_{j} \hat{\overline{Q}}_{k} \rangle \ - \
\langle \hat{\overline{Q}}_{j}\rangle  \langle\hat{\overline{Q}}_{k} \rangle \ .
\end{equation}
These correlations decay exponentially for $\lvert j-k\rvert \gtrsim \xi$ due to confinement and charge screening.
Unlike the chiral condensate, which is a sum of single qubit observables, $\langle \hat{\overline{Q}}_{j} \hat{\overline{Q}}_{k} \rangle_c$ is sensitive to correlations between qubits, i.e., requires measurement of $\langle \hat{Z}_j \hat{Z}_k \rangle$.
The results from {\tt ibm\_cusco}  for $L=50$ are shown in Fig.~\ref{fig:qiqj_40}.
The correlations are symmetric under $j\leftrightarrow k$, 
and only the lower-triangle of the correlation matrix is shown.
Each measured value is within $3\sigma$ of the MPS simulator result, consistent with statistical fluctuations.
Also shown in Fig.~\ref{fig:qiqj_40} are the spatial charge-charge correlations as a function of distance, averaged over the lattice volume,
\begin{equation}
\langle \hat{\overline{Q}} \hat{\overline{Q}} \rangle_c (d) \ = \ \frac{1}{L-4-d}\sum_{k=2}^{L-3-d} \langle \hat{\overline{Q}}_k \hat{\overline{Q}}_{k+d} \rangle_c \ .
\end{equation}
To reduce the effects of the boundaries, this sum omits the first and last two spatial lattice sites.
As anticipated, this correlation function decays exponentially, with a characteristic length scale proportional to $\xi = 1/m_{\text{hadron}}$.
For $d>2$, the correlations are consistent with zero within $2\sigma$ (note that the log scale distorts the error bars),
and 
increased numbers of shots and twirlings 
are needed to distinguish additional points from zero.
The local chiral condensate and charge-charge correlations corresponding to the other values of $L$ are given in App.~\ref{app:qusimDetail}.

\section{Summary and Outlook}
\label{sec:Conclusions}
\noindent
In this chapter, the vacuum of the lattice Schwinger model was prepared 
on up to 100 qubits of IBM's 
127-qubit {\tt Eagle}-processor quantum computers,
{\tt ibm\_brisbane} and {\tt ibm\_cusco}.
This was accomplished with SC-ADAPT-VQE, an algorithm for identifying systematically improvable
state preparation quantum circuits that can be robustly scaled to operate on any number of qubits
The utility of scalable circuits relies on physically relevant systems often having a (discrete) translational symmetry, and a finite correlation length set by the mass gap.
Together, these imply that the state preparation circuits have unique structure over approximately a correlation length~\cite{Klco:2019yrb,Klco:2020aud},
which is replicated across the lattice.
The lattice Schwinger model with OBCs was chosen to explore these ideas as its vacuum has (approximate) translational invariance and, due to confinement, has a mass gap.
By performing SC-ADAPT-VQE on a classical simulator, state preparation circuits for lattices of $L\leq 14$ (28 qubits) were built from an operator pool containing both translationally invariant terms and ones localized to the boundaries.
Exponential convergence in the quality of the prepared state with both system size and circuit depth enabled the extrapolation of circuits that can be scaled to arbitrarily large lattices.
This methodology was successfully demonstrated by preparing the Schwinger model vacuum on up to 100 superconducting qubits of IBM's quantum computers.
Both the charge-charge correlators and the chiral condensate were measured, and were found to agree with results from an MPS simulator, within statistical uncertainty.
Vital to the success of these quantum simulations involving a large number of qubits was the development of an improved error mitigation technique, which we have called Operator Decoherence Renormalization (ODR).

Due to its generality, 
we expect that the scalable circuit framework embodied by SC-ADAPT-VQE can be applied to other gapped theories with translationally-invariant ground states.
Of particular importance is QCD,  
for which the initialization of ground states for quantum simulations continues to be a daunting prospect.
It is likely that many of the ideas used to construct efficient state preparation circuits for the Schwinger model can be applied to the initialization of the ground state of QCD.
Of course, the operator pool that informs the state preparation circuits will be  
more diverse since the  gauge field is no longer completely constrained by Gauss's law.
Local quark-field operators, extended quark operators with associated gauge links, and closed loops of gauge links will need to be included in the pool.
It is also expected that the variational parameters defining the 
ground-state preparation circuits will converge exponentially, 
once the simulation volume can completely contain the pion(s).

The utility of SC-ADAPT-VQE is that it provides a straightforward prescription for determining low-depth quantum circuits that prepare the ground state on systems of arbitrary size with only classical computing overhead.
This not only allows for the quantum simulation of ground state properties, but will be important for future simulations of dynamics, where preparing the initial state is a crucial first step.
In the following chapter hadron dynamics will be simulated by first using SC-ADAPT-VQE to prepare hadron wavepackets on top of the vacuum, and then evolving them forward in time.

\clearpage
\begin{subappendices}

\clearpage
\section{Volume Extrapolation of the Energy Density and Chiral Condensate}
\label{app:LinvScaling}
\begin{figure}[ht!]
    \centering
    \includegraphics[width=0.7\columnwidth]{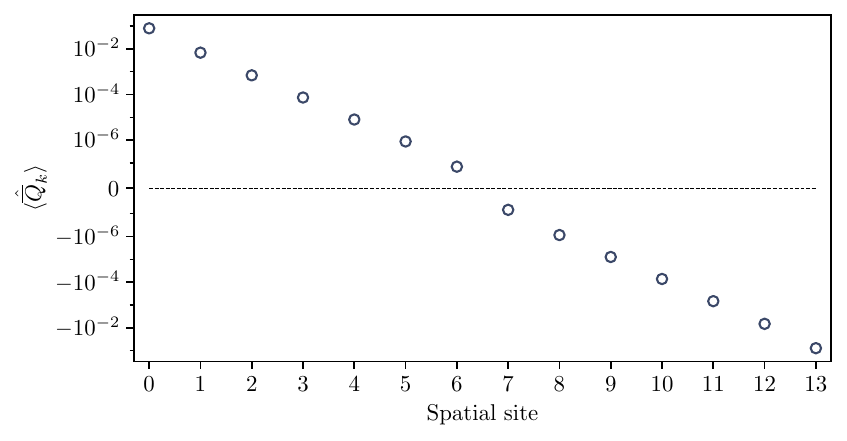}
    \caption{Charge on each spatial site, $\hat{\overline{Q}}_k$, for $m=0.5,g=0.3$ and $L=14$ obtained from exact diagonalization of the Hamiltonian. 
}
    \label{fig:Qboundary}
\end{figure}

\noindent
Here the vacuum energy density and chiral condensate are extrapolated to $L=\infty$.
The results of exact diagonalization and DMRG calculations are considered independently, providing consistent results within uncertainties. For the DMRG calculations, 60 sweeps were performed with a maximum allowed bond dimension of 150 and a truncation of Schmidt coefficients below $10^{-10}$. This showed a convergence of $10^{-10}$ in the energy of the vacuum state.
Discussions in Sec.~\ref{sec:infVolExtra} motivated an inverse-power, $1/L$, dependence 
of the exact vacuum energies as the infinite-volume limit is approached. 
This scaling was argued when $L$ is much larger than the longest correlation length, and with OBCs.
Therefore, for masses and couplings that give rise to the lowest-lying hadron 
being completely contained within the lattice volume,
we anticipate functional forms
\begin{align}
 \varepsilon(L) & = \varepsilon(\infty)\ +\ \frac{e_1}{L}\ +\ {\cal O}\left(\frac{1}{L^2}\right) \ ,\qquad\ 
 \chi(L)  = \chi(\infty)\ +\ \frac{d_1}{L}\ +\ {\cal O}\left(\frac{1}{ L^2}\right)
 \ ,
\end{align}
for $\varepsilon$ and $\chi$.
This is due to the finite penetration depth of boundary effects, and the exponential convergence of both the volume and the surface contributions to their infinite-volume values.
As a result, the surface terms make 
${\cal O}\left(1/L\right)$ contributions to intensive quantities, e.g., densities.
To illustrate this, the expectation value of the charge on each spatial site, $\hat{\overline{Q}}_k$, for $m=0.5,g=0.3$ and $L=14$ is shown in Fig.~\ref{fig:Qboundary}.
This converges exponentially with the distance to the boundary to $\langle \hat{\overline{Q}}_k \rangle = 0$, the expected infinite volume value.

\begin{figure}[ht!]
    \centering
    \includegraphics[width=\columnwidth]{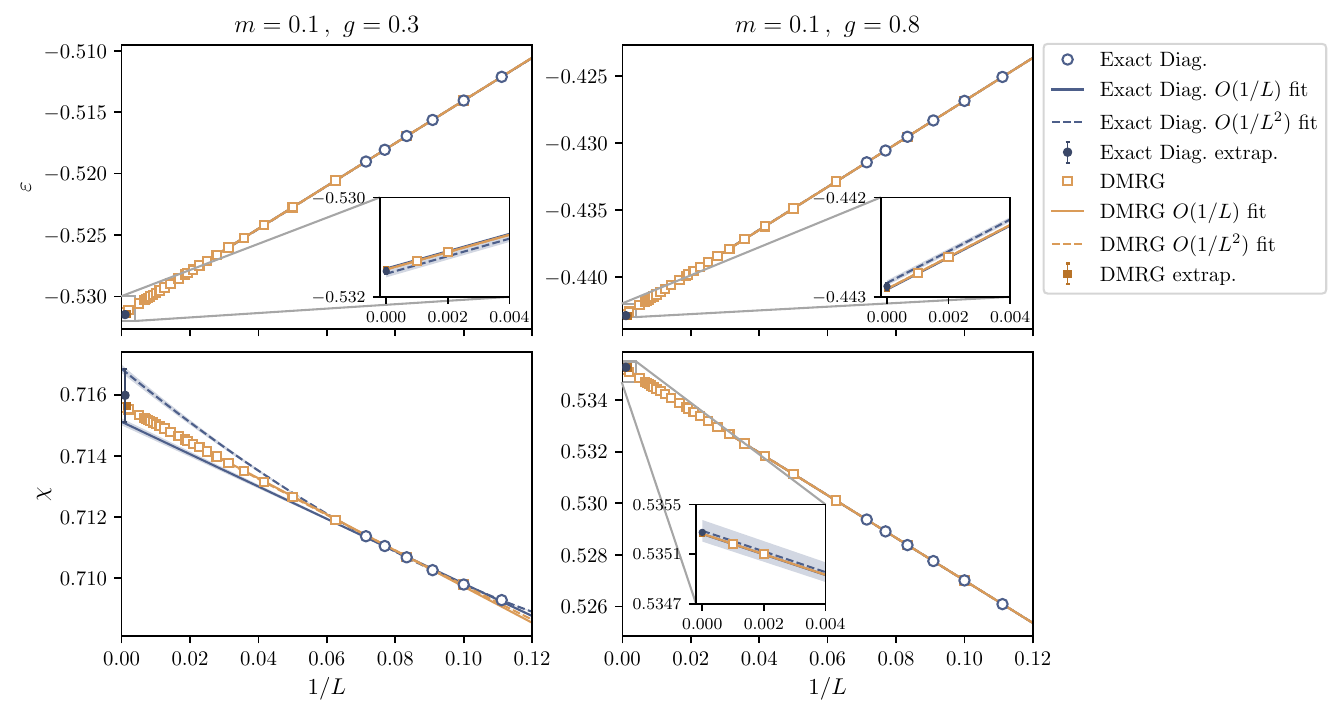}
    \caption{$L$-extrapolations of the vacuum energy density $\varepsilon$ (top) and chiral condensate $\chi$ (bottom) for $m=0.1$ and $g=0.3$ (left) and $m=0.1$ and $g=0.8$ (right).
    Each panel shows extrapolations of the exact values given in Tables~\ref{tab:Echifmp1gp3} and~\ref{tab:Echifmp1gp8} (blue circles) and of the results from DMRG calculations (orange squares) given in Table~\ref{tab:DMRGresults} for $L\ge 9$.
    The solid lines correspond to linear extrapolations and the dashed lines correspond to quadratic extrapolations, and are found to overlap (see the insets).
    The darker points show the $L=\infty$ extrapolated value, with an uncertainty determined by the difference between the linear and quadratic extrapolations.
}
    \label{fig:EdensityLextrap}
\end{figure}
The results of fits to the exact and DMRG results for the energy density and chiral condensate for $m=0.5, g=0.3$ are shown in Fig.~\ref{fig:EdChiLextrap53}, 
and for $m=0.1, g=0.8$ and $m=0.1, g=0.3$ are shown in Fig.~\ref{fig:EdensityLextrap}.
Using polynomials 
that are linear and quadratic in $1/L$,
fits  are performed for $L\geq 9$ and extrapolated to $L=\infty$.
The differences between extrapolations obtained from the two fit forms are used to 
estimate the systematic fitting error, 
corresponding to the black and grey points (and error bars).
The difference between linear and quadratic fits is negligible for the exact results, except 
for the chiral condensate 
in the case of $m=0.1$ and $g=0.3$, which sees a small quadratic dependence.  
When the fit interval is reduced to $L\ge 10$, this dependence once again becomes negligible.

\begin{table}[!ht]
\renewcommand{\arraystretch}{0.9}
\begin{tabularx}{\textwidth}{|c || Y | Y || Y | Y || Y | Y |}
\hline
 & \multicolumn{2}{c||}{$m=0.1, g=0.3$} & \multicolumn{2}{c||}{$m=0.1, g=0.8$} & \multicolumn{2}{c|}{$m=0.5, g=0.3$} \\ \hline
 $L$ & $\varepsilon^{(\text{DMRG})}$ & $\chi^{(\text{DMRG})}$ & $\varepsilon^{(\text{DMRG})}$ & $\chi^{(\text{DMRG})}$ & $\varepsilon^{(\text{DMRG})}$ & $\chi^{(\text{DMRG})}$ \\
 \hline
 \hline
10   & -0.51405 & 0.70979 & -0.42685 & 0.52701 & -0.31707 & 0.33358 \\ \hline
12   & -0.51694 & 0.71068 & -0.42953 & 0.52838 & -0.31936 & 0.33517 \\ \hline
16   & -0.52057 & 0.71190 & -0.43288 & 0.53010 & -0.32221 & 0.33716 \\ \hline
20   & -0.52275 & 0.71265 & -0.43488 & 0.53114 & -0.32393 & 0.33836 \\ \hline
24   & -0.52420 & 0.71315 & -0.43622 & 0.53182 & -0.32508 & 0.33916 \\ \hline
28   & -0.52523 & 0.71350 & -0.43718 & 0.53232 & -0.32589 & 0.33973 \\ \hline
32   & -0.52601 & 0.71377 & -0.43790 & 0.53268 & -0.32650 & 0.34016 \\ \hline
36   & -0.52661 & 0.71398 & -0.43846 & 0.53297 & -0.32698 & 0.34049 \\ \hline
40   & -0.52710 & 0.71414 & -0.43890 & 0.53320 & -0.32736 & 0.34075 \\ \hline
44   & -0.52749 & 0.71428 & -0.43927 & 0.53339 & -0.32768 & 0.34097 \\ \hline
48   & -0.52782 & 0.71439 & -0.43957 & 0.53354 & -0.32794 & 0.34115 \\ \hline
52   & -0.52810 & 0.71449 & -0.43983 & 0.53368 & -0.32816 & 0.34130 \\ \hline
54   & -0.52823 & 0.71453 & -0.43994 & 0.53374 & -0.32825 & 0.34137 \\ \hline
60   & -0.52855 & 0.71464 & -0.44024 & 0.53389 & -0.32851 & 0.34155 \\ \hline
70   & -0.52896 & 0.71479 & -0.44062 & 0.53408 & -0.32883 & 0.34178 \\ \hline
80   & -0.52927 & 0.71489 & -0.44091 & 0.53423 & -0.32908 & 0.34195 \\ \hline
90   & -0.52952 & 0.71498 & -0.44113 & 0.53435 & -0.32927 & 0.34208 \\ \hline
100  & -0.52971 & 0.71504 & -0.44131 & 0.53444 & -0.32942 & 0.34219 \\ \hline
110  & -0.52987 & 0.71510 & -0.44146 & 0.53451 & -0.32955 & 0.34228 \\ \hline
120  & -0.53000 & 0.71514 & -0.44158 & 0.53458 & -0.32965 & 0.34235 \\ \hline
130  & -0.53011 & 0.71518 & -0.44168 & 0.53463 & -0.32974 & 0.34241 \\ \hline
140  & -0.53021 & 0.71521 & -0.44177 & 0.53467 & -0.32981 & 0.34246 \\ \hline
150  & -0.53029 & 0.71524 & -0.44185 & 0.53471 & -0.32988 & 0.34251 \\ \hline
200  & -0.53058 & 0.71534 & -0.44212 & 0.53485 & -0.33011 & 0.34267 \\ \hline
500  & -0.53110 & 0.71552 & -0.44260 & 0.53510 & -0.33052 & 0.34295 \\ \hline
1000 & -0.53128 & 0.71558 & -0.44276 & 0.53518 & -0.33066 & 0.34305 \\ \hline
\end{tabularx}
\caption{Additional results for the energy density $\varepsilon$ and chiral condensate $\chi$ obtained from DMRG calculations, and used in the extrapolations in Figs.~\ref{fig:EdChiLextrap53} and~\ref{fig:EdensityLextrap}.
}
 \label{tab:DMRGresults}
\end{table}

\clearpage
\section{Optimizing Trotterized Circuits for State Preparation}
\label{app:TrotterErrors}
\noindent
As discussed in the main text, even after the operator pool has been chosen for SC-ADAPT-VQE, there remains freedom in how the pool of unitary operators, Eq.~(\ref{eq:TrotOp}), 
is implemented as quantum circuits.
For example, 
instead of leading-order Trotterization,
a higher-order Trotterization could be used to suppress Trotter errors. 
Alternatively, 
different orderings of the terms in the leading-order Trotterization can be considered.
This freedom can be used to optimize the convergence of SC-ADAPT-VQE with circuit depth.
Also, different Trotter orderings can break the CP symmetry.
The circuit orderings in Fig.~\ref{fig:ohd35multicirc} were chosen to minimize the circuit depth, and for $d=1,3,5$ this ordering preserves CP, while for $d=7,9$ it breaks CP.

\begin{figure}[ht!]
    \centering
    \includegraphics[width=\columnwidth]{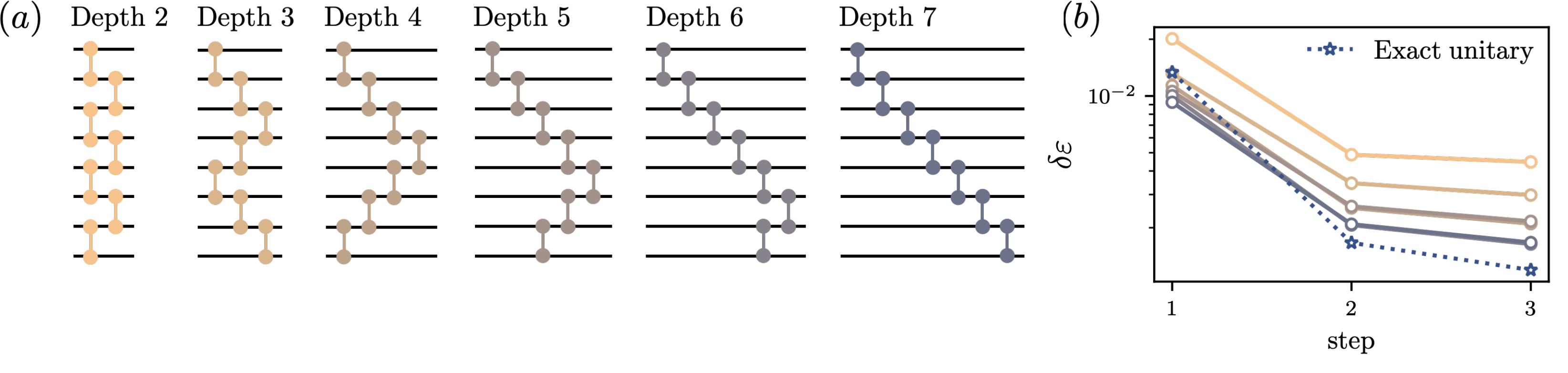}
    \caption{(a) Orderings of Trotterized terms 
    for implementing $\hat{O}^{V}_{mh}(1)$. 
    Circuits of depth 2 to 7 are shown from left to right, with the dumbbells representing the circuit in Fig.~\ref{fig:RpmOh35circ}a.
    (b) Deviations in the energy density of the SC-ADAPT-VQE prepared state for $m=0.5, g=0.3$ and $L=10$
    with different depth implementations of the Trotterization of $\hat{O}^{V}_{mh}(1)$.
    }
    \label{fig:PyramidTrotter}
\end{figure}

Consider the different arrangements of the terms in the Trotterization of $\hat{O}^{V}_{mh}
(1)$, given in Eq.~(\ref{eq:poolComm}), as shown in Fig.~\ref{fig:PyramidTrotter}a.
The depth-2 ordering (left) was used 
to obtain the results presented 
in the main text as it leads to the shallowest circuits.
All the orderings shown
in Fig.~\ref{fig:PyramidTrotter}a
are equivalent up to $\mathcal{O}[(\theta_1)^2]$ 
(where $\theta_1$ is 
the coefficient of the operator in the corresponding unitary operator),
but the deeper circuits allow for the generation of longer-range correlations. 
Note that the deeper circuits can break the CP symmetry; e.g. for $L=10$ depths 2 and 4 preserve CP while depths 3, 4, 5 and 7 break CP.
It is found that this added circuit depth improves the convergence of SC-ADAPT-VQE, as shown in Fig.~\ref{fig:PyramidTrotter}b.
This demonstrates that to minimize circuit depth, for a fixed error threshold, it is preferable to choose a deeper Trotterization of $\hat{O}^{V}_{mh}(1)$, 
instead of 
going to a greater number of SC-ADAPT-VQE steps.
For example, it is more efficient to perform 2 steps of SC-ADAPT-VQE with a depth-3 Trotterization of $\hat{O}^{V}_{mh}(1)$, than to perform 3 steps of SC-ADAPT-VQE with a depth-2 Trotterization of $\hat{O}^{V}_{mh}(1)$.
Also shown in Fig.~\ref{fig:PyramidTrotter}b are results obtained 
from performing SC-ADAPT-VQE with exact unitary operators (no Trotterization).
This is found to always perform better than the Trotterized unitaries, 
except for a single step.
Intriguingly, for a single step, the error is less with a deep first-order Trotterization than with the exact unitary.
This suggests that the optimizer is finding a solution in which the Trotter errors are tuned to improve the overlap with the vacuum.
Note that the deeper Trotterizations of $\hat{O}^{V}_{mh}(1)$ move the recurrence of $\hat{O}^{V}_{mh}(1)$ (e.g., at step 4 for $m=0.5, g=0.3$) to deeper in the SC-ADAPT-VQE ansatz.

\clearpage
\section{Volume Extrapolations of the SC-ADAPT-VQE Variational Parameters: 
An ``effective-\texorpdfstring{$\theta_i^{\infty}$}{theta infinity}''}
\label{app:ThetaScaling}
\noindent
To initialize large quantum registers, the variational parameters defining the state-preparation quantum circuits need to be extrapolated with high precision. 
In volumes large enough to contain the longest correlation length, 
the variational parameters are expected to be exponentially close to their infinite-volume values.
Therefore, we assume that the form of the volume dependence for practical purposes is that given in Eq.~(\ref{eq:thetaextrap}),
\begin{equation}
\theta_i(L) \ = \ \theta_i^{\infty} \ + \  c_1\,e^{-c_2 \, L} 
\ ,
\label{eq:thetaextrapAPP}
\end{equation}
and check the self-consistency of this form.\footnote{For the current paper, due to the small number of parameters, the selection of the points to be fitted was determined by visual inspection (if the points followed an exponential decay or not).}
While there could be a polynomial coefficient of the exponential, 
we find that this is not required.
Fitting exponential functions can be challenging; 
however, with results over a sufficient range of $L$, algebraic techniques, such as effective masses, have proven useful in lattice QCD calculations to eliminate ``uninteresting'' parameters, while at the same time mitigating correlated fluctuations in measurements~\cite{Michael:1985ne,Luscher:1990ck,DeGrand1990FromAT,Fleming:2004hs,Beane:2009kya}.
With the goal of initializing large lattices, it is the
$\theta_i^{\infty}$ that are of particular interest.

Assuming the volume dependence given in Eq.~(\ref{eq:thetaextrapAPP}),
it is useful to form four relations
\begin{align}
y_L & = \theta_i(L) \ - \ \theta_i^{\infty}\ = \ c_1\,e^{-c_2 \, L} 
\ \ ,\ \ 
y_{L+1}\ =\ \theta_i(L+1) \ - \ \theta_i^{\infty}\ = \ c_1\, e^{-c_2} e^{-c_2 \, L} 
\nonumber \\
y_{L+2} & = \theta_i(L+2) \ - \ \theta_i^{\infty}\ = \ c_1\, e^{-2 c_2} e^{-c_2 \, L} 
\ \ ,\ \ 
y_{L+3}\ =\ \theta_i(L+3) \ - \ \theta_i^{\infty}\ = \ c_1\,e^{-3 c_2} e^{-c_2 \, L} 
\ .
\label{eq:LdifsA}
\end{align}
These relations can be combined to isolate $\theta_i^{\infty}$, providing an $L$-dependent ``effective-$\theta_i^{\infty}$'', denoted as $\theta_{i, {\rm eff}}^{\infty}$:
\begin{align}
& y_{L+1} y_{L+2} \  =\  y_{L}  y_{L+3} \ ,
\nonumber \\
& \theta_{i, {\rm eff}}^{\infty}(L) \  =\  
\frac{\theta_i(L) \theta_i(L+3) - \theta_i(L+1)\theta_i(L+2)}{
\theta_i(L) + \theta_i(L+3) - \theta_i(L+1) - \theta_i(L+2)
}
\ .
\label{eq:LdifsB}
\end{align}
For a sufficiently large set of results, 
$\theta_{i, {\rm eff}}^{\infty}(L)$ 
will plateau for large $L$ if the functional form in 
Eq.~(\ref{eq:thetaextrapAPP}) correctly describes the results.
This plateau can be fit by a constant,
over some range of large $L$,
to provide an estimate of $\theta_i^{\infty}$.
This method is similar to using {\tt varpro} (variable projection) in a multi-parameter 
$\chi^2$-minimization.

\begin{figure}[ht!]
    \centering
    \includegraphics[width=0.9\columnwidth]{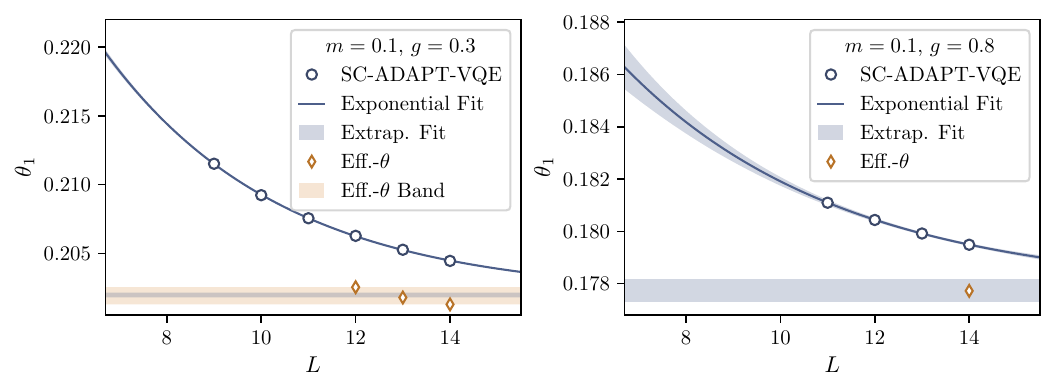}
    \caption{
    Further examples of fitting the asymptotic $L$-dependence of the variational parameters defining the state-preparation quantum circuit, determined from classical simulations using SC-ADAPT-VQE.
    The results for $\theta_1=\hat O_{mh}^{V}(1)$ (blue points) for 
    $m=0.1$ and $g=0.3$ (left panel) given in Table~\ref{tab:AnglesXCircmp1gp3} and
    $m=0.1$ and $g=0.8$ (right panel) given in Table~\ref{tab:AnglesXCircmp1gp8}
    are extrapolated to $L=\infty$ by 
i) use of a 3-parameter fit given in Eq.~(\ref{eq:thetaextrap}), as shown by the blue line and shaded region, 
with an asymptotic value shown by the gray region,
and by
ii) forming of effective-$\theta$ (orange diamonds) 
with the maximum and minimum values shown as the orange shaded region (where possible).
}
    \label{fig:ThetaLextrap}
\end{figure}

As an example, the results for $\theta_1^{\infty}$ 
from a 3-parameter fit of $\theta_1$ to Eq.~\eqref{eq:thetaextrapAPP} are compared with a determination using
$\theta_{1, {\rm eff}}^{\infty}(L)$  from
Eq.~(\ref{eq:LdifsB}).
Results obtained with these two methods for $m=0.1, g=0.3$ and for $m=0.1, g=0.8$ are shown in Fig.~\ref{fig:ThetaLextrap}.
The result obtained 
from fitting a constant to $\theta_{1, {\rm eff}}^{\infty}(L)$ 
is consistent with the asymptotic result from the 3-parameter fit, but with somewhat larger uncertainty.
The current deficiency of this comparison is the small number of points in the plateau region, and results for larger $L$ are required for a more complete comparison.
Analysis of the other variational parameters shows a similar behavior.

The consistency between the two extraction methods is encouraging, and suggests that the selected exponential form may indeed well describe the results.
The fitting method is likely insensitive to polynomial corrections (coefficients), and requires further exploration to fully-quantify uncertainties in these asymptotic values of the variational parameters.
However, as the MPS simulations with these extrapolated angles reproduce the results calculated with DMRG, it appears that, for the systems and parameters we have selected in our analysis, systematic errors introduced by selecting this functional form are small.

\clearpage
\section{Operator Decoherence Renormalization (ODR)}
\label{app:QSimError}
\noindent
To mitigate the effects of noise, the decoherence renormalization technique~\cite{Urbanek_2021,ARahman:2022tkr,Farrell:2022wyt,Ciavarella:2023mfc} 
is modified for use with larger systems. 
In its original form, 
decoherence renormalization assumes that each qubit decoheres at the same rate under a depolarizing noise channel.
When working with a small number of qubits, this is a reasonable approximation, but for larger systems, it is necessary to consider the 
rate of decoherence of each qubit individually.
After Pauli twirling, the qubit errors are well described by a Pauli error channel~\cite{PhysRevA.72.052326}, which maps the $N$ qubit density matrix to
\begin{equation}
\rho \ \to  \ \sum_{i=1}^{4^N} \eta_i \hat{P}_i \rho \hat{P}_i \ , 
\end{equation} 
where $\hat{P}_i$ is a tensor product of Pauli operators ($\hat{I}$, $\hat{X}$, $\hat{Y}$ or $\hat{Z}$) acting on $N$ qubits, and the set of $\eta_i$ characterizes the error channel. It is important to understand the effect of this error channel on observables. Generic observables can be written as a sum over tensor products of Pauli operators, so it suffices to consider an observable, $\hat{O}$, that is a tensor product of Pauli operators. Under a Pauli error channel, the measured (noisy) expectation value, $\langle \hat{O}\rangle_{\text{meas}}$, is given by
\begin{equation}
\langle \hat{O} \rangle_{\text{meas}} \ =  \sum_{i=1}^{4^N} \eta_i\Tr( \hat{P}_i \hat{O} \hat{P}_i \rho) \ .
\end{equation}
Note that $\hat{P}_i \hat{O} \hat{P}_i = \pm \hat{O}$, depending on whether or not $\hat{O}$ and $\hat{P}_i$ commute or anti-commute. Using this fact, the measured (noisy) expectation value $\langle \hat{O}\rangle_{\text{meas}}$, can be seen to be directly proportional to the predicted (noiseless) expectation value, $\langle \hat{O}\rangle_{\text{pred}} = {\Tr}(\hat{O} \rho)$, i.e.,
\begin{equation}
\langle \hat{O} \rangle_{\text{meas}} \ =  \left ( 1-\eta_O\right ) \langle \hat{O }\rangle_{\text{pred}} \ .
\label{eq:etaO}
\end{equation}

The ODR factor $\eta_O$ is, in general, 
distinct for each operator, and can be estimated by running a mitigation circuit that has the same structure as the physics circuit, but where $\langle \hat{O} \rangle_{\text{pred}}$ is already known.
In this work, the mitigation circuit was taken to be the state preparation circuit with variational parameters set to zero, which is the identity in the absence of noise.
This mitigation circuit will have the same noise channel as the physics circuit provided that the noise is dominated by errors in the two-qubit gates and is independent of the single qubit rotation angles in the circuit. Without noise, the mitigation circuit prepares the strong coupling vacuum, where  $\langle \hat{O} \rangle_{\text{pred}}$ is known, 
and therefore $\eta_O$ can be computed.
Once $\eta_O$ is determined, Eq.~\eqref{eq:etaO} is used to estimate the  value of the noiseless observable from the 
results of the physics circuits.

An added benefit of ODR is that it reduces the need for other error mitigation techniques.
For example, readout errors are 
partially mitigated since the measured observables are affected by both gate and measurement errors.
This is convenient as current measurement mitigation techniques require a large classical computing overhead.
It also reduces the need for post-selection, which in our work could have been performed on states with total charge $Q = 0$.
This post-selection removes single-qubit errors, but introduces further correlations between qubits. These correlations effectively increase the size of the 
single-qubit errors (making observables sensitive to errors anywhere on the register).
This reduces the efficacy of the Pauli error model, making post-selection incompatible with ODR.\footnote{This is not true for observables involving the entire qubit register, 
e.g., the vacuum-vacuum persistence probability. 
This is because applying the $Q=0$ constraint when measuring global observables will not introduce any new correlations.}
Another desirable feature of ODR is that it allows simulations to retain the results of 
a much larger fraction of the ensemble. 
This is because 
the probability of a single-qubit error increases with system size, 
and therefore much of the ensemble is lost with naive post selection.
Further, such errors have little effect on local observables that are summed across the entire qubit register.

\clearpage
\section{Additional Results From Classical Simulations}
\label{app:classData}
\noindent
The results corresponding to Fig.~\ref{fig:ClassConvXCirc} for $m=0.1,g=0.3$ are given in Table~\ref{tab:AnglesXCircmp1gp3} and Table~\ref{tab:Echifmp1gp3}, and for
$m=0.1,g=0.8$ are given in Table~\ref{tab:AnglesXCircmp1gp8} and Table~\ref{tab:Echifmp1gp8}.
The $6^{\rm th}$ step of the algorithm is chosen for $m=0.1, g=0.8$ because the operator structure through $L=14$ has converged, allowing a consistent extrapolation of the circuits to large $L$.
This can be seen by comparing the operator structure in Table~\ref{tab:AnglesXCircmp1gp8} (6 steps) and Table~\ref{tab:AnglesXCircmp1gp8Step7} (7 steps). An interesting observation is that the sum of parameters for a particular operator in the ansatz remains approximately unchanged when an additional insertion of the operator is added. 
For example, compare the sum of parameters for $\hat{O}^V_{mh}(1)$ between $L=8$ and $9$ in Table~\ref{tab:AnglesXCircmp1gp3}.
Using the same method as for $m=0.5, g=0.3$ in Sec.~\ref{sec:ClassSim}, scalable circuits for $m=0.1, g=0.3$ and $m=0.1$ and $g=0.8$ were also determined.
The results of running these circuits on {\tt qiskit}'s MPS simulator for $m=0.1, g=0.3$ and $m=0.1$ and $g=0.8$ are given in Table~\ref{tab:SQCResultsApp}.
Due to the longer correlation lengths for these parameters, it was not possible to go to $L=500$ with the available computing resources.
In these MPS simulations, {\tt qiskit}'s default settings were used, where the bond dimension increases until machine precision is achieved.
The details of the qiskit MPS simulator can be found on the qiskit website~\cite{qiskitMPS}.
Again, the energy density and chiral condensate are found to have precision comparable to 
that found on smaller systems.
This shows that, despite the longer correlation lengths for $m=0.1, g=0.3$ and $m=0.1,g=0.8$, it is still possible to accurately extrapolate the state preparation circuits to large lattices. 
Note that stabilization of operator ordering for the different $m$ and $g$ (see Tables \ref{tab:AnglesXCircmp5gp3}, \ref{tab:AnglesXCircmp1gp3} and \ref{tab:AnglesXCircmp1gp8Step7}) does not follow the hierarchy in correlation lengths. 
This is because larger $\xi$ increases both the contribution of the volume $\sim e^{-d/\xi}$ and surface $\sim \xi/L$ terms to the energy density.

To emphasize the advantage of performing SC-ADAPT-VQE using a classical simulator, we give an estimate of the number of shots required to perform SC-ADAPT-VQE on a quantum computer.
For $m=0.5,g=0.3, L=14$ performing 10 steps of SC-ADAPT-VQE required $\sim 6000$ calls to the optimizer, in addition to about $500$ evaluations of $\langle [\hat{H}, \hat{O}_i ] \rangle$ for pool operators $\hat{O}_i$.
Each one of these calls required roughly $10^{-3}$ precision in the measured observable, corresponding to about $10^6$ shots on a noiseless device.
Therefore, SC-ADAPT-VQE for $L=14$ would require $\sim 10^{10}$ shots on a noiseless device. 
Factoring in the effects of device noise would increase this estimate by at least a factor of $10$, and probably close to a trillion shots would be required to perform SC-ADAPT-VQE on a quantum computer. This is infeasible on current hardware.
\begin{table}[!ht]
\renewcommand{\arraystretch}{1.4}
\resizebox{\textwidth}{!}{\begin{tabularx}{1.3\textwidth}{|c || Y | Y | Y | Y | Y ||  Y| Y | Y | Y |}
 \hline
 \diagbox[height=23pt]{$L$}{$\theta_i$} & $\hat O_{mh}^{V}(1)$
 & $\hat O_{mh}^{V}(3)$ & $\hat O_{mh}^{V}(5)$ & $\hat O_{mh}^{V}(1)$ & $\hat O_{mh}^{V}(7)$ &
 $\hat O_{mh}^{V}(9)$  &
 $\hat O_{mh}^{S}(0,1)$ & 
 $\hat O_{mh}^{V}(9)$  &
 $\hat O_{mh}^{V}(1)$  \\
 \hline\hline
 6 & 0.25704 & -0.11697 & 0.04896 & 0.18116 & -0.02664 & -- & 0.19193 & 0.01539 & --\\
 \hline
 7 & 0.25796 & -0.11580 & 0.04776 & 0.18099 & -0.02471 & 0.01250 & 0.18971 & -- & --\\
 \hline
 8 & 0.25862 & -0.11507 & 0.04711 & 0.18087 & -0.02380 & 0.01155 & 0.18832 & -- & --\\
 \hline
 9 & 0.21152 & -0.11560 & 0.04859 & 0.12687 & -0.02419 & 0.01162 & -- & -- & 0.11093\\
 \hline
 10 & 0.20923 & -0.11491 & 0.04809 & 0.12749 & -0.02368 & 0.01122 & -- & -- & 0.11182\\
 \hline
 11 & 0.20755 & -0.11437 & 0.04771 & 0.12792 & -0.02331 & 0.01093 & -- & -- & 0.11244\\
 \hline
 12 & 0.20628 & -0.11393 & 0.04741 & 0.12823 & -0.02303 & 0.01072 & -- & -- & 0.11289\\
 \hline
  13 &0.20526& -0.11357&0.04716&0.12846& -0.02280&0.01056&--&--&0.11324 \\
  \hline
    14 &0.20445& -0.11328&0.04696&0.12863& -0.02262&0.01044&--&--&0.11352\\
 \hline
 \hline
  $\infty$ & 0.202& -0.112&0.046&0.129& -0.022&0.010 & -- & -- & 0.114  \\
 \hline
\end{tabularx}}
\caption{
Same as Table~\ref{tab:AnglesXCircmp5gp3} except for $m=0.1, g=0.3$.
}
 \label{tab:AnglesXCircmp1gp3}
\end{table}
\begin{table}[!ht]
\renewcommand{\arraystretch}{1.4}
\begin{tabularx}{\textwidth}{|c || Y | Y || Y |  Y || Y || c|}
 \hline
 $L$ 
 & $\varepsilon^{\rm (aVQE)}$ 
 & $\varepsilon^{\rm (exact)}$ 
 & $\chi^{\rm (aVQE)}$
  & $\chi^{\rm (exact)}$
 & $\infiL$  & \# CNOTS/\text{qubit}
  \\
 \hline\hline
6 & -0.49927& -0.50256 & 0.68192& 0.70834& 0.00377 & 38\\
\hline
7 & -0.50350& -0.50663  & 0.68442& 0.70837 & 0.00337 & 44.3\\
\hline
8 & -0.50670& -0.50971  & 0.68653& 0.70877 & 0.00309 & 48.8\\
\hline
9 & -0.50838& -0.51212  & 0.68694& 0.70928 & 0.00449 & 53.7\\
\hline
10 & -0.51057& -0.51405  & 0.68902& 0.70979 & 0.00412 & 56.1\\
\hline
11 & -0.51236& -0.51563  & 0.69073& 0.71026 & 0.00382 & 58.1\\
\hline
12 & -0.51385& -0.51694 & 0.69217& 0.71068 & 0.00358 & 59.8\\
\hline
13 & -0.51512& -0.51806 & 0.69340& 0.71105  & 0.00337 & 61.2\\
\hline
14 & -0.51620& -0.51902 &0.69445& 0.71137 & 0.00319 & 62.4\\
 \hline
\end{tabularx}
\caption{
Same as Table~\ref{tab:Echifmp5gp3} except for $m=0.1, g=0.3$.
}
 \label{tab:Echifmp1gp3}
\end{table}
\begin{table}[!ht]
\renewcommand{\arraystretch}{1.4}
\begin{tabularx}{\textwidth}{|c || Y | Y | Y | Y | Y ||  Y| Y |}
 \hline
 \diagbox[height=23pt]{$L$}{$\theta_i$} & $\hat O_{mh}^{V}(1)$
 & $\hat O_{mh}^{V}(3)$ & $\hat O_{mh}^{V}(5)$ & $\hat O_{mh}^{V}(1)$ & $\hat O_{mh}^{V}(7)$ &  $\hat O_{mh}^{S}(0,1)$ & 
 $\hat O_{mh}^{V}(1)$ \\
 \hline\hline
 6 & 0.22698 & -0.06357 & 0.01441 & 0.15594 & -0.00418 & 0.14247 & --\\
 \hline
 7 & 0.22784 & -0.06303 & 0.01416 & 0.15559 & -0.00395 & 0.14111  & --\\
 \hline
 8 & 0.22843 & -0.06267 & 0.01401 & 0.15535 & -0.00382 & 0.14018 & -- \\
 \hline
 9 & 0.22885 & -0.06240 & 0.01390 & 0.15518 & -0.00374 & 0.13951 & --\\
 \hline
 10 & 0.22918 & -0.06219 & 0.01382 & 0.15505 & -0.00368 & 0.13900 & --\\
 \hline
 11 & 0.18110 & -0.06192 & 0.01431 & 0.11095 & -0.00377 & -- & 0.09796 \\
 \hline
 12 & 0.18044 & -0.06169 & 0.01423 & 0.11108 & -0.00372 & -- & 0.09809 \\
 \hline
 13 &0.17992 & -0.06151 & 0.01416 & 0.11116 & -0.00369 & -- & 0.09819 \\
 \hline
 14 &0.17949 & -0.06135 & 0.01410 & 0.11124 & -0.00366 & -- & 0.09825\\
 \hline \hline
 $\infty$ & 0.178 & -0.061 & 0.014 & 0.112 & -0.004 & -- & 0.098 \\
 \hline
\end{tabularx}
\caption{
Same as Table~\ref{tab:AnglesXCircmp5gp3} except with $m=0.1$ and $g=0.8$
and through 6 steps of the SC-ADAPT-VQE algorithm.}
 \label{tab:AnglesXCircmp1gp8}
\end{table}
\begin{table}[!ht]
\renewcommand{\arraystretch}{1.4}
\begin{tabularx}{\textwidth}{|c || Y  | Y || Y |  Y || Y || c |}
 \hline
 $L$ 
 & $\varepsilon^{\rm (aVQE)}$ 
 & $\varepsilon^{\rm (exact)}$ 
 & $\chi^{\rm (aVQE)}$ 
 & $\chi^{\rm (exact)}$ 
 & $\infiL$  & \# CNOTS/\text{qubit}
  \\
 \hline\hline
 6 & -0.41488& -0.41614 & 0.51372 & 0.52154 & 0.00072 & 29.5\\
 \hline
7 & -0.41869& -0.41996 & 0.51579 & 0.52348 & 0.00071 & 32.1 \\
\hline
8 & -0.42156& -0.42283 & 0.51736 & 0.52495 & 0.00071 & 33.9\\
\hline
9 & -0.42379& -0.42506 & 0.51859 & 0.52609 & 0.00071 & 35.2\\
\hline
10 & -0.42557& -0.42685 & 0.51958 & 0.52701 & 0.00071 & 36.3\\
\hline
11 & -0.42669& -0.42831 & 0.51945 & 0.52776 & 0.00129 & 38.9\\
\hline
12 & -0.42799& -0.42953 & 0.52047 & 0.52838 & 0.00121 & 39.7\\
\hline
13 & -0.42909& -0.43056 & 0.52134 & 0.52891 & 0.00115 & 40.3\\
\hline
14 & -0.43003& -0.43144 & 0.52209 & 0.52937 & 0.00109 & 40.9\\
\hline
\end{tabularx}
\caption{
Same as Table~\ref{tab:Echifmp5gp3} except with 
$m=0.1$ and $g=0.8$ and through 6 steps of the SC-ADAPT-VQE algorithm.
}
 \label{tab:Echifmp1gp8}
\end{table}
\begin{table}[!ht]
\renewcommand{\arraystretch}{1.4}
\resizebox{\textwidth}{!}{\begin{tabularx}{1.3\textwidth}{|c || Y | Y | Y | Y | Y ||  Y| Y | Y | Y ||}
 \hline
 \diagbox[height=23pt]{$L$}{$\theta_i$} & $\hat O_{mh}^{V}(1)$
 & $\hat O_{mh}^{V}(3)$ & $\hat O_{mh}^{V}(5)$ & $\hat O_{mh}^{V}(1)$ & $\hat O_{mh}^{V}(7)$ &  $\hat O_{mh}^{S}(0,1)$ & 
 $\hat O_{mh}^{V}(1)$ & $\hat O_{mh}^{S}(0,1)$ & $\hat O_{mh}^{V}(9)$\\
 \hline\hline
 6 & 0.17222 & -0.06236 & 0.01456 & 0.11495 & -0.00409 & 0.07017 & 0.09561 & -- & --\\
 \hline
 7 & 0.17278 & -0.06184 & 0.01433 & 0.11431 & -0.00389 & 0.06947 & 0.09620 & -- & --\\
 \hline
 8 & 0.17316 & -0.06147 & 0.01417 & 0.11388 & -0.00378 & 0.06900 & 0.09659 & -- & -- \\
 \hline
 9 & 0.17344 & -0.06121 & 0.01407 & 0.11357 & -0.00371 & 0.06866 & 0.09688 & -- & --\\
 \hline
 10 & 0.17365 & -0.06101 & 0.01399 & 0.11333 & -0.00366 & 0.06840 & 0.09710 & -- & --\\
 \hline
 11 &0.17300 & -0.06058 & 0.01385 & 0.11216 & -0.00359 & -- & 0.09885 & 0.07047 & --\\
 \hline
 12 & 0.17321 & -0.06048 & 0.01381 & 0.11210 & -0.00357 & -- & 0.09883 & 0.07028 & -- \\
 \hline
 13 & 0.17338 & -0.06039 & 0.01378 & 0.11205 & -0.00355 & -- & 0.09883 & 0.07012 & --\\
 \hline
 14 & 0.17950 & -0.06139 & 0.01417 & 0.11124 & -0.00382 & -- & 0.09825 & -- & 0.00107\\
 \hline 
\end{tabularx}}
\caption{
Same as Table~\ref{tab:Echifmp5gp3} except with 
$m=0.1$ and $g=0.8$ and through 7 steps of the SC-ADAPT-VQE algorithm.}
 \label{tab:AnglesXCircmp1gp8Step7}
\end{table}
\begin{table}[!ht]
\renewcommand{\arraystretch}{1.4}
\begin{tabularx}{\textwidth}{|c || Y | Y || Y | Y |}
\hline
\multicolumn{5}{|c|}{$m=0.1, g=0.3$} \\
\hline
 \hline
 $L$ 
 & $\varepsilon^{\rm (SC-MPS)}$ 
 & $\varepsilon^{\rm (DMRG)}$ 
 & $\chi^{\rm (SC-MPS)}$  
 & $\chi^{\rm (DMRG)}$ \\
 \hline\hline
 50 &-0.52640 & -0.52797 & 0.70967 & 0.71444 \\
 \hline
 100 & -0.52838 & -0.52971 & 0.71359 & 0.71504\\
 \hline
 \hline
 \multicolumn{5}{|c|}{$m=0.1, g=0.8$} \\
 \hline\hline
 50 & -0.43886 & -0.43971 & 0.53339 & 0.53361 \\
 \hline
 100 & -0.44058 & -0.44131 & 0.53604 & 0.53444\\
 \hline
200 & -0.44144 & -0.44212 &0.53737 & 0.53485 \\
 \hline
 300 & -0.44173 & -0.442384 &0.53781 &  0.53499 \\
 \hline
 400 & -0.44187 & -0.44252 &0.53803 & 0.53506 \\
 \hline
\end{tabularx}
\caption{
Same as Table~\ref{tab:SQCResults} except with $m=0.1,g=0.3$ and $m=0.1, g=0.8$.}
 \label{tab:SQCResultsApp}
\end{table}

\clearpage
\section{Additional Details and Results From Simulations using IBM's Quantum Computers}
\label{app:qusimDetail}
\begin{figure}[!ht]
    \centering
    \includegraphics[width=0.42\columnwidth]{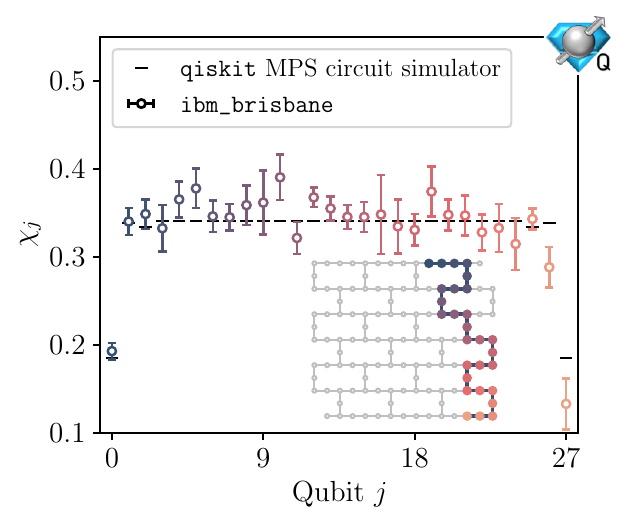}
    \includegraphics[width=0.56\columnwidth]{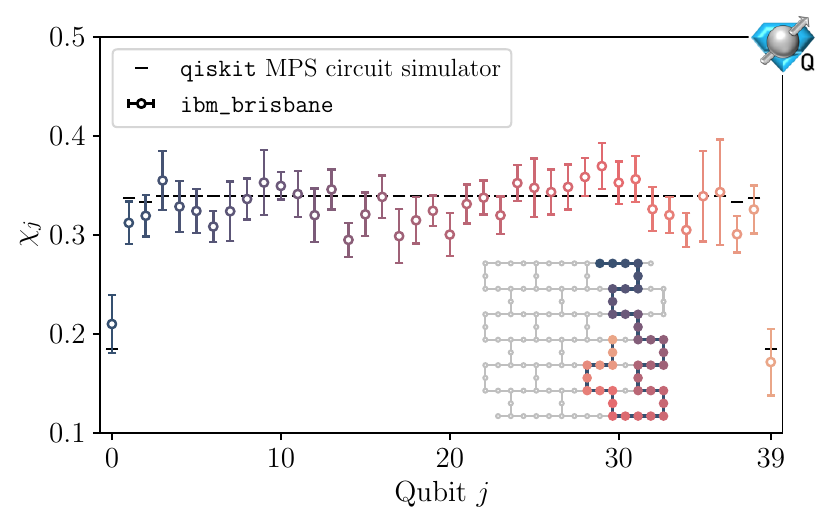}
    \includegraphics[width=0.75\columnwidth]{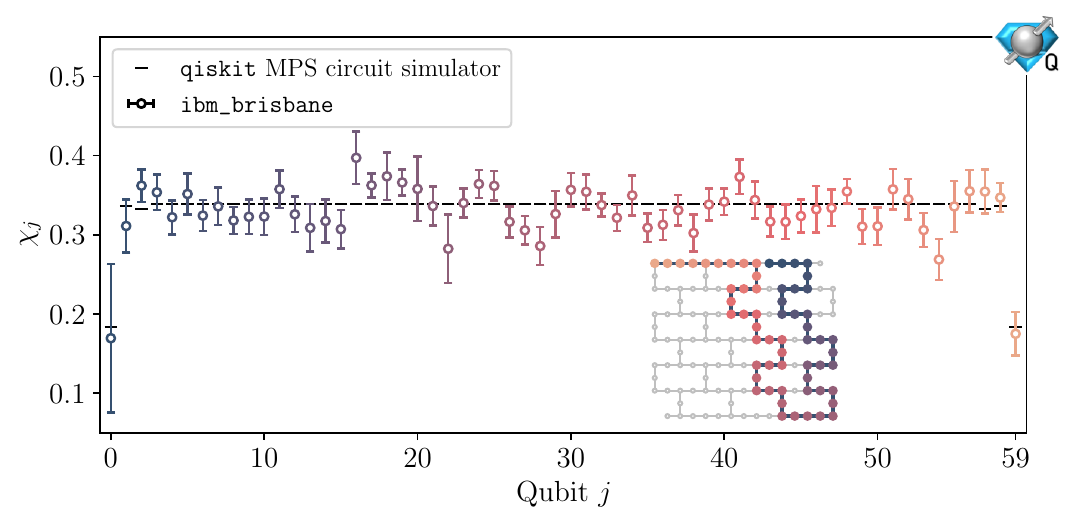}
    \includegraphics[width=0.95\columnwidth]{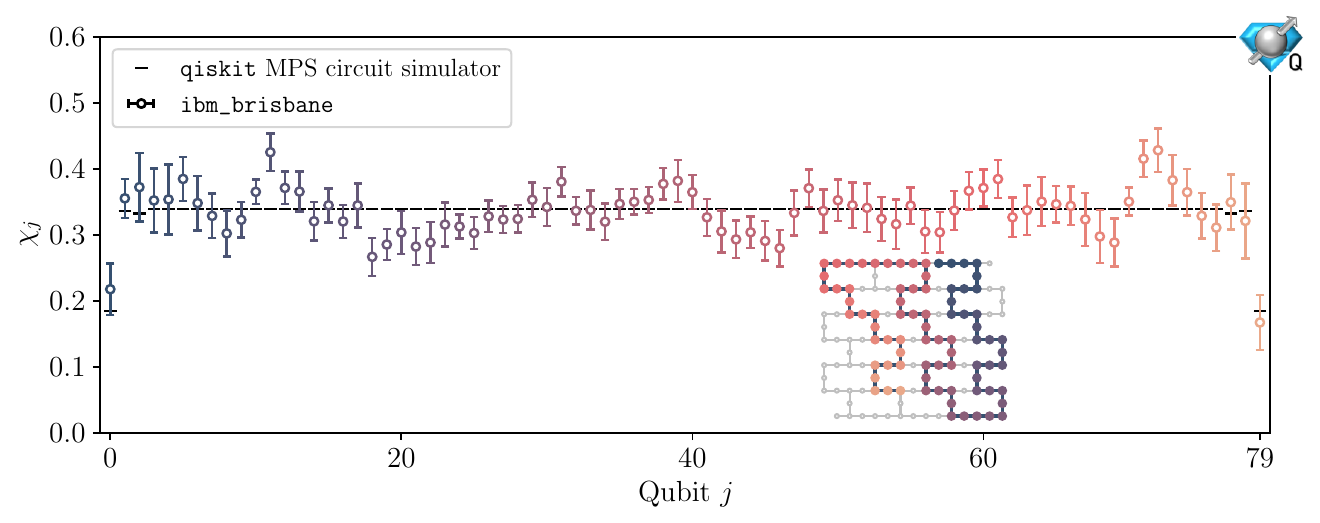}
    \caption{Expectation values of $\chi_j$ for $L=14,20,30$ and $40$ (top to bottom), 
    obtained from simulations using {\tt ibm\_brisbane}. 
    They are compared with the expected results obtained by using {\tt qiskit}'s MPS circuit simulator (black dashes). 
    Averaging $\chi_j$ over all of the 
    qubits provides the chiral condensates presented in Table~\ref{tab:SQCResults_ibm}.
    The layouts of the qubits used on the chip are shown in the insets.}
    \label{fig:chi_14-40}
\end{figure}

\begin{figure}[!ht]
    \centering
    \includegraphics[width=0.37\columnwidth]{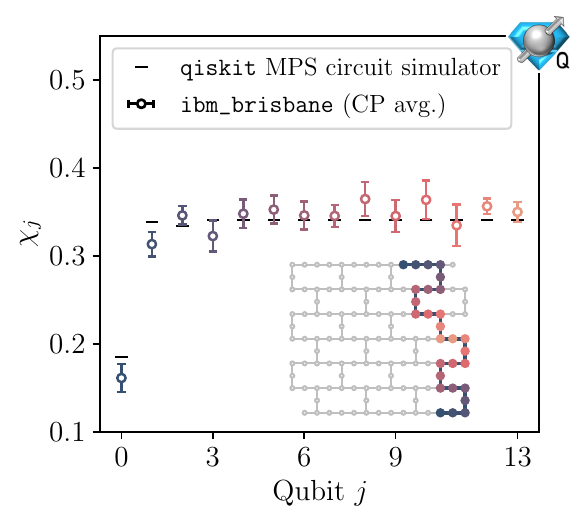}
    \includegraphics[width=0.4\columnwidth]{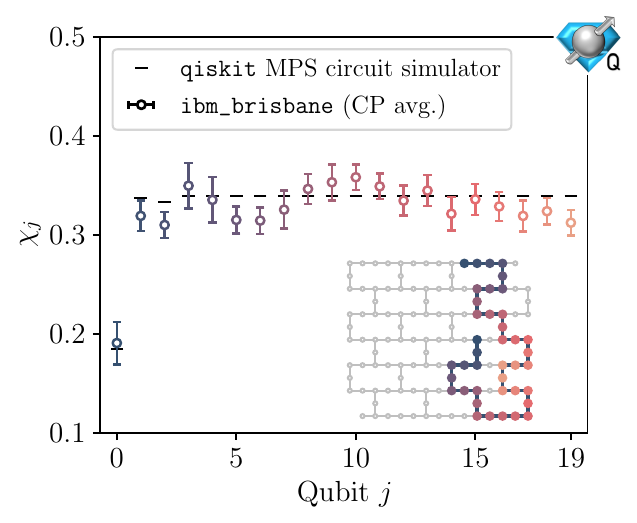}
    \includegraphics[width=0.46\columnwidth]{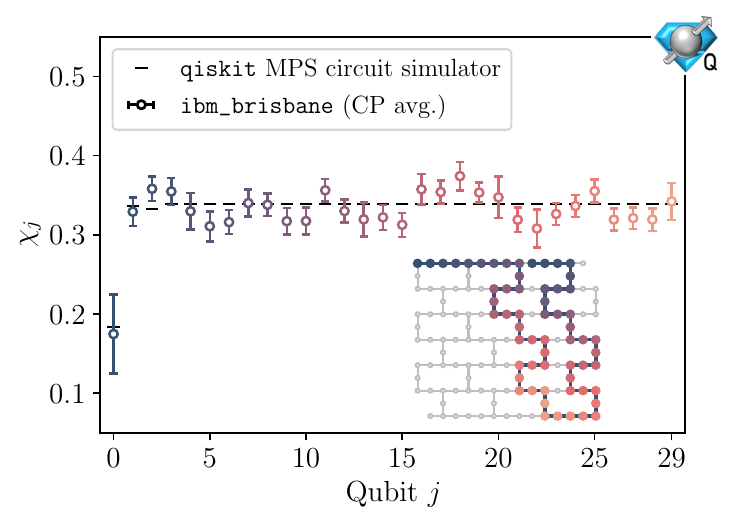}
    \includegraphics[width=0.52\columnwidth]{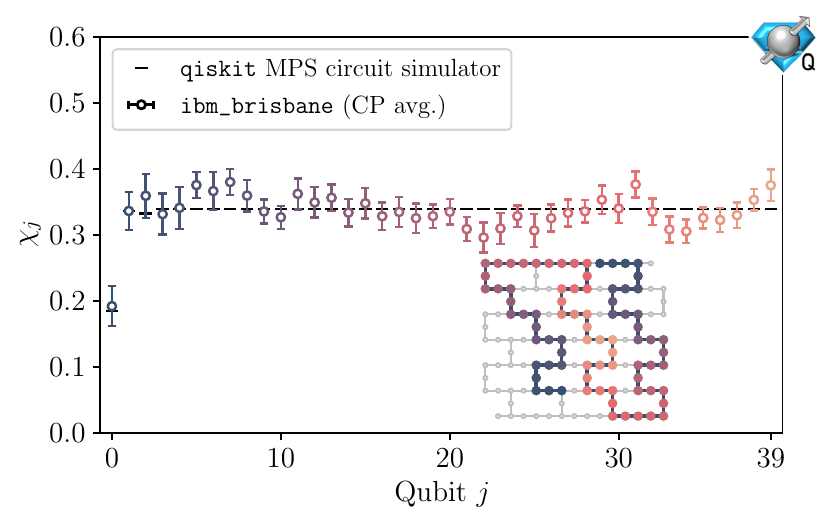}
    \includegraphics[width=0.63\columnwidth]{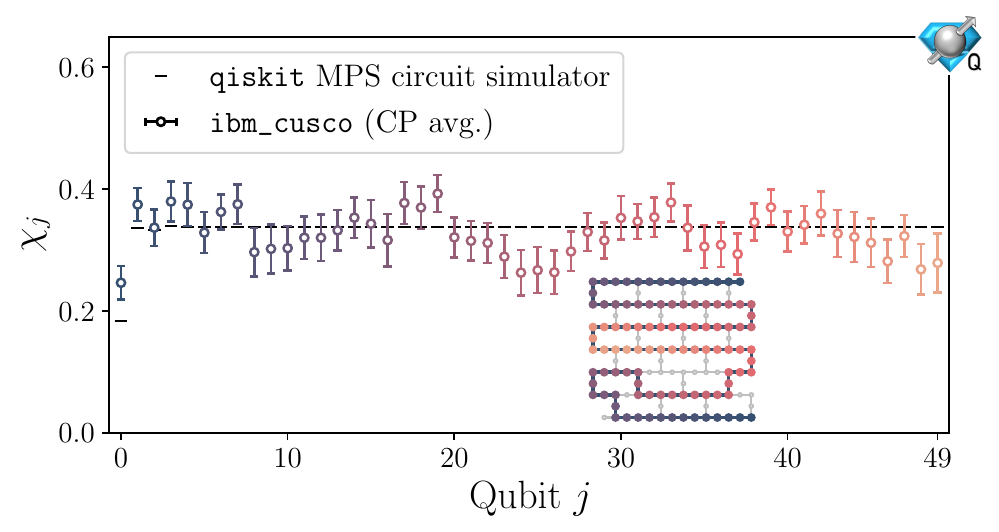}
    \caption{Expectation values of CP averaged $\chi_j$ for $L=14,20,30,40$ and $50$ (top to bottom)
    obtained from simulations using {\tt ibm\_brisbane} and {\tt ibm\_cusco}. 
    They are compared with the expected results obtained by using {\tt qiskit}'s MPS circuit simulator (black dashes). 
    The layouts of the qubits used on the chip are shown in the insets (with same-colored qubits being averaged).}
    \label{fig:chi_14-50_CP}
\end{figure}

\begin{figure}[!ht]
    \centering
    \includegraphics[width=0.92\columnwidth]{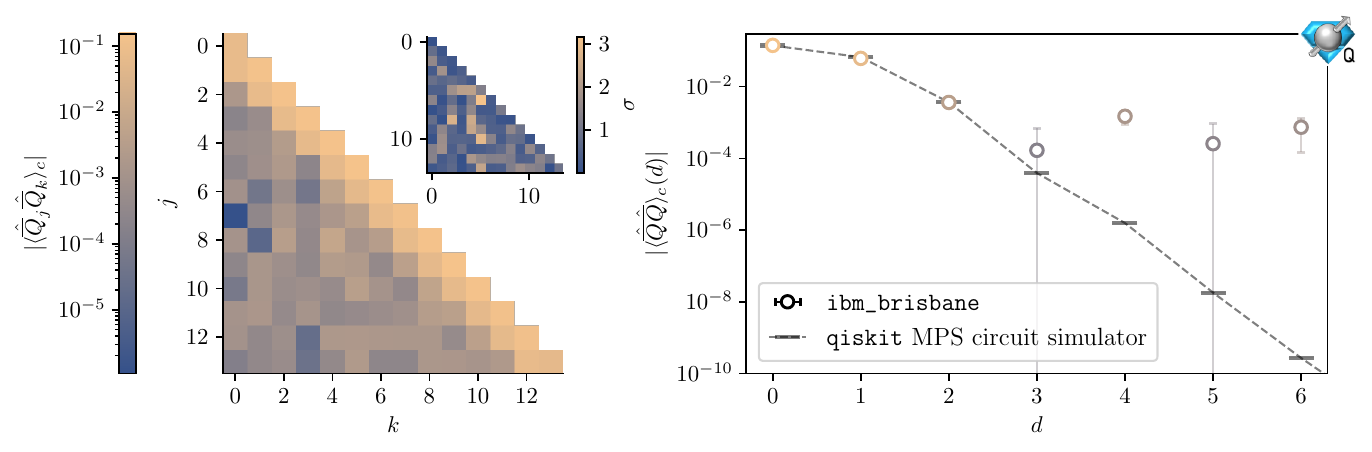}
    \includegraphics[width=0.92\columnwidth]{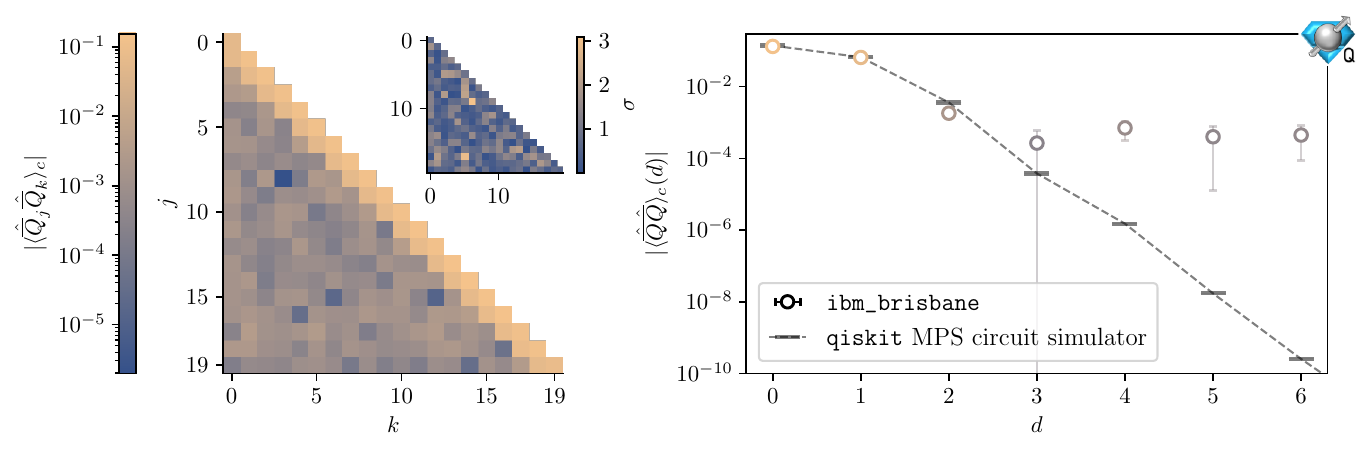}
    \includegraphics[width=0.92\columnwidth]{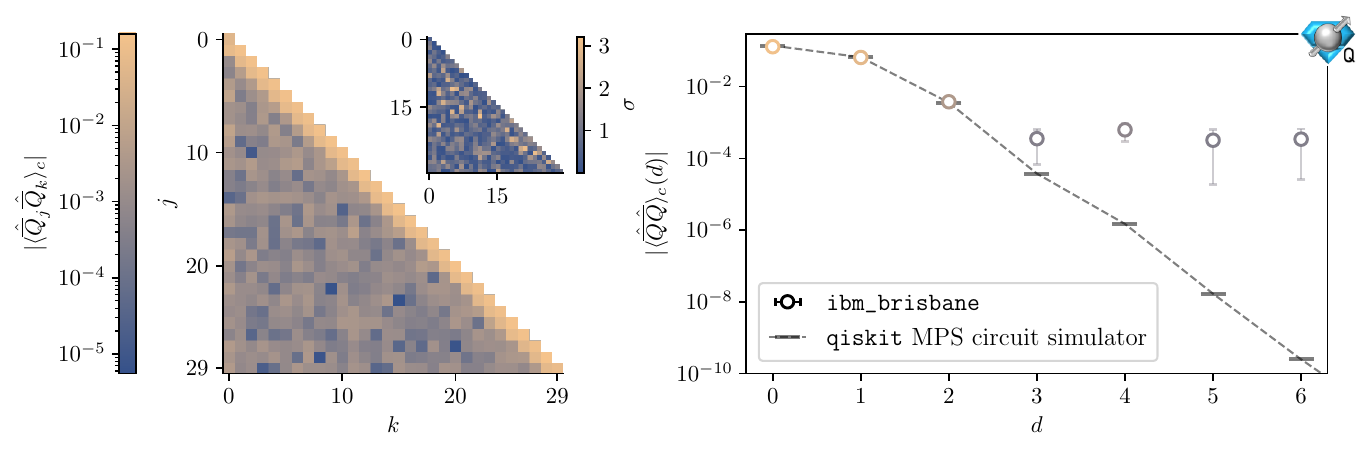}
    \includegraphics[width=0.92\columnwidth]{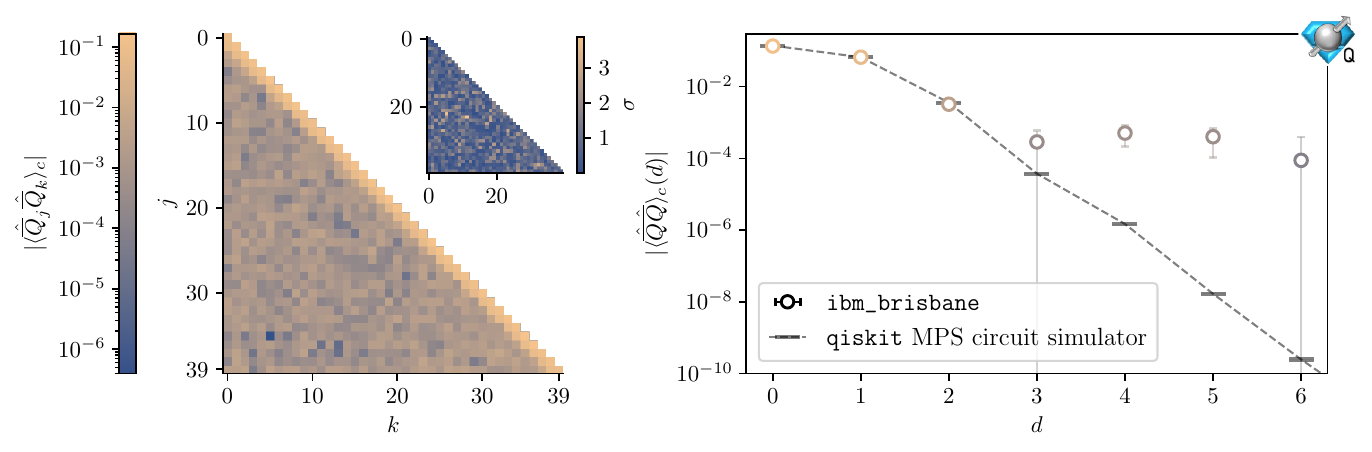}
    \caption{The connected contribution to the spatial charge-charge correlation functions, $\langle \hat{\overline{Q}}_{j} \hat{\overline{Q}}_{k} \rangle_c$ (left)
    and the averaged correlation functions as a function of distance $d$, 
    $\langle \hat{\overline{Q}} \hat{\overline{Q}} \rangle_c (d)$ (right) with the points following the same color map as in the left main panel (error bars show $1\sigma$ standard deviations).
    Results obtained from {\tt ibm\_brisbane} are shown for $L=14,20,30$ and $40$ (top to bottom).}
    \label{fig:qiqj_14-50}
\end{figure}

\noindent
In this appendix, we provide additional details about how our results are obtained from IBM's quantum computers, together with the additional figures not shown in Sec.~\ref{sec:SCQuSim}.
All measurements are performed on {\tt ibm\_brisbane} ($L\leq 40$) and {\tt ibm\_cusco} ($L=50$) by sending the state preparation circuits, 
with measurements in the computational ($z$) basis,
via the {\tt qiskit} Runtime Sampler primitive.
The values of the variational parameters obtained from fitting to the exponential form in Eq.~(\ref{eq:thetaextrap}) for 2 steps of SC-ADAPT-VQE are given in Table~\ref{tab:angles_ibm}. The different qubits used for each lattice size can be seen in the insets in Figs.~\ref{fig:chi_50}
and~\ref{fig:chi_14-40}. 
$\chi_j$, obtained from {\tt ibm\_brisbane} for $L=14,20,30$ and 40, is shown in Fig.~\ref{fig:chi_14-40}, and the charge-charge correlation functions are shown
in Fig.~\ref{fig:qiqj_14-50}.
In Fig.~\ref{fig:chi_14-50_CP}, the CP symmetry relating $\chi_j = \chi_{2L-1-j}$ is used to effectively double the number of shots, resulting in statistical error bars that are smaller by a factor of $\sqrt{2}$.

\begin{figure}[!ht]
    \centering
    \includegraphics[width=0.75\columnwidth]{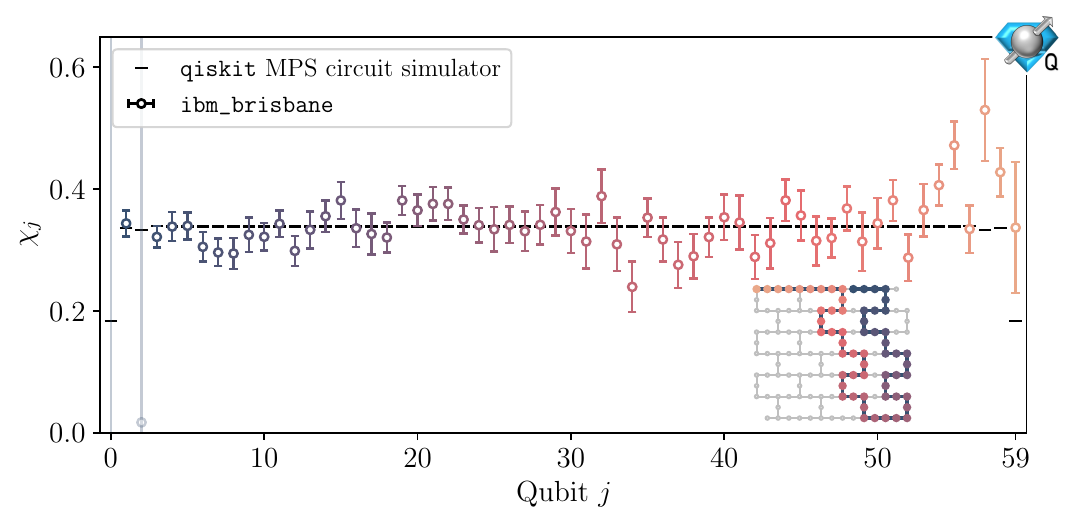}
    \includegraphics[width=0.5\columnwidth]{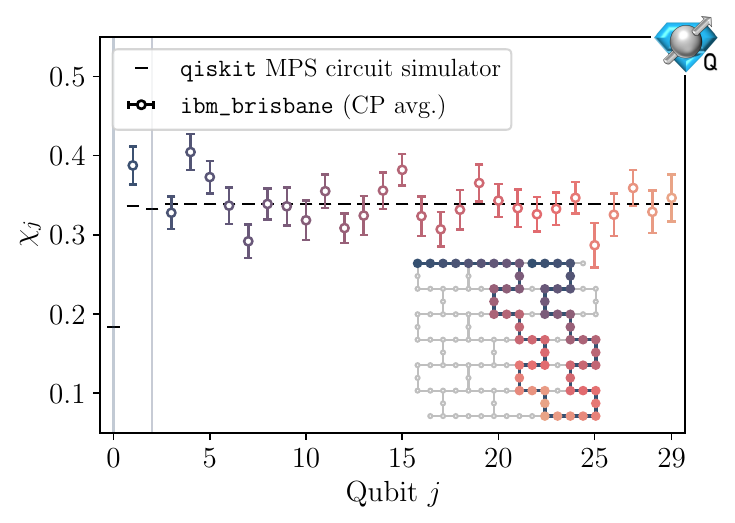}
    \includegraphics[width=0.92\columnwidth]{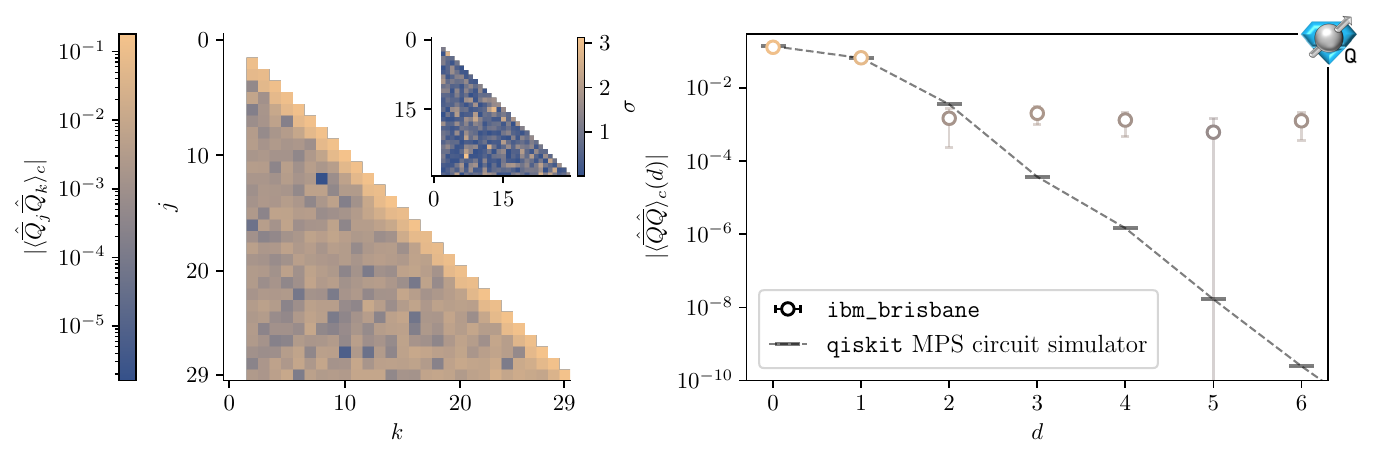}
    \caption{Results for the $L=30$ system obtained with use of three steps of SC-ADAPT-VQE, obtained from simulations using {\tt ibm\_brisbane} with 80 twirled instances. The top panel shows $\chi_j$, the middle panel shows the CP averaged $\chi_j$, and the bottom panels show the connected contribution to the spatial charge-charge correlation functions, $\langle \hat{\overline{Q}}_{j} \hat{\overline{Q}}_{k} \rangle_c$ (the first two spatial sites are not shown due to the errors on qubits 0 and 2),
    and
    the averaged correlation functions as a function of distance $d$, 
    $\langle \hat{\overline{Q}} \hat{\overline{Q}} \rangle_c (d)$, with the points following the same color map as in the left main panel (error bars show $1\sigma$ standard deviations).}
    \label{fig:3layersL30}
\end{figure}

\begin{figure}[!ht]
    \centering
    \includegraphics[width=\columnwidth]{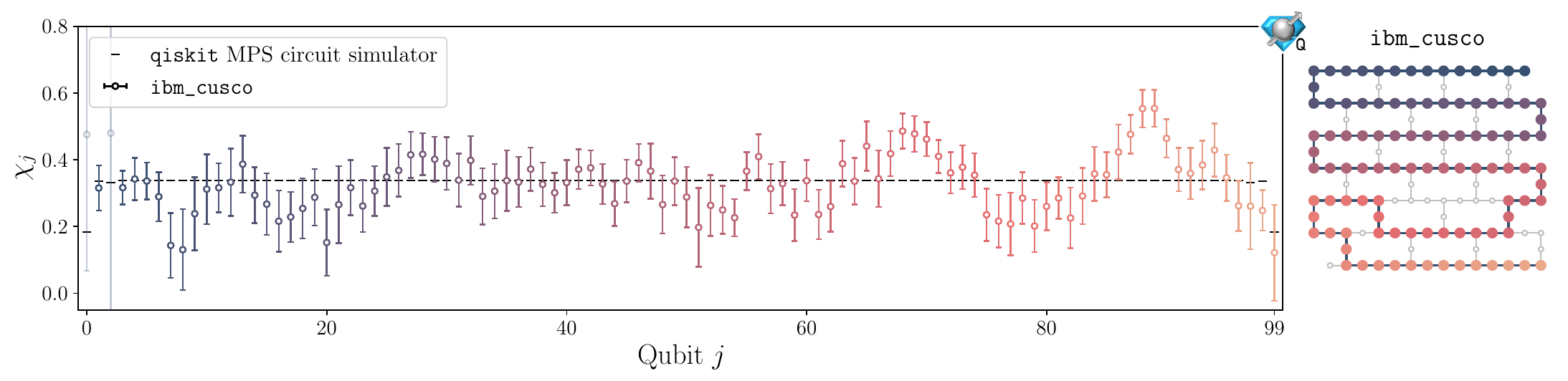}
    \includegraphics[width=0.5\columnwidth]{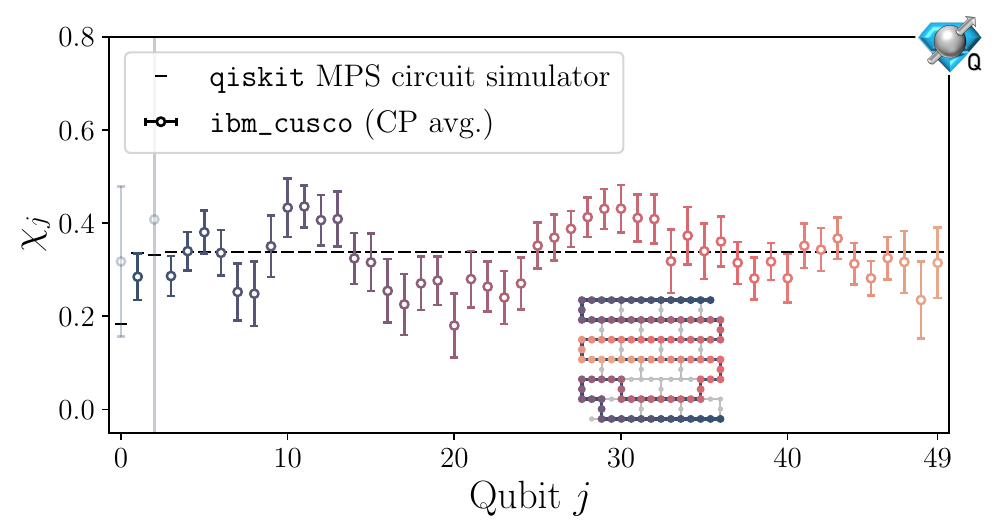}
    \includegraphics[width=0.92\columnwidth]{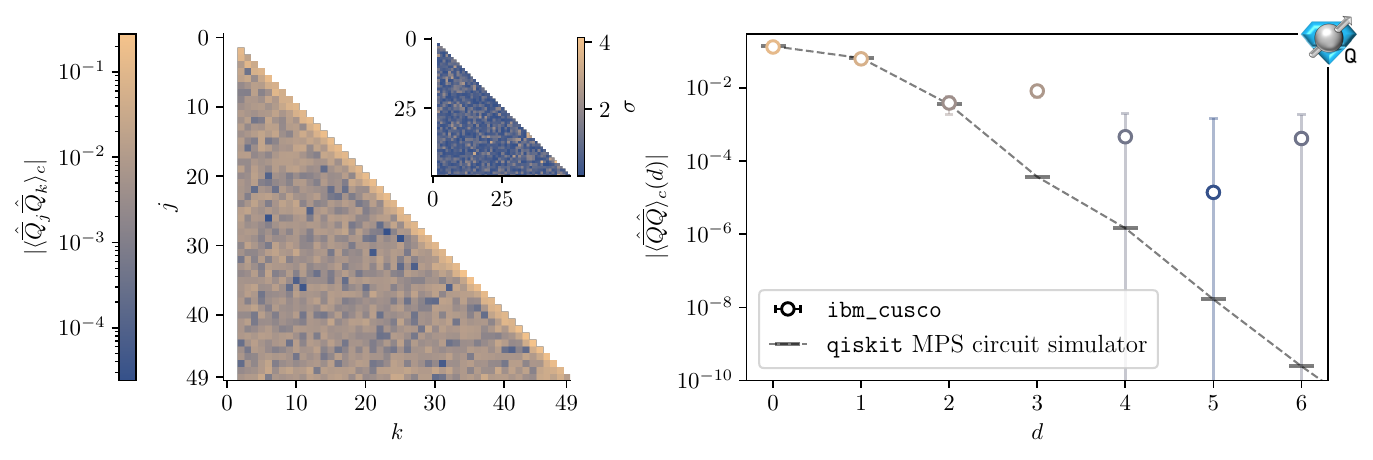}
    \caption{Results for the $L=50$ system obtained with use of three steps of SC-ADAPT-VQE, obtained from simulations using {\tt ibm\_cusco} with 40 twirled instances. The top panel shows $\chi_j$, the middle panel shows the CP averaged $\chi_j$, and the bottom panels show the connected contribution to the spatial charge-charge correlation functions, $\langle \hat{\overline{Q}}_{j} \hat{\overline{Q}}_{k} \rangle_c$ (the first two spatial sites are not shown due to the errors on qubits 0 and 2),
    and
    the averaged correlation functions as a function of distance $d$, 
    $\langle \hat{\overline{Q}} \hat{\overline{Q}} \rangle_c (d)$, with the points following the same color map as in the left main panel (error bars show $1\sigma$ standard deviations).}
    \label{fig:3layersL50}
\end{figure}

In an effort to explore the limitations of the quantum computer, 
the 3-step SC-ADAPT-VQE state preparation circuits for $L=30$ and $L=50$ were implemented on {\tt ibm\_brisbane} and {\tt ibm\_cusco}, respectively. 
The structure of the ansatz wave function and corresponding variational parameters can be found in Table~\ref{tab:angles_ibm}.
The local chiral condensate and charge-charge correlators obtained from 80 ($L=30$) and 40 ($L=50$) twirled instances, each with $8\times 10^3$ shots, are shown in Figs.~\ref{fig:3layersL30} and~\ref{fig:3layersL50}. 
Despite the factor of three increase in the number of
CNOTs relative to 2 layers (1254 versus 468 for $L=30$, and 2134 versus 788 for $L=50$), the results are consistent with those obtained from the {\tt qiskit} MPS circuit simulator.
Note that qubit 0 and 2 have decohered for both volumes, and in principle could be removed
from volume averaged quantities, such as the chiral condensate.

\begin{table}[!ht]
\renewcommand{\arraystretch}{1.4}
\begin{tabularx}{0.8\textwidth}{|c || Y | Y || Y | Y | Y |}
 \hline
 & \multicolumn{2}{c||}{2 steps}  & \multicolumn{3}{c|}{3 steps} \\\hline
 \diagbox[height=23pt]{$L$}{$\theta_i$} 
 & $\hat{O}^{V}_{mh}(1)$ & $\hat{O}^{V}_{mh}(3)$ & $\hat{O}^{V}_{mh}(1)$ & $\hat{O}^{V}_{mh}(3)$ & $\hat{O}^{V}_{mh}(5)$ \\
 \hline\hline
 14 & 0.30699 & -0.04033 & & & \\
 \hline
 20 & 0.30638 & -0.03994 & & & \\
 \hline
 30 & 0.30610 & -0.03978 & 0.30630 & -0.04092 & 0.00671 \\
\hline
 40 & 0.30605 & -0.03975 & & & \\
\hline
 50 & 0.30604 & -0.03975 & 0.30624 & -0.04089 & 0.00670 \\
 \hline
\end{tabularx}
\caption{
Extrapolation of the variational parameters corresponding to 2 and 3 steps of SC-ADAPT-VQE with $m=0.5, g=0.3$.
These parameters were used in the circuits run on {\tt ibm\_brisbane} ($L\leq 40$) and {\tt ibm\_cusco} ($L= 50$). }
 \label{tab:angles_ibm}
\end{table}

By sending the circuits with the Sampler primitive, several error mitigation techniques are applied during runtime, as mentioned in Sec.~\ref{sec:SCQuSim}. Specifically, the readout mitigation technique used (for $L\leq 40$) is M3~\cite{Nation:2021kye}. This method is based on correcting only the subspace of bit-strings observed in the noisy raw counts from the machine (which usually include the ideal ones plus those with short Hamming distance, introduced by the noise in the measurement), and using Krylov subspace methods to avoid having to compute (and store) the full assignment matrix.

Unlike the other works that have utilized $\geq 100$ superconducting qubits~\cite{Yu:2022ivm,Kim:2023bwr,Shtanko:2023tjn}, which used zero-noise extrapolation (ZNE)~\cite{Li:2016vmf,Temme:2016vkz,2020arXiv200510921G} in conjunction with probabilistic error correction (PEC)~\cite{Temme:2016vkz,Berg:2022ugn} to remove incoherent errors, 
we use Operator Decoherence Renormalization (ODR), as explained in App.~\ref{app:QSimError}. 
Both methods require first transforming coherent errors into incoherent errors, which is done via Pauli twirling. However, the overhead in sampling using ZNE and PCE, compared with ODR, is substantial. For ZNE, one has to add two-qubit gates to increase the noise level, and then perform an extrapolation to estimate the noiseless result. In the minimal case, this leads to running only another circuit, like in ODR, but with a circuit depth that is three times as large as the original circuit (e.g., replacing each CNOT with 3 CNOTs). 
However, this leads to a large uncertainty in the functional form of the extrapolation, and ideally the circuit is run with multiple noise levels to have multiple points from which to extrapolate. For PEC, the overhead is even larger, as it involves learning the noise model of the chip, by running multiple random circuits with different depths (see Ref.~\cite{Berg:2022ugn}). For ODR, as explained in App.~\ref{app:QSimError}, only the same ``physics" circuits are run, but with all rotations set to zero, meaning the sampling overhead is only doubled.

To generate the different twirled circuits, the set of two-qubit Pauli gates $G_2$ and $G'_2$ that leave the (noisy) two-qubit gate invariant (up to a global phase) must be identified. For the quantum processors used in this work, the native two-qubit gate is the echoed cross-resonance (ECR) gate, which is equivalent to the CNOT gate via single qubit rotations. Explicitly,
\begin{equation}
ECR = \frac{1}{\sqrt{2}}(\hat{X}\otimes\hat{I}-\hat{Y}\otimes \hat{X})\ , \quad \quad
\begin{quantikz} 
& \gate[wires=2]{ECR} & \qw \\ 
& & \qw
\end{quantikz} =
\begin{quantikz} 
& \qw & \ctrl{1} & \gate{R_z(-\frac{\pi}{2})} & \gate{R_y(\pi)} & \qw \\
& \gate{R_x(\frac{\pi}{2})} & \targ & \qw & \qw & \qw & \qw
\end{quantikz} \ .
\end{equation}
Using the functions from the package {\tt qiskit\_research}~\cite{the_qiskit_research_developers_and_contr_2023_7776174}, together with the two-Pauli gate set shown in Table~\ref{tab:ecrtwirl}, a total of 40 (150) twirled circuits for both mitigation and physics were generated for $L\leq 40$ ($L=50$), each with $8\times 10^3$ shots.
\begin{table}[!t]
\renewcommand{\arraystretch}{1.2}
\resizebox{\textwidth}{!}{\begin{tabularx}{1.6\textwidth}{| Y | Y | Y | Y | Y | Y | Y | Y |}
 \hline
 $(\hat{I}\otimes\hat{I},\hat{I}\otimes\hat{I})$ & $(\hat{I}\otimes\hat{X},\hat{I}\otimes\hat{X})$ & $(\hat{I}\otimes\hat{Y},\hat{Z}\otimes\hat{Z})$ & $(\hat{I}\otimes\hat{Z},\hat{Z}\otimes\hat{Y})$ &
 $(\hat{X}\otimes\hat{I},\hat{Y}\otimes\hat{X})$ & $(\hat{X}\otimes\hat{X},\hat{Y}\otimes\hat{I})$ & $(\hat{X}\otimes\hat{Y},\hat{X}\otimes\hat{Y})$ & $(\hat{X}\otimes\hat{Z},\hat{X}\otimes\hat{Z})$ \\
 \hline
 $(\hat{Y}\otimes\hat{I},\hat{X}\otimes\hat{X})$ & $(\hat{Y}\otimes\hat{X},\hat{X}\otimes\hat{I})$ & $(\hat{Y}\otimes\hat{Y},\hat{Y}\otimes\hat{Y})$ & $(\hat{Y}\otimes\hat{Z},\hat{Y}\otimes\hat{Z})$ &
 $(\hat{Z}\otimes\hat{I},\hat{Z}\otimes\hat{I})$ & $(\hat{Z}\otimes\hat{X},\hat{Z}\otimes\hat{X})$ & $(\hat{Z}\otimes\hat{Y},\hat{I}\otimes\hat{Z})$ & $(\hat{Z}\otimes\hat{Z},\hat{I}\otimes\hat{Y})$ \\
 \hline
\end{tabularx}}
\caption{Two-Pauli gate set $(G_2,G'_2)$ used to generate the twirled ECR gates, $G'_2 \cdot ECR \cdot G_2 = ECR$.}
 \label{tab:ecrtwirl}
\end{table}

From Fig.~\ref{fig:chi_50}, the effects of each error mitigation method can be seen. 
The first set of results shown are semi-raw, obtained directly from the quantum computer. They are not raw since DD is integrated into the circuits that are run on the machine (REM is also included for $L\leq 40$).
To check the effect that DD has, several runs were performed without it, and a degradation of the signal was visible when qubits were idle for long periods (the effects of not using DD were more evident when the deeper 3-step circuit was run). 
Regarding REM, while the final fully-mitigated results for $L=50$ (no REM applied) and $L \leq 40$ (REM applied) systems are similar in quality, a larger statistical sample for $L=50$ was required to achieve an equivalent level of precision ($2.4\times10^6$ vs $6.4\times 10^5$ shots). 
The second set shows the effects of applying PT (the results for no Pauli twirling corresponded to one twirled instance).
It is seen that all the coherent noise on the different qubits has been transformed into uniform incoherent noise.
The last set shown is after ODR has been used to remove the incoherent noise.

\end{subappendices}

\chapter{Quantum simulations of hadron dynamics in the Schwinger model on 112 Qubits}
\label{chap:SchwingerTevol}
\noindent
\textit{This chapter is associated with Ref.~\cite{Farrell:2024fit}: \\
``Quantum simulations of hadron dynamics in the Schwinger model using 112 qubits" by Roland C. Farrell, Anthony N. Ciavarella, Marc Illa and Martin J. Savage.}
\section{Introduction}
\label{sec:intro1}
\noindent
The highest-energy collisions of particles, such as those that take place in colliders and cosmic-ray events, reveal and provide insights into the underlying laws of nature.
They tighten constraints on the content, symmetries and parameters of the Standard Model (SM)~\cite{Glashow:1961tr,Higgs:1964pj,Weinberg:1967tq,Salam:1968rm,Politzer:1973fx,Gross:1973id}, and provide opportunities to discover what may lie beyond.
In searching for new physics and emergent phenomena in exotic states of matter, contributions from known physics must be reliably predicted with a complete quantification of  uncertainties.
The associated complexities, particularly from the strong interactions described by quantum chromodynamics (QCD), provide challenges for phenomenological modeling and classical simulation. 
Many forefront research questions in nuclear and particle physics require simulations of systems of fundamental particles that lie far beyond the capabilities of classical computing.

In principle, the collisions of fundamental and composite particles (hadrons)
could be simulated, from the initial state through to the final state(s), with sufficiently capable quantum computers (for recent reviews, see e.g., Refs.~\cite{Banuls:2019bmf,Guan:2020bdl,Klco:2021lap,Delgado:2022tpc,Bauer:2022hpo,Bauer:2023qgm,Beck:2023xhh,DiMeglio:2023nsa}).
Well before that point, new insights and improvements in predictions for such processes may come from NISQ-era devices~\cite{Preskill:2018jim}.
In this chapter, the real-time dynamics of composite particles, ``hadrons'', in the lattice Schwinger model are simulated using IBM's superconducting-qubit quantum computers. 
This work serves as a proof-of-concept, and builds toward future simulations that will probe highly-inelastic scattering of hadrons and out-of-equilibrium behavior of strongly interacting matter.
Our quantum simulations proceed with the following steps:\footnote{As this work was being completed, similar developments in the Thirring model were reported in Ref.~\cite{Chai:2023qpq}.}
\begin{enumerate}
    \item Prepare the interacting ground state (vacuum);
    \item Establish a localized hadron wavepacket on this vacuum;
    \item Evolve the system forward in time, allowing the hadrons to propagate;
    \item Measure observables in the final state that detect hadron propagation.
\end{enumerate}
Crucial to the success of our quantum simulations is the development of comprehensive suites of scalable techniques that minimize circuit depth and two-qubit entangling gate counts. 
The methods presented here are informed by the symmetries and phenomenological features of the Schwinger model.
They are physics-aware techniques with potential applicability to a broad class of lattice theories.

\begin{figure}[!tb]
    \centering
    \includegraphics[width=0.95\columnwidth]{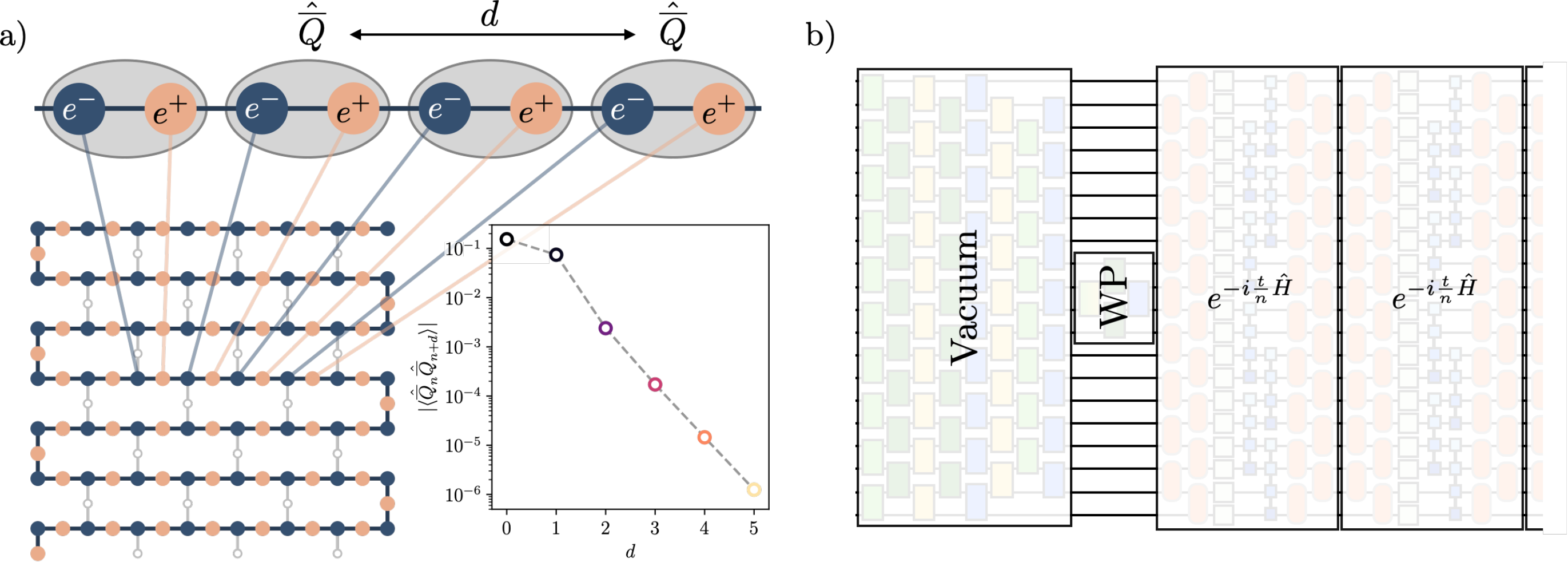}
    \caption{a) Mapping the $L=56$ lattice onto the qubits of IBM's quantum computer {\tt ibm\_torino} (bottom left). 
    The dynamical re-arrangement of charges in the vacuum screens the interactions between electric charges in the Schwinger model, giving rise to an exponential decay of correlations between spatial-site charges, $\langle \hat{\overline{Q}}_n \hat{\overline{Q}}_{n+d}\rangle$ (top and bottom right). 
    b) The charge screening informs an efficient construction of the quantum circuits used to simulate hadron dynamics.
    SC-ADAPT-VQE is used to prepare the vacuum and wavepacket, which are time-evolved using Trotterized circuits implementing $e^{-i t \hat{H} }$ 
    with a truncated electric interaction.
    }
    \label{fig:SimTricks}
\end{figure}

A significant challenge to performing quantum simulations of the Schwinger model is that, in axial gauge ($A_x=0$)~\cite{Farrell:2022wyt}, the electric interaction between fermions is all-to-all.\footnote{Working in Weyl gauge ($A_t=0$) eliminates the need for all-to-all connectivity, but requires additional qubits to encode the gauge field on the links of the lattice.}
This leads to an $\mathcal{O}(L^2)$ scaling in the number of quantum gates required for time evolution, where $L$ is the lattice volume.
It also requires quantum computers to have all-to-all connectivity between qubits for efficient simulation, a native feature in current
trapped-ion devices, but which has a large overhead on superconducting devices.
Fortunately, electric charges are screened in the Schwinger model, causing correlations between distant fermions to decay exponentially with separation; see Fig.~\ref{fig:SimTricks}a).
In Sec.~\ref{sec:HamTruncs}, this screening is used to truncate interactions between fermions beyond a distance, $\overline{\lambda}$, set by the correlation length and the desired level of precision of the simulation.
This improves the scaling of the number of gates required for time evolution to $\mathcal{O}(\overline{\lambda} L)$, with $\mathcal{O}(\overline{\lambda})$-nearest neighbor qubit connectivity.

The construction of low-depth quantum circuits for state preparation is another challenge addressed in this work.
In the previous chapter, we introduced the SC-ADAPT-VQE algorithm, and applied it to the preparation of the Schwinger model vacuum on 100 qubits of {\tt ibm\_cusco}.
SC-ADAPT-VQE uses symmetries and hierarchies in length scales to determine low-depth quantum circuits for state preparation.
Using a hybrid workflow, quantum circuits are determined and optimized on a series of small and modest-sized systems using {\it classical computers}, and then systematically scaled to large systems to be executed on a {\it quantum computer}.
In Sec.~\ref{sec:SCADAPT}, SC-ADAPT-VQE is extended to the preparation of localized states, and used to establish a hadron wavepacket on top of the interacting vacuum; see Fig.~\ref{fig:SimTricks}b).
The wavepacket preparation circuits are optimized on a series of a small lattices by maximizing the overlap with an adiabatically prepared wavepacket.
The locality of the target state ensures that these circuits can be systematically extrapolated to prepare hadron wavepackets on large lattices.

Quantum 
circuits for state preparation and time evolution are developed in Section~\ref{sec:ClSim}.
The circuit design minimizes the two-qubit gate count for implementation on devices with nearest-neighbor connectivity, such as those available from IBM.
A building block for these circuits is a new gate decomposition for $R_{ZZ}$ rotations acting between all pairs of a set of qubits.
This nearest-neighbor decomposition uses the same number of two-qubit gates as decompositions for devices with all-to-all connectivity, at the cost of an increased circuit depth.
Results from classical simulations performed on small lattices are presented in Sec.~\ref{sec:Systematics}.
These simulations quantify the systematic errors originating from the approximations introduced in previous sections: preparation of the hadron wavepacket with SC-ADAPT-VQE, use of a truncated Hamiltonian for time evolution, and  
Trotterization of the time evolution operator.

In Sec.~\ref{sec:Qsim}, the techniques and ideas described in the previous paragraphs
are applied to quantum simulations of hadron dynamics on $L=56$ (112 qubit) lattices using IBM's quantum computer {\tt ibm\_torino}.
The initial state is prepared using SC-ADAPT-VQE, and time evolution is implemented with up to 14 Trotter steps, requiring 13,858 CNOTs (CNOT depth 370).
After applying a suite of error mitigation techniques, measurements of the local chiral condensate show clear signatures of hadron propagation.
The results obtained from {\tt ibm\_torino} are compared to classical simulations using the {\tt cuQuantum} Matrix Product State (MPS) simulator.
In these latter calculations, the bond dimension in the tensor network simulations grows with the simulation time, requiring increased classical computing overhead. 
Appendix~\ref{app:MPSSim} provides details about
the convergence of the MPS simulations, 
and App.~\ref{app:qSimDetails} provides details of
our error mitigation strategy, for our simulations using 112 qubits of IBM's quantum computers.
This work points to quantum simulations of more complex processes, such as inelastic collisions, fragmentation and hadronization, as being strong candidates for a near-term quantum advantage.

\section{Systematic Truncation of the Electric Interactions}
\label{sec:HamTruncs}
\noindent
The Schwinger model Hamiltonian in axial gauge and mapped to spin operators is given in Eq.~\eqref{eq:Hgf}.
Due to the removal of the gauge degrees of freedom, the electric interactions are pair-wise between all of the fermions.
This is problematic for implementing time evolution $e^{-i t \hat{H}}$ on a quantum computer as it implies an $\mathcal{O}(L^2)$ scaling in the number of gates.
In addition, this interaction requires connectivity between every pair of qubits for efficient implementation.
Fortunately, charges are screened in  confining theories like the Schwinger model, and correlation functions decay exponentially between charges separated by more than approximately a correlation length, $\xi$.
The correlation length is a scale that emerges from the solution of the theory, and is naturally related to the hadron mass, $\xi \sim 1/m_{\text{hadron}}$.
This motivates the construction of an effective Hamiltonian where interactions between distant charges are removed.
Such an effective interaction is systematically improvable with exponentially suppressed errors, and only requires $\mathcal{O}(\xi L)$ gates acting between qubits with maximum separation $\sim\xi$.

To form the effective interactions, it is beneficial to first specialize to the $Q=0$ sector with zero background electric field.
There are many equivalent ways to express the interaction due to the freedom of integrating Gauss's law from the left or right side of the lattice 
when constraining the electric field.
However, the desire to preserve CP symmetry in the truncated theory motivates starting from a manifestly CP-symmetric interaction,
\begin{align}
\hat{H}_{el}^{(Q=0)} \ &= \ \frac{g^2}{2} \left \{\sum_{j=0}^{L-2}\left ( \sum_{k=0}^j \hat{Q}_k\right )^2\ + \ \sum_{j=L+1}^{2L-1}\left ( \sum_{k=j}^{2L-1} \hat{Q}_k\right )^2 \ + \ \frac{1}{2}\left [\left (\sum_{j=0}^{L-1} \hat{Q}_j \right )^2 + \left (\sum_{j=L}^{2L-1} \hat{Q}_j \right )^2 \right ] \right \}
\ .
\label{eq:HelCP}
\end{align}
This has decoupled the interactions between charges on different halves of the lattice.
The most straightforward way to form the effective interactions would be to remove $\hat{Q}_j \hat{Q}_{j+d}$ terms with $d\gtrsim \xi$.
However, this is ineffective because it is only the {\it connected} correlations that decay exponentially; on a staggered lattice, $\langle \hat{Q}_j \rangle \neq 0$ and $\langle \hat{Q}_j \hat{Q}_{j+d}\rangle = \langle \hat{Q}_j \rangle \langle \hat{Q}_{j+d} \rangle + {\cal O}(e^{-d/\xi})$.

\begin{figure}[!tb]
    \centering
    \includegraphics[width=\textwidth]{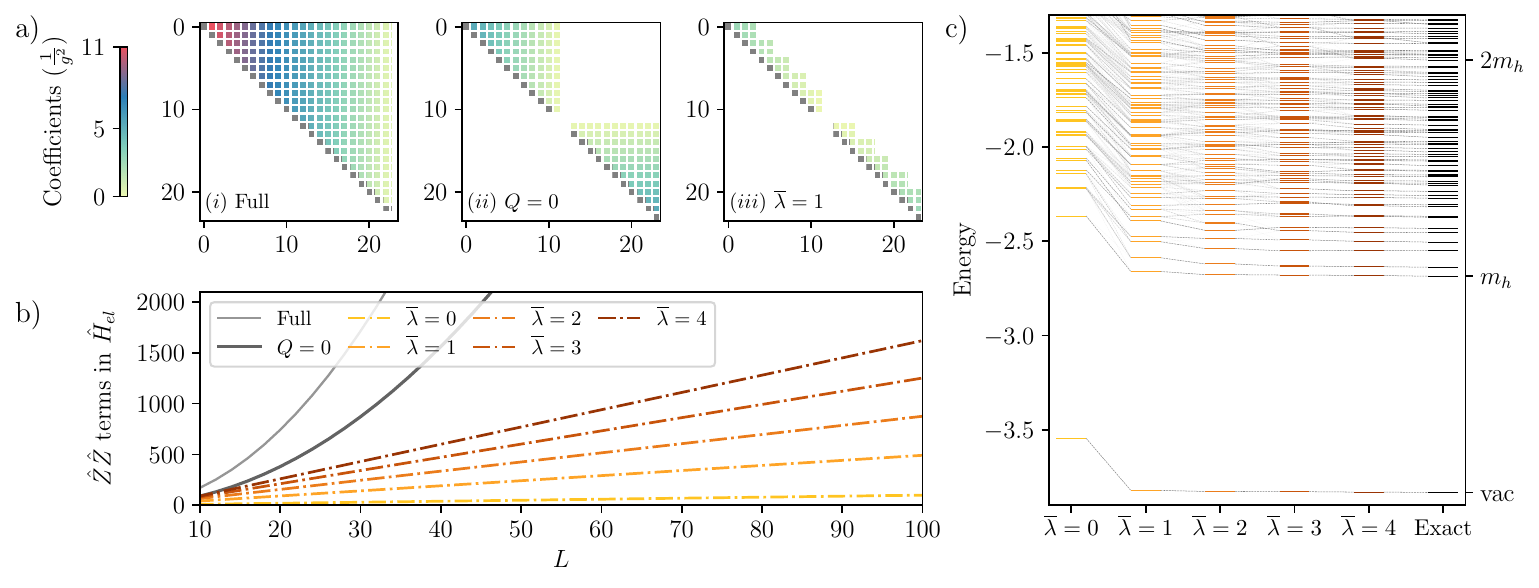}
    \caption{a) The qubit coupling matrix for select electric Hamiltonians with $L=12$:  $(i)$ shows the coupling matrix without truncation [Eq.~\eqref{eq:Hgf}],  
    $(ii)$ shows the impact of restricting to the $Q=0$ sector [Eq.~\eqref{eq:HelCP}], and 
    $(iii)$ corresponds to additionally truncating the interaction between charges separated by more than $\overline{\lambda}=1$ spatial sites [Eq.~\eqref{eq:spatTrun1}]. 
    b) The number of $\hat{Z}\hat{Z}$ terms in  different versions of the electric Hamiltonian as a function of $L$, showing the quadratic $L^2$ and linear $\overline{\lambda}L$ growth. 
    c) The effects of truncating the electric interaction on the low-lying CP-even and $Q=0$ spectrum as a function of $\overline{\lambda}$ for $L=12$. The transparency of the lines connecting energy levels is proportional to the overlap of their corresponding eigenstates.}
    \label{fig:ZZcomparison_2}
\end{figure}

In order to remove the effects of disconnected correlations, consider charges and dipole moments defined on {\it spatial} sites,
\begin{equation}
\hat{\overline{Q}}_n \ = \ \hat{Q}_{2n} + \hat{Q}_{2n+1} \ \ , \ \ \hat{\delta}_n \ = \ \hat{Q}_{2n} - \hat{Q}_{2n+1} \ .
\end{equation}
Unlike charges on staggered sites, the expectation value of a charge on a spatial site is zero, up to exponentially suppressed boundary effects, see App.~B of Ref.~\cite{Farrell:2023fgd}.
Of relevance to constructing the effective Hamiltonian is that correlations between spatial charges, and between spatial charges and dipole moments, decay exponentially,
\begin{equation}
\langle \hat{\overline{Q}}_n \hat{\overline{Q}}_{n+d}\rangle \sim e^{- d/\bar{\xi}} \ \ , \ \ \langle \hat{\overline{Q}}_n \hat{\delta}_{n+d}\rangle \sim e^{-d/\bar{\xi}} \ ,
\end{equation}
for $d\gtrsim \bar{\xi}$,\footnote{Dipole-dipole interactions between spatial sites vanish since the Coulomb potential is linear in one dimension.} where $\bar{\xi} = \xi/2$ is the correlation length in units of spatial sites.
Rewriting $\hat{H}_{el}^{(Q=0)}$ in terms of spatial charges and dipole moments, and truncating interactions beyond $\overline{\lambda}$ spatial sites, it is found that
\begin{align}
\hat{H}_{el}^{(Q=0)} & (\bar{\lambda}) \ = \   \ \frac{g^2}{2}\bigg\{ \sum_{n=0}^{\frac{L}{2}-1} \left[ \left( L - \frac{5}{4} - 2n \right) \hat{\overline{Q}}^2_n + \frac{1}{2} \hat{\overline{Q}}_n \hat{\delta}_n + \frac{1}{4} \hat{\delta}^2_n  \right . \nonumber \\
&+ \left . \left( \frac{3}{4} + 2n \right) \hat{\overline{Q}}^2_{\frac{L}{2}+n} - \frac{1}{2} \hat{\overline{Q}}_{\frac{L}{2}+n} \hat{\delta}_{\frac{L}{2}+n} + \frac{1}{4} \hat{\delta}^2_{\frac{L}{2}+n} \right]   \nonumber \\
&+  2\sum_{n=0}^{\frac{L}{2}-2} \ \ \sum_{m=n+1}^{\min(\frac{L}{2}-1,n+\bar{\lambda})} \left[ \left( L - 1 - 2m \right) \hat{\overline{Q}}_n\hat{\overline{Q}}_m + \frac{1}{2} \hat{\overline{Q}}_{n} \hat{\delta}_{m} \right . \nonumber \\
&+ \left . \left( 1 + 2n \right) \hat{\overline{Q}}_{\frac{L}{2}+n}\hat{\overline{Q}}_{\frac{L}{2}+m} - \frac{1}{2} \hat{\overline{Q}}_{\frac{L}{2}+m} \hat{\delta}_{\frac{L}{2}+n} \right] \bigg\} \ .
\label{eq:spatTrunce}
\end{align}
This expression holds for even $L$, and the analogous expression for odd $L$ can be found in App.~\ref{app:truncHam}.
For $m=0.5,g=0.3$, $\overline{\xi} \sim 0.5$, and $\overline{\lambda}=1$ will be used for demonstration purposes in the remainder of this work. 
Expressed in terms of spin operators, the $\overline{\lambda}=1$ interaction is,
\begin{align}
&\hat{H}_{el}^{(Q=0)}(1) 
\ = \ \frac{g^2}{2}\Bigg\{ \sum_{n=0}^{\frac{L}{2}-1} \left[ \left( \frac{L}{2} - \frac{3}{4} - n \right)\hat{Z}_{2n}\hat{Z}_{2n+1}+\left (n+\frac{1}{4} \right )\hat{Z}_{L+2n}\hat{Z}_{L+2n+1}\right ] 
\nonumber \\
&+ \ \frac{1}{2}
\sum_{n=1}^{\frac{L}{2}-2} \left (2 \hat{Z}_{2n} + \hat{Z}_{2n+1}-\hat{Z}_{L+2n}-2\hat{Z}_{L+2n+1} \right )
\nonumber \\
&+ \ \frac{1}{2}\left (2\hat{Z}_{0}+\hat{Z}_{1}+\hat{Z}_{L-2}-\hat{Z}_{L+1}-\hat{Z}_{2L-2}-2\hat{Z}_{2L-1} \right )  \nonumber \\
&+ \sum_{n=0}^{\frac{L}{2}-2} \bigg[\left (\frac{L}{2}-\frac{5}{4}-n \right )(\hat{Z}_{2n}+\hat{Z}_{2n+1})\hat{Z}_{2n+2} + \left( \frac{L}{2} - \frac{7}{4} - n \right)(\hat{Z}_{2n}+\hat{Z}_{2n+1})\hat{Z}_{2n+3}
\nonumber \\
&+ \  \left (n+\frac{1}{4} \right )(\hat{Z}_{L+2n+2}+\hat{Z}_{L+2n+3})\hat{Z}_{L+2n} +\left (n+\frac{3}{4} \right )(\hat{Z}_{L+2n+2}+\hat{Z}_{L+2n+3})\hat{Z}_{L+2n+1} \bigg] \Bigg\} 
\ .
\label{eq:spatTrun1}
\end{align}
Factors of the identity have been dropped as they do not impact time evolution, and this expression only holds for even $L\geq 4$. 
The effects of these truncations on qubit connectivity, number of two-qubit $\hat{Z}\hat{Z}$ terms, and the low-lying spectrum are illustrated in Fig.~\ref{fig:ZZcomparison_2}.
The number of two-qubit operations required for time evolution now scales linearly with volume ${\cal O}(\overline{\lambda} L)$, and there are only operations between qubits separated by at most $(2\overline{\lambda}+1)$ staggered sites.
This interaction will be used to time evolve a wavepacket of single hadrons, and it is important that the impact of these truncations is small on the low-lying hadron states. 
This is illustrated in panel c) of Fig.~\ref{fig:ZZcomparison_2}, where the low-lying spectrum is shown to rapidly converge with increasing $\overline{\lambda}$.
There is some transient behavior presumably due to tunneling beyond the truncation range.
It is important to stress that the exponentially-converging truncations that are made possible by confinement are not obvious at the level of the spin Hamiltonian in Eq.~\eqref{eq:Hgf} due, in part, to $\hat{\overline{Q}}_n \hat{\overline{Q}}_m$ having conspiring single $\hat{Z}$ and double $\hat{Z}\hat{Z}$ terms.
%

\section{SC-ADAPT-VQE for State Preparation}
\label{sec:SCADAPT}
\noindent
In the previous chapter, the Scalable Circuits-ADAPT-VQE (SC-ADAPT-VQE) algorithm and workflow was introduced, and used to prepare the vacuum of the Schwinger model on 100 qubits of IBM's quantum computers. 
Here, SC-ADAPT-VQE will be detailed in general, and subsequent sections will apply it to prepare both the vacuum and a hadron wavepacket.
The goal of SC-ADAPT-VQE is to determine low-depth circuits for preparing a target wavefunction that are systematically scalable to any lattice size. 
This scalability enables a  hybrid workflow where circuits determined using {\it classical computers} are scaled and executed on a {\it quantum computer}. 
This eliminates the difficult task of optimizing parameterized quantum circuits on a quantum computer that has both statistical noise from a finite number of shots and device errors~\cite{Wang:2020yjh,Scriva:2023sgz,Cerezo:2023nqf}.

The initial steps of SC-ADAPT-VQE parallel those of ADAPT-VQE~\cite{Grimsley_2019}, and can be summarized as:
\begin{itemize}
    \item[1.] Define a pool of operators $\{ \hat{O} \}$ that respect the symmetries of the prepared state. 
    Scalability and phenomenological considerations are used to inform which operators are included in the pool. 
    \item[2.] Initialize a state $\lvert \psi_{{\rm ansatz}}\rangle$ with the quantum numbers of the target state $\lvert \psi_{{\rm target}}\rangle$. 
    \item[3.] Determine a quantity that measures the quality of the ansatz state. 
    For demonstration, consider the infidelity between the ansatz and target states, ${\cal I} = 1 - \vert \langle \psi_{{\rm target}} \vert \psi_{{\rm ansatz}} \rangle \vert ^2$. 
    \item[4.] For each operator in the pool $\hat{O}_i$ determine the gradient of the infidelity between the target and evolved ansatz states, $\frac{\partial}{\partial\theta_i}\left. {\cal I} \right|_{\theta_i=0} = \frac{\partial}{\partial\theta_i} \left.\left( 1 - |\langle\psi_{\rm target} | e^{i \theta_i \hat{O}_i } |\psi_{\rm ansatz}\rangle|^2 \right)\right|_{\theta_i=0}$. 
    This is one way of ranking the relative impact of $\hat{O}_i$ on the infidelity.
    \item[5.] Identify the operator $\hat{O}_n$ with the largest magnitude gradient.
    Update the ansatz with the parameterized evolution of the operator $\lvert \psi_{{\rm ansatz}} \rangle \to e^{i \theta_n \hat{O}_n}\lvert \psi_{{\rm ansatz}} \rangle$.
    \item[6.] Optimize the variational parameters to minimize the infidelity.
    The previously optimized values for $\theta_{1,...,n-1}$ and $\theta_n=0$, are used as initial conditions.
    \item[7.] Return to step 4 until the desired tolerance is achieved.
\end{itemize}
ADAPT-VQE returns an ordered sequence of unitary operators $\{ \hat{U}_i \} = \{ {\exp}(i \theta_i \hat{O}_i) \} $ that prepares the target state up to a desired tolerance.
For use on a quantum computer, the sequence of unitaries can be converted to a sequence of gates through, for example, Trotterization.
If this introduces Trotter errors, the unitaries in steps 4 and 5 should be replaced by their Trotterized versions, ${\exp}(i \theta_i \hat{O}_i) \rightarrow \prod\limits_j \hat{U}_j^{(i)}$.
In SC-ADAPT-VQE, the previous steps are supplemented with the following,
\begin{itemize}
    \item[8.] Repeat ADAPT-VQE for a series of lattice volumes $\{L_1, L_2, \ldots, L_N\}$ using a classical computer (or a small partition of a quantum computer).
    \item[9.] Extrapolate the sequence of unitary operators $\{\{\hat{U}_i\}_{L_1}, \{\hat{U}_i\}_{L_2}, \ldots, \{\hat{U}_i\}_{L_N}\}$ to the desired $L$.
    This sequence is expected to converge for states with localized correlations.
    $L$ can be arbitrarily large and beyond what is accessible
     using a classical computer.
\end{itemize}
The sequence of extrapolated unitaries $\{\hat{U}_i\}_L$ can then be used to prepare the target state on a quantum computer.
This provides an explicit implementation of systematically-localizable~\cite{Klco:2019yrb} and fixed-point~\cite{Klco:2020aud} quantum operators and circuits.

\subsection{Hadron Wavepacket Preparation}
\label{sec:SCADAPTWP}
\begin{figure}[!tb]
    \centering
    \includegraphics[width=\textwidth]{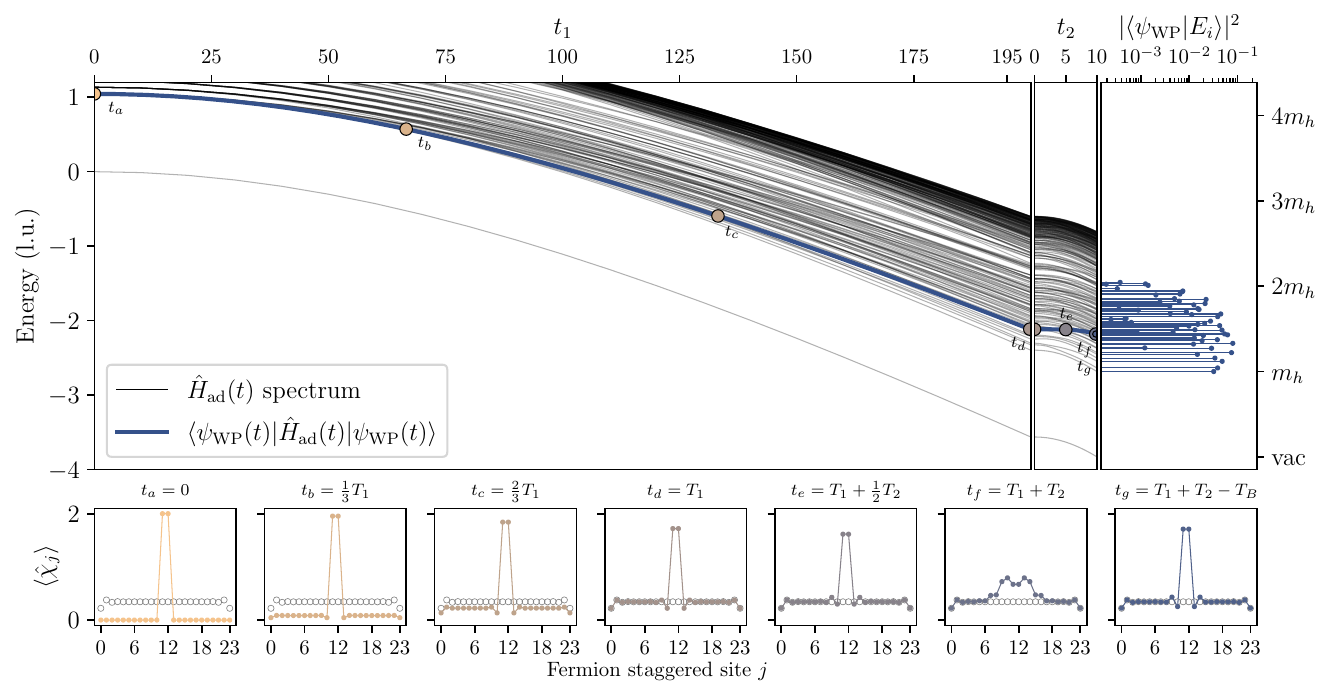}
    \caption{Adiabatic state preparation for $L=12$. Upper panels: the lowest 200 eigenenergies of $\hat{H}_{\text{ad}}(t)$ as a function of adiabatic turn-on time, the energy of the state $|\psi_{\rm WP}(t)\rangle$, and the final overlap between $|\psi_{\rm WP}\rangle$ and the eigenstates of $\hat{H}$, $|E_i\rangle$. Lower panels: evolution of the chiral condensate $\langle\hat{\chi}_j\rangle$, defined in Eq.~\eqref{eq:localCC}, of $|\psi_{\rm WP}(t)\rangle$ for a selection of times, 
    $t_a-t_g$, 
    with the empty markers showing the $\langle \hat{\chi}_j \rangle$ of the vacuum $|\psi_{\rm vac}\rangle$ of $\hat{H}$.}
    \label{fig:adiabatic_stateprep}
\end{figure}

\noindent
SC-ADAPT-VQE can be used to prepare a state that has large overlap with an adiabatically prepared hadron wavepacket.
An alternative method for preparing wavepackets is discussed in App.~\ref{app:chempot}.
In a lattice theory of interacting scalar fields, a complete procedure for preparing single particle wavepackets has been proposed by Jordan, Lee and Preskill~\cite{Jordan_2018,DBLP:journals/qic/JordanLP14}.\footnote{Other proposals for creating initial states and wavepackets can be found in Refs.~\cite{Kuhn:2015zqa,Pichler:2015yqa,Damme:2019rts,Avkhadiev:2019niu,Surace:2020ycc,Karpov:2020pqe,Milsted:2020jmf,Vovrosh:2022bpj,Asaduzzaman:2022bpi,Avkhadiev:2022ttx,Farrell:2022vyh,Atas:2022dqm,Vary:2023ihk,Chai:2023qpq,Roy:2023uil}, including recent work on creating hadronic sources in the bosonized form of the Schwinger model using circuit-QED~\cite{Belyansky:2023rgh}.}
In their method, wavepackets are first prepared in free scalar field theory, and then the $\lambda \phi^4$ interaction is adiabatically ``turned on''.
This method runs into difficulty in the Schwinger model because the single particle states (hadrons) of the interacting theory are non-perturbatively different from the single particle states of the non-interacting theory (electrons).
To overcome this, consider starting in the interacting theory with $m=0.5$ and $g=0.3$, and adiabatically turning on the kinetic term.
The initial Hamiltonian is diagonal in the computational $z$-basis, and the ground state is the same as the infinite coupling (anti-ferromagnetic) vacuum $\vert \Omega_0 \rangle$.
The infinite-coupling vacuum provides a suitable starting configuration upon which to build the wavepacket as it correctly encodes the long-distance correlations that characterize this confining theory.\footnote{The strong-coupling limit has been extensively studied, particularly in the context of lattice QCD. See, for example, Ref.~\cite{Fromm:2009xw} and references therein.}
On this vacuum, a hadron can be excited by creating an $e^-e^+$ pair on adjacent staggered sites.
By preparing a superposition of such hadrons at different locations, an arbitrary wavepacket can be prepared. 
Here, the focus will be on preparing a localized hadron wavepacket that is centered in the middle of the lattice to preserve CP and minimize boundary effects.
A suitable initial state is, 
\begin{equation}
|\psi_{{\rm WP}}\rangle_{\text{init}} = \hat{X}_{L-1}\hat{X}_L\vert \Omega_0\rangle \ .
\end{equation}
To transition to a hadron wavepacket in the full theory, this state is taken through two steps of adiabatic evolution with a time-dependent Hamiltonian (illustrated in Fig.~\ref{fig:adiabatic_stateprep}),
\begin{equation}
    \hat{H}_{\rm ad}(t) = 
    \begin{cases}
    \hat{H}_m + \hat{H}_{el} + \frac{t}{T_1} \left [ \hat{H}_{kin} \ - \ \frac{1}{2} (\sigma^+_{L-2}\sigma^-_{L-1} + \sigma^+_{L}\sigma^-_{L+1} + {\rm h.c.} ) \right ]  \ &0<t\leq T_1  \ , \\[5pt]
    \hat{H}_m + \hat{H}_{el} + \hat{H}_{kin} \ - \ \left (1-\frac{t-T_1}{T_2} \right ) \frac{1}{2} (\sigma^+_{L-2}\sigma^-_{L-1} + \sigma^+_{L}\sigma^-_{L+1} + {\rm h.c.} )   \ &T_1<t\leq T_1+T_2  \ . 
    \end{cases}
\end{equation}
For $t\in (0,T_1 ]$, the kinetic term is adiabatically turned on everywhere except for the links connecting the initial wavepacket to the rest of the lattice.
This mitigates spatial spreading of the initial wavepacket (see times $t_{a,b,c,d}$ in Fig.~\ref{fig:adiabatic_stateprep}).
Next, for $t\in (T_1,T_2 ]$, the remaining two links are adiabatically turned on. 
These remaining links are spatially localized (act over a pair of staggered sites), and therefore primarily couple to high-momentum (energy) states.
This implies that the energy gap relevant for the adiabatic evolution is large, and the second evolution can be performed much faster than the first evolution.
There is a small amount of wavepacket spreading (times $t_{e,f}$), which is undone by evolving backwards in time for a duration $T_B = T_2/2$ with the full Hamiltonian $\hat{H}$ from Eq.~\eqref{eq:Hgf} (time $t_g$).
Explicitly, the hadron wavepacket is given by,
\begin{equation}
\vert \psi_{{\rm WP}} \rangle \ = \  
 e^{i T_B \hat{H}} \, {\cal T} e^{-i \int_{0}^{T_1 + T_2}dt \hat{H}_{\rm ad}(t)} \vert \psi_{{\rm WP}}\rangle_{\text{init}} \ ,
\label{eq:adiabaticWP}
\end{equation}
where ${\cal T}$ denotes time-ordering.
For practical implementation, the evolution of the time-dependent Hamiltonian can be accomplished with Trotterization,
\begin{equation}
{\cal T} e^{-i \int_{0}^{T_1 + T_2}dt \hat{H}_{\rm ad}(t)} \ \approx \ {\cal T} e^{-i \sum_{n=0}^{N_T-1} \delta t \hat{H}_{\rm ad}\left [(n+0.5) \delta t\right ]} \ \approx \ {\cal T} 
 \prod_{n=0}^{N_T-1}e^{-i \left (\delta t \right ) \hat{H}_{\rm ad}\left [(n+0.5) \delta t\right ]} \ ,
\end{equation}
where $N_T$ is the number of Trotter steps and $\delta t = \frac{T_1+T_2}{N_T}$ is the step size.
For the simulation parameters chosen in this work, we find that $T_1 = 200$, $T_2=10$ and $\delta t=0.2$ are sufficient for adiabatic evolution.
The final state is localized (within a few sites), and primarily consists of single-hadron states (see overlaps in Fig.~\ref{fig:adiabatic_stateprep}).

In principle, this adiabatic procedure could be used to prepare a hadronic wavepacket on a quantum computer. 
In practice, the required circuits are too deep to run on current devices.
To address this, SC-ADAPT-VQE is used to find low-depth circuits that prepare an approximation to the adiabatically determined wavepacket.
These low-depth circuits act on the vacuum, whose preparation was outlined in the previous section.
Scalability of the state preparation circuits is expected because the constructed wavepacket is localized away from the boundaries, and is built on top of a vacuum state that has converged exponentially in $L$ to its infinite-volume form~\cite{Farrell:2023fgd}.
As both the initial state (vacuum) and target state (single hadron) are CP even and charge zero, the operators in the pool must conserve charge and CP.
An operator pool that is found to produce a wavefunction that converges exponentially fast in circuit depth is,
\begin{align}
    \{\hat{O} \}_{\text{WP}} \ &= \ 
    \{ \hat{O}_{mh}(n,d), \, \hat{O}_{h}(n,d),\, \hat{O}_{m}(n) \} \ ,\nonumber \\
    \hat{O}_{mh}(n,d) &=\frac{1}{2} \big[ 
    \hat{X}_{L-n} \hat{Z}^{d-1}  \hat{Y}_{L-n+d} 
    - \hat{Y}_{L-n}  \hat{Z}^{d-1} \hat{X}_{L-n+d}  
    \nonumber \\
    &+ \ (-1)^{d+1}\left (1-\delta_{L-n,\gamma} \right  )\left (\hat{X}_{\gamma} \hat{Z}^{d-1}  \hat{Y}_{\gamma+d} - \hat{Y}_{\gamma}  \hat{Z}^{d-1} \hat{X}_{\gamma+d} \right ) \big] 
    \ ,\nonumber \\[4pt]
    \hat{O}_{h}(n,d) &=\frac{1}{2} \big[ 
    \hat{X}_{L-n} \hat{Z}^{d-1}  \hat{X}_{L-n+d} 
    + \hat{Y}_{L-n}  \hat{Z}^{d-1} \hat{Y}_{L-n+d}  
    \nonumber \\
    &+ \ (-1)^{d+1}\left (1-\delta_{L-n,\gamma} \right  )
    \left (\hat{X}_{\gamma} \hat{Z}^{d-1}  \hat{X}_{\gamma+d} + \hat{Y}_{\gamma}  \hat{Z}^{d-1} \hat{Y}_{\gamma+d} \right ) \big] \ ,\nonumber \\[4pt]
    \hat{O}_{m}(n) & = \hat{Z}_{L-n}  \ - \ \hat{Z}_{L-1+n} \ ,
    \label{eq:PacketPool}
\end{align}
where $\gamma=L-1+n-d$, $n\in \{1,\ldots,L \}$, and the $\left (1-\delta_{L-n,\gamma} \right  )$ coefficients prevent double counting operators that are already CP-symmetric.
The pool operators are inspired by the Hamiltonian, with $\hat{O}_m(n)$ being a mass-like operator, $\hat{O}_h(n,d)$ a generalized hopping operator spanning $d$ staggered sites, and $\hat{O}_{mh}(n,d)$ being proportional to their commutator.
Note that unlike the operator pool used to prepare the vacuum, $\{ \hat{O} \}_{\text{WP}}$ is not constrained by time reversal or translational symmetry, and the individual terms in each operator commute.
Thus there are no Trotter errors when the corresponding unitaries are converted to circuits.

\begin{figure}[!tb]
    \centering
    \includegraphics[width=0.55\textwidth]{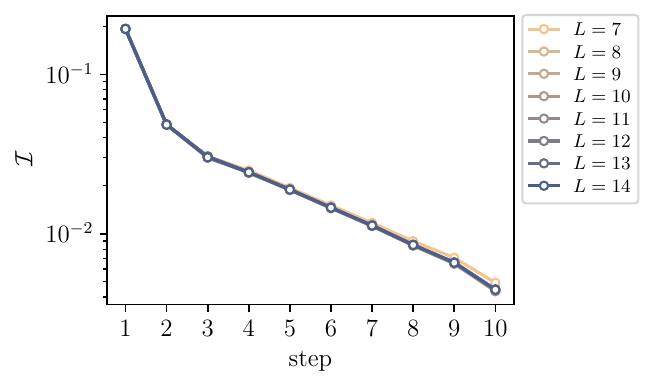}
    \caption{Infidelity of the wavepacket, defined in Eq.~\eqref{eq:inf_wp_scadapt}, prepared with multiple steps of SC-ADAPT-VQE for a range of $L$.}
    \label{fig:WP_adapt_inf}
\end{figure}

The initial state for SC-ADAPT-VQE is chosen to be $\lvert \psi_{\text{ansatz}} \rangle = \lvert \psi_{\text{vac}} \rangle$, as this correctly reproduces the vacuum outside of the support of the hadron wavepacket.
In this section, all calculations are performed with exact diagonalization, and the initial state is the exact vacuum. 
In Secs.~\ref{sec:ClSim} and~\ref{sec:Qsim}, the initial state will be the SC-ADAPT-VQE prepared vacuum. 
Using the exact vacuum instead of the SC-ADAPT-VQE vacuum prevents operators from being chosen that improve the vacuum but do not build out the local profile of the wavepacket.
The quality of the prepared state is determined by the infidelity of the ansatz state with the adiabatically prepared state from Eq.~\eqref{eq:adiabaticWP},
\begin{equation}
{\cal I} \ = \ 1 \ - \ \vert \langle \psi_{\text{WP}} \vert \psi_{\text{ansatz}}\rangle\vert^2 \ .
\label{eq:inf_wp_scadapt}
\end{equation}
Results obtained from performing the steps in SC-ADAPT-VQE (outlined in the introduction of Sec.~\ref{sec:SCADAPT}) for $L=7-14$ are shown in Fig.~\ref{fig:WP_adapt_inf} and Table~\ref{tab:AnglesWP10}.\footnote{The vacuum maximizes the infidelity (has ${\cal I}=1$) with the adiabatically determined state as there is no overlap between the vacuum and the single-hadron states that make up the wavepacket.
This presents a problem in step 4 of SC-ADAPT-VQE since $\frac{\partial}{\partial \theta_i} {\cal I}$ is zero for all operators in the pool.
To overcome this, for the first iteration of SC-ADAPT-VQE, the parameterized evolution of the ansatz with each operator is determined separately.
The operator that minimizes the infidelity is chosen for the first operator in the SC-ADAPT-VQE ansatz.}
Up to the tolerance of the optimizer, the variational parameters have converged in $L$, and therefore the $L=14$ parameters and operator ordering can be used 
to prepare a hadron wavepacket for any $L>14$.
Initially, short-range operators localized around the center of the wavepacket are 
selected by SC-ADAPT-VQE.\footnote{It is interesting to note the similarities between this wavepacket construction, and the construction of hadronic sources and sinks in Euclidean-space lattice QCD calculation.
Here, the initial interpolating operator for the hadronic wavepacket is being ``dressed'' by an increasing number of operators with exponentially improving precision. 
In Euclidean-space lattice QCD, a matrix of correlation functions between a set of sources and sinks is diagonalized to provide a set of correlators with extended plateaus toward shorter times, corresponding to the lowest-lying levels in the spectrum that have overlap with the operator set.
This ``variational method'', e.g., Refs.~\cite{Michael:1982gb,Luscher:1990ck,blossier2008efficient}, provides upper bounds to the energies of the states in the spectrum.
The sources and sinks for hadrons are operators constructed in terms of quark and gluon fields, and correlation functions are formed by contracting field operators of the sinks with those of the sources (or with themselves when both quark and anti-quark operators are present).
This becomes computationally challenging with increasingly complex operator structures, as required, for instance, to study nuclei, see for example Refs.~\cite{Beane:2009gs,Beane:2009py,Aoki:2009yy,Beane:2010em,NPLQCD:2012mex,Yamazaki:2012hi,Yamazaki:2015vjn,Yamazaki:2015asa,Drischler:2019xuo,Davoudi:2020ngi,Amarasinghe:2021lqa}.}
This is as expected for a wavepacket composed of single hadron states with short correlation lengths,  that is approximately a delta function in position space.\footnote{The variational parameters change sign between even- and odd-values of $L$ if $d$ is odd (even) in $\hat{O}_{mh}$ ($\hat{O}_h$). 
Also, note that $\hat{O}_m$ is not chosen until after step 10 in the SC-ADAPT-VQE ansatz.}
The convergence of the infidelity is found to be exponential in the 
step of the algorithm (circuit depth), and independent of $L$.
This is in agreement with previous discussions on localized states being built on top of an exponentially converged vacuum.
Note that the convergence in $L$ is smoother for the SC-ADAPT-VQE wavepacket than for the vacuum as the boundary effects are much smaller (see Fig.~5 in Ref.~\cite{Farrell:2023fgd}).
Two steps of SC-ADAPT-VQE reaches an infidelity of 0.05, and will be used in the remainder of the work to prepare the wavepacket.
\begin{table}[!tb]
\renewcommand{\arraystretch}{1.4}
\resizebox{\textwidth}{!}{\begin{tabularx}{1.4\textwidth}{|c || Y | Y | Y | Y | Y | Y | Y | Y | Y | Y |}
 \hline
 \diagbox[height=23pt]{$L$}{$\theta_i$} & $\hat O_{mh}(1,1)$ & $\hat O_{mh}(2,2)$ & $\hat O_{mh}(3,2)$ & $\hat O_{mh}(3,1)$ & $\hat O_{mh}(5,4)$ &  $\hat O_{h}(2,2)$ & $\hat O_{mh}(4,4)$ & $\hat O_{mh}(4,5)$ & $\hat O_{h}(4,4)$ & $\hat O_{mh}(2,3)$ \\
 \hline\hline
 7 & 1.6370 & -0.3154 & -0.0978 & 0.0590 & -0.0513 & -0.0494 & -0.0518 & -0.0389 & 0.0359 & 0.0528\\
 \hline
 8 & -1.6371 & -0.3157 & -0.0976 & -0.0615 & -0.0499 & 0.0493 & -0.0515 & 0.0391 & -0.0360 & -0.0529\\
 \hline
 9 & 1.6370 & -0.3155 & -0.0980 & 0.0609 & -0.0509 & -0.0493 & -0.0515 & -0.0390 & 0.0361 & 0.0527\\
 \hline
 10 & -1.6370 & -0.3154 & -0.0984 & -0.0598 & -0.0501 & 0.0493 & -0.0515 & 0.0389 & -0.0360 & -0.0527\\
 \hline
 11 & 1.6370 & -0.3155 & -0.0984 & 0.0598 & -0.0507 & -0.0492 & -0.0515 & -0.0390 & 0.0360 & 0.0527\\
 \hline
 12 & -1.6371 & -0.3156 & -0.0975 & -0.0616 & -0.0505 & 0.0493 & -0.0516 & 0.0391 & -0.0361 & -0.0528\\
 \hline
 13 & 1.6371 & -0.3157 & -0.0973 &  0.0617 & -0.0506 & -0.0494 & -0.0516 & -0.0391 &  0.0359 &  0.0529\\
 \hline
 14 & -1.6370 & -0.3155 & -0.0981 & -0.0602 & -0.0506 & 0.0493 & -0.0515 &  0.0390 & -0.0359 & -0.0527\\
 \hline
\end{tabularx}}
\caption{The operator ordering and variational parameters that prepare the 10 step SC-ADAPT-VQE hadron wavepacket. Results are shown for $L=7-14$, and were obtained from a classical simulation using exact exponentiation.}
 \label{tab:AnglesWP10}
\end{table}
%

\section{Quantum Circuits}
\label{sec:ClSim}
\noindent
In this section, the quantum circuits that prepare hadron wavepackets and implement time evolution are developed.
These circuits are constructed to minimize CNOT count and circuit depth in order to reduce the effects of device errors. 
In addition, with the goal of running on {\tt IBM}'s quantum computers, the circuits are optimized for nearest-neighbor connectivity.
These circuits are verified using the {\tt qiskit} classical simulator, and the systematic errors arising from the approximations used in this work are quantified.

\subsection{Quantum Circuits for Vacuum and Hadron Wavepacket Preparation}
\begin{figure}[!b]
    \centering
    \includegraphics[width=0.45\textwidth]{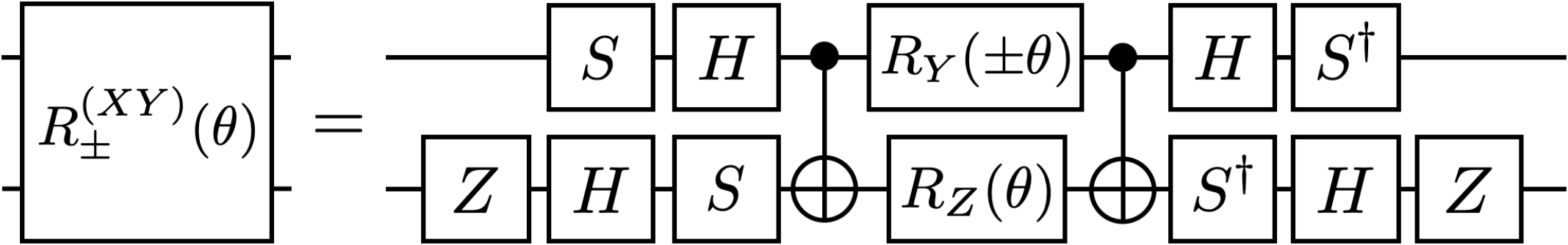}\hfill
    \includegraphics[width=0.45\textwidth]{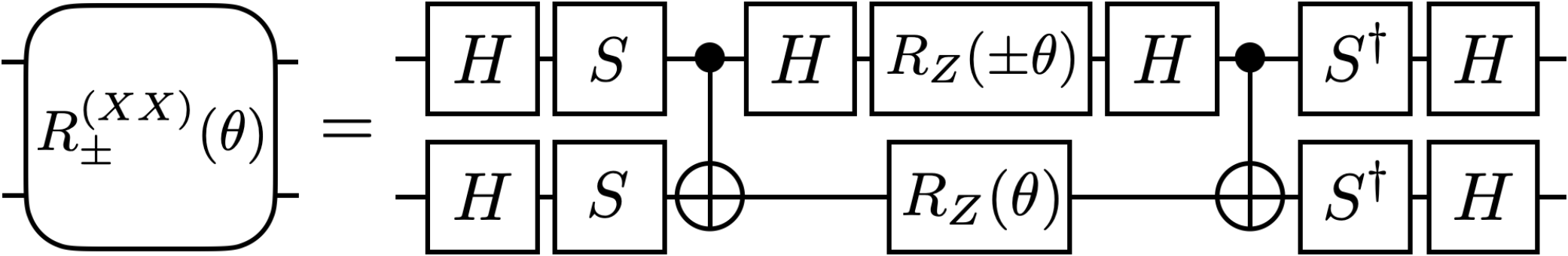}
    \caption{Circuits implementing $R^{(XY)}_{\pm}(\theta)={\rm exp}[-i\tfrac{\theta}{2}(\hat{Y}\hat{X}\pm \hat{X}\hat{Y})]$ (left) and $R^{(XX)}_{\pm}(\theta)={\rm exp}[-i\tfrac{\theta}{2}(\hat{X}\hat{X}\pm \hat{Y}\hat{Y})]$ (right).}
    \label{fig:xy_xx_circuits}
\end{figure}

\begin{figure}[!tb]
    \centering
    \includegraphics[width=0.8\textwidth]{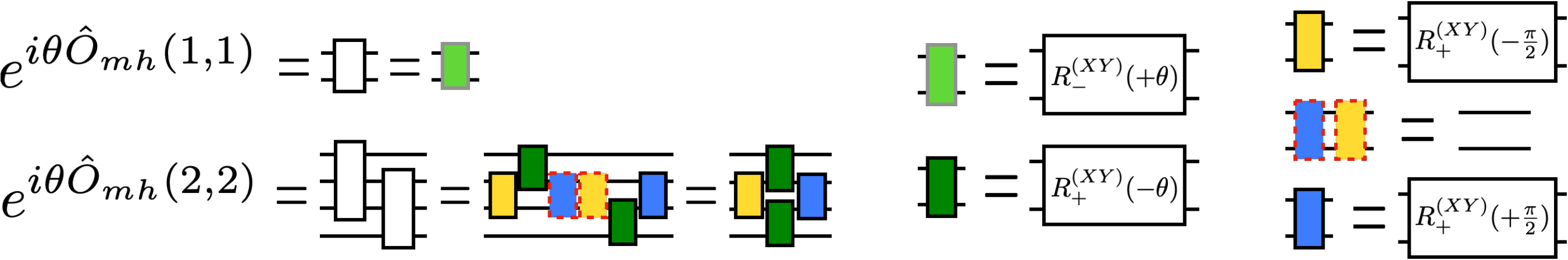}
    \caption{Circuits implementing the unitaries that prepare the 2-step SC-ADAPT-VQE wavepacket. The circuits for the individual blocks $R^{(XY)}_{\pm}(\theta)$ and $R^{(XX)}_{\pm}(\theta)$ are shown in Fig.~\ref{fig:xy_xx_circuits}.}
    \label{fig:2step_wp_adapt}
\end{figure}

\noindent
In order to prepare the SC-ADAPT-VQE vacuum on a quantum computer, the circuits presented in the previous chapter can be used.
The circuit building technique follows the strategy of Ref.~\cite{Algaba:2023enr}, where an ``X''-shaped construction is used to minimize circuit depth and CNOT gate count.
Preparing the SC-ADAPT-VQE hadron wavepacket requires converting the exponential of the pool operators in Eq.~\eqref{eq:PacketPool} to sequences of gates. 
The individual terms in each operator in the wavepacket pool commute, and therefore first-order Trotterization is exact.
The corresponding circuits extend those used for preparing the vacuum, and are shown in Figs.~\ref{fig:xy_xx_circuits} and~\ref{fig:2step_wp_adapt} for the 
2-step SC-ADAPT-VQE wavepacket used in subsequent sections 
(see App.~\ref{app:circuitDetails} for the 10-step SC-ADAPT-VQE circuits). 
These circuits are arranged to maximize cancellations between CNOTs, and minimize the circuit depth.

%

\subsection{Quantum Circuits for Time Evolution }
\noindent
To perform time evolution, a second-order Trotterization of the time-evolution operator with the $\overline{\lambda}=1$ truncated electric interaction will be used,
\begin{equation}
\hat{U}^{\text{(Trot)}}_2(t) \ = \  e^{-i \frac{t}{2} \hat{H}_{kin\text{-1}}} e^{-i \frac{t}{2} \hat{H}_{kin\text{-0}}} e^{-i t  \hat{H}_{m}} e^{-i t \hat{H}_{el}^{(Q=0)}(1)} e^{-i \frac{t}{2} \hat{H}_{kin\text{-0}}}  e^{-i \frac{t}{2} \hat{H}_{kin\text{-1}}} \ ,
\end{equation}
where $\hat{H}_{kin\text{-0}}$ ($\hat{H}_{kin\text{-1}}$) are the hopping terms between even (odd) staggered sites.
This ordering was chosen to maximize the cancellations between neighboring CNOTs. 
A second-order Trotterization is used as it provides a good balance between minimizing both circuit depth and Trotter errors.
In addition, the property of second-order Trotterization $\hat{U}^{(\text{Trot})}_2(t) \hat{U}^{(\text{Trot})}_2(-t) = \hat{1}$ enables a powerful error-mitigation technique~\cite{ARahman:2022tkr}, see Sec.~\ref{sec:Qsim}.

\begin{figure}[!htb]
    \centering
    \includegraphics[width=0.7\columnwidth]{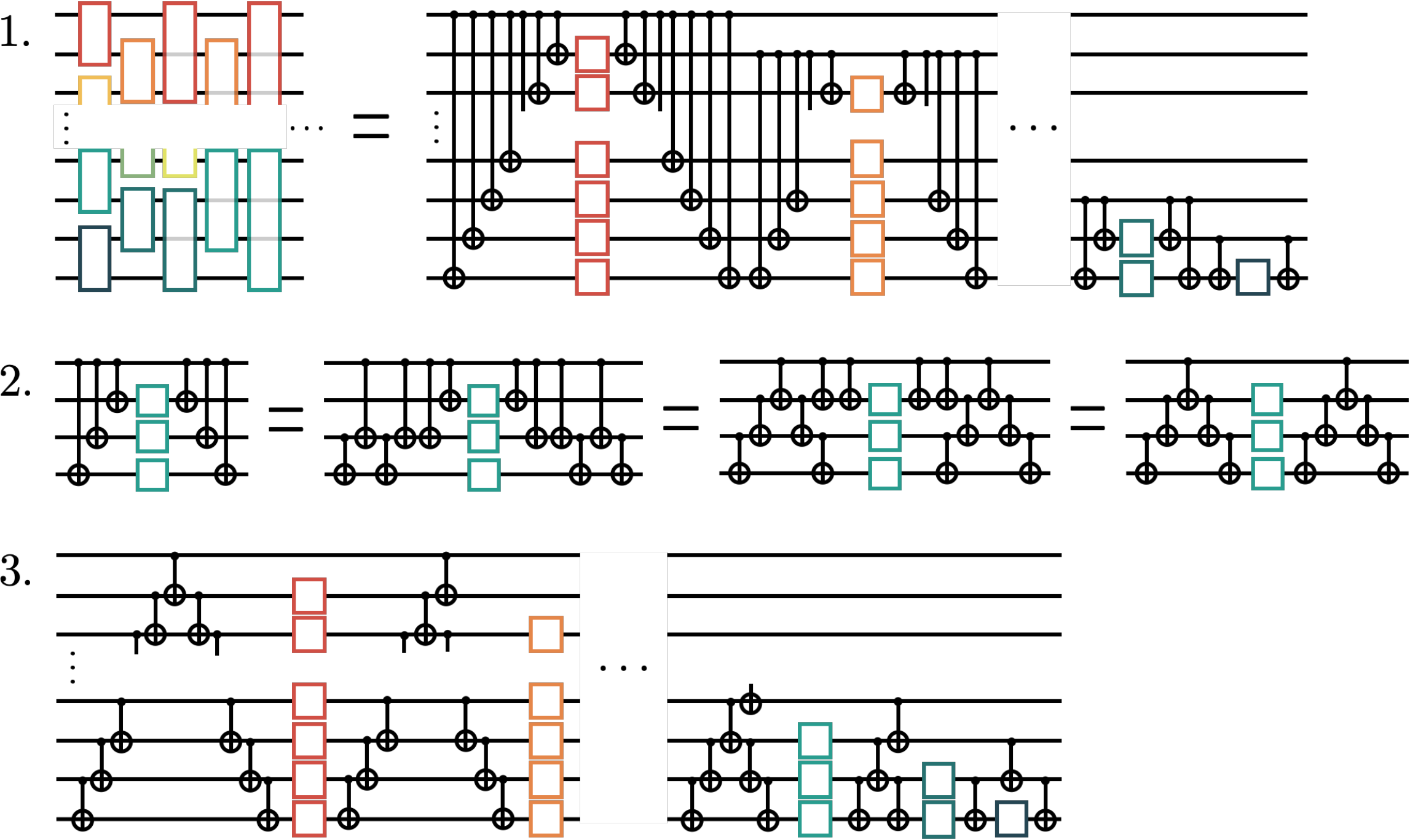}
    \caption{An efficient nearest-neighbor CNOT decomposition for $R_{ZZ}$s between all pairs of $N_q$ qubits.}
    \label{fig:ZZsimplifications}
\end{figure}

\begin{figure}[!htb]
    \centering
    \includegraphics[width=0.85\columnwidth]{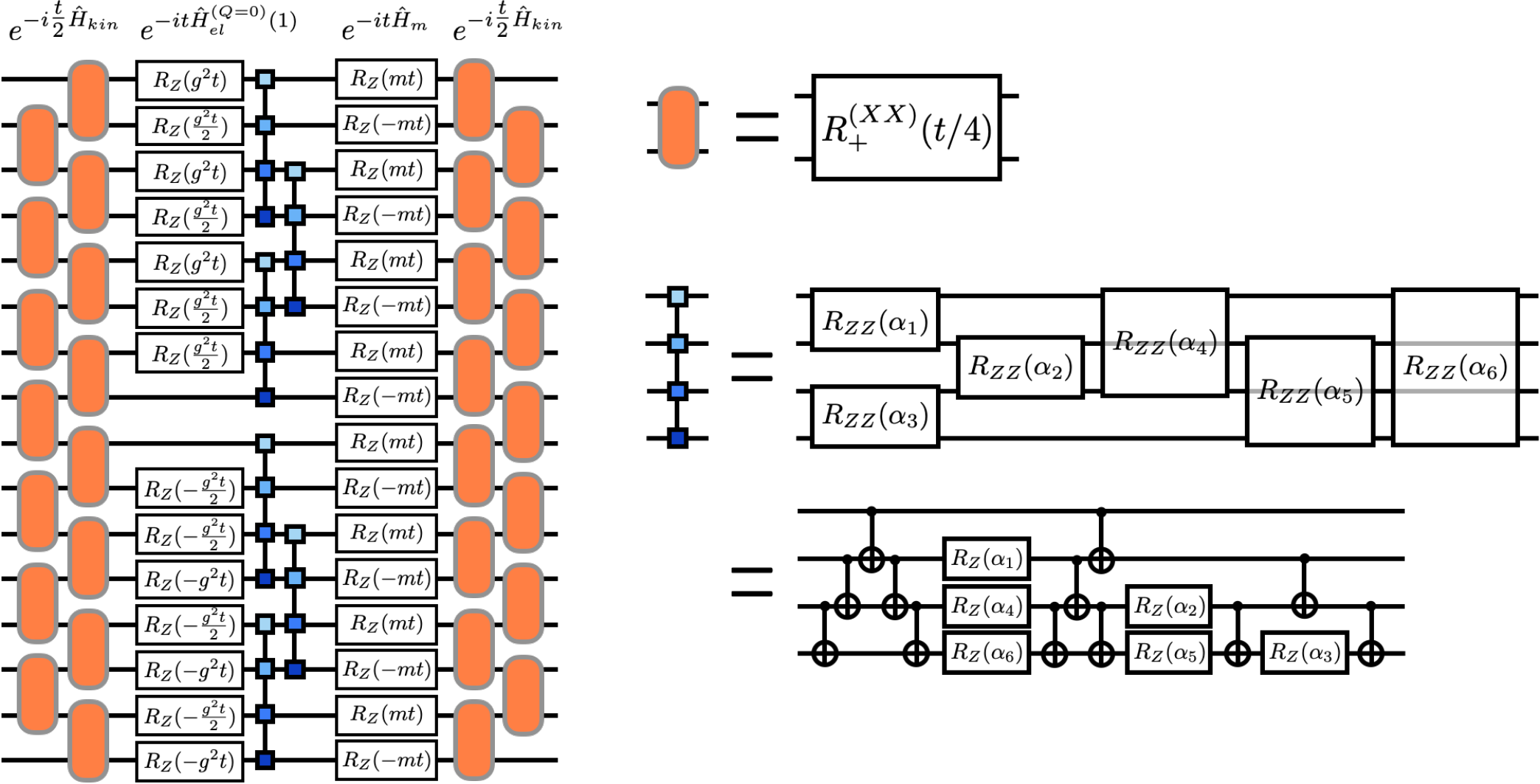}
    \caption{A quantum circuit that implements a single second-order Trotter step associated with the $\overline{\lambda}=1$ truncated Hamiltonian in Eq.~\eqref{eq:spatTrun1} for $L=8$. The orange boxes implement the kinetic term (the right circuit in Fig.~\ref{fig:xy_xx_circuits}) and the blue ``barbells'' are $\hat{Z}\hat{Z}$ rotations. 
    With this ordering, some of the CNOTs in the barbells can be combined with the ones in the kinetic terms.
    The $\alpha_i$ angles can be derived from Eq.~\eqref{eq:spatTrun1} and are given in App.~\ref{app:circuitDetails}.}
    \label{fig:StagCirc}
\end{figure}

The Trotterization of $\hat{H}_m$ only involves single qubit $\hat{Z}$ rotations, which has a straightforward circuit implementation.
The Trotterization of the kinetic terms uses the right circuit in Fig.~\ref{fig:xy_xx_circuits} arranged in a brickwall pattern to minimize circuit depth, and requires $4(2L-1)$ CNOTs per second-order Trotter step.
The Trotterization of $\hat{H}_{el}^{(Q=0)}(1)$ in Eq.~\eqref{eq:spatTrun1} requires nearest-neighbor, next-to-nearest-neighbor and next-to-next-to-nearest-neighbor entangling $R_{ZZ}= e^{-i \frac{\theta}{2} \hat Z \hat Z}$ operations acting between qubits on adjacent spatial sites.
Organizing into blocks of adjacent spatial sites, the problem is to find a nearest-neighbor CNOT decomposition for $R_{ZZ}$s between all pairs of $N_q=4$ qubits.
Generalizing to any $N_q\geq3$, a strategy for constructing these circuits, depicted in Fig.~\ref{fig:ZZsimplifications}, is
\begin{enumerate}
    \item Group all the rotations that share the top qubit.
    \item For each block of grouped rotations, 
    use the bridge decomposition to convert the long-range CNOTs into nearest neighbor ones. Simplify the CNOTs within each block.
    \item Simplify the  CNOTs from neighboring blocks.
\end{enumerate}
These circuits have a total number of CNOTs $N$ and circuit depth $D$ given by,
\begin{equation}
    N=2\, \binom{N_q}{2} \ , \ \ D=N_q(N_q-2)+3 \ .
\end{equation}
Compared to the circuits before the nearest-neighbor decomposition (e.g., using the circuits in step 1.), this does not introduce any additional CNOTs, but has a depth that scales as ${\cal O}(N_q^2)$ compared to ${\cal O}(N_q)$.
The $N_q = 4$ circuit used for the $\overline{\lambda}=1$ interaction contributes $12(L-2)$ CNOTs per second-order Trotter step.
Circuits implementing a full second-order Trotter steps are shown in Fig.~\ref{fig:StagCirc}.
Taking into account the CNOT cancellations between the electric and kinetic terms, as well as between adjacent Trotter steps, the total number of CNOTs required is
\begin{equation}
\# \ \text{of CNOTs for }N_T \ 2^\text{nd} \ \text{order Trotter steps with $\overline{\lambda}=1$ :} \ \ 19L-28+(17L-26)(N_T-1) \ .
\end{equation}
For $L=56$, this is 926 CNOTs per additional second-order Trotter step, comparable to the 890 CNOTs required for the 2-step SC-ADAPT-VQE vacuum and hadron wavepacket preparation.
%

\section{Quantifying the Systematic Errors of the Approximations}
\label{sec:Systematics}
\noindent
The systematic errors that are introduced by the approximations we have employed can be analyzed and quantified by performing end-to-end classical simulations using {\tt qiskit}.
The approximations are:
\begin{enumerate}
\item The vacuum is prepared using the 2-step SC-ADAPT-VQE circuits. 
This furnishes an infidelity density of ${\cal I}_L \ = \ {\cal I}/L \ = \ 0.01$ with the exact vacuum.\footnote{The infidelity density ${\cal I}_L$ is a relevant measure for the vacuum as the state is being established across the whole lattice, whereas the infidelity is a relevant figure of merit for the (localized) hadron wavepacket.}
\item A hadron wavepacket is prepared using the 2-step SC-ADAPT-VQE circuits. 
This furnishes an infidelity of ${\cal I}\ = \ 0.05$ with an adiabatically prepared wavepacket.
\item A Hamiltonian with the electric interactions truncated beyond $\overline{\lambda} = 1$ spatial sites is used to evolve the prepared wavepacket forward in time.
\item The time-evolution operator is implemented in quantum circuits using a second-order Trotterization.
\end{enumerate}
This section will focus on a system size of $L=12$, where the classical simulations can be performed exactly. 
The circuit structure and variational parameters for the 
2-step SC-ADAPT-VQE vacuum and wavepacket preparation are given in Table~\ref{tab:AnglesVACWP}.
Note that the (2-step) wavepacket parameters differ slightly from those in Table~\ref{tab:AnglesWP10}, which are for the 10-step SC-ADAPT-VQE ansatz.
\begin{table}[!htb]
\renewcommand{\arraystretch}{1.4}
\begin{tabularx}{\textwidth}{| Y || Y | Y || Y | Y |}
 \hline
   & \multicolumn{2}{c||}{Vacuum}&  \multicolumn{2}{c|}{Wavepacket} \\
   \hline
   \hline
   & $\hat{O}^V_{mh}(1)$ & $\hat{O}^V_{mh}(3)$ & $\hat{O}_{mh}(1,1)$ & $\hat{O}_{mh}(2,2)$ \\
   \hline
    $L=12$ & 0.30738 & -0.04059 & -1.6492 & -0.3281 \\
   \hline
   $L=56$ & 0.30604 & -0.03975 & -1.6494 &  -0.3282 \\
   \hline
\end{tabularx}
\caption{The structure of the SC-ADAPT-VQE preparation circuits for the vacuum and wavepacket. The pool operators in the second row are defined in Eqs.~\eqref{eq:poolComm} and~\eqref{eq:PacketPool}. The parameters for the $L=12$ vacuum were determined in Ref.~\cite{Farrell:2023fgd}, and for the $L=12$ wavepacket in Sec.~\ref{sec:SCADAPTWP}. The wavepacket parameters for $L=56$ are the same as those for $L=14$, and the vacuum parameters are extrapolated via an exponential fit (in $L$).}
\label{tab:AnglesVACWP}
\end{table}

To identify the propagation of hadrons, we choose to measure the local chiral condensate,
\begin{align}
\hat{\chi}_j  = (-1)^j 
\hat{Z}_j + \hat{I} \ ,
\label{eq:localCC}
\end{align}
with eigenvalues of 0 (staggered site $j$ is empty) and $2$ (staggered site $j$ is occupied by a fermion).
It is useful to define the expectation value of the local chiral condensate relative to its vacuum expectation value,
\begin{equation}
{\cal X}_j(t) 
\ = \ \langle \psi_{\text{WP}}\vert
\ \hat{\chi}_j(t)\  \vert \psi_{\text{WP}} \rangle  
\ - \ \langle \psi_{\text{vac}} \vert\  \hat{\chi}_j(t) \  \vert\psi_{\text{vac}} \rangle 
\ .
\label{eq:chij}
\end{equation}
Here, $\hat{\chi}_j(t)$ is the time evolved observable; with exact exponentiation of the full Hamiltonian this would be $\hat{\chi}_j(t) = e^{i t \hat{H}} \hat{\chi}_j e^{-i t \hat{H}}$. 
When using a truncated interaction and/or Trotterization, the time-evolution operator changes.
The states $\vert \psi_{\text{vac}} \rangle $ and $\vert \psi_{\text{WP}} \rangle $ represent the prepared vacuum and wavepacket, either exact or using the SC-ADAPT-VQE approximation.
The subtraction of the vacuum expectation value is also time dependent because, for example, the SC-ADAPT-VQE prepared vacuum is not an eigenstate of the truncated Hamiltonian.
This time-dependent subtraction removes systematic errors that are present in both the wavepacket and vacuum time evolution.
It also proves to be an effective way to mitigate some effects of device errors, see Sec.~\ref{sec:Qsim}.

\begin{figure}[!htb]
    \centering
    \includegraphics[width=0.75\columnwidth]{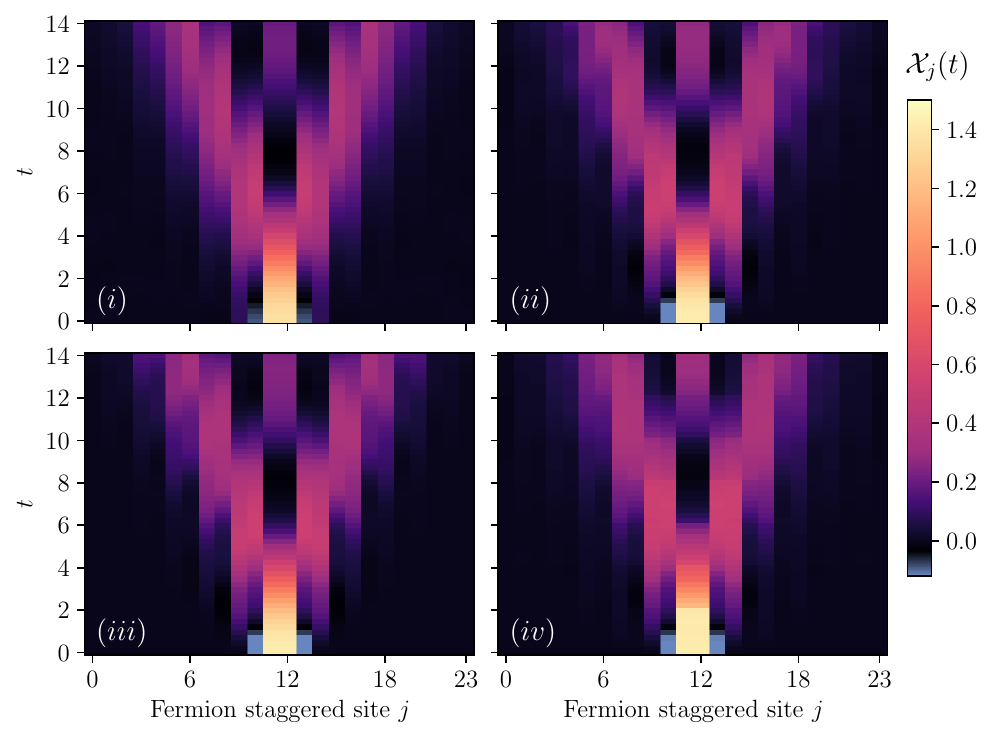}
    \caption{The effects of the approximations introduced in this work on the time evolution of the local chiral condensate ${\cal X}_j(t)$, given in Eq.~\eqref{eq:chij}.
    $(i)$ Without approximations: time evolution of the adiabatically-prepared hadron wavepacket with exact exponentiation of the full Hamiltonian, $\hat{H}$ in Eq.~\eqref{eq:Hgf}.
    $(ii)$ Approximate initial state preparation: time evolution of the 
    2-step SC-ADAPT-VQE hadron wavepacket built on top of the 
    2-step SC-ADAPT-VQE vacuum with exact exponentiation of the full Hamiltonian.
    $(iii)$ The same as $(ii)$, but with the electric interaction replaced with the $\overline{\lambda} = 1$ truncated interaction in Eq.~\eqref{eq:spatTrun1}.
    $(iv)$ The same as $(iii)$, but with time evolution implemented with $2\lceil\frac{t}{2} \rceil$ 
    second-order Trotter steps (maximum step size of $\delta t = 1$).}
    \label{fig:SystematicErr}
\end{figure}
Results obtained for the time evolved chiral condensate are shown in Fig.~\ref{fig:SystematicErr} with four different levels of approximation. 
Small errors are introduced with each approximation, but the results are found to recover expectations
within the uncertainties of the approximations.
Panel $(iv)$ in Fig.~\ref{fig:SystematicErr} shows the time-evolution operator approximated with $2\lceil\frac{t}{2}\rceil$ second-order Trotter steps, giving a maximum step size of $\delta t = 1$.
These step sizes introduce minimal (Trotter) errors, and will be used 
for the time evolution using a digital quantum computer presented in the next section.
The propagation of hadrons outward from an initially localized wavepacket is clearly identified in deviations of the local chiral condensate from its vacuum expectation value.
The oscillations of the condensate at the center of the wavepacket are consistent with expectations, and are discussed further in App.~\ref{app:SFTsrc}.
Due to the symmetry of the initial state, the hadron has equal amplitude to propagate in either direction, with a profile that is bounded by the speed of light ($1$ staggered site per unit time).

The (composite) hadrons that make up the wavepacket are 
(bosonic) vector particles, and some features of the hadron dynamics can be qualitatively understood in the simpler setting of non-interacting 1+1D scalar field theory.
In particular, the light-cone structure of propagating hadrons, the damped oscillations at the origin of the wavepacket and the effects of OBCs in both theories are similar.
This is treated in detail in App.~\ref{app:SFTsrc}, where the (textbook) example of a localized classical source coupled to a scalar field in 1+1D is treated in the continuum and on the lattice, and in App.~\ref{app:OBC}, where OBCs are compared to periodic boundary conditions (PBCs).

\section{Real-time Simulations using IBM's Digital Quantum Computers}
\label{sec:Qsim}
\noindent
The end-to-end simulations performed in the previous section using {\tt qiskit} and classical computers are scaled up to $L=56$ (112 qubits) 
and executed on IBM's 133-qubit {\tt ibm\_torino} Heron processor.
The scalability of the SC-ADAPT-VQE vacuum preparation circuits was demonstrated in Ref.~\cite{Farrell:2023fgd}, where it was shown that the variational parameters are reproduced well by an exponential in $L$.
This enables the extrapolation of the state preparation circuits, determined for $L\leq 14$, to arbitrarily large $L$.
In principle, a similar exponential convergence of parameters for the hadronic wavepacket preparation circuits is expected. 
However, as shown in Sec.~\ref{sec:SCADAPTWP}, the operator ordering and variational parameters of the SC-ADAPT-VQE wavepacket preparation have converged up to the tolerance of the optimizer by $L=14$. 
Therefore, the circuit structure and parameters determined for $L=14$ can be used to initialize the $L=56$ hadron wavepacket.
The operator ordering and parameters used to prepare the 
2-step SC-ADAPT-VQE vacuum and 2-step SC-ADAPT-VQE hadron wavepacket 
for $L=56$ are given in Table~\ref{tab:AnglesVACWP}.

Error mitigation is essential for successful simulations utilizing large quantum volumes~\cite{Kim:2023bwr}.
Here, our error mitigation methods are outlined, and a more detailed discussion can be found in App.~\ref{app:qSimDetails}. 
Through cloud-access, the circuits are sent to {\tt ibm\_torino} using the {\tt qiskit} sampler primitive, which includes both dynamical decoupling~\cite{Viola:1998dsd,2012RSPTA.370.4748S,Ezzell:2022uat} and {\tt M3} measurement mitigation~\cite{Nation:2021kye}.
To mitigate coherent two-qubit gate errors, Pauli twirling~\cite{Wallman:2016nts} is used on the native two-qubit gates, control-$Z$ for {\tt ibm\_torino}.
After twirling, we assume that the coherent two-qubit gate errors are transformed into statistically independent and unbiased incoherent errors, which can be modeled by a Pauli noise channel. 
Observables are then estimated using Operator Decoherence Renormalization (ODR)~\cite{Farrell:2023fgd}, which extends decoherence renormalization~\cite{Urbanek_2021,ARahman:2022tkr,Farrell:2022wyt,Ciavarella:2023mfc} to large systems.\footnote{Instead of setting the single-qubit rotations to zero in the mitigation circuits~\cite{Farrell:2022wyt,Ciavarella:2023mfc,Farrell:2023fgd}, 
they could be replaced by Clifford gates~\cite{Qin:2021jpm,Robbiati:2023eyl}.
}
To implement ODR, two kinds of circuits are run on the device: a ``physics'' circuit, and a ``mitigation'' circuit.
For a simulation of wavepacket dynamics, the physics circuit implements the time evolution of either the wavepacket or the vacuum (to compute ${\cal X}_j(t)$ in Eq.~\eqref{eq:chij}).
The mitigation circuit(s), with a priori known error-free (predicted) results, 
and the physics circuits have similar structures and similar error profiles.
From the mitigation circuits, deviations of measured observables ${\langle \hat{O} \rangle_{\text{meas}}}$ from their predicted values ${\langle \hat{O} \rangle_{\text{pred}}}$ are used to compute the depolarizing noise parameters,
\begin{equation}
\eta_{O} \ = \ 1 \ - \ \frac{\langle \hat{O} \rangle_{\text{meas}}}{\langle \hat{O} \rangle_{\text{pred}}}
\ .
\end{equation}
These $\eta_{O}$ are used to estimate the expectation values from the physics circuits (using the same relation).
For wavepacket (vacuum) time evolution, we choose a mitigation circuit that creates the wavepacket (vacuum), time evolves with half of the Trotter steps until $t/2$ and then evolves for $-t/2$ with the remaining Trotter steps~\cite{ARahman:2022tkr}.
This forwards-backwards time evolution corresponds to the identity operator in the absence of device errors, and restricts our simulations to an even number of Trotter steps.
To determine the $\eta_{O}$, the prediction of a desired observable from the mitigation circuit must be known.
In our case, this requires classically computing $\langle \hat{\chi}_j \rangle$ in both the SC-ADAPT-VQE vacuum and wavepacket.
This can be accomplished even for large systems using the {\tt qiskit} or {\tt cuQuantum} MPS simulator, as was demonstrated in Ref.~\cite{Farrell:2023fgd} for the SC-ADAPT-VQE vacuum up to $L=500$.
Interestingly, our numerical calculations highlight that it is the time evolution, and not the state preparation, that is difficult for classical MPS techniques.

\begin{table}[!tb]
\renewcommand{\arraystretch}{1.4}
\resizebox{\textwidth}{!}{\begin{tabularx}{1.4\textwidth}{| c || c | Y | Y | Y | Y | Y || Y | Y |}
 \hline
 $t$ & $N_T$ & \# of CNOTs (per~$t$) & CNOT depth (per~$t$)& \# of distinct circuits (per~$t$)
 &   \# of twirls (per circuit) & \# of shots (per twirl) 
 &  Executed CNOTs  ($\scalemath{0.9}{\times 10^9}$) & Total \# of shots ($\scalemath{0.9}{\times 10^6}$) \\
   \hline
   \hline
    1 \& 2 & 2 &  2,746    & 70     & 4 & 480 & 8,000  & $4\times 2\times 10.5$ & $4\times 2\times 3.8$\\
    \hline
    3 \& 4 & 4 & 4,598 & 120 & 4 & 480 & 8,000 & $4\times 2\times 17.7$ & $4\times 2\times 3.8$\\
    \hline
    5 \& 6 & 6 & 6,450 & 170  & 4 & 480 & 8,000 & $4\times 2\times 24.8$ & $4\times 2\times 3.8$\\
    \hline
    7 \& 8 & 8 & 8,302 & 220 & 4  & 480 & 8,000 & $4\times 2\times 31.9$ & $4\times 2\times 3.8$\\
    \hline
    9 \& 10 & 10 & 10,154 & 270 & 4  & 160 & 8,000 & $4\times 2\times 13.0$  & $4\times 2\times 1.3$\\
    \hline
    11 \& 12 & 12 & 12,006 & 320 & 4  &  160 & 8,000 & $4\times 2\times 15.4$ & $4\times 2\times 1.3$\\
    \hline
    13 \& 14 & 14 & 13,858 & 370 & 4  & 160 & 8,000 & $4\times 2\times 17.7$ & $4\times 2\times 1.3$ \\
    \hline \hline
    {\bf Totals} & & & & & & & $1.05 \times 10^{12}$ & $1.54\times 10^8$ \\
    \hline
\end{tabularx}}
\caption{Details of our quantum simulations performed using 112 qubits of IBM's {\tt ibm\_torino} Heron processor. 
For a given simulation time, $t$ (first column), the second column gives the number of employed Trotter steps $N_T$.
The third and fourth columns give the number of CNOTs and corresponding CNOT depth.
The CNOT totals given in the third column include the cancellations that occur during transpilation, and the CNOT depth should be compared to the minimum depth that is equal to twice the number of CNOTs/qubit (49, 82, 115, 148, 181, 214, 247 for increasing $N_T$) to assess the sparsity of the circuits. 
The fifth column gives the number of distinct circuits per $t$ (this number does not include the circuits needed for readout mitigation) and the sixth column gives the number of Pauli-twirls executed per distinct circuit.
For each twirl, 8,000 shots are performed (seventh column).
The total number of executed CNOT gates are given in the eighth column, and the total number of shots are given in the ninth column.
The total number of CNOT gates applied in this production is one trillion, and the total number of shots is 154 million.} 
 \label{tab:QsimCNOT}
\end{table}
\begin{figure}[!t]
    \centering
    \includegraphics[width=\columnwidth]{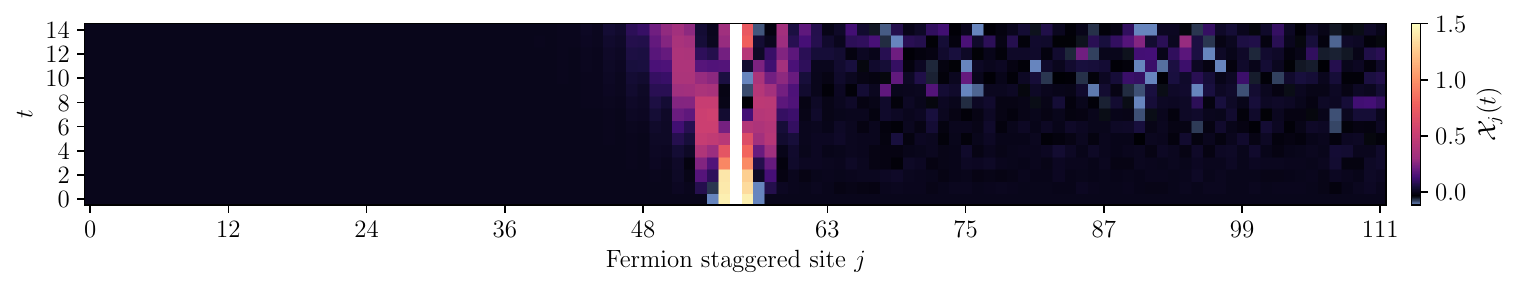}
    \caption{The time evolution of the vacuum subtracted chiral condensate $\mathcal{X}_j(t)$, defined in Eq.~\eqref{eq:chij}, for a $L=56$ (112 qubits) spatial-site lattice. 
    The initial state is prepared using the 
    2-step SC-ADAPT-VQE vacuum and wavepacket preparation circuits.
    Time evolution is implemented using a second-order Trotterization of the Hamiltonian with the $\overline{\lambda}=1$ truncated electric interaction.
    The left side shows the results of error-free classical simulations from the {\tt cuQuantum} MPS simulator, while the right side shows the CP-averaged results obtained using IBM's superconducting-qubit digital quantum computer {\tt ibm\_torino} (both sides show the MPS result for $t=0$).
    Due to CP symmetry, the right and left halves would be mirror images of each other in the absence of device errors.
    A more detailed view for each time slice is given in Fig.~\ref{fig:srcmvac_evol}, and discussions of the error-mitigation techniques are presented in the main text and App.~\ref{app:qSimDetails}.}
    \label{fig:IBMresultsMPS}
\end{figure}
We implement time evolution for $t=\{1,2,\ldots,14\}$ with $2\lceil\frac{t}{2} \rceil $ second-order Trotter steps (a maximum step size of $\delta t = 1$).
As shown in the previous section, this step size does not introduce significant Trotter errors.
The number of CNOTs and corresponding CNOT depth for each simulation time are given in Table~\ref{tab:QsimCNOT}, and range from 2,746 CNOTs (depth 70) for 2 Trotter steps to 13,858 CNOTs (depth 370) for 14 Trotter steps.
\begin{figure}[!htb]
    \centering
    \includegraphics[width=0.9\columnwidth]{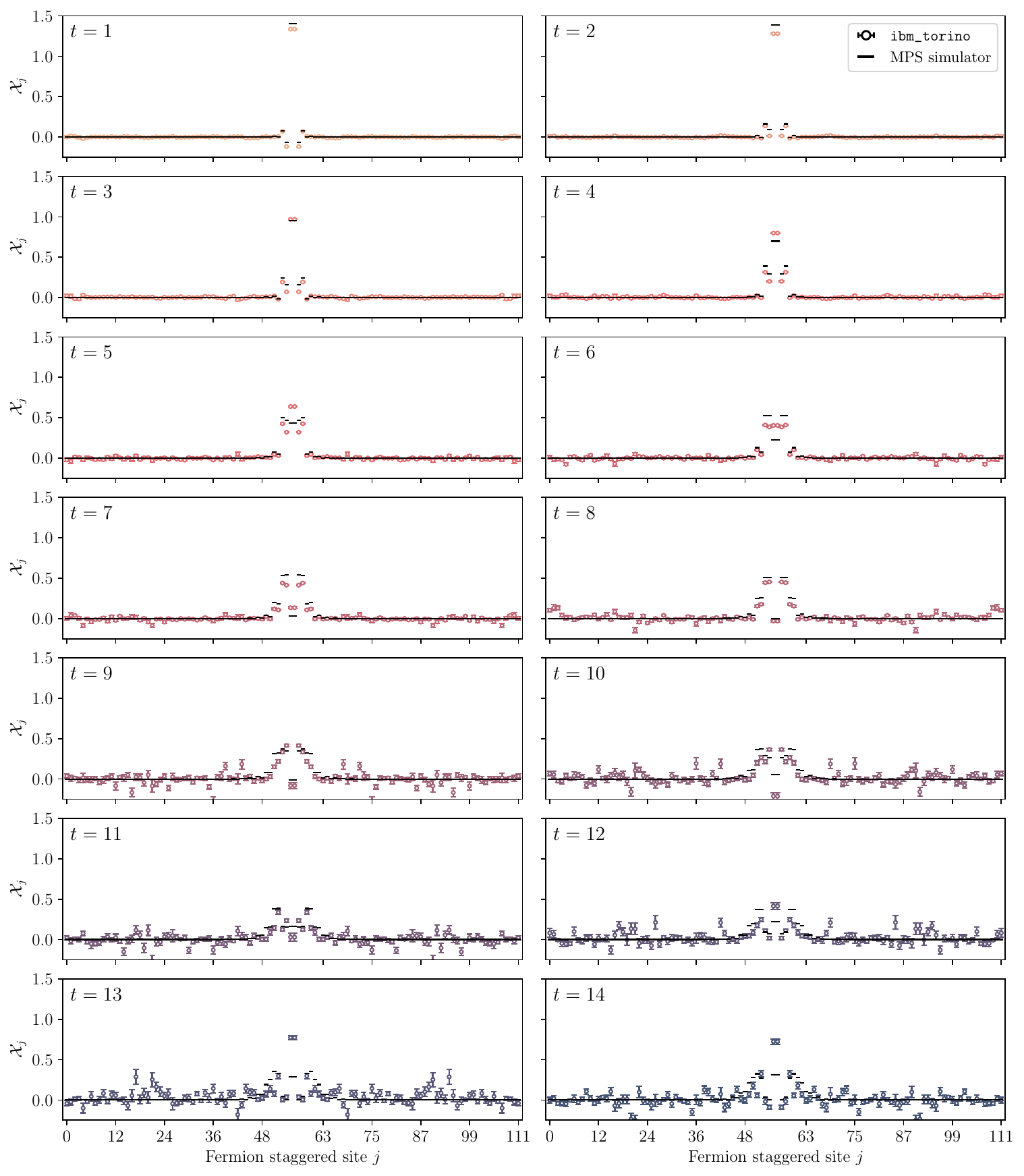}
    \caption{A detailed view of the time evolution of the vacuum subtracted chiral condensate shown in Fig.~\ref{fig:IBMresultsMPS} for each simulation time.
    The open circles are CP averaged results obtained using IBM's superconducting-qubit digital quantum computer {\tt ibm\_torino}.
    The black dashes are the error-free expectations obtained from the {\tt cuQuantum} MPS classical simulator.
    A complete discussion of the error-mitigation techniques, post processing and statistical uncertainties is presented in the main text and App.~\ref{app:qSimDetails}.}
    \label{fig:srcmvac_evol}
\end{figure}
The results for ${\cal X}_j(t)$ obtained from {\tt ibm\_torino} and the MPS simulator are shown in Fig.~\ref{fig:IBMresultsMPS}, with a breakdown of each $t$ given in Fig.~\ref{fig:srcmvac_evol} (the separate evolutions of the wavepacket and vacuum are shown in Fig.~\ref{fig:src_vac_evol}).
For each time, four circuits are run: time evolution of the wavepacket, time evolution of the vacuum, forward-backward time evolution of the wavepacket and forward-backward time evolution of the vacuum.
For $t=1-8$, 480 twirled instances of each circuit are run, and for $t=9-14$, 160 twirled instances are run. 
Each twirled instance has 8,000 shots, using a total of $\sim 1.5 \times 10^8$ shots for the complete production.
We have estimated the uncertainties in the results from the quantum computer using bootstrap-mean resampling.\footnote{Due to the noisy nature of the device, the utility of the Hodges-Lehmann (HL) estimator was studied, and consistent results were obtained.
The HL estimator has been considered in lattice QCD studies to mitigate the impact of outliers in nuclear correlation functions~\cite{Beane:2014oea,Orginos:2015aya,NPLQCD:2020lxg}.}
The expected results are determined by using the {\tt cuQuantum} MPS simulator with maximum bond dimension 200.
The run time and convergence of the MPS simulations are discussed in App.~\ref{app:MPSSim}.

The individual time evolutions of the wavepacket and vacuum, used to compute ${\cal X}_j(t)$, are shown in Fig.~\ref{fig:src_vac_evol} of App.~\ref{app:qSimDetails}.
A systematic error in the chiral condensate away from the center of the lattice is seen to increase with simulation time.
Fortunately, it is similar for the wavepacket and vacuum evolution, and largely cancels in the subtraction to form ${\cal X}_j(t)$, as shown in Fig.~\ref{fig:srcmvac_evol}.
The origin of this systematic error is currently unknown to us, and either stems from a deficiency in our error-mitigation techniques, or from insufficient convergence in the MPS simulations.
Without the approximations in the state preparation and time evolution, the chiral condensate would not evolve in regions that are locally the vacuum.
This qualitatively holds for smaller systems with $L\leq 14$ that can be simulated exactly.
The results from the quantum computer agree with these expectations, showing little evolution of the chiral condensate in the vacuum (right column of Fig.~\ref{fig:src_vac_evol}).
The MPS simulations, on the other hand, show significant evolution of the vacuum chiral condensate.
For the range of maximum bond dimensions we have been able to explore, it appears that the chiral condensate has converged at the level of $10^{-2}$ for late times.
However, these results are not exact, and at this point we cannot rule out systematic errors being present in the MPS simulations.
From preliminary investigations, it appears that the vacuum evolution is due to $\overline{\lambda}=1$ being too small for exponential convergence.
This is not surprising since the relevant ratio for exponential convergence is $\propto \overline{\lambda}/\overline{\xi}$, with possibly a prefactor proportional to, for example, $2\pi$.
However, the maximum bond dimension required for convergence becomes significantly larger with increasing $\overline{\lambda}$, and it is unclear if this conclusion is consistent.
A future detailed study of the effects of increasing the precision of the state preparation, increasing $\overline{\lambda}$, and increasing the number of Trotter steps will be needed to determine if this discrepancy is due to errors in the MPS simulation or from imperfect error mitigation.

The results shown in Figs.~\ref{fig:IBMresultsMPS} and \ref{fig:srcmvac_evol} demonstrate that, by implementing a series of exponentially convergent approximations (beyond Trotterization), wavepackets of hadrons can be prepared and evolved forward in time with available quantum computers.
Propagating hadrons are clearly identified as a disturbance in the chiral condensate, with random fluctuations due to device errors outside of the hadron's light-cone.
It should be emphasized that obtaining ${\cal X}_n(t) = 0$ outside of the light-cone using IBM's device is a non-trivial result, as it requires cancellations between the wavepacket and vacuum evolutions.
The simulations performed using {\tt ibm\_torino} show qualitative agreement with classical MPS results, but degrade with increasing number of Trotter steps (circuit depth).
The simulations highlight that device errors dominate over the systematic errors due to approximate state preparation and time evolution.
The results qualitatively recover expectations, but often differ by many standard deviations from classical expectations, indicating that we do not have a complete quantification of uncertainties.
This is not surprising given the simplicity and limitations of the assumed error model.
Despite the device errors, it is clear that current hardware is capable of creating and possibly colliding (composite) hadrons over a meaningful time interval.
Such simulations could provide first glimpses of inelastic hadron scattering and fragmentation in the Schwinger model that are beyond present capabilities of classical computing.

\section{Summary and Outlook}
\label{sec:Summary}
\noindent
Quantum computing offers the potential of reliably simulating the collisions of high-energy hadrons and nuclei directly from quantum chromodynamics, the quantum field theory describing the strong interactions.
First steps are being taken to develop scalable techniques and algorithms for QCD simulations by working with the Schwinger model defined in 1+1D. 
Towards these goals, this work develops
protocols for quantum simulations of hadron dynamics that are demonstrated on a $L=56$ (112 qubit) lattice using IBM's superconducting-qubit digital quantum computer, {\tt ibm\_torino}.
These simulations start with establishing a wavepacket of hadrons in the center of the lattice on top of the vacuum.
The necessary quantum circuits for the creation of this wavepacket are determined using the SC-ADAPT-VQE algorithm that was recently introduced by the authors in Ref.~\cite{Farrell:2023fgd}.
In SC-ADAPT-VQE, low-depth circuits for state preparation are determined on a series of small lattices using {\it classical computers}, and then systematically scaled up to prepare states on a {\it quantum computer}.
For the present purposes, the SC-ADAPT-VQE circuits are variationally optimized to have maximal overlap with an adiabatically prepared hadron wavepacket.
The vacuum and hadronic wavepacket that are initialized on the quantum computer are then time evolved using a second-order Trotterization of the time evolution operator.
Naively, the electric interaction between fermions is all-to-all, giving rise to a prohibitive ${\cal O}(L^2)$ scaling in the number of two-qubit gates needed for time evolution.
Motivated by confinement, an approximation that truncates the electric interaction between distant charges is introduced.
This interaction converges exponentially with increasing interaction distance, and improves the scaling of the number of two-qubit gates required for time evolution to ${\cal O}(\overline{\lambda} L)$, where $\overline{\lambda}$ is proportional to the confinement length scale.
These new methods for state preparation are verified on small systems using a classical simulator, and then applied to time evolve hadron wavepackets on a $L=56$ (112 qubit) lattice using {\tt ibm\_torino}.
Our digital quantum simulations utilize some of the largest quantum volumes to date~\cite{Yu:2022ivm,Kim:2023bwr,Shtanko:2023tjn,Farrell:2023fgd,Baumer:2023vrf,Chen:2023tfg,Liao:2023eug,Chowdhury:2023vbk}, with up to 13,858 two-qubit entangling gates applied (CNOT depth of 370).
A large number of shots with which to implement the error mitigation techniques is found to be essential to the success of our simulations. Our results show clear signatures of hadron propagation through modifications of the local chiral condensate.

Real-time dynamics typically explore highly-entangled regions of Hilbert space and, as a result, classical methods scale unfavorably with simulation time $t$, lattice volume $L$, and energy.
To explore this in more detail, our quantum simulations have been compared to classical MPS circuit simulations using {\tt qiskit} and {\tt cuQuantum}.
We have found that our initial state preparation circuits can be simulated relatively easily with these simulators.
However, the bond dimension needed for proper convergence grows rapidly as more steps of Trotterized time evolution are added to the quantum circuit.
All of this points to a potential near-term quantum advantage for the simulation of hadronic dynamics.
In particular, it is likely that the simulation of high-energy hadronic collisions will exceed the capabilities of classical computing for simulation times and volumes that are not excessively large.
Exactly where such a quantum advantage can be realized remains to be established.

On this path, future work will use the hadron wavepacket preparation and time evolution circuits that we have presented here to simulate hadron scattering. 
Evolving out to later times will require time-evolution methods that improve upon Trotterization.
A promising direction is to use SC-ADAPT-VQE to find low-depth circuits for simulating over the early times.
The light-cone restricts early-time dynamics to only a modest number of qubits, and scalable low-depth circuits can likely be found with classical computing.
Another direction worth pursuing  is to approach the continuum by taking $m$ and $g$ smaller, increasing the correlation length.
These longer correlation lengths will require deeper state-preparation circuits and larger truncations of the electric interaction to reach a target simulation quality.
Further into the future, improved methods for hadron detection will also be needed. 
Finally, it will be necessary to extend these techniques to non-Abelian gauge theories and higher dimensions to perform more realistic simulations of QCD.

\clearpage
\begin{subappendices}

\section{The Classical Dynamics of a Sourced Non-Interacting Scalar Field}
\label{app:SFTsrc}
\noindent
The spectrum of the Schwinger model consists of composite hadrons due to confinement.
Unlike the underlying electron and positron degrees of freedom, which are fermions, the hadrons are bosonic scalar and vector particles.
Important features of the hadronic dynamics simulated in this work can be understood in the simpler setting of a non-interacting scalar field evolving from a localized source.
The framework for the latter is well known, and can be found in quantum field theory textbooks, for example, Ref.~\cite{Peskin:1995ev}.
The spatial and temporal extents of the hadron wavepacket (in the Schwinger model) that we work with are approximately determined by the correlation length, $\xi$, and we model this by a Gaussian source for the scalar field (describing the Schwinger model vector hadron). The Klein-Gordon equation in the presence of a classical source,
\begin{equation}
    \left( \partial^\mu\partial_\mu + m^2 \right)\phi(x,t) = j(x,t)
    \ ,
    \label{eq:KG1}
\end{equation}
is solved in 1+1D in infinite volume and with vanishing lattice spacing by
\begin{equation}
    \phi_j(x,t) = \phi_{j=0}(x,t) 
    \ +\ 
    i \int dy\ dt^\prime \ G_R(x-y, t-t^\prime) j(y,t^\prime)
    \ ,
    \label{eq:KG2}
\end{equation}
where $G_R(a,b)$ is the retarded Green's function and $\phi_{j=0}(x,t)$ is the field in the absence of the source.
The effective source we consider is 
\begin{equation}
    j(x,t) = J_0\ \sqrt{\frac{\alpha}{\pi}}\ e^{-\alpha x^2}\ \delta(t)
    \ \ ,\ \ 
    \int dx\ dt\ j(x,t) = J_0
    \ .
    \label{eq:KG3}
\end{equation}
After writing the propagator in momentum space, and using the spatial symmetry of the source, the field in the presence of the source is given by
\begin{equation}
    \phi_j(x,t) = \phi_{j=0}(x,t) 
    \ +\ 
2 J_0 \int_0^\infty \frac{dp}{2\pi} \frac{e^{-p^2/(4 \alpha)}}{\omega_p}\ 
\cos p x\ \sin \omega_p t \ ,
    \label{eq:KG4}
\end{equation}
where $\omega_p = \sqrt{p^2+m^2}$.
In a finite volume with a discrete set of uniformly spaced lattice points, it is straightforward to derive the appropriate analogous relation.
Spatial integrals are replaced by a discrete sum over the finite number lattice sites, and momentum integrals are replaced by sums over momentum modes within the first Brillouin zone (the exact set of modes are determined by the selected boundary conditions imposed on the field).
\begin{figure}[!b]
    \centering
    \includegraphics[width=0.95\columnwidth]{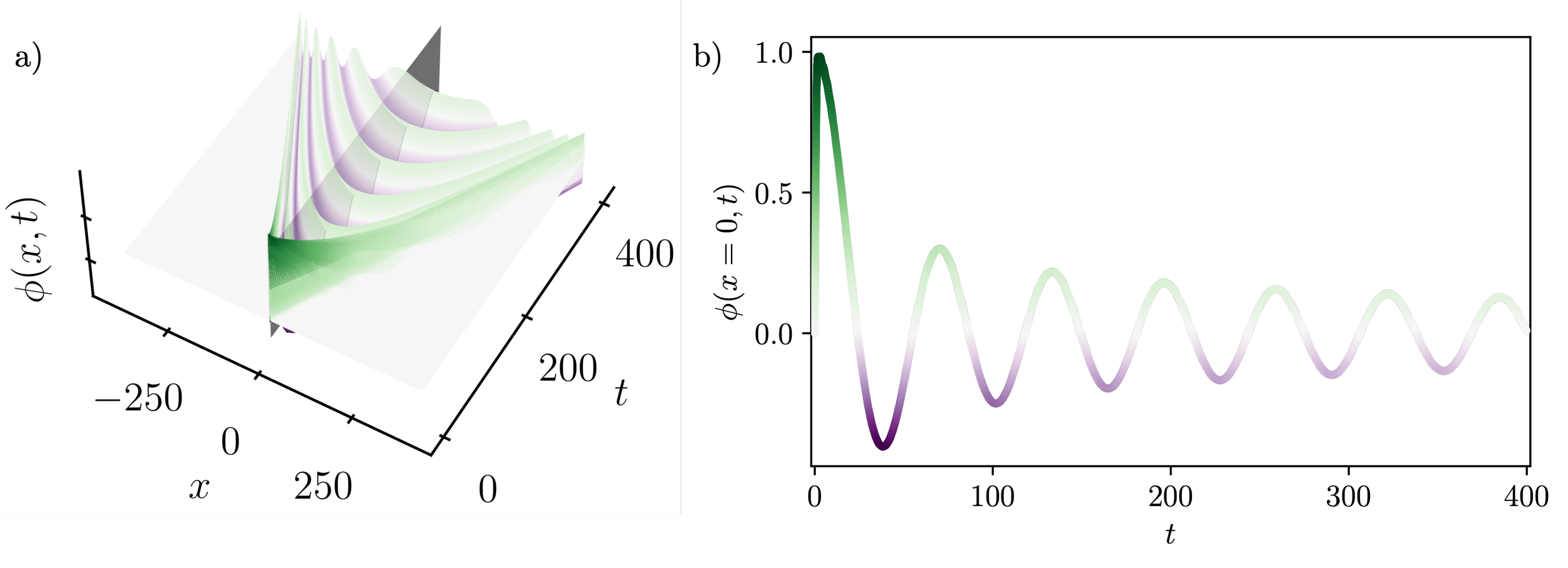}
    \caption{a) The free scalar field downstream from a Gaussian source given in Eq.~\eqref{eq:KG3}
    with $m=0.1$ and $\alpha=J_0=1$, determined by Eq.~\eqref{eq:KG4}. b) Profile of the scalar field at $x=0$.}
    \label{fig:KGdynamics}
\end{figure}
Figure~\ref{fig:KGdynamics} shows the downstream field in spacetime from the source given in Eq.~\eqref{eq:KG3}, with parameters $m=0.1$ and $\alpha=J_0=1$.
The light cone at $x=t$ is clear, with the field decaying exponentially beyond these lines. 
Importantly, the field near the origin is seen to ``ring down'', continuing to emit particles until the initially localized energy density is dispersed via particle production.

The total energy injected into the field by the source is
\begin{equation}
\langle \hat H \rangle = \sqrt{\frac{\alpha}{8 \pi}}\ J_0^2 \ ,
    \label{eq:KG5}
\end{equation}
where $\hat{H}$ is the free Hamiltonian without the source, and the energy of the vacuum has been set to zero.
The probability of creating a particle in the $|p\rangle$ momentum state, ${\rm Prob}(|p\rangle)$, and the expectation value of the total number of particles produced in such an event, $N_\phi$, are
\begin{equation}
{\rm Prob}(|p\rangle) = 
\frac{J_0^2}{2 \omega_p} e^{-\frac{p^2}{2 \alpha}}
\ \ ,\ \ 
N_\phi  =  
\frac{J_0^2}{4\pi} \ e^{\frac{m^2}{4\alpha}}\ K_0\left(\frac{m^2}{4\alpha}\right) \ ,
    \label{eq:KG6}
\end{equation}
with $K_0$ being the modified Bessel function of the second kind of order zero.

\section{Aspects of Open Boundary Conditions}
\label{app:OBC}
\noindent
Ideally, quantum simulations of lattice field theories would utilize periodic boundary conditions (PBCs) in order to maintain the translation invariance of free space (in the continuum limit).
However, without connectivity between the initial and final lattice sites, as is the case in some quantum computers, simulations can be performed with open boundary conditions (OBCs).
In this appendix, we demonstrate some key features of OBCs in the context of scalar field theory, and make connections to the Schwinger model.

The Hamiltonian describing non-interacting lattice scalar field theory with continuous fields at each lattice site and with OBCs is given by 
\begin{equation}
\hat H_{\rm lsft} = 
\frac{1}{2} \sum_{j=0}^{L-1} \hat \Pi_j^2
+ \frac{1}{2} \sum_{j=0}^{L-1} m_0^2 \hat \phi_j^2
-\frac{1}{2} \sum_{\substack{j=0\\ j-1\ge 0\\ j+1\le L-1}}^{L-1} 
\hat\phi_j (\hat\phi_{j+1}+\hat\phi_{j-1}-2\hat\phi_j)
= 
\frac{1}{2}
\hat \Pi^2 
\ +\ 
\frac{1}{2}
\Phi^T \left[\ 
m_0^2 \hat I
\ +\ 
{\cal G}
\right] \Phi \ ,
    \label{eq:OBC1}
\end{equation}
where 
\begin{equation}
{\cal G} = 
\left(
\begin{array}{cccccc}
2 & -1 & 0 & 0 & \cdots & 0 \\
-1 & 2 & -1 & 0 & \cdots & 0 \\
0 & -1 & 2 & -1 & \cdots & 0 \\
\vdots & & & & & \vdots \\
0 & 0 & 0 & 0 & \cdots & -1 \\
0 & 0 & 0 & 0 & \cdots & 2 \\
\end{array}
\right)
\ \ ,\ \ 
\Phi^T = \left(\phi_0, \phi_1, \cdots ,\phi_{L-1} \right) \ ,
    \label{eq:OBC2}
\end{equation}
and where $\hat\Pi$ is the conjugate-momentum operator.
The only difference between this expression and that for PBCs is the absence of terms in the extreme anti-diagonal entries in ${\cal G}$, which renders the matrix non-circulant, reflecting the lack of discrete translational invariance.
An orthogonal transformation can be applied to the fields to diagonalize the Hamiltonian matrix,
\begin{equation}
\Phi = V \Psi
\ \ ,\ \
\hat H \ = \ 
\frac{1}{2}
\hat \Pi^2
\ +\ 
\frac{1}{2}
\Psi^T  
\ 
\Omega^2 \ \Psi 
\ ,
    \label{eq:OBC3}
\end{equation}
where $\Omega$ is a $L\times L$ diagonal matrix with eigenvalues $\omega_i$.
Therefore, the $L$ towers of single-particle energy eigenvalues of these systems are 
\begin{equation}
E_i = \left(n_i + \frac{1}{2} \right) \omega_i \ ,
    \label{eq:OBC4}
\end{equation}
where $n_i$ are the number of bosons with energy $\omega_i$, with a vacuum energy that is the sum of zero-point energies,
\begin{equation}
E_{\rm vac} = \frac{1}{2} \sum_i \omega_i \ .
    \label{eq:OBC5}
\end{equation}
%

\subsection{OBCs and PBCs for \texorpdfstring{$L=4$}{L=4}}
\label{app:OBCL4}
\noindent
It is instructive to consider the similarities and differences 
between OBCs and PBCs for non-interacting scalar field theory on $L=4$ lattice sites.
It is well known that the structure of the Hamiltonian in Eq.~\eqref{eq:OBC1} indicates that this (and other such systems) can be diagonalized by the eigenvectors of ${\cal G}$, and are hence independent of the mass and conjugate momentum (as these are both local operators).

For OBCs, the $\omega_i$ are
\begin{align}
&\omega_i  =
\left\{ \sqrt{m_0^2 + \frac{1}{2}(3-\sqrt{5})} , 
\sqrt{m_0^2 + \frac{1}{2}(5-\sqrt{5})} ,
\sqrt{m_0^2 + \frac{1}{2}(3+\sqrt{5})} , 
\sqrt{m_0^2 + \frac{1}{2}(5+\sqrt{5})} \right\}
\nonumber\\
& = 
\left\{ \sqrt{m_0^2 + \frac{1}{2}(3-\sqrt{5})} , 
\sqrt{m_0^2 + 2 + \frac{1}{2}(1-\sqrt{5})} , 
\sqrt{m_0^2 + 2 - \frac{1}{2}(1-\sqrt{5})} , 
\sqrt{m_0^2 + 4 - \frac{1}{2}(3-\sqrt{5})} \right\}
\nonumber\\
& = 
\left\{ \sqrt{m_0^2 + 0.3819} , 
\sqrt{m_0^2 + 1.3819} , 
\sqrt{m_0^2 + 2.6180} , 
\sqrt{m_0^2 + 3.6180} \right\}
\ ,
\label{eq:OBCL4}
\end{align}
which are to be compared with those from PBCs,
\begin{align}
\omega_i & =
\left\{ m_0 , 
\sqrt{m_0^2 + 4\sin^2 \frac{\pi}{4}} ,
\sqrt{m_0^2 + 4\sin^2 \frac{\pi}{4}} , 
\sqrt{m_0^2 + 4\sin^2 \frac{\pi}{2}} \right\}
\nonumber\\
& =
\left\{ m_0 ,
\sqrt{m_0^2 + 2} , 
\sqrt{m_0^2 + 2} , 
\sqrt{m_0^2 + 4} \right\}
\ .
\label{eq:PBCL4}
\end{align}
The kinetic contributions to the energies in Eq.~\eqref{eq:OBCL4} correspond to ``momentum modes'' with $k=n \pi/5$ with $n=\{1,2,3,4\}$, and generalizes to $k=n \pi/(L+1)$ with $n=\{1,\ldots, L\}$.\footnote{A more direct comparison between  
Eq.~(\ref{eq:OBCL4}) and Eq.~(\ref{eq:PBCL4}) can be made using relations such as $4 \sin^2 \frac{\pi}{10} = (3-\sqrt{5})/2$.}
The energies of the OBC states are split around the energies of the PBC states, 
with the lowest is raised, and highest lowered.
This splits the degeneracies of the left- and right-moving momentum eigenstates associated with PBCs.
These features extend to larger values of $L$, with the splittings reducing with increasing $L$.

\begin{figure}[!ht]
	\centering
	\includegraphics[width=\columnwidth]{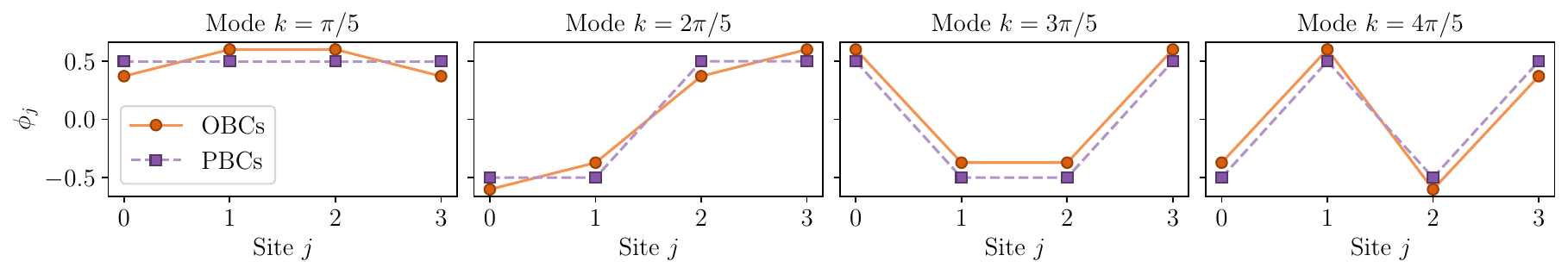}
	\caption{The eigenvectors of the $L=4$ lattice scalar field theory with PBCs (purple) and OBCs (orange).}
	\label{fig:OPBCevecsL4}
\end{figure}

The eigenstates can all be made real by global phase rotations, and identification of these states with the associated states with PBCs can be made by forming linear combinations of the degenerate PBC states.
Figure~\ref{fig:OPBCevecsL4} shows the eigenstates for PBCs and OBCs.
Even for $L=4$, the difference between the eigenstates is not large, and diminishes with increasing $L$.
%

\subsection{Matching the Schwinger Model to Non-Interacting Scalar Field Theory for \texorpdfstring{$L=8$}{L=8} and \texorpdfstring{$L=14$}{L=14}
with OBCs}
\label{app:OBCSMmatch}
\noindent
In large enough spatial volumes, 
it is expected that the low-lying continuum states of the Schwinger model will be approximately recovered by an effective field theory (EFT) of scalar and vector particles~\cite{Luscher:1986pf,Gasser:1987zq,Luscher:1990ux,2001afpp.book..683V,Beane:2003da,PhysRevD.70.074029,Colangelo_2005}. 
To explore this more with OBC simulations, the mass of the scalar particle needs to be determined from the spectrum of the Schwinger model.
As the energies of the states of the scalar field depend in a non-trivial way on the mass of the scalar particle, this is accomplished numerically.

In the Schwinger model, fermions are discretized on a lattice with $2L$ staggered sites, corresponding to $L$ spatial sites.
To match to the spectrum of lattice scalar field theory, a conversion must be performed to switch from units of staggered lattice spacing, $a_{st}$, to units of spatial lattice spacing, $a_{sp}$.
A dimensionless energy, $\Delta_{st}$, in the Schwinger model is related to a physical energy by $\Delta_{st} = a_{st} E /(\hbar c)$, where $\hbar c=197.32$ MeV fm, and $E$ is an energy in units of MeV.
The corresponding quantity in terms of the spatial lattice spacing is $\Delta_{sp} = a_{sp} E /(\hbar c) = 2 a_{st} E /(\hbar c) = 2 \Delta_{st}$.

Exact diagonalization of the Schwinger model Hamiltonian with parameters $m=0.5,g=0.3,L=8$ gives a gap to the first excited state (vector hadron mass) of $a_{st} E_1 = a_{st} m_{\text{hadron}} = 1.15334$.
In the $L=8$ non-interacting scalar field theory, this corresponds to an excitation of $a_{sp}\  \omega_1= 2 m_{\text{hadron}} = 2.30668$.
Fitting the bare scalar field mass $m_0$ to this value gives $m_{0}^{({\rm fit})} = 2.28039$ in spatial lattice units, which can be then used to predict higher-lying states in the Schwinger model spectrum.
Converting back to the staggered lattice spacing gives the values of $a_{st}\omega_i$ to be compared with the exact results from the Schwinger model, $a_{st}E_i$, shown in Table~\ref{tab:L8comp}.
\begin{table}[!t]
\renewcommand{\arraystretch}{1.4}
\resizebox{\textwidth}{!}{\begin{tabularx}{1.3\textwidth}{|c || Y | Y | Y | Y | Y | Y | Y | Y | Y | Y | }
 \hline
Quantity & State 1 & State 2 & State 3 & State 4 & State 5 & State 6 & State 7 & State 8 & State 9 & State 10  \\
 \hline\hline
$a_{st} \ E_i$ &1.15334& 1.19133& 1.25209& 1.33035& 1.33728& 1.38401& 1.41968& 1.44693& 1.47535& 1.51249\\
\hline
$a_{st} \ \omega_i$ &1.15334& 1.19039& 1.24501& 1.30890 & - & 1.37363& 1.43180 & - & 1.47752& 1.50662\\
 \hline
\end{tabularx}}
\caption{The lowest-lying energies, $a_{st} \ E_i$,  of the Schwinger model with Hamiltonian parameters $m=0.5, g=0.3$ and  $L=8$. These are compared with the lowest-lying eigenvalues of a non-interacting scalar field theory with OBCs, $a_{st} \ \omega_i$, with a scalar mass parameter fit to reproduce $a_{st} \ E_1$.}
 \label{tab:L8comp}
\end{table}
Each of the energies $a_{st} \ \omega_i$ can be identified with an energy in the Schwinger model, within $\sim 2\%$, indicating that the low-lying spectrum is largely from the motion of a single hadron on the lattice.
We assume that the two states that do not correspond to states in the scalar theory result from internal excitations of the single particle state in the Schwinger model.
This analysis can be repeated for $L=14$ where it is found that $a_{st} E_1 = a_{st} m_{\text{hadron}} =  1.1452$ and $m_{0}^{({\rm fit})} = 2.28096$ (spatial lattice units). 
These quantities are very similar to the $L=8$ ones, as expected since $m_{\text{hadron}} \ll L$ and finite-size effects are small.
Table~\ref{tab:L14comp} shows the energy levels in the Schwinger model compared with those predicted from non-interacting scalar field theory fit to the lowest level.
Good agreement is again found, supporting the identification of the excited states in the Schwinger model with OBC momentum modes.
\begin{table}[!t]
\renewcommand{\arraystretch}{1.4}
\begin{tabularx}{\textwidth}{|c || Y | Y | Y | Y | Y | }
 \hline
Quantity & State 1 & State 2 & State 3 & State 4 & State 5 \\
 \hline\hline
$a_{st} \ E_i$ &1.1452 &1.1588&1.1812&1.2118&1.2496\\
\hline
$a_{st} \ \omega_i$ &1.1452&1.1592&1.1816&1.2108&1.2452\\
 \hline
\end{tabularx}
\caption{The same as Table~\ref{tab:L8comp} but for $L=14$.}
 \label{tab:L14comp}
\end{table}
%

\subsection{Sources with OBCs}
\label{app:OBCSMsrcs}
\noindent
The analysis in App.~\ref{app:SFTsrc} related to source dynamics in non-interacting scalar field theory is performed in infinite volume and in the continuum limit.
To better understand the impact of finite-volume and OBCs, it is helpful to consider the retarded-Green's function on such lattices.
The Green's function in Eq.~\eqref{eq:KG2} in 3+1D is given by 
\begin{align}
D_R({\bf x},{\bf y},t,0) & =
\theta(t)\ 
\int\ \frac{d^3{\bf k}}{(2\pi)^3}\ \frac{1}{2\omega_k}\ 
\left(
e^{-i \left( \omega_k t - {\bf k}\cdot ({\bf x}-{\bf y}) \right)}
\ -\ 
e^{+i \left( \omega_k t - {\bf k}\cdot ({\bf x}-{\bf y}) \right)}
\right)
\nonumber\\
& = - i \theta(t)\ 
\int\ \frac{d^3{\bf k}}{(2\pi)^3}\ \frac{1}{\omega_k}\ 
\sin \omega_k t\ 
e^{i  {\bf k}\cdot ({\bf x}-{\bf y}) } \ .
    \label{eq:GROBC1}
\end{align}
In a 3+1D finite volume with PBCs, this becomes
\begin{equation}
D_R({\bf x},{\bf y},t,0) \rightarrow 
- i \theta(t)\ 
\frac{1}{L^3} \sum_{\bf k}
\ \frac{1}{\omega_{\bf k}}\ 
\sin \omega_{\bf k} t\ 
e^{i  {\bf k}\cdot ({\bf x}-{\bf y}) }
 =
- i \theta(t)\ 
\sum_{\bf n}
\ \frac{1}{\omega_{\bf n}}\ 
\sin \omega_{\bf n} t\ 
\psi_{\bf n}^\dagger({\bf y})
\psi_{\bf n}({\bf x}) \ ,
    \label{eq:GROBC2PBC}
\end{equation}
where $\psi_{\bf n}({\bf x})$ is an appropriately normalized lattice eigenstate subject to PBCs, defined by a triplet of integers ${\bf n}$,
\begin{equation}
\psi_{\bf n}({\bf x}) = \frac{1}{L^{3/2}}\ 
e^{i  {\bf k}\cdot {\bf x} }
\ \ ,\ \ 
\sum_{\bf x}
\psi_{\bf n}^\dagger({\bf x}) \psi_{\bf m}({\bf x}) = \delta^{(3)}_{{\bf n},{\bf m}}
\ \ ,\ \ 
\sum_{\bf n}
\psi_{\bf n}^\dagger({\bf y}) \psi_{\bf n}({\bf x}) = \delta^{(3)}_{{\bf x},{\bf y}} \ ,
    \label{eq:GROBC2PBCb}
\end{equation}
with ${\bf k}=2 \pi {\bf n}/L$.
To transition to OBCs, the OBC eigenstates $\psi_{\bf m}({\bf x})$ are used.
For simulations in 1+1D with OBCs, the relevant retarded Green's function is
\begin{equation}
D_R(x,y,t,0) = 
- i \theta(t)\ 
\sum_n
\ \frac{\sin \omega_n t}{\omega_n}\ 
\psi_n^\dagger(y)
\psi_n(x) \ ,
    \label{eq:GROBC2OBCa}
\end{equation}
with appropriately orthonormalized wavefunctions, such as those shown in Fig.~\ref{fig:OPBCevecsL4}.

Consider a source with a Gaussian profile, as was considered earlier, 
on a lattice of length $L$,
\begin{equation}
j_L(y) = \eta\ \sum_{n=0}^{L-1}\ \delta (y-n)\ e^{-\alpha (y - \frac{L-1}{2})^2} \ .
    \label{eq:jy}
\end{equation}
where $\eta$ is the appropriate normalization factor determined by requiring,
\begin{equation}
\int_{-\infty}^{+\infty}\ dy\ j_L(y) = J_0
 \ = \ \
 \eta \sum_{n=0}^{L-1} e^{-\alpha \left (\frac{L-1-2n}{2}\right )^2 } 
 \ \approx \ \eta \sqrt{\frac{\pi}{\alpha}}\ \left[
1\ +\ 2\sum_{p=1}^\infty\ (-)^p e^{-\pi^2 p^2/\alpha} \right]
 \equiv 
\eta \sqrt{\frac{\pi}{\alpha}}\  S(\alpha) \ .
    \label{eq:jyb}
\end{equation}
The approximate equality holds for a well-localized source with large $L$ and small $\alpha$ (in which case the bounds of the sum can extended to $\pm\infty$ with exponentially-suppressed errors, and the Poisson resummation formula can be used).
The function $S(\alpha)$ rapidly approaches the continuum result of unity, for decreasing $\alpha$.
Therefore, the sources can be written as
\begin{equation}
j_L(y) = J_0\ \sqrt{\frac{\alpha}{\pi}}\ \frac{1}{S(\alpha)}\ 
\sum_{n=0}^{L-1}\ \delta (y-n)\ e^{-\alpha (y - \frac{L-1}{2})^2} \ ,
    \label{eq:jyc}
\end{equation}
which is the discrete version of Eq.~\eqref{eq:KG3}.
The expression for the downstream field from the source is given by Eq.~\eqref{eq:KG2}, and can be written as
\begin{equation}
    \phi_j(x,t) = \phi_{j=0}(x,t) 
    \ +\ 
    J_0\ \sqrt{\frac{\alpha}{\pi}}\ \frac{1}{S(\alpha)}\ 
    \sum_{n=1}^L\ 
    \left(
    \sum_{y=0}^{L-1} \psi_n^\dagger(y)
    e^{-\alpha (y - \frac{L-1}{2})^2}
    \right)
    \ \frac{\sin \omega_n t}{\omega_n}\ 
    \psi_n(x) \ .
    \label{eq:phiDS}
\end{equation}
The expression in Eq.~(\ref{eq:phiDS}) is the corresponding result to Eq.~(\ref{eq:KG4}) but in a finite volume with OBCs.
Numerically, evaluating the field evolution from the source are the same until boundary effects become important.

\section{Truncated Electric Interactions for Odd \texorpdfstring{$L$}{L}}
\label{app:truncHam}
\noindent
The Hamiltonian corresponding to Eq.~\eqref{eq:spatTrunce} for odd $L$ is,
\begin{align}
&\hat{H}_{el}^{(Q=0)}(\bar{\lambda}) \ =  \ \frac{g^2}{2}\left\{ \sum_{n=0}^{\frac{L-3}{2}} \bigg[ \left( L - \frac{5}{4} - 2n \right) \hat{\overline{Q}}^2_n + \frac{1}{2} \hat{\overline{Q}}_n \hat{\delta}_n + \frac{1}{4} \hat{\delta}^2_n + \left( \frac{7}{4} + 2n \right) \hat{\overline{Q}}^2_{\frac{L+1}{2}+n}   \right. \nonumber \\
& - \frac{1}{2} \hat{\overline{Q}}_{\frac{L+1}{2}+n} \hat{\delta}_{\frac{L+1}{2}+n} + \frac{1}{4} \hat{\delta}^2_{\frac{L+1}{2}+n}\bigg ] + \frac{1}{4}(\hat{\overline{Q}}^2_{\frac{L-1}{2}}+\hat{\delta}^2_{\frac{L-1}{2}}) \nonumber \\
&+ 2\sum_{n=0}^{\frac{L-5}{2}}\ \ \sum_{m=n+1}^{\min(\frac{L-3}{2},n+\bar{\lambda})} \bigg[ \left( L-1 - 2m \right) \hat{\overline{Q}}_n\hat{\overline{Q}}_m + \frac{1}{2} \hat{\overline{Q}}_{n} \hat{\delta}_{m} + \left( 2 + 2n \right) \hat{\overline{Q}}_{\frac{L+1}{2}+n}\hat{\overline{Q}}_{\frac{L+1}{2}+m}
\nonumber \\
&- \frac{1}{2} \hat{\overline{Q}}_{\frac{L+1}{2}+m} \hat{\delta}_{\frac{L+1}{2}+n} \bigg ]  \nonumber \\
&+ \left. \frac{1}{2} \sum_{n=1}^{\min(\frac{L-1}{2},\bar{\lambda})} \left[ \hat{\overline{Q}}_{\frac{L-1}{2}-n} \hat{\overline{Q}}_{\frac{L-1}{2}} + \hat{\overline{Q}}_{\frac{L-1}{2}-n} \hat{\delta}_{\frac{L-1}{2}} +  \hat{\overline{Q}}_{\frac{L-1}{2}+n} \hat{\overline{Q}}_{\frac{L-1}{2}} - \hat{\overline{Q}}_{\frac{L-1}{2}+n} \hat{\delta}_{\frac{L-1}{2}} \right] \right\} \ .
\end{align}
%

\clearpage
\section{Further Details on Circuit Construction}
\label{app:circuitDetails}
\noindent
\begin{figure}[!ht]
    \centering
    \includegraphics[width=\textwidth]{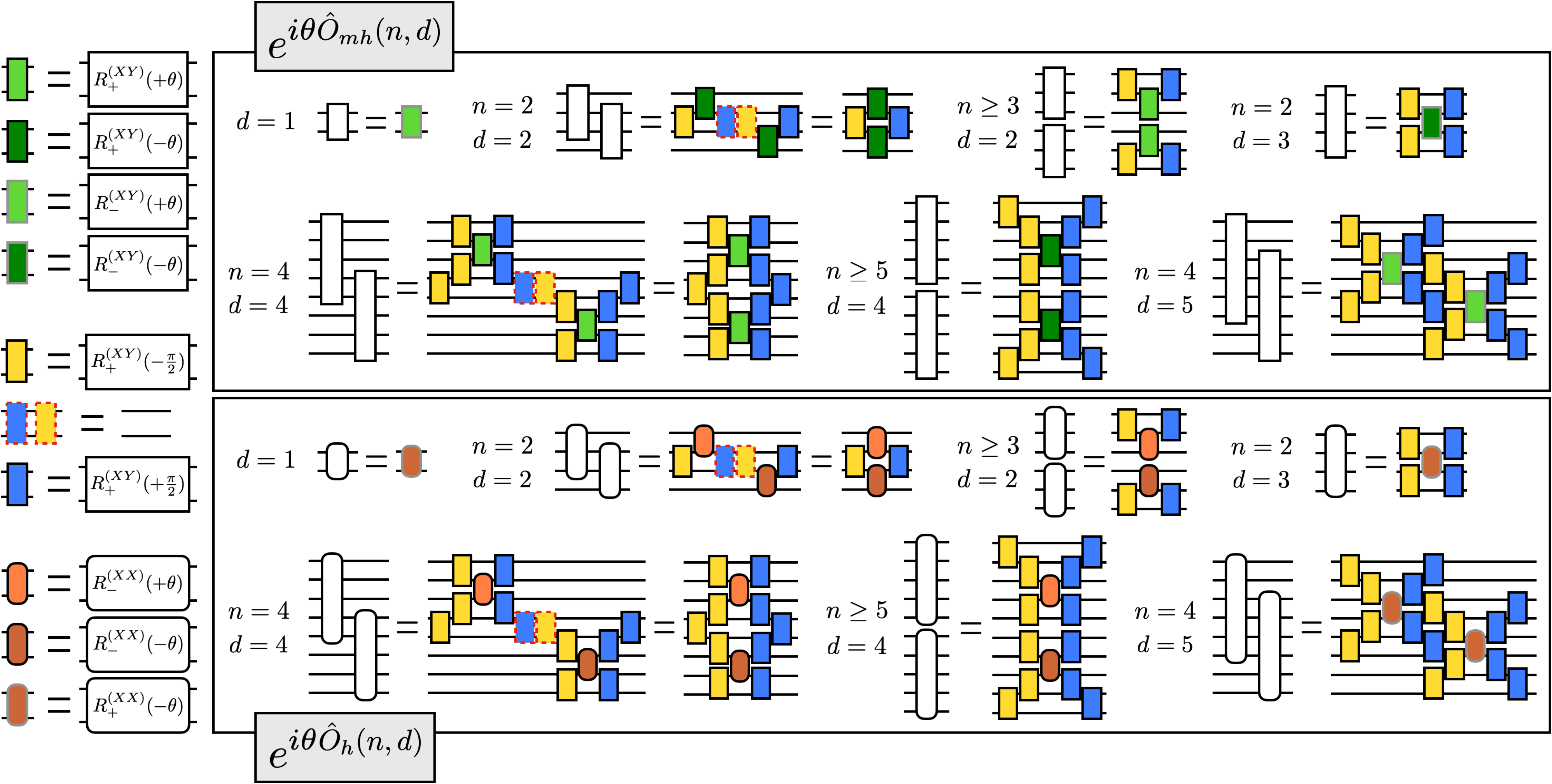}
    \caption{Efficient circuits implementing the unitaries corresponding to the wavepacket pool operators in Eq.~\eqref{eq:PacketPool}. 
    The circuits for the individual blocks $R^{(XY)}_{\pm}(\theta)$ and $R^{(XX)}_{\pm}(\theta)$ are shown in Fig.~\ref{fig:xy_xx_circuits}.}
    \label{fig:blocks_wp_adapt}
\end{figure}
The circuit implementation of the operators from the wavepacket pool in Eq.~\eqref{eq:PacketPool} for $d\leq 5$ is shown in Fig.~\ref{fig:blocks_wp_adapt}.

\section{Another Way to Create Hadron Wavepackets}
\label{app:chempot}
\noindent
In the main text, circuits are constructed that optimize the overlap with an adiabatically prepared hadron wavepacket.
Here, an alternative method for preparing hadron wavepackets is presented based on minimizing the energy in the single-hadron sector.
Desirable features of a hadronic wavepacket are that it is localized (i.e., outside of the wavepacket profile, the system is locally in the vacuum), and that it is composed of single hadrons.
When establishing a wavepacket on top of the interacting vacuum, as is done in the main text, localizability can be implemented at the level of the operator pool.
For example, by only including operators in the pool that have support over a predefined spatial interval, $l$, it is guaranteed that outside of $l$ is vacuum.
To ensure that the wavepacket is composed of single hadron states, consider adding a vacuum chemical potential, $\mu$, to the Hamiltonian,
\begin{equation}
\hat{H}_{\text{1-hadron}} \ = \ \hat{H} \ + \ \mu \vert \psi_{\text{vac}} \rangle \langle \psi_{\text{vac}}\vert \ .
\label{eq:H1chem}
\end{equation}
For $\mu > m_{\text{hadron}}$, the ground state of $\hat{H}_{\text{1-hadron}}$ 
in the $Q=0$ sector is the lowest-energy state of a single hadron.
The strategy for building a wavepacket is to minimize $\langle \psi_{\text{ansatz}} \vert \hat{H}_{\text{1-hadron}}\vert  \psi_{\text{ansatz}} \rangle$, where $\vert  \psi_{\text{ansatz}} \rangle$ is adaptively built using a localized operator pool. 
The resulting state will be the lowest energy configuration orthogonal to the vacuum that is localized within the interval $l$. 
The prepared state will primarily be a superposition of single hadrons, with multi-hadron contributions decreasing as $l$ increases.

As an example, consider using this procedure to construct a single-hadron wavepacket with an operator pool localized to $l=2$ sites on either side of the midpoint of the lattice.
Starting from the operator pool in Eq.~\eqref{eq:PacketPool}, the $l=2$ pool consists of $\hat{O}_m(1)$, $\hat{O}_m(2)$, $\hat{O}_{mh}(1,1)$, $\hat{O}_{h}(1,1)$, $\hat{O}_{mh}(2,d)$ and $\hat{O}_{h}(2,d)$ with $d=\{1,2,3\}$ .
Choosing $\mu = 2.5 \, m_{\text{hadron}}$ pushes the energy of the vacuum above two-particle threshold (which is slightly below $2 \, m_{\text{hadron}}$ due to the presence of a two-hadron bound state), and is found to be effective for our purposes.
To update the SC-ADAPT-VQE ansatz, the gradient can be computed with 
\begin{align}
&\frac{\partial}{\partial\theta_i}\left. \langle \psi_{\text{ansatz}}\vert e^{-i \theta_i \hat{O}_i}\hat{H}_{\text{1-hadron}} e^{i \theta_i \hat{O}_i}\vert \psi_{\text{ansatz}}\rangle \right|_{\theta_i=0}
   \nonumber \\
   &= - \text{Im} \left [\langle \psi_{\text{ansatz}}\vert \left ( [ \hat{H},\hat{O}_i ] \ + \ 2 \mu \, \vert \psi_{\text{vac}} \rangle \langle \psi_{\text{vac}} \vert \hat{O}_i \right ) \vert \psi_{\text{ansatz}} \rangle \right ] 
    \ .
\end{align}
Note that it can be necessary to bias the initial parameters to avoid the optimizer choosing $\theta_i=0$ because the initial state is a local maxima of energy (and second derivatives are then required).
Due to the limited size of the operator pool, the SC-ADAPT-VQE algorithm converges relatively well after 4 steps, with the optimal operators and associated variational parameters shown in Table~\ref{tab:AnglesWPchem}.
\begin{table}[!tb]
\renewcommand{\arraystretch}{1.4}
\begin{tabularx}{\textwidth}{|c || Y | Y | Y | Y ||}
 \hline
 \diagbox[height=23pt]{$L$}{$\theta_i$} & $\hat O_{mh}(1,1)$ & $\hat O_{mh}(2,3)$ & $\hat O_{mh}(2,2)$ & $\hat O_{mh}(2,1)$  \\
 \hline\hline
 7  &  -2.4342  & -0.9785  & 0.0819  & 0.2599  \\
 \hline
8  &  2.4343  & 0.9778  & 0.0808  & -0.2591  \\
\hline
9  &  -2.4342  & -0.9780  & 0.0812  & 0.2594  \\
\hline
10  &  2.4340  & 0.9778  & 0.0811  & -0.2595  \\
\hline
11  &  -2.4339  & -0.9776  & 0.0810  & 0.2593  \\
\hline
12  &  2.4321  & 0.9781  & 0.0837  & -0.2605  \\
\hline
13  &  -2.4343  & -0.9780  & 0.0810  & 0.2593  \\
 \hline
\end{tabularx}
\caption{The operator ordering and variational parameters that minimize $\hat{H}_{\text{1-hadron}}$, and prepare a hadron wavepacket for $L=7-13$.}
 \label{tab:AnglesWPchem}
\end{table}
The resulting state has an $L$-independent energy expectation value of $\langle \psi_{\text{ansatz}} \vert \hat{H} \vert \psi_{\text{ansatz}} \rangle = 1.18 \, m_{\text{hadron}}$, and overlap onto the vacuum state of $\vert\langle \psi_{\text{vac}} \vert \psi_{\text{ansatz}} \rangle \vert^2 = 8.5\times 10^{-5}$.
These results show that the prepared wavepacket is primarily composed of single-hadron states, and both $\langle \psi_{\text{ansatz}} \vert \hat{H} \vert \psi_{\text{ansatz}} \rangle$ and $\vert\langle \psi_{\text{vac}} \vert \psi_{\text{ansatz}} \rangle \vert^2$ can be further reduced by increasing $l$, i.e., de-localizing the prepared wavepacket.

\section{Details on the 112-qubit MPS Simulations}
\label{app:MPSSim}
\noindent
The 112-qubit quantum simulations in Sec.~\ref{sec:Qsim} are compared to the expected, error-free, results determined using the {\tt qiskit} and {\tt cuQuantum} MPS circuit simulators.
MPS techniques are approximations that can be improved by increasing the bond dimension in the MPS ansatz. 
A higher bond dimension increases the maximum amount of entanglement in the ansatz state, at the cost of longer run-time on a classical computer.
As a result, simulations that explore highly-entangled states are promising candidates for a near-term quantum advantage.
Our numerical investigations have found a large contrast between the bond dimension needed for state preparation and time evolution.
The initial hadron wavepacket coincides with the vacuum state outside of the few sites where the wavepacket has support.
This state has a low amount of entanglement as the ground states of gapped 1D systems have area-law entanglement~\cite{Hastings_2007,arad2013area,Brand_o_2014}. 
Therefore, a relatively small bond dimension can be used in the MPS simulations to faithfully reproduce the preparation of the vacuum and initial hadron wavepacket.
Time evolution, on the other hand, involves a superposition of many single-hadron states, which disturb the vacuum as they propagate.
This produces a significant amount of entanglement, and subsequently requires a larger bond dimension.

\begin{figure}[!ht]
    \centering
    \includegraphics[width=0.9\columnwidth]{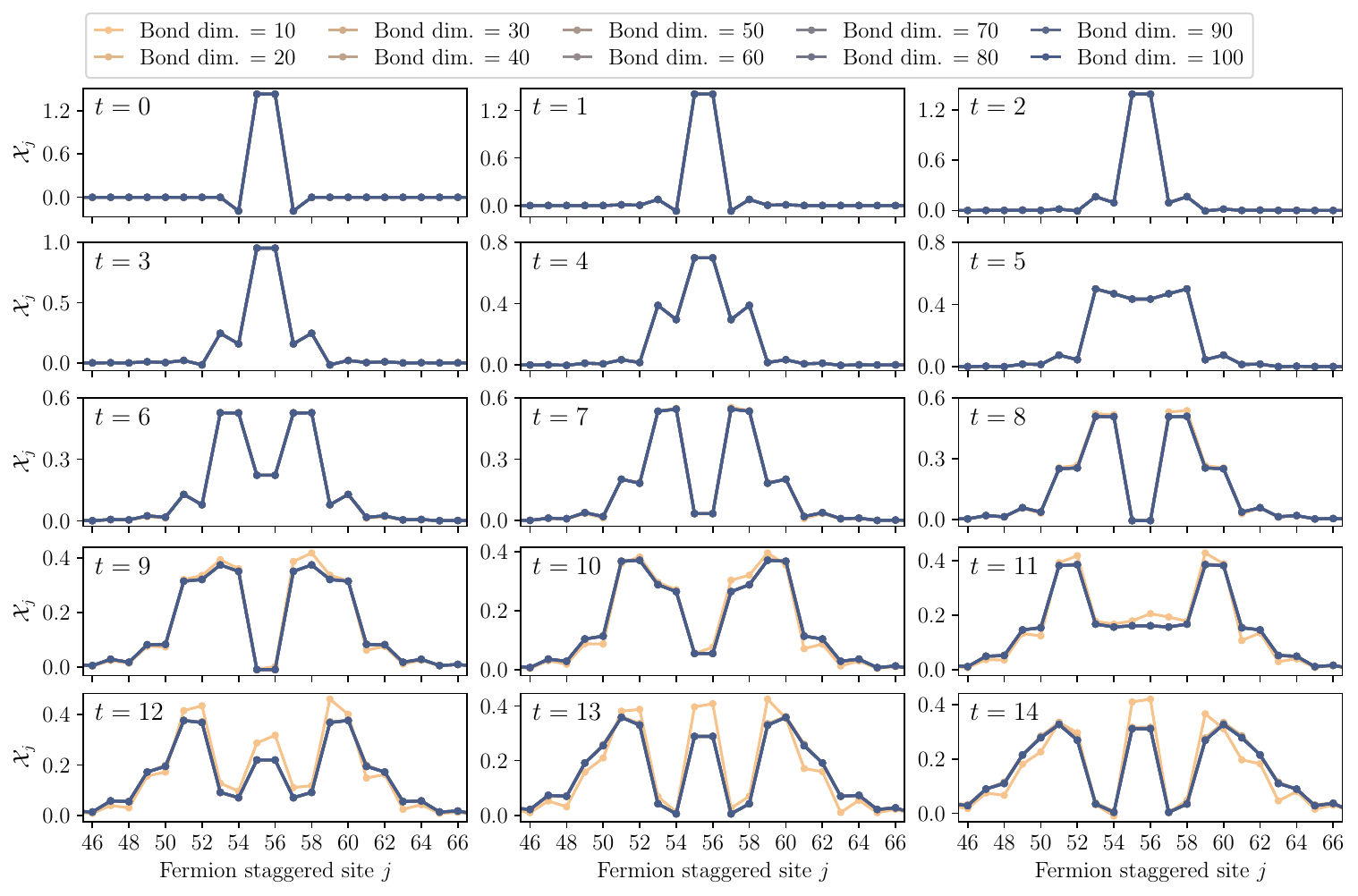}
    \caption{Convergence of the vacuum subtracted local chiral condensate, ${\cal X}_j(t)$, with different maximum bond dimension in the {\tt cuQuantum} MPS simulator.
    Results are shown for $L=56$, and are focused around the center of the lattice where the convergence is the slowest.}
    \label{fig:MPS_Convergence}
\end{figure}

\begin{figure}[!ht]
    \centering
    \includegraphics[width=0.9\columnwidth]{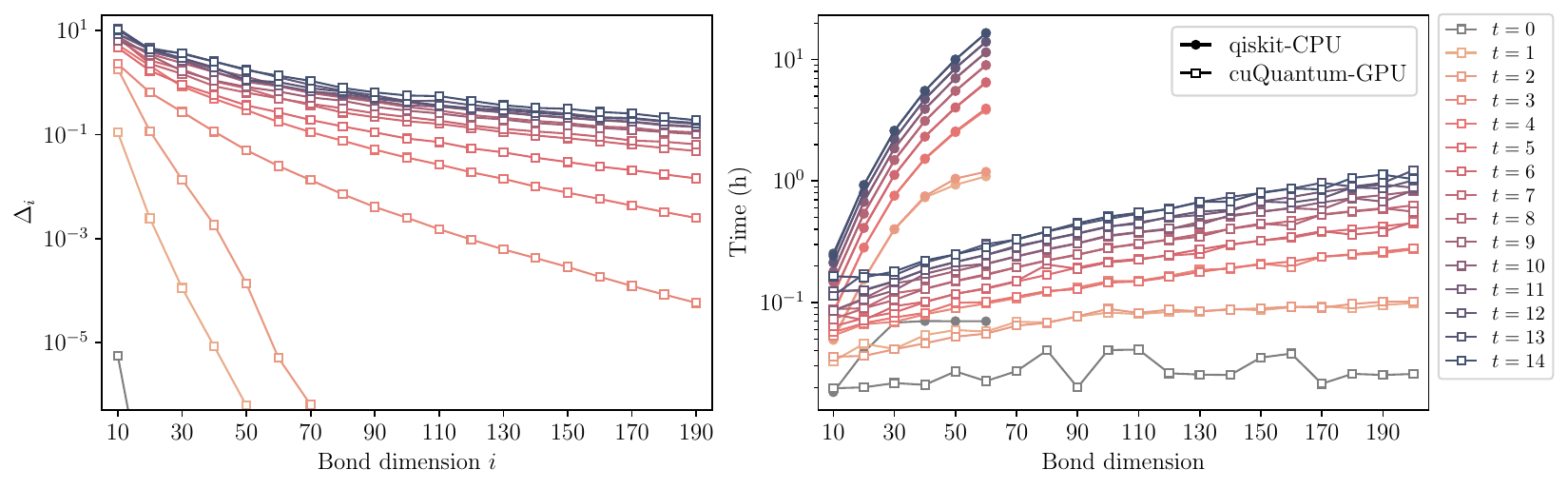}
    \caption{The left panel shows the relative convergence of the chiral condensate, $\Delta_i$, defined in Eq.~\eqref{eq:deltai}, for a selection of times as a function of maximum bond dimension, while the right panel shows the computational run time using the {\tt qiskit} MPS simulator on a single 40-core CPU-node and the {\tt cuQuantum} MPS simulator on a single NVIDIA RTX A5000, for $t=\{0,1,\ldots,14\}$ with different maximum bond dimension.}
    \label{fig:MPS_Runtime}
\end{figure}

The bond dimension needed for convergence of the chiral condensate for different simulation times is shown in Fig.~\ref{fig:MPS_Convergence}.
It is seen that a relatively small bond dimension is sufficient for convergence, even out to late times. 
This should be compared to the convergence of $\langle \psi_{\text{WP}} \vert \hat{\chi}_j \vert\psi_{\text{WP}} \rangle$
in the left panel of Fig.~\ref{fig:MPS_Runtime}, where the quantity
\begin{equation}
    \Delta_i \ = \ \sum_j \ | \ \langle \psi^{MPS_i}_{\rm WP} \vert \ \hat{\chi}_j(t) \ \vert \psi^{MPS_i}_{\rm WP} \rangle \ - \ \langle \psi^{MPS_{i+10}}_{\rm WP} \vert \ \hat{\chi}_j(t) \ \vert \psi^{MPS_{i+10}}_{\rm WP} \rangle \ | \ ,
\label{eq:deltai}
\end{equation}
is computed for different bond dimensions.
This quantity determines how much the local chiral condensate of the evolved wavepacket changes as the maximum bond dimension is increased from $i$ to $i+10$.
This reveals that MPS calculation of the chiral condensate of 
(a) the initial state can be done very efficiently (results with a maximum bond dimension of 10 have already converged below a $10^{-5}$ precision), and 
(b) the evolved wavepacket converges  slowly, especially at late times. 
Indeed, the quick convergence of ${\cal X}_n(t)$ in Fig.~\ref{fig:MPS_Convergence} is due to the cancellations of errors between the MPS simulation of the wavepacket and vacuum evolution.
These MPS simulations take increasingly more compute run-time as the bond dimension increases.
This is illustrated in the right panel of Fig.~\ref{fig:MPS_Runtime}, where the run-time for a selection of times  and various bond dimensions are shown. In this panel, we compare the performance of the CPU-based {\tt qiskit} MPS simulator, run on a single 40-core CPU-node on Hyak, and the GPU-based {\tt cuQuantum} MPS simulator, run on a single NVIDIA RTX A5000 through the OSG Pool.
%

\section{Further Details about the Error Mitigation and Analysis}
\label{app:qSimDetails}
\begin{figure}[!b]
    \centering
    \includegraphics[width=0.5\columnwidth]{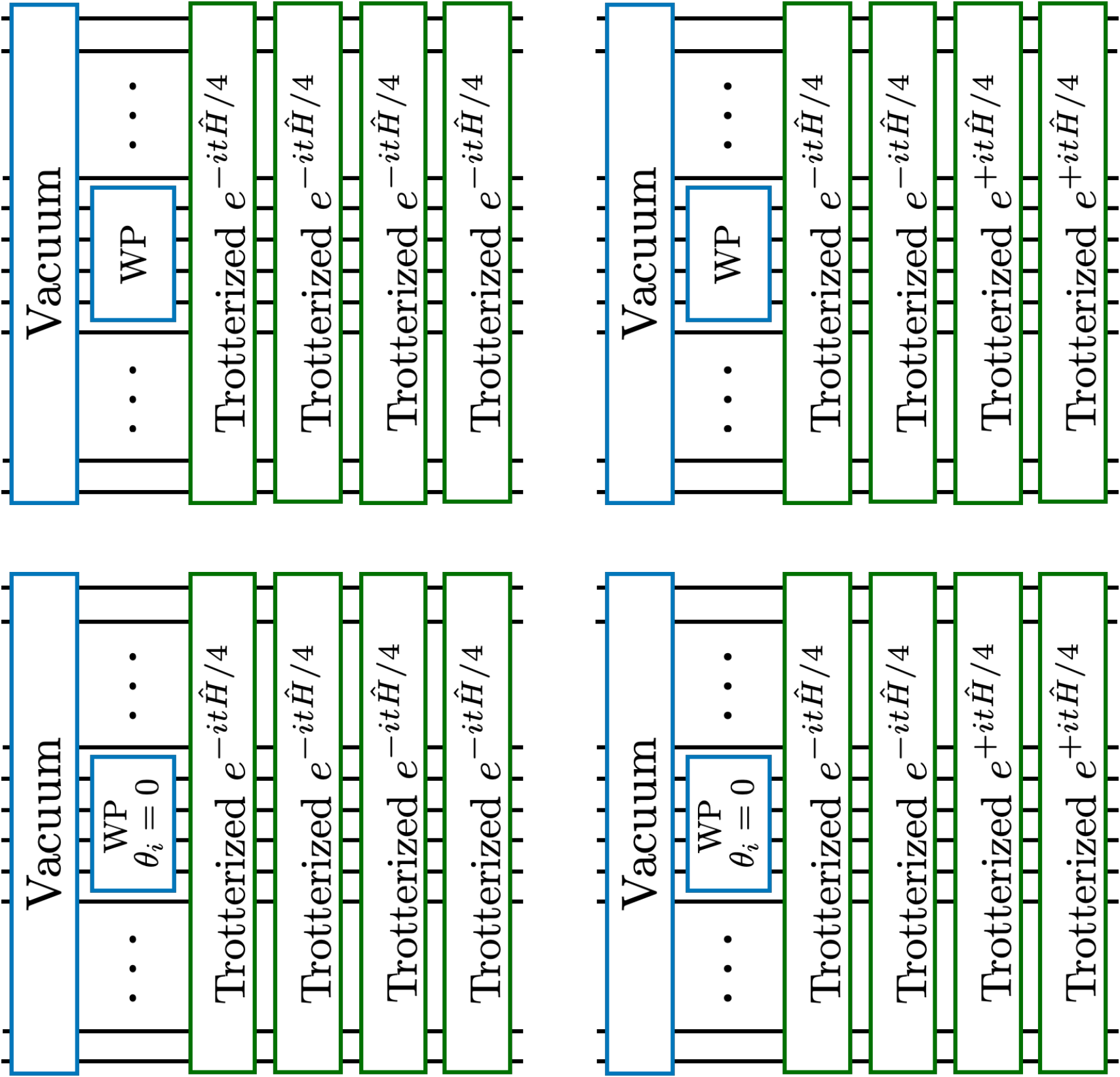}
    \caption{The four types of circuits run on {\tt ibm\_torino}.
    Blue boxes denote SC-ADAPT-VQE circuits, and green boxes denote Trotterized 
    time-evolution circuits.
    Shown are examples for 4 Trotter steps of time evolution, with straightforward extension to other even number of Trotter steps.
    (Lower) upper-left show the circuits used for the time evolution of the (vacuum) wavepacket. 
    (Lower) upper-right are the forwards-backwards time-evolution circuits used for ODR error-mitigation of the wavepacket (vacuum).
    The $\theta_i=0$ in the wavepacket circuit box denotes that the SC-ADAPT-VQE parameters are set to zero, i.e., it is the identity operator in the absence of device errors.}
    \label{fig:FourCircuits}
\end{figure}

\noindent
For each time $t=\{1,2,\ldots,14\}$, four kinds of circuits are run on the quantum computer: time evolution of the vacuum, time evolution of the wavepacket, and the corresponding forward-backward evolution for ODR error mitigation, see Fig.~\ref{fig:FourCircuits}.
Each circuit for $t=1-8$ is run with 480 twirls and each circuit for $t=9-14$ is run with 160 twirls; each twirl with 8,000 shots, as displayed in Table~\ref{tab:QsimCNOT}.
The longest continuous one-dimensional chain on {\tt ibm\_torino} that we utilize is 112 qubits, corresponding to a $L=56$ lattice (see layout in Fig.~\ref{fig:SimTricks}). 
We use two lattice-to-qubit mappings to minimize the effects of poorly performing qubits.
Half of the twirls assign staggered site 0 to the top-right device qubit, and the other half assign staggered site 0 to the bottom left device qubit.
Averaging over multiple layouts mitigates some of the effects of qubit-specific noise.
Indeed, in our simulations there are twirled instances where qubits perform poorly, either due to decoherence or to readout errors.
Such errors can be identified and removed from analysis by filtering out measurements where $\langle \hat{Z}_j \rangle_{\text{meas}}/\langle \hat{Z}_j \rangle_{\text{pred}} < \epsilon$ in the mitigation runs, with $\epsilon$ some predetermined threshold.\footnote{This type of event post-selection, requiring device performance to exceed a specified level in interleaved calibration circuits, has been employed previously, for example, Ref.~\cite{Klco:2019xro}.}
If this ratio is negative, then the qubit has flipped, and if it is 0 then the qubit has completely decohered, i.e., it has become a maximally mixed state.
We choose $\epsilon = 0.01$, and do not see much difference varying up to $\epsilon = 0.05$.
Our scheduling of  jobs interleaves physics and mitigation circuits with the same twirl.
Poorly performing qubits, identified from measuring the mitigation circuit, are cut from both the ensemble of mitigation and associated physics measurements.\footnote{
Note that 160 of the 480 twirls for $t=1-7$ do not interleave physics and mitigation. Instead, they are sent in batches of 40 circuits with uncorrelated twirls between mitigation and physics circuits. 
In this case a qubit measurement of physics circuit $n$ in the batch is cut if the corresponding qubit measurement in mitigation circuit $n$ is cut.
Surprisingly, no improvement is found when correlating the twirls and interleaving mitigation and physics circuits.}

\begin{figure}[!htb]
    \centering
    \includegraphics[width=\columnwidth]{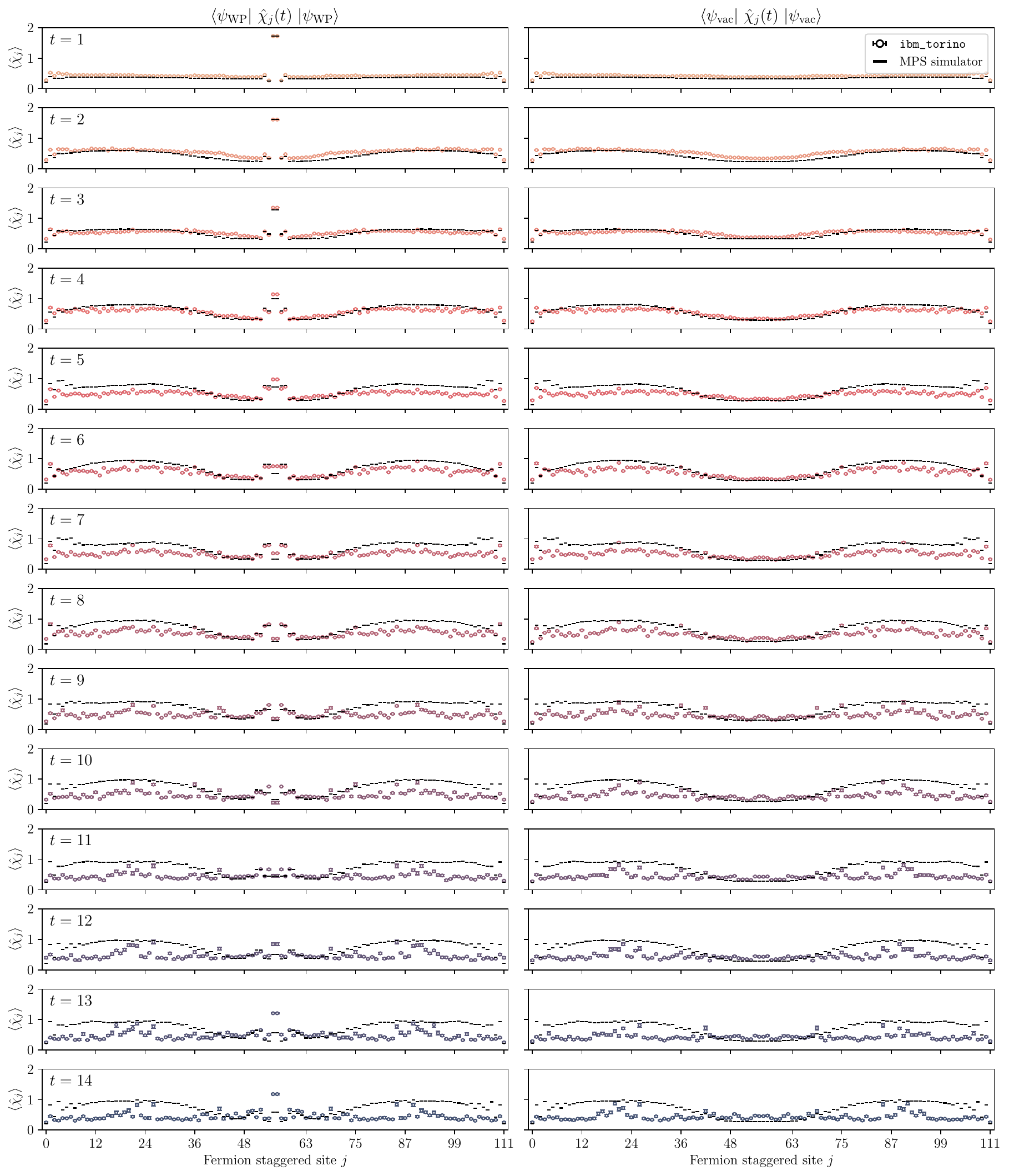}
    \caption{CP-averaged local chiral condensate for the time-evolved wavepacket (left subpanels) and vacuum (right subpanels). The points are obtained from {\tt ibm\_torino}, and the black lines from the {\tt cuQuantum} MPS circuit simulator.}
    \label{fig:src_vac_evol}
\end{figure}

The results of measurements related by CP symmetry are combined.
For $\hat{Z}_j$, this means combining $\langle \hat{Z}_j \rangle$ and $-\langle \hat{Z}_{2L-1-j} \rangle$ (for runs with 480 twirls, this can lead to up to 960 independent measurements for $\langle \hat{Z}_j \rangle$).
The central values and corresponding uncertainties are determined from bootstrap re-sampling over twirls. 
Due to the filtering  procedure, $\langle \hat{Z}_j\rangle$ for each qubit can have a different number of contributing twirls, $N^{(\text{meas})}_j$.
For each sample in the bootstrap ensemble, $N^{(\text{meas})}_j$ random integers with replacement $\{ x \} \in \{1,2,\ldots, N^{(\text{meas})}_j\}$ are generated, with the prediction for the error-free physics expectation value for that sample given by
\begin{equation}
\left.{\overline{\langle \hat{Z}_j \rangle}_{\text{pred}}}\right\rvert_{\text{phys}} \ = \ \left( \sum_{i \in \{ x \}} \left.{\langle \hat{Z}_j \rangle^{(i)}_{\text{meas}}} \right\rvert_{\text{phys}} \right) \times  \left ( \sum_{i \in \{ x \}} \left . \frac{\langle \hat{Z}_j \rangle_{\text{pred}}}{\langle \hat{Z}_j \rangle^{(i)}_{\text{meas}}}\right \rvert_{\text{mit}} \right )\ ,
\label{eq:bootsrap}
\end{equation}
where the superscript $(i)$ labels the twirl.
This is performed for the wavepacket and for the vacuum evolution, with the vacuum subtracted chiral condensate given by
\begin{equation}
\overline{ {\cal X}_j } \ = \ (-1)^j \left ( \left.\overline{\langle \hat{Z}_j \rangle}_{\text{pred}} \right\rvert_{\text{phys}}^{(\text{WP})} \ - \ \left. \overline{\langle \hat{Z}_j \rangle}_{\text{pred}} \right\rvert_{\text{phys}}^{(\text{Vac})}  \right ) \ .
\end{equation}
This process is repeated $N_{\text{Boot}}$ times, with $N_{\text{Boot}}$ large enough for the mean and standard deviation of the bootstrap ensemble $\{ \ \overline{ {\cal X}_j }\ \}$ to have converged.
This mean and standard deviation are used to produce the points with error bars in Figs.~\ref{fig:srcmvac_evol} and~\ref{fig:src_vac_evol}.\footnote{The two sums in Eq.~\eqref{eq:bootsrap} compute the mean of the bootstrap sample. 
If instead the median is used, larger error bars are found.
This is likely due to there being correlations in the tails of both the ensembles of physics and mitigation measurements that are captured by the mean, but suppressed 
by the median.}

We have found that larger angles in the circuits lead to larger systematic errors, independent of circuit depth.
This is likely due to cross-talk errors between gates acting on neighboring qubits when large rotations are applied.
These kinds of errors are not corrected by ODR.
Thus, there is a trade-off between increased number of Trotter steps with smaller angles, and the associated increased circuit depth.
A full determination of this trade off remains to be explored.

\begin{figure}[!htb]
    \centering
    \includegraphics[width=0.8\columnwidth]{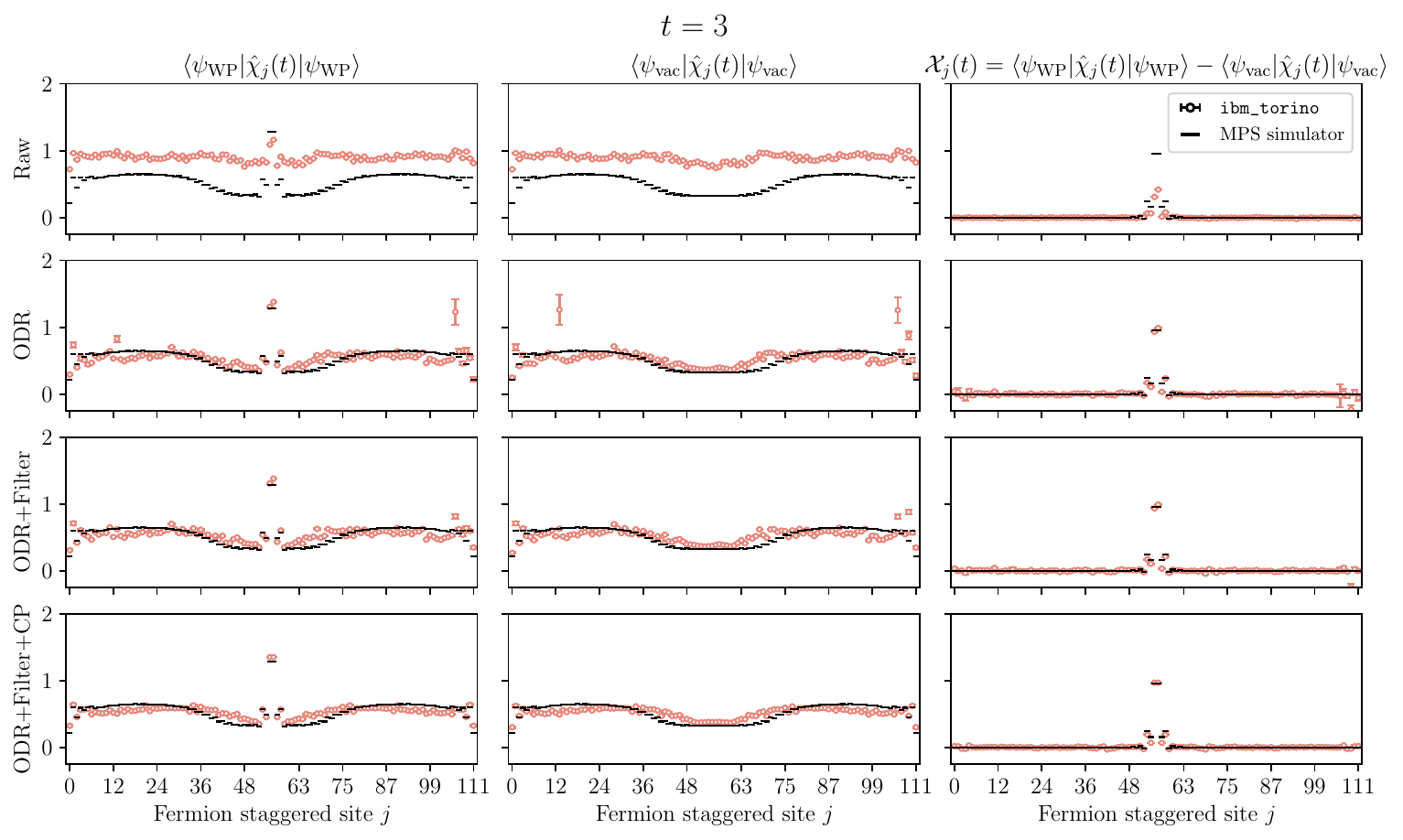}
    \includegraphics[width=0.8\columnwidth]{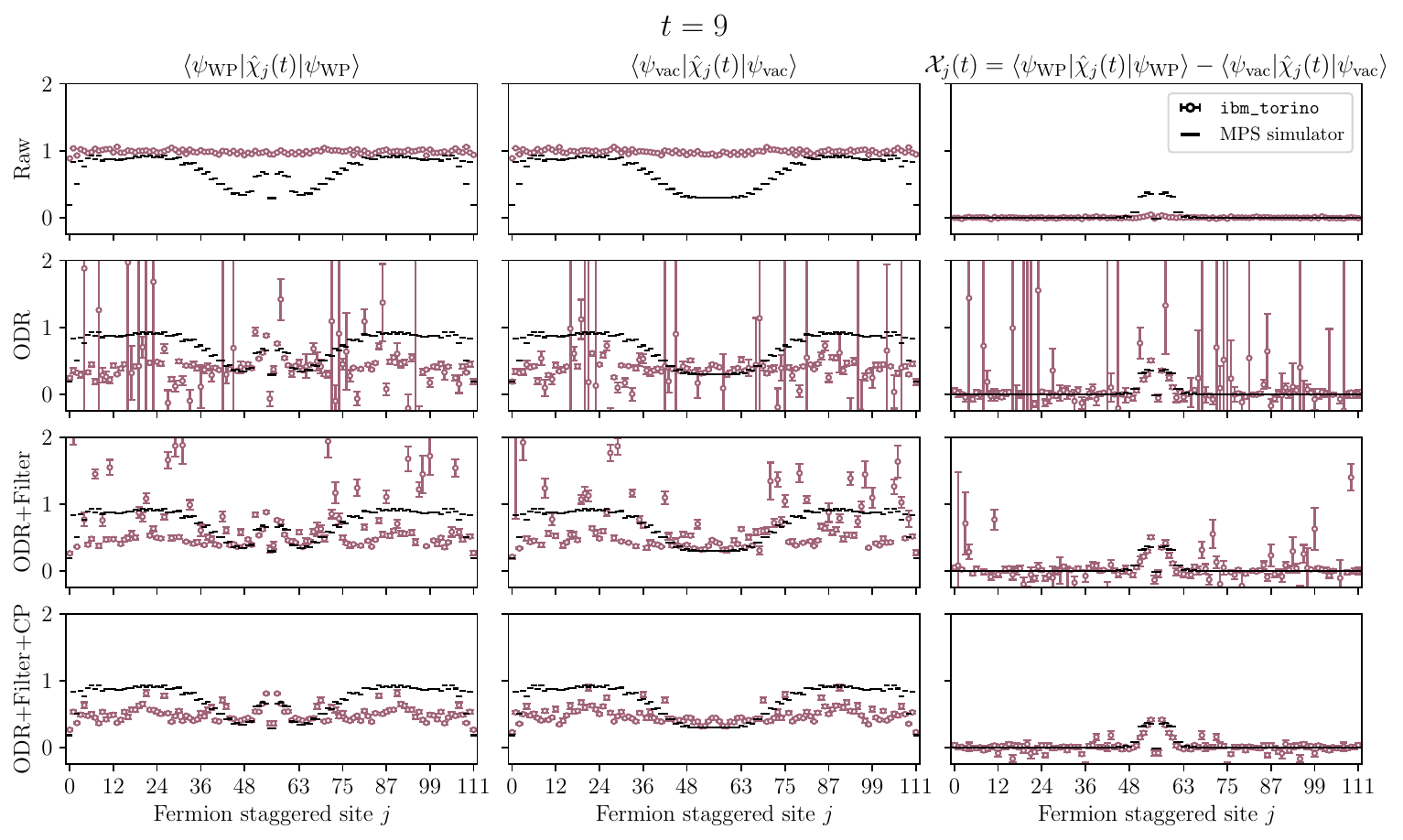}
    \caption{The results obtained from {\tt ibm\_torino} after different stages of error mitigation. Results for $t=3$ and $t=9$ are shown for the chiral condensate of the wavepacket and vacuum evolution, and their subtraction. The four rows give: raw results, after applying ODR, after filtering out the decohered qubits, and after CP averaging. Note that after the filtering procedure, different qubits have different numbers of twirls contributing (and hence different sized errors bars).}
    \label{fig:mitigation_stages_IBM}
\end{figure}

The different stages of error mitigation are displayed in Fig.~\ref{fig:mitigation_stages_IBM}.
Two times, $t=3$ (CNOT depth 120) and $t=9$ (CNOT depth 270), are chosen for the purpose of demonstration. 
Note that in these plots, the decohered value of the chiral condensate is $\langle \hat{\chi}_j\rangle =1$ (with $\mathcal{X}_j =0$).
The first row of Fig.~\ref{fig:mitigation_stages_IBM} shows the ``raw'' results obtained from the device (with dynamical decoupling and readout error mitigation) after averaging over all Pauli twirls.
The device errors for the wavepacket and vacuum evolution outside of the wavepacket region are very similar, and cancel to a large degree in forming the subtraction in ${\cal X}_j(t)$.
It is striking that, for $t=9$, there is no discernible sign of the presence of a wavepacket in the raw results.
The second row of Fig.~\ref{fig:mitigation_stages_IBM} shows the effect of applying ODR.
This helps recover the chiral condensate, being more effective for $t=3$ than $t=9$, but can also lead to large error bars when the qubit is close to being completely decohered ($\langle\hat{Z}_j\rangle_{\text{meas}}\vert_{\text{mit}}$ close to zero).
The third row of Fig.~\ref{fig:mitigation_stages_IBM} shows the effects of filtering out runs where $\langle \hat{Z}_j \rangle_{\text{meas}}/\langle \hat{Z}_j \rangle_{\text{pred}}\vert_{\text{mit}} < 0.01$.
This removes most of the runs contributing to the large error bars, and is more significant for $t=9$ than $t=3$.
It also leads to different numbers of twirls surviving the filtering for different qubits.
Sometimes only a small number survive, compromising the assumption of a depolarizing channel for ODR (and also compromising the error estimates from bootstrap re-sampling).
The fourth row of Fig.~\ref{fig:mitigation_stages_IBM} shows the effects of using the CP symmetry to combine the measurements of $\langle \hat{Z}_j \rangle$ and $-\langle \hat{Z}_{2L-1-j} \rangle$.
This reduces the effects of poorly performing qubits, and gives the final results presented in Figs.~\ref{fig:IBMresultsMPS}, \ref{fig:srcmvac_evol}, and~\ref{fig:src_vac_evol}.

\end{subappendices}

\chapter{Steps Toward Quantum simulations of Hadronization and Energy-Loss in Dense Matter}
\label{chap:deDx}
\noindent
\textit{This chapter is associated with Ref.~\cite{Farrell:2024mgu}: \\
``Steps Toward Quantum Simulations of Hadronization and Energy-Loss in Dense Matter" by Roland C. Farrell, Marc Illa and Martin J. Savage.}
\section{Introduction}
\noindent
An improved understanding of the transport of energy, momentum, flavor, and other quantum numbers 
in non-equilibrium strongly-interacting dense matter is needed to  
refine predictive capabilities for 
the extreme matter created in the early universe and in astrophysical environments.
Motivation and inspiration comes from ongoing and planned experiments of heavy-ions collisions~\cite{Dusling:2015gta,Busza:2018rrf} and from astronomical observations of multi-messenger signals~\cite{LIGOScientific:2017ync,Troja:2018uns,KAGRA:2021vkt,LIGOScientific:2020aai}.
The state-of-the-art predictions for the structure and dynamics of extreme matter 
integrate the experimental results from these programs with analytic and computational techniques, 
and phenomenological modeling~\cite{Lovato:2022vgq,Fevre:2023swv,Jacobi:2023olu,Brandes:2023hma}.

Knowledge of energy-loss and stopping ranges of electrically charged particles, from electrons to nuclei, in ordinary materials is essential to many scientific, technological, societal, and therapeutic endeavors, including in protecting humans and scientific equipment from cosmic rays and background radiation.
It is also critical to designing ultra-sensitive experiments to search for new physics that interacts weakly with ordinary matter, such as searches for Dark Matter~\cite{Trickle:2019nya,Baxter:2022dkm} and $0\nu\beta\beta$-decay of nuclei~\cite{Giuliani:2012zu,Dolinski:2019nrj}.
Long-term experimental programs have measured the energy loss for particles penetrating a wide selection of materials over a broad spectrum of energies.
Theoretical descriptions of the underlying mechanisms are well-established for electrically-charged particles and $\gamma$-rays over a significant energy regime, including, for example, the effects of elastic and inelastic collisions, pair-production and bremsstrahlung radiation~\cite{Tsai:1973py,SELTZER1984665}.
Interactions with nuclei at higher energies remain the focus of work at current and future high-energy colliders and fixed-target experiments~\cite{Meehan:2017cum,Maurice:2017iom,Hadjidakis:2018ifr,Barschel:2020drr,AbdulKhalek:2021gbh}, e.g., experiments at the Thomas Jefferson National Accelerator Facility (TJNAF), Brookhaven National Laboratory (BNL), or at the Large Hadron Collider (LHC).
Together, experiment, theory, and computation have established an extensive catalogue enabling predictions for the behavior of charged particles moving through a variety of materials (a brief summary can be found in the Particle Data Group~\cite{Workman:2022ynf}).
In contrast, energy loss mechanisms in dense nuclear matter in (and out of) equilibrium are much less well understood. 
There has been significant theoretical progress on this topic 
(e.g., see Ref.~\cite{zhao2023hadronization}), 
largely driven by input from heavy-ion collision experiments at BNL~\cite{PHENIX:2004vcz,BRAHMS:2004adc,PHOBOS:2004zne,STAR:2005gfr,STAR:2010vob,STAR:2017sal} and the LHC~\cite{Loizides:2016tew,Foka:2016zdb,Foka:2016vta}.
As an example, because of their mass and compactness, the transport properties, yields, and distributions of 
heavy quarks and
quarkonia systems provide insights into the nature of the  matter created in heavy-ion collisions (for recent reviews on different theoretical approaches, see Refs.~\cite{Rothkopf:2019ipj,Akamatsu:2020ypb,Chapon:2020heu,Yao:2021lus,Montana:2023sft}).
However, much remains unknown, particularly in processes where quantum coherence and entanglement play a significant role, and conventional methods such as Monte-Carlo sampling over event probability distributions, break down.
Previous approaches have often neglected particular elements of such processes, e.g., the effects of coherent scattering in parton shower simulations~\cite{Bauer:2019qxa,Bepari:2020xqi,Bepari:2021kwv,Macaluso:2021ngq,Chigusa:2022act,Gustafson:2022dsq,Bauer:2023ujy}.
Generally, a more
complete understanding of the dynamics of transport, fragmentation, color screening, and hadronization in non-equilibrium quantum chromodynamics  (QCD) matter remains a forefront challenge.

In this chapter, classical simulations of the Schwinger model are performed to determine the energy-loss and other observables associated with particles moving through dense matter.
An extensive study of heavy-hadrons moving through the 
lattice vacuum is performed in order to 
isolate lattice artifacts that arise from the breaking of Lorentz symmetry, 
and will not survive in the continuum.
These lattice artifacts are magnified in certain entanglement measures, 
and lead to energy loss and light-hadron production 
even for a heavy-hadron moving through the vacuum at constant velocity.
Once parameters are found where these artifacts are minimized, the propagation of heavy-hadrons through a medium of static heavy-hadrons is considered.
Energy loss due to the production of hadrons (hadronization) and internal excitations are identified, and the crucial role of quantum coherence is emphasized.
These classical simulations are limited to system sizes of $L=12$ spatial sites (24 staggered sites), and we present scalable quantum circuits for state preparation and estimate the resources required for large-scale quantum simulations of these phenomena.

\section{The Simulation Strategy}
\label{sec:framework}
\noindent
To investigate how charges move through dense strongly-correlated matter,
heavy ``external'' charges are introduced into the Schwinger model, 
with positions specified by classical trajectories.
Heavy charges with fixed positions define regions of dense matter 
within the lattice, 
and additional heavy charges moving across these regions probe energy-loss, 
fragmentation, hadronization, and entanglement arising from propagation through a dense medium. 
These heavy charges emulate the heavy fields 
in heavy-quark effective theory (HQET)~\cite{Isgur:1989vq,Isgur:1990yhj,Eichten:1989zv,Georgi:1990um,Grinstein:1990mj} that are used to define 
a systematic expansion about the heavy-quark limit.
In this limit, analogous to a B-meson,
a heavy-hadron in the Schwinger model is composed of a single heavy charge 
that is electrically neutralized by a ``cloud'' of light charges.
Important to the current treatment
is that the position, velocity, and acceleration of the moving heavy charge 
are well-defined throughout its motion.
To access the desired physics, we
choose a classical trajectory where the heavy charge accelerates to a constant velocity, 
moves through the dense medium, and then decelerates to rest.

\subsection{The Lattice Schwinger Model Hamiltonian with Heavy Charges}
\label{sec:SM}
\noindent
The Hamiltonian used in this chapter is the same as in Eq.~\ref{eq:Hgf}, except that now we allow for heavy background charges.
This leads to 
\begin{align}
 \hat H_{el} 
\ \rightarrow \  \frac{g^2}{ 2}\sum_{j=0}^{2L-2}\bigg (\sum_{k\leq j} \hat{q}_k +Q_k \bigg )^2 
\label{eq:Hgf1}
\end{align}
The electric charge operator $\hat q_k$ acts on the $k^{\rm th}$ staggered site, and $Q_k$ is a heavy background charge.
The heavy charges have been included as discontinuities in Gauss's law, which
(on the lattice) is,
\begin{equation}
{\bf E}_k - {\bf E}_{k-1} = q_k + Q_k \ ,
\end{equation}
where ${\bf E}_k$ is the electric field on the link between staggered sites $k$ and $k+1$.
Due to confinement and without background charges, the low-energy excitations are charge-neutral bound states of electrons and positrons. 
The hadron mass $m_{\text{hadron}}$ and confinement length
$\xi \sim m_{\text{hadron}}^{-1}$ depend non-perturbatively on $m$ and $g$.
The range of $\{m,g\}$ values used in this work give rise to $1.3 \leq m_{\text{hadron}}^{-1} \leq 1.8$ 
that is well contained inside of the volumes accessible to classical simulation, 
$\xi \ll L$.
All lengths are measured in units of the staggered lattice spacing that has been set to
$a_{\text{staggered}} = a_{\text{spatial}}/2=1$.
It is important to keep in mind that
this Hamiltonian measures the energy in the light degrees of freedom, as there is no mass 
or kinetic term for the heavy charges.
An explicit expression for $\hat{H}_{el}$ with the charge operator $\hat{q}_k$ expanded in terms of $\hat{Z}_k$ can be found in App.~\ref{app:hamZ}.

\subsection{Heavy-\texorpdfstring{$Q$}{} Trajectories}
\label{sec:trajsQ}
\begin{figure}[!t]
\centering
\includegraphics[width=\textwidth]{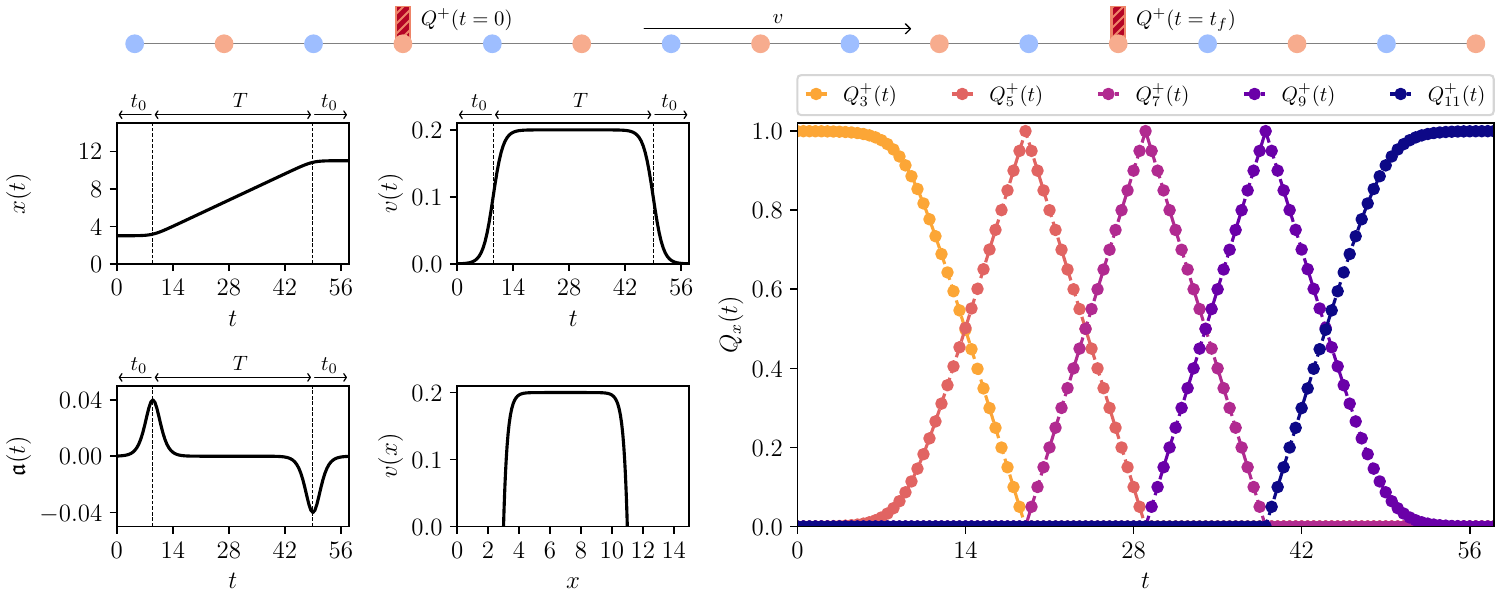}
\caption{
An example of a
trajectory 
moving a heavy-$Q^+$ from $x_0=3$ to $x_f=11$ on a $L=8$ lattice.
The panels on the left show
$x(t)$, $v(t)$, $\mathfrak{a}(t)$, and $v(x)$ for
$v_{\rm max}=0.2$ and $\mathfrak{a}_{\rm max}=v_{\rm max}/5$,
using
$t_0=9$ and $T=40$, as defined in Eq.~\eqref{eq:xva_softer_rels}. 
The panel on the right shows 
the distribution of the heavy charge between the two nearest positron sites
as a function of time, 
computed via the procedure described in Eq.~\eqref{eq:chargePart}.
}

\label{fig:TRAJv0p2a0p2}
\end{figure}

\noindent
In our treatment, moving heavy-$Q^+$s follow
 a trajectory with
a smooth acceleration up to a uniform velocity, 
followed by a smooth deceleration to rest.
The trajectory is parameterized by the following equations of motion,
\begin{align}
x(t) & \ =
\ \frac{v_{\text{max}}^2}{4 {\mathfrak a}_{\text{max}}}\ 
\log \frac{\cosh \left[ \beta (t-t_0) \right]}{\cosh \left[ \beta (t-t_0-T) \right]} 
\ + \ \frac{x_f + x_0}{2}
\ ,
\nonumber\\[4pt]
v(t) & \ = \  \frac{v_{\text{max}}}{2}\ \left( 
\tanh \left[ \beta (t-t_0) \right] -  \tanh \left[ \beta (t-t_0-T)  \right] 
\right)
\ ,
\nonumber\\[4pt]
{\mathfrak a}(t) & \ = \ {\mathfrak a}_{\text{max}} \  \left( 
\sech^2 \left[ \beta (t-t_0) \right] -  \sech^2 \left[ \beta (t-t_0-T)  \right] 
\right)
\ ,
\label{eq:xva_softer}
\end{align}
where 
\begin{equation}
\beta\ =\ 2\frac{{\mathfrak a}_{\rm max}}{v_{\text{max}}}
\ ,\quad 
t_0\ =\ \left\lfloor \frac{\text{arccosh}[\sqrt{{\mathfrak a}_{\text{max}}/{\mathfrak a}(0)}]}{\beta} \right\rceil
\ ,\quad 
T\ =\ \frac{x_f-x_0}{v_{\text{max}}}
\ .
\label{eq:xva_softer_rels}
\end{equation}
The trajectories are defined by the maximum velocity ($v_{\text{max}}$) and acceleration ($\mathfrak{a}_{\text{max}}$), as well as the initial position
($x_0$) and final position ($x_f$) of the heavy charge. 
The variable $t_0$ is fixed by the initial acceleration ${\mathfrak a}(0)$, which for our numerical studies is set to ${\mathfrak a}(0) = 10^{-4}$, and 
$\lfloor \cdot\cdot\cdot \rceil$ denotes the round function to the nearest integer.
The continuous position $x(t)$ is distributed among the two nearest odd numbered (positron) staggered sites to match the positive heavy charge.
Defining $x_{q1}$ to be the smaller numbered site and $x_{q2}=x_{q1+2}$ to be the larger numbered 
site, the charge is distributed as,
\begin{equation}
Q_{q1} \ = \ 
\frac{Q}{2} \left(x_{q2}-x(t)\right)
\ , \quad 
Q_{q2} \ = \ 
\frac{Q}{2}  \left(x(t)-x_{q1}\right)
    \ ,
\label{eq:chargePart}
\end{equation}
where $Q$ is the value of the heavy charge.
To minimize boundary effects, the initial and final positions of a heavy-$Q^+$ are placed as far as practical from the edges of the lattice, but in such a way to have an extended period of constant velocity. 
An example trajectory is shown in Fig.~\ref{fig:TRAJv0p2a0p2},
where a heavy-$Q^+$ moves
from $x_0=3$ to $x_f=11$ with the constraints that $v_{\rm max}=0.2$ and ${\mathfrak a}_{\rm max}=0.04$.
Throughout this work, we will set ${\mathfrak a}_{\rm max}=v_{\rm max}/5$, 
which we have verified does not lead to an appreciable energy loss due to the 
radiation from an accelerating charge.
%

\subsection{Maximum Lattice Velocity}
\label{sec:DispRels}
\noindent
Features of the dynamical simulations that will be
presented in the following sections 
can be understood from the lattice dispersion relations.
The spectrum of the Schwinger model has been extensively studied in the literature, 
with a focus on the first excitations in the charge $q_{tot}=0$ sector (scalar and vector mesons)~\cite{Hamer:1982mx,Berruto:1997jv,Sriganesh:1999ws,Byrnes:2002nv,Cichy:2012rw,Banuls:2013jaa}.
It is convenient to first consider the theory with $g=0$, 
corresponding to non-interacting electrons and positrons.
The lattice dispersion relation for the electrons (in the charge $q_{tot}=-1$ sector)
resulting from the Hamiltonian given in Eq.~\eqref{eq:Hgf}, 
subject to OBCs, is
\begin{equation}
    E^2 \ = \ m^2 + \sin^2\left( K/2 \right) 
    \ ,\ \ 
    K \ = \ \frac{(\overline{n}+\frac{1}{2}) \pi}{L+\frac{1}{2}}
    \ ,
\label{eq:dispF}
\end{equation}
where $\overline{n}=\{0,1,\ldots, L-1\}$.
These energies are the gap above the vacuum, and
there are more energy levels in the $g=0$ spectrum beyond single-particle states
corresponding to multi-particle excitations.
The form of this (electron) dispersion relation is from, in part, the spatial lattice spacing, relevant for the distance between adjacent lattice sites, being twice the staggered lattice spacing, $a_{\text{staggered}}=1$ and $a_{\text{spatial}}=2$. 
The electron dispersion is relevant for simulations with a moving heavy-$Q^+$, whose charge is neutralized by 
electrons.
The group velocity of electrons is  
\begin{equation}
v \ = \ \frac{dE}{dK} \ = \ \frac{\sin K}{4\sqrt{m^2 + \sin^2 \left (K/2 \right )}} \ ,
\label{eq:dispBv}
\end{equation}
where the limit $L\to\infty$ has been taken in order
to define the derivative.
Unlike in the continuum, there is a maximum group velocity, 
\begin{equation}
v_\star \ = \ \frac{1}{2}\left (\sqrt{m^2+1}-m \right )
    \ ,
\label{eq:Vstar}
\end{equation}
which reduces to the speed of light, $c=\frac{1}{2}$ 
(in spatial lattice units),
in the $m\rightarrow 0$ (continuum) limit. 
This maximum velocity will persist in the interacting theory, with a value that is shifted away from $v_{\star}$.
Therefore, on the lattice, there is a critical velocity of the heavy charge that exceeds the maximum group velocity of the light degrees of freedom. 
This critical velocity is lower than the speed of light, which is 1 staggered site per unit time.
This indicates that particle production can occur 
on the lattice even when the heavy charge is moving at constant velocity
because some or all of the light degrees of freedom are unable to ``keep up'' with the charge for sufficiently high velocity.
Conceptually, the moving heavy charge will separate from the light degrees of freedom that were initially screening it, exposing the vacuum to an electric field, 
which will create hadrons in its wake.
When determining the energy loss of a 
charged particle
in medium at a non-zero lattice spacing, 
the energy loss into the vacuum will be subtracted.

\section{Classical Simulations}
\label{sec:CSims}
\noindent
Classical simulations of a selection of heavy-$Q^+$ trajectories with different ``dense mediums'' are performed.
These simulations determine the state at time $t$, $\vert \psi(t) \rangle$, from 
the ground  state in a particular charge sector,
$\vert \psi(0) \rangle$,
via a Trotterized time evolution associated with the time-dependent Hamiltonian,
\begin{align}
\vert \psi(t) \rangle \ = \ \mathcal{T} \prod_{j=1}^{t/\Delta t}\, e^{- i \, \Delta t \, \hat{H}(j \, \Delta t)} \vert \psi(0) \rangle \ ,
\label{eq:tevol}
\end{align}
where $\mathcal{T}$ denotes the time-ordered product.
It was found that a (minimal) time step of $\Delta t=0.25$ was sufficient for the convergence of the observables considered.\footnote{For small values of $v$, the time step $\Delta t$ can be increased. Explicitly, for $v\leq 0.05$, $\Delta t=2.0$; for $0.05 < v\leq 0.1$, $\Delta t=1.5$; for $0.1 < v\leq 0.2$, $\Delta t=1.0$; for $0.2 < v\leq 0.4$, $\Delta t=0.5$; for $0.4 < v\leq 0.99$, $\Delta t=0.25$.} 
The time-dependent energy, charge distribution and various entanglement measures 
are determined from $\vert \psi(t) \rangle$.

\subsection{A Heavy-\texorpdfstring{$Q^+$}{} Moving Through the Lattice Vacuum}
\label{sec:HQVAC}
\noindent
Lorentz invariance is broken down to discrete translational invariance
in simulations using a spatial lattice.
The lattice dispersion relation
allows processes to occur that  are forbidden in the continuum by energy-momentum conservation, 
such as pair production below threshold.
This leads to {\it increasing energy} in the light degrees of freedom
as a heavy-$Q^+$ moves with constant velocity across the lattice vacuum,
which we connect to the more standard framework of {\it energy-loss} by the moving heavy-$Q^+$.
The workflow that we employ to simulate the dynamics of
a neutralized heavy-$Q^+$ moving across the lattice vacuum is the following:
\begin{enumerate}
    \item 
    Determine the vacuum state, $|\psi_{\rm vac}\rangle$,  
    and low-lying excited states without background charges.
    This defines the vacuum energy ($E_{\rm vac}$), the mass of the hadronic excitations
    in the light sector, the chiral condensate, and other vacuum observables.
    \item 
    Determine the ground state with a neutralized
    heavy-$Q^+$ at rest at site $x_0$, $|\psi_{\rm vac}\rangle_{Q^+_{\{x_0\}}}$. 
    The energy gap above $E_{\rm vac}$ defines the mass of the heavy hadron 
    (analogous to the B-meson), or more precisely the lattice evaluation of $\overline{\Lambda}$ in HQET~\cite{Falk:1992ws,Luke:1993za,Kronfeld:2000gk,Gambino:2017vkx,FermilabLattice:2018est}.
     \item 
    Time evolve the state $|\psi (0)\rangle=|\psi_{\rm vac}\rangle_{Q^+_{\{x_0,v\}}}$ using Eq.~\eqref{eq:tevol}, with the heavy-$Q^+$ trajectory, $x(t)$, defined above in Eqs.~\eqref{eq:xva_softer} and \eqref{eq:chargePart}.
    At each time step, the relevant observables are computed, 
    including the total energy, given in Eq.~\eqref{eq:Etottx}.
\end{enumerate}
It is convenient to define observables as functions of the position of the moving heavy-charge, instead of time. For example, the total energy is
\begin{equation}
E_{Q^+}(x) \ = \ \langle \  \psi[t(x)] \  \vert  \ \hat{H}[t(x)] \  \vert \  \psi[t(x)] \ \rangle_{Q^+_{\{x_0,v\}}} 
\ ,
\label{eq:Etottx}
\end{equation}
where $t(x)$ can be determined from inverting the heavy-$Q^+$ trajectory $x(t)$ from Eq.~\eqref{eq:xva_softer}. 
All the quantities displayed
in the rest of the paper will depend on the position of the heavy-$Q^+$.

\begin{figure}[!ht]
\centering
\includegraphics[width=\textwidth]{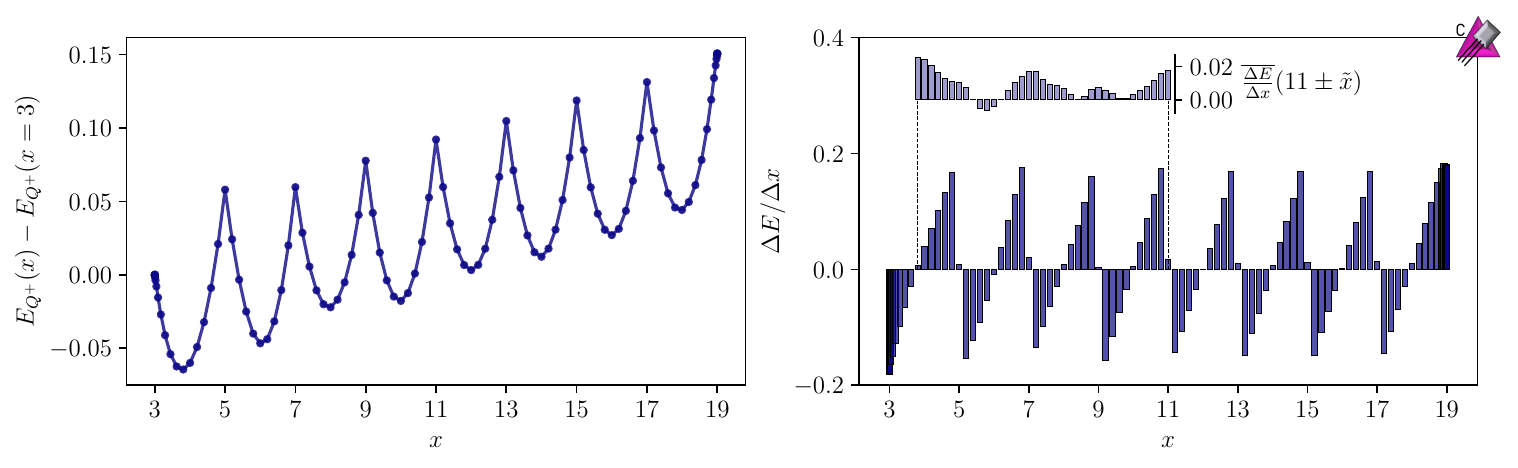}
\caption{
The energy as a function of the position of the heavy-$Q^+$ 
with the initial energy subtracted (left) and rate of energy loss $\Delta E/ \Delta x$, as defined in Eq.~\eqref{eq:dedxFD} (right).
The inset on the right panel shows the symmetrized average of $\Delta E/ \Delta x$ around $x=11$.
This simulation was performed with $L=12$, $m=0.1$, and $g=0.8$.
The heavy-$Q^+$ 
is initially at staggered site $x_0=3$ and moves to $x_f=19$, using a velocity profile with $v_{\rm max}=0.2$.
A purple icon in the upper right indicates that this calculation was performed using a classical computer~\cite{Klco:2019xro}.}
\label{fig:dEdxVAC}
\end{figure}
Figure~\ref{fig:dEdxVAC} shows the change in the total energy,
$E_{Q^+}(x)$
with parameters $m=0.1$ and $g=0.8$, using the same classical trajectory as in Fig.~\ref{fig:TRAJv0p2a0p2}, 
but with $L=12$ and $x_f = 19$.
Also shown is the instantaneous energy-loss $\Delta E/\Delta x$ 
defined by a 
finite-difference approximation to the energy loss 
at position $x$ at time $t$ during a Trotter step of size $\Delta t$,
\begin{equation}
    \frac{dE}{dx}(x) \ \rightarrow \  \frac{\Delta E}{\Delta x} \ = \ 
\frac{E_{Q^+}(t+\Delta t) - E_{Q^+}(t-\Delta t)}{x(t+\Delta t) - x(t-\Delta t)}
    \ .
\label{eq:dedxFD}
\end{equation}
The saw-tooth structure of the energy is due to the staggering of charges.
Energy decreases as the heavy-$Q^+$ moves toward an even-numbered (electron) site and away from an odd-numbered (positron) site, due to the Coulomb interaction.
Similarly, the energy increases as the heavy-$Q^+$ moves away from an electron site and toward a positron site.
This structure likely results from our implementation of motion across the lattice, 
and could be mitigated by smoothing the charge evolution over more than two lattice sites,
and by  decreasing the lattice spacing.
The inset of the right panel shows the sum of the contributions symmetrized around the midpoint of the lattice ($x=11$), $\overline{\frac{\Delta E}{\Delta x}}(11\pm \tilde{x})=\frac{1}{2}[\frac{\Delta E}{\Delta x} (11+\tilde{x})+\frac{\Delta E}{\Delta x} (11-\tilde{x})]$, with $0\leq\tilde{x}\leq 7$.
The strong cancellations between the positive and negative contributions indicates that the net 
$\Delta E / \Delta x$ is small when averaged across  lattice sites, 
much smaller than the magnitude of typical instantaneous values,
but importantly not equal to zero.  
This demonstrates that there is a net energy 
gain per unit length in the light degrees of freedom
as the heavy charge moves across the lattice 
vacuum, corresponding
to the production of light hadrons, and a net energy loss of heavy charge.

Lorentz breaking operators in the Hamiltonian
will, in general, contribute terms that are suppressed by powers of the lattice spacing, 
${\cal O}(a^n)$, with the lowest contribution at
${\cal O}(a^2)$ for the KS Hamiltonian~\cite{Kogut:1974ag,Banks:1975gq}.
Matrix elements of the Lorentz-breaking operators with a moving 
heavy-$Q^+$ are expected to give rise to contributions to observables that scale as 
${\cal O}\left(a^2 v^2\right)$ 
at low velocities
after 
parity considerations and renormalization of the Lorentz-preserving contributions.
As the velocity of the heavy charge approaches the speed of light, $v\rightarrow 1$, 
higher order terms will become increasingly important. 
\begin{figure}[!t]
\centering
\includegraphics[width=\textwidth]{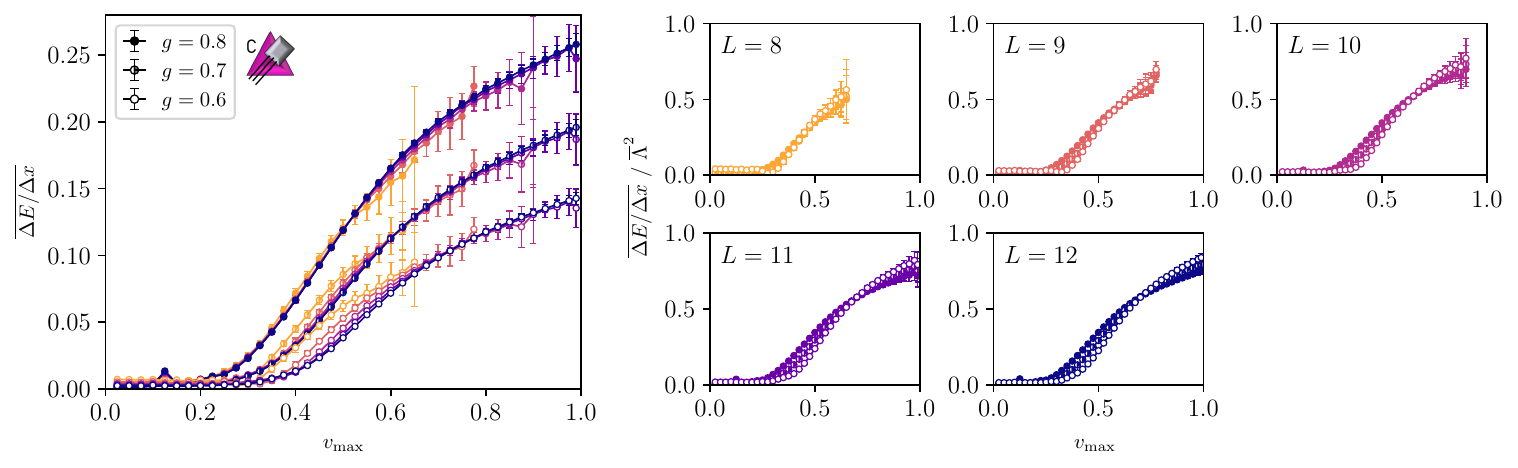}
\caption{
Left:
the lattice-averaged rate of energy loss $\overline{\Delta E/\Delta x}$, for a heavy-hadron
moving from $x_0=3$ to $x_f=2(L-3)+1$ across the vacuum as a function of velocity.
Right: the same as the left panel
but rescaled by the square of the heavy-hadron mass $\overline{\Lambda}^2$.
Results are shown for $L=\{8,\ldots,12\}$ and $g=\{0.8,0.7,0.6\}$ with $m/g=0.125$. 
Decreasing $g$ with $m/g$ fixed effectively decreases the lattice spacing.
}
\label{fig:AVdEdxVsvelocity}
\end{figure}
Figure~\ref{fig:AVdEdxVsvelocity} shows a 
lattice-averaged rate of energy loss, 
$\overline{\Delta E/\Delta x}$, as a function of the velocity of the heavy-$Q^+$.
$\overline{\Delta E/\Delta x}$ is determined by a linear fit to $E_{Q^+}(x)$ 
in the region of constant velocity, 
defined by $v \geq (v_{\rm max}-0.01)$.
Using our trajectories, there is an upper bound on $v_{\text{max}}$ 
for a given simulation volume, leading to incomplete curves for the smaller $L$ in Fig.~\ref{fig:AVdEdxVsvelocity}.
The results are consistent with
expected quadratic dependence on $v$ for low velocity.
The lattice-spacing dependence can also be probed by decreasing the coupling $g$ while keeping $m/g$ fixed (effectively decreasing $a$).
The results in Fig.~\ref{fig:AVdEdxVsvelocity} verify that the energy loss decreases as the lattice spacing decreases. 
This can be made more manifest by forming a dimensionless quantity between two physical quantities, that vanishes in the continuum.
The right panels of Fig.~\ref{fig:AVdEdxVsvelocity}
show $\overline{\Delta E/\Delta x}$ rescaled by the square of the heavy hadron mass, $\overline{\Lambda}^2$.
Keeping the physical heavy hadron mass fixed gives $\overline{\Lambda}\sim a$, 
whereas $\overline{\Delta E/\Delta x}$ is expected to scale at least as $\sim a^3$.
The rescaled energy loss is indeed seen to decrease for smaller lattice spacing up to a $v_{\text{max}} \sim 0.7$,  
where this analysis appears to break down.

\begin{figure}[!t]
\centering
\includegraphics[width=\textwidth]{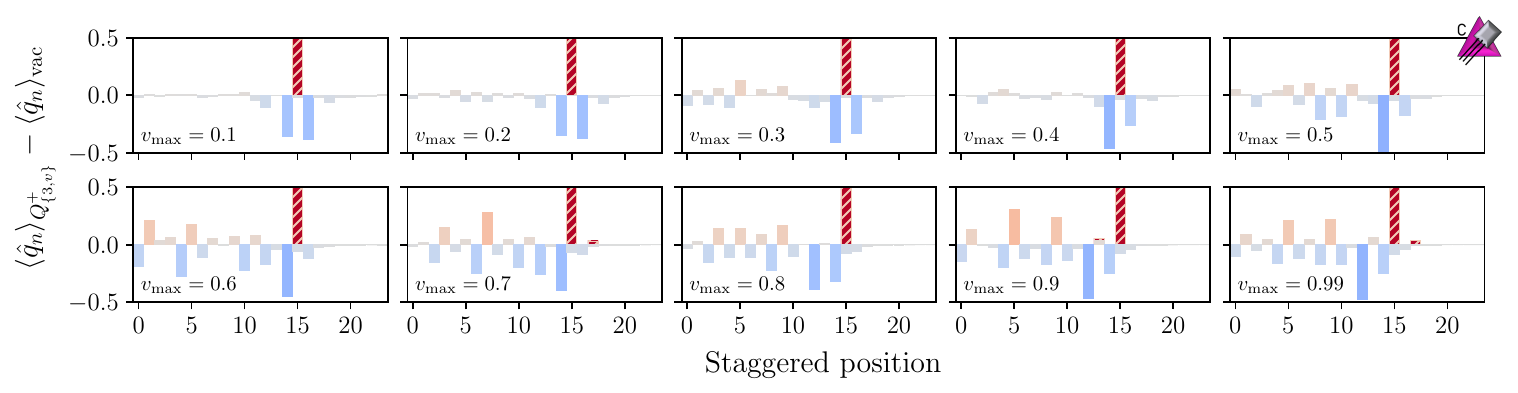}
\caption{
The vacuum subtracted charge density in the light degrees of freedom (solid bars) 
as a function of velocity, 
when the heavy-$Q^+$ (dashed bar) is at (or near) $x=15$.
These simulations were performed with $L=12$, $m=0.1$, and $g=0.8$.
}
\label{fig:QplusVac}
\end{figure}

As the heavy charge moves through the vacuum, the light charges re-arrange to dynamically screen the $Q^+$.
The charge density in the Lorentz-invariant continuum remains localized around the heavy charge, with a symmetric profile that is increasingly Lorentz-contracted with increasing velocity.
On the lattice, this picture changes due to Lorentz symmetry being broken, and particles having a modified dispersion relation. 
Importantly, the light charges have a maximum velocity $v_{\star}$ 
(see Eq.~\eqref{eq:Vstar} for $g=0$) that is less than the speed of light.
These lattice effects are illustrated in Fig.~\ref{fig:QplusVac}, which shows the charge density when the heavy-charge is at $x=15$ for a range of velocities.\footnote{This is not an ideal 
observable because the charge density in the wake of the heavy charge will fluctuate with time.
This is why the charge density behind the heavy charge is unusually small for $v_{\text{max}}=0.4$ for that specific time (compared to $v_{\text{max}}=0.3$ or $v_{\text{max}}=0.5$).}
At $t=0$ in our simulations, there is a symmetric distribution of charges screening the $Q^+$ 
(up to boundary effects).
For $v_{\text{max}} \lesssim 0.3$, 
this screening cloud largely
travels with the heavy charge, reproducing continuum expectations.
However, as $v_{\text{max}}$ becomes comparable to $v_{\star}$, the light charges cannot keep up with the heavy charge.
The light charges are dragged behind the heavy charge, and the profile becomes more asymmetric with increasing velocity.
This asymmetric charge distribution exposes the vacuum to a strong electric field, causing particle (hadron) production in the wake of the moving charge.
This is seen in the fluctuations in the light degrees of freedom on the opposite (left) side of the lattice, where light hadrons have a non-zero probability of being produced during the motion.

\begin{figure}[!t]
\centering
\includegraphics[width=\textwidth]{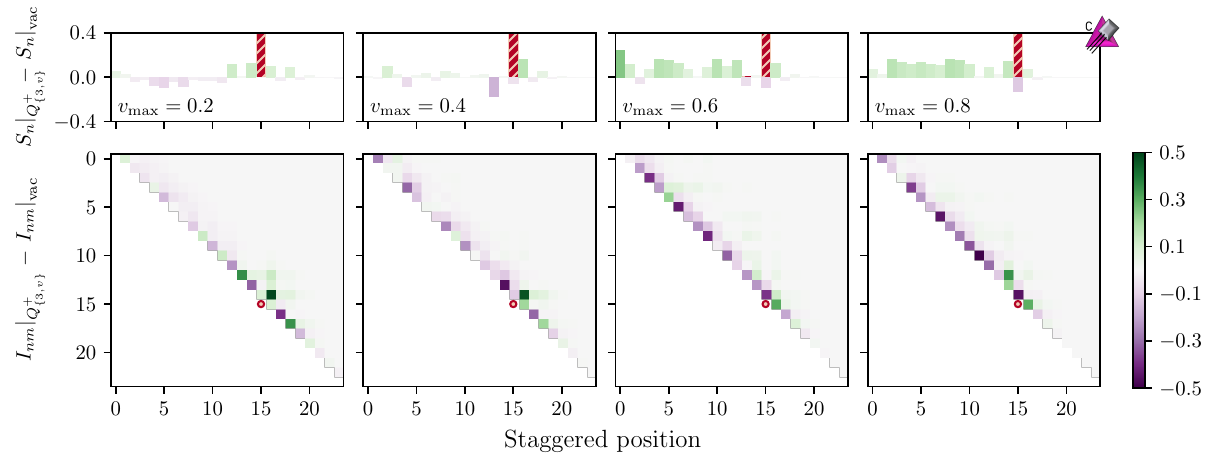}
\caption{
The vacuum subtracted single-site entanglement entropy in the light degrees of freedom (top) and mutual information (bottom) when the moving heavy-$Q^+$ is at $x=15$ (dashed red bar) 
Being a symmetric matrix, only the upper triangular part of the mutual information is shown.
The simulation parameters are the same as those used to generate Fig.~\ref{fig:QplusVac}.
}
\label{fig:QplusSiInm}
\end{figure}

The role of quantum correlations in strongly interacting systems is an area of active research, with pioneering work connecting entanglement to the confinement and chiral phase transitions in QCD~\cite{Beane:2018oxh,Beane:2019loz,Beane:2020wjl,Beane:2021zvo,Liu:2022grf,Miller:2023ujx,Liu:2023bnr}, with parallel works on low-energy nuclear systems~\cite{Robin:2020aeh,Bulgac:2022cjg,Johnson:2022mzk,Gu:2023aoc,Hengstenberg:2023ryt,Perez-Obiol:2023wdz}, high-energy processes~\cite{Kharzeev:2017qzs,Cervera-Lierta:2017tdt,Baker:2017wtt,Kharzeev:2021yyf,Gong:2021bcp,Florio:2024aix}, and quantum field theories~\cite{Srednicki:1993im,Marcovitch:2008sxc,Klco:2020rga,Klco:2021biu,Klco:2021cxq,Klco:2023ojt,Parez:2023uxu,Florio:2023mzk}.
In the continuum, and similar to the charge density,
it is expected that disturbances in the entanglement above the vacuum will be localized around the position of the heavy charge.
However, on the lattice, the production of hadrons in the wake of the $Q^+$ alter the localized entanglement signatures.
The single-site entanglement entropy $S_n=- {\rm Tr}(\rho_n \log_2 \rho_n)$ 
is related to the purity of the reduced density matrix, $\rho_n$, 
on site $n$, 
and is shown in Fig.~\ref{fig:QplusSiInm} 
when the heavy-$Q^+$ is at $x=15$ for a selection of $v_{\text{max}}$ (same situation as in Fig.~\ref{fig:QplusVac}).
For small
$v_{\text{max}}$, the continuum expectation of entanglement entropy localized around the heavy-$Q^+$ is recovered.
For larger $v_{\text{max}}$, 
considerable entanglement entropy is generated in the wake of the moving charge, consistent with a lattice artifact that scales as ${\mathcal O}(a^2 v^2)$ at low velocities.
To investigate correlations between sites, the bottom panels of Fig.~\ref{fig:QplusSiInm} 
show the mutual information $I_{nm}$.
The mutual information between sites $n$ and $m$ is defined as 
$I_{nm}=S_n+S_m-S_{nm}$, where $S_{nm} = - {\rm Tr}(\rho_{nm} \log_2 \rho_{nm})$ 
is the entanglement entropy of the two-site reduced density matrix.
This quantity shows that the correlations produced by the moving charge are 
short range, with a scale naturally set by confinement.
For this particular system, these entanglement measures are providing a qualitative picture that is consistent with the charge densities in Fig.~\ref{fig:QplusVac}.\footnote{The difference in sign between $I_{nm}$ and $S_n$ in Fig.~\ref{fig:QplusSiInm} (especially in the last column) is due to $S_{nm}$ being much larger with the moving charge than in the vacuum.}

\begin{figure}[!t]
\centering
\includegraphics[width=\textwidth]{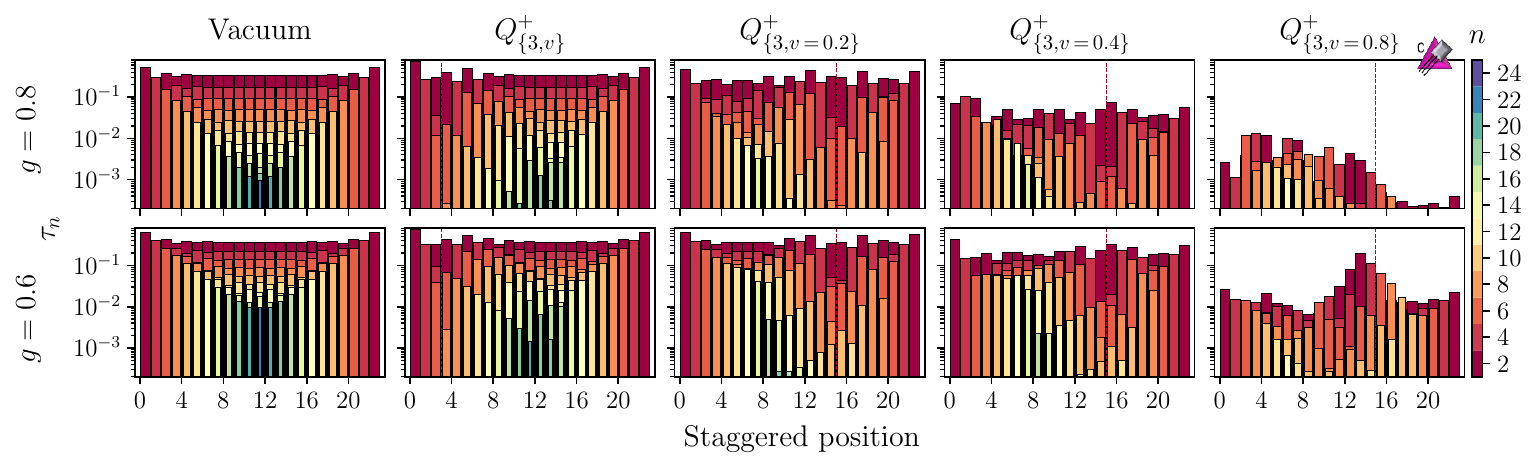}
\caption{
The $n$-tangles for $n=\{2,\ldots,24\}$
associated with the vacuum state (first column), with a moving heavy-$Q^+$ at its starting position $x_0=3$ (second column), and near position $x=15$ (on its way to $x_f = 19$) with velocity $v_{\rm max}=0.2$ (third column), $v_{\rm max}=0.4$ (fourth column), and $v_{\rm max}=0.8$ (fifth column). 
The simulations were performed using $g=0.8$ (top row) and $g=0.6$ (bottom row) with $m/g=0.125$ and $L=12$.
The bars representing $\tau_n(|\psi\rangle)^{i_1,... i_n}$ are centered around the mid $\{i_1,\ldots,i_n\}$ point, with $i_k=i_1+(k-1)$.}
\label{fig:ntangs}
\end{figure}

Quantum correlations beyond two sites can be characterized by
the $n$-tangle~\cite{Wong:2001} of a pure state $\vert \psi \rangle$, defined by
\begin{equation}
\tau_n(|\psi\rangle)^{i_1,... i_n} \ = \
|  \langle \psi |\tilde \psi \rangle  |^2
\ ,\ \ 
|\tilde \psi \rangle \ = \ \hat Y_{i_1}\hat Y_{i_2}\cdots\hat Y_{i_n}
\ |\psi\rangle^*
\ .
\label{eq:ntang}
\end{equation}
The $n$-tangle for odd-$n$ is only defined for $n=3$, and 
vanishes in this system by charge conservation.
Confinement suggests that the $n$-tangles should
fall off exponentially when the number of contributing lattice sites exceeds the confinement length. 
This can be seen in the first column of Fig.~\ref{fig:ntangs}, 
where the non-zero $n$-tangles of the vacuum state $| \psi_{\rm vac}\rangle$ are shown.
The second column of Fig.~\ref{fig:ntangs} shows the $n$-tangles with a heavy-$Q^+$ at its initial position $x_0=3$.
They deviate from those in the vacuum in a local region around the heavy charge, as expected for a system with a finite correlation length.
If there were no lattice discretization effects, deviations from the vacuum $n$-tangles would remain localized around the moving charge.
The lattice effects are illustrated in columns three through five, which show the $n$-tangles when the heavy-charge is at $x=15$ for a selection of velocities.
Relative to the initial state (column two), a moving charge modifies the $n$-tangles across the whole lattice. 
These effects are magnified for larger velocities or coarser lattice spacings, compare $g=0.8$ (top row) and $g=0.6$ (bottom row), as expected for a lattice artifact.
However, these lattice artifacts behave noticeably different than the local observables examined in the previous paragraphs.
The $n$-tangles are more fragile than other observables; with $g=0.8$, the $2$-tangle is diminished by more than $10\times$ for a velocity of $v=0.4$.\footnote{Note that the finite time step in our time evolution, 
$\Delta t$ in Eq.~\eqref{eq:tevol}, 
induces errors on the order of $\sim 10^{-3}$ in the $n$-tangles. 
While decreasing the value of $\Delta t$ modifies the values of the $n$-tangles, it does not change the qualitative features observed.
This is much more severe than, for example, the deviation of the charge density shown in Fig.~\ref{fig:QplusVac}.
In addition, while the deviations in the charge density are restricted to the region behind the moving charge, the $n$-tangle is significantly destroyed across the entire lattice.
These differences are likely because the $n$-tangle is not a local observable 
due to the complex conjugation in Eq.~\eqref{eq:ntang}.}

The suppression of the $n$-tangles is a striking result, compared to the single-site 
and two-site entropy.
A possible explanation for this difference is that the $n$-tangles are not capturing all
of the entanglement in the system 
(e.g., when evaluated in the GHZ and W states, $S_n$ and $I_{mn}$ are non-zero while certain $n$-tangles are zero~\cite{Wong:2001,Illa:2022zgu}). 
The results from these entanglement measures point to 
observables of the system evolving toward those of a classically mixed ensemble,
as predicted for pure state evolution consistent with the 
{\it Eigenstate Thermalization Hypothesis}
(ETH)~\cite{Srednicki:1994mfb} 
(a similar connection was found in Ref.~\cite{Florio:2024aix}).
In order to further understand how the entanglement structure evolves, 
other measures, such as the negativity~\cite{Zyczkowski:1998yd,Vidal:2002zz} or non-stabilizerness entanglement 
(magic)~\cite{Bravyi:2012,Leone:2021rzd,Tirrito:2023fnw}, could be studied.
We remind the reader that velocity dependence (aside from Lorentz contraction) of the observables calculated in this section are lattice artifacts that will vanish in the continuum.

\subsection{A Heavy-\texorpdfstring{$Q^+$}{} Moving Through a Dense Medium}
\label{sec:HQmed}
\noindent
A main objective of this work is to develop machinery for quantum simulations of dynamics in 
strongly-interacting dense matter.
The previous subsection quantified the energy loss and other lattice artifacts that are already present for a heavy charge moving across the vacuum.
This provides a benchmark to compare with the results of in-medium simulations.
Matter is introduced into our simulations by including one or more static heavy-$Q$s, 
whose positions are fixed
in time.
For well separated static charges, the ground state consists of a grid of heavy hadrons at rest.
For tightly packed static charges, the screening clouds merge together, analogous to the electron sea in a metal.
The parameters used in this work give rise to screening 
with high fermion occupation numbers localized over a couple of staggered sites.
Because of this, Pauli blocking plays a significant role in the dynamics.
Combined with the kinematic restrictions of one dimension, 
evolution within the medium leads
to interesting phenomena, 
such as significant distortions to the screening profiles that is 
more pronounced in the leading edge of the collision.
In the continuum, collisions between hadrons 
are inelastic above a given
threshold invariant mass, depending on the hadronic spectrum.
In our simulations, with the hadron velocity fixed to $v_{\text{max}}$ throughout 
the collision, 
hadron production is possible for all kinematics.
The simulations in this section are all performed
with $L=12$ and a relatively low heavy-$Q^+$ velocity of $v_{\text{max}}=0.2$ to minimize lattice artifacts.
The rest of the parameters defining the classical trajectory of the heavy-$Q^+$ are the same as in the vacuum simulations of the previous section.

\subsubsection{A Heavy-\texorpdfstring{$Q^+$}{} Incident upon One Static-\texorpdfstring{$Q^+$}{} }
\label{sec:QonQ}
\noindent
The simplest system to begin studies of energy loss in matter is that of 
a neutralized heavy-$Q^+$ moving past another neutralized heavy-$Q^+$ that is fixed in place.
Initially, we prepare the ground state of the system in the presence of two heavy-$Q^+$s: the moving charge at $x_0=3$ and the static charge at $x=11$. 
The total charge of the light 
sector in the
ground state is $q_{\text{tot}}=-2$.
This initial wavefunction is time-evolved 
using Eq.~\eqref{eq:tevol}, and the energy loss and charge density are compared to the results from the vacuum simulations in the previous section.

\begin{figure}[!t]
\centering
\includegraphics[width=\textwidth]{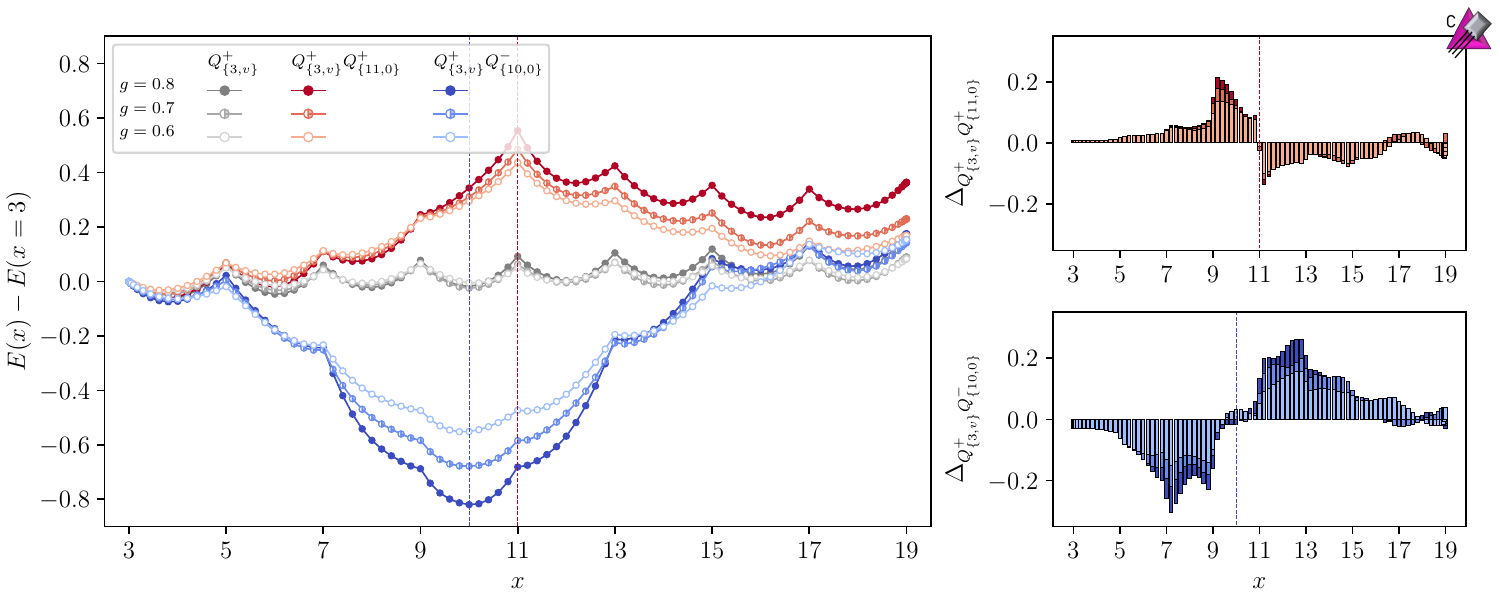}
\caption{
The energy as a function of the position  of the heavy-$Q^+$ with the initial energy subtracted (left) and rate of energy loss, as defined in Eq.~\eqref{eq:DeltaMed} (right).
The heavy-$Q^+$ velocity is $v_{\rm max}=0.2$ and the static heavy-$Q^+$ ($Q^-$) is located at $x=11$ ($x=10$).
These simulations were performed with $L=12$ for  $g=\{0.8,0.7,0.6\}$ and $m/g=0.125$.
The curve labeled as $Q^+_{\{3,v\}}$ denotes a heavy-$Q^+$ initially at $x_0=3$ moving though the vacuum.
The curve denoted by $Q^+_{\{3,v\}} Q^+_{\{10,0\}} $ denotes a heavy-$Q^+$ initially at $x_0=3$ moving past a static heavy-$Q^+$ at $x=10$, and similarly for the curve  $Q^+_{\{3,v\}} Q^-_{\{11,0\}}$.
The dashed red (blue) vertical lines mark the position of the positive (negative) static charge.
}
\label{fig:QplusQplusminus}
\end{figure}

\begin{figure}[!t]
\centering
\includegraphics[width=\textwidth]{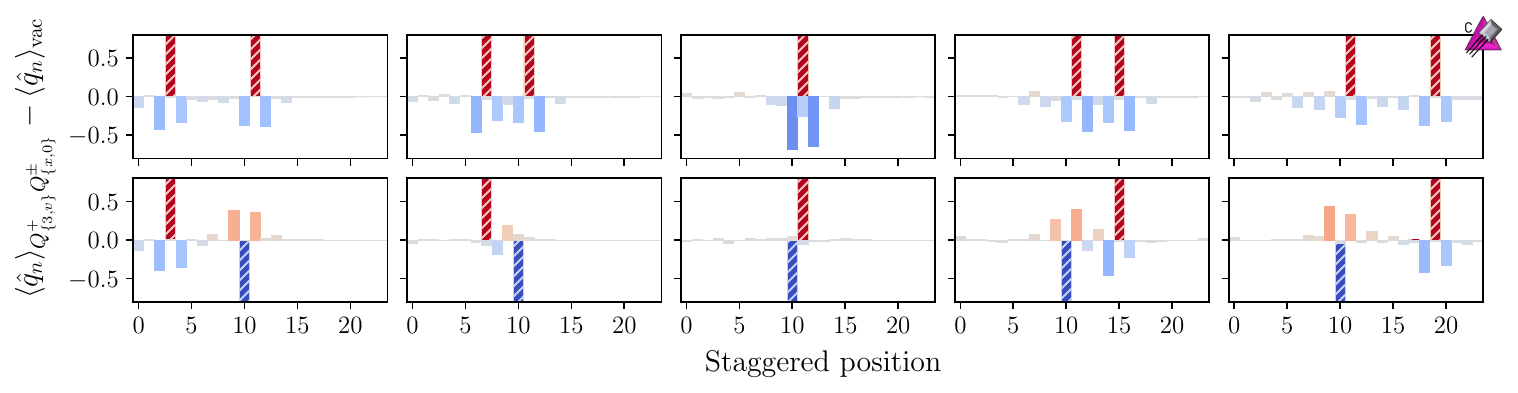}
\caption{
The vacuum subtracted charge distribution 
in the light degrees of freedom (solid bars) shown for when the moving heavy-$Q^+$ is at $x=\{3,7,11,15,19\}$ (left-to-right).
The results in the upper (bottom) panels are for a static heavy-$Q^+$ ($Q^-$) located at $x=11$ ($x=10$).
The dashed bars denote the location of the heavy-charges, and simulation parameters $v_{\text{max}}=0.2$, $m=0.1$, $g=0.8$, and $L=12$ have been used.
}
\label{fig:QplusQplusQminus_charge}
\end{figure}

Figure~\ref{fig:QplusQplusminus} shows the energy loss as a function of the 
position of the moving heavy-$Q^+$ in the presence of the static heavy-$Q^+$ (red points).
To remove some of the lattice artifacts, it is useful to define the vacuum-subtracted quantity $\Delta_{Q^+_{\{3,v\}}Q^\pm_{\{x',0\}}}$,
\begin{equation}
    \Delta_{Q^+_{\{3,v\}}Q^\pm_{\{x',0\}}} \ = \ \left. \frac{\Delta E}{\Delta x} \right|_{Q^+_{\{3,v\}} Q^\pm_{\{x',0\}}}
    - \left. \frac{\Delta E}{\Delta x} \right|_{Q^+_{\{3,v\}}}
    \ ,
    \label{eq:DeltaMed}
\end{equation}
where $x'$ is the staggered site of the static charge.
As expected, the energy loss receives its largest contributions when the screening 
of the two heavy-charges overlap.
The change in energy is more rapid
on the leading edge of the collision than the trailing edge.
This suggests that the initial collision is similar to a violent quench whereas the trailing interactions are occurring 
closer to equilibrium.
Further insight into the mechanisms involved in this process can be gathered from Fig.~\ref{fig:QplusQplusQminus_charge}, which shows the evolution of the charge density.
Particularly striking is the charge density around the static charge after the collision (top-right panel).
Compared to the 
results in vacuum, Fig.~\ref{fig:QplusVac}, the charge density in-medium is more de-localized.
This is indicative of excitations of the static hadron, and of light hadron production in the collision.
Note that the difference in the charge distributions of the static and dynamic charge at $t=0$ (left panel) are due to boundary effects.
%

\subsubsection{A Heavy-\texorpdfstring{$Q^+$}{} Incident upon One Static-\texorpdfstring{$Q^-$}{}}
\label{sec:QonQbar}
\noindent
By placing a static $Q^-$ in the volume instead of a static $Q^+$, 
collisions of a heavy-meson with a heavy-anti-meson can  also be studied.
One difference between the simulations involving two equal charges is that the effects of Pauli blocking are less significant; the electrons surrounding the moving heavy-$Q^+$ are not Pauli blocked by the positrons surrounding the static $Q^-$.
In addition, there is now a particle-antiparticle annihilation channel open during the collision.
The energies during these simulations are shown in Fig.~\ref{fig:QplusQplusminus} (blue points).
As expected, they are seen to be lower for the oppositely charged heavy-$Q$s than for the same
charged heavy-$Q$s.
The right panels show that the energy changes much more rapidly than for the two colliding $Q^+$s, due to the lack of Pauli blocking.
Also, the net change in energy in this process is noticeably less than 
when a  heavy-$Q^+$ passes a static heavy-$Q^+$.

The charge density is shown in the lower panels of Fig.~\ref{fig:QplusQplusQminus_charge}.
For this relatively low velocity,
the screening vanishes when the charges are on top of each other (middle column), because the total net 
heavy charge is zero.
It is interesting to look 
at the charge distribution surrounding the static-$Q^-$ after the moving charge has passed.
When the positive charge is at $x=15$ (fourth column) the charge distribution surrounding the $Q^-$ has a dipole moment pointing to the right.
However, when the positive charge is at $x=19$ (fifth column), the dipole moment is 
pointing to the left.
This suggests that the heavy-$Q^-$ hadron  is left 
in an excited state characterized by a time-dependent dipole moment.
Such excitations have recently been identified in dynamical simulations of nuclei moving through dense neutron matter~\cite{Pecak:2024zqq}.

\subsubsection{A Heavy-\texorpdfstring{$Q^+$}{} Incident upon Two Static-\texorpdfstring{$Q^+$}{}s}
\label{sec:QonQQ}
\noindent
A medium of multiple neutralized static $Q^+$s allows for an exploration of in-medium quantum coherence, beyond those involved in heavy-hadron  collision.
Limited by the sizes of lattice volumes available for classical simulation, we consider two static $Q^+$s located in the middle of the lattice and separated by one or two spatial sites.
Quantum correlations between tightly packed static charges are expected to have a large effect on the energy loss, which will not simply be an incoherent sum of the energy lost to each static charge separately.
\begin{figure}[!t]
\centering
\includegraphics[width=\textwidth]{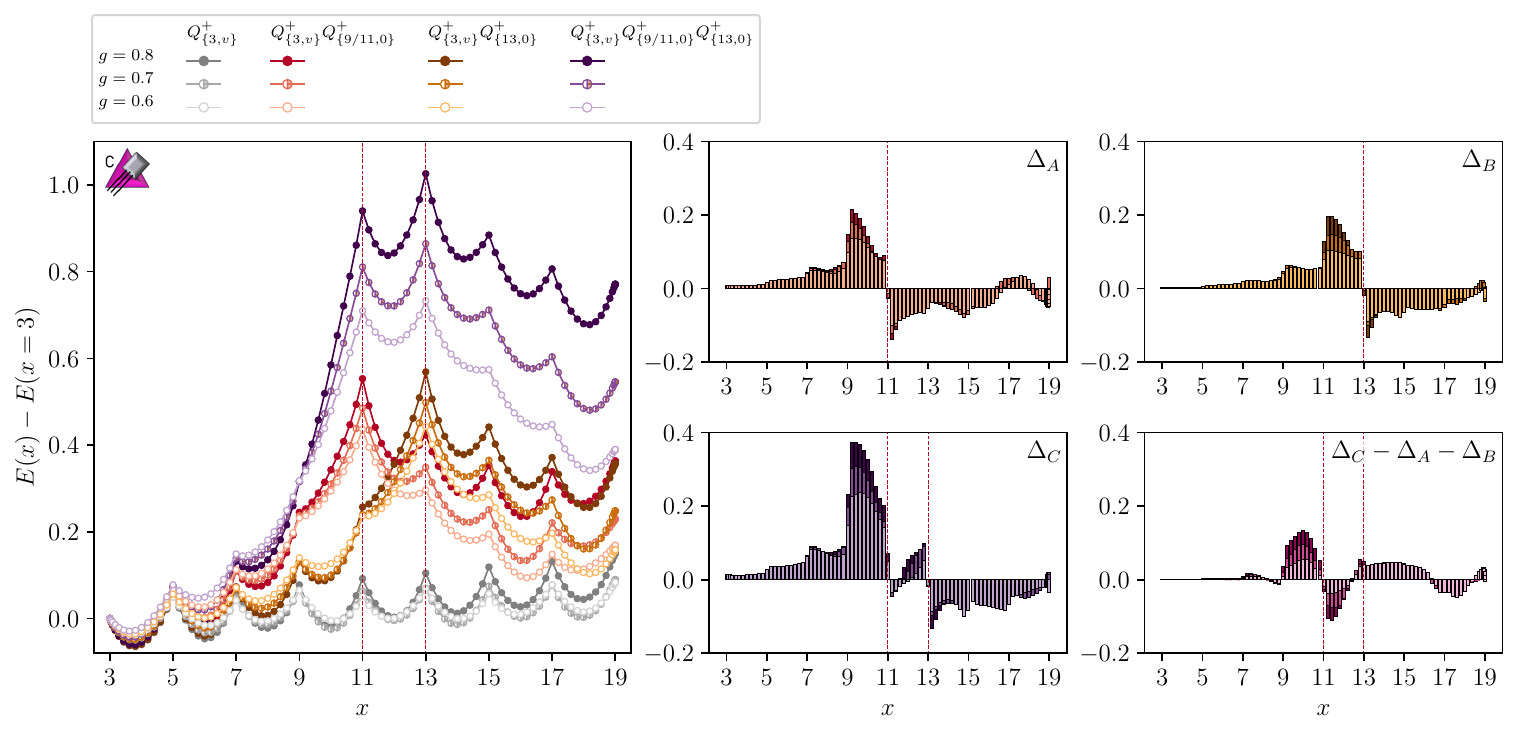}
\includegraphics[width=\textwidth]{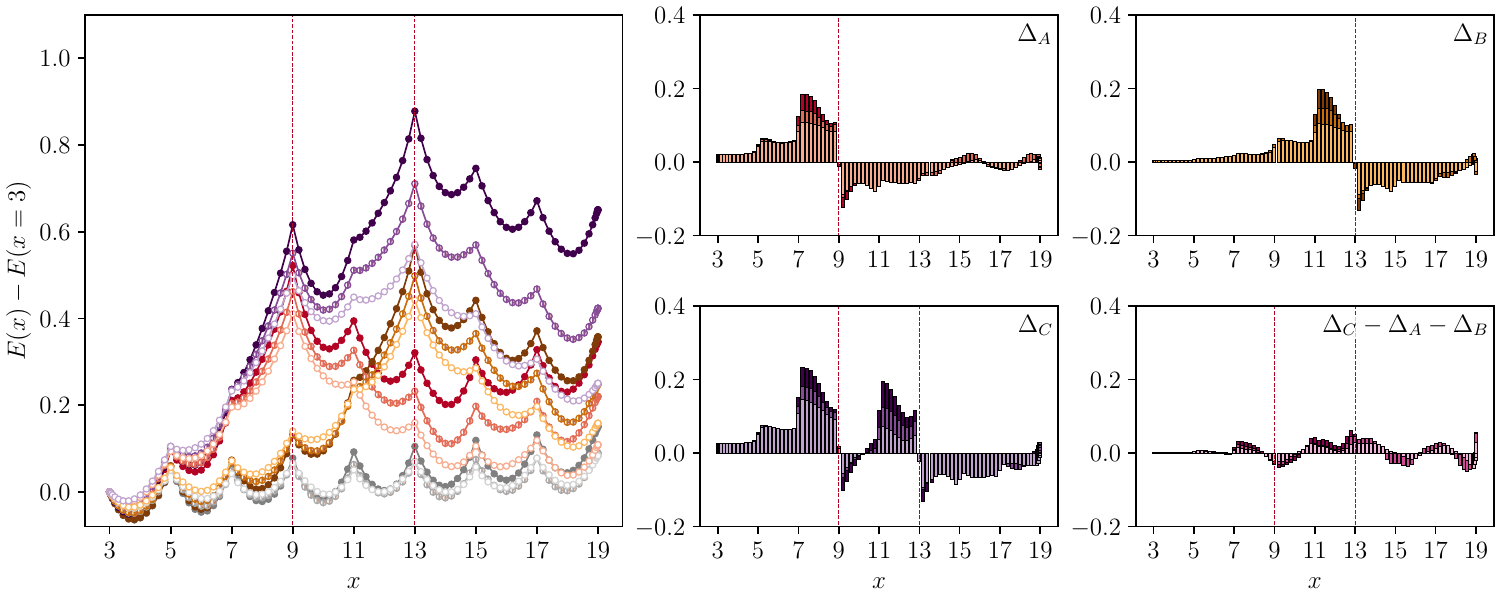}
\caption{
The energies as a function of position of the heavy-$Q^+$ moving 
through two static heavy-$Q^+$s located at $x=\{11,13\}$ (top) and $x=\{9,13\}$ (bottom)
after the initial-state energy has been subtracted.
The left panel shows the total energy, $E(x)$, for the 
$Q^+_v$-$Q^+Q^+$ systems (purple), 
the individual $Q^+_v$-$Q^+$ systems (red and orange), 
and the $Q^+_v$-vacuum system (gray).
The right panels show $\frac{\Delta E}{\Delta x}$ for these same systems
as defined in  Eq.~\eqref{eq:Deltadef}.
The simulations were  performed with $L=12$ for different values of $g=\{0.8,0.7,0.6\}$ and $m/g=0.125$.
The heavy-$Q^+$ started at staggered site $x=3$ and moved to $x=19$, using a velocity profile 
with $v_{\text{max}}=0.2$.
}
\label{fig:QplusONQplusQplus}
\end{figure}
Figure~\ref{fig:QplusONQplusQplus} shows the energy loss as a function of position of the moving heavy-$Q^+$ in the presence of two static-$Q^+$ 
(for separations of one (upper) and two (lower) spatial sites).
To isolate the effects of in-medium quantum coherence, 
it is useful to define the following quantities,
\begin{align}
    \Delta_A & \ = \ \left. \frac{\Delta E}{\Delta x} \right|_{Q^+_{\{3,v\}} Q^+_{\{x',0\}}}
    - \  \left. \frac{\Delta E}{\Delta x} \right|_{Q^+_{\{3,v\}}}
     , \quad \quad \Delta_B \ = \ \left. \frac{\Delta E}{\Delta x} \right|_{Q^+_{\{3,v\}} Q^+_{\{x'',0\}}}
    - \  \left. \frac{\Delta E}{\Delta x} \right|_{Q^+_{\{3,v\}}}  ,
    \nonumber\\
    \Delta_C & \ = \ \left. \frac{\Delta E}{\Delta x} \right|_{Q^+_{\{3,v\}} Q^+_{\{x',0\}} Q^+_{\{x'',0\}}}
    - \  \left. \frac{\Delta E}{\Delta x} \right|_{Q^+_{\{3,v\}}}
     ,
    \label{eq:Deltadef}
\end{align}
where $x', x''$ indicate the lattice sites of the 
static-$Q^+$s, 
and these quantities are shown in the right panels of Fig.~\ref{fig:QplusONQplusQplus}.
The combination $\Delta_C - \Delta_A - \Delta_B$ is a measure of in-medium quantum coherence, and is seen to be more significant for static-$Q^+$s separated by one spatial site compared to two.
Even at low velocities, the energy loss function is sensitive to the increased fermion occupancy and quantum coherence present in dense systems.
As in the case of a single static heavy hadron,
there is a clear asymmetry between the interactions of the leading light degrees of freedom and the trailing ones.


\section{Quantum Simulations}
\label{sec:Qsims}
\noindent
The initial state for the simulations performed in this work is the ground state 
in the presence of   background charges, $\vert \psi_{\text{vac}} \rangle_{Q_{\{x\}}}$. 
Without these background  charges, the total charge of the vacuum is $q_{\text{tot}} = 0$.
In the presence of the charges, the ground state of the system re-arranges in such a way that 
$q_{\text{tot}} +  Q_{\text{tot}} = 0$  for sufficiently large lattices.\footnote{
For a finite lattice size, 
there is a regime of large $m/g$ for which $q_{\text{tot}} = 0$ 
when $Q_{\rm tot}\neq 0$, and the ground state of the system is charged.
For the relatively small $m/g=0.125$ used in this work, the ground state has 
$q_{\text{tot}} +  Q_{\text{tot}} = 0$.}
In this phase, the charges are completely screened over the scale of a confinement length $\xi \sim m_{\text{hadron}}^{-1}$.
Outside of this screening length, the system is locally in the vacuum {\it without} static charges, $\vert \psi_{\text{vac}} \rangle$.
These observations inform an efficient and scalable method for preparing ground states  
in the presence of background  charges on a quantum computer.
A key ingredient in the method for state preparation is the Scalable-Circuit-ADAPT-VQE (SC-ADAPT-VQE) algorithm that is detailed in Sec.~\ref{sec:SCADAPT}.

\subsection{Preparing Ground States  with Background Charges}
\label{sec:StatVac}
\noindent
Consider preparing the ground state  
in the presence of a single positive background charge in the middle of the lattice $Q_{L-1}=+1$.
As argued above, the ground state 
has $q_{\text{tot}}=-1$, with the charge density localized around staggered site $L-1$.
The method that we will use to prepare $\vert \psi_{\text{vac}} \rangle_{Q^+_{L-1}}$ will have two steps:
\begin{itemize}
    \item[1.] Prepare a state $\vert \psi_{\text{init}}\rangle$ which has the qualitative features of $\vert \psi_{\text{vac}} \rangle_{Q^+_{L-1}}$ correct. 
     This state will be $\vert \psi_{\text{vac}} \rangle$ far from the background charge, and possess a local integrated charge of $q_{\text{tot}}=-1$ around the background  charge. 
     $\vert \psi_{\text{init}}\rangle$ is quantitatively correct everywhere except for a few correlation lengths around staggered site $L-1$.
    \item[2.] Modify the wavefunction around staggered site $L-1$. This builds the correct profile of the screening charges, and can be done with circuits that act locally around site $L-1$.
\end{itemize}

To construct $\vert \psi_{\text{init}}\rangle$, first initialize the strong coupling ground state 
in the presence of the background charge,
\begin{align}
\vert \Omega_0\rangle_{Q^+_{L-1}}  \ = \ \frac{1}{\sqrt{2}}\left (\hat{X}_{L-2}\vert \Omega_0\rangle \ + \ \hat{X}_{L}\vert \Omega_0\rangle \right ) \ ,
\end{align}
where $\vert \Omega_0\rangle = \vert 01\rangle^{\otimes L}$ is the strong-coupling vacuum without a background charge.
The $\hat{X}$ operators leads to electron occupation on the staggered sites next to the 
background charge.\footnote{
The state with an electron occupied on site $L-2$ and site $L$ are degenerate with the kinetic term turned off.
By time reversal symmetry, the state can be taken to be a real superposition, and it is found that the equal superposition with a $(+)$ has the lowest energy when the kinetic term is turned on.}
$\vert \Omega_0\rangle_{Q^+_{L-1}}$ has the desired property of charge $(-1)$ localized around the position of the background  charge.
Next, $\vert \psi_{\text{vac}}\rangle$ is prepared far away from the background  charge.
One way to accomplish this is to act with a unitary $\hat{U}^{\text{aVQE}}$, that prepares the vacuum when acting on the strong coupling vacuum, $\hat{U}^{\text{aVQE}}\, \vert \Omega_0 \rangle = \vert \psi_{\text{vac}} \rangle$. 
The problem of determining such a unitary with an efficient circuit implementation was recently addressed by the authors~\cite{Farrell:2023fgd}, 
and is an application of the SC-ADAPT-VQE algorithm outlined in Sec.~\ref{sec:SCADAPT}.
The use of SC-ADAPT-VQE 
to determine $\hat{U}^{\text{aVQE}}$ is reviewed in App.~\ref{app:SCADAPTvac}.
Acting this vacuum preparation unitary on $\vert \Omega_0\rangle_{Q^+_{L-1}}$,
\begin{align}
\vert \psi_{\text{init}}\rangle  \ = \ \hat{U}^{\text{aVQE}}\vert \Omega_0\rangle_{Q^+_{L-1}} \ ,
\label{eq:StatInit}
\end{align}
furnishes 
an initial state with the desired properties of having charge $(-1)$ localized around the 
background  charge and being $\vert\psi_{\text{vac}}\rangle$ away from the position of the 
background  charge.

For step 2, $\vert \psi_{\text{init}}\rangle$ is used as the initial state for another application of SC-ADAPT-VQE. 
The goal of this round of SC-ADAPT-VQE is to determine localized circuits that build the correct wavefunction in the region around the background  charge.
The target state $\vert \psi_{\text{vac}} \rangle_{Q^+_{L-1}}$ is the ground state of the Hamiltonian with a background charge $Q_{L-1}=+1$ and, since both the initial state and target state have $q_{\text{tot}}=-1$, the operators in the pool should conserve charge. 
In addition, as the Hamiltonian is real, the operators are also constrained by time-reversal invariance (operators with an odd number of $\hat{Y}$ in the Pauli string decomposition).
These constraints imply that there are no single-qubit operators, 
and a similar pool to 
that
used for preparing a hadron wavepacket in our previous work~\cite{Farrell:2024fit} is found to be effective,
\begin{align}
\{ \hat{O}  \}_{Q^+_{L-1}} \ &= \  \{ \hat{O}_{mh}(n,d)  \} \ , \nonumber \\
\hat{O}_{mh}(n,d) \ & \equiv \ \frac{i}{4}\left [ \hat{\sigma}^+_{L-1-n}\hat{Z}^{d-1}\hat{\sigma}^-_{L-1-n+d} \,  + \, {\rm h.c.}\ , \ \hat{Z}_{L-1-n}  \right ] \nonumber \\
&= \ \frac{1}{2}\left (\hat{X}_{L-1-n} \hat{Z}^{d-1}\hat{Y}_{L-1-n+d}  -  \hat{Y}_{L-1-n} \hat{Z}^{d-1}\hat{X}_{L-1-n+d} \right ) ,
\label{eq:statPool}
\end{align}
where $n$ measures the staggered distance from the background  charge, with $n \in \{-L+1,-L+2,\ldots,L-1\}$, and $d \in \{1,2,\ldots,N-n-1\}$.
This pool satisfies the desired symmetry constraints, 
as $e^{i \theta \hat{O}_{mh}}$ is real and conserves charge.
The two terms in the RHS of the second line of Eq.~\eqref{eq:statPool} commute, and their exponentials can be converted to circuits without Trotter errors.

The convergence of the SC-ADAPT-VQE prepared ground state  $\vert \psi_{\text{ansatz}} \rangle $ 
to the true ground state  can be quantified with the deviation in the energy of the 
ansatz state $E_{\text{ans}}$ compared to the true  ground state  energy $E_{\text{gs}}$,
\begin{align}
\delta E = \frac{E_{\text{gs}} - E_{\text{ans}}}{E_{\text{gs}}} \ ,
\label{eq:EnergyDeviation}
\end{align}
as well as the infidelity density of the ansatz wavefunction with respect to the exact 
ground state,\footnote{
The average infidelity density is not an optimal metric to use as it asymptotes to 
the infidelity of the vacuum prepared with $\hat{U}^{\text{aVQE}}$ for large system sizes,
i.e., the deviation from the vacuum infidelity density will scale as $1/L$.
A better measure of infidelity would be the overlap of partially-reduced density matrices over a region of the lattice localized about the background charge.
Even with this, requiring a precision exceeding that of the vacuum state is not helpful.
}
\begin{align}
{\cal I}_L =\frac{1}{L}\left (1 - \vert \langle \psi_{\text{ansatz}} \vert  \psi_{\text{vac}} \rangle_{Q^+_{L-1}} \vert^2  \right )\ .
\label{eq:Infidelity}
\end{align}
The deviation in the energy and infidelity obtained from performing SC-ADAPT-VQE for $m=0.1$, $g=0.8$, and $L=\{8,10,12\}$ are given in Table~\ref{tab:AdaptEF}. 
The number of steps used to prepare the ground state  (and its convergence) can be found in App.~\ref{app:SCADAPTvac}.
The sequence of operators and the corresponding variational parameters are given in Table~\ref{tab:AdaptAng}.
It is surprising the first set of operators that are chosen have $d=4$, indicating that correlations of separation 4 are more important than separation 2, which come later in the ansatz.
One explanation is that the initial state has already included some of the short-range correlations (the vacuum prepared with $\hat{U}^{\text{avQE}}$  has $d=1$ and $d=3$ correlations). 
The initial state, $\vert \psi_{\text{init}} \rangle$ in Eq.~\eqref{eq:StatInit} is labelled as step 0 in Table~\ref{tab:AdaptEF}, and already has pretty good overlap with the desired state. 
After 4 steps, a deviation of the energy density of $\delta E \approx 0.012$ is reached which is sufficiently converged for our purposes.
The operators that are chosen, e.g., $\hat{O}_{mh}(3,4)$ and $\hat{O}_{mh}(1,4)$ in steps 1 and 2, and $\hat{O}_{mh}(3,2)$ and $\hat{O}_{mh}(-1,2)$ in steps 3 and 4, are related by a reflection about the position of the background  charge.
The optimal variational parameters are equal with
opposite signs, and the wavefunction that is being 
established has a version of the CP symmetry, but which is broken by boundary effects beyond 4 steps of SC-ADAPT-VQE.
The operator sequence is stable with increasing $L$, and the variational parameters are converging rapidly. 
This indicates that the extrapolation and scaling of these state preparation circuits should be robust.

Due to CP symmetry, the circuits for establishing the vacuum with a negative  background charge at site $L$ are identical to those for a positive charge, but with the variational parameters negated.
This technique can be generalized to prepare the ground state  in the presence of multiple background  charges located within the simulation volume.
As long as background  charges are well separated from each other and the boundaries, then the circuits (presented in the following section) simply need to be repeated around the location of each additional background  charge.
\begin{table}[!t]
\renewcommand{\arraystretch}{1.4}
\resizebox{\textwidth}{!}{\begin{tabularx}{1.2\textwidth}{|c || Y | Y | Y | Y | Y || Y |  Y| Y | Y | Y |}
 \hline
  & \multicolumn{5}{c||}{$\delta E $ } & \multicolumn{5}{c|}{${\cal I}_L $} \\
  \hline
 \diagbox[height=23pt]{$L$}{\text{step}} & 0 & 1 & 2 & 3 & 4 & 0 & 1 & 2 & 3 & 4 \\
 \hline\hline
 8 & 0.0412 & 0.0356 & 0.0300 & 0.0242 & 0.0184 & 0.0164 & 0.0136 & 0.0109 & 0.0085 & 0.0062 \\
 \hline
 10 & 0.0317 & 0.0275 & 0.0232 & 0.0188 & 0.0145 & 0.0131 & 0.0109 & 0.0087 & 0.0069 & 0.0050 \\
 \hline
 12 & 0.0259 & 0.0226 & 0.0192 & 0.0157 & 0.0122& 0.0110 & 
0.0092 & 0.0073 & 0.0058 & 0.0043 \\
 \hline
\end{tabularx}}
\caption{
The deviation in the energy $\delta E$, and infidelity density ${\cal I}_L$, of the SC-ADAPT-VQE prepared ground state  in the presence of a background  charge $\hat{Q}_{L-1} = +1$.
Results are shown through 4 steps of SC-ADAPT-VQE for $L=8,10,12$ and $m=0.1,g=0.8$. 
Step 0 corresponds to the initial state $\vert \psi_{\text{init}}\rangle$, defined in Eq.~\eqref{eq:StatInit}.}
 \label{tab:AdaptEF}
\end{table}
\begin{table}[!t]
\renewcommand{\arraystretch}{1.4}
\begin{tabularx}{0.8\textwidth}{|c || Y | Y | Y | Y |}
  \hline
 \diagbox[height=23pt]{$L$}{$\theta_i$} & $\hat{O}_{mh}(3,4)$ & $\hat{O}_{mh}(1,4)$ &  $\hat{O}_{mh}(3,2)$ &  $\hat{O}_{mh}(-1,2)$  \\
 \hline\hline
 8 & 0.1375 & -0.1375 & -0.1409 &  0.1409 \\
 \hline
 10 &  0.1367 & -0.1367  & -0.1404 & 0.1404 \\
 \hline
 12 & 0.1363  &  -0.1363  &  -0.1400 & 0.1400 \\
 \hline
\end{tabularx}
\caption{
The order of the operators $\hat{O}$ and variational parameters $\theta_i$ used to prepare the SC-ADAPT-VQE ground state  in the presence of a background  charge $\hat{Q}_{L-1} = +1$.
Results are shown through 4 steps of SC-ADAPT-VQE for $L=8,10,12$ and $m=0.1,g=0.8$. }
 \label{tab:AdaptAng}
\end{table}
%

\subsection{Quantum Circuits and Resource requirements}
\label{sec:Qcirqs}
\begin{figure}[!t]
    \centering
    \includegraphics[width=0.7\textwidth]{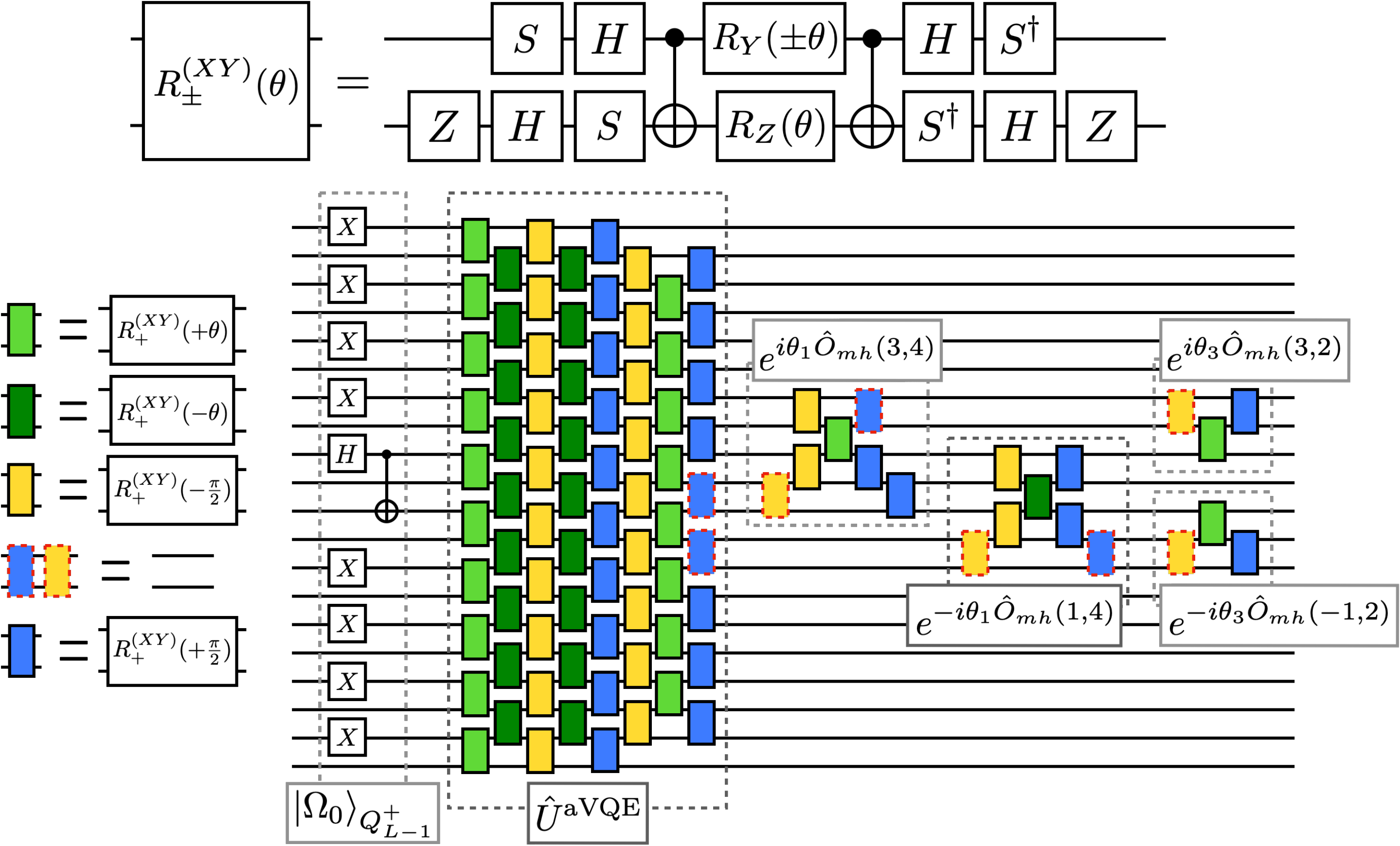}
    \caption{
    Circuits that prepare $\vert \psi_{\text{vac}}\rangle_{Q^+_{L-1}}$ with $L=10$ and a 
    background  charge at $Q_{L-1}=+1$. Initially, the strong coupling vacuum in the presence of the background charge $\vert \Omega_0 \rangle_{Q^+_{L-1}}$ is prepared. Next, the circuits that prepare the two-step SC-ADAPT-VQE vacuum without background  charges $\hat{U}^{\text{aVQE}}$, is applied. 
    Lastly, the four-step SC-ADAPT-VQE circuits for modifying the wavefunction around the position of the background  charge are applied.
    The parameters $\theta_i$ for $\hat{U}^{\text{aVQE}}$ are 
    given
    in App.~\ref{app:SCADAPTvac} and the angles
    for the second application of SC-ADAPT-VQE are given in Table~\ref{tab:AdaptAng}.
    The colored rectangles defining the circuits are defined in the left and top of the figure, with $R_{\pm}^{(XY)}(\theta) \equiv e^{-i \frac{\theta}{2}\left (\hat{Y}\hat{X} \pm \hat{X} \hat{Y} \right )}$. 
    }
    \label{fig:VacPrepCirc}
\end{figure}

\noindent
In the previous section, a sequence of 
unitary operations
that prepares the ground state  in the presence of a single background  charge was presented.
In order to perform simulations on a quantum computer, these 
unitary operations  
must be converted to a sequence of gates. 
Our circuit design is tailored toward  devices with linear nearest-neighbor connectivity, such as is native on IBM's quantum computers~\cite{ibmquantum},
and aims to minimize the circuit depth and two-qubit gate count.
The 
unitary operators forming
the operator pool in Eq.~\eqref{eq:statPool} are of the form $e^{i \theta\left ( \hat{Y}\hat{Z}^{d-1} \hat{X} - \hat{X}\hat{Z}^{d-1} \hat{Y}\right )}$,\footnote{The convention is that the operator on the far left acts on the lower numbered qubit, e.g. $\hat{Y}\hat{X} = \hat{Y}_n \hat{X}_{n+1}$.} 
and we will use the circuit  design introduced in our recent work~\cite{Farrell:2023fgd,Farrell:2024fit} that extends the techniques in Ref.~\cite{Algaba:2023enr}.
These circuits have an ``X'' shape, and are arranged in such a way to cancel the maximum number of CNOT gates.
An example of the circuit that prepares the ground state  in the presence of a background
charge $Q_{L-1}=+1$ for $L=10$ is shown in Fig.~\ref{fig:VacPrepCirc}.
This circuit has been decomposed into three parts. 
First, the strong coupling vacuum in the presence of the heavy charge $\vert \Omega_0\rangle_{Q^+_{L-1}}$ is prepared.
Next, the circuits that prepare the two step SC-ADAPT-VQE vacuum without a background charge are applied. 
These circuits are collectively denoted as $\hat{U}^{\text{aVQE}}$, and were treated in detail in Ref.~\cite{Farrell:2023fgd}.
Lastly, the circuits that implement the four step SC-ADAPT-VQE unitaries $e^{i \theta_i \hat{O}_i}$ in Sec.~\ref{sec:StatVac} are applied.
These circuits are localized and only modify the wavefunction around the position of the 
background charge.

As discussed in the previous section, preparing the ground state  in the presence of multiple 
background  charges is a straightforward extension of these circuits, provided the charges are well separated from each other and the boundaries. 
The first modification is to the preparation of $\vert \Omega_0\rangle_Q$: there is a hadamard-CNOT sequence centered around each heavy charge.
The second change is that the $e^{i \theta \hat{O}}$ are repeated around the center of each additional background  charge.
In total, the resources required for this state preparation are
\begin{align}
\text{\# of CNOTs = } 16L-12+25N_Q \ , \quad \text{CNOT depth = }35 \ ,
\end{align}
where $N_Q$ is the number of background  charges.
This circuit depth is well within the capabilities of current devices.
Note that the number of SC-ADAPT-VQE steps to maintain a constant quality of the prepared state will scale linearly with the confinement length $\xi$.
The circuit depth for each step of SC-ADAPT-VQE also scales with $\xi$: as ${\cal O}(\xi^2)$ for devices with nearest-neighbor connectivity and ${\cal O}(\xi)$ for devices with all-to-all connectivity.
As a result, for devices with nearest-neighbor connectivity, the circuit depth is expected to scale as ${\cal O}(\xi^3)$ for state preparation.

Once the initial state is prepared, time evolution can be implemented with the time-dependent Hamiltonian defined by the classical trajectory of the heavy charge, $Q(t)$ in Eq.~\eqref{eq:xva_softer}.
As shown in our previous work, time evolution in systems without background  charges can be reproduced up to exponentially small errors using a truncated electric interaction~\cite{Farrell:2024fit}. 
In Sec.~\ref{sec:SM}, it was argued that a similar procedure should also be possible for systems with heavy charges, provided the charge operators are suitably averaged over the extent of the heavy hadrons.
To get an estimate of the scaling, we assume a truncation of interactions between spatial charges separated by more than $\overline{\lambda} \approx \xi/2$  spatial sites. 
The resources required for one second-order Trotter step of time evolution can be estimated using the circuits in Ref.~\cite{Farrell:2024fit}, with a CNOT gate count of, 
\begin{equation}
\text{\# of CNOTs = }4(2L-1)+(2L-4\overline{\lambda})(\overline{\lambda}+1)(2\overline{\lambda}+1) -(L-2\overline{\lambda}+2) \ .
\end{equation}
Taking $\overline{\lambda} \sim \xi$, this gives a scaling of ${\mathcal O}(L \, \xi^2)$ for the number of two-qubit gates and a corresponding circuit depth of $\mathcal{O}(\xi^2)$.
A proper determination of the minimum $\overline{\lambda}$ required to reach a predetermined error threshold is left for future work.

To approach the continuum limit, $\xi$ is held fixed in physical units, while the lattice spacing is decreased, i.e., $\xi \sim a^{-1}$.
The number of Trotter steps must also grow as $\xi \sim a^{-1}$ 
as can be seen with the following argument. 
Trotterization of the kinetic term with a brickwork ordering only allows for correlations to spread two staggered sites per Trotter step.
Keeping the lattice volume traversed by the moving charge fixed in physical units implies that the number of staggered sites traversed scales as ${\mathcal O}(a^{-1})$.\footnote{It is possible that there is a Trotter ordering, different from brickwork, 
that improves this scaling.}
Therefore, the number of Trotter steps 
also scales as ${\mathcal O}(a^{-1})$ and time evolution is estimated to have a circuit depth that scales as ${\mathcal O}(a^{-3})$.
This is the same scaling as for the initial state preparation, giving a total circuit depth for simulating dynamics in dense matter in the Schwinger model to be ${\cal O}(a^{-3})$.
This depth would improve to ${\cal O}(a^{-2})$ on devices with all-to-all connectivity.
Of course, actual simulations that approach the continuum will need to be performed to validate these scaling arguments.

\section{Summary and Outlook}
\noindent
The mechanisms responsible for energy-loss and transport in dense matter are key to understanding the evolution of
matter under extreme conditions: from high-energy collisions of large nuclei, to high-energy cosmic-ray penetrating ordinary matter, to the dynamics of core-collapse supernova.
There has been a long history of successfully using
classical techniques, such as Monte-Carlo simulation, to determine the 
electromagnetic responses when charged particles move through matter.
In contrast, 
the dynamics of high-energy quarks and gluons in dense matter is much less understood, in part due to the non-perturbative phenomena of confinement and hadronization.
With an eye toward understanding such processes in QCD,
we have performed real-time simulations of energy-loss and hadronization in 
the simpler setting of the Schwinger model at finite density.
In particular, we have performed classical simulations of 
heavy-hadrons moving through regions of dense matter characterized by static heavy-hadrons.
These simulations have provided insight into internal excitations of hadrons, and the crucial role of quantum coherence between the particles that make up the dense medium.
The effects of quantum coherence between the constituents of matter are visible 
in the energy-loss as a function of incident velocity
in the highest density systems we have prepared.  
By subtracting the individual contributions, the remaining energy loss is attributed 
to quantum correlations  in the matter wavefunction with increasing density.
Further, we have provided scalable quantum circuits for preparing ground states with a finite density of heavy hadrons.
In combination with the time evolution circuits presented in our previous work~\cite{Farrell:2024fit}, we estimated the circuit depths required for large-scale quantum simulations of energy loss in the 
Schwinger model.
The outlook looks promising, and simulations of the dynamics of dense matter in the Schwinger model will be possible in the near-term.

While it is no surprise, 
present-day simulations are significantly effected by relatively large lattice spacings,
restricted by the number of qubits (or qudits) that can be assembled into a quantum register 
to form a spatial lattice volume
that is large enough to contain more than a few confinement length scales.
The hadronic wavefunctions
have support only over a few lattice sites, rendering an obvious discretization of their wavefunctions, with large lattice spacing artifacts.  
One consequence is that when hadrons pass each other, there are relatively large differences between the incident and outgoing fields
at the lattice spacing scale.
In addition, the dispersion relation is such that the velocity of momentum modes has a maximum that is less than the speed of light. 
This occurs around the scale of the inverse lattice spacing 
(depending on mass and electric charge), and causes high-velocity hadrons 
to partially disintegrate as they move, even in the vacuum,
leaving behind a wake of low-energy hadronic excitations. 
This corresponds to fragmentation at fixed velocity, and is entirely a lattice spacing artifact.
These effects are mitigated by forming differences between propagation in matter and vacuum, but nonetheless present an unwelcome background from which to extract the physical fragmentation and hadronization.
These differences have a well-defined continuum limit, reflecting the target physics observables, 
and quantum simulations using multiple lattice spacings, tuned to known physics observables, 
are required
in order to make robust predictions with a complete quantification of uncertainties. 
An important result to highlight is that while lattice 
discretization effects are seen (and understood) in the energy loss of a single heavy-hadron moving through the vacuum, new effects are seen in the modification of the entanglement structure. This is pointing in the direction that quantum correlations are more sensitive to lattice artifacts than classical correlations.

On top of the discussion in the previous paragraph, the impact of lattice-spacing artifacts in high-energy processes cannot be under-estimated.  
One concern is that when colliding high-energy wavepackets together, the non-zero lattice spacing will induce scattering and fragmentation through the modified dispersion relation and beyond.
Care must be taken in such quantum simulations to ensure that the observed inelasticities are coming from physics, and not from the underlying lattice upon which the simulation is being performed.
Alternative formulations to Kogut-Susskind  where discretization errors are suppressed, 
such as improved-KS~\cite{Carlsson:2001wp,Carena:2022kpg,Dempsey:2022nys,Gustafson:2023aai} or improved-Wilson~\cite{Zache:2018jbt,Mathis:2020fuo,Mazzola:2021hma,Hayata:2023skf} 
Hamiltonians, are starting to be pursued. More development is required, leveraging knowledge from classical Euclidean lattice QCD
calculations.

A limitation of working in one spatial dimension is that peripheral collisions are not possible, and all collisions between the constituent electrons and positrons (partons)
that make up the hadrons are ``head-on''.  
In addition, there are no soft momentum-transfer processes, like bremsstrahlung radiation, due to the absence of dynamical gauge fields and the finite spatial extent of the lattice.
More realistic simulations of QCD will require advancing from a $U(1)$ to a $SU(3)$ lattice gauge theory, first in 1+1D and then in higher dimensions.
Development of  these more realistic simulations are underway, and will enable a study of the explicit role of 
non-Abelian color charges in the dynamics of dense matter.

\clearpage

\begin{subappendices}

\section{Spin Hamiltonian with External Charges}
\label{app:hamZ}
\noindent
The electric part of the Hamiltonian in Eq.~\eqref{eq:Hgf1} can be expanded as,
\begin{equation}
    \frac{2}{g^2}\hat{H}_{el} = \sum_{j=0}^{2L-2}\bigg (\sum_{k\leq j} \hat q_k +Q_k\bigg )^2 = \sum_{j=0}^{2L-2}\bigg (\sum_{k\leq j} \hat q_k \bigg )^2 + 2 \sum_{j=0}^{2L-2} \bigg( \sum_{k\leq j}  \hat q_k   \bigg) \bigg( \sum_{l\leq j}  Q_l   \bigg) + \sum_{j=0}^{2L-2}\ \bigg( \sum_{k\leq j}  Q_k   \bigg)^2 \ ,
    \label{eq:backQ}
\end{equation}
The first term is unaffected by the presence of external charges and is given by,
\begin{equation}
\sum_{j=0}^{2L-2}\bigg (\sum_{k\leq j} \hat q_k \bigg )^2 = \frac{L^2}{2}+\frac{1}{4}\sum_{j=0}^{2L-2}\left(2L-j-\frac{1}{2}[1+(-1)^{j+1}]\right)\hat{Z}_j+\sum_{j=0}^{2L-3}\sum_{k=j+1}^{2L-2}\frac{2L-1-k}{2}\hat{Z}_j\hat{Z}_k \ .
\end{equation}
The terms that couple to the external charge are 
\begin{align}
2 \sum_{j=0}^{2L-2} \bigg( \sum_{k\leq j}  \hat q_k   \bigg) 
&\bigg( \sum_{l\leq j}  Q_l   \bigg)
\ +\ 
\sum_{j=0}^{2L-2}\ \bigg( \sum_{k\leq j}  Q_k   \bigg)^2 \nonumber \\
& = \sum_{j=0}^{2L-2}\left[ \bigg( \sum_{k\leq j}  Q_k   \bigg)^2 - \bigg( \sum_{l\leq j}  (-1)^l   \bigg)\bigg( \sum_{k\leq j}  Q_k   \bigg)\right] - \sum_{j=0}^{2L-2}\bigg(\sum_{l=j}^{2L-2}\sum_{m\leq l}Q_m\bigg) \hat{Z}_j \ ,
\end{align}
and contains terms proportional to the identity
and single $\hat{Z}_j$.

\section{SC-ADAPT-VQE for Preparation of the Vacuum without Static Charges}
\label{app:SCADAPTvac}
\noindent
This appendix provides an overview of
the use of SC-ADAPT-VQE to prepare the vacuum without background charges.
The operator pool is given in Eq.~\ref{eq:poolComm} and for $m=0.1,g=0.8$, it was found that  two steps of SC-ADAPT-VQE, was sufficient to achieve an infidelity density of ${\cal I}_L \approx 0.004$ and a deviation in the energy density of $\delta E \approx 0.006$.
These quantities are defined in Eq.~\eqref{eq:Infidelity} and Eq.~\eqref{eq:EnergyDeviation}, respectively.
The two-step vacuum preparation is used as a proof of principle for this work, and defines the vacuum state preparation unitary $\hat{U}^{\text{aVQE}}$ used in Sec.~\ref{sec:StatVac} to prepare the vacuum with background charges.
The operator sequencing and corresponding variational parameters for $m=0.1,g=0.8$, and several different $L$, are given in Table~\ref{tab:AdaptAngNoStat}.

\begin{table}[!t]
\renewcommand{\arraystretch}{1.4}
\begin{tabularx}{0.5\textwidth}{|c || Y | Y |}
  \hline
 \diagbox[height=23pt]{$L$}{$\theta_i$} & $\hat{O}_{mh}^V(1)$ & $\hat{O}_{mh}^V(3)$ \\
 \hline\hline
 10 & 0.3902 &  -0.0676  \\
 \hline
 11 & 0.3896 &  -0.0671  \\
 \hline
 12 & 0.3892 &  -0.0667  \\
 \hline
 13 & 0.3888 &  -0.0664  \\
 \hline
 14 & 0.3884 &  -0.0662  \\
 \hline
 \hline
 $\infty$ & 0.387 & -0.065 \\
 \hline
\end{tabularx}
\caption{
The order of the operators $\hat{O}$ and variational parameters $\theta_i$ used to prepare the two-step SC-ADAPT-VQE vacuum without background  charges.
Shown are results for $L=10-14$ and $m=0.1,g=0.8$. 
The extrapolation to $L=\infty$ is 
performed using the results from 
$L=11-14$, assuming an exponential dependence on $L$; 
see Appendix E of Ref.~\cite{Farrell:2023fgd} for details.}
 \label{tab:AdaptAngNoStat}
\end{table}

\end{subappendices}

\chapter{Brief Reflections on Quantum Simulation}
\noindent
It is a very exciting time to be involved in the field of quantum simulation and quantum information more generally.
There is a lot that is unexplored, and a motivated student can reach the frontier and begin doing research without a lot of overhead. 
I first started really thinking about quantum simulation over Thanksgiving break 2021, initially struggling to reproduce results from 4-qubit classical simulations of the Schwinger model in Ref.~\cite{Klco:2018kyo}.
Persistence payed off, and by January 2022 I was beginning to work toward quantum simulations of $1+1$D QCD.
Due to wonderful collaborators, we were able to make swift progress, and on July 4 we posted the first paper on quantum simulations of $1+1$D QCD to arXiv~\cite{Farrell:2022wyt}.
Three days later, researchers from Canada posted the second paper on quantum simulations of $1+1$D QCD to arXiv~\cite{Atas:2022dqm}.
The takeaway is that progress in this field happens quickly and that, with good mentors and collaborators, even a novice can contribute to the research community if they work hard. 

I have been astounded by the rapid growth of the capabilities of quantum computers over the last two years, and am very optimistic for the future of digital quantum simulation.
In Summer 2022 we were running circuits on 6 qubits with 34 two-qubit gates. 
By Fall 2022 we were running circuits with 17 qubits and 212 two-qubit gates. 
In Summer 2023 we reached 100 qubits and were able to extract sensible results from circuits with 2,134 two-qubit gates.
And in Winter 2023 we increased this to 112 qubits and 13,858 two-qubit gates.
There are no signs that the capabilities of quantum hardware have reached a ceiling, and I am looking forward to see what the quantum simulation community will achieve in the next 3-5 years.
I believe that we are on the verge of performing quantum simulations that are beyond the capabilities of even the best approximate algorithms run on supercomputers, and that will provide qualitative insight into dynamical processes relevant to nuclear and particle physics.
The amazing thing is that it is still not clear which quantum computing architecture will offer the greatest utility moving forward.
Table~\ref{tab:specs} has been included for posterity, and is a current snapshot of the state-of-the-art.
\begin{table}[!t]
\renewcommand{\arraystretch}{1.4}
\resizebox{\textwidth}{!}{\begin{tabularx}{1.1\textwidth}{||c | Y | Y | Y | Y | Y ||}
  \hline
 Device & \# of qubits & Connectivity & Gate fidelity & Gate duration & Coherence time \\
 \hline\hline
 IBM {\tt ibm\_fez} & 156 & heavy-hex & 99.72 & 68ns & 127$\mu$s \\
 \hline
 Google Sycamore & 105 & square-grid & $\sim$ 99.7 & 42ns & 89$\mu$s \\
 \hline
 Quantinuum {\tt H-2} & 56 & all-to-all & 99.87 & 70ms & minutes \\
 \hline
 IonQ {\tt Forte} & 35 & all-to-all & 99.6 & $\sim$1ms & seconds \\
 \hline
 Lukin  
 neutral atom & $>$250 & all-to-all & $\sim$ 99.5 & $\sim$3ms & seconds \\
 \hline
\end{tabularx}}
\caption{The specifications for the current (August 2024) state-of-the-art quantum computers~\cite{queraweb,quantinuumweb,ibmweb,ionqweb,Bluvstein:2023zmt,Acharya:2024btg}.
Some of the specifications have been estimated from the publicly available information.
}
 \label{tab:specs}
\end{table}

\printendnotes

%
%
\bibliographystyle{unsrt}
\bibliography{uwthesis}

\begin{thebibliography}{100}

\bibitem{NNOnline}
Radboud~University Nijmegen.
\newblock {NN}-online.
\newblock \url{http://nn-online.org/}, 2005.
\newblock Accessed: 2018-12-01.

\bibitem{Beane:2020wjl}
Silas~R. Beane and Roland~C. Farrell.
\newblock {Geometry and entanglement in the scattering matrix}.
\newblock {\em Annals Phys.}, 433:168581, 2021.

\bibitem{Grimsley_2019}
Harper~R. {Grimsley}, Sophia~E. {Economou}, Edwin {Barnes}, and Nicholas~J. {Mayhall}.
\newblock An adaptive variational algorithm for exact molecular simulations on a quantum computer.
\newblock {\em Nat. Commun.}, 10:3007, 7 2019.

\bibitem{Farrell:2023fgd}
Roland~C. Farrell, Marc Illa, Anthony~N. Ciavarella, and Martin~J. Savage.
\newblock {Scalable Circuits for Preparing Ground States on Digital Quantum Computers: The Schwinger Model Vacuum on 100 Qubits}.
\newblock {\em PRX Quantum}, 5(2):020315, 2024.

\bibitem{Ananthanarayan:2000ht}
B.~Ananthanarayan, G.~Colangelo, J.~Gasser, and H.~Leutwyler.
\newblock Roy equation analysis of pi pi scattering.
\newblock {\em Phys. Rept.}, 353:207--279, 2001.

\bibitem{Colangelo:2001df}
G.~Colangelo, J.~Gasser, and H.~Leutwyler.
\newblock $\pi \pi$ scattering.
\newblock {\em Nucl. Phys. B}, 603:125--179, 2001.

\bibitem{Hoferichter_2016}
Martin Hoferichter, Jacobo Ruiz~de Elvira, Bastian Kubis, and Ulf-G. Mei\ss{}ner.
\newblock Roy\textendash{}steiner-equation analysis of pion\textendash{}nucleon scattering.
\newblock {\em Phys. Rept.}, 625:1--88, 2016.

\bibitem{Klco:2019xro}
Natalie Klco and Martin~J. Savage.
\newblock {Minimally entangled state preparation of localized wave functions on quantum computers}.
\newblock {\em Phys. Rev. A}, 102(1):012612, 2020.

\bibitem{gadi_aleksandrowicz_2019_2562111}
Matthew Treinish, Jay Gambetta, Paul Nation, qiskit bot, Paul Kassebaum, Diego~M. Rodríguez, Salvador de~la Puente~González, Shaohan Hu, Kevin Krsulich, Jim Garrison, Laura Zdanski, Jake Lishman, Jessie Yu, Julien Gacon, David McKay, Juan Gomez, Lauren Capelluto, Travis-S-IBM, Manoel Marques, Ashish Panigrahi, lerongil, Rafey~Iqbal Rahman, Steve Wood, Luciano Bello, Toshinari Itoko, Christopher~J. Wood, Divyanshu Singh, Drew, Eli Arbel, and Joachim Schwarm.
\newblock {Qiskit/qiskit: Qiskit 0.36.2}, May 2022.

\bibitem{queraweb}
QuEra.
\newblock {QuEra}'s quantum roadmap, 2024.

\bibitem{quantinuumweb}
Quantinuum.
\newblock Quantinuum hardware, 2024.

\bibitem{ibmweb}
IBM.
\newblock {IBM} quantum summit {2023}, 2024.

\bibitem{ionqweb}
IonQ.
\newblock {IonQ} technical roadmap, 2024.

\bibitem{Bluvstein:2023zmt}
Dolev Bluvstein et~al.
\newblock {Logical quantum processor based on reconfigurable atom arrays}.
\newblock {\em Nature}, 626(7997):58--65, 12 2024.

\bibitem{Acharya:2024btg}
Rajeev Acharya et~al.
\newblock {Quantum error correction below the surface code threshold}.
\newblock 8 2024.

\bibitem{Kim:2023bwr}
Youngseok Kim et~al.
\newblock {Evidence for the utility of quantum computing before fault tolerance}.
\newblock {\em Nature}, 618(7965):500--505, 2023.

\bibitem{Shinjo:2024vci}
Kazuya Shinjo, Kazuhiro Seki, Tomonori Shirakawa, Rong-Yang Sun, and Seiji Yunoki.
\newblock {Unveiling clean two-dimensional discrete time quasicrystals on a digital quantum computer}.
\newblock 3 2024.

\bibitem{Andersen:2024aob}
Trond~I. Andersen et~al.
\newblock {Thermalization and Criticality on an Analog-Digital Quantum Simulator}.
\newblock 5 2024.

\bibitem{DeCross:2024tmi}
Matthew DeCross et~al.
\newblock {The computational power of random quantum circuits in arbitrary geometries}.
\newblock 6 2024.

\bibitem{Arute:2019zxq}
Frank Arute et~al.
\newblock {Quantum supremacy using a programmable superconducting processor}.
\newblock {\em Nature}, 574(7779):505--510, 2019.

\bibitem{Farrell:2024fit}
Roland~C. Farrell, Marc Illa, Anthony~N. Ciavarella, and Martin~J. Savage.
\newblock {Quantum simulations of hadron dynamics in the Schwinger model using 112 qubits}.
\newblock {\em Phys. Rev. D}, 109(11):114510, 1 2024.

\bibitem{Valentini:2024fly}
Marco Valentini et~al.
\newblock {Demonstration of two-dimensional connectivity for a scalable error-corrected ion-trap quantum processor architecture}.
\newblock 6 2024.

\bibitem{Guo:2023xza}
S.~A. Guo et~al.
\newblock {A site-resolved two-dimensional quantum simulator with hundreds of trapped ions}.
\newblock {\em Nature}, 630(8017):613--618, 2024.

\bibitem{Manetsch:2024lwl}
Hannah~J. Manetsch, Gyohei Nomura, Elie Bataille, Kon~H. Leung, Xudong Lv, and Manuel Endres.
\newblock {A tweezer array with 6100 highly coherent atomic qubits}.
\newblock 3 2024.

\bibitem{Zalivako:2024bjm}
Ilia~V. Zalivako et~al.
\newblock {Towards multiqudit quantum processor based on a $^{171}$Yb$^{+}$ ion string: Realizing basic quantum algorithms}.
\newblock 2 2024.

\bibitem{Vazquez:2024qmo}
Almudena~Carrera Vazquez, Caroline Tornow, Diego Riste, Stefan Woerner, Maika Takita, and Daniel~J. Egger.
\newblock {Scaling quantum computing with dynamic circuits}.
\newblock 2 2024.

\bibitem{Li_2008}
Hui Li and F.~D.~M. Haldane.
\newblock Entanglement spectrum as a generalization of entanglement entropy: Identification of topological order in non-abelian fractional quantum hall effect states.
\newblock {\em Physical Review Letters}, 101(1), July 2008.

\bibitem{Kitaev:2005dm}
Alexei Kitaev and John Preskill.
\newblock {Topological entanglement entropy}.
\newblock {\em Phys. Rev. Lett.}, 96:110404, 2006.

\bibitem{Beane:2018oxh}
Silas~R. Beane, David~B. Kaplan, Natalie Klco, and Martin~J. Savage.
\newblock {Entanglement Suppression and Emergent Symmetries of Strong Interactions}.
\newblock {\em Phys. Rev. Lett.}, 122(10):102001, 2019.

\bibitem{Low:2021ufv}
Ian Low and Thomas Mehen.
\newblock {Symmetry from entanglement suppression}.
\newblock {\em Phys. Rev. D}, 104(7):074014, 2021.

\bibitem{Bai:2022hfv}
Dong Bai and Zhongzhou Ren.
\newblock {Entanglement generation in few-nucleon scattering}.
\newblock {\em Phys. Rev. C}, 106(6):064005, 2022.

\bibitem{Liu:2023bnr}
Qiaofeng Liu and Ian Low.
\newblock {Hints of Entanglement Suppression in Hyperon-Nucleon Scattering}.
\newblock 12 2023.

\bibitem{Liu:2022grf}
Qiaofeng Liu, Ian Low, and Thomas Mehen.
\newblock {Minimal entanglement and emergent symmetries in low-energy QCD}.
\newblock {\em Phys. Rev. C}, 107(2):025204, 2023.

\bibitem{Kirchner:2023dvg}
Tanja Kirchner, Wael Elkamhawy, and Hans-Werner Hammer.
\newblock {Entanglement in Few-Nucleon Scattering Events}.
\newblock {\em Few Body Syst.}, 65(2):29, 2024.

\bibitem{Carena:2023vjc}
Marcela Carena, Ian Low, Carlos E.~M. Wagner, and Ming-Lei Xiao.
\newblock {Entanglement suppression, enhanced symmetry, and a standard-model-like Higgs boson}.
\newblock {\em Phys. Rev. D}, 109(5):L051901, 2024.

\bibitem{Beane:2021zvo}
Silas~R. Beane, Roland~C. Farrell, and Mira Varma.
\newblock Entanglement minimization in hadronic scattering with pions.
\newblock {\em Int. J. Mod. Phys. A}, 36(30):2150205, 2021.

\bibitem{Aoude_2020}
Rafael Aoude, Ming-Zhi Chung, Yu-tin Huang, Camila~S. Machado, and Man-Kuan Tam.
\newblock Silence of binary kerr black holes.
\newblock {\em Phys. Rev. Lett.}, 125(18):181602, 2020.

\bibitem{PhysRevLett.69.2863}
Steven~R. White.
\newblock Density matrix formulation for quantum renormalization groups.
\newblock {\em Phys. Rev. Lett.}, 69:2863--2866, Nov 1992.

\bibitem{Verstraete_2008}
F.~Verstraete, V.~Murg, and J.I. Cirac.
\newblock Matrix product states, projected entangled pair states, and variational renormalization group methods for quantum spin systems.
\newblock {\em Advances in Physics}, 57(2):143–224, March 2008.

\bibitem{Bauer:2020epk}
Bela Bauer, Sergey Bravyi, Mario Motta, and Garnet Kin-Lic Chan.
\newblock {Quantum Algorithms for Quantum Chemistry and Quantum Materials Science}.
\newblock {\em Chem. Rev.}, 120(22):12685--12717, 2020.

\bibitem{Banuls:2019bmf}
Mari~Carmen Ba{\~{n}}uls, Rainer Blatt, Jacopo Catani, Alessio Celi, Juan~Ignacio Cirac, Marcello Dalmonte, Leonardo Fallani, Karl Jansen, Maciej Lewenstein, Simone Montangero, Christine~A. Muschik, Benni Reznik, Enrique Rico, Luca Tagliacozzo, Karel~Van Acoleyen, Frank Verstraete, Uwe-Jens Wiese, Matthew Wingate, Jakub Zakrzewski, and Peter Zoller.
\newblock {Simulating Lattice Gauge Theories within Quantum Technologies}.
\newblock {\em Eur. Phys. J. D}, 74(8):165, Aug 2020.

\bibitem{Bauer:2022hpo}
Christian~W. Bauer et~al.
\newblock {Quantum Simulation for High-Energy Physics}.
\newblock {\em PRX Quantum}, 4(2):027001, 4 2023.

\bibitem{Bauer:2023qgm}
Christian~W. Bauer, Zohreh Davoudi, Natalie Klco, and Martin~J. Savage.
\newblock Quantum simulation of fundamental particles and forces.
\newblock {\em Nature Rev. Phys.}, 5(7):420--432, 2023.

\bibitem{Glashow:1961tr}
S.L. Glashow.
\newblock {Partial Symmetries of Weak Interactions}.
\newblock {\em Nucl. Phys.}, 22:579--588, 1961.

\bibitem{Higgs:1964pj}
Peter~W. Higgs.
\newblock {Broken Symmetries and the Masses of Gauge Bosons}.
\newblock {\em Phys. Rev. Lett.}, 13:508--509, 1964.

\bibitem{Weinberg:1967tq}
Steven Weinberg.
\newblock {A Model of Leptons}.
\newblock {\em Phys. Rev. Lett.}, 19:1264--1266, 1967.

\bibitem{Salam:1968rm}
Abdus Salam.
\newblock {Weak and Electromagnetic Interactions}.
\newblock {\em Conf. Proc. C}, 680519:367--377, 1968.

\bibitem{Politzer:1973fx}
H.David Politzer.
\newblock {Reliable Perturbative Results for Strong Interactions?}
\newblock {\em Phys. Rev. Lett.}, 30:1346--1349, 1973.

\bibitem{Gross:1973id}
David~J. Gross and Frank Wilczek.
\newblock {Ultraviolet Behavior of Nonabelian Gauge Theories}.
\newblock {\em Phys. Rev. Lett.}, 30:1343--1346, 1973.

\bibitem{Fan:2022eto}
X.~Fan, T.~G. Myers, B.~A.~D. Sukra, and G.~Gabrielse.
\newblock {Measurement of the Electron Magnetic Moment}.
\newblock {\em Phys. Rev. Lett.}, 130(7):071801, 2023.

\bibitem{ParticleDataGroup:2022pth}
R.~L. Workman et~al.
\newblock {Review of Particle Physics}.
\newblock {\em PTEP}, 2022:083C01, 2022.

\bibitem{Tiesinga:2021myr}
Eite Tiesinga, Peter~J. Mohr, David~B. Newell, and Barry~N. Taylor.
\newblock {CODATA} recommended values of the fundamental physical constants: {2018}.
\newblock {\em Rev. Mod. Phys.}, 93(2):025010, 2021.

\bibitem{Troyer_2005}
Matthias Troyer and Uwe-Jens Wiese.
\newblock Computational complexity and fundamental limitations to fermionic quantum monte carlo simulations.
\newblock {\em Physical Review Letters}, 94(17):170201, May 2005.

\bibitem{Schafer:2005ff}
Thomas Sch\"afer.
\newblock {Phases of QCD}.
\newblock In {\em {20th Annual Hampton University Graduate Studies Program}}, 9 2005.

\bibitem{Andersson:1983ia}
Bo~Andersson, G.~Gustafson, G.~Ingelman, and T.~Sjostrand.
\newblock {Parton Fragmentation and String Dynamics}.
\newblock {\em Phys. Rept.}, 97:31--145, 1983.

\bibitem{Fries:2003vb}
R.~J. Fries, Berndt Muller, C.~Nonaka, and S.~A. Bass.
\newblock {Hadronization in heavy ion collisions: Recombination and fragmentation of partons}.
\newblock {\em Phys. Rev. Lett.}, 90:202303, 2003.

\bibitem{Fries:2003kq}
R.~J. Fries, Berndt Muller, C.~Nonaka, and S.~A. Bass.
\newblock {Hadron production in heavy ion collisions: Fragmentation and recombination from a dense parton phase}.
\newblock {\em Phys. Rev. C}, 68:044902, 2003.

\bibitem{Srednicki:1994mfb}
Mark Srednicki.
\newblock {Chaos and Quantum Thermalization}.
\newblock {\em Phys. Rev. E}, 50:888--901, Aug 1994.

\bibitem{Burgio:2021vgk}
G.~F. Burgio, H.~J. Schulze, I.~Vidana, and J.~B. Wei.
\newblock {Neutron stars and the nuclear equation of state}.
\newblock {\em Prog. Part. Nucl. Phys.}, 120:103879, 2021.

\bibitem{Shor:1994jg}
Peter~W. Shor.
\newblock {Polynomial time algorithms for prime factorization and discrete logarithms on a quantum computer}.
\newblock {\em SIAM J. Sci. Statist. Comput.}, 26:1484, 1997.

\bibitem{Bennett:2014rmv}
Charles~H. Bennett and Gilles Brassard.
\newblock {Quantum cryptography: Public key distribution and coin tossing}.
\newblock {\em Theor. Comput. Sci.}, 560:7--11, 2014.

\bibitem{Wigner:1936dx}
E.~Wigner.
\newblock {On the Consequences of the Symmetry of the Nuclear Hamiltonian on the Spectroscopy of Nuclei}.
\newblock {\em Phys. Rev.}, 51:106--119, 1937.

\bibitem{Wigner:1937zz}
E.~Wigner.
\newblock {On the Structure of Nuclei Beyond Oxygen}.
\newblock {\em Phys. Rev.}, 51:947--958, 1937.

\bibitem{Wigner:1939zz}
E.~P. Wigner.
\newblock On coupling conditions in light nuclei and the lifetimes of beta-radioactivities.
\newblock {\em Phys. Rev.}, 56:519--527, 9 1939.

\bibitem{NPLQCD:2020lxg}
Marc Illa, Silas~R. Beane, Emmanuel Chang, Zohreh Davoudi, William Detmold, David~J. Murphy, Kostas Orginos, Assumpta Parre\~no, Martin~J. Savage, Phiala~E. Shanahan, Michael~L. Wagman, and Frank Winter.
\newblock Low-energy scattering and effective interactions of two baryons at $m_{\pi}\sim {450}$ {MeV} from lattice quantum chromodynamics.
\newblock {\em Phys. Rev. D}, 103(5):054508, 2021.

\bibitem{Wagman:2017tmp}
Michael~L. Wagman, Frank Winter, Emmanuel Chang, Zohreh Davoudi, William Detmold, Kostas Orginos, Martin~J. Savage, and Phiala~E. Shanahan.
\newblock {Baryon-Baryon Interactions and Spin-Flavor Symmetry from Lattice Quantum Chromodynamics}.
\newblock {\em Phys. Rev. D}, 96(11):114510, 2017.

\bibitem{PhysRev.124.246}
Y.~Nambu and G.~Jona-Lasinio.
\newblock Dynamical model of elementary particles based on an analogy with superconductivity. ii.
\newblock {\em Phys. Rev.}, 124:246--254, Oct 1961.

\bibitem{PhysRev.122.345}
Y.~Nambu and G.~Jona-Lasinio.
\newblock Dynamical model of elementary particles based on an analogy with superconductivity. i.
\newblock {\em Phys. Rev.}, 122:345--358, Apr 1961.

\bibitem{PhysRevLett.4.380}
Yoichiro Nambu.
\newblock Axial vector current conservation in weak interactions.
\newblock {\em Phys. Rev. Lett.}, 4:380--382, Apr 1960.

\bibitem{PhysRev.117.648}
Yoichiro Nambu.
\newblock Quasi-particles and gauge invariance in the theory of superconductivity.
\newblock {\em Phys. Rev.}, 117:648--663, Feb 1960.

\bibitem{tHooft:1973alw}
Gerard 't~Hooft.
\newblock {A Planar Diagram Theory for Strong Interactions}.
\newblock {\em Nucl. Phys. B}, 72:461, 1974.

\bibitem{tHooft:1974pnl}
Gerard 't~Hooft.
\newblock {A Two-Dimensional Model for Mesons}.
\newblock {\em Nucl. Phys. B}, 75:461--470, 1974.

\bibitem{RevModPhys.38.447}
JOHN~S. BELL.
\newblock On the problem of hidden variables in quantum mechanics.
\newblock {\em Rev. Mod. Phys.}, 38:447--452, Jul 1966.

\bibitem{Weinberg:1978kz}
Steven Weinberg.
\newblock Phenomenological lagrangians.
\newblock {\em Physica A}, 96(1-2):327--340, 1979.

\bibitem{Beane:2000fx}
Silas~R. Beane, Paulo~F. Bedaque, Wick~C. Haxton, Daniel~R. Phillips, and Martin~J. Savage.
\newblock {From hadrons to nuclei: Crossing the border}.
\newblock pages 133--269, 8 2000.

\bibitem{Hammer:2019poc}
H.~W. Hammer, S.~K\"onig, and U.~van Kolck.
\newblock Nuclear effective field theory: status and perspectives.
\newblock {\em Rev. Mod. Phys.}, 92(2):025004, 2020.

\bibitem{PhysRev.76.38}
H.~A. Bethe.
\newblock Theory of the effective range in nuclear scattering.
\newblock {\em Phys. Rev.}, 76:38--50, Jul 1949.

\bibitem{Weinberg:1990rz}
Steven Weinberg.
\newblock {Nuclear forces from chiral Lagrangians}.
\newblock {\em Phys. Lett. B}, 251:288--292, 1990.

\bibitem{Kaplan:1998we}
David~B. Kaplan, Martin~J. Savage, and Mark~B. Wise.
\newblock {Two nucleon systems from effective field theory}.
\newblock {\em Nucl. Phys. B}, 534:329--355, 1998.

\bibitem{deSwart:1995ui}
J.~J. de~Swart, C.~P.~F. Terheggen, and V.~G.~J. Stoks.
\newblock The low-energy n p scattering parameters and the deuteron.
\newblock In {\em {3rd International Symposium on Dubna Deuteron 95}}, 9 1995.

\bibitem{Yamaguchi:1954mp}
Yoshio Yamaguchi.
\newblock Two nucleon problem when the potential is nonlocal but separable. {1}.
\newblock {\em Phys. Rev.}, 95:1628--1634, 1954.

\bibitem{FlavourLatticeAveragingGroupFLAG:2021npn}
Y.~Aoki et~al.
\newblock {FLAG Review 2021}.
\newblock {\em Eur. Phys. J. C}, 82(10):869, 2022.

\bibitem{Davoudi:2022bnl}
Zohreh Davoudi et~al.
\newblock {Report of the Snowmass 2021 Topical Group on Lattice Gauge Theory}.
\newblock In {\em {Snowmass 2021}}, 9 2022.

\bibitem{USQCD:2022mmc}
Andreas~S. Kronfeld et~al.
\newblock Lattice {QCD} and particle physics, 7 2022.

\bibitem{Davoudi:2020ngi}
Zohreh Davoudi, William Detmold, Kostas Orginos, Assumpta Parre\~no, Martin~J. Savage, Phiala Shanahan, and Michael~L. Wagman.
\newblock {Nuclear matrix elements from lattice QCD for electroweak and beyond-Standard-Model processes}.
\newblock {\em Phys. Rept.}, 900:1--74, 2021.

\bibitem{BMW:2008jgk}
S.~Durr et~al.
\newblock {Ab-Initio Determination of Light Hadron Masses}.
\newblock {\em Science}, 322:1224--1227, 2008.

\bibitem{BMW:2014pzb}
Sz. Borsanyi et~al.
\newblock {Ab initio calculation of the neutron-proton mass difference}.
\newblock {\em Science}, 347:1452--1455, 2015.

\bibitem{Catterall:2022wjq}
Simon Catterall et~al.
\newblock {Report of the Snowmass 2021 Theory Frontier Topical Group on Quantum Information Science}.
\newblock In {\em {Snowmass 2021}}, 9 2022.

\bibitem{Humble:2022klb}
Travis~S. Humble, Gabriel~N. Perdue, and Martin~J. Savage.
\newblock {Snowmass Computational Frontier: Topical Group Report on Quantum Computing}.
\newblock 9 2022.

\bibitem{Benioff1980}
Paul Benioff.
\newblock The computer as a physical system: A microscopic quantum mechanical hamiltonian model of computers as represented by turing machines.
\newblock {\em Journal of Statistical Physics}, 22(5):563--591, May 1980.

\bibitem{Feynman1982}
Richard~P. Feynman.
\newblock Simulating physics with computers.
\newblock {\em International Journal of Theoretical Physics}, 21:467--488, 1982.

\bibitem{Feynman1986}
Richard~P. Feynman.
\newblock Quantum mechanical computers.
\newblock {\em Foundations of Physics}, 16:507--531, 1986.

\bibitem{Lloyd1073}
Seth Lloyd.
\newblock Universal quantum simulators.
\newblock {\em Science}, 273(5278):1073--1078, 1996.

\bibitem{farhi2000quantum}
Edward Farhi, Jeffrey Goldstone, Sam Gutmann, and Michael Sipser.
\newblock Quantum computation by adiabatic evolution, 2000.

\bibitem{van_Dam_2001}
W.~van Dam, M.~Mosca, and U.~Vazirani.
\newblock How powerful is adiabatic quantum computation?
\newblock In {\em Proceedings 42nd {IEEE} Symposium on Foundations of Computer Science}. {IEEE}, 2001.

\bibitem{Ebadi2021}
Sepehr Ebadi, Tout~T. Wang, Harry Levine, Alexander Keesling, Giulia Semeghini, Ahmed Omran, Dolev Bluvstein, Rhine Samajdar, Hannes Pichler, Wen~Wei Ho, Soonwon Choi, Subir Sachdev, Markus Greiner, Vladan Vuleti{\'{c}}, and Mikhail~D. Lukin.
\newblock Quantum phases of matter on a 256-atom programmable quantum simulator.
\newblock {\em Nature}, 595(7866):227--232, Jul 2021.

\bibitem{Semeghini:2021wls}
Giulia Semeghini et~al.
\newblock {Probing topological spin liquids on a programmable quantum simulator}.
\newblock {\em Science}, 374(6572):abi8794, 2021.

\bibitem{Preskill:2018jim}
John Preskill.
\newblock {Quantum Computing in the NISQ era and beyond}.
\newblock {\em Quantum}, 2:79, 2018.

\bibitem{Klco:2018kyo}
N.~Klco, E.~F. Dumitrescu, A.~J. McCaskey, T.~D. Morris, R.~C. Pooser, M.~Sanz, E.~Solano, P.~Lougovski, and M.~J. Savage.
\newblock {Quantum-classical computation of Schwinger model dynamics using quantum computers}.
\newblock {\em Phys. Rev. A}, 98(3):032331, 2018.

\bibitem{Mazzola:2021hma}
Giulia Mazzola, Simon~V. Mathis, Guglielmo Mazzola, and Ivano Tavernelli.
\newblock Gauge-invariant quantum circuits for {U}({1}) and yang-mills lattice gauge theories.
\newblock {\em Phys. Rev. Res.}, 3(4):043209, 2021.

\bibitem{deJong:2021wsd}
Wibe~A. de~Jong, Kyle Lee, James Mulligan, Mateusz P\l{}osko\'n, Felix Ringer, and Xiaojun Yao.
\newblock Quantum simulation of nonequilibrium dynamics and thermalization in the schwinger model.
\newblock {\em Phys. Rev. D}, 106(5):054508, 2022.

\bibitem{Gong:2021bcp}
Wenjie Gong, Ganesh Parida, Zhoudunming Tu, and Raju Venugopalan.
\newblock Measurement of bell-type inequalities and quantum entanglement from \ensuremath{\Lambda}-hyperon spin correlations at high energy colliders.
\newblock {\em Phys. Rev. D}, 106(3):L031501, 2022.

\bibitem{Mildenberger:2022jqr}
Julius Mildenberger, Wojciech Mruczkiewicz, Jad~C. Halimeh, Zhang Jiang, and Philipp Hauke.
\newblock {Probing confinement in a $\mathbb{Z}_2$ lattice gauge theory on a quantum computer}.
\newblock 3 2022.

\bibitem{Charles:2023zbl}
Clement Charles, Erik~J. Gustafson, Elizabeth Hardt, Florian Herren, Norman Hogan, Henry Lamm, Sara Starecheski, Ruth~S. Van~de Water, and Michael~L. Wagman.
\newblock Simulating $\mathbb{Z}_{2}$ lattice gauge theory on a quantum computer, 5 2023.

\bibitem{Pomarico:2023png}
Domenico Pomarico, Leonardo Cosmai, Paolo Facchi, Cosmo Lupo, Saverio Pascazio, and Francesco~V. Pepe.
\newblock Dynamical quantum phase transitions of the schwinger model: Real-time dynamics on {IBM} quantum.
\newblock {\em Entropy}, 25(4):608, 2023.

\bibitem{Lu:2018pjk}
Hsuan-Hao Lu, Natalie Klco, Joseph~M. Lukens, Titus~D. Morris, Aaina Bansal, Andreas Ekstr\"om, Gaute Hagen, Thomas Papenbrock, Andrew~M. Weiner, Martin~J. Savage, and Pavel Lougovski.
\newblock Simulations of subatomic many-body physics on a quantum frequency processor.
\newblock {\em Phys. Rev. A}, 100(1):012320, Jul 2019.

\bibitem{Mil:2019pbt}
Alexander Mil, Torsten~V. Zache, Apoorva Hegde, Andy Xia, Rohit~P. Bhatt, Markus~K. Oberthaler, Philipp Hauke, J\"urgen Berges, and Fred Jendrzejewski.
\newblock {A scalable realization of local U(1) gauge invariance in cold atomic mixtures}.
\newblock {\em Science}, 367(6482):1128--1130, 2020.

\bibitem{Yang:2020yer}
Bing Yang, Hui Sun, Robert Ott, Han-Yi Wang, Torsten~V. Zache, Jad~C. Halimeh, Zhen-Sheng Yuan, Philipp Hauke, and Jian-Wei Pan.
\newblock {Observation of gauge invariance in a 71-site Bose\textendash{}Hubbard quantum simulator}.
\newblock {\em Nature}, 587(7834):392--396, 2020.

\bibitem{Zhou:2021kdl}
Zhao-Yu Zhou, Guo-Xian Su, Jad~C. Halimeh, Robert Ott, Hui Sun, Philipp Hauke, Bing Yang, Zhen-Sheng Yuan, J\"urgen Berges, and Jian-Wei Pan.
\newblock Thermalization dynamics of a gauge theory on a quantum simulator.
\newblock {\em Science}, 377(6603):311--314, 2022.

\bibitem{PhysRevResearch.5.023010}
Guo-Xian Su, Hui Sun, Ana Hudomal, Jean-Yves Desaules, Zhao-Yu Zhou, Bing Yang, Jad~C. Halimeh, Zhen-Sheng Yuan, Zlatko Papi\'c, and Jian-Wei Pan.
\newblock Observation of many-body scarring in a bose-hubbard quantum simulator.
\newblock {\em Phys. Rev. Res.}, 5(2):023010, Apr 2023.

\bibitem{zhang2023observation}
Wei-Yong Zhang, Ying Liu, Yanting Cheng, Ming-Gen He, Han-Yi Wang, Tian-Yi Wang, Zi-Hang Zhu, Guo-Xian Su, Zhao-Yu Zhou, Yong-Guang Zheng, Hui Sun, Bing Yang, Philipp Hauke, Wei Zheng, Jad~C. Halimeh, Zhen-Sheng Yuan, and Jian-Wei Pan.
\newblock Observation of microscopic confinement dynamics by a tunable topological {$\theta$}-angle, 2023.

\bibitem{Farrell:2022vyh}
Roland~C. Farrell, Ivan~A. Chernyshev, Sarah J.~M. Powell, Nikita~A. Zemlevskiy, Marc Illa, and Martin~J. Savage.
\newblock {Preparations for quantum simulations of quantum chromodynamics in 1+1 dimensions. II. Single-baryon \ensuremath{\beta}-decay in real time}.
\newblock {\em Phys. Rev. D}, 107(5):054513, 2023.

\bibitem{Farrell:2022wyt}
Roland~C. Farrell, Ivan~A. Chernyshev, Sarah J.~M. Powell, Nikita~A. Zemlevskiy, Marc Illa, and Martin~J. Savage.
\newblock Preparations for quantum simulations of quantum chromodynamics in {1}+{1} dimensions: ({I}) axial gauge, 2022.

\bibitem{Ciavarella:2024fzw}
Anthony~N. Ciavarella and Christian~W. Bauer.
\newblock {Quantum Simulation of SU(3) Lattice Yang Mills Theory at Leading Order in Large N}, 2 2024.

\bibitem{h1-1}
Quantinuum.
\newblock System model {H1} product data sheet, version {5}.{00}, 6 2022.

\bibitem{De_Raedt_2019}
Hans De~Raedt, Fengping Jin, Dennis Willsch, Madita Willsch, Naoki Yoshioka, Nobuyasu Ito, Shengjun Yuan, and Kristel Michielsen.
\newblock Massively parallel quantum computer simulator, eleven years later.
\newblock {\em Computer Physics Communications}, 237:47–61, April 2019.

\bibitem{Wang:2020yjh}
Samson Wang, Enrico Fontana, M.~Cerezo, Kunal Sharma, Akira Sone, Lukasz Cincio, and Patrick~J. Coles.
\newblock Noise-induced barren plateaus in variational quantum algorithms.
\newblock {\em Nat. Commun.}, 12:6961, 2021.

\bibitem{Scriva:2023sgz}
Giuseppe Scriva, Nikita Astrakhantsev, Sebastiano Pilati, and Guglielmo Mazzola.
\newblock Challenges of variational quantum optimization with measurement shot noise, 7 2023.

\bibitem{Hastings_2007}
M~B Hastings.
\newblock An area law for one-dimensional quantum systems.
\newblock {\em Journal of Statistical Mechanics: Theory and Experiment}, 2007(08):P08024--P08024, aug 2007.

\bibitem{arad2013area}
Itai Arad, Alexei Kitaev, Zeph Landau, and Umesh Vazirani.
\newblock An area law and sub-exponential algorithm for {1D} systems, 2013.

\bibitem{Brand_o_2014}
Fernando G. S.~L. Brandão and Michal Horodecki.
\newblock Exponential decay of correlations implies area law.
\newblock {\em Commun. Math. Phys.}, 333(2):761--798, 2015.

\bibitem{PHENIX:2004vcz}
K.~Adcox et~al.
\newblock {Formation of dense partonic matter in relativistic nucleus-nucleus collisions at RHIC: Experimental evaluation by the PHENIX collaboration}.
\newblock {\em Nucl. Phys. A}, 757:184--283, 2005.

\bibitem{BRAHMS:2004adc}
I.~Arsene et~al.
\newblock {Quark gluon plasma and color glass condensate at RHIC? The Perspective from the BRAHMS experiment}.
\newblock {\em Nucl. Phys. A}, 757:1--27, 2005.

\bibitem{PHOBOS:2004zne}
B.~B. Back et~al.
\newblock {The PHOBOS perspective on discoveries at RHIC}.
\newblock {\em Nucl. Phys. A}, 757:28--101, 2005.

\bibitem{STAR:2005gfr}
John Adams et~al.
\newblock {Experimental and theoretical challenges in the search for the quark gluon plasma: The STAR Collaboration's critical assessment of the evidence from RHIC collisions}.
\newblock {\em Nucl. Phys. A}, 757:102--183, 2005.

\bibitem{STAR:2010vob}
M.~M. Aggarwal et~al.
\newblock {An Experimental Exploration of the QCD Phase Diagram: The Search for the Critical Point and the Onset of De-confinement}, 7 2010.

\bibitem{STAR:2017sal}
L.~Adamczyk et~al.
\newblock {Bulk Properties of the Medium Produced in Relativistic Heavy-Ion Collisions from the Beam Energy Scan Program}.
\newblock {\em Phys. Rev. C}, 96(4):044904, 2017.

\bibitem{Loizides:2016tew}
Constantin Loizides.
\newblock {Experimental overview on small collision systems at the LHC}.
\newblock {\em Nucl. Phys. A}, 956:200--207, 2016.

\bibitem{Foka:2016zdb}
Panagiota Foka and Ma\l{}gorzata~Anna Janik.
\newblock {An overview of experimental results from ultra-relativistic heavy-ion collisions at the CERN LHC: Hard probes}.
\newblock {\em Rev. Phys.}, 1:172--194, 2016.

\bibitem{Foka:2016vta}
Panagiota Foka and Ma\l{}gorzata~Anna Janik.
\newblock {An overview of experimental results from ultra-relativistic heavy-ion collisions at the CERN LHC: Bulk properties and dynamical evolution}.
\newblock {\em Rev. Phys.}, 1:154--171, 2016.

\bibitem{Klco:2021lap}
Natalie Klco, Alessandro Roggero, and Martin~J. Savage.
\newblock {Standard model physics and the digital quantum revolution: thoughts about the interface}.
\newblock {\em Rept. Prog. Phys.}, 85(6):064301, 2022.

\bibitem{Witten:1979kh}
Edward Witten.
\newblock {Baryons in the $1/N$ Expansion}.
\newblock {\em Nucl. Phys. B}, 160:57--115, 1979.

\bibitem{Witten:1979pi}
Edward Witten.
\newblock The {1}/{N} expansion in atomic and particle physics.
\newblock {\em NATO Sci. Ser. B}, 59:403--419, 1980.

\bibitem{Kaiser:2000gs}
Roland Kaiser and H.~Leutwyler.
\newblock Large-{N} in chiral perturbation theory.
\newblock {\em Eur. Phys. J. C}, 17:623--649, 2000.

\bibitem{Yaffe:1981vf}
Laurence~G. Yaffe.
\newblock Large {N} limits as classical mechanics.
\newblock {\em Rev. Mod. Phys.}, 54:407, 4 1982.

\bibitem{PhysRevA.63.040304}
Paolo Zanardi.
\newblock Entanglement of quantum evolutions.
\newblock {\em Phys. Rev. A}, 63:040304, 3 2001.

\bibitem{mahdavi2011cross}
Aaron~David Ballard and Yong-Shi Wu.
\newblock {\em Cross Disciplinary Advances in Quantum Computing}, chapter Cartan Decomposition and Entangleing Power of Braiding Quantum Gates.
\newblock Contemporary mathematics - American Mathematical Society. American Mathematical Society, 2011.

\bibitem{Kaplan:1995yg}
David~B. Kaplan and Martin~J. Savage.
\newblock {The Spin flavor dependence of nuclear forces from large n QCD}.
\newblock {\em Phys. Lett. B}, 365:244--251, 1996.

\bibitem{Kaplan:1996rk}
David~B. Kaplan and Aneesh~V. Manohar.
\newblock {The Nucleon-nucleon potential in the 1/N(c) expansion}.
\newblock {\em Phys. Rev. C}, 56:76--83, 1997.

\bibitem{CalleCordon:2008cz}
A.~Calle~Cordon and E.~Ruiz~Arriola.
\newblock Wigner symmetry, large {N}(c) and renormalized one boson exchange potential.
\newblock {\em Phys. Rev.}, C78:054002, 2008.

\bibitem{lanz2018determination}
Stefan Lanz.
\newblock Determination of the quark mass ratio $q$ from $\eta \to 3 \pi$.
\newblock 2018.

\bibitem{Brody_2001}
Dorje~C. Brody and Lane~P. Hughston.
\newblock Geometric quantum mechanics.
\newblock {\em Journal of Geometry and Physics}, 38(1):19–53, 4 2001.

\bibitem{Bengtsson:2001yd}
I.~Bengtsson, J.~Braennlund, and K.~Zyczkowski.
\newblock ${CP}^n$, or, entanglement illustrated.
\newblock {\em Int. J. Mod. Phys. A}, 17(31):4675--4696, 12 2002.

\bibitem{bengtsson_zyczkowski_2006}
Ingemar Bengtsson and Karol Zyczkowski.
\newblock {\em Geometry of Quantum States: An Introduction to Quantum Entanglement}.
\newblock Cambridge University Press, 2006.

\bibitem{Weinberg:2013cfa}
Steven Weinberg.
\newblock Tetraquark mesons in large-{N} quantum chromodynamics.
\newblock {\em Physical Review Letters}, 110(26), 6 2013.

\bibitem{Roy:1971tc}
S.~M. Roy.
\newblock Exact integral equation for pion pion scattering involving only physical region partial waves.
\newblock {\em Phys. Lett. B}, 36:353--356, 1971.

\bibitem{Gasser:1983yg}
J.~Gasser and H.~Leutwyler.
\newblock Chiral perturbation theory to one loop.
\newblock {\em Annals Phys.}, 158:142, 1984.

\bibitem{Beane:2019loz}
Silas~R. Beane and Peter Ehlers.
\newblock {Chiral symmetry breaking, entanglement, and the nucleon spin decomposition}.
\newblock {\em Mod. Phys. Lett. A}, 35(08):2050048, 2019.

\bibitem{Dashen:1993as}
Roger~F. Dashen and Aneesh~V. Manohar.
\newblock Baryon - pion couplings from large {N} {QCD}.
\newblock {\em Phys. Lett. B}, 315:425--430, 1993.

\bibitem{Dashen:1993ac}
Roger~F. Dashen and Aneesh~V. Manohar.
\newblock {1}/{N} corrections to the baryon axial currents in {QCD}.
\newblock {\em Phys. Lett. B}, 315:438--440, 1993.

\bibitem{Dashen:1993jt}
Roger~F. Dashen, Elizabeth~Ellen Jenkins, and Aneesh~V. Manohar.
\newblock The {1}/{N} expansion for baryons.
\newblock {\em Phys. Rev. D}, 49:4713, 1994.
\newblock [Erratum: Phys.Rev.D 51, 2489 (1995)].

\bibitem{Dashen:1994qi}
Roger~F. Dashen, Elizabeth~Ellen Jenkins, and Aneesh~V. Manohar.
\newblock Spin flavor structure of large {N} baryons.
\newblock {\em Phys. Rev. D}, 51:3697--3727, 1995.

\bibitem{Fettes_1998}
Nadia Fettes, Ulf-G. Meissner, and Sven Steininger.
\newblock Pion - nucleon scattering in chiral perturbation theory. {1}. isospin symmetric case.
\newblock {\em Nucl. Phys. A}, 640:199--234, 1998.

\bibitem{Yao_2016}
De-Liang Yao, D.~Siemens, V.~Bernard, E.~Epelbaum, A.~M. Gasparyan, J.~Gegelia, H.~Krebs, and Ulf-G. Mei\ss{}ner.
\newblock Pion-nucleon scattering in covariant baryon chiral perturbation theory with explicit delta resonances.
\newblock {\em JHEP}, 05:038, 2016.

\bibitem{Moj_i__1998}
Martin Mojzis.
\newblock Elastic $\pi ${N} scattering to {O}$(p^{3})$ in heavy baryon chiral perturbation theory.
\newblock {\em Eur. Phys. J. C}, 2:181--195, 1998.

\bibitem{Scherer2012}
Stefan Scherer and Matthias~R. Schindler.
\newblock {\em Chiral Perturbation Theory for Baryons}, pages 145--214.
\newblock Springer Berlin Heidelberg, Berlin, Heidelberg, 2012.

\bibitem{Fettes_2001}
Nadia Fettes and Ulf~G. Meissner.
\newblock Pion - nucleon scattering in an effective chiral field theory with explicit spin {3}/{2} fields.
\newblock {\em Nucl. Phys. A}, 679:629--670, 2001.

\bibitem{PhysRevLett.17.616}
Steven Weinberg.
\newblock Pion scattering lengths.
\newblock {\em Phys. Rev. Lett.}, 17:616--621, 9 1966.

\bibitem{Epelbaum_2002}
Evgeny Epelbaum, Ulf~G. Meissner, Walter Gloeckle, and Charlotte Elster.
\newblock Resonance saturation for four nucleon operators.
\newblock {\em Phys. Rev. C}, 65:044001, 2002.

\bibitem{Beane:2021B}
Silas~R. Beane and Roland~C. Farrell.
\newblock Causality and dimensionality in geometric scattering.
\newblock {\em Annals Phys.}, 440:168841, 12 2022.

\bibitem{Cohen:1998jr}
Thomas~D. Cohen and James~M. Hansen.
\newblock Low-energy theorems for nucleon-nucleon scattering.
\newblock {\em Phys. Rev. C}, 59:13--20, 1999.

\bibitem{PhysRev.74.131}
Ning Hu.
\newblock On the application of {H}eisenberg's theory of ${S}$-matrix to the problems of resonance scattering and reactions in nuclear physics.
\newblock {\em Phys. Rev.}, 74:131--140, 7 1948.

\bibitem{PhysRev.83.249}
Walter Sch\"utzer and J.~Tiomno.
\newblock On the connection of the scattering and derivative matrices with causality.
\newblock {\em Phys. Rev.}, 83:249--251, 7 1951.

\bibitem{PhysRev.91.1267}
N.~G. van Kampen.
\newblock ${S}$- matrix and causality condition. {{II}.} nonrelativistic particles.
\newblock {\em Phys. Rev.}, 91:1267--1276, 9 1953.

\bibitem{Wigner:1955zz}
Eugene~P. Wigner.
\newblock Lower limit for the energy derivative of the scattering phase shift.
\newblock {\em Phys. Rev.}, 98:145--147, 1955.

\bibitem{Phillips:1996ae}
Daniel~R. Phillips and Thomas~D. Cohen.
\newblock How short is too short? constraining contact interactions in nucleon-nucleon scattering.
\newblock {\em Phys. Lett. B}, 390:7--12, 1997.

\bibitem{Hammer:2009zh}
H.~W. Hammer and Dean Lee.
\newblock Causality and universality in low-energy quantum scattering.
\newblock {\em Phys. Lett. B}, 681:500--503, 2009.

\bibitem{Hammer:2010fw}
H.-W. Hammer and Dean Lee.
\newblock Causality and the effective range expansion.
\newblock {\em Annals Phys.}, 325:2212--2233, 2010.

\bibitem{garay2019classical}
I\~naki Garay and Salvador Robles-P\'erez.
\newblock {Classical geodesics from the canonical quantisation of spacetime coordinates}.
\newblock 1 2019.

\bibitem{blau2020}
Matthias Blau.
\newblock Lecture notes on general relativity, 2020.
\newblock \url{http://www.blau.itp.unibe.ch/GRLecturenotes.html}.

\bibitem{Beck:2019abp}
Saar Beck, Betzalel Bazak, and Nir Barnea.
\newblock Removing the wigner bound in non-perturbative effective field theory.
\newblock {\em Phys. Lett. B}, 806:135485, 2020.

\bibitem{Habashi:2020qgw}
J.~Balal Habashi, S.~Sen, S.~Fleming, and U.~van Kolck.
\newblock Effective field theory for two-body systems with shallow {S}-wave resonances.
\newblock {\em Annals Phys.}, 422:168283, 2020.

\bibitem{PhysRevA.64.012706}
D.~S. Petrov and G.~V. Shlyapnikov.
\newblock Interatomic collisions in a tightly confined bose gas.
\newblock {\em Phys. Rev. A}, 64:012706, 6 2001.

\bibitem{PhysRevA.76.063610}
J.~P. Kestner and L.-M. Duan.
\newblock Effective low-dimensional hamiltonian for strongly interacting atoms in a transverse trap.
\newblock {\em Phys. Rev. A}, 76:063610, 12 2007.

\bibitem{PhysRevA.85.061604}
Stefan~K. Baur, Bernd Fr\"ohlich, Michael Feld, Enrico Vogt, Daniel Pertot, Marco Koschorreck, and Michael K\"ohl.
\newblock Radio-frequency spectra of feshbach molecules in quasi-two-dimensional geometries.
\newblock {\em Phys. Rev. A}, 85:061604, 6 2012.

\bibitem{PhysRevA.98.051603}
Pawe\l{} Zin, Maciej Pylak, Tomasz Wasak, Mariusz Gajda, and Zbigniew Idziaszek.
\newblock Quantum bose-bose droplets at a dimensional crossover.
\newblock {\em Phys. Rev. A}, 98:051603, 11 2018.

\bibitem{PhysRevLett.94.210401}
Henning Moritz, Thilo St\"oferle, Kenneth G\"unter, Michael K\"ohl, and Tilman Esslinger.
\newblock Confinement induced molecules in a {1D} fermi gas.
\newblock {\em Phys. Rev. Lett.}, 94:210401, 6 2005.

\bibitem{PhysRevA.93.063631}
Soeren Lammers, Igor Boettcher, and Christof Wetterich.
\newblock Dimensional crossover of nonrelativistic bosons.
\newblock {\em Phys. Rev. A}, 93:063631, 6 2016.

\bibitem{Beane:2018huc}
Silas~R. Beane and Murtaza Jafry.
\newblock Dimensional crossover in non-relativistic effective field theory.
\newblock {\em J. Phys. B}, 52(3):035001, 2019.

\bibitem{Braaten:2004rn}
Eric Braaten and H.~W. Hammer.
\newblock Universality in few-body systems with large scattering length.
\newblock {\em Phys. Rept.}, 428:259--390, 2006.

\bibitem{Beane:2010ny}
Silas~R. Beane.
\newblock Ground state energy of the interacting bose gas in two dimensions: An explicit construction.
\newblock {\em Phys. Rev. A}, 82:063610, 2010.

\bibitem{Kaplan:2005es}
David~B. Kaplan.
\newblock Five lectures on effective field theory.
\newblock 10 2005.

\bibitem{Beane:2021A}
Silas~R. Beane and Roland~C. Farrell.
\newblock {UV/IR symmetries of the S-matrix and RG flow}.
\newblock {\em Nucl. Phys. A}, 1024:122478, 12 2022.

\bibitem{VanRaamsdonk:2010pw}
Mark Van~Raamsdonk.
\newblock {Building up spacetime with quantum entanglement}.
\newblock {\em Gen. Rel. Grav.}, 42:2323--2329, 2010.

\bibitem{Pastawski:2015qua}
Fernando Pastawski, Beni Yoshida, Daniel Harlow, and John Preskill.
\newblock {Holographic quantum error-correcting codes: Toy models for the bulk/boundary correspondence}.
\newblock {\em JHEP}, 06:149, 2015.

\bibitem{Headrick:2014cta}
Matthew Headrick, Veronika~E. Hubeny, Albion Lawrence, and Mukund Rangamani.
\newblock {Causality \& holographic entanglement entropy}.
\newblock {\em JHEP}, 12:162, 2014.

\bibitem{Beane:2021xrk}
Silas~R. Beane and Roland~C. Farrell.
\newblock {UV/IR symmetries of the S-matrix and RG flow}.
\newblock {\em Nucl. Phys. A}, 1024:122478, 2022.

\bibitem{Weinberg:1991um}
Steven Weinberg.
\newblock {Effective chiral Lagrangians for nucleon - pion interactions and nuclear forces}.
\newblock {\em Nucl. Phys. B}, 363:3--18, 1991.

\bibitem{SanchezSanchez:2017tws}
M.~S\'anchez~S\'anchez, C.~J. Yang, Bingwei Long, and U.~van Kolck.
\newblock Two-nucleon $^1s_{0}$ amplitude zero in chiral effective field theory.
\newblock {\em Phys. Rev. C}, 97(2):024001, 2018.

\bibitem{Peng:2021pvo}
Rui Peng, Songlin Lyu, Sebastian K\"onig, and Bingwei Long.
\newblock {Constructing chiral effective field theory around unnatural leading-order interactions}.
\newblock {\em Phys. Rev. C}, 105(5):054002, 2022.

\bibitem{Mishra:2021luw}
Chinmay Mishra, A.~Ekstr\"om, G.~Hagen, T.~Papenbrock, and L.~Platter.
\newblock {Two-pion exchange as a leading-order contribution in chiral effective field theory}.
\newblock {\em Phys. Rev. C}, 106(2):024004, 2022.

\bibitem{Ebert:2021epn}
M.~Ebert, H.~W. Hammer, and A.~Rusetsky.
\newblock An alternative scheme for effective range corrections in pionless {EFT}.
\newblock {\em Eur. Phys. J. A}, 57(12):332, 2021.

\bibitem{Habashi:2020ofb}
J.~B. Habashi, S.~Fleming, and U.~van Kolck.
\newblock Nonrelativistic effective field theory with a resonance field.
\newblock {\em Eur. Phys. J. A}, 57(5):169, 2021.

\bibitem{Kaplan:1998tg}
David~B. Kaplan, Martin~J. Savage, and Mark~B. Wise.
\newblock {A New expansion for nucleon-nucleon interactions}.
\newblock {\em Phys. Lett. B}, 424:390--396, 1998.

\bibitem{vanKolck:1998bw}
U.~van Kolck.
\newblock {Effective field theory of short range forces}.
\newblock {\em Nucl. Phys. A}, 645:273--302, 1999.

\bibitem{Birse:1998dk}
Michael~C. Birse, Judith~A. McGovern, and Keith~G. Richardson.
\newblock {A} renormalization group treatment of two-body scattering.
\newblock {\em Phys. Lett. B}, 464:169--176, 1999.

\bibitem{Mehen:1999nd}
Thomas Mehen, Iain~W. Stewart, and Mark~B. Wise.
\newblock Conformal invariance for nonrelativistic field theory.
\newblock {\em Phys. Lett. B}, 474:145--152, 2000.

\bibitem{Beane:2021C}
Silas~R. Beane and Roland~C. Farrell.
\newblock {Symmetries of the Nucleon\textendash{}Nucleon S-Matrix and Effective Field Theory Expansions}.
\newblock {\em Few Body Syst.}, 63(2):45, 2022.

\bibitem{Hagen:1972pd}
C.~R. Hagen.
\newblock Scale and conformal transformations in galilean-covariant field theory.
\newblock {\em Phys. Rev. D}, 5:377--388, 1972.

\bibitem{Niederer:1972zz}
U.~Niederer.
\newblock The maximal kinematical invariance group of the free schrodinger equation.
\newblock {\em Helv. Phys. Acta}, 45:802--810, 1972.

\bibitem{Nishida:2007pj}
Yusuke Nishida and Dam~T. Son.
\newblock Nonrelativistic conformal field theories.
\newblock {\em Phys. Rev. D}, 76:086004, 2007.

\bibitem{Kaplan:1996nv}
David~B. Kaplan.
\newblock More effective field theory for nonrelativistic scattering.
\newblock {\em Nucl. Phys. B}, 494:471--484, 1997.

\bibitem{Phillips:1997xu}
Daniel~R. Phillips, Silas~R. Beane, and Thomas~D. Cohen.
\newblock Nonperturbative regularization and renormalization: Simple examples from nonrelativistic quantum mechanics.
\newblock {\em Annals Phys.}, 263:255--275, 1998.

\bibitem{Beane:1997pk}
S.~R. Beane, T.~D. Cohen, and Daniel~R. Phillips.
\newblock The potential of effective field theory in {N} {N} scattering.
\newblock {\em Nucl. Phys. A}, 632:445--469, 1998.

\bibitem{Phillips:1999bf}
Daniel~R. Phillips, I.~R. Afnan, and A.~G. Henry-Edwards.
\newblock Numerical renormalization using dimensional regularization: {A} simple test case in the lippmann-schwinger equation.
\newblock {\em Phys. Rev. C}, 61:044002, 2000.

\bibitem{5392446}
R.~Landauer.
\newblock Irreversibility and heat generation in the computing process.
\newblock {\em IBM Journal of Research and Development}, 5(3):183--191, 1961.

\bibitem{5391327}
C.~H. Bennett.
\newblock Logical reversibility of computation.
\newblock {\em IBM Journal of Research and Development}, 17(6):525--532, 1973.

\bibitem{Manin1980}
Yuri Manin.
\newblock {Computable and Uncomputable}.
\newblock {\em Sovetskoye Radio, Moscow"}, 128, 1980.

\bibitem{Fredkin1982}
Edward Fredkin and Tommaso Toffoli.
\newblock Conservative logic.
\newblock {\em International Journal of Theoretical Physics}, 21(3-4):219--253, April 1982.

\bibitem{doi:10.1063/1.881299}
Rolf Landauer.
\newblock Information is physical.
\newblock {\em Physics Today}, 44(5):23--29, 1991.

\bibitem{williamsNASAconference}
Colin~P. Williams.
\newblock {Quantum Computing and Quantum Communications: First NASA International Conference, QCQC’98 Palm Springs, California, USA February 17–20, Selected Papers}, 1998.

\bibitem{Preskill:2021apy}
John Preskill.
\newblock {Quantum computing 40 years later}.
\newblock 6 2021.

\bibitem{Preskill2018quantumcomputingin}
John Preskill.
\newblock Quantum {C}omputing in the {NISQ} era and beyond.
\newblock {\em {Quantum}}, 2:79, August 2018.

\bibitem{Hauke:2013jga}
Philipp Hauke, David Marcos, Marcello Dalmonte, and Peter Zoller.
\newblock {Quantum simulation of a lattice Schwinger model in a chain of trapped ions}.
\newblock {\em Phys. Rev. X}, 3(4):041018, 2013.

\bibitem{Banuls:2013jaa}
M.~C. Ba\~nuls, K.~Cichy, K.~Jansen, and J.~I. Cirac.
\newblock {The mass spectrum of the Schwinger model with Matrix Product States}.
\newblock {\em JHEP}, 11:158, 2013.

\bibitem{Zohar:2016iic}
Erez Zohar, Alessandro Farace, Benni Reznik, and J.~Ignacio Cirac.
\newblock {Digital lattice gauge theories}.
\newblock {\em Phys. Rev. A}, 95(2):023604, 2017.

\bibitem{Muschik:2016tws}
Christine Muschik, Markus Heyl, Esteban Martinez, Thomas Monz, Philipp Schindler, Berit Vogell, Marcello Dalmonte, Philipp Hauke, Rainer Blatt, and Peter Zoller.
\newblock {U(1) Wilson lattice gauge theories in digital quantum simulators}.
\newblock {\em New J. Phys.}, 19(10):103020, 2017.

\bibitem{Martinez:2016yna}
Esteban~A. Martinez, Christine~A. Muschik, Philipp Schindler, Daniel Nigg, Alexander Erhard, Markus Heyl, Philipp Hauke, Marcello Dalmonte, Thomas Monz, Peter Zoller, and Rainer Blatt.
\newblock {Real-time dynamics of lattice gauge theories with a few-qubit quantum computer}.
\newblock {\em Nature}, 534:516--519, 2016.

\bibitem{Buyens:2016hhu}
Boye Buyens, Jutho Haegeman, Florian Hebenstreit, Frank Verstraete, and Karel Van~Acoleyen.
\newblock {Real-time simulation of the Schwinger effect with Matrix Product States}.
\newblock {\em Phys. Rev. D}, 96(11):114501, 2017.

\bibitem{Banuls:2016lkq}
Mari~Carmen Ba\~nuls, Krzysztof Cichy, Karl Jansen, and Hana Saito.
\newblock {Chiral condensate in the Schwinger model with Matrix Product Operators}.
\newblock {\em Phys. Rev. D}, 93(9):094512, 2016.

\bibitem{Gonzalez-Cuadra:2017lvz}
Daniel Gonz\'alez-Cuadra, Erez Zohar, and J.~Ignacio Cirac.
\newblock {Quantum Simulation of the Abelian-Higgs Lattice Gauge Theory with Ultracold Atoms}.
\newblock {\em New J. Phys.}, 19(6):063038, 2017.

\bibitem{Dumitrescu:2018njn}
E.~F. Dumitrescu, A.~J. McCaskey, G.~Hagen, G.~R. Jansen, T.~D. Morris, T.~Papenbrock, R.~C. Pooser, D.~J. Dean, and P.~Lougovski.
\newblock {Cloud Quantum Computing of an Atomic Nucleus}.
\newblock {\em Phys. Rev. Lett.}, 120(21):210501, 2018.

\bibitem{PhysRevA.98.032331}
N.~Klco, E.~F. Dumitrescu, A.~J. McCaskey, T.~D. Morris, R.~C. Pooser, M.~Sanz, E.~Solano, P.~Lougovski, and M.~J. Savage.
\newblock Quantum-classical computation of schwinger model dynamics using quantum computers.
\newblock {\em Phys. Rev. A}, 98(3):032331, Sep 2018.

\bibitem{Kaplan:2018vnj}
David~B. Kaplan and Jesse~R. Stryker.
\newblock {Gauss\textquoteright{}s law, duality, and the Hamiltonian formulation of U(1) lattice gauge theory}.
\newblock {\em Phys. Rev. D}, 102(9):094515, 2020.

\bibitem{Stryker:2018efp}
Jesse~R. Stryker.
\newblock {Oracles for Gauss's law on digital quantum computers}.
\newblock {\em Phys. Rev. A}, 99(4):042301, 2019.

\bibitem{Yeter-Aydeniz:2018mix}
Kubra Yeter-Aydeniz, Eugene~F. Dumitrescu, Alex~J. McCaskey, Ryan~S. Bennink, Raphael~C. Pooser, and George Siopsis.
\newblock Scalar quantum field theories as a benchmark for near-term quantum computers.
\newblock {\em Phys. Rev. A}, 99(3):032306, 2019.

\bibitem{PhysRevD.101.074512}
Natalie Klco, Martin~J. Savage, and Jesse~R. Stryker.
\newblock {SU(2) non-Abelian gauge field theory in one dimension on digital quantum computers}.
\newblock {\em Phys. Rev. D}, 101(7):074512, Apr 2020.

\bibitem{Avkhadiev:2019niu}
A.~Avkhadiev, P.~E. Shanahan, and R.~D. Young.
\newblock {Accelerating Lattice Quantum Field Theory Calculations via Interpolator Optimization Using Noisy Intermediate-Scale Quantum Computing}.
\newblock {\em Phys. Rev. Lett.}, 124(8):080501, 2020.

\bibitem{Bauer:2019qxa}
Christian~W. Bauer, Wibe~A. de~Jong, Benjamin Nachman, and Davide Provasoli.
\newblock {Quantum Algorithm for High Energy Physics Simulations}.
\newblock {\em Phys. Rev. Lett.}, 126(6):062001, 2021.

\bibitem{Klco:2019yrb}
Natalie Klco and Martin~J. Savage.
\newblock {Systematically Localizable Operators for Quantum Simulations of Quantum Field Theories}.
\newblock {\em Phys. Rev. A}, 102(1):012619, 2020.

\bibitem{Luo:2019vmi}
Di~Luo, Jiayu Shen, Michael Highman, Bryan~K. Clark, Brian DeMarco, Aida~X. El-Khadra, and Bryce Gadway.
\newblock {Framework for simulating gauge theories with dipolar spin systems}.
\newblock {\em Phys. Rev. A}, 102(3):032617, 2020.

\bibitem{Funcke:2019zna}
Lena Funcke, Karl Jansen, and Stefan K\"uhn.
\newblock {Topological vacuum structure of the Schwinger model with matrix product states}.
\newblock {\em Phys. Rev. D}, 101(5):054507, 2020.

\bibitem{Davoudi:2019bhy}
Zohreh Davoudi, Mohammad Hafezi, Christopher Monroe, Guido Pagano, Alireza Seif, and Andrew Shaw.
\newblock {Towards analog quantum simulations of lattice gauge theories with trapped ions}.
\newblock {\em Phys. Rev. Research}, 2(2):023015, 2020.

\bibitem{Magnifico:2019kyj}
Giuseppe Magnifico, Marcello Dalmonte, Paolo Facchi, Saverio Pascazio, Francesco~V. Pepe, and Elisa Ercolessi.
\newblock {Real Time Dynamics and Confinement in the $\mathbb{Z}_{n}$ Schwinger-Weyl lattice model for 1+1 QED}.
\newblock {\em Quantum}, 4:281, 2020.

\bibitem{Mishra:2019xbh}
Chinmay Mishra, Shane Thompson, Raphael Pooser, and George Siopsis.
\newblock Quantum computation of the massive thirring model.
\newblock {\em Quantum Sci. Technol.}, 5(3):035010, 2020.

\bibitem{Shehab:2019gfn}
Omar Shehab, Kevin~A. Landsman, Yunseong Nam, Daiwei Zhu, Norbert~M. Linke, Matthew~J. Keesan, Raphael~C. Pooser, and Christopher~R. Monroe.
\newblock {Toward convergence of effective field theory simulations on digital quantum computers}.
\newblock {\em Phys. Rev. A}, 100(6):062319, 2019.

\bibitem{Yang_2020}
Bing Yang, Hui Sun, Robert Ott, Han-Yi Wang, Torsten~V. Zache, Jad~C. Halimeh, Zhen-Sheng Yuan, Philipp Hauke, and Jian-Wei Pan.
\newblock Observation of gauge invariance in a 71-site bose–hubbard quantum simulator.
\newblock {\em Nature}, 587(7834):392–396, Nov 2020.

\bibitem{Kharzeev:2020kgc}
Dmitri~E. Kharzeev and Yuta Kikuchi.
\newblock Real-time chiral dynamics from a digital quantum simulation.
\newblock {\em Phys. Rev. Research}, 2:023342, Jun 2020.

\bibitem{Shaw2020quantumalgorithms}
Alexander~F. Shaw, Pavel Lougovski, Jesse~R. Stryker, and Nathan Wiebe.
\newblock Quantum {A}lgorithms for {S}imulating the {L}attice {S}chwinger {M}odel.
\newblock {\em {Quantum}}, 4:306, August 2020.

\bibitem{PhysRevD.103.094501}
Anthony Ciavarella, Natalie Klco, and Martin~J. Savage.
\newblock {Trailhead for quantum simulation of SU(3) Yang-Mills lattice gauge theory in the local multiplet basis}.
\newblock {\em Phys. Rev. D}, 103(9):094501, May 2021.

\bibitem{Halimeh:2020ecg}
Jad~C. Halimeh, Haifeng Lang, Julius Mildenberger, Zhang Jiang, and Philipp Hauke.
\newblock {Gauge-Symmetry Protection Using Single-Body Terms}.
\newblock {\em PRX Quantum}, 2:040311, 2021.

\bibitem{Halimeh:2020djb}
Jad~C. Halimeh, Valentin Kasper, and Philipp Hauke.
\newblock {Fate of Lattice Gauge Theories Under Decoherence}, 9 2020.

\bibitem{VanDamme:2020rur}
Maarten Van~Damme, Jad~C. Halimeh, and Philipp Hauke.
\newblock {Gauge-Symmetry Violation Quantum Phase Transition in Lattice Gauge Theories}, 10 2020.

\bibitem{Haase:2020kaj}
Jan~F. Haase, Luca Dellantonio, Alessio Celi, Danny Paulson, Angus Kan, Karl Jansen, and Christine~A. Muschik.
\newblock {A} resource efficient approach for quantum and classical simulations of gauge theories in particle physics.
\newblock {\em Quantum}, 5:393, 2021.

\bibitem{Yeter-Aydeniz:2020jte}
K\"ubra Yeter-Aydeniz, George Siopsis, and Raphael~C. Pooser.
\newblock Scattering in the ising model with the quantum lanczos algorithm.
\newblock {\em New J. Phys.}, 23(4):043033, 2021.

\bibitem{Davoudi:2021ney}
Zohreh Davoudi, Norbert~M. Linke, and Guido Pagano.
\newblock {Toward simulating quantum field theories with controlled phonon-ion dynamics: A hybrid analog-digital approach}.
\newblock {\em Phys. Rev. Research}, 3(4):043072, 2021.

\bibitem{ARahman:2021ktn}
Sarmed A~Rahman, Randy Lewis, Emanuele Mendicelli, and Sarah Powell.
\newblock {SU(2) lattice gauge theory on a quantum annealer}.
\newblock {\em Phys. Rev. D}, 104(3):034501, 2021.

\bibitem{PhysRevLett.122.050403}
T.~V. Zache, N.~Mueller, J.~T. Schneider, F.~Jendrzejewski, J.~Berges, and P.~Hauke.
\newblock {Dynamical Topological Transitions in the Massive Schwinger Model with a $\theta$ Term}.
\newblock {\em Phys. Rev. Lett.}, 122(5):050403, Feb 2019.

\bibitem{Stryker:2021asy}
Jesse~R. Stryker.
\newblock {Shearing approach to gauge invariant Trotterization}, 5 2021.

\bibitem{aidelsburger2021cold}
Monika Aidelsburger, Luca Barbiero, Alejandro Bermudez, Titas Chanda, Alexandre Dauphin, Daniel González-Cuadra, Przemysław~R. Grzybowski, Simon Hands, Fred Jendrzejewski, Johannes Jünemann, Gediminas Juzeliunas, Valentin Kasper, Angelo Piga, Shi-Ju Ran, Matteo Rizzi, Gérman Sierra, Luca Tagliacozzo, Emanuele Tirrito, Torsten~V. Zache, Jakub Zakrzewski, Erez Zohar, and Maciej Lewenstein.
\newblock Cold atoms meet lattice gauge theory.
\newblock {\em Phil. Trans. Roy. Soc. Lond. A}, 380:20210064, 2021.

\bibitem{Bauer:2021gup}
Christian~W. Bauer, Marat Freytsis, and Benjamin Nachman.
\newblock {Simulating Collider Physics on Quantum Computers Using Effective Field Theories}.
\newblock {\em Phys. Rev. Lett.}, 127(21):212001, 2021.

\bibitem{VanDamme:2021njp}
Maarten Van~Damme, Julius Mildenberger, Fabian Grusdt, Philipp Hauke, and Jad~C. Halimeh.
\newblock {Suppressing nonperturbative gauge errors in the thermodynamic limit using local pseudogenerators}.
\newblock 10 2021.

\bibitem{Halimeh:2021lnv}
Jad~C. Halimeh, Lukas Homeier, Christian Schweizer, Monika Aidelsburger, Philipp Hauke, and Fabian Grusdt.
\newblock Stabilizing lattice gauge theories through simplified local pseudo generators.
\newblock {\em Phys. Rev. Res.}, 4(3):033120, 2022.

\bibitem{Knaute:2021xna}
Johannes Knaute and Philipp Hauke.
\newblock Relativistic meson spectra on ion-trap quantum simulators.
\newblock {\em Phys. Rev. A}, 105(2):022616, 2022.

\bibitem{Halimeh:2021vzf}
Jad~C. Halimeh, Haifeng Lang, and Philipp Hauke.
\newblock Gauge protection in non-abelian lattice gauge theories.
\newblock {\em New J. Phys.}, 24(3):033015, 2022.

\bibitem{Thompson:2021eze}
Shane Thompson and George Siopsis.
\newblock Quantum computation of phase transition in the massive schwinger model.
\newblock {\em Quantum Sci. Technol.}, 7(3):035001, 2022.

\bibitem{Yeter-Aydeniz:2021mol}
K\"ubra Yeter-Aydeniz, Eleftherios Moschandreou, and George Siopsis.
\newblock Quantum imaginary-time evolution algorithm for quantum field theories with continuous variables.
\newblock {\em Phys. Rev. A}, 105(1):012412, 2022.

\bibitem{Yeter-Aydeniz:2021olz}
K\"ubra Yeter-Aydeniz, Shikha Bangar, George Siopsis, and Raphael~C. Pooser.
\newblock Collective neutrino oscillations on a quantum computer.
\newblock {\em Quant. Inf. Proc.}, 21(3):84, 2022.

\bibitem{Funcke:2021aps}
Lena Funcke, Tobias Hartung, Karl Jansen, Stefan K\"uhn, Manuel Schneider, Paolo Stornati, and Xiaoyang Wang.
\newblock Towards quantum simulations in particle physics and beyond on noisy intermediate-scale quantum devices.
\newblock {\em Phil. Trans. A. Math. Phys. Eng. Sci.}, 380(2216):20210062, 2021.

\bibitem{Zhang:2021bjq}
Jinglei Zhang, Ryan Ferguson, Stefan K\"uhn, Jan~F. Haase, C.~M. Wilson, Karl Jansen, and Christine~A. Muschik.
\newblock {Simulating gauge theories with variational quantum eigensolvers in superconducting microwave cavities}.
\newblock {\em Quantum}, 7:1148, 8 2023.

\bibitem{Rahman:2022rlg}
Sarmed A~Rahman, Randy Lewis, Emanuele Mendicelli, and Sarah Powell.
\newblock Self-mitigating trotter circuits for {SU}({2}) lattice gauge theory on a quantum computer.
\newblock {\em Phys. Rev. D}, 106(7):074502, 2022.

\bibitem{Deliyannis:2022uyh}
Plato Deliyannis, James Sud, Diana Chamaki, Zo\"e Webb-Mack, Christian~W. Bauer, and Benjamin Nachman.
\newblock Improving quantum simulation efficiency of final state radiation with dynamic quantum circuits.
\newblock {\em Phys. Rev. D}, 106(3):036007, 2022.

\bibitem{Bauer:2021gek}
Christian~W. Bauer and Dorota~M. Grabowska.
\newblock Efficient representation for simulating {U}({1}) gauge theories on digital quantum computers at all values of the coupling.
\newblock {\em Phys. Rev. D}, 107(3):L031503, 2023.

\bibitem{Illa:2022jqb}
Marc Illa and Martin~J. Savage.
\newblock Basic elements for simulations of standard-model physics with quantum annealers: Multigrid and clock states.
\newblock {\em Phys. Rev. A}, 106(5):052605, 2022.

\bibitem{Milsted:2020jmf}
Ashley Milsted, Junyu Liu, John Preskill, and Guifre Vidal.
\newblock {Collisions of False-Vacuum Bubble Walls in a Quantum Spin Chain}.
\newblock {\em PRX Quantum}, 3(2):020316, 2022.

\bibitem{Kokail:2018eiw}
Christian Kokail et~al.
\newblock Self-verifying variational quantum simulation of lattice models.
\newblock {\em Nature}, 569(7756):355--360, 2019.

\bibitem{Nguyen:2021hyk}
Nhung~H. Nguyen, Minh~C. Tran, Yingyue Zhu, Alaina~M. Green, C.~Huerta Alderete, Zohreh Davoudi, and Norbert~M. Linke.
\newblock Digital quantum simulation of the schwinger model and symmetry protection with trapped ions.
\newblock {\em PRX Quantum}, 3(2):020324, 2022.

\bibitem{Zohar:2011cw}
Erez Zohar and Benni Reznik.
\newblock {Confinement and lattice QED electric flux-tubes simulated with ultracold atoms}.
\newblock {\em Phys. Rev. Lett.}, 107:275301, 2011.

\bibitem{Zohar:2012ay}
Erez Zohar, J.~Ignacio Cirac, and Benni Reznik.
\newblock {Simulating Compact Quantum Electrodynamics with ultracold atoms: Probing confinement and nonperturbative effects}.
\newblock {\em Phys. Rev. Lett.}, 109:125302, 2012.

\bibitem{Tagliacozzo:2012vg}
L.~Tagliacozzo, A.~Celi, A.~Zamora, and M.~Lewenstein.
\newblock {Optical Abelian Lattice Gauge Theories}.
\newblock {\em Annals Phys.}, 330:160--191, 2013.

\bibitem{Zohar:2012ts}
Erez Zohar, J.~Ignacio Cirac, and Benni Reznik.
\newblock {Simulating (2+1)-Dimensional Lattice QED with Dynamical Matter Using Ultracold Atoms}.
\newblock {\em Phys. Rev. Lett.}, 110(5):055302, 2013.

\bibitem{Wiese:2013uua}
Uwe-Jens Wiese.
\newblock {Ultracold Quantum Gases and Lattice Systems: Quantum Simulation of Lattice Gauge Theories}.
\newblock {\em Annalen Phys.}, 525:777--796, 2013.

\bibitem{Marcos:2014lda}
D.~Marcos, P.~Widmer, E.~Rico, M.~Hafezi, P.~Rabl, U.~J. Wiese, and P.~Zoller.
\newblock {Two-dimensional Lattice Gauge Theories with Superconducting Quantum Circuits}.
\newblock {\em Annals Phys.}, 351:634--654, 2014.

\bibitem{Kuno:2014npa}
Yoshihito Kuno, Kenichi Kasamatsu, Yoshiro Takahashi, Ikuo Ichinose, and Tetsuo Matsui.
\newblock {Real-time dynamics and proposal for feasible experiments of lattice gauge\textendash{}Higgs model simulated by cold atoms}.
\newblock {\em New J. Phys.}, 17(6):063005, 2015.

\bibitem{Bazavov:2015kka}
Alexei Bazavov, Yannick Meurice, Shan-Wen Tsai, Judah Unmuth-Yockey, and Jin Zhang.
\newblock {Gauge-invariant implementation of the Abelian Higgs model on optical lattices}.
\newblock {\em Phys. Rev. D}, 92(7):076003, 2015.

\bibitem{Kasper:2015cca}
V.~Kasper, F.~Hebenstreit, M.~Oberthaler, and J.~Berges.
\newblock {Schwinger pair production with ultracold atoms}.
\newblock {\em Phys. Lett. B}, 760:742--746, 2016.

\bibitem{Brennen:2015pgn}
G.~K. Brennen, G.~Pupillo, E.~Rico, T.~M. Stace, and D.~Vodola.
\newblock {Loops and Strings in a Superconducting Lattice Gauge Simulator}.
\newblock {\em Phys. Rev. Lett.}, 117(24):240504, 2016.

\bibitem{Kuno:2016xbf}
Yoshihito Kuno, Shinya Sakane, Kenichi Kasamatsu, Ikuo Ichinose, and Tetsuo Matsui.
\newblock {Atomic quantum simulation of a three-dimensional U(1) gauge-Higgs model}.
\newblock {\em Phys. Rev. A}, 94(6):063641, 2016.

\bibitem{Kasper:2016mzj}
V.~Kasper, F.~Hebenstreit, F.~Jendrzejewski, M.~K. Oberthaler, and J.~Berges.
\newblock {Implementing quantum electrodynamics with ultracold atomic systems}.
\newblock {\em New J. Phys.}, 19(2):023030, 2017.

\bibitem{Ott:2020ycj}
Robert Ott, Torsten~V. Zache, Fred Jendrzejewski, and J\"urgen Berges.
\newblock {Scalable Cold-Atom Quantum Simulator for Two-Dimensional QED}.
\newblock {\em Phys. Rev. Lett.}, 127(13):130504, 2021.

\bibitem{Paulson:2020zjd}
Danny Paulson, Luca Dellantonio, Jan~F. Haase, Alessio Celi, Angus Kan, Andrew Jena, Christian Kokail, Rick van Bijnen, Karl Jansen, Peter Zoller, and Christine~A. Muschik.
\newblock {Towards simulating 2D effects in lattice gauge theories on a quantum computer}.
\newblock {\em PRX Quantum}, 2:030334, 2021.

\bibitem{Kan:2021nyu}
Angus Kan, Lena Funcke, Stefan K\"uhn, Luca Dellantonio, Jinglei Zhang, Jan~F. Haase, Christine~A. Muschik, and Karl Jansen.
\newblock {Investigating a 3+1D Topological $\theta$-Term in the Hamiltonian Formulation of Lattice Gauge Theories for Quantum and Classical Simulations}.
\newblock {\em Phys. Rev. D}, 104:034504, Aug 2021.

\bibitem{Wilson:1994fk}
Kenneth~G. Wilson, Timothy~S. Walhout, Avaroth Harindranath, Wei-Min Zhang, Robert~J. Perry, and Stanislaw~D. Glazek.
\newblock {Nonperturbative QCD: A Weak coupling treatment on the light front}.
\newblock {\em Phys. Rev. D}, 49:6720--6766, 1994.

\bibitem{Heinzl:1995jn}
T.~Heinzl.
\newblock {Hamiltonian formulations of Yang-Mills quantum theory and the Gribov problem}.
\newblock In {\em {5th Meeting on Light Cone Quantization and Nonperturbative QCD}}, 6 1995.

\bibitem{PhysRevD.31.2020}
John~B. Bronzan.
\newblock {Explicit Hamiltonian for SU(2) lattice gauge theory}.
\newblock {\em Phys. Rev. D}, 31:2020--2028, Apr 1985.

\bibitem{LIGTERINK2000983c}
N.E. Ligterink, N.R. Walet, and R.F. Bishop.
\newblock A many-body treatment of hamiltonian lattice gauge theory.
\newblock {\em Nucl. Phys. A}, 663:983--986, 2000.

\bibitem{LIGTERINK2000215}
N.E. Ligterink, N.R. Walet, and R.F. Bishop.
\newblock {Toward a Many-Body Treatment of Hamiltonian Lattice SU(N) Gauge Theory}.
\newblock {\em Annals Phys.}, 284(2):215--262, 2000.

\bibitem{PhysRevD.11.395}
John Kogut and Leonard Susskind.
\newblock Hamiltonian formulation of wilson's lattice gauge theories.
\newblock {\em Phys. Rev. D}, 11:395--408, Jan 1975.

\bibitem{PhysRevA.73.022328}
Tim Byrnes and Yoshihisa Yamamoto.
\newblock Simulating lattice gauge theories on a quantum computer.
\newblock {\em Phys. Rev. A}, 73:022328, Feb 2006.

\bibitem{Zohar:2012xf}
Erez Zohar, J.~Ignacio Cirac, and Benni Reznik.
\newblock {Cold-Atom Quantum Simulator for SU(2) Yang-Mills Lattice Gauge Theory}.
\newblock {\em Phys. Rev. Lett.}, 110(12):125304, 2013.

\bibitem{PhysRevLett.110.125303}
D.~Banerjee, M.~B\"ogli, M.~Dalmonte, E.~Rico, P.~Stebler, U.-J. Wiese, and P.~Zoller.
\newblock {Atomic Quantum Simulation of $\mathbf{U}(N)$ and $\mathrm{SU}(N)$ Non-Abelian Lattice Gauge Theories}.
\newblock {\em Phys. Rev. Lett.}, 110(12):125303, Mar 2013.

\bibitem{Banuls:2017ena}
Mari~Carmen Ba\~nuls, Krzysztof Cichy, J.~Ignacio Cirac, Karl Jansen, and Stefan K\"uhn.
\newblock {Efficient basis formulation for 1+1 dimensional SU(2) lattice gauge theory: Spectral calculations with matrix product states}.
\newblock {\em Phys. Rev. X}, 7(4):041046, 2017.

\bibitem{Atas:2021ext}
Yasar~Y. Atas, Jinglei Zhang, Randy Lewis, Amin Jahanpour, Jan~F. Haase, and Christine~A. Muschik.
\newblock {SU}({2}) hadrons on a quantum computer via a variational approach.
\newblock {\em Nature Commun.}, 12(1):6499, 2021.

\bibitem{Frishman:2010tc}
Yitzhak Frishman and Jacob Sonnenschein.
\newblock {Non-Perturbative Field Theory -- From Two Dimensional Conformal field Theory to QCD in Four Dimensions}.
\newblock 4 2010.

\bibitem{Frishman2014book}
Yitzhak Frishman and Jacob Sonnenschein.
\newblock {\em Non-Perturbative Field Theory: From Two Dimensional Conformal Field Theory to {QCD} in Four Dimensions}.
\newblock Cambridge Monographs on Mathematical Physics. Cambridge University Press; 1st edition (July 1, 2014), 4 2014.

\bibitem{Joo:2019byq}
B\'alint Jo\'o, Chulwoo Jung, Norman~H. Christ, William Detmold, Robert Edwards, Martin Savage, and Phiala Shanahan.
\newblock Status and future perspectives for lattice gauge theory calculations to the exascale and beyond.
\newblock {\em Eur. Phys. J. A}, 55(11):199, 2019.

\bibitem{Aoki:2021kgd}
Y.~Aoki et~al.
\newblock {FLAG} review {2021}.
\newblock {\em Eur. Phys. J. C}, 82(10):869, 2022.

\bibitem{osti_1369223}
Joseph Carlson, Martin~J. Savage, Richard Gerber, Katie Antypas, Deborah Bard, Richard Coffey, Eli Dart, Sudip Dosanjh, James Hack, Inder Monga, Michael~E. Papka, Katherine Riley, Lauren Rotman, Tjerk Straatsma, Jack Wells, Harut Avakian, Yassid Ayyad, Steffen~A. Bass, Daniel Bazin, Amber Boehnlein, Georg Bollen, Leah~J. Broussard, Alan Calder, Sean Couch, Aaron Couture, Mario Cromaz, William Detmold, Jason Detwiler, Huaiyu Duan, Robert Edwards, Jonathan Engel, Chris Fryer, George~M. Fuller, Stefano Gandolfi, Gagik Gavalian, Dali Georgobiani, Rajan Gupta, Vardan Gyurjyan, Marc Hausmann, Graham Heyes, W.~Ralph Hix, Mark ito, Gustav Jansen, Richard Jones, Balint Joo, Olaf Kaczmarek, Dan Kasen, Mikhail Kostin, Thorsten Kurth, Jerome Lauret, David Lawrence, Huey-Wen Lin, Meifeng Lin, Paul Mantica, Peter Maris, Bronson Messer, Wolfgang Mittig, Shea Mosby, Swagato Mukherjee, Hai~Ah Nam, Petr navratil, Witek Nazarewicz, Esmond Ng, Tommy O'Donnell, Konstantinos Orginos, Frederique Pellemoine, Peter Petreczky,
  Steven~C. Pieper, Christopher~H. Pinkenburg, Brad Plaster, R.~Jefferson Porter, Mauricio Portillo, Scott Pratt, Martin~L. Purschke, Ji~Qiang, Sofia Quaglioni, David Richards, Yves Roblin, Bjorn Schenke, Rocco Schiavilla, Soren Schlichting, Nicolas Schunck, Patrick Steinbrecher, Michael Strickland, Sergey Syritsyn, Balsa Terzic, Robert Varner, James Vary, Stefan Wild, Frank Winter, Remco Zegers, He~Zhang, Veronique Ziegler, and Michael Zingale.
\newblock Nuclear physics exascale requirements review: An office of science review sponsored jointly by advanced scientific computing research and nuclear physics, june 15 - 17, 2016, gaithersburg, maryland.
\newblock 2 2017.

\bibitem{Habib:2016sce}
Salman Habib et~al.
\newblock {ASCR/HEP Exascale Requirements Review Report}.
\newblock 3 2016.

\bibitem{Sala:2018dui}
P.~Sala, T.~Shi, S.~K\"uhn, M.~C. Ba\~nuls, E.~Demler, and J.~I. Cirac.
\newblock Variational study of {U}({1}) and {SU}({2}) lattice gauge theories with gaussian states in {1}+{1} dimensions.
\newblock {\em Phys. Rev. D}, 98(3):034505, Aug 2018.

\bibitem{Jordan:1928wi}
Pascual Jordan and Eugene~P. Wigner.
\newblock {About the Pauli exclusion principle}.
\newblock {\em Z. Phys.}, 47:631--651, 1928.

\bibitem{Peruzzo_2014}
Alberto Peruzzo, Jarrod McClean, Peter Shadbolt, Man-Hong Yung, Xiao-Qi Zhou, Peter~J. Love, Alán Aspuru-Guzik, and Jeremy~L. O’Brien.
\newblock A variational eigenvalue solver on a photonic quantum processor.
\newblock {\em Nat Commun}, 5(1):4213, Jul 2014.

\bibitem{IBMQ}
IBM.
\newblock {IBM Quantum}.
\newblock \url{https://quantum-computing.ibm.com/}, 2022, 2023.

\bibitem{Klco:2019evd}
Natalie Klco, Jesse~R. Stryker, and Martin~J. Savage.
\newblock {SU(2) non-Abelian gauge field theory in one dimension on digital quantum computers}.
\newblock {\em Phys. Rev. D}, 101(7):074512, 2020.

\bibitem{Ciavarella:2021nmj}
Anthony Ciavarella, Natalie Klco, and Martin~J. Savage.
\newblock {Trailhead for quantum simulation of SU(3) Yang-Mills lattice gauge theory in the local multiplet basis}.
\newblock {\em Phys. Rev. D}, 103(9):094501, 2021.

\bibitem{Ciavarella:2021lel}
Anthony~N. Ciavarella and Ivan~A. Chernyshev.
\newblock Preparation of the {SU}({3}) lattice yang-mills vacuum with variational quantum methods.
\newblock {\em Phys. Rev. D}, 105(7):074504, 2022.

\bibitem{Brower:1997ha}
R.~Brower, S.~Chandrasekharan, and U.~J. Wiese.
\newblock {QCD as a quantum link model}.
\newblock {\em Phys. Rev. D}, 60:094502, 1999.

\bibitem{Banerjee:2012xg}
D.~Banerjee, M.~B\"ogli, M.~Dalmonte, E.~Rico, P.~Stebler, U.~J. Wiese, and P.~Zoller.
\newblock {Atomic Quantum Simulation of U(N) and SU(N) Non-Abelian Lattice Gauge Theories}.
\newblock {\em Phys. Rev. Lett.}, 110(12):125303, 2013.

\bibitem{Tagliacozzo:2012df}
L.~Tagliacozzo, A.~Celi, P.~Orland, and M.~Lewenstein.
\newblock {Simulations of non-Abelian gauge theories with optical lattices}.
\newblock {\em Nat Commun}, 4:2615, 2013.

\bibitem{Alexandru:2019nsa}
Andrei Alexandru, Paulo~F. Bedaque, Siddhartha Harmalkar, Henry Lamm, Scott Lawrence, and Neill~C. Warrington.
\newblock {Gluon Field Digitization for Quantum Computers}.
\newblock {\em Phys. Rev. D}, 100(11):114501, 2019.

\bibitem{Ji:2020kjk}
Yao Ji, Henry Lamm, and Shuchen Zhu.
\newblock {Gluon Field Digitization via Group Space Decimation for Quantum Computers}.
\newblock {\em Phys. Rev. D}, 102(11):114513, 2020.

\bibitem{Wiese:2021djl}
Uwe-Jens Wiese.
\newblock From quantum link models to {D}-theory: a resource efficient framework for the quantum simulation and computation of gauge theories.
\newblock {\em Phil. Trans. A. Math. Phys. Eng. Sci.}, 380(2216):20210068, 2021.

\bibitem{Caspar:2022llo}
Stephan Caspar and Hersh Singh.
\newblock From asymptotic freedom to {$\theta$} vacua: Qubit embeddings of the {O}({3}) nonlinear {$\sigma$} model.
\newblock {\em Phys. Rev. Lett.}, 129(2):022003, 2022.

\bibitem{Reinhardt:1996dy}
H.~Reinhardt.
\newblock Yang-mills theory in axial gauge.
\newblock {\em Phys. Rev. D}, 55:2331--2346, 1997.

\bibitem{Shaw:2020udc}
Alexander~F. Shaw, Pavel Lougovski, Jesse~R. Stryker, and Nathan Wiebe.
\newblock {Quantum Algorithms for Simulating the Lattice Schwinger Model}.
\newblock {\em Quantum}, 4:306, 2020.

\bibitem{Kan:2021xfc}
Angus Kan and Yunseong Nam.
\newblock {Lattice Quantum Chromodynamics and Electrodynamics on a Universal Quantum Computer}.
\newblock 7 2021.

\bibitem{PhysRevD.107.054513}
Roland~C. Farrell, Ivan~A. Chernyshev, Sarah J.~M. Powell, Nikita~A. Zemlevskiy, Marc Illa, and Martin~J. Savage.
\newblock Preparations for quantum simulations of quantum chromodynamics in {$1+1$} dimensions. {II}. single-baryon $\ensuremath{\beta}$-decay in real time.
\newblock {\em Phys. Rev. D}, 107(5):054513, 3 2023.

\bibitem{Andrade:2021pil}
B\'arbara Andrade, Zohreh Davoudi, Tobias Gra\ss{}, Mohammad Hafezi, Guido Pagano, and Alireza Seif.
\newblock Engineering an effective three-spin hamiltonian in trapped-ion systems for applications in quantum simulation.
\newblock {\em Quantum Sci. Technol.}, 7(3):034001, 2022.

\bibitem{Katz:2022czu}
Or~Katz, Marko Cetina, and Christopher Monroe.
\newblock {$N$}-body interactions between trapped ion qubits via spin-dependent squeezing.
\newblock {\em Phys. Rev. Lett.}, 129(6):063603, 2022.

\bibitem{Beane:2002xf}
Silas~R. Beane and Martin~J. Savage.
\newblock The quark mass dependence of two nucleon systems.
\newblock {\em Nucl. Phys. A}, 717:91--103, 2003.

\bibitem{Epelbaum:2012iu}
Evgeny Epelbaum, Hermann Krebs, Timo~A. L\"ahde, Dean Lee, and Ulf-G. Mei\ss{}ner.
\newblock Viability of carbon-based life as a function of the light quark mass.
\newblock {\em Phys. Rev. Lett.}, 110(11):112502, 2013.

\bibitem{Berengut:2013nh}
J.~C. Berengut, E.~Epelbaum, V.~V. Flambaum, C.~Hanhart, U.~G. Meissner, J.~Nebreda, and J.~R. Pelaez.
\newblock Varying the light quark mass: impact on the nuclear force and big bang nucleosynthesis.
\newblock {\em Phys. Rev. D}, 87(8):085018, 2013.

\bibitem{Chen:1999tn}
Jiunn-Wei Chen, Gautam Rupak, and Martin~J. Savage.
\newblock Nucleon-nucleon effective field theory without pions.
\newblock {\em Nucl. Phys. A}, 653:386--412, 1999.

\bibitem{Kogut:1974ag}
John~B. Kogut and Leonard Susskind.
\newblock {Hamiltonian Formulation of Wilson's Lattice Gauge Theories}.
\newblock {\em Phys. Rev. D}, 11:395--408, 1975.

\bibitem{Banks:1975gq}
Tom Banks, Leonard Susskind, and John~B. Kogut.
\newblock Strong coupling calculations of lattice gauge theories: ({1}+{1})-dimensional exercises.
\newblock {\em Phys. Rev. D}, 13:1043, 1976.

\bibitem{PhysRev.127.1821}
R.~L. Arnowitt and S.~I. Fickler.
\newblock Quantization of the yang-mills field.
\newblock {\em Phys. Rev.}, 127:1821--1829, 9 1962.

\bibitem{weinberg1995quantum}
Steven Weinberg.
\newblock {\em The quantum theory of fields}, volume~2.
\newblock Cambridge university press, 1995.

\bibitem{1928ZPhy...47..631J}
P.~{Jordan} and E.~{Wigner}.
\newblock {{\"U}ber das Paulische {\"A}quivalenzverbot}.
\newblock {\em Zeitschrift fur Physik}, 47(9-10):631--651, September 1928.

\bibitem{PhysRev.51.106}
E.~Wigner.
\newblock On the consequences of the symmetry of the nuclear hamiltonian on the spectroscopy of nuclei.
\newblock {\em Phys. Rev.}, 51:106--119, Jan 1937.

\bibitem{PhysRev.51.947}
E.~Wigner.
\newblock On the structure of nuclei beyond oxygen.
\newblock {\em Phys. Rev.}, 51:947--958, Jun 1937.

\bibitem{PhysRev.56.519}
E.~P. Wigner.
\newblock On coupling conditions in light nuclei and the lifetimes of $\ensuremath{\beta}$-radioactivities.
\newblock {\em Phys. Rev.}, 56:519--527, Sep 1939.

\bibitem{LongRangePlan}
Donald Geesaman, Vincenzo Cirigliano, Abhay Deshpande, Frederic Fahey, John Hardy, Karsten Heeger, David Hobart, Suzanne Lapi, Jamie Nagle, Filomena Nunes, Erich Ormand, Jorge Piekarewicz, Patrizia Rossi, Jurgen Schukraft, Kate Scholberg, Matthew Shepherd, Raju Venugopalan, Michael Wiescher, and John Wilkerson.
\newblock Reaching for the horizon: The {2015} long range plan for nuclear science, 2015.

\bibitem{deFlorian:2009vb}
Daniel de~Florian, Rodolfo Sassot, Marco Stratmann, and Werner Vogelsang.
\newblock Extraction of spin-dependent parton densities and their uncertainties.
\newblock {\em Phys. Rev. D}, 80:034030, 2009.

\bibitem{Nocera:2014gqa}
Emanuele~R. Nocera, Richard~D. Ball, Stefano Forte, Giovanni Ridolfi, and Juan Rojo.
\newblock {A} first unbiased global determination of polarized {PDFs} and their uncertainties.
\newblock {\em Nucl. Phys. B}, 887:276--308, 2014.

\bibitem{COMPASS:2015mhb}
C.~Adolph et~al.
\newblock The spin structure function $g_{1}^{\rm p}$ of the proton and a test of the bjorken sum rule.
\newblock {\em Phys. Lett. B}, 753:18--28, 2016.

\bibitem{Yang:2018nqn}
Yi-Bo Yang, Jian Liang, Yu-Jiang Bi, Ying Chen, Terrence Draper, Keh-Fei Liu, and Zhaofeng Liu.
\newblock Proton mass decomposition from the {QCD} energy momentum tensor.
\newblock {\em Phys. Rev. Lett.}, 121(21):212001, 2018.

\bibitem{Alexandrou:2020sml}
C.~Alexandrou, S.~Bacchio, M.~Constantinou, J.~Finkenrath, K.~Hadjiyiannakou, K.~Jansen, G.~Koutsou, H.~Panagopoulos, and G.~Spanoudes.
\newblock Complete flavor decomposition of the spin and momentum fraction of the proton using lattice {QCD} simulations at physical pion mass.
\newblock {\em Phys. Rev. D}, 101(9):094513, 2020.

\bibitem{Ji:2021mtz}
Xiangdong Ji.
\newblock Proton mass decomposition: naturalness and interpretations.
\newblock {\em Front. Phys. (Beijing)}, 16(6):64601, 2021.

\bibitem{Wang:2021vqy}
Gen Wang, Yi-Bo Yang, Jian Liang, Terrence Draper, and Keh-Fei Liu.
\newblock Proton momentum and angular momentum decompositions with overlap fermions.
\newblock {\em Phys. Rev. D}, 106(1):014512, 2022.

\bibitem{Lorce:2021xku}
C\'edric Lorc\'e, Andreas Metz, Barbara Pasquini, and Simone Rodini.
\newblock Energy-momentum tensor in {QCD}: nucleon mass decomposition and mechanical equilibrium.
\newblock {\em JHEP}, 11:121, 2021.

\bibitem{Boer:2011fh}
Daniel Boer et~al.
\newblock {Gluons and the quark sea at high energies: Distributions, polarization, tomography}.
\newblock 8 2011.

\bibitem{Accardi:2012qut}
A.~Accardi et~al.
\newblock {Electron Ion Collider: The Next {QCD} Frontier}: {Understanding the glue that binds us all}.
\newblock {\em Eur. Phys. J. A}, 52(9):268, 2016.

\bibitem{DwaveLeap}
{D-Wave Systems Inc.}
\newblock {D}-wave leap, 2022.

\bibitem{PhysRevB.80.052506}
R.~Harris, T.~Lanting, A.~J. Berkley, J.~Johansson, M.~W. Johnson, P.~Bunyk, E.~Ladizinsky, N.~Ladizinsky, T.~Oh, and S.~Han.
\newblock Compound josephson-junction coupler for flux qubits with minimal crosstalk.
\newblock {\em Phys. Rev. B}, 80:052506, 8 2009.

\bibitem{doi:10.1021/acs.jctc.9b00402}
Alexander Teplukhin, Brian~K. Kendrick, and Dmitri Babikov.
\newblock {Calculation of Molecular Vibrational Spectra on a Quantum Annealer}.
\newblock {\em J. Chem. Theory Comput.}, 15(8):4555--4563, 2019.

\bibitem{Chang:2019}
Chia~Cheng Chang, Arjun Gambhir, Travis~S. Humble, and Shigetoshi Sota.
\newblock Quantum annealing for systems of polynomial equations.
\newblock {\em Sci. Rep.}, 9(1):10258, Jul 2019.

\bibitem{Stetina:2020abi}
Torin~F. Stetina, Anthony Ciavarella, Xiaosong Li, and Nathan Wiebe.
\newblock Simulating effective {QED} on quantum computers.
\newblock {\em Quantum}, 6:622, 1 2022.

\bibitem{Qchem2014}
M.~B. Hastings, D.~Wecker, B.~Bauer, and M.~Troyer.
\newblock Improving quantum algorithms for quantum chemistry.
\newblock {\em Quantum Inf. Comput.}, 15:1--21, 2015.

\bibitem{Kim2021ScalableEM}
Youngseok Kim, Christopher~J. Wood, Theodore~J. Yoder, Seth~T. Merkel, Jay~M. Gambetta, Kristan Temme, and Abhinav Kandala.
\newblock {Scalable error mitigation for noisy quantum circuits produces competitive expectation values}.
\newblock {\em Nature Phys.}, 19(5):752--759, 2023.

\bibitem{Symanzik:1983dc}
K.~Symanzik.
\newblock {Continuum limit and improved action in lattice theories: (I). Principles and $\phi^4$ theory}.
\newblock {\em Nucl. Phys.}, B226:187--204, 1983.

\bibitem{Symanzik:1983gh}
K.~Symanzik.
\newblock {Continuum limit and improved action in lattice theories: (II). O(N) non-linear sigma model in perturbation theory}.
\newblock {\em Nucl. Phys.}, B226:205--227, 1983.

\bibitem{Zohar:2015hwa}
Erez Zohar, J.~Ignacio Cirac, and Benni Reznik.
\newblock Quantum simulations of lattice gauge theories using ultracold atoms in optical lattices.
\newblock {\em Rept. Prog. Phys.}, 79(1):014401, 2016.

\bibitem{Dalmonte:2016alw}
M.~Dalmonte and S.~Montangero.
\newblock Lattice gauge theory simulations in the quantum information era.
\newblock {\em Contemp. Phys.}, 57(3):388--412, 2016.

\bibitem{Halimeh:2019svu}
Jad~C. Halimeh and Philipp Hauke.
\newblock {Reliability of lattice gauge theories}.
\newblock {\em Phys. Rev. Lett.}, 125(3):030503, 2020.

\bibitem{PhysRevLett.112.120406}
K.~Stannigel, P.~Hauke, D.~Marcos, M.~Hafezi, S.~Diehl, M.~Dalmonte, and P.~Zoller.
\newblock Constrained dynamics via the zeno effect in quantum simulation: Implementing non-abelian lattice gauge theories with cold atoms.
\newblock {\em Phys. Rev. Lett.}, 112:120406, Mar 2014.

\bibitem{Kasper:2020owz}
Valentin Kasper, Torsten~V. Zache, Fred Jendrzejewski, Maciej Lewenstein, and Erez Zohar.
\newblock {Non-Abelian gauge invariance from dynamical decoupling}.
\newblock {\em Phys. Rev. D}, 107:014506, 2023.

\bibitem{Feynman:1981tf}
Richard~P. Feynman.
\newblock {Simulating physics with computers}.
\newblock {\em Int. J. Theor. Phys.}, 21:467--488, 1982.

\bibitem{DiVincenzo2000ThePI}
David~P. DiVincenzo.
\newblock {The Physical Implementation of Quantum Computation}.
\newblock {\em Fortschritte der Physik: Progress of Physics}, 48:771--783, 2000.

\bibitem{Childs_2021}
Andrew~M. Childs, Yuan Su, Minh~C. Tran, Nathan Wiebe, and Shuchen Zhu.
\newblock Theory of trotter error with commutator scaling.
\newblock {\em Physical Review X}, 11(1):011020, 2 2021.

\bibitem{https://doi.org/10.48550/arxiv.quant-ph/0701004}
Mark~R. Dowling and Michael~A. Nielsen.
\newblock The geometry of quantum computation, 2007.

\bibitem{doi:10.1126/science.1121541}
Michael~A. Nielsen, Mark~R. Dowling, Mile Gu, and Andrew~C. Doherty.
\newblock Quantum computation as geometry.
\newblock {\em Science}, 311(5764):1133--1135, 2006.

\bibitem{https://doi.org/10.48550/arxiv.quant-ph/0502070}
Michael~A. Nielsen.
\newblock {A} geometric approach to quantum circuit lower bounds, 2005.

\bibitem{Jefferson:2017sdb}
Ro~Jefferson and Robert~C. Myers.
\newblock Circuit complexity in quantum field theory.
\newblock {\em JHEP}, 10:107, 2017.

\bibitem{PhysRevA.94.052325}
Joel~J. Wallman and Joseph Emerson.
\newblock Noise tailoring for scalable quantum computation via randomized compiling.
\newblock {\em Phys. Rev. A}, 94:052325, 11 2016.

\bibitem{Urbanek:2021oej}
Miroslav Urbanek, Benjamin Nachman, Vincent~R. Pascuzzi, Andre He, Christian~W. Bauer, and Wibe~A. de~Jong.
\newblock Mitigating depolarizing noise on quantum computers with noise-estimation circuits.
\newblock {\em Phys. Rev. Lett.}, 127:270502, 12 2021.

\bibitem{PhysRevA.58.2733}
Lorenza Viola and Seth Lloyd.
\newblock Dynamical suppression of decoherence in two-state quantum systems.
\newblock {\em Phys. Rev. A}, 58:2733--2744, 10 1998.

\bibitem{DUAN1999139}
Lu-Ming Duan and Guang-Can Guo.
\newblock Suppressing environmental noise in quantum computation through pulse control.
\newblock {\em Phys. Lett. A}, 261(3):139--144, 1999.

\bibitem{ZANARDI199977}
Paolo Zanardi.
\newblock Symmetrizing evolutions.
\newblock {\em Phys. Lett. A}, 258(2):77--82, 1999.

\bibitem{PhysRevLett.82.2417}
Lorenza Viola, Emanuel Knill, and Seth Lloyd.
\newblock Dynamical decoupling of open quantum systems.
\newblock {\em Phys. Rev. Lett.}, 82:2417--2421, 3 1999.

\bibitem{Yeter-Aydeniz:2022vuy}
K\"ubra Yeter-Aydeniz, Zachary Parks, Aadithya Nair, Erik Gustafson, Alexander~F. Kemper, Raphael~C. Pooser, Yannick Meurice, and Patrick Dreher.
\newblock Measuring qubit stability in a gate-based {NISQ} hardware processor.
\newblock {\em Quantum Inf Process}, 22(2):96, 2023.

\bibitem{Wagman:2016bam}
Michael~L. Wagman and Martin~J. Savage.
\newblock Statistics of baryon correlation functions in lattice {QCD}.
\newblock {\em Phys. Rev. D}, 96(11):114508, 2017.

\bibitem{Wagman:2017gqi}
Michael~L. Wagman.
\newblock {\em Statistical Angles on the Lattice {QCD} Signal-to-Noise Problem}.
\newblock PhD thesis, U. Washington, Seattle (main), 2017.

\bibitem{RevModPhys.54.407}
Laurence~G. Yaffe.
\newblock Large $n$ limits as classical mechanics.
\newblock {\em Rev. Mod. Phys.}, 54:407--435, Apr 1982.

\bibitem{Aoude:2020mlg}
Rafael Aoude, Ming-Zhi Chung, Yu-tin Huang, Camila~S. Machado, and Man-Kuan Tam.
\newblock {Silence of Binary Kerr Black Holes}.
\newblock {\em Phys. Rev. Lett.}, 125(18):181602, 2020.

\bibitem{Cervera-Lierta:2017tdt}
Alba Cervera-Lierta, Jos\'e~I. Latorre, Juan Rojo, and Luca Rottoli.
\newblock {Maximal Entanglement in High Energy Physics}.
\newblock {\em SciPost Phys.}, 3(5):036, 2017.

\bibitem{Zewail:1999}
{The Nobel Prize Organization}.
\newblock The nobel prize in chemistry {1999}, 1999.
\newblock Accessed = 2022-09-05.

\bibitem{Durrani:2020}
{Jamie Durrani}.
\newblock Caught on camera the transition state is giving up its secrets, 2020.
\newblock Accessed = 2022-09-05.

\bibitem{Cabibbo:1963yz}
Nicola Cabibbo.
\newblock Unitary symmetry and leptonic decays.
\newblock {\em Phys. Rev. Lett.}, 10:531--533, 1963.

\bibitem{Kobayashi:1973fv}
Makoto Kobayashi and Toshihide Maskawa.
\newblock {CP} violation in the renormalizable theory of weak interaction.
\newblock {\em Prog. Theor. Phys.}, 49:652--657, 1973.

\bibitem{Gonz_lez_Alonso_2019}
M.~Gonz{\'{a}}lez-Alonso, O.~Naviliat-Cuncic, and N.~Severijns.
\newblock New physics searches in nuclear and neutron beta-decay.
\newblock {\em Prog. Part. Nucl. Phys.}, 104:165--223, 1 2019.

\bibitem{Hassan:2020hrj}
M.~T. Hassan et~al.
\newblock Measurement of the neutron decay electron-antineutrino angular correlation by the {aCORN} experiment.
\newblock {\em Phys. Rev. C}, 103(4):045502, 2021.

\bibitem{Algora:2021}
A.~Algora, J.~L. Tain, B.~Rubio, M.~Fallot, and W.~Gelletly.
\newblock Beta-decay studies for applied and basic nuclear physics.
\newblock {\em Eur. Phys. J. A}, 57(3):85, 2021.

\bibitem{PhysRevC.102.045501}
J.~C. Hardy and I.~S. Towner.
\newblock Superallowed ${0}^{+}\ensuremath{\rightarrow}{0}^{+}$ nuclear $\ensuremath{\beta}$ decays: {2020} critical survey, with implications for ${V}_{\mathit{ud}}$ and {CKM} unitarity.
\newblock {\em Phys. Rev. C}, 102:045501, 10 2020.

\bibitem{Feynman:1958ty}
R.~P. Feynman and Murray Gell-Mann.
\newblock Theory of fermi interaction.
\newblock {\em Phys. Rev.}, 109:193--198, 1958.

\bibitem{Baroni_2016}
A.~Baroni, L.~Girlanda, S.~Pastore, R.~Schiavilla, and M.~Viviani.
\newblock Nuclear axial currents in chiral effective field theory.
\newblock {\em Phys. Rev. C}, 93(1):015501, 2016.
\newblock [Erratum: Phys.Rev.C 93, 049902 (2016), Erratum: Phys.Rev.C 95, 059901 (2017)].

\bibitem{Krebs:2016rqz}
H.~Krebs, E.~Epelbaum, and U.~G. Mei\ss{}ner.
\newblock Nuclear axial current operators to fourth order in chiral effective field theory.
\newblock {\em Annals Phys.}, 378:317--395, 2017.

\bibitem{Gysbers:2019uyb}
P.~Gysbers, G.~Hagen, J.~D. Holt, G.~R. Jansen, T.~D. Morris, P.~Navr{\'{a}}til, T.~Papenbrock, S.~Quaglioni, A.~Schwenk, S.~R. Stroberg, and K.~A. Wendt.
\newblock Discrepancy between experimental and theoretical {$\beta$}-decay rates resolved from first principles.
\newblock {\em Nature Phys.}, 15(5):428--431, 2019.

\bibitem{Baroni_2021}
Alessandro Baroni, Garrett~B. King, and Saori Pastore.
\newblock Electroweak currents from chiral effective field theory.
\newblock {\em Few-Body Syst.}, 62(4):114, 2021.

\bibitem{Butler:1999sv}
Malcolm Butler and Jiunn-Wei Chen.
\newblock Elastic and inelastic neutrino deuteron scattering in effective field theory.
\newblock {\em Nucl. Phys. A}, 675:575--600, 2000.

\bibitem{Butler:2002cw}
Malcolm Butler, Jiunn-Wei Chen, and Petr Vogel.
\newblock Constraints on two-body axial currents from reactor anti-neutrino deuteron breakup reactions.
\newblock {\em Phys. Lett. B}, 549:26--31, 2002.

\bibitem{Baroni:2016xll}
A.~Baroni, L.~Girlanda, A.~Kievsky, L.~E. Marcucci, R.~Schiavilla, and M.~Viviani.
\newblock Tritium {$\beta$}-decay in chiral effective field theory.
\newblock {\em Phys. Rev. C}, 94(2):024003, 2016.
\newblock [Erratum: Phys.Rev.C 95, 059902 (2017)].

\bibitem{Li:2017udr}
Yan-Ling Li, Yong-Liang Ma, and Mannque Rho.
\newblock Nuclear axial currents from scale-chiral effective field theory.
\newblock {\em Chin. Phys. C}, 42(9):094102, 2018.

\bibitem{Baroni:2018fdn}
A.~Baroni et~al.
\newblock Local chiral interactions, the tritium gamow-teller matrix element, and the three-nucleon contact term.
\newblock {\em Phys. Rev. C}, 98(4):044003, 2018.

\bibitem{Grinstein:1997gv}
Benjamin Grinstein and Ira~Z. Rothstein.
\newblock {Effective field theory and matching in nonrelativistic gauge theories}.
\newblock {\em Phys. Rev. D}, 57:78--82, 1998.

\bibitem{Luke:1997ys}
Michael~E. Luke and Martin~J. Savage.
\newblock {Power counting in dimensionally regularized NRQCD}.
\newblock {\em Phys. Rev. D}, 57:413--423, 1998.

\bibitem{Parreno:2021ovq}
Assumpta Parre\~no, Phiala~E. Shanahan, Michael~L. Wagman, Frank Winter, Emmanuel Chang, William Detmold, and Marc Illa.
\newblock Axial charge of the triton from lattice {QCD}.
\newblock {\em Phys. Rev. D}, 103(7):074511, 2021.

\bibitem{Dolinski_2019}
Michelle~J. Dolinski, Alan W.~P. Poon, and Werner Rodejohann.
\newblock Neutrinoless double-beta decay: Status and prospects.
\newblock {\em Ann. Rev. Nucl. Part. Sci.}, 69(1):219--251, 10 2019.

\bibitem{Savage:1998yh}
Martin~J. Savage.
\newblock Pionic matrix elements in neutrinoless double-beta decay.
\newblock {\em Phys. Rev. C}, 59:2293--2296, 1999.

\bibitem{Shanahan:2017bgi}
Phiala~E. Shanahan, Brian~C. Tiburzi, Michael~L. Wagman, Frank Winter, Emmanuel Chang, Zohreh Davoudi, William Detmold, Kostas Orginos, and Martin~J. Savage.
\newblock {Isotensor Axial Polarizability and Lattice QCD Input for Nuclear Double-$\beta$ Decay Phenomenology}.
\newblock {\em Phys. Rev. Lett.}, 119(6):062003, 2017.

\bibitem{Tiburzi:2017iux}
Brian~C. Tiburzi, Michael~L. Wagman, Frank Winter, Emmanuel Chang, Zohreh Davoudi, William Detmold, Kostas Orginos, Martin~J. Savage, and Phiala~E. Shanahan.
\newblock {Double-$\beta$ Decay Matrix Elements from Lattice Quantum Chromodynamics}.
\newblock {\em Phys. Rev. D}, 96(5):054505, 2017.

\bibitem{Cirigliano:2018hja}
Vincenzo Cirigliano, Wouter Dekens, Jordy De~Vries, Michael~L. Graesser, Emanuele Mereghetti, Saori Pastore, and Ubirajara Van~Kolck.
\newblock {New Leading Contribution to Neutrinoless Double-\ensuremath{\beta} Decay}.
\newblock {\em Phys. Rev. Lett.}, 120(20):202001, 2018.

\bibitem{Cirigliano:2019vdj}
V.~Cirigliano, W.~Dekens, J.~De~Vries, M.~L. Graesser, E.~Mereghetti, S.~Pastore, M.~Piarulli, U.~Van~Kolck, and R.~B. Wiringa.
\newblock {Renormalized approach to neutrinoless double- $\beta$ decay}.
\newblock {\em Phys. Rev. C}, 100(5):055504, 2019.

\bibitem{Monge-Camacho:2019nby}
Henry Monge-Camacho, Evan Berkowitz, David Brantley, Chia~Cheng Chang, M.~A. Clark, Arjun Gambhir, Nicolas Garrón, Bálint Joó, Thorsten Kurth, Amy Nicholson, Enrico Rinaldi, Brian Tiburzi, Pavlos Vranas, and André Walker-Loud.
\newblock Short range operator contributions to {$0\nu\beta\beta$} decay from {LQCD}.
\newblock {\em PoS}, LATTICE2018:263, 2019.

\bibitem{Davoudi:2021noh}
Zohreh Davoudi and Saurabh~V. Kadam.
\newblock Extraction of low-energy constants of single- and double-\ensuremath{\beta} decays from lattice {QCD}: {A} sensitivity analysis.
\newblock {\em Phys. Rev. D}, 105(9):094502, 2022.

\bibitem{Cirigliano:2022oqy}
A.~van Kolck, {V. Cirigliano, Z. Davoudi, W. Dekens, J. de Vries, J. Engel, X. Feng, J. Gehrlein, M. L. Graesser, L. Gráf, H. Hergert, L. Jin, E. Mereghetti, A. Nicholson, S. Pastore, M. J. Ramsey-Musolf, R. Ruiz, M. Spinrath, U. and Walker-Loud}.
\newblock Neutrinoless double-beta decay: {A} roadmap for matching theory to experiment, 3 2022.

\bibitem{Detmold:2022jwu}
William Detmold, William~I. Jay, David~J. Murphy, Patrick~R. Oare, and Phiala~E. Shanahan.
\newblock Neutrinoless double beta decay from lattice {QCD}: The short-distance $\pi^-\rightarrow\pi^+ e^- e^-$ amplitude, 8 2022.

\bibitem{Cirigliano:2022rmf}
V.~Cirigliano et~al.
\newblock Towards precise and accurate calculations of neutrinoless double-beta decay: Project scoping workshop report.
\newblock {\em J. Phys. G}, 49(12):120502, 2022.

\bibitem{Sudarshan:1958vf}
E.~C.~G. Sudarshan and R.~e. Marshak.
\newblock Chirality invariance and the universal fermi interaction.
\newblock {\em Phys. Rev.}, 109:1860--1860, 1958.

\bibitem{Atas:2022dqm}
Yasar~Y. Atas, Jan~F. Haase, Jinglei Zhang, Victor Wei, Sieglinde M.~L. Pfaendler, Randy Lewis, and Christine~A. Muschik.
\newblock Real-time evolution of {SU}({3}) hadrons on a quantum computer, 7 2022.

\bibitem{Weinberg:1979sa}
Steven Weinberg.
\newblock Baryon and lepton nonconserving processes.
\newblock {\em Phys. Rev. Lett.}, 43:1566--1570, 1979.

\bibitem{QuantHoney}
Quantinuum.
\newblock Quantinuum.
\newblock \url{https://www.quantinuum.com/}, 2022.

\bibitem{h1-1e}
Quantinuum.
\newblock Quantinuum system model {H1} emulator product data sheet, version {5}.{00}, 6 2022.

\bibitem{Ghosh:2015iwa}
Sudip Ghosh, Ronak~M Soni, and Sandip~P. Trivedi.
\newblock On the entanglement entropy for gauge theories.
\newblock {\em JHEP}, 09:069, 2015.

\bibitem{Soni:2015yga}
Ronak~M Soni and Sandip~P. Trivedi.
\newblock Aspects of entanglement entropy for gauge theories.
\newblock {\em JHEP}, 01:136, 2016.

\bibitem{Panizza:2022gvd}
Veronica Panizza, Ricardo~Costa de~Almeida, and Philipp Hauke.
\newblock Entanglement witnessing for lattice gauge theories.
\newblock {\em JHEP}, 09:196, 2022.

\bibitem{Rigobello:2021fxw}
Marco Rigobello, Simone Notarnicola, Giuseppe Magnifico, and Simone Montangero.
\newblock Entanglement generation in ({1}+{1}){D} {QED} scattering processes.
\newblock {\em Phys. Rev. D}, 104(11):114501, 2021.

\bibitem{Pelofske:2022vyy}
Elijah Pelofske, Andreas B\"artschi, and Stephan Eidenbenz.
\newblock Quantum volume in practice: What users can expect from {NISQ} devices.
\newblock {\em IEEE Trans. Quantum Eng.}, 3:1--19, 2022.

\bibitem{Kamakari:2021nmf}
Hirsh Kamakari, Shi-Ning Sun, Mario Motta, and Austin~J. Minnich.
\newblock Digital quantum simulation of open quantum systems using quantum imaginary time evolution.
\newblock {\em PRX Quantum}, 3:010320, 2 2022.

\bibitem{Hubisz:2020vhx}
Jay Hubisz, Bharath Sambasivam, and Judah Unmuth-Yockey.
\newblock Quantum algorithms for open lattice field theory.
\newblock {\em Phys. Rev. A}, 104(5):052420, 2021.

\bibitem{Turro:2021vbk}
Francesco Turro, Alessandro Roggero, Valentina Amitrano, Piero Luchi, Kyle~A. Wendt, Jonathan~L. DuBois, Sofia Quaglioni, and Francesco Pederiva.
\newblock Imaginary-time propagation on a quantum chip.
\newblock {\em Phys. Rev. A}, 105(2):022440, 2022.

\bibitem{Workman:2022ynf}
R.~L. Workman et~al.
\newblock Review of particle physics.
\newblock {\em Prog. Theor. Exp. Phys.}, 2022:083C01, 2022.

\bibitem{Luscher:1985dn}
M.~Luscher.
\newblock {Volume Dependence of the Energy Spectrum in Massive Quantum Field Theories. 1. Stable Particle States}.
\newblock {\em Commun. Math. Phys.}, 104:177, 1986.

\bibitem{Luscher:1986pf}
M.~L{\"u}scher.
\newblock {Volume Dependence of the Energy Spectrum in Massive Quantum Field Theories. 2. Scattering States}.
\newblock {\em Commun. Math. Phys.}, 105:153--188, 1986.

\bibitem{Luscher:1990ux}
M.~Lüscher.
\newblock Two particle states on a torus and their relation to the scattering matrix.
\newblock {\em Nucl. Phys. B}, 354:531, 1991.

\bibitem{Lellouch:2000pv}
Laurent Lellouch and Martin L\"{u}scher.
\newblock Weak transition matrix elements from finite volume correlation functions.
\newblock {\em Commun. Math. Phys.}, 219:31--44, 2001.

\bibitem{Detmold:2004qn}
William Detmold and Martin~J. Savage.
\newblock Electroweak matrix elements in the two nucleon sector from lattice {QCD}.
\newblock {\em Nucl. Phys. A}, 743:170--193, 2004.

\bibitem{Christ:2005gi}
Norman~H. Christ, Changhoan Kim, and Takeshi Yamazaki.
\newblock Finite volume corrections to the two-particle decay of states with non-zero momentum.
\newblock {\em Phys. Rev. D}, 72:114506, 2005.

\bibitem{Kim:2005gf}
C.~h. Kim, C.~T. Sachrajda, and Stephen~R. Sharpe.
\newblock Finite-volume effects for two-hadron states in moving frames.
\newblock {\em Nucl. Phys. B}, 727:218--243, 2005.

\bibitem{Hansen:2012tf}
Maxwell~T. Hansen and Stephen~R. Sharpe.
\newblock Multiple-channel generalization of lellouch-lüscher formula.
\newblock {\em Phys. Rev. D}, 86:016007, 2012.

\bibitem{Meyer:2011um}
Harvey~B. Meyer.
\newblock Lattice {QCD} and the timelike pion form factor.
\newblock {\em Phys. Rev. Lett.}, 107:072002, 2011.

\bibitem{Briceno:2012yi}
Ra\'ul~A. Brice\~no and Zohreh Davoudi.
\newblock Moving multichannel systems in a finite volume with application to proton-proton fusion.
\newblock {\em Phys. Rev. D}, 88(9):094507, 2013.

\bibitem{Feng:2014gba}
Xu~Feng, Sinya Aoki, Shoji Hashimoto, and Takashi Kaneko.
\newblock Timelike pion form factor in lattice {QCD}.
\newblock {\em Phys. Rev. D}, 91(5):054504, 2015.

\bibitem{Meyer:2012wk}
Harvey~B. Meyer.
\newblock Photodisintegration of a bound state on the torus, 2 2012.

\bibitem{Bernard:2012bi}
V.~Bernard, D.~Hoja, U.~G. Mei{\ss}ner, and A.~Rusetsky.
\newblock Matrix elements of unstable states.
\newblock {\em JHEP}, 09:023, 2012.

\bibitem{Agadjanov:2014kha}
A.~Agadjanov, V.~Bernard, U.~G. Mei{\ss}ner, and A.~Rusetsky.
\newblock {A} framework for the calculation of the $\delta {N} \gamma^*$ transition form factors on the lattice.
\newblock {\em Nucl. Phys. B}, 886:1199--1222, 2014.

\bibitem{Briceno:2014uqa}
Ra\'ul~A. Brice\~no, Maxwell~T. Hansen, and Andr\'e Walker-Loud.
\newblock Multichannel {1} {$\rightarrow$} {2} transition amplitudes in a finite volume.
\newblock {\em Phys. Rev. D}, 91(3):034501, 2015.

\bibitem{Briceno:2015csa}
Ra\'ul~A. Brice\~no and Maxwell~T. Hansen.
\newblock Multichannel {0} {$\to$} {2} and {1} {$\to$} {2} transition amplitudes for arbitrary spin particles in a finite volume.
\newblock {\em Phys. Rev. D}, 92(7):074509, 2015.

\bibitem{Briceno:2015tza}
Ra\'ul~A. Brice\~no and Maxwell~T. Hansen.
\newblock Relativistic, model-independent, multichannel ${2}\to {2}$ transition amplitudes in a finite volume.
\newblock {\em Phys. Rev. D}, 94(1):013008, 2016.

\bibitem{Briceno:2019opb}
Ra\'ul~A. Brice\~no, Zohreh Davoudi, Maxwell~T. Hansen, Matthias~R. Schindler, and Alessandro Baroni.
\newblock Long-range electroweak amplitudes of single hadrons from euclidean finite-volume correlation functions.
\newblock {\em Phys. Rev. D}, 101(1):014509, 2020.

\bibitem{Briceno:2020vgp}
Ra\'ul~A. Brice\~no, Andrew~W. Jackura, Felipe~G. Ortega-Gama, and Keegan~H. Sherman.
\newblock On-shell representations of two-body transition amplitudes: Single external current.
\newblock {\em Phys. Rev. D}, 103(11):114512, 2021.

\bibitem{Beane:2003da}
S.~R. Beane, P.~F. Bedaque, A.~Parre\~no, and M.~J. Savage.
\newblock Two nucleons on a lattice.
\newblock {\em Phys. Lett. B}, 585:106--114, 2004.

\bibitem{Beane:2007qr}
Silas~R. Beane, William Detmold, and Martin~J. Savage.
\newblock n-boson energies at finite volume and three-boson interactions.
\newblock {\em Phys. Rev. D}, 76:074507, 2007.

\bibitem{Beane:2007es}
Silas~R. Beane, William Detmold, Thomas~C. Luu, Kostas Orginos, Martin~J. Savage, and Aaron Torok.
\newblock Multi-pion systems in lattice {QCD} and the three-pion interaction.
\newblock {\em Phys. Rev. Lett.}, 100:082004, 2008.

\bibitem{Luu:2010hw}
Thomas Luu, Martin~J. Savage, Achim Schwenk, and James~P. Vary.
\newblock Nucleon-nucleon scattering in a harmonic potential.
\newblock {\em Phys. Rev. C}, 82:034003, 2010.

\bibitem{Luu:2011ep}
Thomas Luu and Martin~J. Savage.
\newblock Extracting scattering phase-shifts in higher partial-waves from lattice {QCD} calculations.
\newblock {\em Phys. Rev. D}, 83:114508, 2011.

\bibitem{Davoudi:2011md}
Zohreh Davoudi and Martin~J. Savage.
\newblock Improving the volume dependence of two-body binding energies calculated with lattice {QCD}.
\newblock {\em Phys. Rev. D}, 84:114502, 2011.

\bibitem{Briceno:2012rv}
Ra\'ul~A. Brice\~no and Zohreh Davoudi.
\newblock Three-particle scattering amplitudes from a finite volume formalism.
\newblock {\em Phys. Rev. D}, 87(9):094507, 2013.

\bibitem{Briceno:2013lba}
Ra\'ul~A. Brice\~no, Zohreh Davoudi, and Thomas~C. Luu.
\newblock Two-nucleon systems in a finite volume: ({I}) quantization conditions.
\newblock {\em Phys. Rev. D}, 88(3):034502, 2013.

\bibitem{Briceno:2013bda}
Ra\'ul~A. Brice\~no, Zohreh Davoudi, Thomas Luu, and Martin~J. Savage.
\newblock Two-nucleon systems in a finite volume. {II}. $^3s_{1}-^3d_{1}$ coupled channels and the deuteron.
\newblock {\em Phys. Rev. D}, 88(11):114507, 2013.

\bibitem{Briceno:2013hya}
Ra\'ul~A. Brice\~no, Zohreh Davoudi, Thomas~C. Luu, and Martin~J. Savage.
\newblock Two-baryon systems with twisted boundary conditions.
\newblock {\em Phys. Rev. D}, 89(7):074509, 2014.

\bibitem{Briceno:2014oea}
Ra\'ul~A. Brice\~no.
\newblock Two-particle multichannel systems in a finite volume with arbitrary spin.
\newblock {\em Phys. Rev. D}, 89(7):074507, 2014.

\bibitem{Grabowska:2021qqz}
Dorota~Maria Grabowska and Maxwell~T. Hansen.
\newblock Analytic expansions of two- and three-particle excited-state energies.
\newblock {\em PoS}, LATTICE2021:203, 2022.

\bibitem{Maiani:1990ca}
L.~Maiani and M.~Testa.
\newblock Final state interactions from euclidean correlation functions.
\newblock {\em Phys. Lett. B}, 245:585, 1990.

\bibitem{PhysRevD.103.014506}
Ra\'ul~A. Brice\~no, Juan~V. Guerrero, Maxwell~T. Hansen, and Alexandru~M. Sturzu.
\newblock Role of boundary conditions in quantum computations of scattering observables.
\newblock {\em Phys. Rev. D}, 103:014506, Jan 2021.

\bibitem{Urbanowski_2017}
K.~Urbanowski.
\newblock True quantum face of the \textquotedblleft{}exponential\textquotedblright{} decay law.
\newblock {\em Eur. Phys. J. D}, 71(5):118, 2017.

\bibitem{Giacosa_2017}
Francesco Giacosa.
\newblock Time evolution of an unstable quantum system.
\newblock {\em Acta Phys. Polon. B}, 48:1831, 2017.

\bibitem{Katz:2022ajk}
Or~Katz, Lei Feng, Andrew Risinger, Christopher Monroe, and Marko Cetina.
\newblock Demonstration of three- and four-body interactions between trapped-ion spins, 9 2022.

\bibitem{sivarajah2020t}
Seyon Sivarajah, Silas Dilkes, Alexander Cowtan, Will Simmons, Alec Edgington, and Ross Duncan.
\newblock t{$\vert$}ket{$\rangle$}: a retargetable compiler for {NISQ} devices.
\newblock {\em Quantum Sci. Technol.}, 6(1):014003, 2020.

\bibitem{Guan:2020bdl}
Wen Guan, Gabriel Perdue, Arthur Pesah, Maria Schuld, Koji Terashi, Sofia Vallecorsa, and Jean-Roch Vlimant.
\newblock {Quantum Machine Learning in High Energy Physics}.
\newblock {\em Mach. Learn. Sci. Tech.}, 2:011003, 2021.

\bibitem{Delgado:2022tpc}
Andrea Delgado et~al.
\newblock Quantum computing for data analysis in high-energy physics.
\newblock In {\em {Snowmass 2021}}, 3 2022.

\bibitem{Beck:2023xhh}
Douglas Beck et~al.
\newblock {Quantum Information Science and Technology for Nuclear Physics. Input into U.S. Long-Range Planning, 2023}.
\newblock 2 2023.

\bibitem{DiMeglio:2023nsa}
Alberto Di~Meglio et~al.
\newblock Quantum computing for high-energy physics: State of the art and challenges. summary of the {QC4HEP} working group, 7 2023.

\bibitem{10.1007/978-3-540-30538-5_31}
Julia Kempe, Alexei Kitaev, and Oded Regev.
\newblock The complexity of the local hamiltonian problem.
\newblock In Kamal Lodaya and Meena Mahajan, editors, {\em FSTTCS 2004: Foundations of Software Technology and Theoretical Computer Science}, pages 372--383, Berlin, Heidelberg, 2005. Springer Berlin Heidelberg.

\bibitem{Chakraborty:2020uhf}
Bipasha Chakraborty, Masazumi Honda, Taku Izubuchi, Yuta Kikuchi, and Akio Tomiya.
\newblock Classically emulated digital quantum simulation of the schwinger model with a topological term via adiabatic state preparation.
\newblock {\em Phys. Rev. D}, 105(9):094503, 2022.

\bibitem{Schwinger:1962tp}
Julian~S. Schwinger.
\newblock Gauge invariance and mass. {II}.
\newblock {\em Phys. Rev.}, 128:2425--2429, 1962.

\bibitem{Mueller:2022xbg}
Niklas Mueller, Joseph~A. Carolan, Andrew Connelly, Zohreh Davoudi, Eugene~F. Dumitrescu, and K\"ubra Yeter-Aydeniz.
\newblock Quantum computation of dynamical quantum phase transitions and entanglement tomography in a lattice gauge theory, 10 2022.

\bibitem{Surace:2019dtp}
Federica~M. Surace, Paolo~P. Mazza, Giuliano Giudici, Alessio Lerose, Andrea Gambassi, and Marcello Dalmonte.
\newblock {Lattice gauge theories and string dynamics in Rydberg atom quantum simulators}.
\newblock {\em Phys. Rev. X}, 10(2):021041, 2020.

\bibitem{Riechert:2021ink}
Hannes Riechert, Jad~C. Halimeh, Valentin Kasper, Landry Bretheau, Erez Zohar, Philipp Hauke, and Fred Jendrzejewski.
\newblock Engineering a {U}({1}) lattice gauge theory in classical electric circuits.
\newblock {\em Phys. Rev. B}, 105(20):205141, 2022.

\bibitem{Banerjee:2012pg}
D.~Banerjee, M.~Dalmonte, M.~Muller, E.~Rico, P.~Stebler, U.~J. Wiese, and P.~Zoller.
\newblock Atomic quantum simulation of dynamical gauge fields coupled to fermionic matter: From string breaking to evolution after a quench.
\newblock {\em Phys. Rev. Lett.}, 109:175302, 2012.

\bibitem{Marcos:2013aya}
D.~Marcos, P.~Rabl, E.~Rico, and P.~Zoller.
\newblock Superconducting circuits for quantum simulation of dynamical gauge fields.
\newblock {\em Phys. Rev. Lett.}, 111(11):110504, 2013.

\bibitem{Yang:2016hjn}
Dayou Yang, Gouri~Shankar Giri, Michael Johanning, Christof Wunderlich, Peter Zoller, and Philipp Hauke.
\newblock Analog quantum simulation of ({1}+{1})-dimensional lattice {QED} with trapped ions.
\newblock {\em Phys. Rev. A}, 94(5):052321, 2016.

\bibitem{Ferguson:2020qyf}
R.~R. Ferguson, L.~Dellantonio, A.~Al~Balushi, K.~Jansen, W.~D\"ur, and C.~A. Muschik.
\newblock {Measurement-Based Variational Quantum Eigensolver}.
\newblock {\em Phys. Rev. Lett.}, 126(22):220501, 2021.

\bibitem{Yamamoto:2021vxp}
Arata Yamamoto.
\newblock Quantum variational approach to lattice gauge theory at nonzero density.
\newblock {\em Phys. Rev. D}, 104(1):014506, 2021.

\bibitem{Honda:2021aum}
Masazumi Honda, Etsuko Itou, Yuta Kikuchi, Lento Nagano, and Takuya Okuda.
\newblock Classically emulated digital quantum simulation for screening and confinement in the schwinger model with a topological term.
\newblock {\em Phys. Rev. D}, 105(1):014504, 2022.

\bibitem{Bennewitz:2021jqi}
Elizabeth~R. Bennewitz, Florian Hopfmueller, Bohdan Kulchytskyy, Juan Carrasquilla, and Pooya Ronagh.
\newblock Neural error mitigation of near-term quantum simulations.
\newblock {\em Nat. Mach. Intell.}, 4:618--624, 2022.

\bibitem{Honda:2021ovk}
Masazumi Honda, Etsuko Itou, Yuta Kikuchi, and Yuya Tanizaki.
\newblock Negative string tension of a higher-charge schwinger model via digital quantum simulation.
\newblock {\em PTEP}, 2022(3):033B01, 2022.

\bibitem{Halimeh:2022pkw}
Jad~C. Halimeh, Ian~P. McCulloch, Bing Yang, and Philipp Hauke.
\newblock Tuning the topological \ensuremath{\theta}-angle in cold-atom quantum simulators of gauge theories.
\newblock {\em PRX Quantum}, 3(4):040316, 2022.

\bibitem{Xie:2022jgj}
Xu-Dan Xie, Xingyu Guo, Hongxi Xing, Zheng-Yuan Xue, Dan-Bo Zhang, and Shi-Liang Zhu.
\newblock Variational thermal quantum simulation of the lattice schwinger model.
\newblock {\em Phys. Rev. D}, 106(5):054509, 2022.

\bibitem{Davoudi:2022uzo}
Zohreh Davoudi, Niklas Mueller, and Connor Powers.
\newblock Toward quantum computing phase diagrams of gauge theories with thermal pure quantum states, 8 2022.

\bibitem{Nagano:2023uaq}
Lento Nagano, Aniruddha Bapat, and Christian~W. Bauer.
\newblock Quench dynamics of the schwinger model via variational quantum algorithms.
\newblock {\em Phys. Rev. D}, 108:034501, 2 2023.

\bibitem{Popov:2023xft}
Pavel~P. Popov, Michael Meth, Maciej Lewenstein, Philipp Hauke, Martin Ringbauer, Erez Zohar, and Valentin Kasper.
\newblock Variational quantum simulation of {U}({1}) lattice gauge theories with qudit systems, 7 2023.

\bibitem{Nagano:2023kge}
Lento Nagano, Alexander Miessen, Tamiya Onodera, Ivano Tavernelli, Francesco Tacchino, and Koji Terashi.
\newblock Quantum data learning for quantum simulations in high-energy physics, 6 2023.

\bibitem{Notarnicola:2019wzb}
Simone Notarnicola, Mario Collura, and Simone Montangero.
\newblock Real-time-dynamics quantum simulation of ({1}+{1})-dimensional lattice {QED} with rydberg atoms.
\newblock {\em Phys. Rev. Res.}, 2(1):013288, 2020.

\bibitem{Tran:2020azk}
Minh~C. Tran, Yuan Su, Daniel Carney, and Jacob~M. Taylor.
\newblock Faster digital quantum simulation by symmetry protection.
\newblock {\em PRX Quantum}, 2:010323, 2021.

\bibitem{Shen:2021zrg}
Jiayu Shen, Di~Luo, Chenxi Huang, Bryan~K. Clark, Aida~X. El-Khadra, Bryce Gadway, and Patrick Draper.
\newblock Simulating quantum mechanics with a \ensuremath{\theta}-term and an \textquoteright{}t hooft anomaly on a synthetic dimension.
\newblock {\em Phys. Rev. D}, 105(7):074505, 2022.

\bibitem{Jensen:2022hyu}
Rasmus~Berg Jensen, Simon~Panyella Pedersen, and Nikolaj~Thomas Zinner.
\newblock Dynamical quantum phase transitions in a noisy lattice gauge theory.
\newblock {\em Phys. Rev. B}, 105(22):224309, 2022.

\bibitem{Florio:2023dke}
Adrien Florio, David Frenklakh, Kazuki Ikeda, Dmitri Kharzeev, Vladimir Korepin, Shuzhe Shi, and Kwangmin Yu.
\newblock Real-time nonperturbative dynamics of jet production in schwinger model: Quantum entanglement and vacuum modification.
\newblock {\em Phys. Rev. Lett.}, 131(2):021902, 2023.

\bibitem{Ikeda:2023zil}
Kazuki Ikeda, Dmitri~E. Kharzeev, Ren\'e Meyer, and Shuzhe Shi.
\newblock Detecting the critical point through entanglement in the schwinger model.
\newblock {\em Phys. Rev. D}, 108(9):L091501, 5 2023.

\bibitem{Byrnes:2002nv}
T.~Byrnes, P.~Sriganesh, R.~J. Bursill, and C.~J. Hamer.
\newblock Density matrix renormalization group approach to the massive schwinger model.
\newblock {\em Phys. Rev. D}, 66:013002, 2002.

\bibitem{Rico:2013qya}
E.~Rico, T.~Pichler, M.~Dalmonte, P.~Zoller, and S.~Montangero.
\newblock Tensor networks for lattice gauge theories and atomic quantum simulation.
\newblock {\em Phys. Rev. Lett.}, 112:201601, 2014.

\bibitem{Buyens:2013yza}
Boye Buyens, Jutho Haegeman, Karel Van~Acoleyen, Henri Verschelde, and Frank Verstraete.
\newblock Matrix product states for gauge field theories.
\newblock {\em Phys. Rev. Lett.}, 113:091601, 2014.

\bibitem{Kuhn:2014rha}
Stefan K\"uhn, J.~Ignacio Cirac, and Mari-Carmen Ba\~nuls.
\newblock {Quantum simulation of the Schwinger model: A study of feasibility}.
\newblock {\em Phys. Rev. A}, 90(4):042305, 2014.

\bibitem{Banuls:2015sta}
M.~C. Ba\~nuls, K.~Cichy, J.~I. Cirac, K.~Jansen, and H.~Saito.
\newblock Thermal evolution of the schwinger model with matrix product operators.
\newblock {\em Phys. Rev. D}, 92(3):034519, 2015.

\bibitem{Pichler:2015yqa}
T.~Pichler, M.~Dalmonte, E.~Rico, P.~Zoller, and S.~Montangero.
\newblock {Real-time Dynamics in U(1) Lattice Gauge Theories with Tensor Networks}.
\newblock {\em Phys. Rev. X}, 6(1):011023, 2016.

\bibitem{Buyens:2015tea}
Boye Buyens, Jutho Haegeman, Henri Verschelde, Frank Verstraete, and Karel Van~Acoleyen.
\newblock Confinement and string breaking for {QED}{$_2$} in the hamiltonian picture.
\newblock {\em Phys. Rev. X}, 6(4):041040, 2016.

\bibitem{Buyens:2016ecr}
Boye Buyens, Frank Verstraete, and Karel Van~Acoleyen.
\newblock {Hamiltonian simulation of the Schwinger model at finite temperature}.
\newblock {\em Phys. Rev. D}, 94(8):085018, 2016.

\bibitem{Zapp:2017fcr}
Kai Zapp and Roman Orus.
\newblock Tensor network simulation of {QED} on infinite lattices: Learning from $({1}+{1})\text{ }\mathrm{d}$, and prospects for $({2}+{1})\text{ }\mathrm{d}$.
\newblock {\em Phys. Rev. D}, 95(11):114508, 2017.

\bibitem{Ercolessi:2017jbi}
Elisa Ercolessi, Paolo Facchi, Giuseppe Magnifico, Saverio Pascazio, and Francesco~V. Pepe.
\newblock Phase transitions in $z_{n}$ gauge models: Towards quantum simulations of the schwinger-weyl {QED}.
\newblock {\em Phys. Rev. D}, 98(7):074503, 2018.

\bibitem{Butt:2019uul}
Nouman Butt, Simon Catterall, Yannick Meurice, Ryo Sakai, and Judah Unmuth-Yockey.
\newblock {Tensor network formulation of the massless Schwinger model with staggered fermions}.
\newblock {\em Phys. Rev. D}, 101(9):094509, 2020.

\bibitem{Okuda:2022hsq}
Takuya Okuda.
\newblock Schwinger model on an interval: Analytic results and {DMRG}.
\newblock {\em Phys. Rev. D}, 107(5):054506, 2023.

\bibitem{Honda:2022edn}
Masazumi Honda, Etsuko Itou, and Yuya Tanizaki.
\newblock {DMRG} study of the higher-charge schwinger model and its \textquoteright{}t hooft anomaly.
\newblock {\em JHEP}, 11:141, 2022.

\bibitem{Desaules:2023ghs}
Jean-Yves Desaules, Guo-Xian Su, Ian~P. McCulloch, Bing Yang, Zlatko Papi\'c, and Jad~C. Halimeh.
\newblock Ergodicity breaking under confinement in cold-atom quantum simulators, 1 2023.

\bibitem{Angelides:2023bme}
Takis Angelides, Lena Funcke, Karl Jansen, and Stefan K\"uhn.
\newblock Computing the mass shift of wilson and staggered fermions in the lattice schwinger model with matrix product states.
\newblock {\em Phys. Rev. D}, 108(1):014516, 2023.

\bibitem{Belyansky:2023rgh}
Ron Belyansky, Seth Whitsitt, Niklas Mueller, Ali Fahimniya, Elizabeth~R. Bennewitz, Zohreh Davoudi, and Alexey~V. Gorshkov.
\newblock High-energy collision of quarks and hadrons in the schwinger model: From tensor networks to circuit {QED}, 7 2023.

\bibitem{Banuls:2019rao}
Mari~Carmen Ba\~nuls and Krzysztof Cichy.
\newblock Review on novel methods for lattice gauge theories.
\newblock {\em Rept. Prog. Phys.}, 83(2):024401, 2020.

\bibitem{Ciavarella:2023mfc}
Anthony~N. Ciavarella.
\newblock Quantum simulation of lattice {QCD} with improved hamiltonians, 7 2023.

\bibitem{Kuhn:2015zqa}
Stefan K\"uhn, Erez Zohar, J.~Ignacio Cirac, and Mari~Carmen Ba\~nuls.
\newblock Non-abelian string breaking phenomena with matrix product states.
\newblock {\em JHEP}, 07:130, 2015.

\bibitem{Ciavarella:2022zhe}
Anthony Ciavarella, Natalie Klco, and Martin~J. Savage.
\newblock Some conceptual aspects of operator design for quantum simulations of non-abelian lattice gauge theories.
\newblock 3 2022.

\bibitem{ARahman:2022tkr}
Sarmed A~Rahman, Randy Lewis, Emanuele Mendicelli, and Sarah Powell.
\newblock {Self-mitigating Trotter circuits for SU(2) lattice gauge theory on a quantum computer}.
\newblock {\em Phys. Rev. D}, 106(7):074502, 2022.

\bibitem{Klco:2020aud}
Natalie Klco and Martin~J. Savage.
\newblock {Fixed-point quantum circuits for quantum field theories}.
\newblock {\em Phys. Rev. A}, 102(5):052422, 2020.

\bibitem{Dempsey:2022nys}
Ross Dempsey, Igor~R. Klebanov, Silviu~S. Pufu, and Bernardo Zan.
\newblock Discrete chiral symmetry and mass shift in the lattice hamiltonian approach to the schwinger model.
\newblock {\em Phys. Rev. Res.}, 4(4):043133, 2022.

\bibitem{VanDyke:2022ffj}
John~S. Van~Dyke, Karunya Shirali, George~S. Barron, Nicholas~J. Mayhall, Edwin Barnes, and Sophia~E. Economou.
\newblock Scaling adaptive quantum simulation algorithms via operator pool tiling, 6 2022.

\bibitem{Tang:2021PRX}
Ho~Lun {Tang}, V.~O. {Shkolnikov}, George~S. {Barron}, Harper~R. {Grimsley}, Nicholas~J. {Mayhall}, Edwin {Barnes}, and Sophia~E. {Economou}.
\newblock Qubit-{ADAPT}-{VQE}: An adaptive algorithm for constructing hardware-efficient ans{\"a}tze on a quantum processor.
\newblock {\em PRX Quantum}, 2(2):020310, 4 2021.

\bibitem{Yordanov:2021}
Yordan~S. Yordanov, Crispin H.~W. Barnes, and David R.~M. Arvidsson-Shukur.
\newblock Molecular-excited-state calculations with the qubit-excitation-based adaptive variational quantum eigensolver protocol.
\newblock {\em Phys. Rev. A}, 106:032434, 2022.

\bibitem{Shkolnikov:2021btx}
V.~O. Shkolnikov, Nicholas~J. Mayhall, Sophia~E. Economou, and Edwin Barnes.
\newblock Avoiding symmetry roadblocks and minimizing the measurement overhead of adaptive variational quantum eigensolvers.
\newblock {\em Quantum}, 7:1040, 2023.

\bibitem{Bertels:2022oga}
Luke~W. Bertels, Harper~R. Grimsley, Sophia~E. Economou, Edwin Barnes, and Nicholas~J. Mayhall.
\newblock Symmetry breaking slows convergence of the {ADAPT} variational quantum eigensolver.
\newblock {\em J. Chem. Theor. Comput.}, 18(11):6656--6669, 2022.

\bibitem{Anastasiou:2022swg}
Panagiotis~G. Anastasiou, Yanzhu Chen, Nicholas~J. Mayhall, Edwin Barnes, and Sophia~E. Economou.
\newblock {TETRIS}-{ADAPT}-{VQE}: An adaptive algorithm that yields shallower, denser circuit ans\"atze, 9 2022.

\bibitem{Feniou:2023gvo}
C\'esar Feniou, Muhammad Hassan, Diata Traor\'e, Emmanuel Giner, Yvon Maday, and Jean-Philip Piquemal.
\newblock Overlap-{ADAPT}-{VQE}: Practical quantum chemistry on quantum computers via overlap-guided compact ans\"atze.
\newblock {\em Commun. Phys.}, 6(1):192, 1 2023.

\bibitem{Romero:2022blx}
A.~M. Romero, J.~Engel, Ho~Lun Tang, and Sophia~E. Economou.
\newblock Solving nuclear structure problems with the adaptive variational quantum algorithm.
\newblock {\em Phys. Rev. C}, 105(6):064317, 2022.

\bibitem{Perez-Obiol:2023vod}
A.~P\'erez-Obiol, A.~M. Romero, J.~Men\'endez, A.~Rios, A.~Garc\'\i{}a-S\'aez, and B.~Juli\'a-D\'\i{}az.
\newblock Nuclear shell-model simulation in digital quantum computers.
\newblock {\em Sci. Rep.}, 13(1):12291, 2023.

\bibitem{Bell:1955djs}
J.~S. Bell.
\newblock Time reversal in field theory.
\newblock {\em Proc. Roy. Soc. Lond. A}, 231:479--495, 1955.

\bibitem{Schwinger:1951xk}
Julian~S. Schwinger.
\newblock The theory of quantized fields. {1}.
\newblock {\em Phys. Rev.}, 82:914--927, 1951.

\bibitem{Luders:1954zz}
Gerhart Luders.
\newblock On the equivalence of invariance under time reversal and under particle-antiparticle conjugation for relativistic field theories.
\newblock {\em Kong. Dan. Vid. Sel. Mat. Fys. Med.}, 28N5(5):1--17, 1954.

\bibitem{RevModPhys.79.291}
Rodney~J. Bartlett and Monika Musia\l{}.
\newblock Coupled-cluster theory in quantum chemistry.
\newblock {\em Rev. Mod. Phys.}, 79:291--352, 2 2007.

\bibitem{Hagen:2013nca}
G.~Hagen, T.~Papenbrock, M.~Hjorth-Jensen, and D.~J. Dean.
\newblock Coupled-cluster computations of atomic nuclei.
\newblock {\em Rept. Prog. Phys.}, 77(9):096302, 2014.

\bibitem{qiskit}
IBM.
\newblock Qiskit: An open-source framework for quantum computing, July 2021.

\bibitem{Algaba:2023enr}
Manuel~G. Algaba, P.~V. Sriluckshmy, Martin Leib, and Fedor Simkovic.
\newblock Low-depth simulations of fermionic systems on square-grid quantum hardware, 2 2023.

\bibitem{Cowtan:2019}
Alexander {Cowtan}, Silas {Dilkes}, Ross {Duncan}, Will {Simmons}, and Seyon {Sivarajah}.
\newblock Phase gadget synthesis for shallow circuits.
\newblock {\em EPTCS}, 318:213--228, 2020.

\bibitem{Nation:2021kye}
Paul~D. Nation, Hwajung Kang, Neereja Sundaresan, and Jay~M. Gambetta.
\newblock Scalable mitigation of measurement errors on quantum computers.
\newblock {\em PRX Quantum}, 2:040326, 2021.

\bibitem{Viola:1998dsd}
Lorenza Viola and Seth Lloyd.
\newblock Dynamical suppression of decoherence in two-state quantum systems.
\newblock {\em Phys. Rev. A}, 58:2733--2744, Oct 1998.

\bibitem{2012RSPTA.370.4748S}
A.~M. {Souza}, G.~A. {Álvarez}, and D.~{Suter}.
\newblock Robust dynamical decoupling.
\newblock {\em Phil. Trans. R. Soc.}, 370(1976):4748--4769, 10 2012.

\bibitem{Ezzell:2022uat}
Nic Ezzell, Bibek Pokharel, Lina Tewala, Gregory Quiroz, and Daniel~A. Lidar.
\newblock Dynamical decoupling for superconducting qubits: {A} performance survey.
\newblock {\em Phys. Rev. Applied}, 20(6):064027, 7 2023.

\bibitem{Urbanek_2021}
Miroslav Urbanek, Benjamin Nachman, Vincent~R. Pascuzzi, Andre He, Christian~W. Bauer, and Wibe~A. de~Jong.
\newblock Mitigating depolarizing noise on quantum computers with noise-estimation circuits.
\newblock {\em Phys. Rev. Lett.}, 127:270502, Dec 2021.

\bibitem{Wallman:2016nts}
Joel~J. Wallman and Joseph Emerson.
\newblock Noise tailoring for scalable quantum computation via randomized compiling.
\newblock {\em Phys. Rev. A}, 94:052325, Nov 2016.

\bibitem{Yu:2022ivm}
Hongye Yu, Yusheng Zhao, and Tzu-Chieh Wei.
\newblock Simulating large-size quantum spin chains on cloud-based superconducting quantum computers.
\newblock {\em Phys. Rev. Res.}, 5(1):013183, 2023.

\bibitem{Shtanko:2023tjn}
Oles Shtanko, Derek~S. Wang, Haimeng Zhang, Nikhil Harle, Alireza Seif, Ramis Movassagh, and Zlatko Minev.
\newblock Uncovering local integrability in quantum many-body dynamics, 7 2023.

\bibitem{Michael:1985ne}
Christopher Michael.
\newblock Adjoint sources in lattice gauge theory.
\newblock {\em Nucl. Phys. B}, 259:58--76, 1985.

\bibitem{Luscher:1990ck}
Martin L{\"u}scher and Ulli Wolff.
\newblock How to calculate the elastic scattering matrix in two-dimensional quantum field theories by numerical simulation.
\newblock {\em Nucl. Phys. B}, 339:222--252, 1990.

\bibitem{DeGrand1990FromAT}
Thomas~A. DeGrand and Doug Toussaint.
\newblock From actions to answers : proceedings of the {1989} theoretical advanced study institute in elementary particle physics, {5}-{30} june, {1989}, university of colorado, boulder.
\newblock 1990.

\bibitem{Fleming:2004hs}
George~Tamminga Fleming.
\newblock What can lattice {QCD} theorists learn from {NMR} spectroscopists?
\newblock In {\em {3rd International Workshop on Numerical Analysis and Lattice QCD}}, pages 143--152, 3 2004.

\bibitem{Beane:2009kya}
Silas~R. Beane, William Detmold, Thomas~C. Luu, Kostas Orginos, Assumpta Parre\~no, Martin~J. Savage, Aaron Torok, and Andre Walker-Loud.
\newblock High statistics analysis using anisotropic clover lattices: ({I}) single hadron correlation functions.
\newblock {\em Phys. Rev. D}, 79:114502, 2009.

\bibitem{PhysRevA.72.052326}
W.~D\"ur, M.~Hein, J.~I. Cirac, and H.-J. Briegel.
\newblock Standard forms of noisy quantum operations via depolarization.
\newblock {\em Phys. Rev. A}, 72:052326, 11 2005.

\bibitem{qiskitMPS}
IBM.
\newblock Qiskit aer {12}.{0} {MPS} documentation, 2023.

\bibitem{Li:2016vmf}
Ying Li and Simon~C. Benjamin.
\newblock {Efficient Variational Quantum Simulator Incorporating Active Error Minimization}.
\newblock {\em Phys. Rev. X}, 7(2):021050, 2017.

\bibitem{Temme:2016vkz}
Kristan Temme, Sergey Bravyi, and Jay~M. Gambetta.
\newblock {Error Mitigation for Short-Depth Quantum Circuits}.
\newblock {\em Phys. Rev. Lett.}, 119(18):180509, 2017.

\bibitem{2020arXiv200510921G}
Tudor Giurgica-Tiron, Yousef Hindy, Ryan LaRose, Andrea Mari, and William~J. Zeng.
\newblock Digital zero noise extrapolation for quantum error mitigation.
\newblock In {\em 2020 IEEE International Conference on Quantum Computing and Engineering (QCE)}, pages 306--316, 2020.

\bibitem{Berg:2022ugn}
Ewout van~den Berg, Zlatko~K. Minev, Abhinav Kandala, and Kristan Temme.
\newblock Probabilistic error cancellation with sparse pauli\textendash{}lindblad models on noisy quantum processors.
\newblock {\em Nat. Phys.}, 19(8):1116--1121, 2023.

\bibitem{the_qiskit_research_developers_and_contr_2023_7776174}
{The Qiskit Research and contributors} developers.
\newblock Qiskit research, 3 2023.

\bibitem{Chai:2023qpq}
Yahui Chai, Arianna Crippa, Karl Jansen, Stefan K\"uhn, Vincent~R. Pascuzzi, Francesco Tacchino, and Ivano Tavernelli.
\newblock Entanglement production from scattering of fermionic wave packets: a quantum computing approach, 12 2023.

\bibitem{Cerezo:2023nqf}
M.~Cerezo et~al.
\newblock Does provable absence of barren plateaus imply classical simulability? or, why we need to rethink variational quantum computing, 12 2023.

\bibitem{Jordan_2018}
Stephen~P. Jordan, Hari Krovi, Keith S.~M. Lee, and John Preskill.
\newblock Bqp-completeness of scattering in scalar quantum field theory.
\newblock {\em Quantum}, 2:44, Jan 2018.

\bibitem{DBLP:journals/qic/JordanLP14}
Stephen~P. Jordan, Keith S.~M. Lee, and John Preskill.
\newblock Quantum computation of scattering in scalar quantum field theories.
\newblock {\em Quantum Inf. Comput.}, 14(11-12):1014--1080, 2014.

\bibitem{Damme:2019rts}
Maarten Van~Damme, Laurens Vanderstraeten, Jacopo De~Nardis, Jutho Haegeman, and Frank Verstraete.
\newblock Real-time scattering of interacting quasiparticles in quantum spin chains.
\newblock {\em Phys. Rev. Res.}, 3:013078, 1 2021.

\bibitem{Surace:2020ycc}
Federica~Maria Surace and Alessio Lerose.
\newblock Scattering of mesons in quantum simulators.
\newblock {\em New J. Phys.}, 23(6):062001, 2021.

\bibitem{Karpov:2020pqe}
P.~I. Karpov, G.~Y. Zhu, M.~P. Heller, and M.~Heyl.
\newblock Spatiotemporal dynamics of particle collisions in quantum spin chains.
\newblock {\em Phys. Rev. Res.}, 4(3):L032001, 2022.

\bibitem{Vovrosh:2022bpj}
Joseph Vovrosh, Rick Mukherjee, Alvise Bastianello, and Johannes Knolle.
\newblock Dynamical hadron formation in long-range interacting quantum spin chains.
\newblock {\em PRX Quantum}, 3(4):040309, 2022.

\bibitem{Asaduzzaman:2022bpi}
Muhammad Asaduzzaman, Simon Catterall, Goksu~Can Toga, Yannick Meurice, and Ryo Sakai.
\newblock Quantum simulation of the {N}-flavor gross-neveu model.
\newblock {\em Phys. Rev. D}, 106(11):114515, 2022.

\bibitem{Avkhadiev:2022ttx}
A.~Avkhadiev, P.~E. Shanahan, and R.~D. Young.
\newblock Strategies for quantum-optimized construction of interpolating operators in classical simulations of lattice quantum field theories, 9 2022.

\bibitem{Vary:2023ihk}
Michael Kreshchuk, James~P. Vary, and Peter~J. Love.
\newblock Simulating scattering of composite particles, 10 2023.

\bibitem{Roy:2023uil}
Ananda Roy, Sameer Erramilli, and Robert~M. Konik.
\newblock Efficient quantum circuits based on the quantum natural gradient, 10 2023.

\bibitem{Fromm:2009xw}
Michael Fromm and Philippe de~Forcrand.
\newblock Nuclear physics from strong coupling {QCD}.
\newblock {\em PoS}, LAT2009:193, 2009.

\bibitem{Michael:1982gb}
Christopher Michael and I.~Teasdale.
\newblock Extracting glueball masses from lattice {QCD}.
\newblock {\em Nucl. Phys. B}, 215:433--446, 1983.

\bibitem{blossier2008efficient}
B.~Blossier, G.~von Hippel, T.~Mendes, R.~Sommer, and M.~Della Morte.
\newblock Efficient use of the generalized eigenvalue problem.
\newblock {\em PoS}, LAT2008:135, 2008.

\bibitem{Beane:2009gs}
Silas~R. Beane, William Detmold, Thomas~C Luu, Kostas Orginos, Assumpta Parre\~{n}o, Martin~J. Savage, Aaron Torok, and Andre Walker-Loud.
\newblock High statistics analysis using anisotropic clover lattices. {II}. three-baryon systems.
\newblock {\em Phys. Rev. D}, 80:074501, 2009.

\bibitem{Beane:2009py}
Silas~R. Beane, William Detmold, Huey-Wen Lin, Thomas~C. Luu, Kostas Orginos, Martin~J. Savage, Aaron Torok, and Andre Walker-Loud.
\newblock High statistics analysis using anisotropic clover lattices: ({III}) baryon-baryon interactions.
\newblock {\em Phys. Rev. D}, 81:054505, 2010.

\bibitem{Aoki:2009yy}
S.~Aoki.
\newblock From quarks to nuclei: Challenges of lattice {QCD}.
\newblock {\em Nucl. Phys. B Proc. Suppl.}, 195:281--287, 2009.

\bibitem{Beane:2010em}
S.~R. Beane, W.~Detmold, K.~Orginos, and M.~J. Savage.
\newblock {Nuclear Physics from Lattice QCD}.
\newblock {\em Prog. Part. Nucl. Phys.}, 66:1--40, 2011.

\bibitem{NPLQCD:2012mex}
S.~R. Beane, E.~Chang, S.~D. Cohen, William Detmold, H.~W. Lin, T.~C. Luu, K.~Orginos, A.~Parre{\~n}o, M.~J. Savage, and A.~Walker-Loud.
\newblock {Light Nuclei and Hypernuclei from Quantum Chromodynamics in the Limit of SU(3) Flavor Symmetry}.
\newblock {\em Phys. Rev. D}, 87(3):034506, 2013.

\bibitem{Yamazaki:2012hi}
Takeshi Yamazaki, Ken-ichi Ishikawa, Yoshinobu Kuramashi, and Akira Ukawa.
\newblock Helium nuclei, deuteron and dineutron in {2}+{1} flavor lattice {QCD}.
\newblock {\em Phys. Rev. D}, 86:074514, 2012.

\bibitem{Yamazaki:2015vjn}
Takeshi Yamazaki.
\newblock Light nuclei and nucleon form factors in {$N_f=2+1$} lattice {QCD}.
\newblock {\em PoS}, LATTICE2015:081, 2016.

\bibitem{Yamazaki:2015asa}
Takeshi Yamazaki, Ken-ichi Ishikawa, Yoshinobu Kuramashi, and Akira Ukawa.
\newblock Study of quark mass dependence of binding energy for light nuclei in {2}+{1} flavor lattice {QCD}.
\newblock {\em Phys. Rev. D}, 92(1):014501, 2015.

\bibitem{Drischler:2019xuo}
Christian Drischler, Wick Haxton, Kenneth McElvain, Emanuele Mereghetti, Amy Nicholson, Pavlos Vranas, and Andr\'e Walker-Loud.
\newblock {Towards grounding nuclear physics in QCD}.
\newblock {\em Prog. Part. Nucl. Phys.}, 121:103888, 2021.

\bibitem{Amarasinghe:2021lqa}
Saman Amarasinghe, Riyadh Baghdadi, Zohreh Davoudi, William Detmold, Marc Illa, Assumpta Parre{\~n}o, Andrew~V. Pochinsky, Phiala~E. Shanahan, and Michael~L. Wagman.
\newblock Variational study of two-nucleon systems with lattice {QCD}.
\newblock {\em Phys. Rev. D}, 107(9):094508, 2023.

\bibitem{Qin:2021jpm}
Dayue Qin, Yanzhu Chen, and Ying Li.
\newblock Error statistics and scalability of quantum error mitigation formulas.
\newblock {\em npj Quantum Inf.}, 9(1):35, 2023.

\bibitem{Robbiati:2023eyl}
Matteo Robbiati, Alejandro Sopena, Andrea Papaluca, and Stefano Carrazza.
\newblock Real-time error mitigation for variational optimization on quantum hardware, 11 2023.

\bibitem{Beane:2014oea}
Silas~R. Beane, William Detmold, Kostas Orginos, and Martin~J. Savage.
\newblock {Uncertainty Quantification in Lattice QCD Calculations for Nuclear Physics}.
\newblock {\em J. Phys. G}, 42(3):034022, 2015.

\bibitem{Orginos:2015aya}
Kostas Orginos, Assumpta Parre\~{n}o, Martin~J. Savage, Silas~R. Beane, Emmanuel Chang, and William Detmold.
\newblock Two nucleon systems at $m_\pi\sim {450}~{\rm {mev}}$ from lattice {QCD}.
\newblock {\em Phys. Rev. D}, 92(11):114512, 2015.
\newblock [Erratum: Phys.Rev.D 102, 039903 (2020)].

\bibitem{Baumer:2023vrf}
Elisa B\"aumer, Vinay Tripathi, Derek~S. Wang, Patrick Rall, Edward~H. Chen, Swarnadeep Majumder, Alireza Seif, and Zlatko~K. Minev.
\newblock Efficient long-range entanglement using dynamic circuits, 8 2023.

\bibitem{Chen:2023tfg}
Edward~H. Chen et~al.
\newblock Realizing the nishimori transition across the error threshold for constant-depth quantum circuits, 9 2023.

\bibitem{Liao:2023eug}
Haoran Liao, Derek~S. Wang, Iskandar Sitdikov, Ciro Salcedo, Alireza Seif, and Zlatko~K. Minev.
\newblock Machine learning for practical quantum error mitigation, 9 2023.

\bibitem{Chowdhury:2023vbk}
Talal~Ahmed Chowdhury, Kwangmin Yu, Mahmud~Ashraf Shamim, M.~L. Kabir, and Raza~Sabbir Sufian.
\newblock Enhancing quantum utility: simulating large-scale quantum spin chains on superconducting quantum computers, 12 2023.

\bibitem{Peskin:1995ev}
Michael~Edward Peskin and Daniel~V. Schroeder.
\newblock {\em An Introduction to Quantum Field Theory}.
\newblock Westview Press, 1995.
\newblock Reading, USA: Addison-Wesley (1995) 842 p.

\bibitem{Gasser:1987zq}
J.~Gasser and H.~Leutwyler.
\newblock Spontaneously broken symmetries: Effective lagrangians at finite volume.
\newblock {\em Nucl. Phys. B}, 307:763--778, 1988.

\bibitem{2001afpp.book..683V}
Pierre {van Baal}.
\newblock {QCD} in a finite volume.
\newblock In {\em At the Frontier of Particle Physics: Handbook of QCD}, pages 683--760. {World Scientific Publishing Co. Pte. Ltd}, 8 2001.

\bibitem{PhysRevD.70.074029}
S.~R. Beane and M.~J. Savage.
\newblock Baryon axial charge in a finite volume.
\newblock {\em Phys. Rev. D}, 70:074029, 10 2004.

\bibitem{Colangelo_2005}
Gilberto Colangelo, Stephan Dürr, and Christoph Haefeli.
\newblock Finite volume effects for meson masses and decay constants.
\newblock {\em Nucl. Phys. B}, 721(1–3):136–174, 8 2005.

\bibitem{Farrell:2024mgu}
Roland~C. Farrell, Marc Illa, and Martin~J. Savage.
\newblock {Steps Toward Quantum Simulations of Hadronization and Energy-Loss in Dense Matter}.
\newblock 5 2024.

\bibitem{Dusling:2015gta}
Kevin Dusling, Wei Li, and Bj\"orn Schenke.
\newblock {Novel collective phenomena in high-energy proton\textendash{}proton and proton\textendash{}nucleus collisions}.
\newblock {\em Int. J. Mod. Phys. E}, 25(01):1630002, 2016.

\bibitem{Busza:2018rrf}
Wit Busza, Krishna Rajagopal, and Wilke van~der Schee.
\newblock {Heavy Ion Collisions: The Big Picture, and the Big Questions}.
\newblock {\em Ann. Rev. Nucl. Part. Sci.}, 68:339--376, 2018.

\bibitem{LIGOScientific:2017ync}
B.~P. Abbott et~al.
\newblock {Multi-messenger Observations of a Binary Neutron Star Merger}.
\newblock {\em Astrophys. J. Lett.}, 848(2):L12, 2017.

\bibitem{Troja:2018uns}
E.~Troja, H.~van Eerten, G.~Ryan, R.~Ricci, J.~M. Burgess, M.~H. Wieringa, L.~Piro, S.~B. Cenko, and T.~Sakamoto.
\newblock {A year in the life of GW 170817: the rise and fall of a structured jet from a binary neutron star merger}.
\newblock {\em Mon. Not. Roy. Astron. Soc.}, 489(2):1919--1926, 2019.

\bibitem{KAGRA:2021vkt}
R.~Abbott et~al.
\newblock {GWTC-3: Compact Binary Coalescences Observed by LIGO and Virgo during the Second Part of the Third Observing Run}.
\newblock {\em Phys. Rev. X}, 13(4):041039, 2023.

\bibitem{LIGOScientific:2020aai}
B.~P. Abbott et~al.
\newblock {GW190425: Observation of a Compact Binary Coalescence with Total Mass $\sim 3.4 M_{\odot}$}.
\newblock {\em Astrophys. J. Lett.}, 892(1):L3, 2020.

\bibitem{Lovato:2022vgq}
Alessandro Lovato et~al.
\newblock {Long Range Plan: Dense matter theory for heavy-ion collisions and neutron stars}.
\newblock 11 2022.

\bibitem{Fevre:2023swv}
Arnaud~Le F\`evre et~al.
\newblock {Constraining Neutron-Star Matter \textemdash{} Combination of heavy-ion experiments, astronomy and theory}.
\newblock {\em EPJ Web Conf.}, 290:05006, 2023.

\bibitem{Jacobi:2023olu}
Maximilian Jacobi, Federico~Maria Guercilena, Sabrina Huth, Giacomo Ricigliano, Almudena Arcones, and Achim Schwenk.
\newblock {Effects of nuclear matter properties in neutron star mergers}.
\newblock {\em Mon. Not. Roy. Astron. Soc.}, 527:8812, 2023.

\bibitem{Brandes:2023hma}
Len Brandes, Wolfram Weise, and Norbert Kaiser.
\newblock {Evidence against a strong first-order phase transition in neutron star cores: Impact of new data}.
\newblock {\em Phys. Rev. D}, 108(9):094014, 2023.

\bibitem{Trickle:2019nya}
Tanner Trickle, Zhengkang Zhang, Kathryn~M. Zurek, Katherine Inzani, and Sin\'ead~M. Griffin.
\newblock {Multi-Channel Direct Detection of Light Dark Matter: Theoretical Framework}.
\newblock {\em JHEP}, 03:036, 2020.

\bibitem{Baxter:2022dkm}
Daniel Baxter et~al.
\newblock {Snowmass2021 Cosmic Frontier White Paper: Calibrations and backgrounds for dark matter direct detection}.
\newblock 3 2022.

\bibitem{Giuliani:2012zu}
Andrea Giuliani and Alfredo Poves.
\newblock {Neutrinoless Double-Beta Decay}.
\newblock {\em Adv. High Energy Phys.}, 2012:857016, 2012.

\bibitem{Dolinski:2019nrj}
Michelle~J. Dolinski, Alan W.~P. Poon, and Werner Rodejohann.
\newblock {Neutrinoless Double-Beta Decay: Status and Prospects}.
\newblock {\em Ann. Rev. Nucl. Part. Sci.}, 69:219--251, 2019.

\bibitem{Tsai:1973py}
Yung-Su Tsai.
\newblock {Pair Production and Bremsstrahlung of Charged Leptons}.
\newblock {\em Rev. Mod. Phys.}, 46:815, 1974.
\newblock [Erratum: Rev.Mod.Phys. 49, 421--423 (1977)].

\bibitem{SELTZER1984665}
Stephen~M. Seltzer and Martin~J. Berger.
\newblock Improved procedure for calculating the collision stopping power of elements and compounds for electrons and positrons.
\newblock {\em Int. J. Appl. Radiat. Isot.}, 35(7):665--676, 1984.

\bibitem{Meehan:2017cum}
Kathryn Meehan.
\newblock {STAR Results from Au + Au Fixed-Target Collisions at $\sqrt{s_{NN}} = 4.5$ GeV}.
\newblock {\em Nucl. Phys. A}, 967:808--811, 2017.

\bibitem{Maurice:2017iom}
Emilie Maurice.
\newblock {Fixed-target physics at LHCb}.
\newblock In {\em {5th Large Hadron Collider Physics Conference}}, 8 2017.

\bibitem{Hadjidakis:2018ifr}
C.~Hadjidakis et~al.
\newblock {A fixed-target programme at the LHC: Physics case and projected performances for heavy-ion, hadron, spin and astroparticle studies}.
\newblock {\em Phys. Rept.}, 911:1--83, 2021.

\bibitem{Barschel:2020drr}
Colin Barschel et~al.
\newblock {\em {LHC fixed target experiments : Report from the LHC Fixed Target Working Group of the CERN Physics Beyond Colliders Forum}}, volume 4/2020 of {\em CERN Yellow Reports: Monographs}.
\newblock CERN, Geneva, 2020.

\bibitem{AbdulKhalek:2021gbh}
R.~Abdul~Khalek et~al.
\newblock {Science Requirements and Detector Concepts for the Electron-Ion Collider}: {EIC Yellow Report}.
\newblock {\em Nucl. Phys. A}, 1026:122447, 2022.

\bibitem{zhao2023hadronization}
Jiaxing Zhao, Jörg Aichelin, Pol~Bernard Gossiaux, Andrea Beraudo, Shanshan Cao, Wenkai Fan, Min He, Vincenzo Minissale, Taesoo Song, Ivan Vitev, Ralf Rapp, Steffen Bass, Elena Bratkovskaya, Vincenzo Greco, and Salvatore Plumari.
\newblock Hadronization of heavy quarks, 2023.

\bibitem{Rothkopf:2019ipj}
Alexander Rothkopf.
\newblock {Heavy Quarkonium in Extreme Conditions}.
\newblock {\em Phys. Rept.}, 858:1--117, 2020.

\bibitem{Akamatsu:2020ypb}
Yukinao Akamatsu.
\newblock {Quarkonium in quark\textendash{}gluon plasma: Open quantum system approaches re-examined}.
\newblock {\em Prog. Part. Nucl. Phys.}, 123:103932, 2022.

\bibitem{Chapon:2020heu}
Emilien Chapon et~al.
\newblock {Prospects for quarkonium studies at the high-luminosity LHC}.
\newblock {\em Prog. Part. Nucl. Phys.}, 122:103906, 2022.

\bibitem{Yao:2021lus}
Xiaojun Yao.
\newblock {Open quantum systems for quarkonia}.
\newblock {\em Int. J. Mod. Phys. A}, 36(20):2130010, 2021.

\bibitem{Montana:2023sft}
Gloria Monta\~{n}a, Angels Ramos, Laura Tolos, and Juan~M. Torres-Rincon.
\newblock {Recent progress on in-medium properties of heavy mesons from finite-temperature EFTs}.
\newblock {\em Front. in Phys.}, 11:1250939, 2023.

\bibitem{Bepari:2020xqi}
Khadeejah Bepari, Sarah Malik, Michael Spannowsky, and Simon Williams.
\newblock {Towards a quantum computing algorithm for helicity amplitudes and parton showers}.
\newblock {\em Phys. Rev. D}, 103(7):076020, 2021.

\bibitem{Bepari:2021kwv}
Khadeejah Bepari, Sarah Malik, Michael Spannowsky, and Simon Williams.
\newblock {Quantum walk approach to simulating parton showers}.
\newblock {\em Phys. Rev. D}, 106(5):056002, 2022.

\bibitem{Macaluso:2021ngq}
Sebastian Macaluso and Kyle Cranmer.
\newblock {The Quantum Trellis: A classical algorithm for sampling the parton shower with interference effects}, 12 2021.

\bibitem{Chigusa:2022act}
So~Chigusa and Masahito Yamazaki.
\newblock {Quantum simulations of dark sector showers}.
\newblock {\em Phys. Lett. B}, 834:137466, 2022.

\bibitem{Gustafson:2022dsq}
G\"osta Gustafson, Stefan Prestel, Michael Spannowsky, and Simon Williams.
\newblock {Collider events on a quantum computer}.
\newblock {\em JHEP}, 11:035, 2022.

\bibitem{Bauer:2023ujy}
Christian~W. Bauer, So~Chigusa, and Masahito Yamazaki.
\newblock {Quantum parton shower with kinematics}.
\newblock {\em Phys. Rev. A}, 109(3):032432, 2024.

\bibitem{Isgur:1989vq}
Nathan Isgur and Mark~B. Wise.
\newblock {Weak Decays of Heavy Mesons in the Static Quark Approximation}.
\newblock {\em Phys. Lett. B}, 232:113--117, 1989.

\bibitem{Isgur:1990yhj}
Nathan Isgur and Mark~B. Wise.
\newblock {Weak transition form factors between heavy mesons}.
\newblock {\em Phys. Lett. B}, 237:527--530, 1990.

\bibitem{Eichten:1989zv}
Estia Eichten and Brian~Russell Hill.
\newblock {An Effective Field Theory for the Calculation of Matrix Elements Involving Heavy Quarks}.
\newblock {\em Phys. Lett. B}, 234:511--516, 1990.

\bibitem{Georgi:1990um}
Howard Georgi.
\newblock {An Effective Field Theory for Heavy Quarks at Low-energies}.
\newblock {\em Phys. Lett. B}, 240:447--450, 1990.

\bibitem{Grinstein:1990mj}
Benjamin Grinstein.
\newblock {The Static Quark Effective Theory}.
\newblock {\em Nucl. Phys. B}, 339:253--268, 1990.

\bibitem{Hamer:1982mx}
C.~J. Hamer, John~B. Kogut, D.~P. Crewther, and M.~M. Mazzolini.
\newblock {The Massive Schwinger Model on a Lattice: Background Field, Chiral Symmetry and the String Tension}.
\newblock {\em Nucl. Phys. B}, 208:413--438, 1982.

\bibitem{Berruto:1997jv}
F.~Berruto, G.~Grignani, G.~W. Semenoff, and P.~Sodano.
\newblock {Chiral symmetry breaking on the lattice: A Study of the strongly coupled lattice Schwinger model}.
\newblock {\em Phys. Rev. D}, 57:5070--5083, 1998.

\bibitem{Sriganesh:1999ws}
P.~Sriganesh, R.~Bursill, and C.~J. Hamer.
\newblock {A New finite lattice study of the massive Schwinger model}.
\newblock {\em Phys. Rev. D}, 62:034508, 2000.

\bibitem{Cichy:2012rw}
Krzysztof Cichy, Agnieszka Kujawa-Cichy, and Marcin Szyniszewski.
\newblock {Lattice Hamiltonian approach to the massless Schwinger model: Precise extraction of the mass gap}.
\newblock {\em Comput. Phys. Commun.}, 184:1666--1672, 2013.

\bibitem{Falk:1992ws}
Adam~F. Falk and Matthias Neubert.
\newblock {Second order power corrections in the heavy quark effective theory. 2. Baryon form-factors}.
\newblock {\em Phys. Rev. D}, 47:2982--2990, 1993.

\bibitem{Luke:1993za}
Michael~E. Luke and Martin~J. Savage.
\newblock {Extracting $|V_{bc}|$, $m_{c}$ and $m_{b}$ from inclusive D and B decays}.
\newblock {\em Phys. Lett. B}, 321:88--94, 1994.

\bibitem{Kronfeld:2000gk}
Andreas~S. Kronfeld and James~N. Simone.
\newblock {Computation of $\overline{\Lambda}$ and $\lambda_1$ with lattice QCD}.
\newblock {\em Phys. Lett. B}, 490(3):228--235, 2000.

\bibitem{Gambino:2017vkx}
P.~Gambino, A.~Melis, and S.~Simula.
\newblock {Extraction of heavy-quark-expansion parameters from unquenched lattice data on pseudoscalar and vector heavy-light meson masses}.
\newblock {\em Phys. Rev. D}, 96(1):014511, 2017.

\bibitem{FermilabLattice:2018est}
A.~Bazavov et~al.
\newblock {Up-, down-, strange-, charm-, and bottom-quark masses from four-flavor lattice QCD}.
\newblock {\em Phys. Rev. D}, 98(5):054517, 2018.

\bibitem{Miller:2023ujx}
Gerald~A. Miller.
\newblock {Entanglement maximization in low-energy neutron-proton scattering}.
\newblock {\em Phys. Rev. C}, 108(3):L031002, 2023.

\bibitem{Robin:2020aeh}
Caroline Robin, Martin~J. Savage, and Nathalie Pillet.
\newblock {Entanglement Rearrangement in Self-Consistent Nuclear Structure Calculations}.
\newblock {\em Phys. Rev. C}, 103(3):034325, 2021.

\bibitem{Bulgac:2022cjg}
Aurel Bulgac, Matthew Kafker, and Ibrahim Abdurrahman.
\newblock {Measures of complexity and entanglement in many-fermion systems}.
\newblock {\em Phys. Rev. C}, 107(4):044318, 2023.

\bibitem{Johnson:2022mzk}
Calvin~W. Johnson and Oliver~C. Gorton.
\newblock {Proton-neutron entanglement in the nuclear shell model}.
\newblock {\em J. Phys. G}, 50(4):045110, 2023.

\bibitem{Gu:2023aoc}
Chenyi Gu, Z.~H. Sun, G.~Hagen, and T.~Papenbrock.
\newblock {Entanglement entropy of nuclear systems}.
\newblock {\em Phys. Rev. C}, 108(5):054309, 2023.

\bibitem{Hengstenberg:2023ryt}
S.~Momme Hengstenberg, Caroline E.~P. Robin, and Martin~J. Savage.
\newblock {Multi-body entanglement and information rearrangement in nuclear many-body systems: a study of the Lipkin\textendash{}Meshkov\textendash{}Glick model}.
\newblock {\em Eur. Phys. J. A}, 59(10):231, 2023.

\bibitem{Perez-Obiol:2023wdz}
A.~P\'erez-Obiol, S.~Masot-Llima, A.~M. Romero, J.~Men\'endez, A.~Rios, A.~Garc\'\i{}a-S\'aez, and B.~Juli\'a-D\'\i{}az.
\newblock {Quantum entanglement patterns in the structure of atomic nuclei within the nuclear shell model}.
\newblock {\em Eur. Phys. J. A}, 59(10):240, 2023.

\bibitem{Kharzeev:2017qzs}
Dmitri~E. Kharzeev and Eugene~M. Levin.
\newblock {Deep inelastic scattering as a probe of entanglement}.
\newblock {\em Phys. Rev. D}, 95(11):114008, 2017.

\bibitem{Baker:2017wtt}
O.~K. Baker and D.~E. Kharzeev.
\newblock {Thermal radiation and entanglement in proton-proton collisions at energies available at the CERN Large Hadron Collider}.
\newblock {\em Phys. Rev. D}, 98(5):054007, 2018.

\bibitem{Kharzeev:2021yyf}
Dmitri~E. Kharzeev and Eugene Levin.
\newblock {Deep inelastic scattering as a probe of entanglement: confronting experimental data}.
\newblock {\em Phys. Rev. D}, 104(3):L031503, 2 2021.

\bibitem{Florio:2024aix}
Adrien Florio, David Frenklakh, Kazuki Ikeda, Dmitri~E. Kharzeev, Vladimir Korepin, Shuzhe Shi, and Kwangmin Yu.
\newblock {Quantum simulation of entanglement and hadronization in jet production: lessons from the massive Schwinger model}, 3 2024.

\bibitem{Srednicki:1993im}
Mark Srednicki.
\newblock {Entropy and area}.
\newblock {\em Phys. Rev. Lett.}, 71:666--669, 1993.

\bibitem{Marcovitch:2008sxc}
S.~Marcovitch, A.~Retzker, M.B. Plenio, and B.~Reznik.
\newblock {Critical and noncritical long-range entanglement in Klein-Gordon fields}.
\newblock {\em Phys. Rev. A}, 80(1):012325, 2009.

\bibitem{Klco:2020rga}
Natalie Klco and Martin~J. Savage.
\newblock {Geometric quantum information structure in quantum fields and their lattice simulation}.
\newblock {\em Phys. Rev. D}, 103(6):065007, 2021.

\bibitem{Klco:2021biu}
Natalie Klco and Martin~J. Savage.
\newblock {Entanglement Spheres and a UV-IR Connection in Effective Field Theories}.
\newblock {\em Phys. Rev. Lett.}, 127(21):211602, 2021.

\bibitem{Klco:2021cxq}
Natalie Klco, D.~H. Beck, and Martin~J. Savage.
\newblock {Entanglement Structures in Quantum Field Theories: Negativity Cores and Bound Entanglement in the Vacuum}, 10 2021.

\bibitem{Klco:2023ojt}
Natalie Klco and D.~H. Beck.
\newblock {Entanglement structures in quantum field theories. II. Distortions of vacuum correlations through the lens of local observers}.
\newblock {\em Phys. Rev. A}, 108(1):012429, 2023.

\bibitem{Parez:2023uxu}
Gilles Parez and William Witczak-Krempa.
\newblock {Entanglement negativity between separated regions in quantum critical systems}.
\newblock {\em Phys. Rev. Res.}, 6(2):023125, 2024.

\bibitem{Florio:2023mzk}
Adrien Florio.
\newblock {Two-fermion negativity and confinement in the Schwinger model}.
\newblock {\em Phys. Rev. D}, 109(7):L071501, 2024.

\bibitem{Wong:2001}
Alexander Wong and Nelson Christensen.
\newblock Potential multiparticle entanglement measure.
\newblock {\em Phys. Rev. A}, 63:044301, Mar 2001.

\bibitem{Illa:2022zgu}
Marc Illa and Martin~J. Savage.
\newblock {Multi-Neutrino Entanglement and Correlations in Dense Neutrino Systems}.
\newblock {\em Phys. Rev. Lett.}, 130(22):221003, 2023.

\bibitem{Zyczkowski:1998yd}
Karol Zyczkowski, Pawel Horodecki, Anna Sanpera, and Maciej Lewenstein.
\newblock {On the volume of the set of mixed entangled states}.
\newblock {\em Phys. Rev. A}, 58:883, 1998.

\bibitem{Vidal:2002zz}
G.~Vidal and R.~F. Werner.
\newblock {Computable measure of entanglement}.
\newblock {\em Phys. Rev. A}, 65:032314, 2002.

\bibitem{Bravyi:2012}
Sergey {Bravyi} and Jeongwan {Haah}.
\newblock {Magic-state distillation with low overhead}.
\newblock {\em PRA}, 86(5):052329, November 2012.

\bibitem{Leone:2021rzd}
Lorenzo Leone, Salvatore F.~E. Oliviero, and Alioscia Hamma.
\newblock {Stabilizer R\'enyi Entropy}.
\newblock {\em Phys. Rev. Lett.}, 128(5):050402, 2022.

\bibitem{Tirrito:2023fnw}
Emanuele Tirrito, Poetri~Sonya Tarabunga, Gugliemo Lami, Titas Chanda, Lorenzo Leone, Salvatore F.~E. Oliviero, Marcello Dalmonte, Mario Collura, and Alioscia Hamma.
\newblock {Quantifying nonstabilizerness through entanglement spectrum flatness}.
\newblock {\em Phys. Rev. A}, 109(4):L040401, 2024.

\bibitem{Pecak:2024zqq}
Daniel P\c{e}cak, Agata Zdanowicz, Nicolas Chamel, Piotr Magierski, and Gabriel Wlaz\l{}owski.
\newblock {Time-dependent nuclear energy-density functional theory toolkit for neutron star crust: Dynamics of a nucleus in a neutron superfluid}, 3 2024.

\bibitem{ibmquantum}
{IBM Quantum}.
\newblock \url{https://quantum.ibm.com}, 2024.

\bibitem{Carlsson:2001wp}
Jesse Carlsson and Bruce H.~J. McKellar.
\newblock {Direct improvement of Hamiltonian lattice gauge theory}.
\newblock {\em Phys. Rev. D}, 64:094503, 2001.

\bibitem{Carena:2022kpg}
Marcela Carena, Henry Lamm, Ying-Ying Li, and Wanqiang Liu.
\newblock Improved hamiltonians for quantum simulations of gauge theories.
\newblock {\em Phys. Rev. Lett.}, 129(5):051601, 2022.

\bibitem{Gustafson:2023aai}
Erik Gustafson and Ruth Van~de Water.
\newblock {Improved Fermion Hamiltonians for Quantum Simulation}.
\newblock {\em PoS}, LATTICE2023:215, 2024.

\bibitem{Zache:2018jbt}
T.~V. Zache, F.~Hebenstreit, F.~Jendrzejewski, M.~K. Oberthaler, J.~Berges, and P.~Hauke.
\newblock {Quantum simulation of lattice gauge theories using Wilson fermions}.
\newblock {\em Quantum Sci. Technol.}, 3(3):034010, 2018.

\bibitem{Mathis:2020fuo}
Simon~V. Mathis, Guglielmo Mazzola, and Ivano Tavernelli.
\newblock {Toward scalable simulations of Lattice Gauge Theories on quantum computers}.
\newblock {\em Phys. Rev. D}, 102(9):094501, 2020.

\bibitem{Hayata:2023skf}
Tomoya Hayata, Katsumasa Nakayama, and Arata Yamamoto.
\newblock {Dynamical chirality production in one dimension}.
\newblock {\em Phys. Rev. D}, 109(3):034501, 2024.

\end{thebibliography}

\end{document}